\documentclass[usenatbib]{mnras}
\usepackage{epsfig}
\usepackage{amsmath,amssymb}
\usepackage{color}
\def\apj{ApJ}                 % Astrophysical Journal
\def\apjl{ApJ}                % Astrophysical Journal, Letters
\def\apjs{ApJS}               % Astrophysical Journal, Supplement
\def\aap{A\&A}                % Astronomy and Astrophysics
\def\mnras{MNRAS}             % Monthly Notices of the RAS
\newcommand{\Msolar}{\mbox{\,$\rm M_{\odot}$}}        % solar mass
\newcommand{\Lsolar}{\mbox{\,$\rm L_{\odot}$}}        % solar luminosity
\title[Pulsation instability]{Radial pulsation as a function of hydrogen abundance}

\author[C. S. Jeffery \& H. Saio]
       {C. S. Jeffery\thanks{E-mail: csj@arm.ac.uk}
\& H. Saio\thanks{E-mail: saio@astr.tohoku.ac.jp} \\
Armagh Observatory, College Hill, Armagh BT61 9DG, Northern Ireland\\
Astronomical Institute, School of Science, Tohoku University, Sendai 980-8578, Japan\\
}

\date{Accepted .....
      Received 2016 April 1;
      in original form 2016 April 1}

\pagerange{\pageref{firstpage}--\pageref{lastpage}}
\pubyear{2016}

\begin{document}

\maketitle

\label{firstpage}

\begin{abstract}
Using linear non-adabatic pulsation analysis, we explore the radial-mode (p-mode) stability of stars  across a  
wide range of mass ($0.2 \leq M \leq 50\Msolar$), composition ($0 \leq X \leq 0.7$, $Z=0.001, 0.02$),  
effective temperature ($3\,000 \leq T_{\rm eff} \leq 40\,000$\,K), and luminosity ($0.01 \leq L/M \leq 100,000$ solar units).
We  identify the instability boundaries associated with low- to  high-order  radial oscillations ($0\le n\le16$). 
The instability boundaries are a strong function of both composition and  radial order ($n$).
With decreasing hydrogen abundance we find that i) the classical blue edge of the Cepheid instability strip shifts to higher effective 
temperature and luminosity, and ii) high-order modes are more easily 
excited and small islands of high radial-order instability develop, some of
which  correspond with real stars.  
Driving in all cases is by the classical $\kappa$-mechanism and/or strange modes.
We identify regions of parameter space where new classes of pulsating variable 
may, in future, be discovered. The majority of these are associated with reduced 
hydrogen abundance in the envelope; one has not  been identified previously. 
\end{abstract}


\section{Introduction}              %%%%%%%%%%%%%%%%%%  INTRO  %%%%%%%%%%%%
\label{intro}

\begin{figure}
        \centering
                \includegraphics[width=0.47\textwidth]{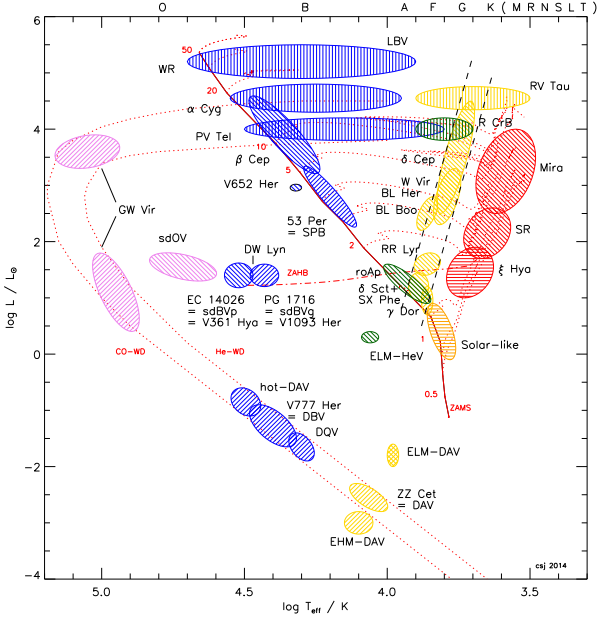}
\caption{Luminosity-effective temperature (or Hertzsprung-Russell) diagram showing the 
approximate locations of major pulsating variables coloured roughly by spectral type, the zero-age
main sequence and horizontal branch, the Cepheid instability strip, and evolution tracks for model
stars of various masses, indicated by small numbers (\Msolar). Shadings represent opacity-driven
p-modes ($\backslash\backslash\backslash$), g-modes (///) and strange modes ($|||$) and
acoustically-driven modes ($\equiv$). Approximate spectral types are indicated on the top axis.
Based on figures by J. Christensen-Dalsgaard and subsequently by \citet{jeffery08.coast}.  }
        \label{f:puls_hrd}
\end{figure}

Since the discovery of periodic light variations in the luminous giant $\delta$ Cephei, the study of
stellar pulsations has transformed our understanding of how stars work, as well as establishing 
a distance scale whereby the cosmos can be measured. The fact that the light variations in 
$\delta$ Cep represented a major discovery testifies to the fact that not all stars are variable. 
However, as telescopes and detectors have become more sensitive, pulsations have been identified in 
diverse groups of stars of all masses and across the Hertzsprung-Russell diagram. Such
discoveries continue to the present day, with pulsations in low-mass white dwarfs and pre-white 
dwarfs being the latest additions to the pulsating star zoo (Fig.~\ref{f:puls_hrd}) \citep{maxted13,hermes13a}. 

\citet{jeffery13b} demonstrated that pulsation instability in the low-mass pre-white dwarf J0247-25B
would arise in a high-order overtone if the envelope was depleted in hydrogen. The principal 
reason for this, demonstrated previously by \citet{saio88b} and \citet{jeffery07}, is that hydrogen acts as a poison, 
suppressing the positive opacity gradient   
around an opacity peak which would otherwise drive pulsations  if located at an appropriate depth beneath the stellar surface\footnote{The condition for driving 
an oscillation by the $\kappa$-mechanism in such a region is generally understood to be that the spatial derivative 
${\rm d} (\kappa_T + \kappa_{\rho}/(\Gamma_3-1))/{\rm d}r > 0$, where $\kappa_T$ and $\kappa_{\rho}$ 
are the temperature and density derivatives of the opacity $\kappa$, and $\Gamma_3$ is the usual adiabatic exponent \citep[p. 243]{unno89}.}.

The question therefore arose whether it would be possible to predict the properties of other 
hitherto undiscovered pulsating variables, especially those in which hydrogen has been depleted
as a consequence of  prior evolution. Consequently, we have carried out a parametric survey
 in order to identify locations where new classes of variable star await discovery.

\section{Radial Pulsation Models}

The investigation commenced by computing a grid
of 258,000 models of stellar envelopes covering a range of chemical mixtures and for masses on the range $0.2 \le M/\Msolar \le50$, 
effective temperaures $\log T_{\rm eff}/{\rm K} = 3.50 (0.02) 4.60 $, and luminosity-to-mass ratios 
$\log (L/\Lsolar)/(M/\Msolar) = -2.0 (0.2) 5.0$.  The linear nonadiabatic analysis of stability against pulsation was 
carried out following methods described by \citet{saio83b} and \citet{jeffery06a,jeffery06b}.  

The OPAL95 \citep{iglesias96} opacities were adopted, except at low temperatures, where  \citet{alexander94} 
opacites were used.  As a test of sensitivity to opacity, additional calculations were made with OP opacities \citep{badnell05}. 
Convection is treated assuming a standard mixing-length theory with the ratio mixing-length to pressure scale height $l/H_p = 1.5$. 
Any convection/pulsation interaction is neglected by setting the divergence of the convective flux perturbation 
to zero.  Therefore results for $T_{\rm eff}<4\,000$\,K should be treated with caution.
 
The outer boundary for the envelope model is set at the Rosseland mean optical depth $\tau = 10^{-3}$, 
The integration is carried out with pressure $\log_{10} P$ as the independent variable, with initial stepsize $\delta \log_{10} P = 0.02$, 
which is adjusted to maintain increments in radius $\delta r/r < 0.01$, density $\delta \rho/\rho < 0.1$ and electron pressure 
$\delta P_{\rm e}/P_{\rm e} < 0.08$ at each step. The integration is halted at a fractional mass  $m = M_\star/10$ or 
fractional radius $r=R_\star/100$, whichever occurs first.

For each model envelope,  
the first 17 eigenfrequencies were located and stored, including the real and imaginary components
$\omega_{\rm r}$ and $\omega_{\rm i}$, the period $\Pi$ and
the number of nodes in the eigensolution. Modes with $\omega_{\rm i} < 0 $ were deemed to
be unstable, {\it i.e.} pulsations could be excited. 

We considered a range of abundances with  hydrogen-mass fraction 
$X=0.002, 0.1, 0.30$ and $0.70$ and  metal mass fraction $Z=0.001$ and $0.02$. 
We have assumed that the iron and nickel abundances are scaled to solar values 
\citep[metal mixture GN93:][]{grevesse93} for all values of $Z$. 
We have {\it not} considered any additional enhancements to iron-group or other elements. 

\begin{figure*}
\begin{center}
\epsfig{file=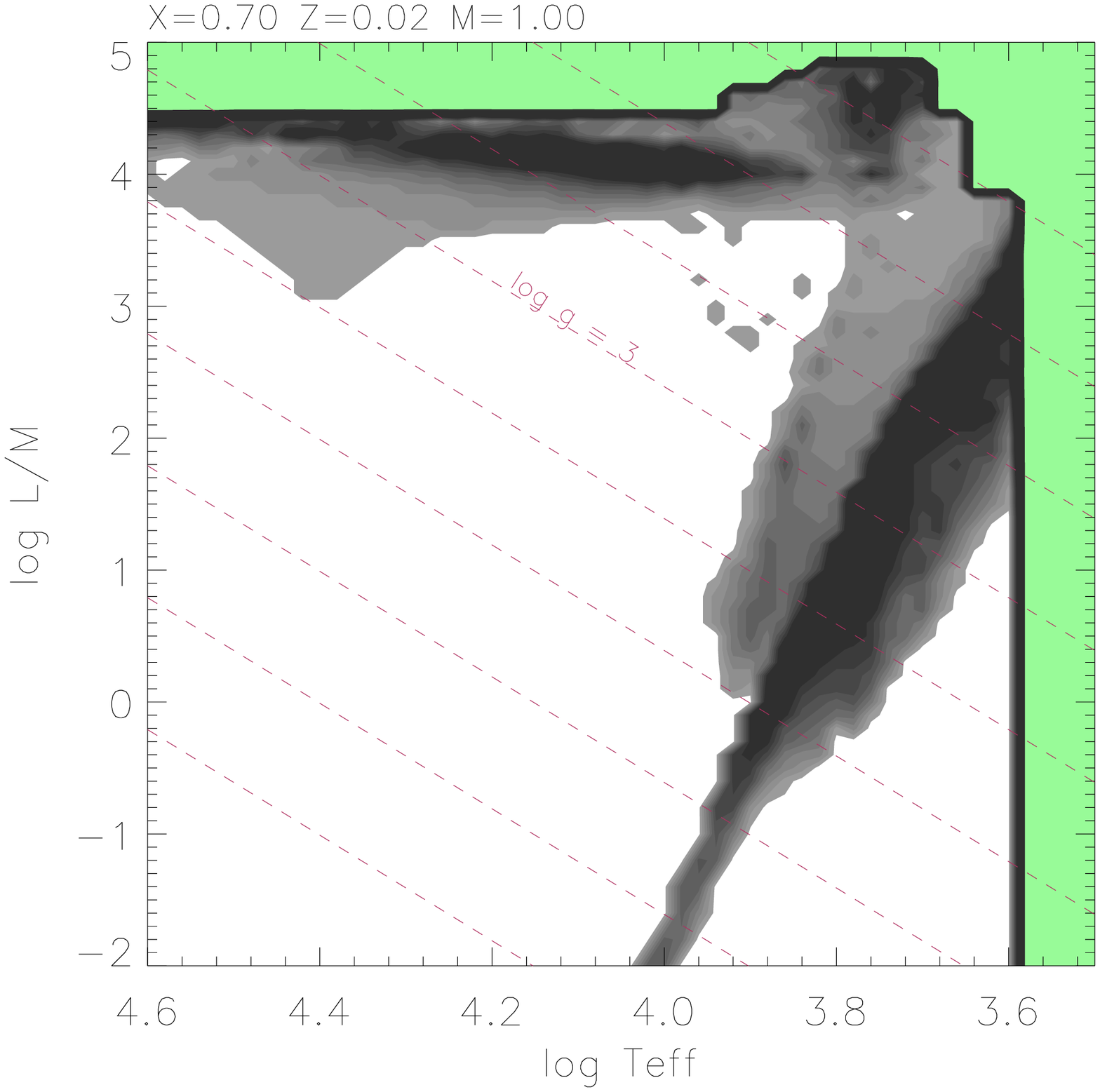,width=4.3cm,angle=0}
\epsfig{file=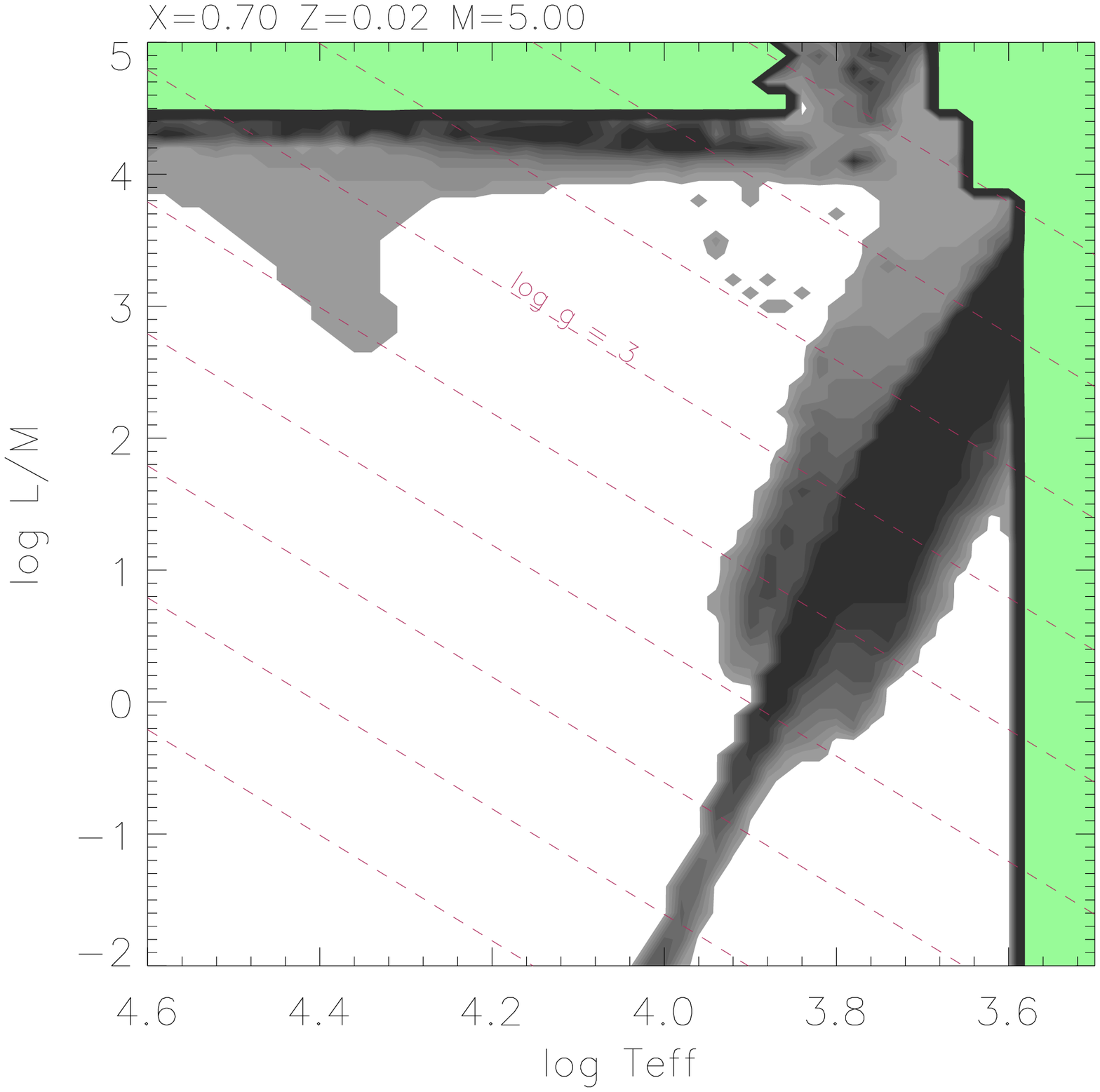,width=4.3cm,angle=0}
\epsfig{file=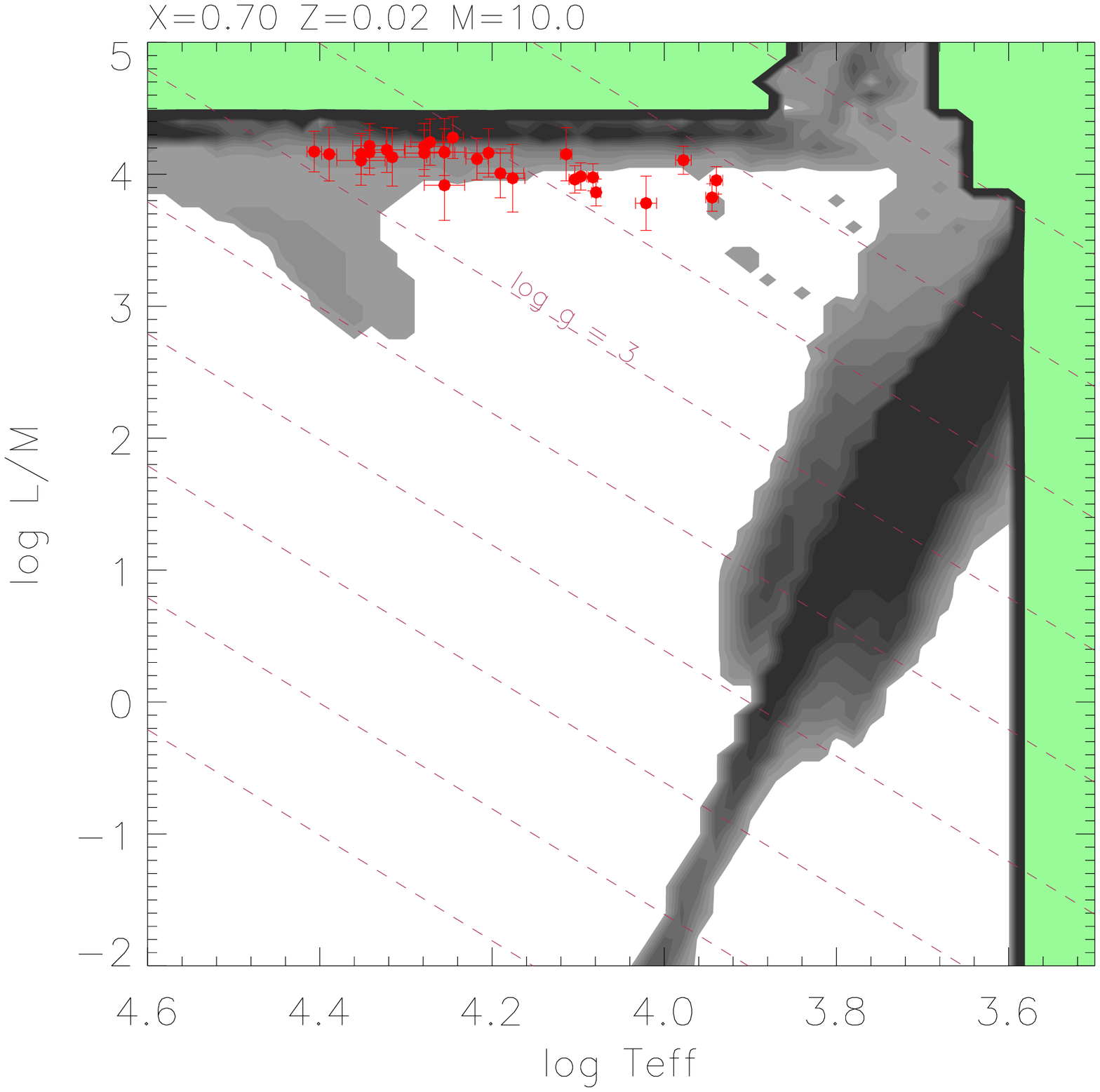,width=4.3cm,angle=0}
\epsfig{file=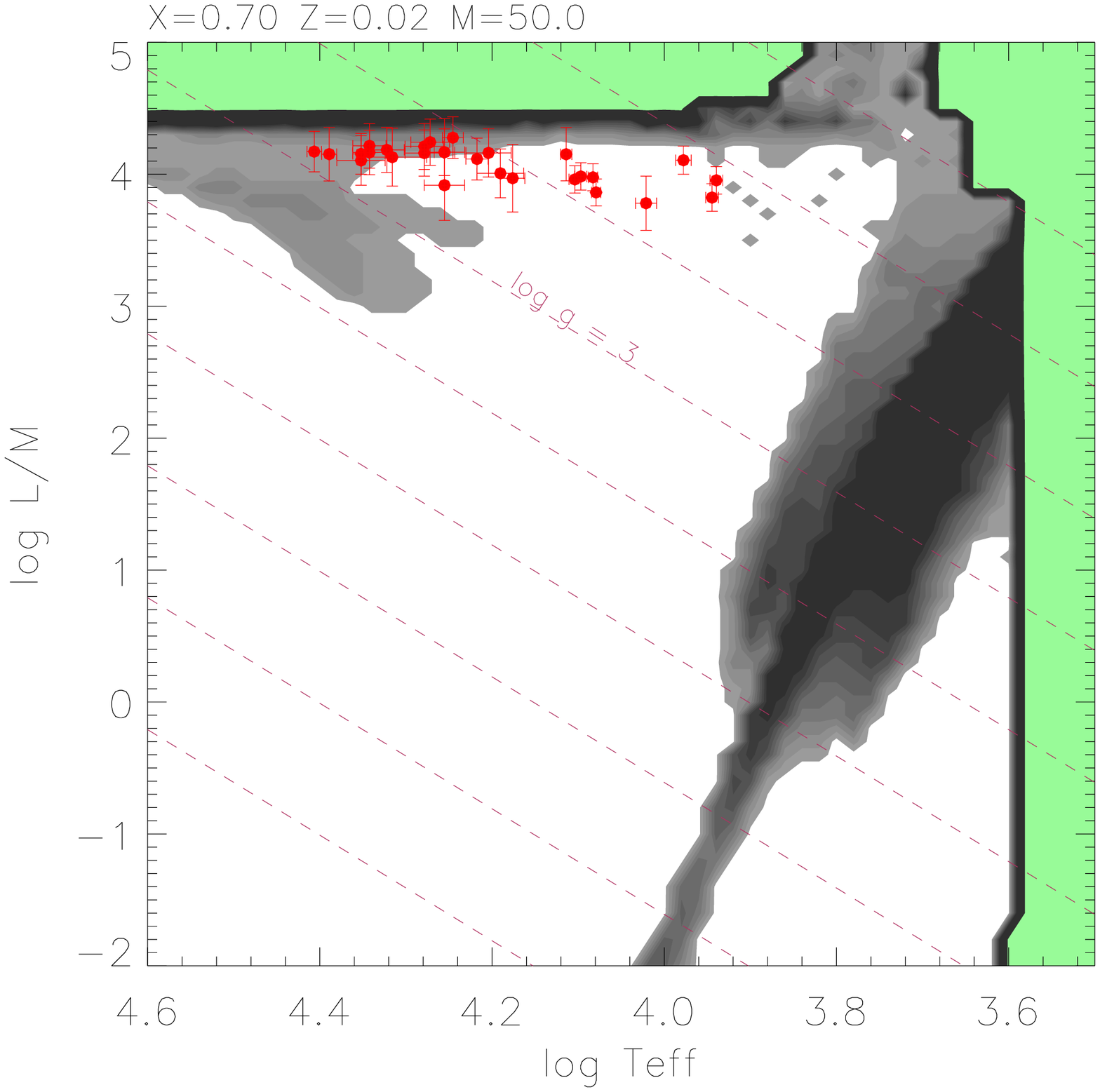,width=4.3cm,angle=0}\\
\epsfig{file=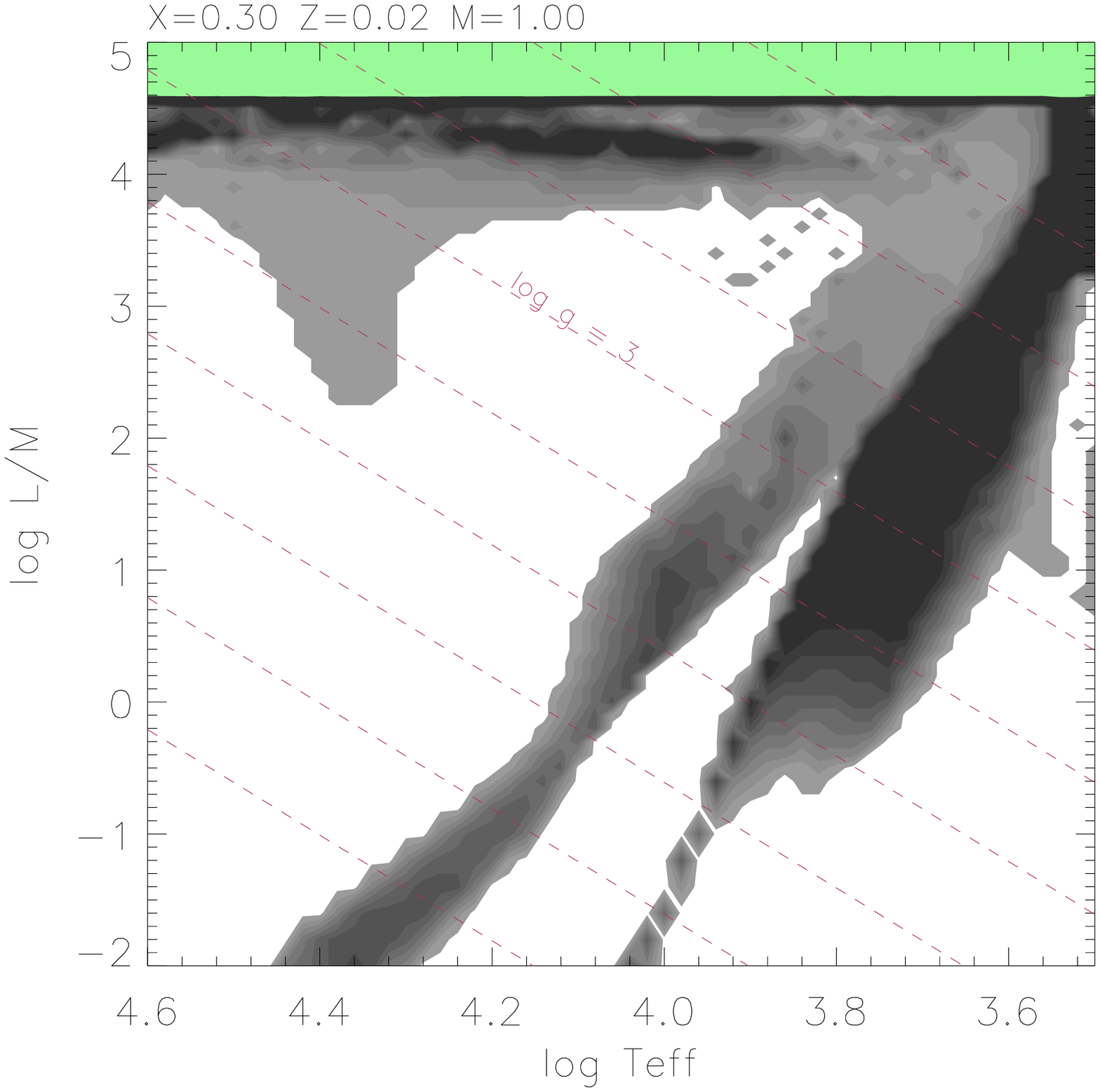,width=4.3cm,angle=0}
\epsfig{file=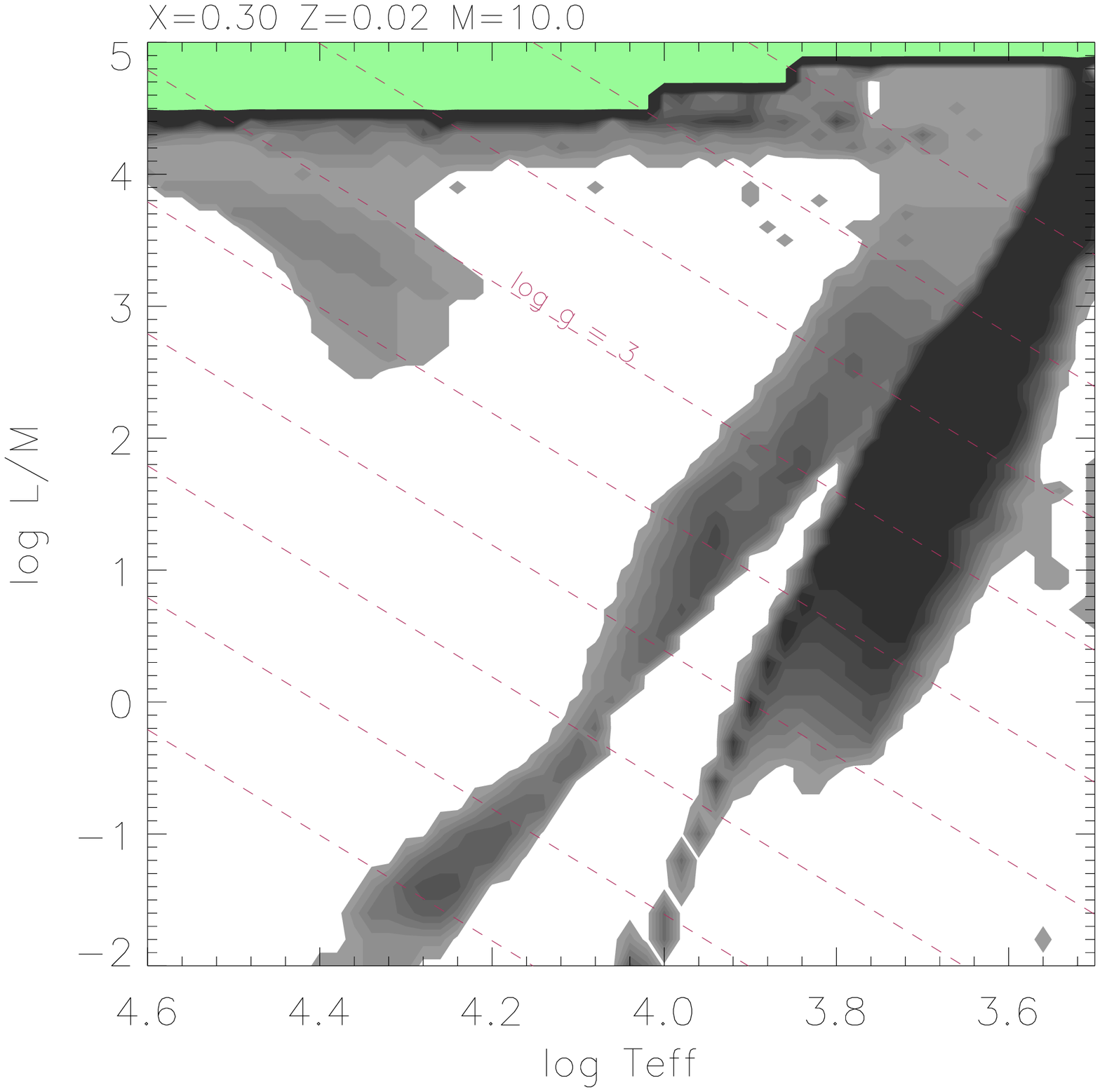,width=4.3cm,angle=0}
\epsfig{file=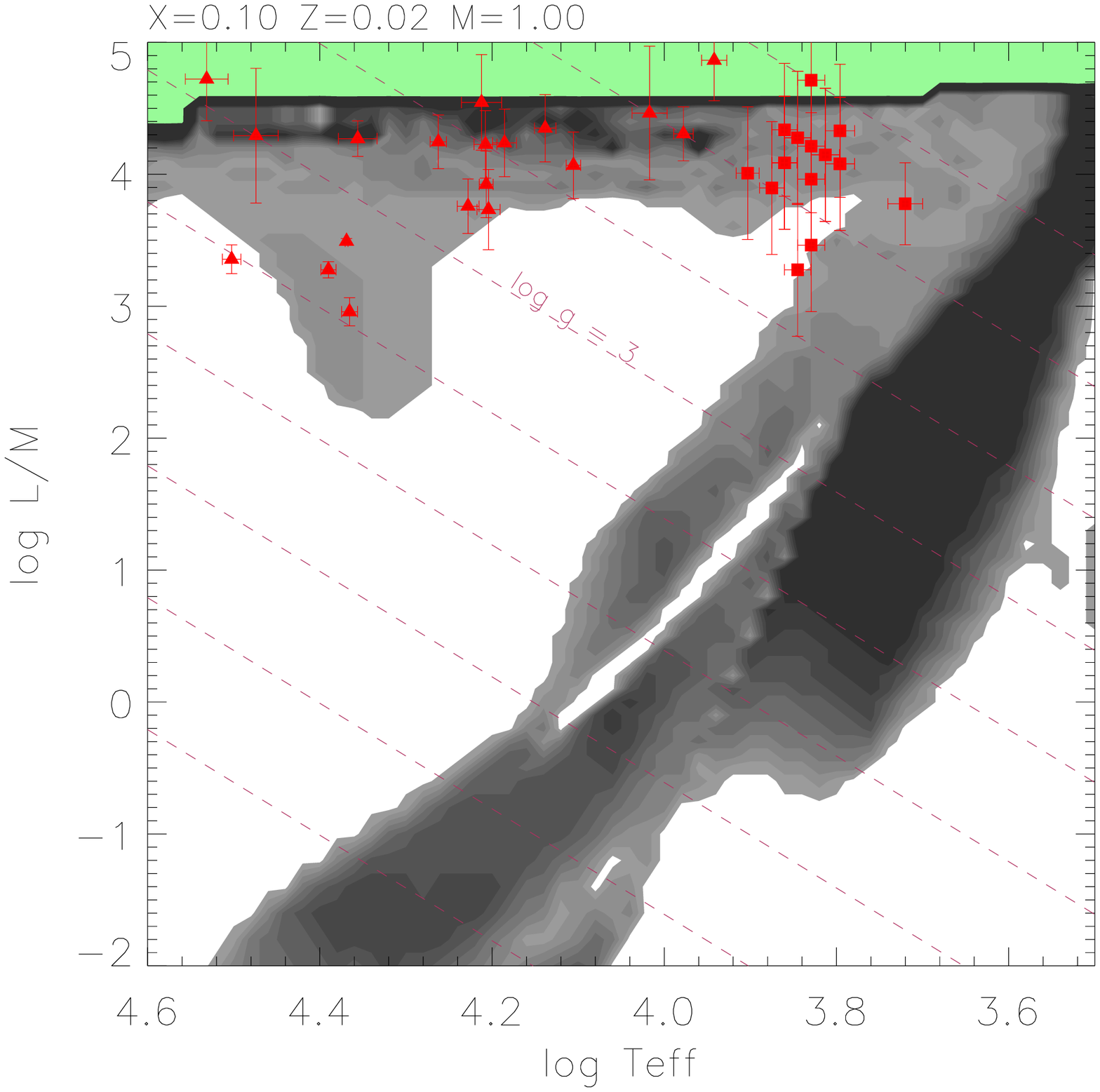,width=4.3cm,angle=0}
\epsfig{file=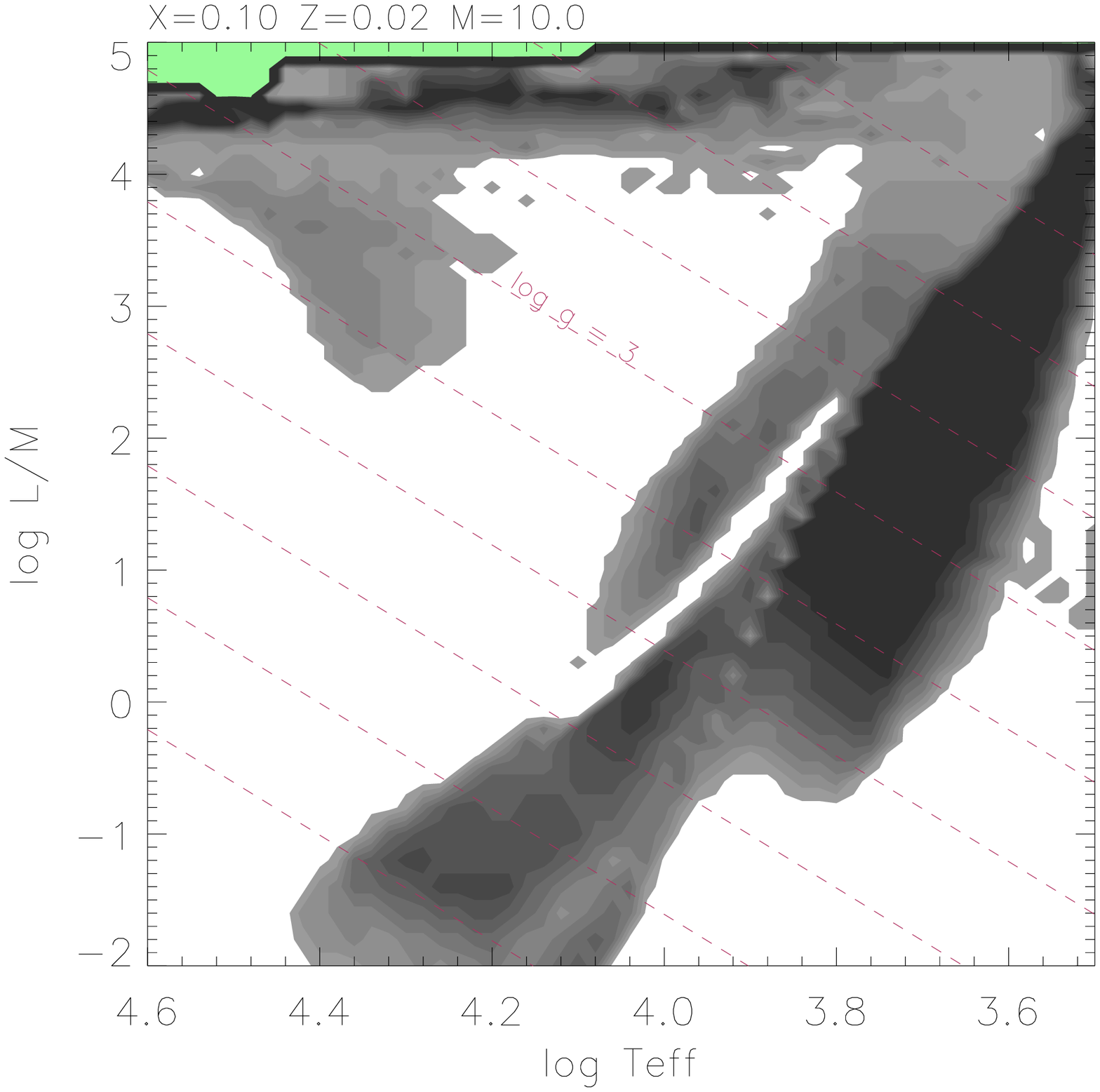,width=4.3cm,angle=0}\\
\epsfig{file=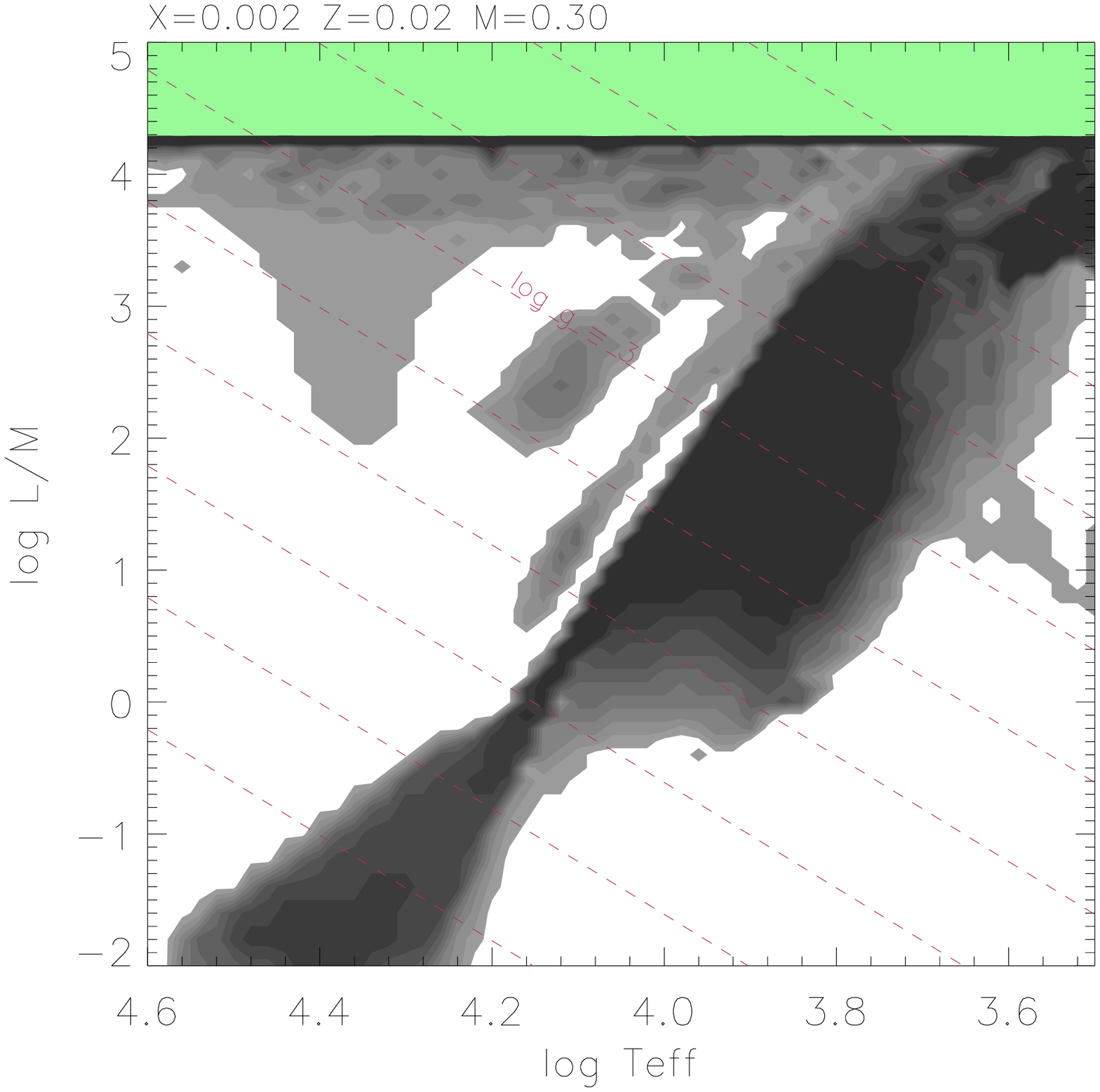,width=4.3cm,angle=0}
\epsfig{file=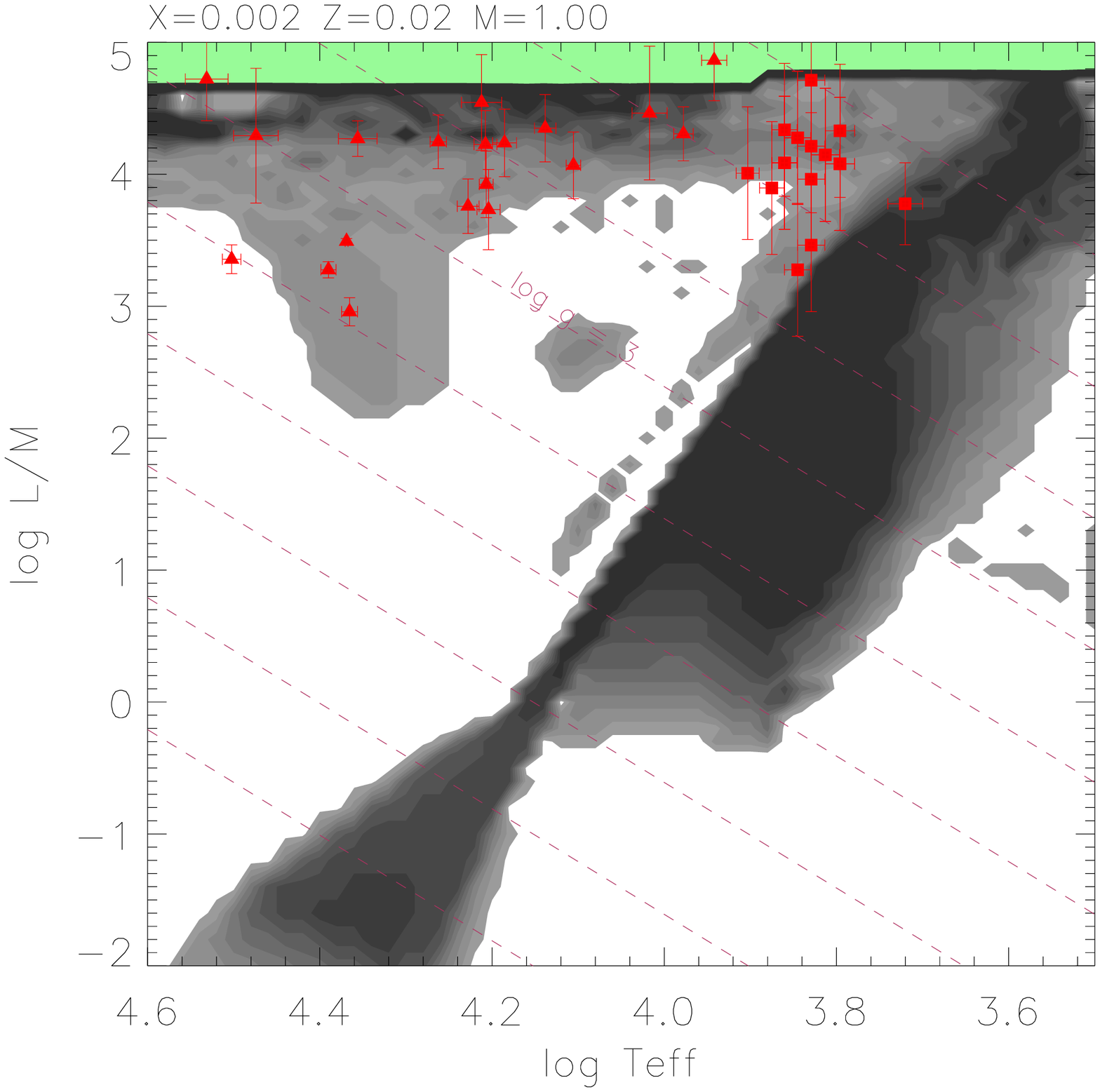,width=4.3cm,angle=0}
\epsfig{file=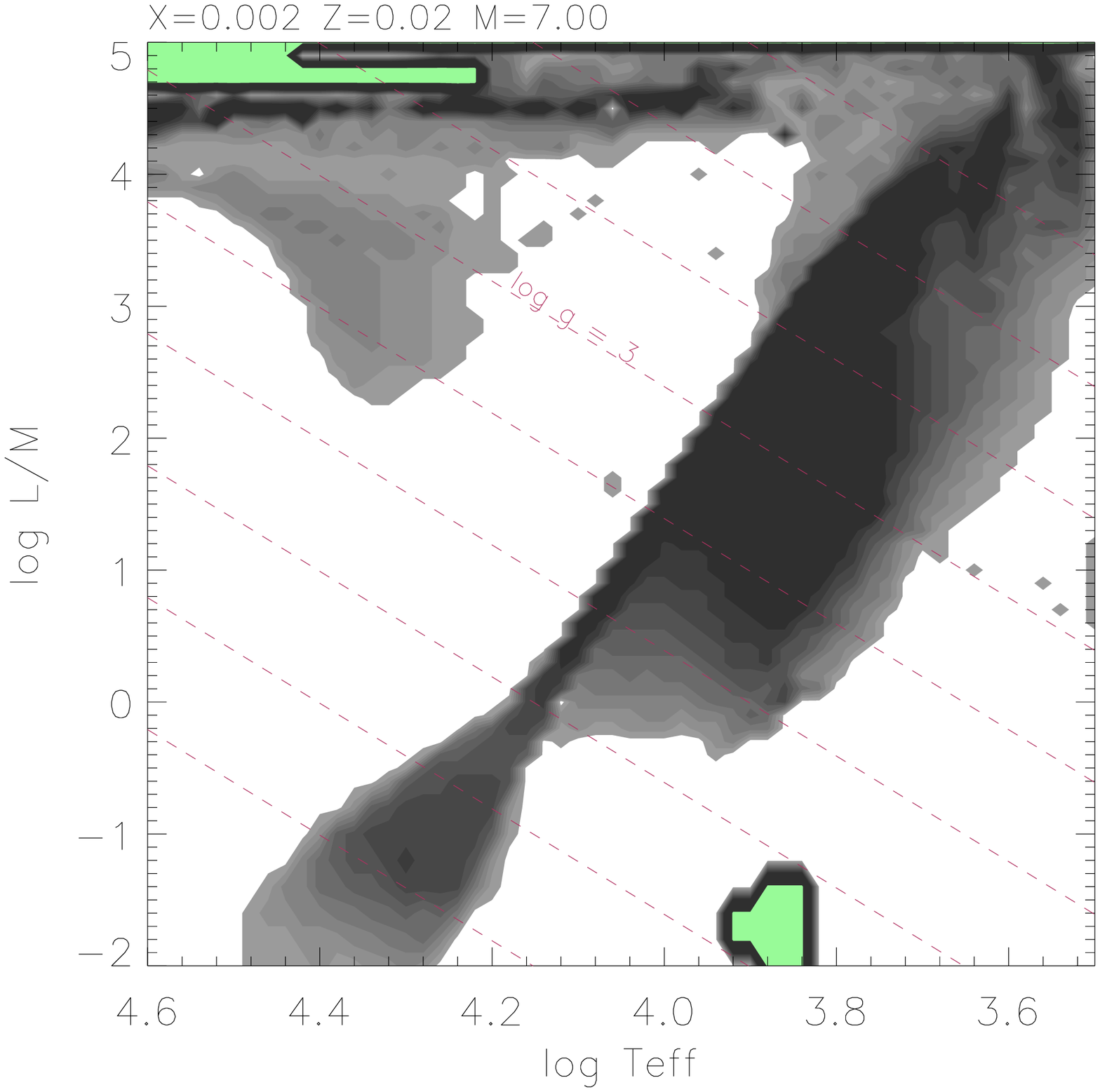,width=4.3cm,angle=0}
\epsfig{file=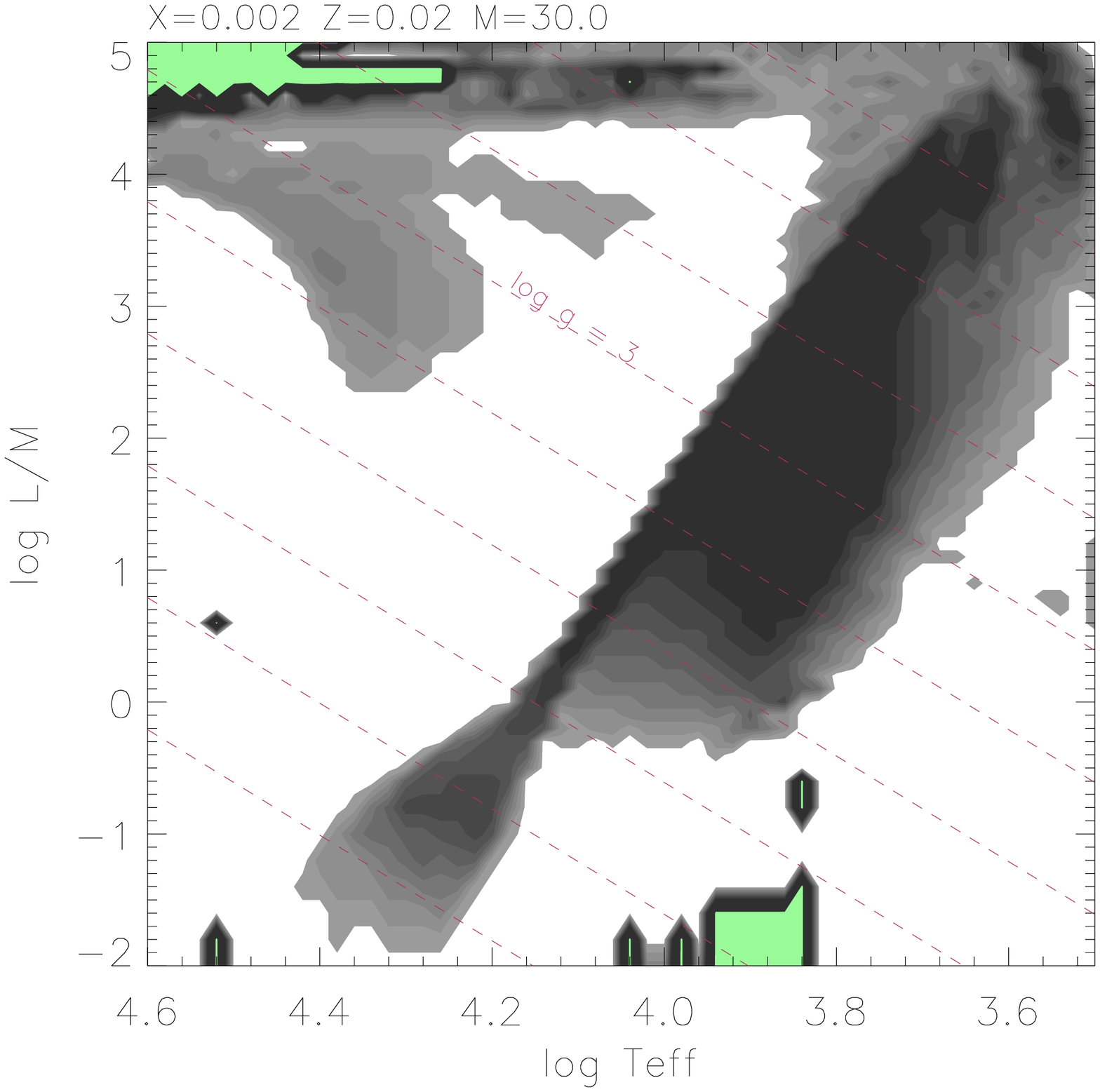,width=4.3cm,angle=0}\\
\epsfig{file=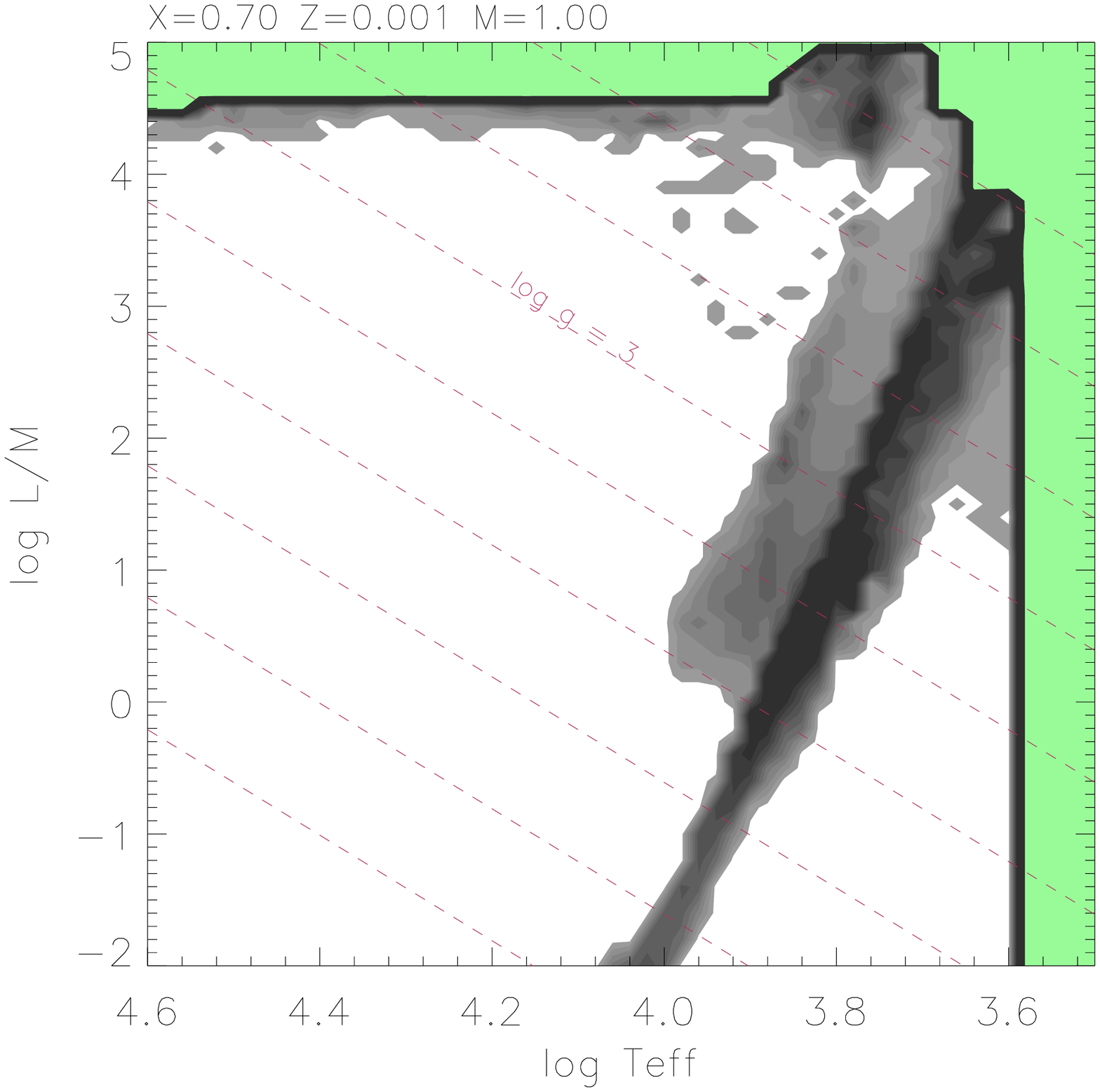,width=4.3cm,angle=0}
\epsfig{file=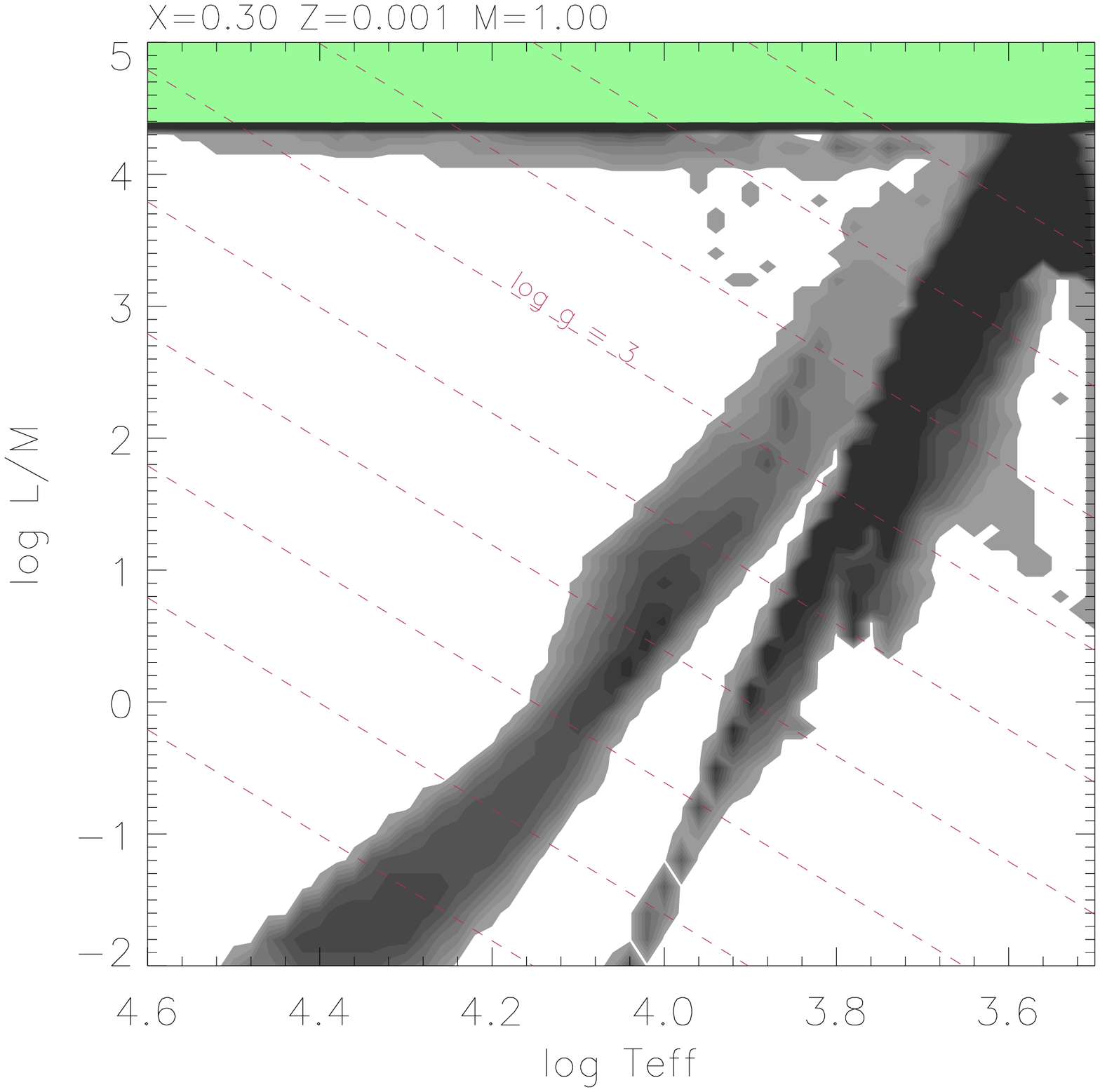,width=4.3cm,angle=0}
\epsfig{file=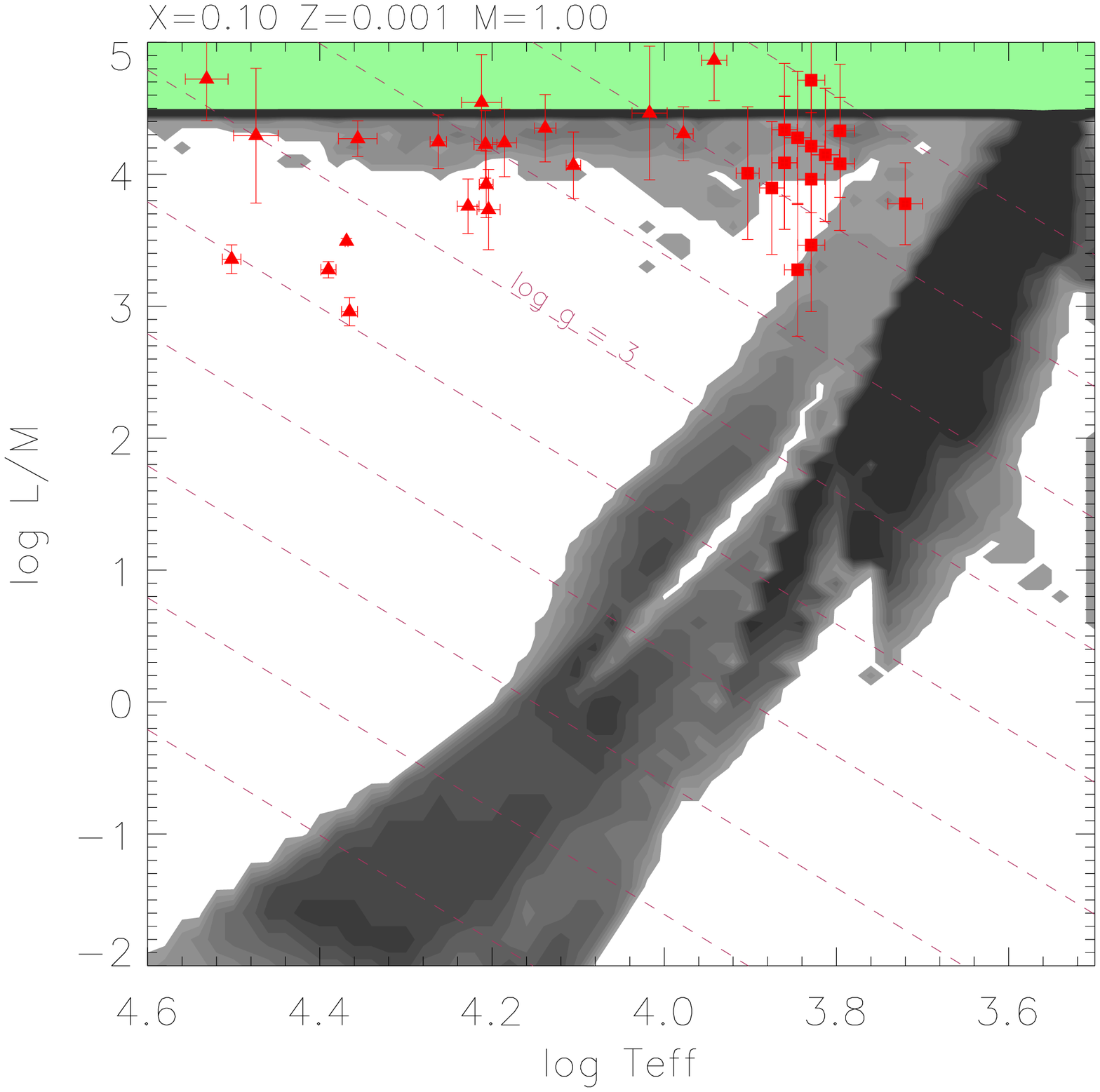,width=4.3cm,angle=0}
\epsfig{file=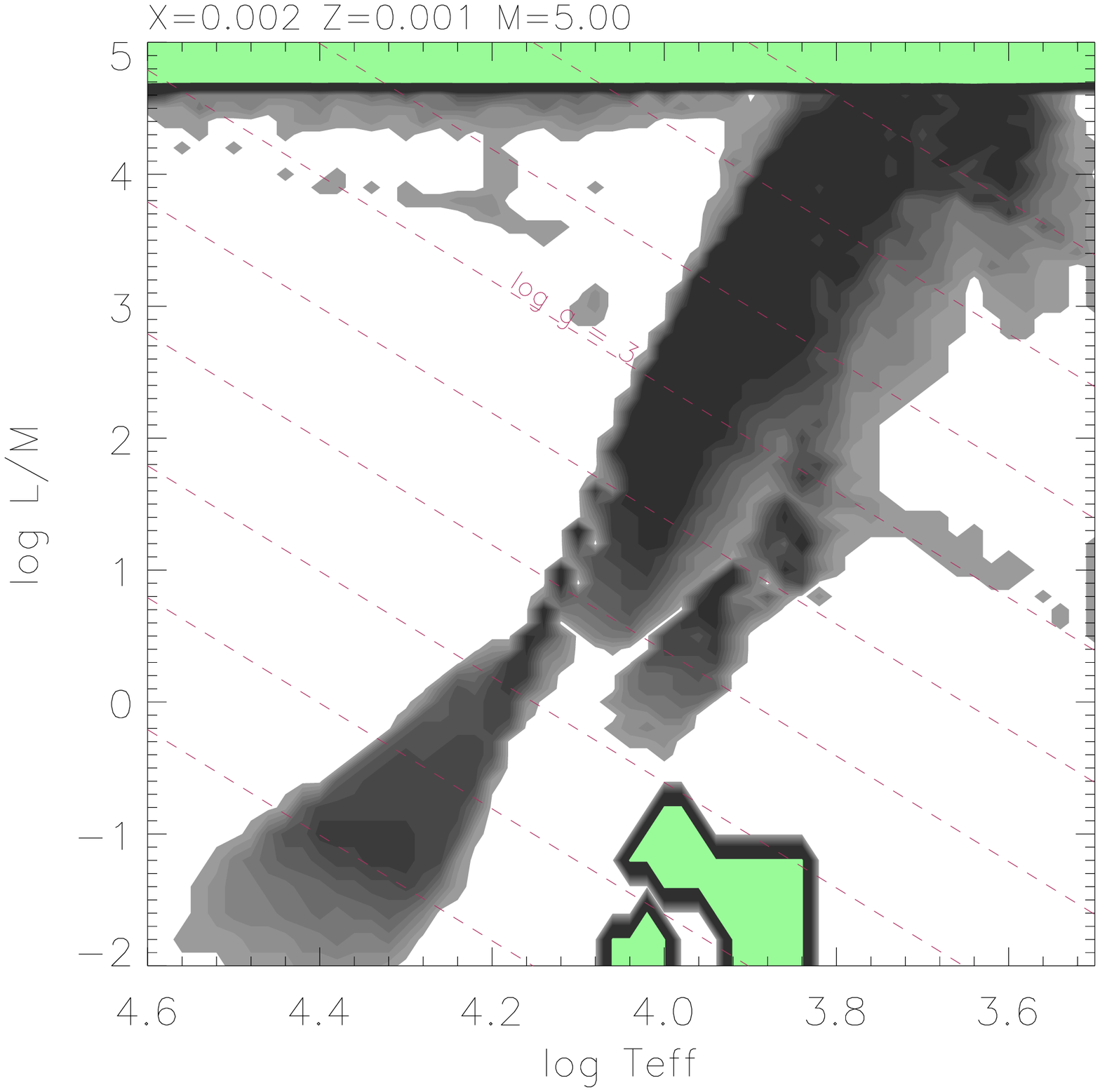,width=4.3cm,angle=0}
\caption[Unstable modes numbers]
{Unstable pulsation modes in stars with homogeneous
envelopes for selected compositions and masses, as labelled. 
Full grids are shown in Appendix A. 
The {\it number} of unstable radial modes is represented by grey scale contours, with the lightest shade marking the instability boundary
(one unstable mode), and the darkest shade representing ten or more more unstable modes. 
 Broken (maroon online) diagonal lines represent contours  of constant surface gravity at  $\log g = 8, 7, 6, \ldots, 1$. 
Pale green areas denote regions where envelope models encountered convergende difficulties.   
Red symbols with error bars  shown on selected panels  
represent the observed positions of pulsating low-mass hydrogen-deficient stars, including extreme
helium stars and R Coronae Borealis variables \citep{jeffery08.ibvs}, shown on panels with $X \leq 0.1$ and $M=1.0 \Msolar$, 
and $\alpha$\,Cyg variables \citep{crowther06,searle08,firnstein12}, shown on panels with  $X=0.7, Z=0.02$ and $M \geq 10 \Msolar$. 
Figs.\,\ref{f:nmodes} and \ref{f:harmonics} are best viewed online and expanded; one grid is enlarged in 
Fig.\,\ref{f:enlarged}.
}
\label{f:nmodes}
\end{center}
\end{figure*}

The results are presented primarily as contour plots representing
the number of unstable modes as a function of 
$(T_{\rm eff}, L/M)$   for each composition (Figs.~\ref{f:nmodes} and \ref{f:nx70} -- \ref{f:nx002z001} ).
This provides an overall instability boundary since it includes 
pulsations in both low- and high-order modes. 
In some cases, envelope models with very high $L/M$ ratios and/or very low $T_{\rm eff}$ were difficult
to integrate due to very low densities in the equation of state; these appear as voids on the contour plots. 

\begin{figure*}
\begin{center}
\epsfig{file=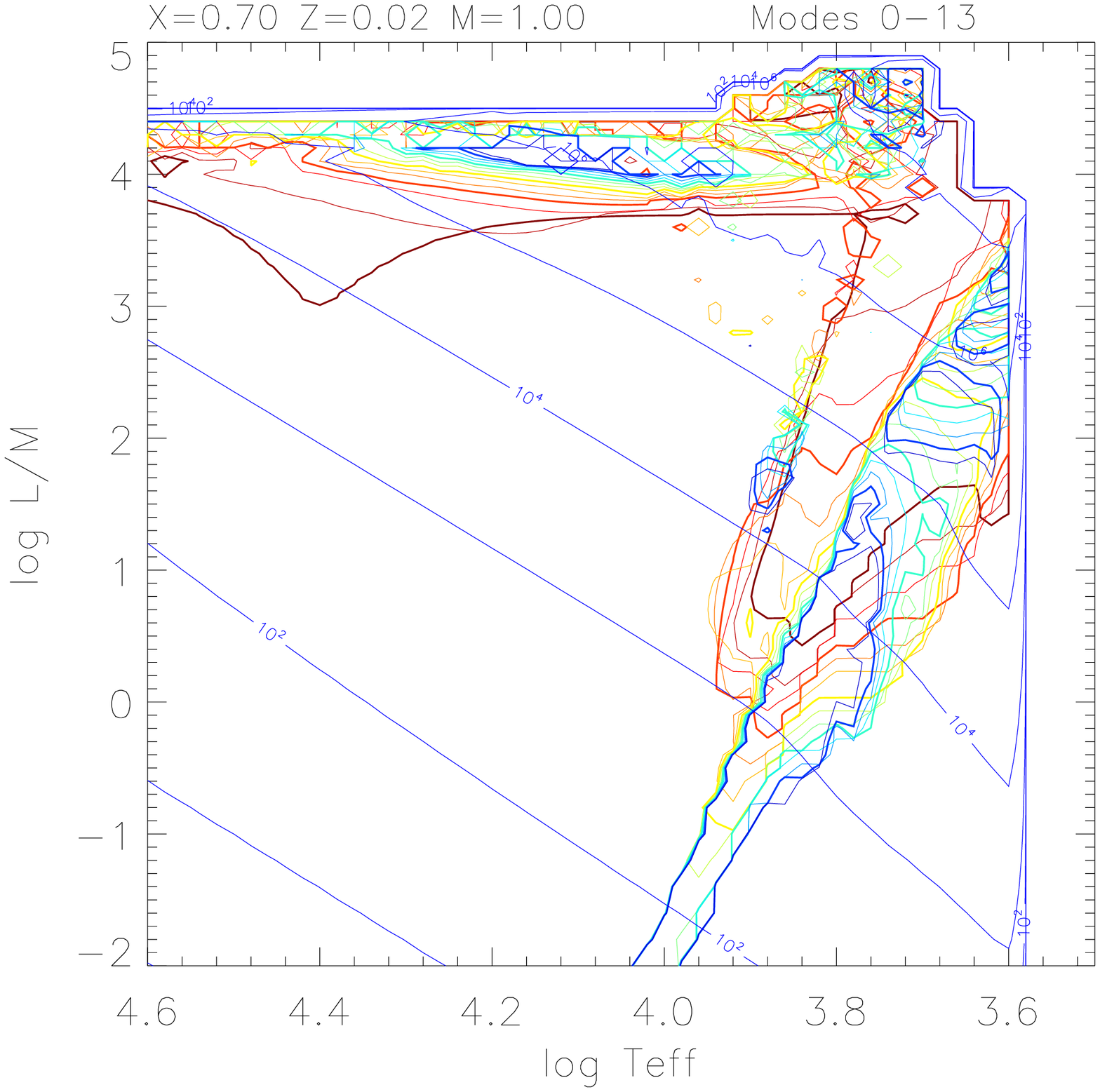,width=4.3cm,angle=0}
\epsfig{file=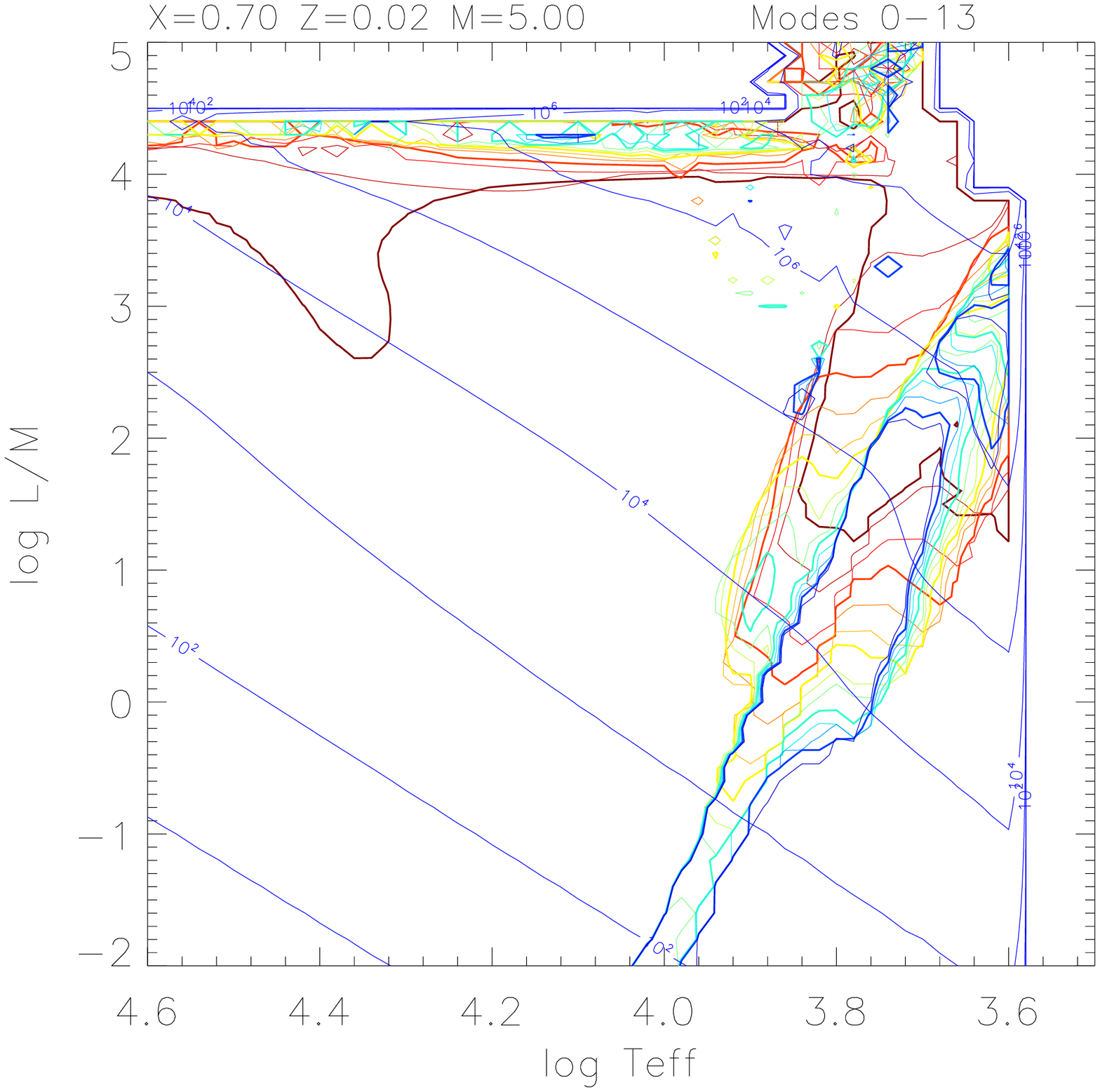,width=4.3cm,angle=0}
\epsfig{file=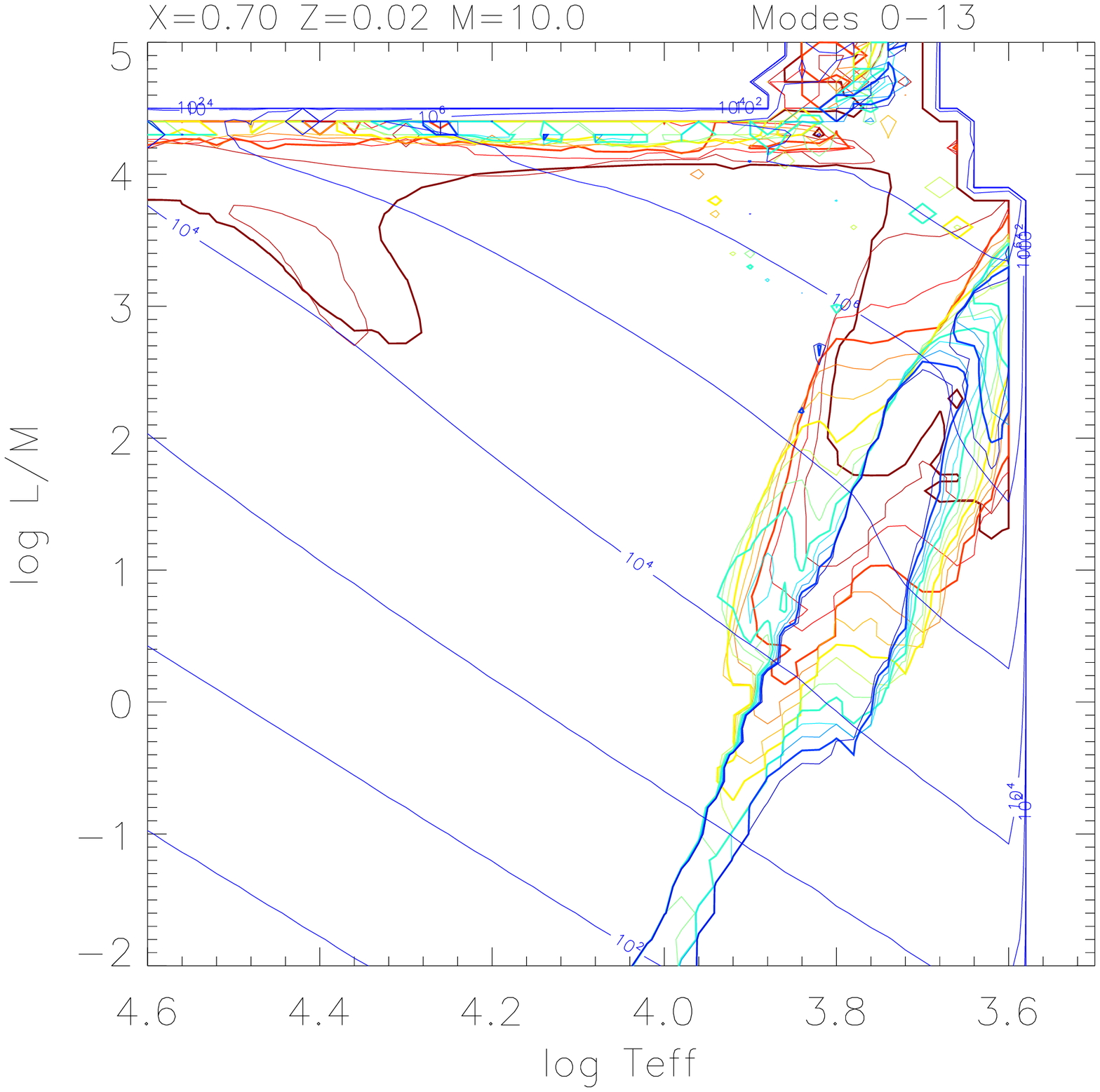,width=4.3cm,angle=0}
\epsfig{file=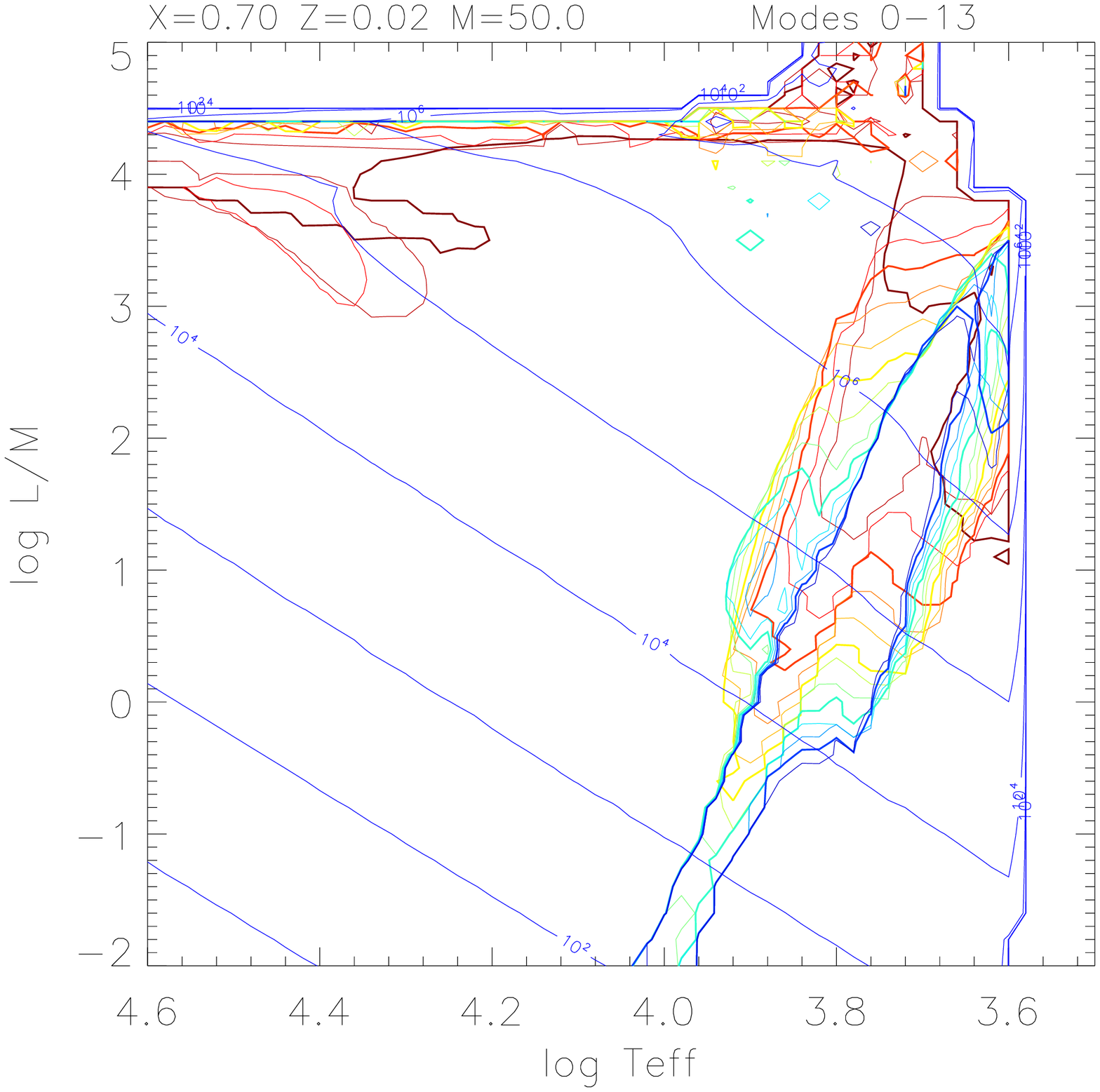,width=4.3cm,angle=0}\\
\epsfig{file=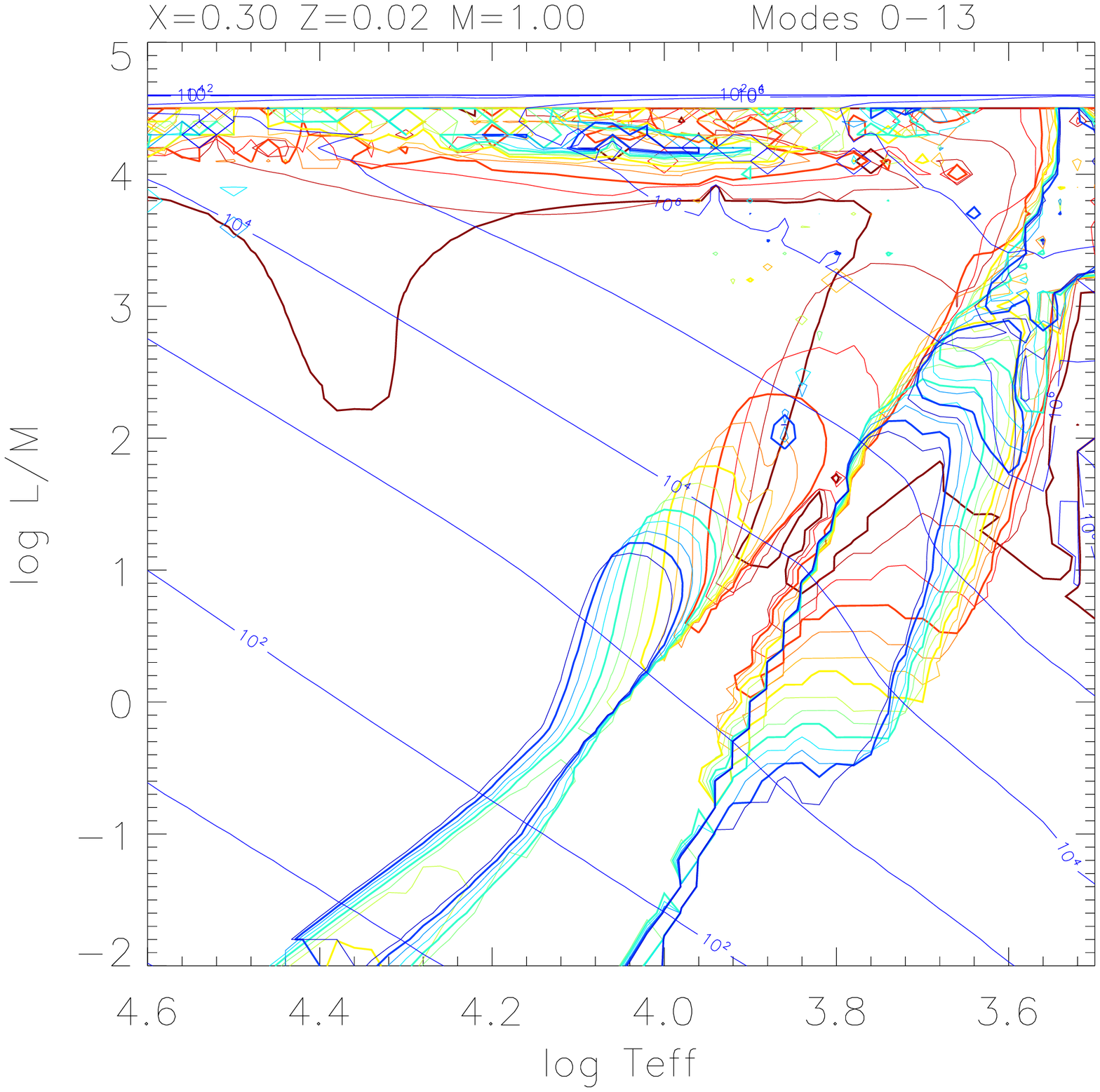,width=4.3cm,angle=0}
\epsfig{file=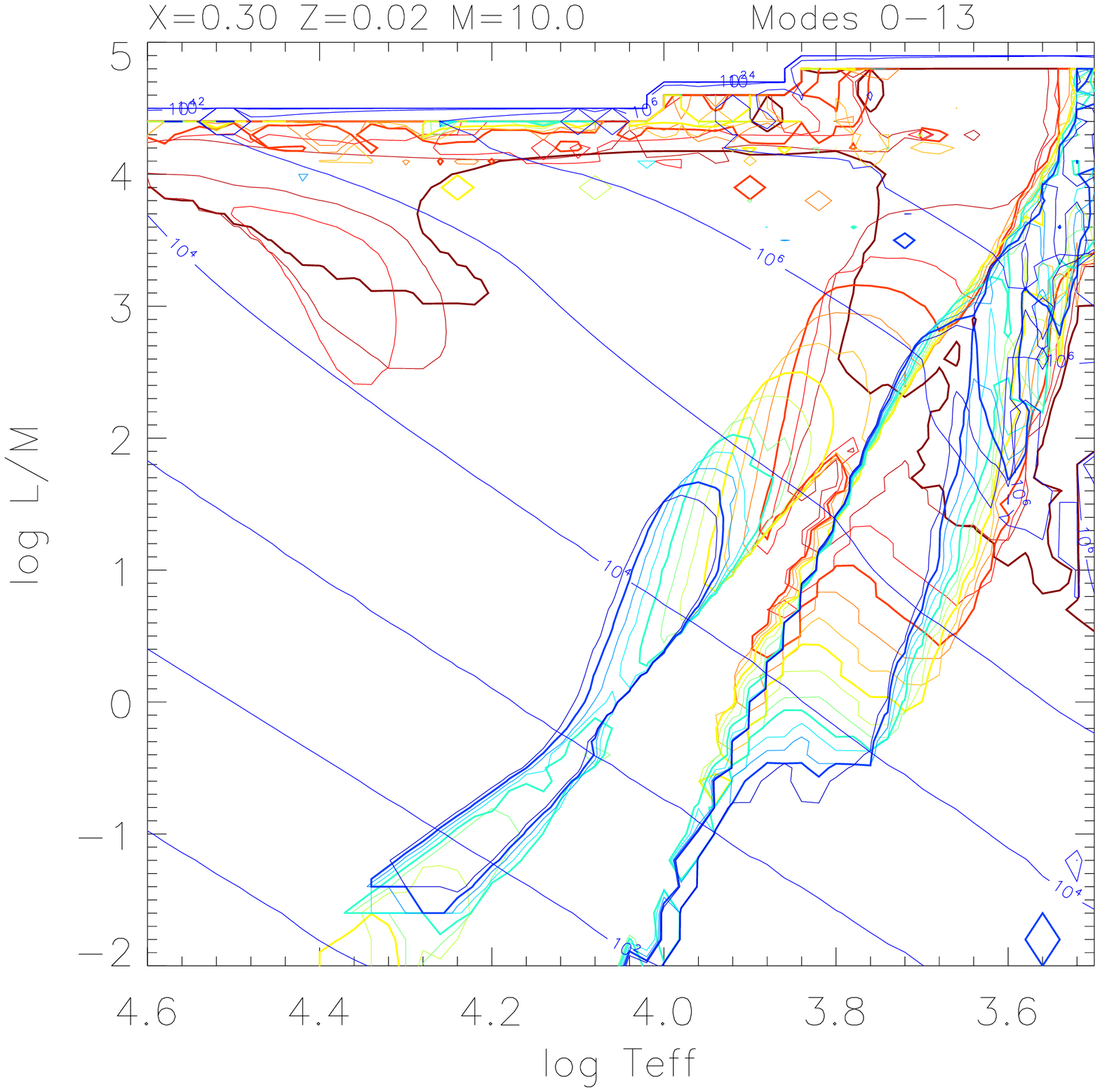,width=4.3cm,angle=0}
\epsfig{file=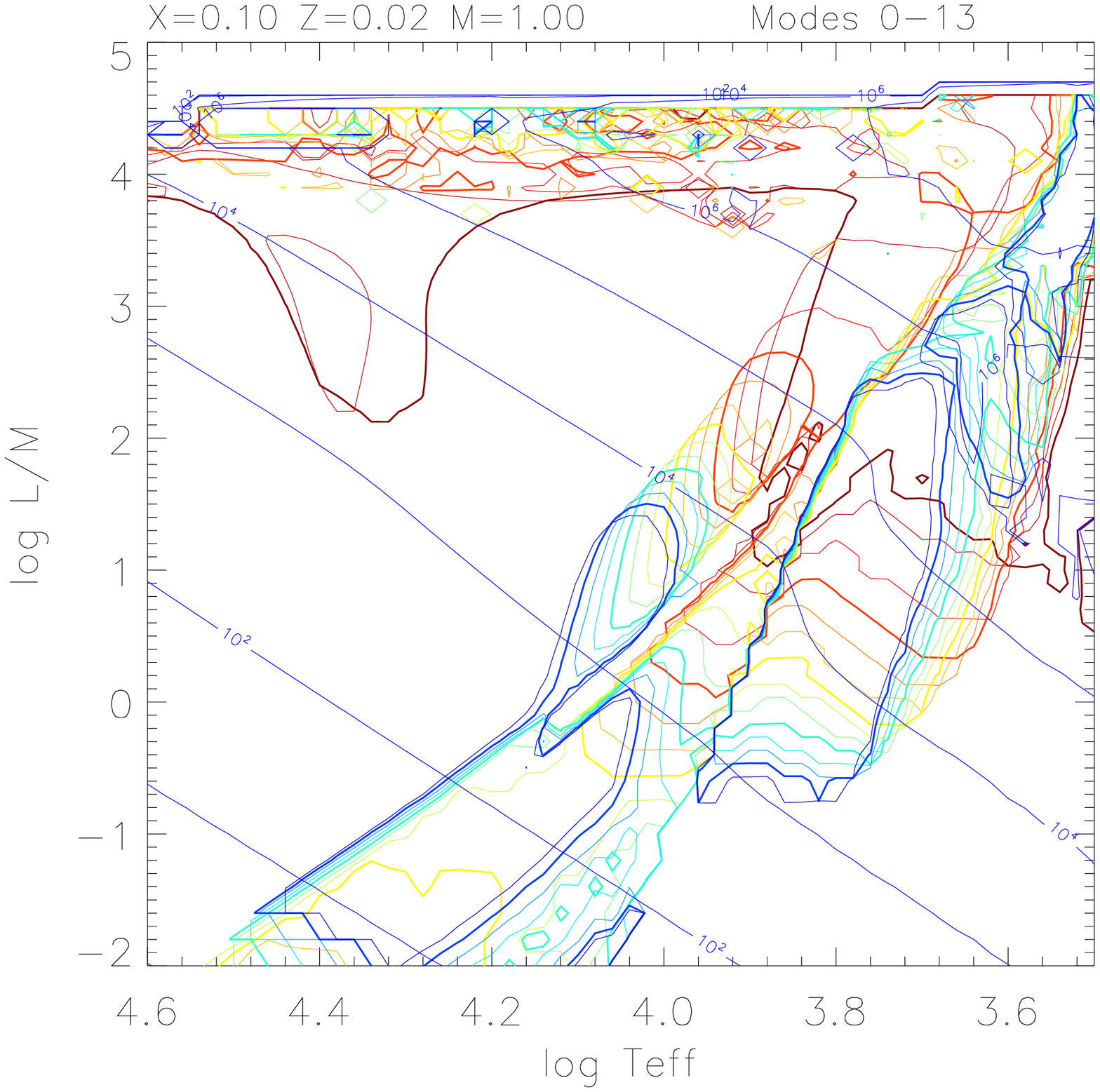,width=4.3cm,angle=0}
\epsfig{file=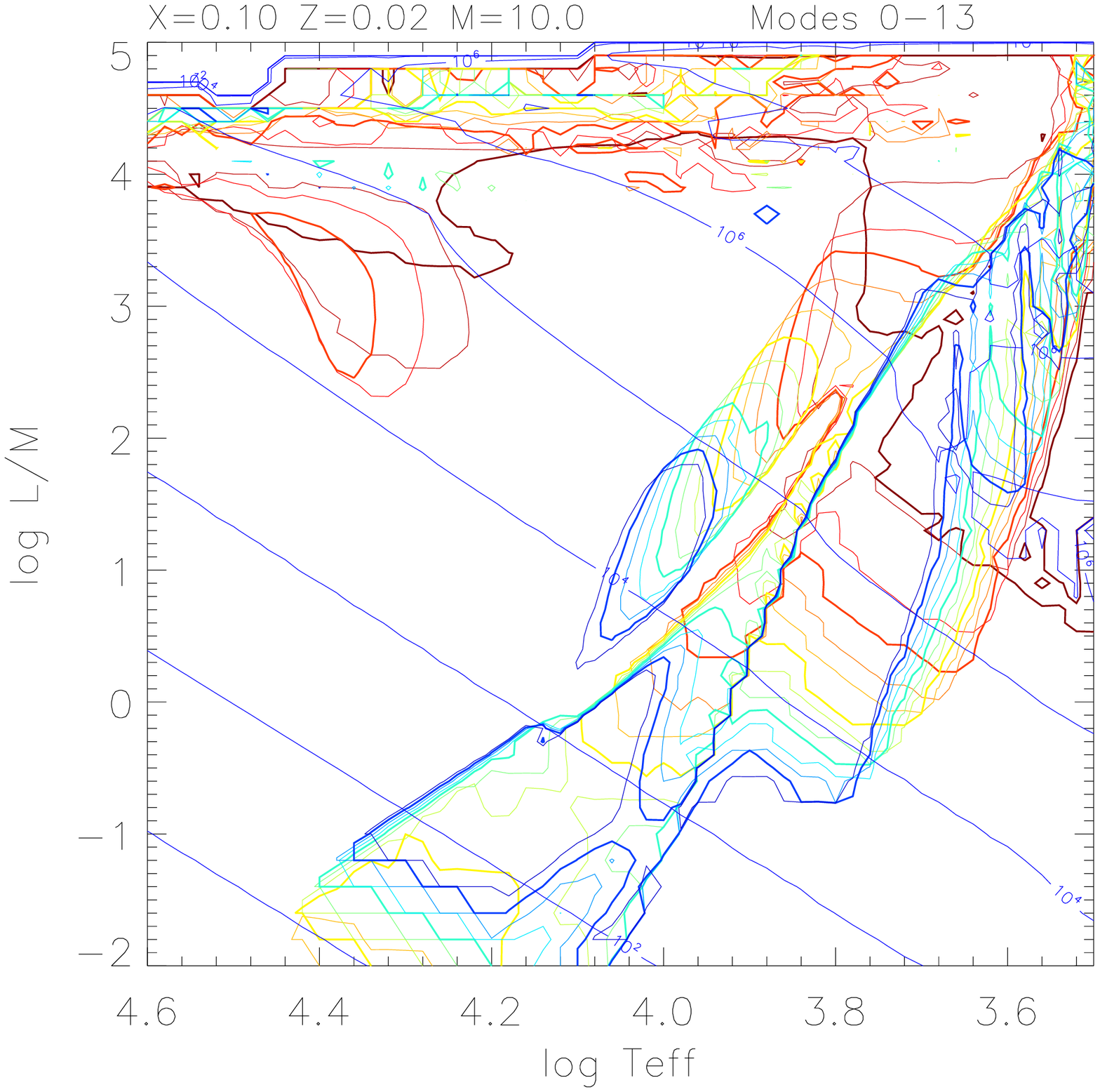,width=4.3cm,angle=0}\\
\epsfig{file=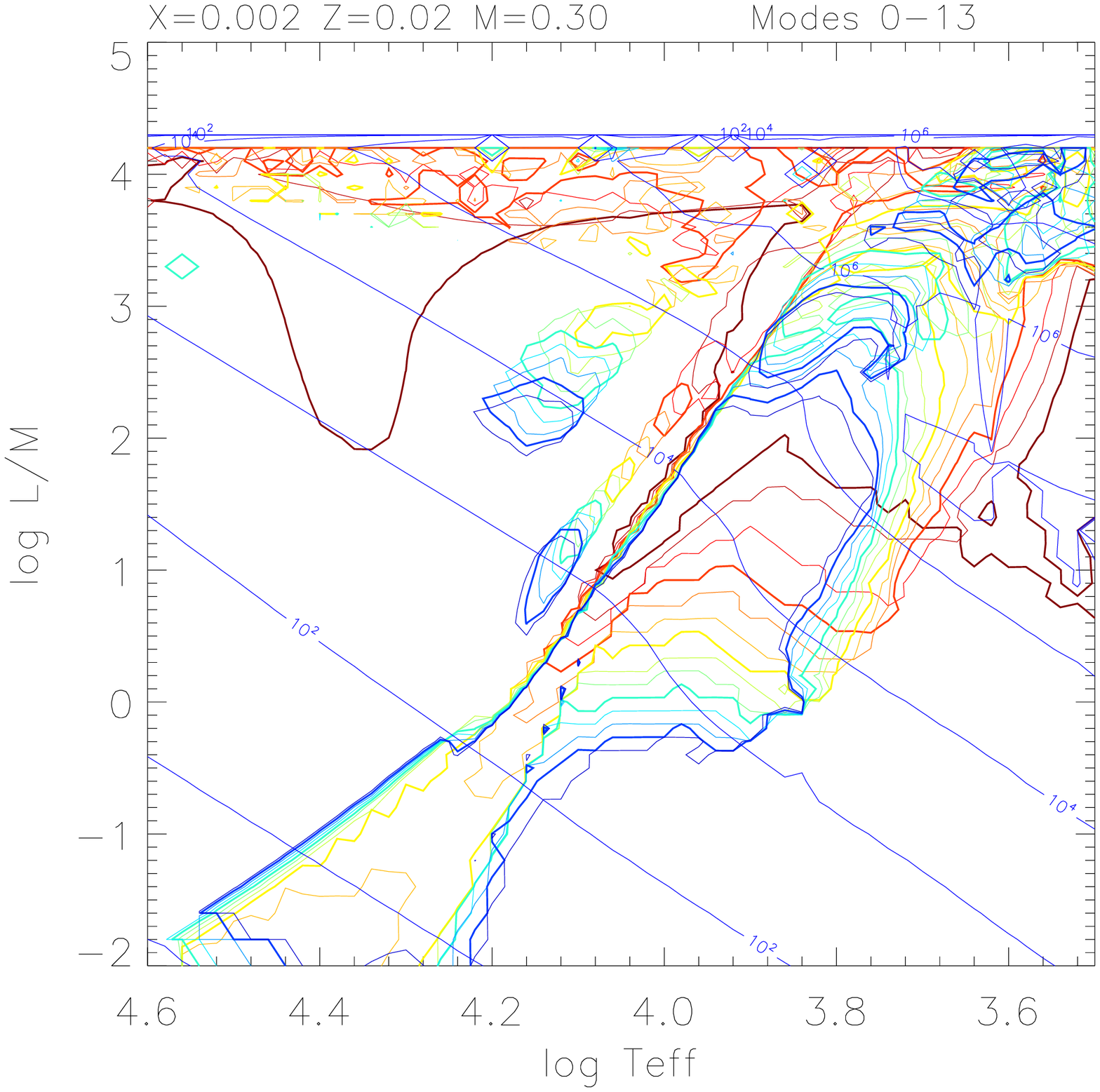,width=4.3cm,angle=0}
\epsfig{file=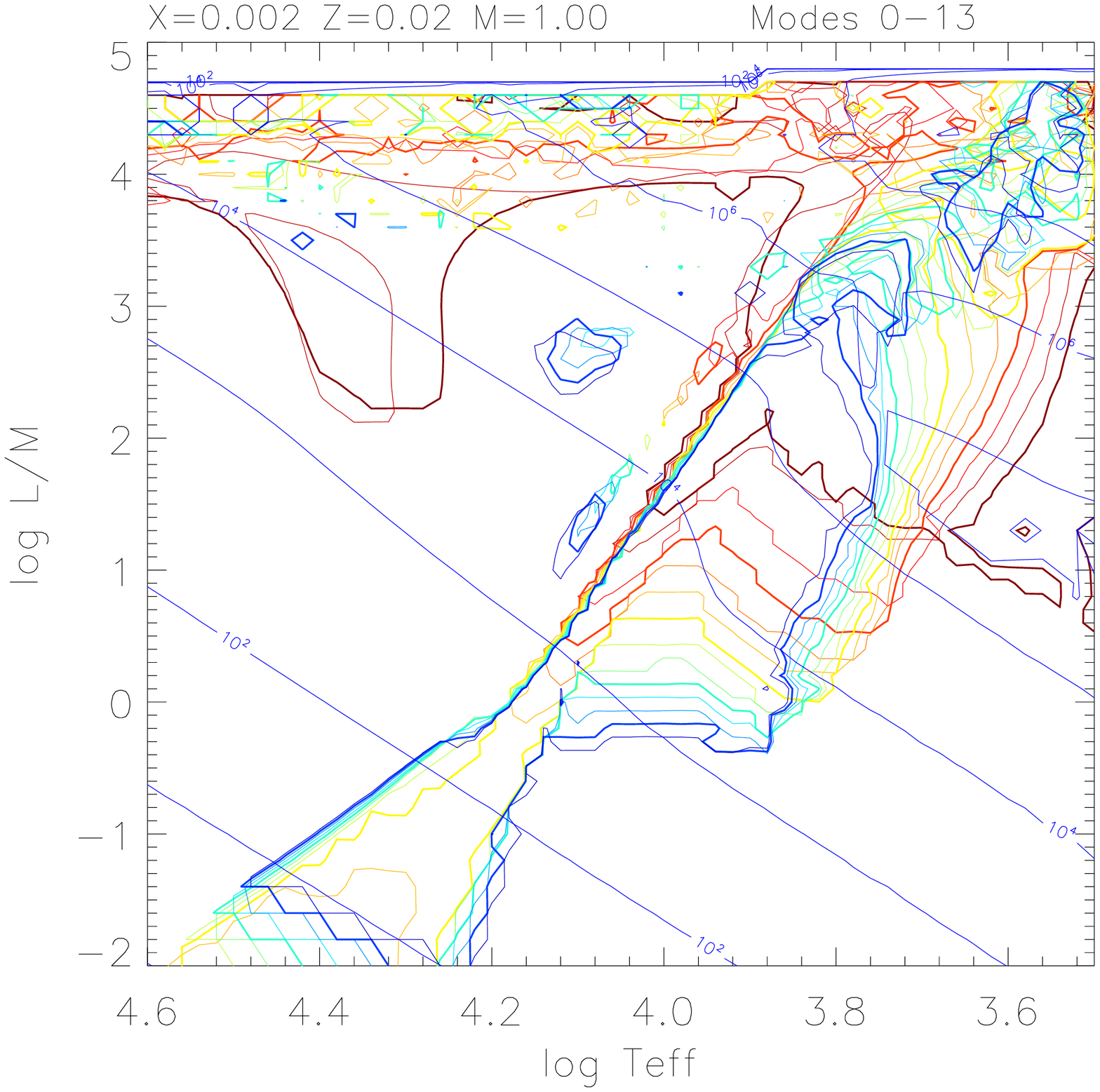,width=4.3cm,angle=0}
\epsfig{file=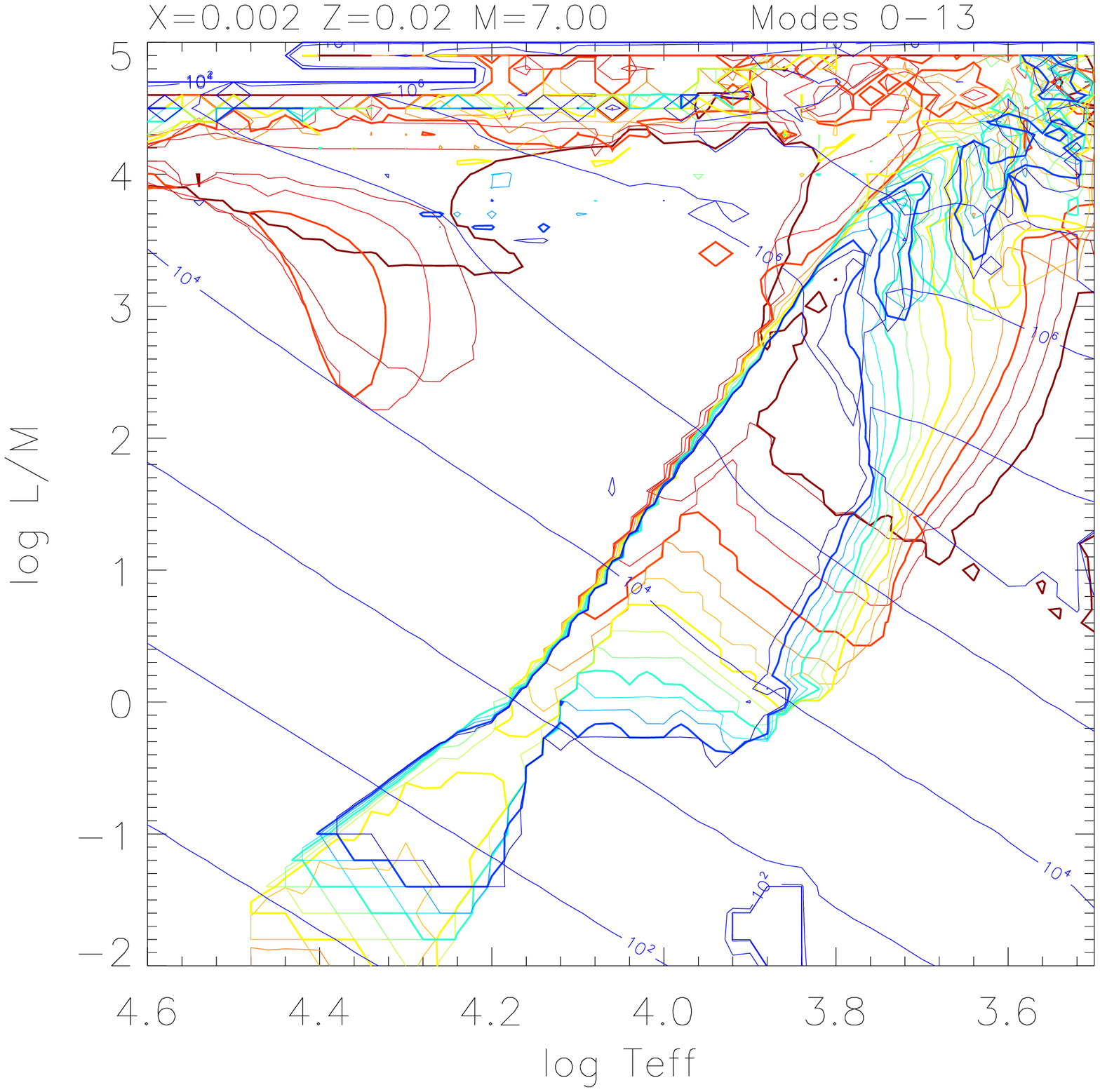,width=4.3cm,angle=0}
\epsfig{file=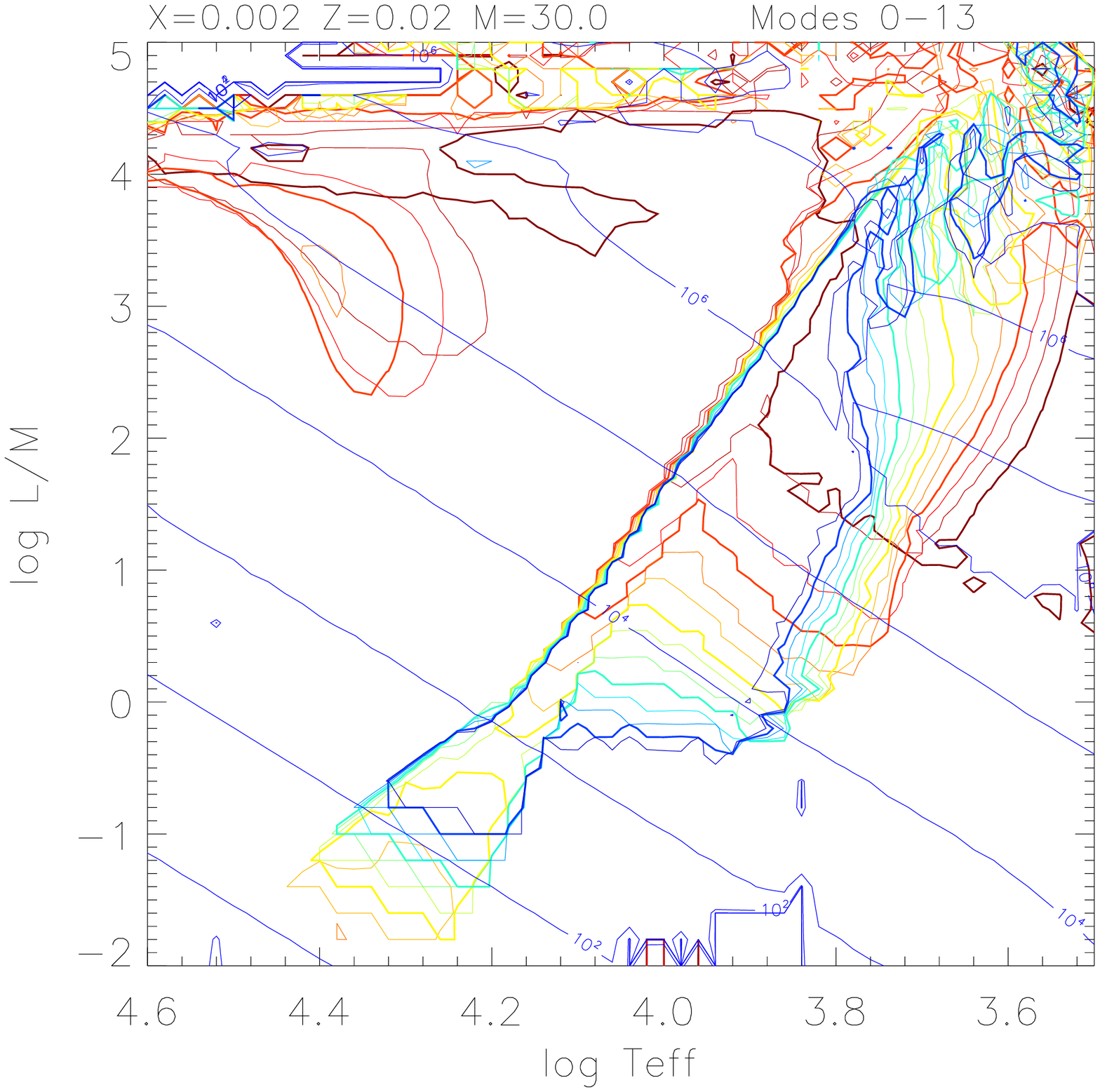,width=4.3cm,angle=0}\\
\epsfig{file=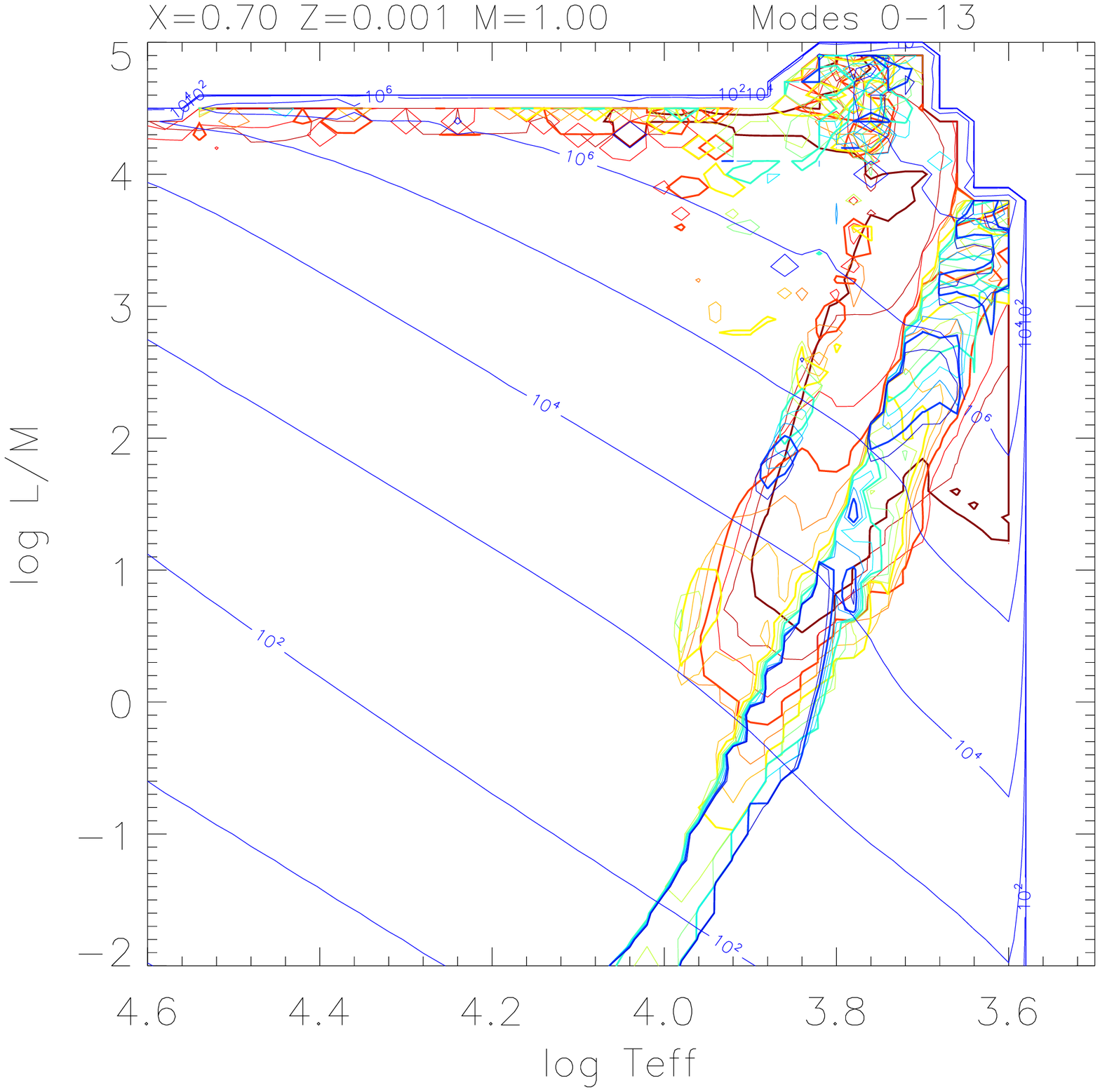,width=4.3cm,angle=0}
\epsfig{file=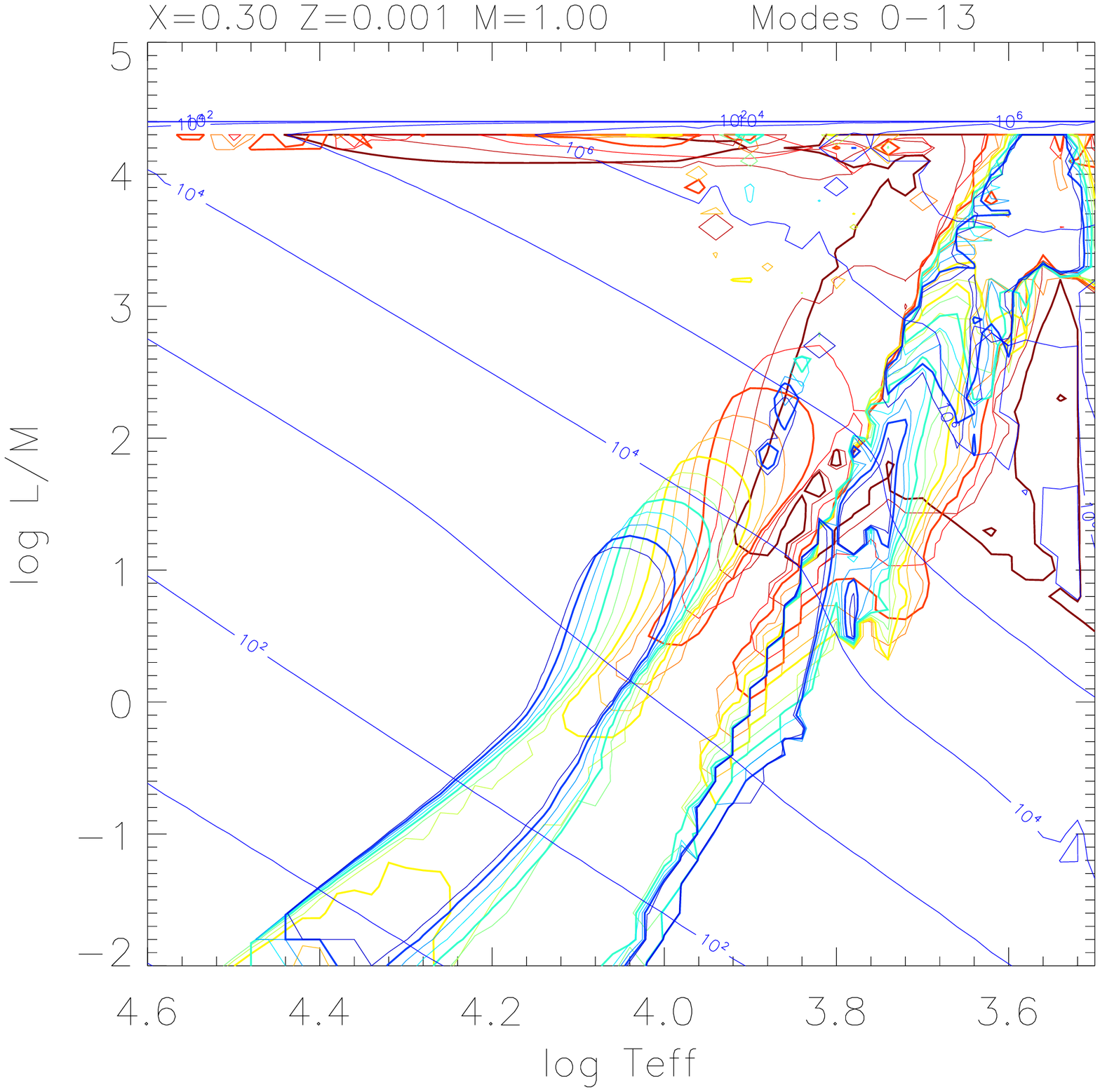,width=4.3cm,angle=0}
\epsfig{file=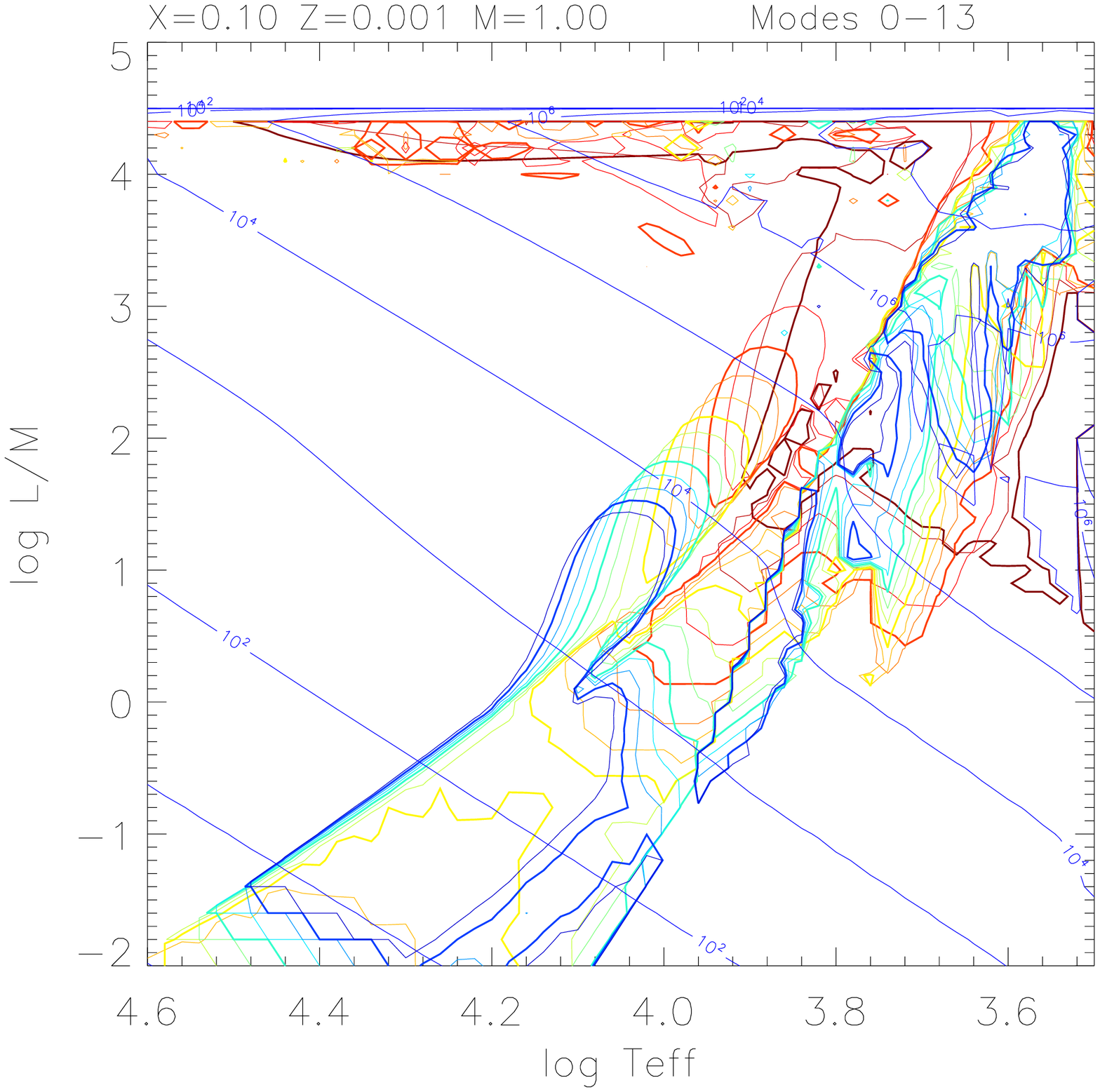,width=4.3cm,angle=0}
\epsfig{file=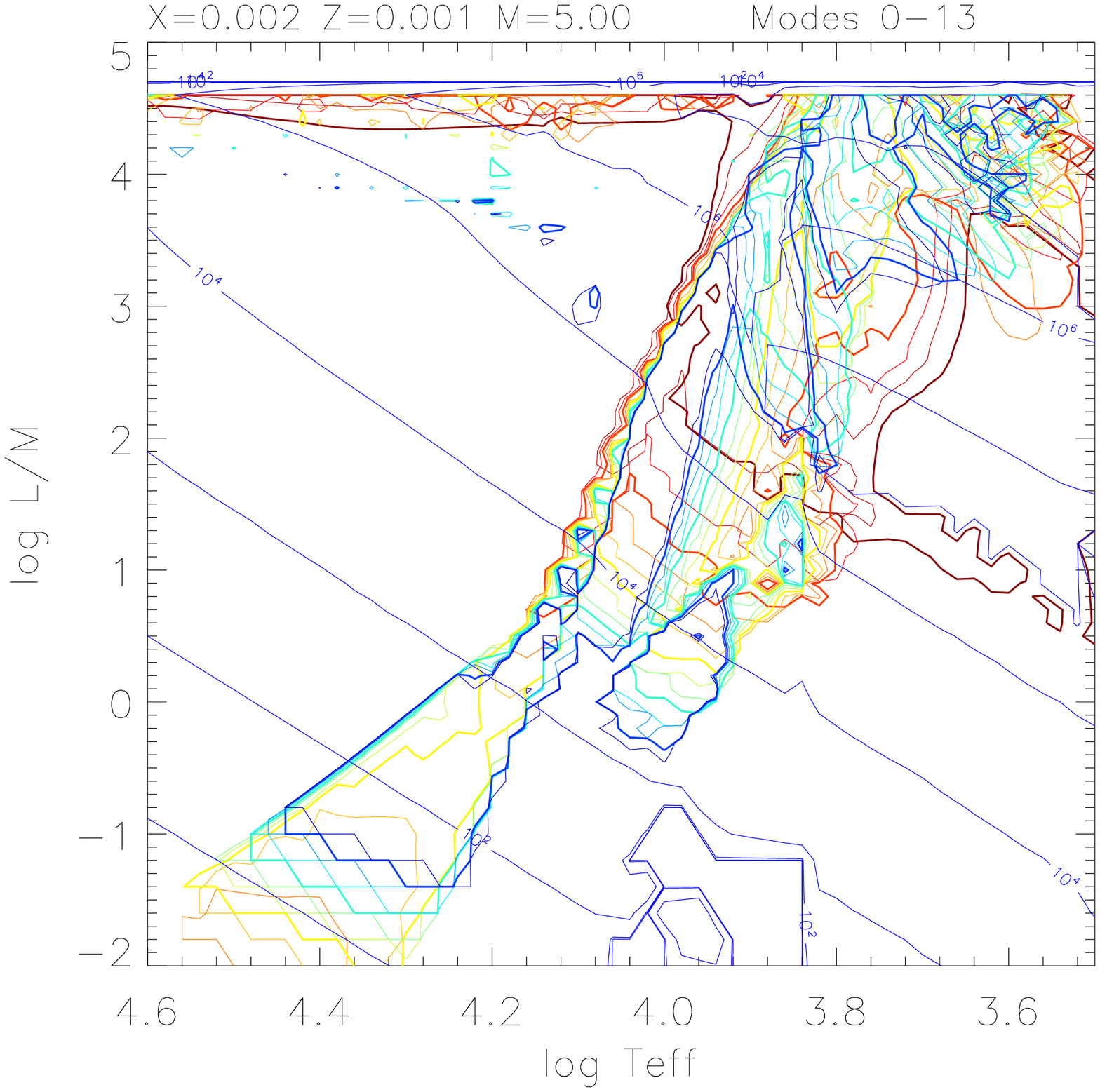,width=4.3cm,angle=0}
\caption[Unstable mode boundaries]
{As Fig.\,\ref{f:nmodes}, but showing the 
{\it boundaries} for individual radial modes as coloured contours, with the darkest red representing
the boundary of the fundamental ($n=0$) mode, with increasing higher orders ($n=1 - 10$) represented progressively by colours
of increasing frequency (orange, yellow, green, blue \ldots). 
Solid blue lines represent contours of equal fundamental radial-mode period in seconds spaced at decadal intervals. 
}
\label{f:harmonics}
\end{center}
\end{figure*}

Second,  the  instability boundaries for modes with $n = 0, 1$, or more
nodes, {\it i.e.} the instability boundaries for the fundamental
radial, and for the first and higher overtone pulsations are shown, also 
as a function of $(T_{\rm eff}, L/M)$ 
(Figs. ~\ref{f:harmonics} and \ref{f:px70} -- \ref{f:px002z001}). 

It is noted that many models may not represent any known stars ({\it e.g.} 
high $M$ models with very low $L/M$. ) Nevertheless, exploring such
models provides a systematic insight into the pulsation properties of stars 
in general.

\begin{figure*}
\begin{center}
\epsfig{file=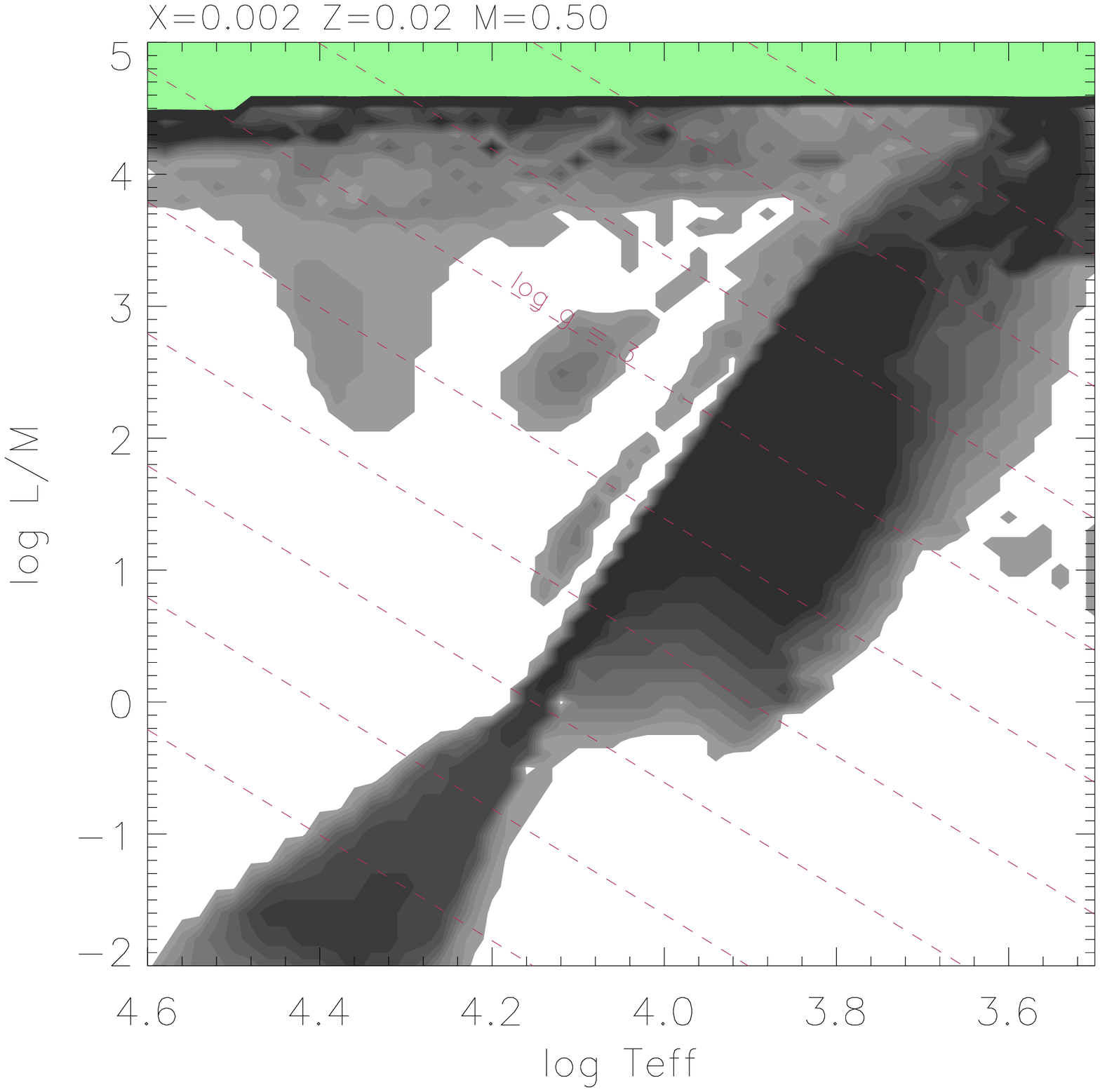,width=8.6cm,angle=0}
\epsfig{file=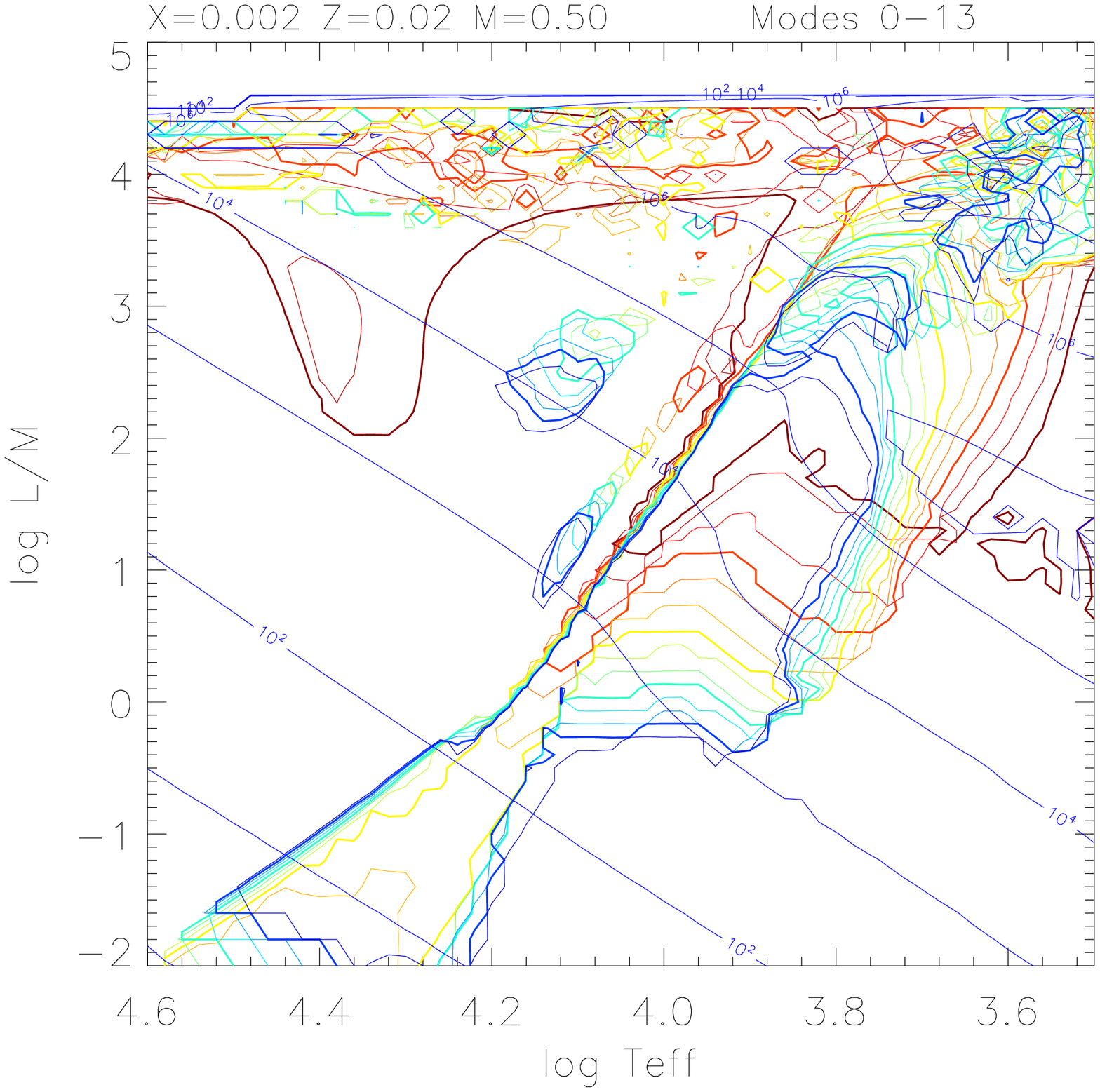,width=8.6cm,angle=0}
\caption[Enlargement]{As Figs.\,\ref{f:nmodes} and \ref{f:harmonics}, for composition $X=0.002, Z=0.02$, $M=0.50\Msolar$,
enlarged for clarity.}  
\label{f:enlarged}
\end{center}
\end{figure*}

\begin{figure*}
\begin{center}
\epsfig{file=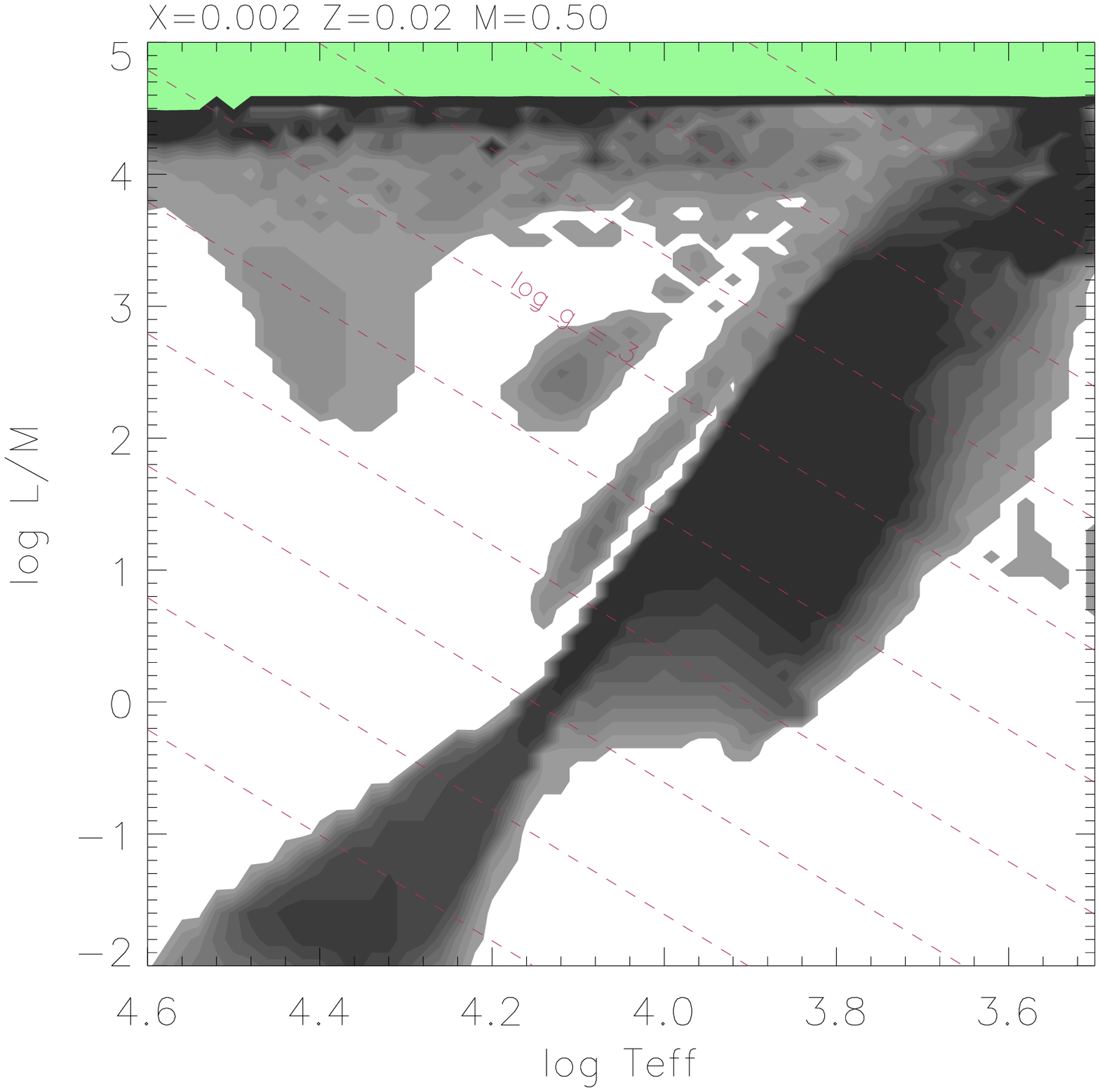,width=8.6cm,angle=0}
\epsfig{file=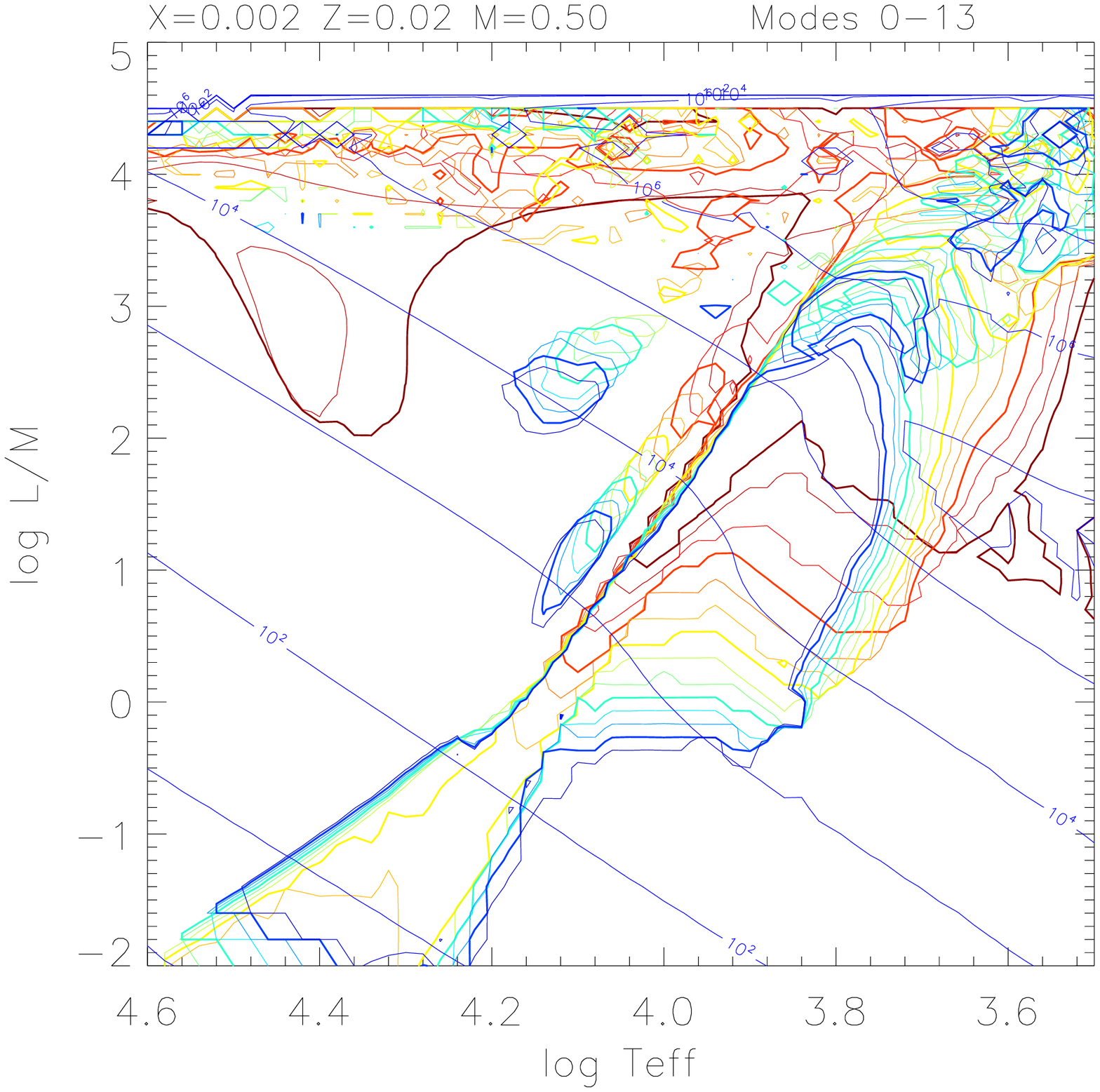,width=8.6cm,angle=0}
\caption[Opacity effects]{As Fig.\,\ref{f:enlarged} with OP opacities rather than OPAL95 opacities.}  
\label{f:opacity}
\end{center}
\end{figure*}

\section{Radial-Mode Instability}

The use of $L/M$ to parameterize the model grid exploits the
fact that the radial pulsation properties of stellar envelopes are
only slowly dependent on the mass of the star  
over a range from 0.2 to 50 \Msolar (cf. Fig.~\ref{f:nmodes}: top row). 
Nevertheless there are significant features to note; details may be verified from 
figures in Appendix~\ref{s:app1}. \\

\noindent At $X=0.70, Z=0.02$ (Figs.~\ref{f:nx70} and \ref{f:px70}): \\
i) The most familiar feature is the classical Cepheid instability strip running diagonally toward high $L/M$ and low $T_{\rm eff}$
at $\log L/M>0$. The principal driving is provided by the second ionization of singly-ionized helium (He$^+$).  \\  
ii) This is supplemented by an adjacent and parallel strip at lower 
$T_{\rm eff}$  in  which higher-order modes are excited (cf. Fig.~\ref{f:harmonics}: top row). 
This may be related to the solar-like oscillations detected by {\it Corot} and {\it Kepler} in many less luminous red giants \citep{miglio09,bedding10,huber10}
and marked as $\xi$\,Hya variables in Fig.~\ref{f:puls_hrd}.
Although the latter are generally thought to be excited stochastically by turbulence, 
the {\it order} of the p-modes detected in red giants decreases as the luminosity increases, which is consistent with 
our prediction from the kappa-mechanism.  
\citet{xiong07}  report from a non-adiabatic analysis including coupling between convection and pulsations  
that these modes are excited.  We hence find that evidence exists for 
a contribution of the kappa-mechanism from ionization of H and He$^0$ to the driving of small-amplitude oscillations in red giants.\\
iii) The low $T_{\rm eff}$ strip extends to very low $L/M$ values and is responsible in low-mass main-sequence stars 
for $\delta$ Scuti variables. The principal driving comes from H and He$^0$ ionisation,  although the 
high overtone pulsations in hotter $\delta$ Sct stars are probably  associated with He$^+$ ionization 
({\it i.e.} the classical instability strip), which extends to the main sequence with $\log g < \sim 5$, $\log L/M > \sim 0$.\\
iv) At high $L/M$, models show instability in one or more modes; these
are the so-called {\it strange} modes \citep{gautschy90,saio98b}. 
The general confusion in the mode boundary diagram  (Fig.~\ref{f:px70})
is accounted for by a breakdown in the strict 1-1 correspondence between mode 
eigenfrequency and node number for strange modes\footnote{Strange modes in high $L/M$ 
stars are essentially as defined by \citet{gautschy90,saio98b}, and first identified by \citet{wood76}. 
Other modes of somewhat different character have also been described as {\it strange}, {\it e.g.} 
by \citet{buchler97,buchler01,smolec16}. The current analysis makes no distinction between {\it strange} 
and {\it normal} modes; it only identifies which modes are stable or unstable.}  
$\alpha$ Cygni variables 
can be explained by strange modes, if considerable mass has been lost \citep{saio13}. \\
v) At $\log L/M>3$ and $\log T_{\rm eff}\approx 4.4$ there is a 
weak instability finger caused by iron-group opacities at temperatures $T\approx 2\times10^5$\,K. 
This can be identified with $\beta$ Cepheid variables on the upper main sequence. The shape of this finger
is sensitive to mass, and also to the iron and nickel abundances \citep{jeffery07}. \\

\noindent At $X=0.30, Z=0.02$ (Figs.~\ref{f:nx30} and \ref{f:px30}), 
a modest reduction in hydrogen abundance has the effect of increasing the driving effect of helium
in the classicial instability strip and iron in the `Z-bump' instability finger. The chief consequences are:\\
i) As the mean molecular weight in the envelope increases, the He$^+$ instability strip shifts to
higher $T_{\rm eff}$. With the excitation of higher-order radial modes, the strip also extends to 
lower $L/M$.  \\
ii) For $\log L/M < -1$ the high-order strip narrows significantly. \\
iii) With  increasing contrast between iron-group and hydrogen opacity, the  `Z-bump' finger 
becomes significantly stronger. \\
iv) At high mass, `Z-bump' excited fundamental modes extend to $\log T_{\rm eff}\approx 4.0$,
forming a `spur' to the `Z-bump' finger. Close inspection of these modes show an unrealistically 
large amplitude propagating deep into the interior (to fractional mass $m/M < 0.001$);  
we do not consider these to be real. 
 \\

\noindent At $X=0.10, Z=0.02$ (Figs.~\ref{f:nx10} and \ref{f:px10}), 
the consequences of reducing the hydrogen abundance seen at $X=0.3$ continue. 
In addition: \\
i) The width of the high-order He$^+$ instability strip at low $L/M$ increases. As mass increases, the 
region becomes fragmented.  \\
ii)  For $\log L/M < -1$, there are virtually no unstable modes in the high-order strip. \\
iii) The Z-bump finger extends to $\log L/M>2.0$; the high-mass spur persists, but see above.  \\

\begin{figure}
        \centering
                \includegraphics[width=0.47\textwidth]{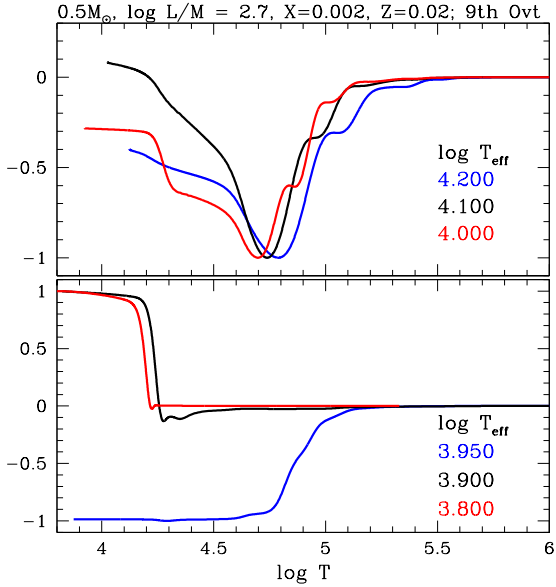}
\caption{Work integrals $W$ as a function of temperature for the 9th overtone in the neighbourhood 
of the instability island at $X=0.002$, $Z=0.02$, with  $M=0.5\Msolar$, $\log L/M = 2.7$, and  $\log T_{\rm eff}$ as shown. 
Positive values of $W$ at the surface indicate instability. At $\log T_{\rm eff} \leq 3.9$, driving is due to He$^0$ ionization -- essentially the classical Cepheid mechanism.  
For $\log T_{\rm eff} = 3.95$, 4.0 and 4.2, the mode is stable.  For  $\log T_{\rm eff} = 4.1$, the 
mode is unstable due to driving by both  He$^0$ and He$^+$ ionization.  }
        \label{f:work1}
\end{figure}

\noindent At $X=0.002, Z=0.02$ (Figs.~\ref{f:nx002} and \ref{f:px002}), 
the consequences of reducing the hydrogen abundance seen at $X=0.1$ continue. 
In addition: \\
i) At $M<5\Msolar$, an additional narrow strip of high-overtone instability is seen immediately to the blue 
of the He$^+$ instability strip at $\log L/M\approx1$. This strip has been identified with pulsations observed in
J0247-25B \citep{jeffery13b}. \\
ii) The Z-bump finger is very strong and broad, but only low-order ($n<4$) modes are excited; the high mass $n=0$ Z-bump spur persists. \\
iii) The width of the high-order He$^+$ instability strip at low $L/M$ remains large. At low mass, the models give evidence that radial modes in 
some DB white dwarfs may be unstable, as previously shown by \citet{kawaler93}.
iv) At $M<2\Msolar$, an island of high-overtone instability is seen around $\log T_{\rm eff}\approx 4.1$ and $\log L/M\approx2 - 3$ (Fig.~\ref{f:enlarged}),
diminishing in extent with increasing mass. This island has appeared in previous calculations \citep{jeffery13b}, but only here
identified as a persistent feature over a range of masses. Driving is due to a combination of He$^0$ and He$^+$ ionization (Fig.~\ref{f:work1}).
Since pulsation instability boundaries are sensitive to details of the precise chemical mixture and to the overall opacity calculation, 
a few of these models were recalculated using OP opacities \citep{badnell05} instread of OPAL95 opacities  \citep{iglesias96}. Minor differences 
may be seen (Fig.\,\ref{f:opacity}), but the new instability island persists and the differences are small compared with the large variations in $X$ which are of primary interest.  
{\it We suggest that stars having these characteristics, should they actually occur in nature, would be good candidates to 
show high-order radial (or non-radial) p-mode pulsations.} \\

\noindent At $X=0.70, Z=0.001$ (Figs.~\ref{f:nx70z001} and \ref{f:px70z001}), 
the principal consequence of reducing the metallicity is that the Z-bump finger disappears.  
In addition, the classical instability strip, and the high-overtone strip to the red are both narrowed. \\

\noindent At $X=0.002, Z=0.001$ (Figs.~\ref{f:nx002z001} and \ref{f:px002z001}), 
the  high-overtone instability island identified at $X=0.002, Z=0.02$ persists but, as there were substantial difficulties 
computing envelope models with $M<2\Msolar$ at this composition for other ranges of $L/M$ and $T_{\rm eff}$,
these models are not shown.  \\

\noindent At $X=0.70$, $Z=0.02$ and $\log L/M \leq0$  (Fig.~\ref{f:nx70}), 
 unstable radial modes  initially  form a single narrow strip, extending the classical Cepheid instability strip 
to very low $L/M$. 
With  $X=0.30$ (Fig.~\ref{f:nx30}),  a second strip develops substantially
to the blue of the first. Again, this is the low $L/M$ extension of the He$^+$-driven strip already seen. We note that 
the excited modes are dominated by high-order modes (Fig.~\ref{f:px30}). 
At  $X=0.10$ both strips broaden to form a single region with a complicated mode structure 
 (Figs.~\ref{f:nx10},\ref{f:px10}). 
Finally, at $X=0.002$, the redward strip stabilizies, to leave only a broad blue instability strip  (Fig.~\ref{f:px002}).
Similar behaviour is replicated at  $Z=0.001$. 
Predictions of radial instability in both DA and DB  white dwarfs are well established \citep{saio83b,kawaler93}. 
However there has so far been no  successful detection  of p-mode pulsations in any white dwarfs \citep{silvotti11,kilkenny14}.  

\section{Conclusion}

We have made an extensive survey of the stability against {\it radial} pulsations for the  envelopes of stars 
having masses in the range $0.2 -- 50 \Msolar$ and  hydrogen abundances (by mass fraction) 
 from $X=0.70$ to 0.002,  considering both metal-rich ($Z=0.02$) and metal-poor ($Z=0.001$) mixtures. 
The grid of models ranges in effective temperature from $T_{\rm eff} = 3\,000 -  40\,000$\,K, and in luminosity-to-mass 
ratio from $L/M = 0.01 - 100,000$ (in solar units), covering most of parameter space occupied by stars, 
excepting only the hottest and coolest supergiants, the hottest subdwarfs, the most massive white dwarfs and 
the coolest supergiants and dwarfs. By considering overtones up to $n=16$, we identify nearly all stars likely to
be unstable to p-mode oscillations driven by the $\kappa-$mechanism. 

{\it We demonstrate that the Hertszprung-Russell diagram for pulsation instability} expressed as a plot of
$L/M$ versus $T_{\rm eff}$ {\it is only slowly variant with mass $M$,} but is {\it much more sensitive to composition}, 
especially the hydrogen abundance since the latter normally acts to damp pulsations. 
 
Within a single computational framework, {\it we recover all hitherto known regions of
radial and non-radial p-mode instability} due to the $\kappa-$mechanism and/or strange modes. 
The detailed boundaries are likely to vary for specific cases owing to other properties such as internal composition
gradients and long-term evolution effects. 

{\it We identify one new region of pulsation instability} for low-mass hydrogen-deficient stars with 
$\log L/M \approx 2 - 3$ and $\log T_{\rm eff} \approx 4.0 - 4.2$;  $\kappa$-mechanism driving is 
by He$^0$ and He$^+$ ionization. No stars are currently known to exhibit these pulsations. 

In addition, we conclude that solar-like oscillations in red giants may be at least partially driven by the $\kappa-$mechanism, 
supporting \citet{xiong07}. To investigate this and other questions, a more detailed study of the relative r\^oles of convection 
and the $\kappa$-mechanism in exciting pulsations across the H-R diagram will follow.

\section*{Acknowledgments}
The Armagh Observatory is funded by direct grant from the Northern Ireland 
Department of Culture, Arts and Leisure. 

\bibliographystyle{mnras}
\bibliography{ehe}

\appendix
\renewcommand\thefigure{A.\arabic{figure}} 
\renewcommand\thetable{A.\arabic{table}} 
%\counterwithin{figure}{section}

\section[]{Radial Pulsation Model Grids}
\label{s:app1}
\begin{figure*}
\begin{center}
\epsfig{file=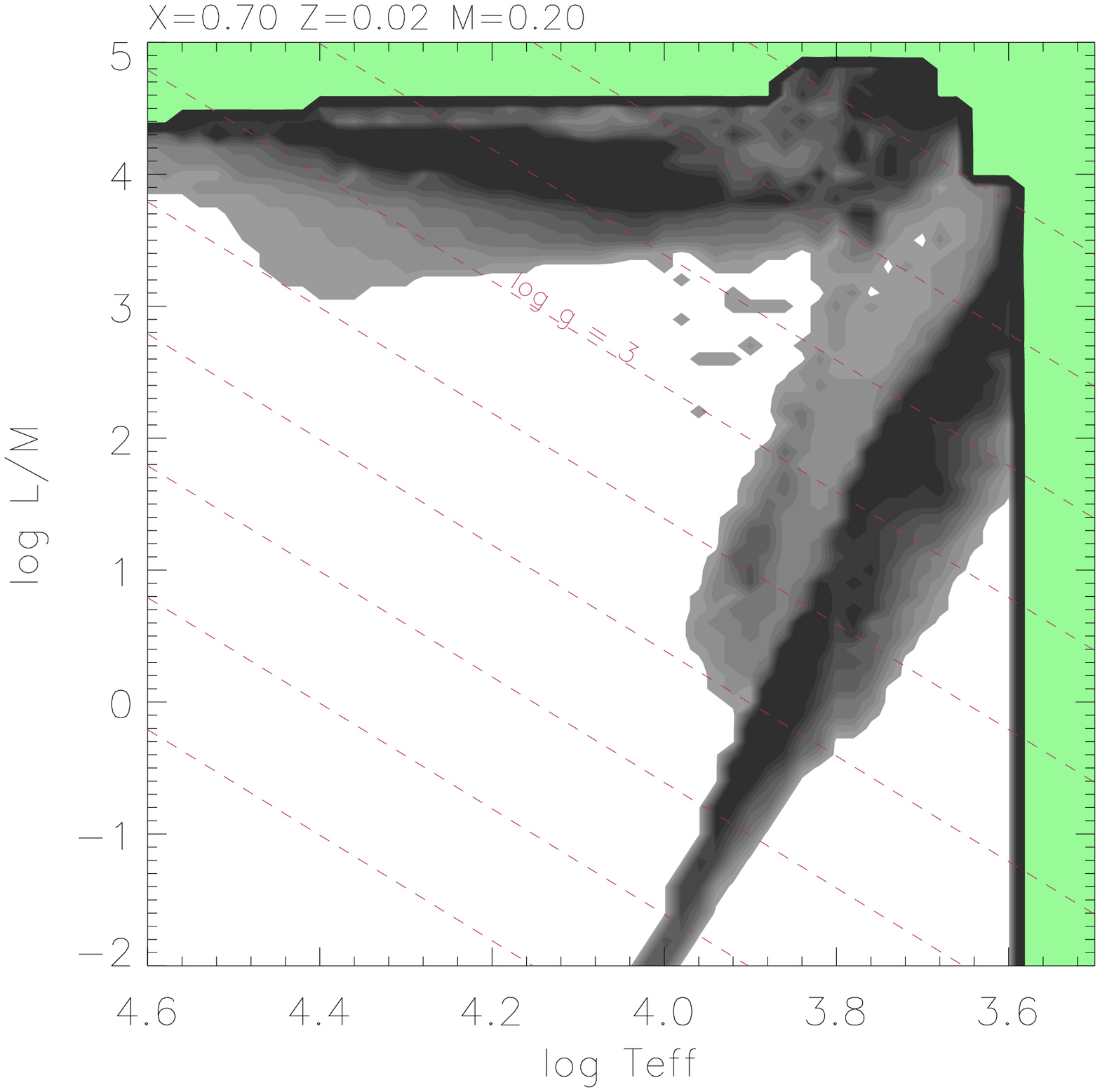,width=4.3cm,angle=0}
\epsfig{file=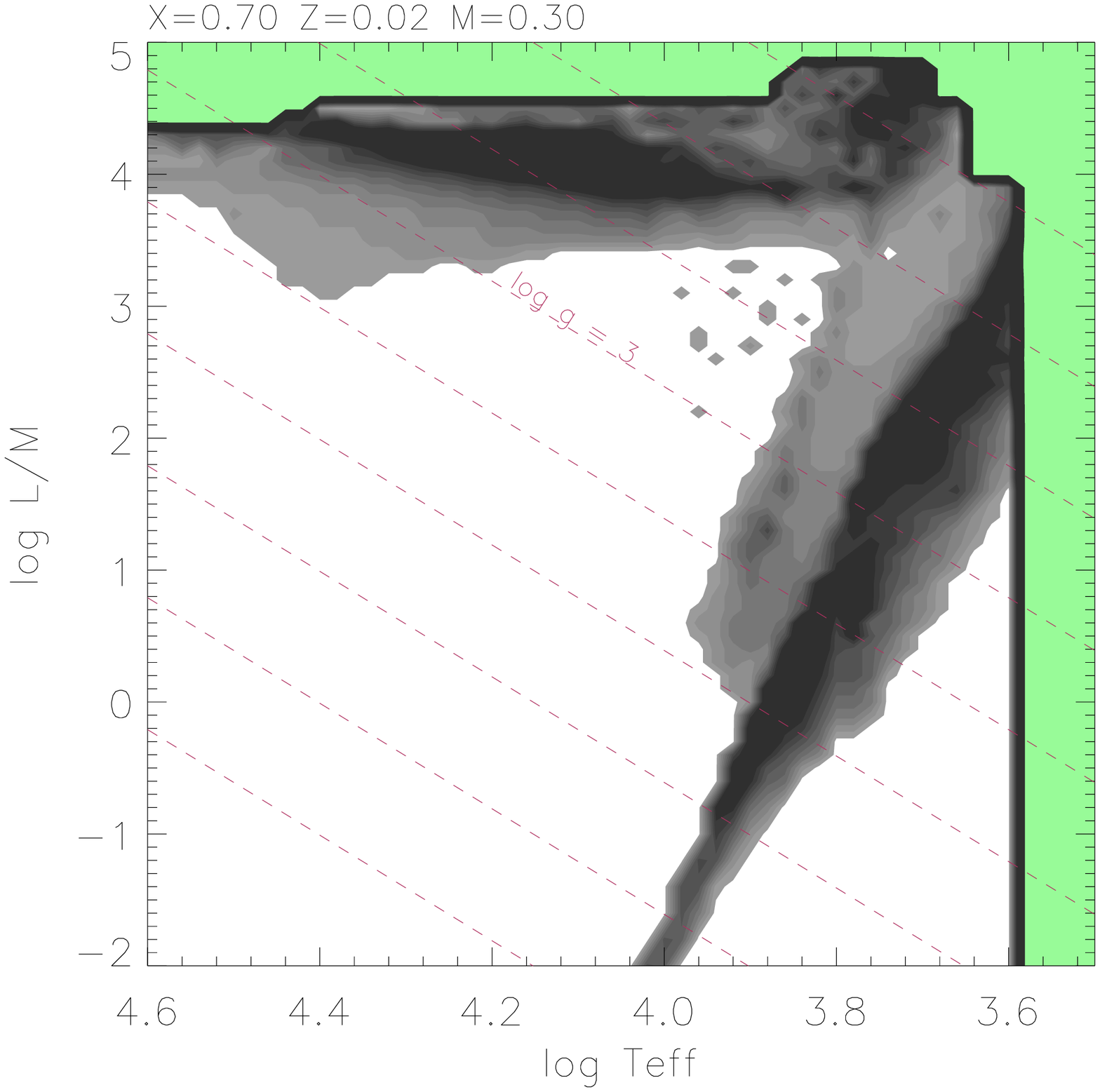,width=4.3cm,angle=0}
\epsfig{file=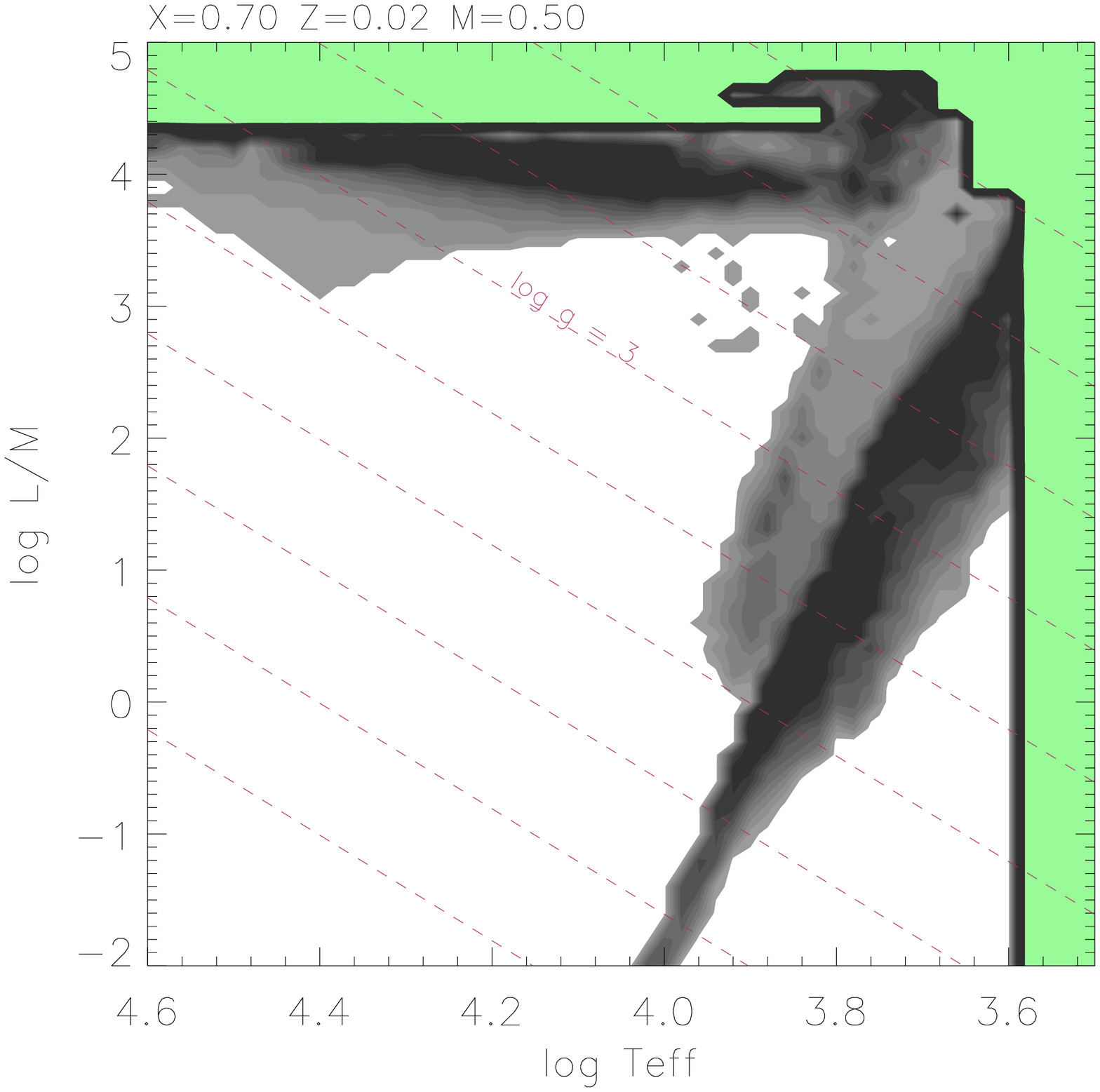,width=4.3cm,angle=0}
\epsfig{file=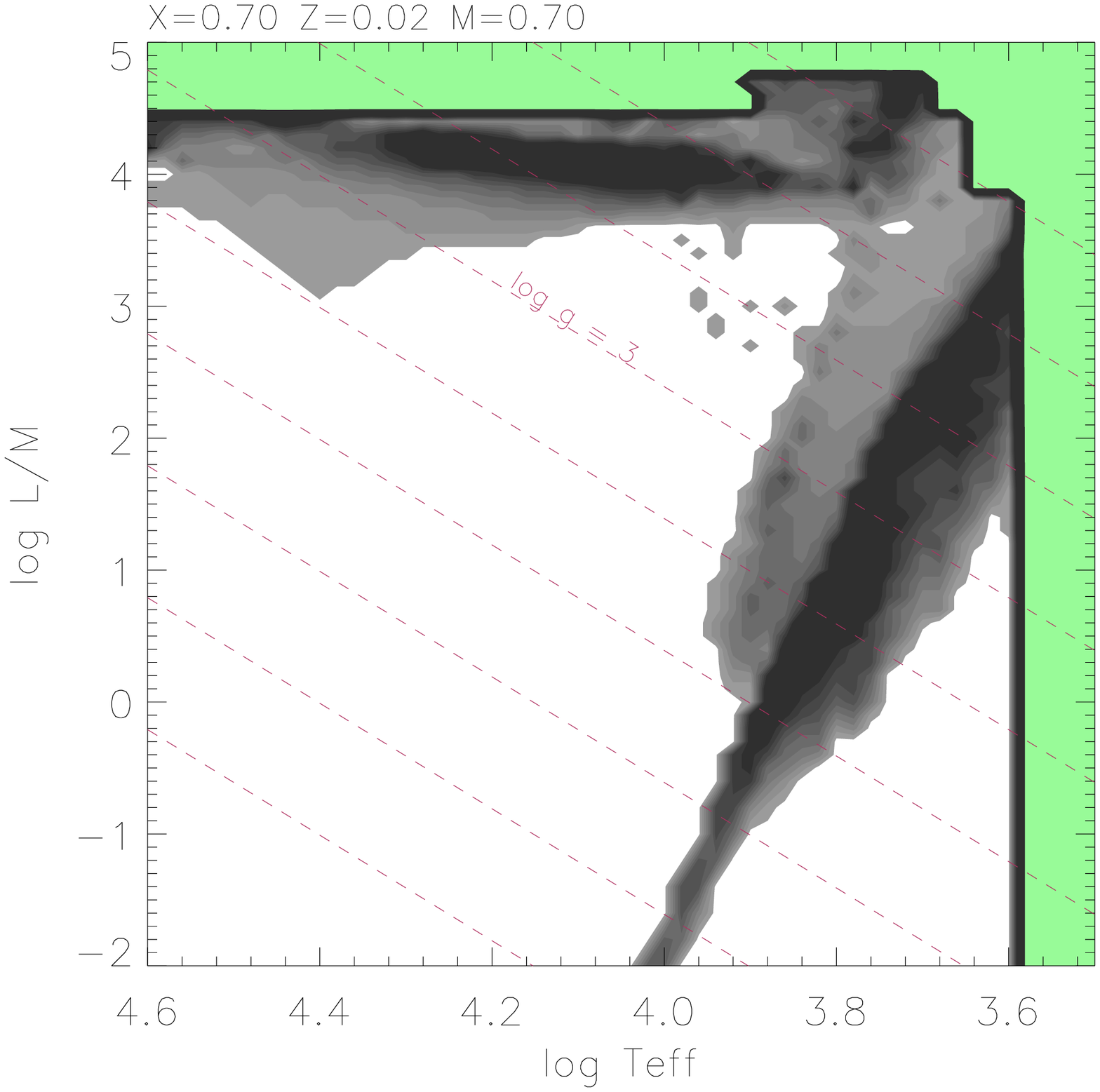,width=4.3cm,angle=0}\\
\epsfig{file=figs/nmodes_x70z02m01.0_00_opal.eps,width=4.3cm,angle=0}
\epsfig{file=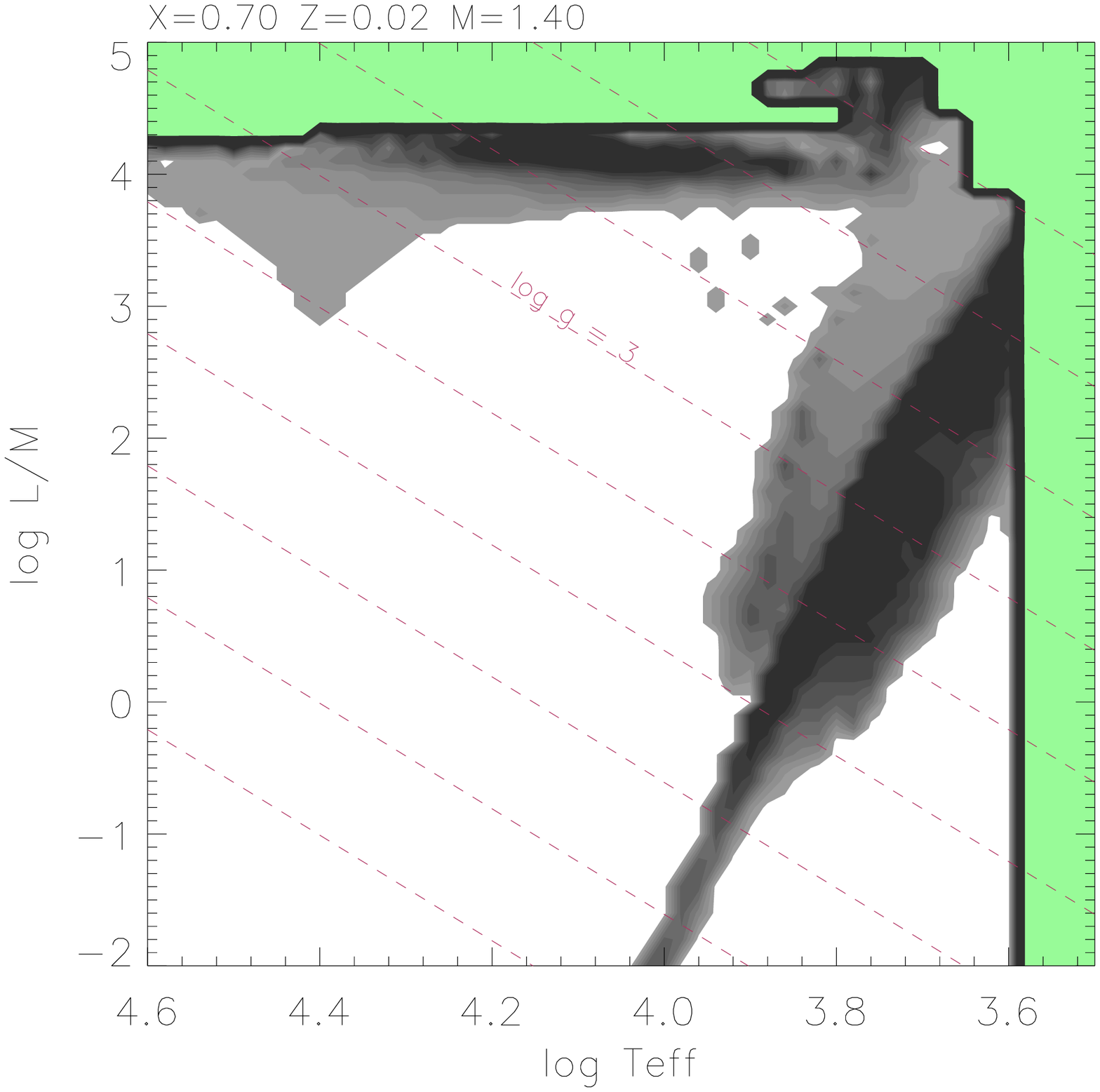,width=4.3cm,angle=0}
\epsfig{file=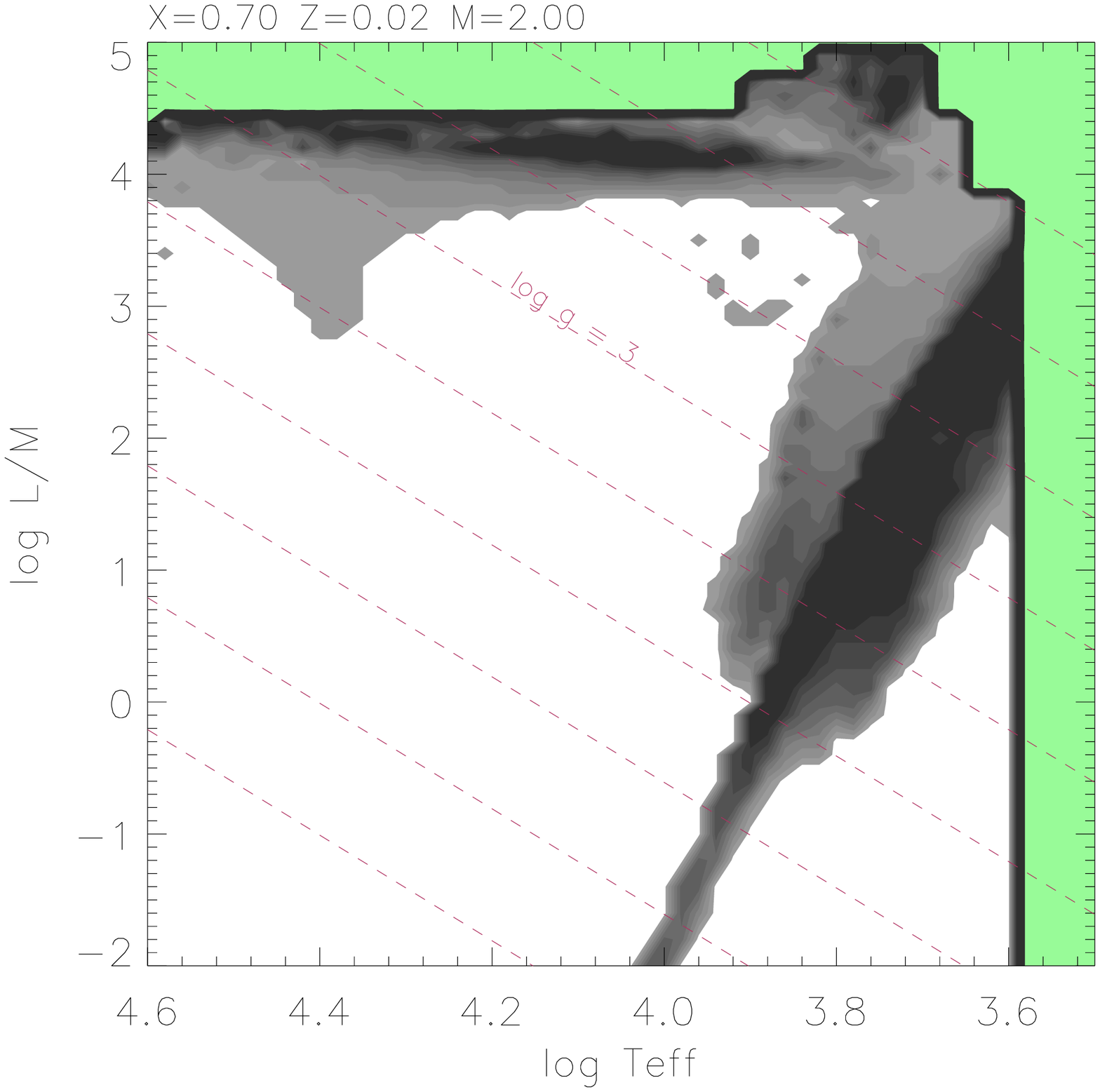,width=4.3cm,angle=0}
\epsfig{file=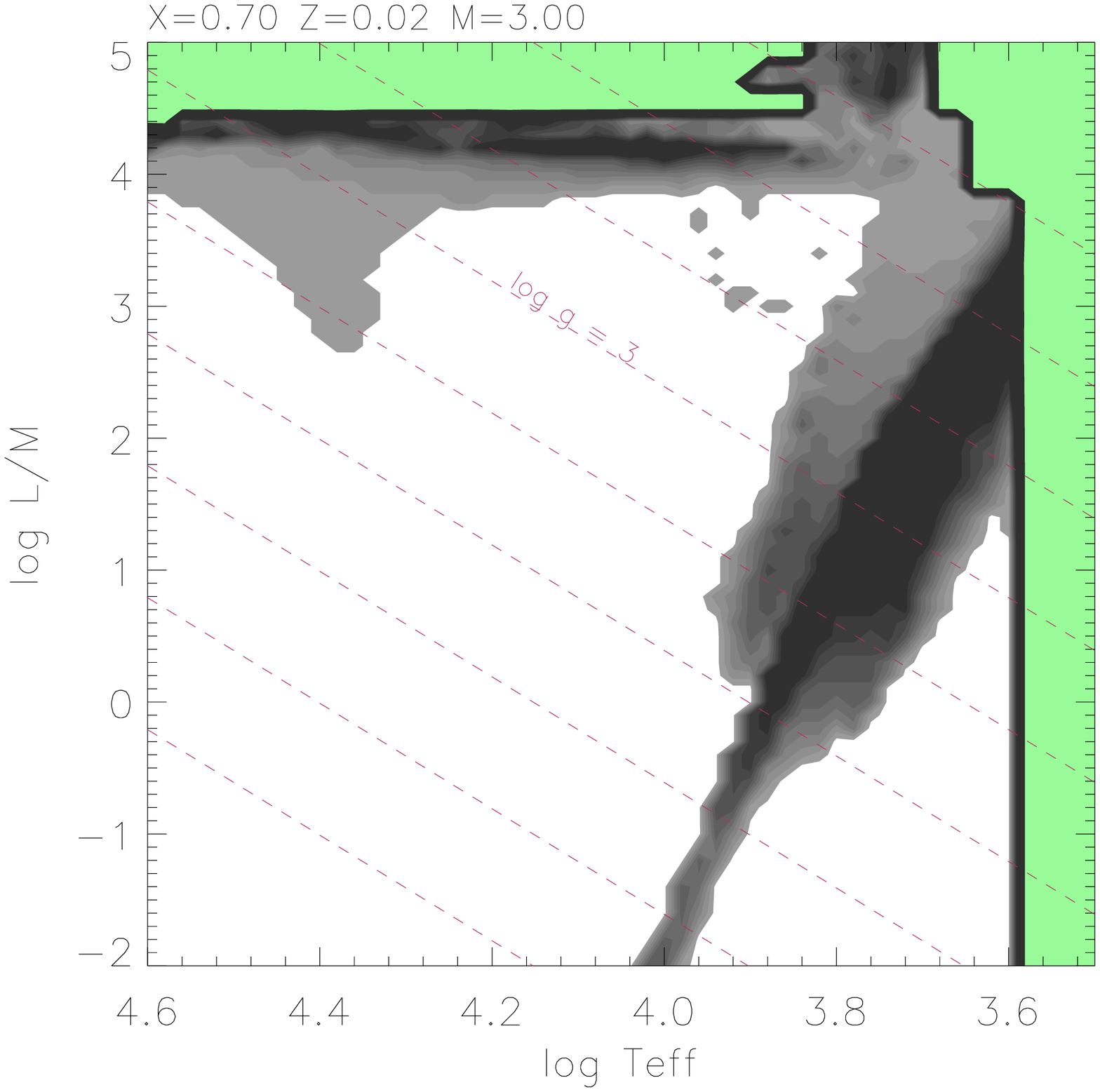,width=4.3cm,angle=0}\\
\epsfig{file=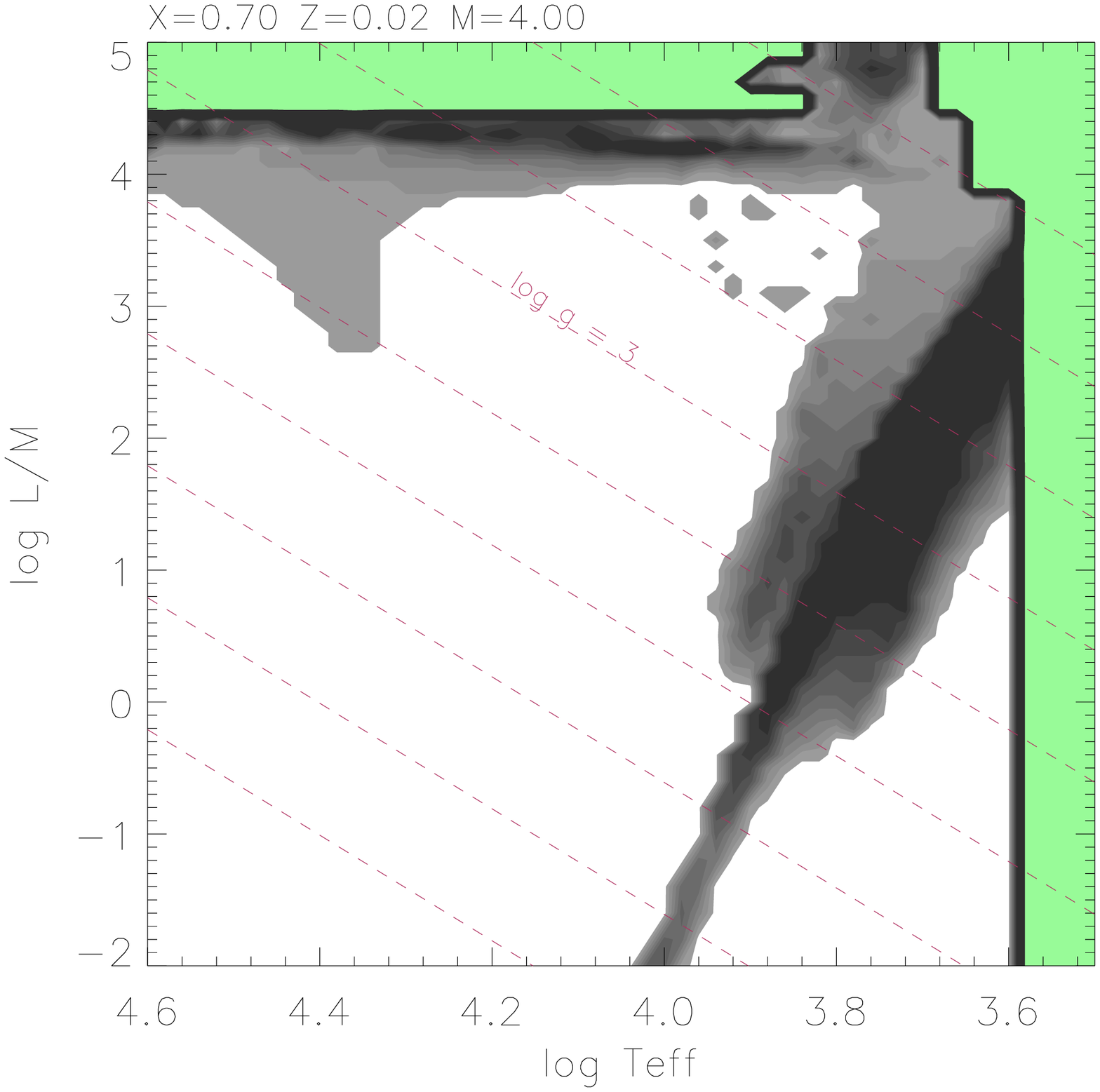,width=4.3cm,angle=0}
\epsfig{file=figs/nmodes_x70z02m05.0_00_opal.eps,width=4.3cm,angle=0}
\epsfig{file=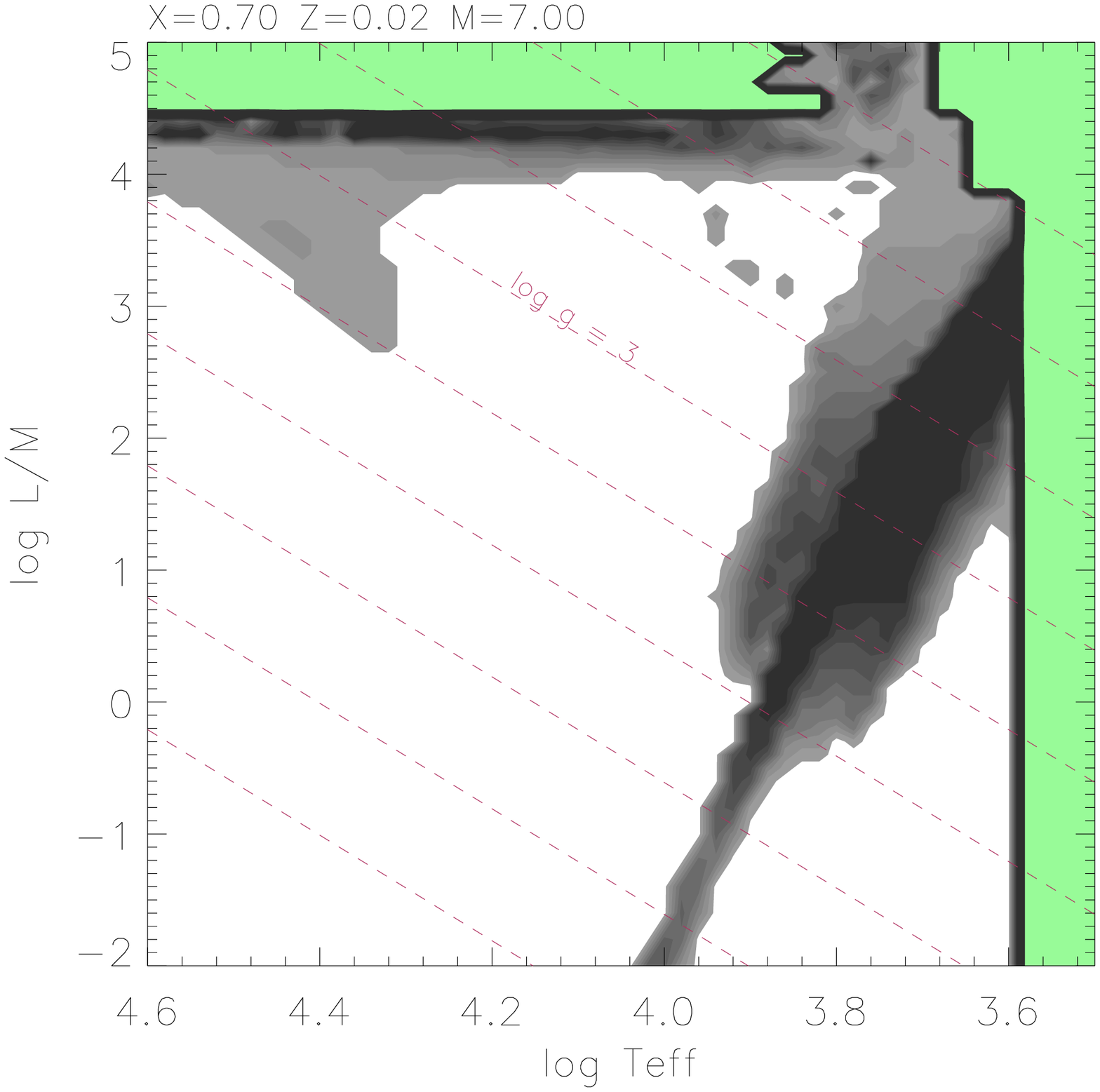,width=4.3cm,angle=0}
\epsfig{file=figs/nmodes_x70z02m10.0_00_opal.eps,width=4.3cm,angle=0}\\
\epsfig{file=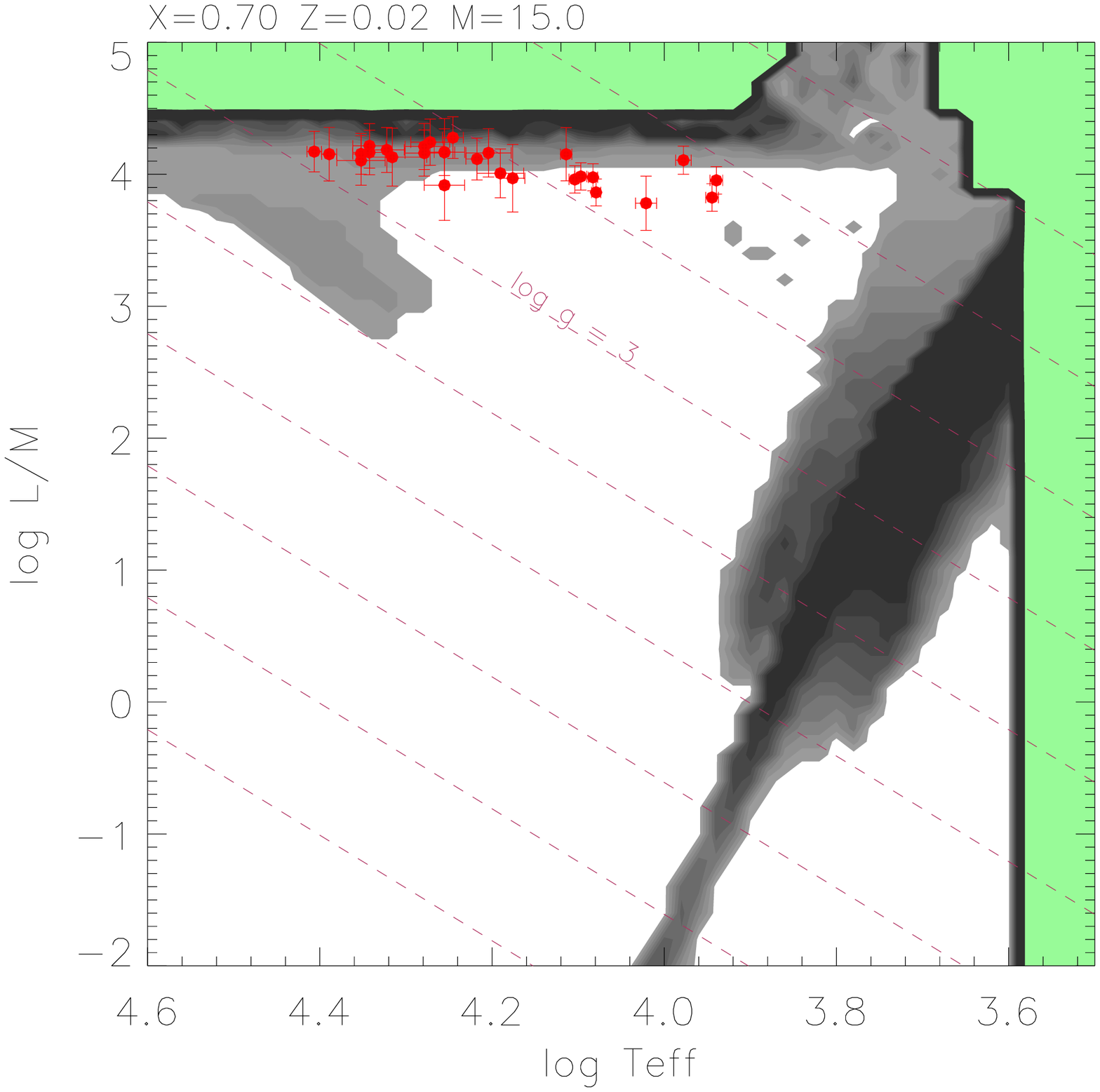,width=4.3cm,angle=0}
\epsfig{file=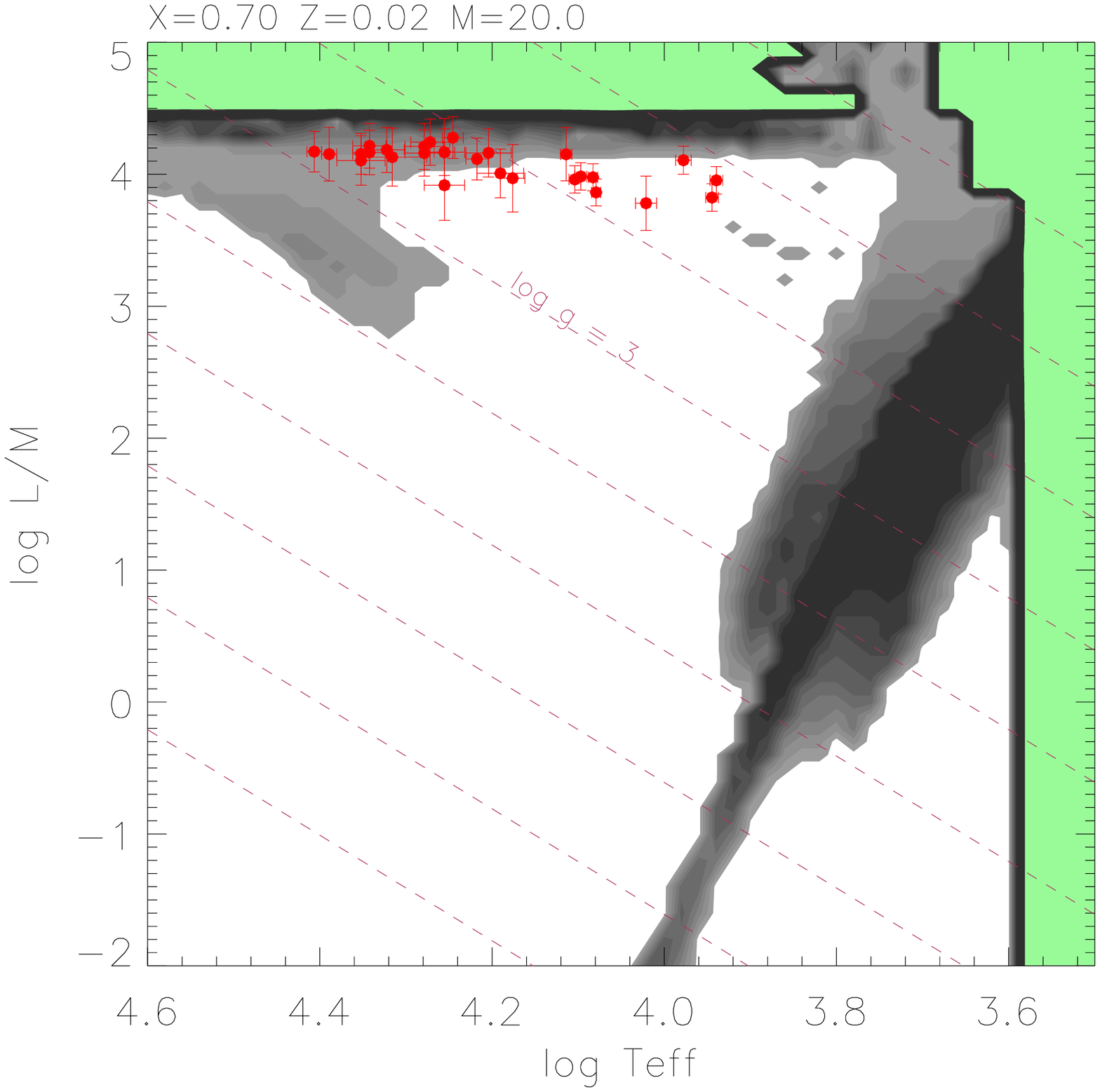,width=4.3cm,angle=0}
\epsfig{file=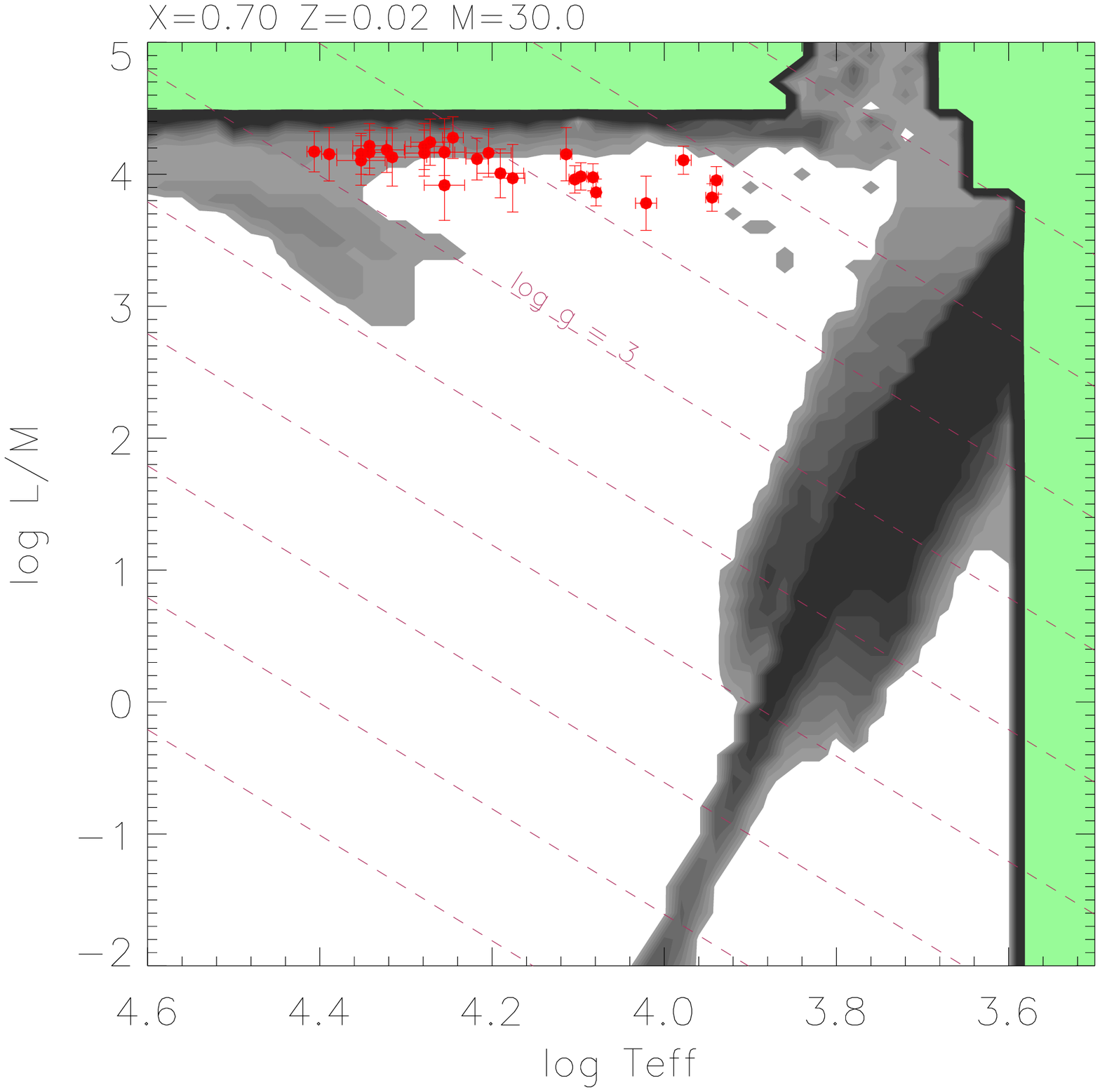,width=4.3cm,angle=0}
\epsfig{file=figs/nmodes_x70z02m50.0_00_opal.eps,width=4.3cm,angle=0}
\caption[Unstable modes: $X=0.70, Z=0.02$]
{Unstable pulsation modes in stars with homogeneous
envelopes with hydrogen content $X=0.70, Z=0.02$,  
 and OPAL opacities, and mass $0.20 < M/\Msolar < 50$, as labelled. 
The number of unstable radial modes is represented by grey scale contours, with the lightest shade marking the instability boundary
(one unstable mode), and the darkest shade representing ten or more more unstable modes. 
 Broken (maroon online) diagonal lines represent contours  of constant surface gravity at  $\log g = 8, 7, 6, \ldots, 1$. 
Models with $X=0.7$ and $\log T_{\rm eff} < 3.6$ were excluded. 
Red symbols with error bars  shown on selected high-mass panels  
represent the observed parameters of pulsating $\alpha$ Cygni variables \citep{crowther06,searle08,firnstein12}.
}
\label{f:nx70}
\end{center}
\end{figure*}

\begin{figure*}
\begin{center}
\epsfig{file=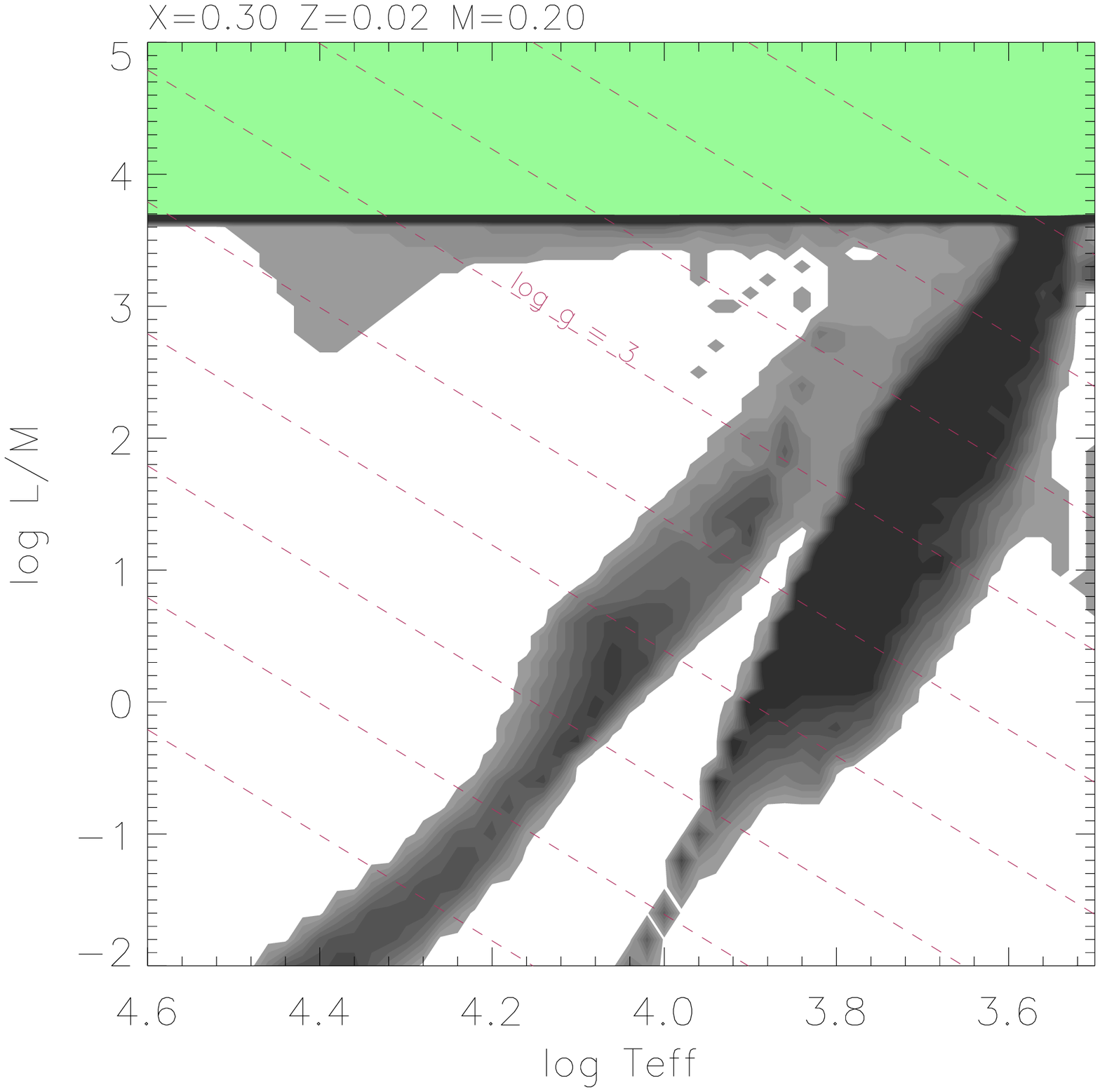,width=4.3cm,angle=0}
\epsfig{file=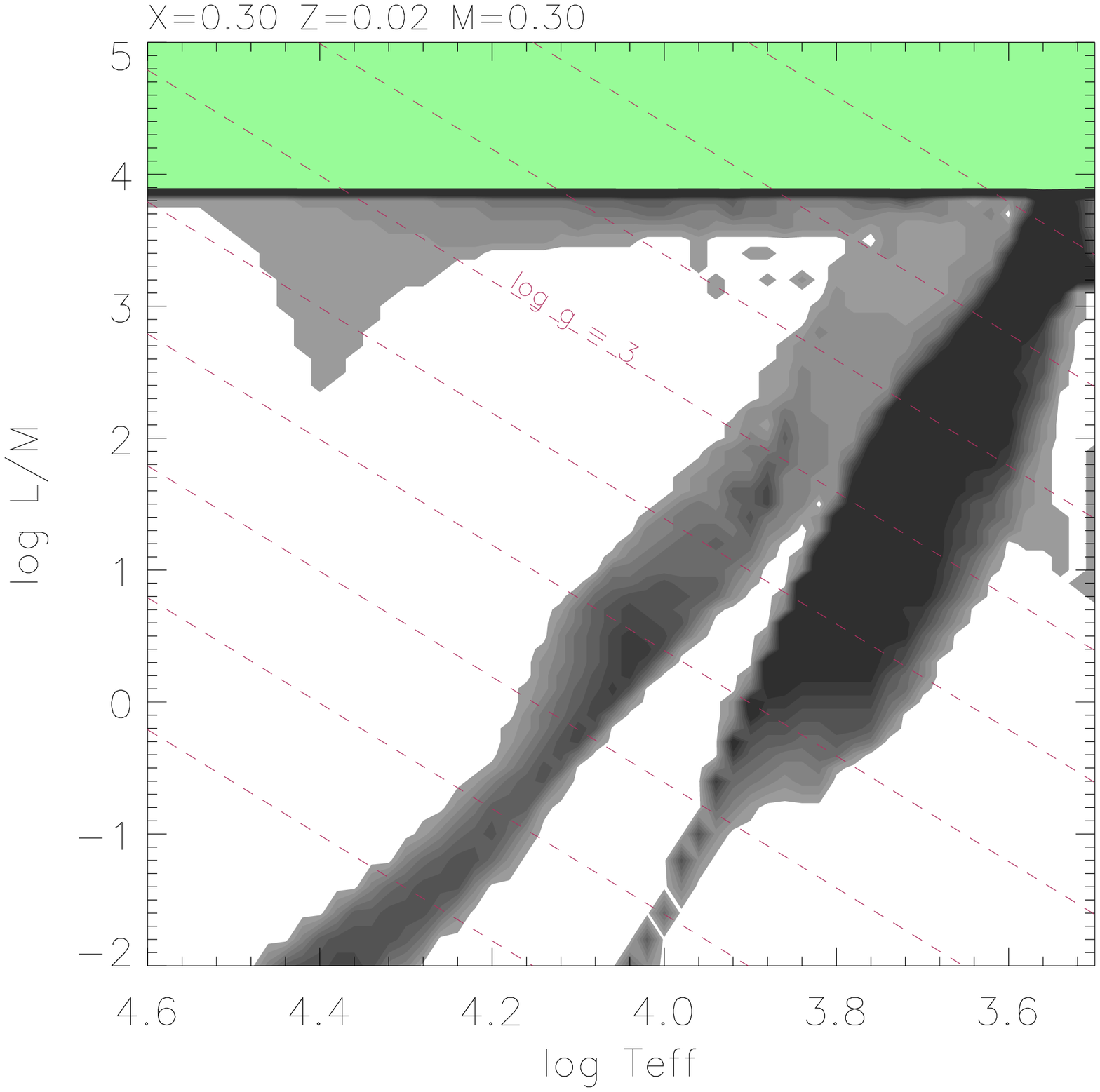,width=4.3cm,angle=0}
\epsfig{file=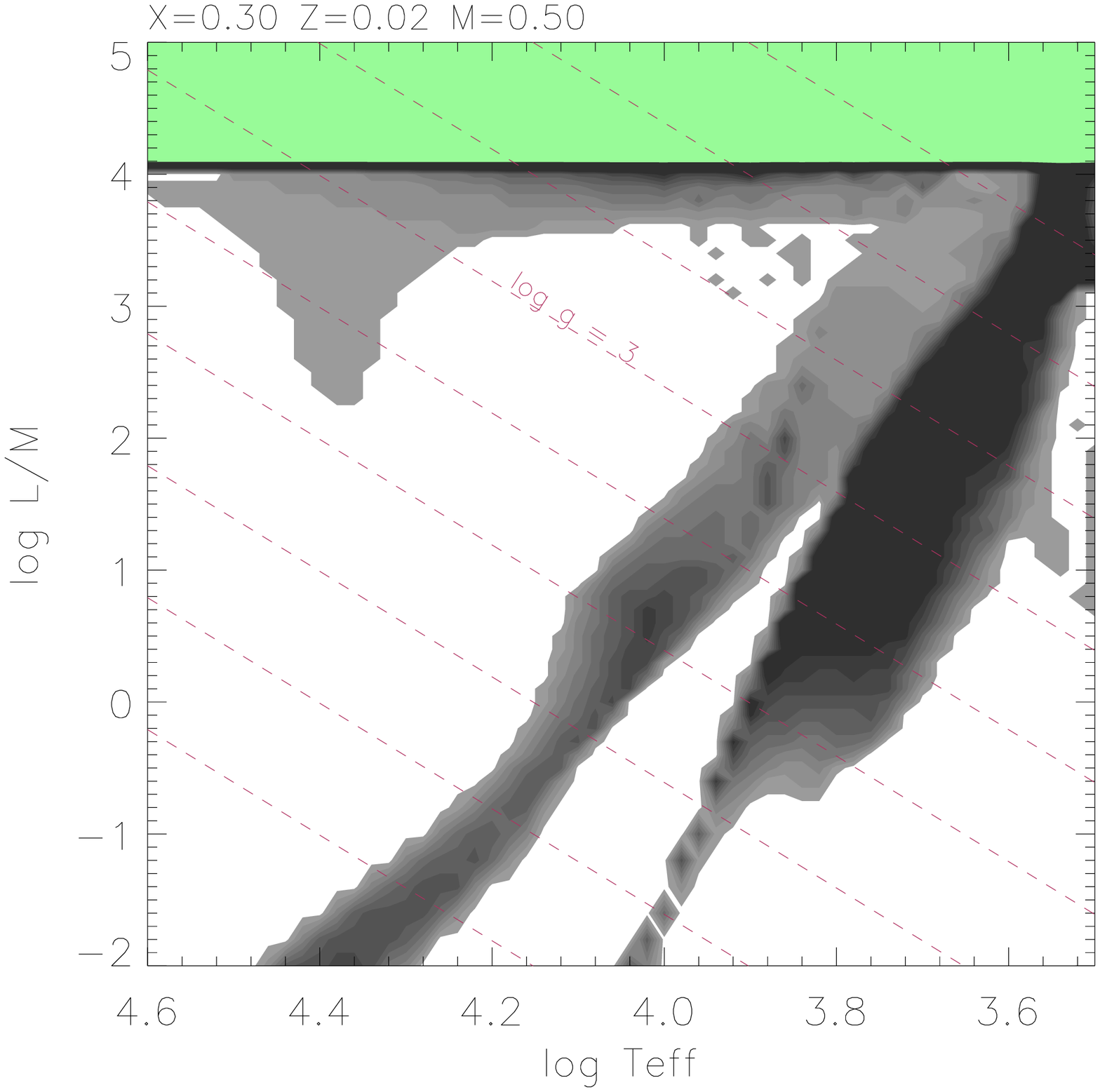,width=4.3cm,angle=0}
\epsfig{file=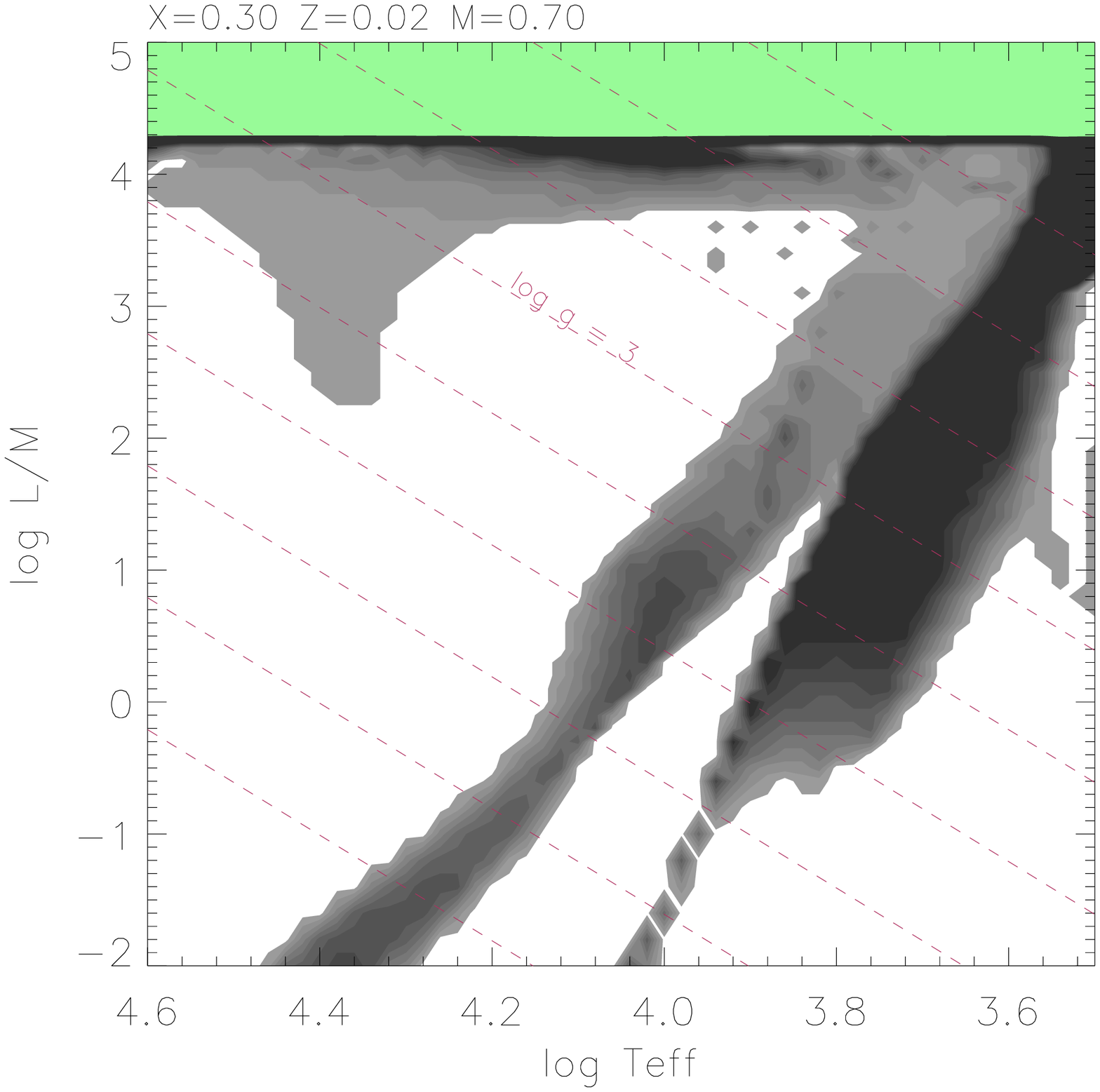,width=4.3cm,angle=0}\\
\epsfig{file=figs/nmodes_x30z02m01.0_00_opal.eps,width=4.3cm,angle=0}
\epsfig{file=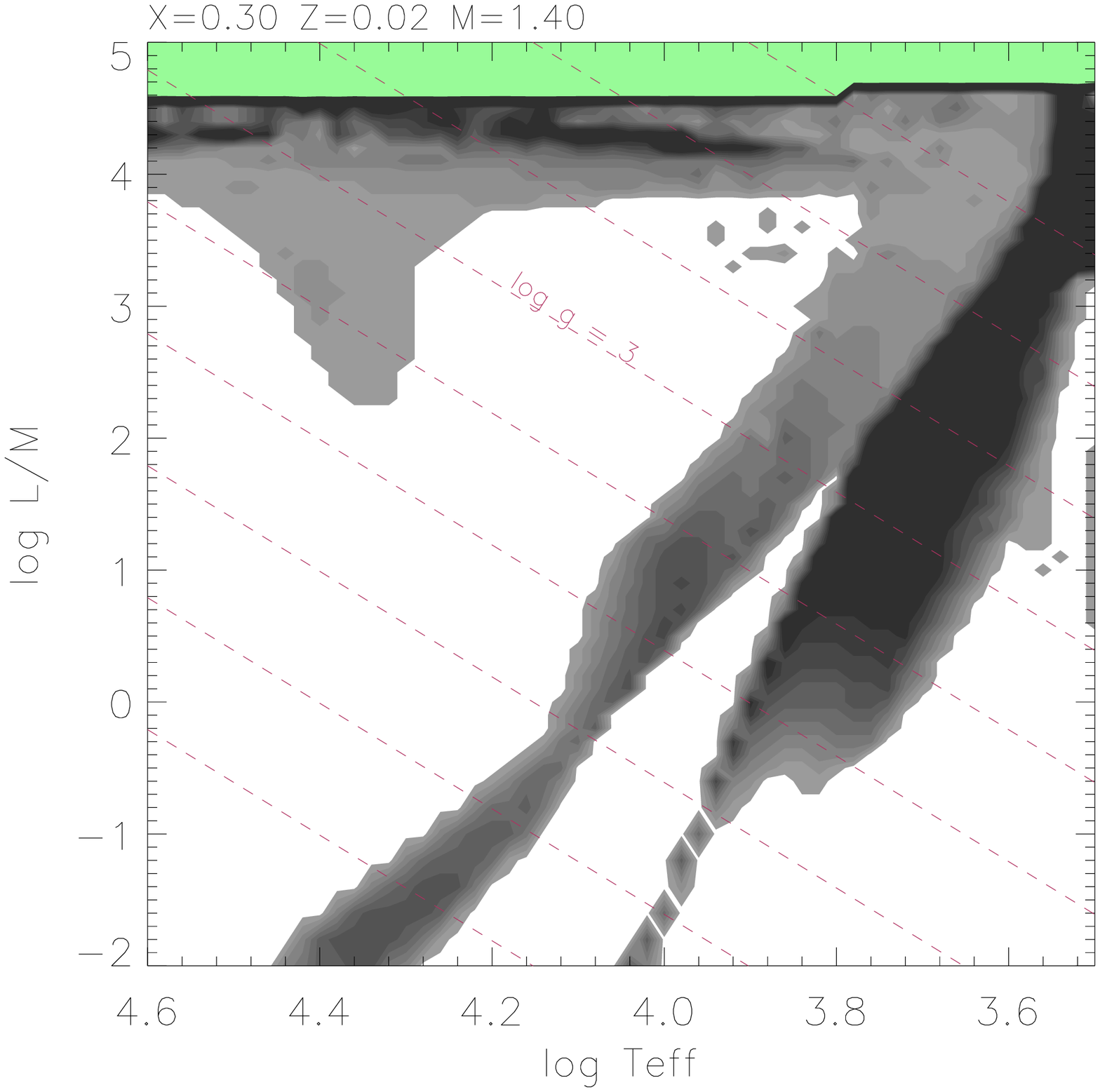,width=4.3cm,angle=0}
\epsfig{file=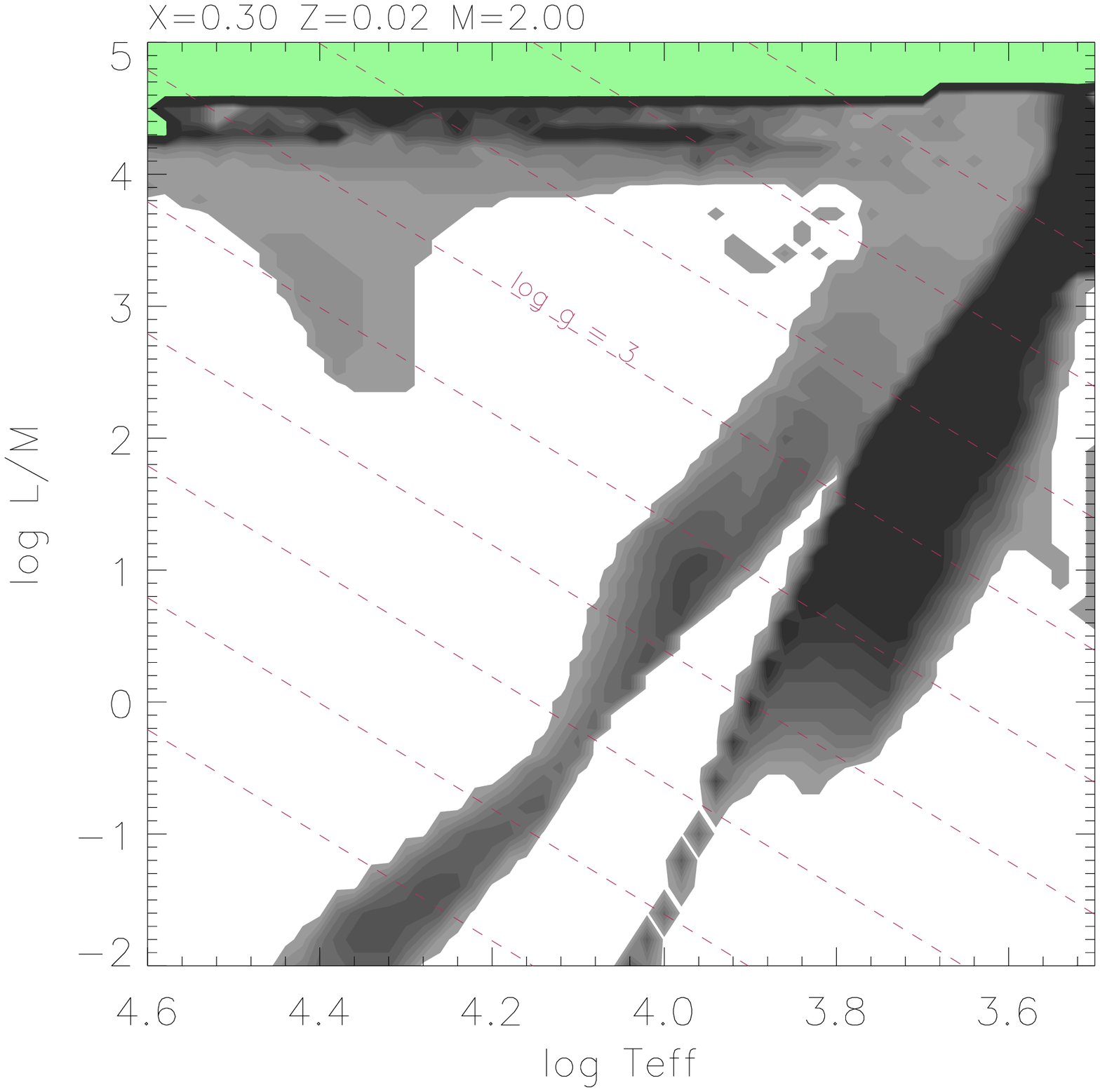,width=4.3cm,angle=0}
\epsfig{file=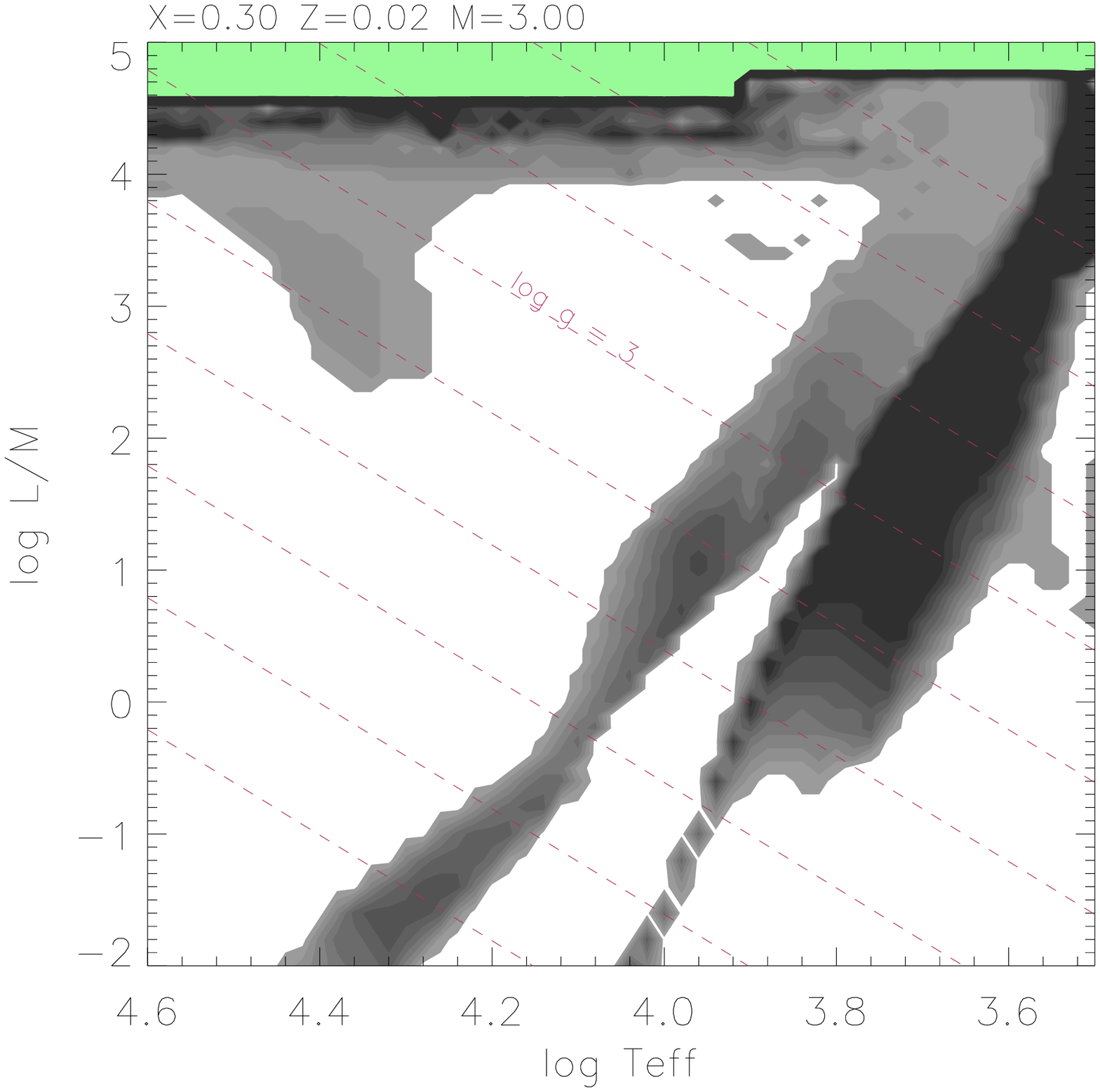,width=4.3cm,angle=0}\\
\epsfig{file=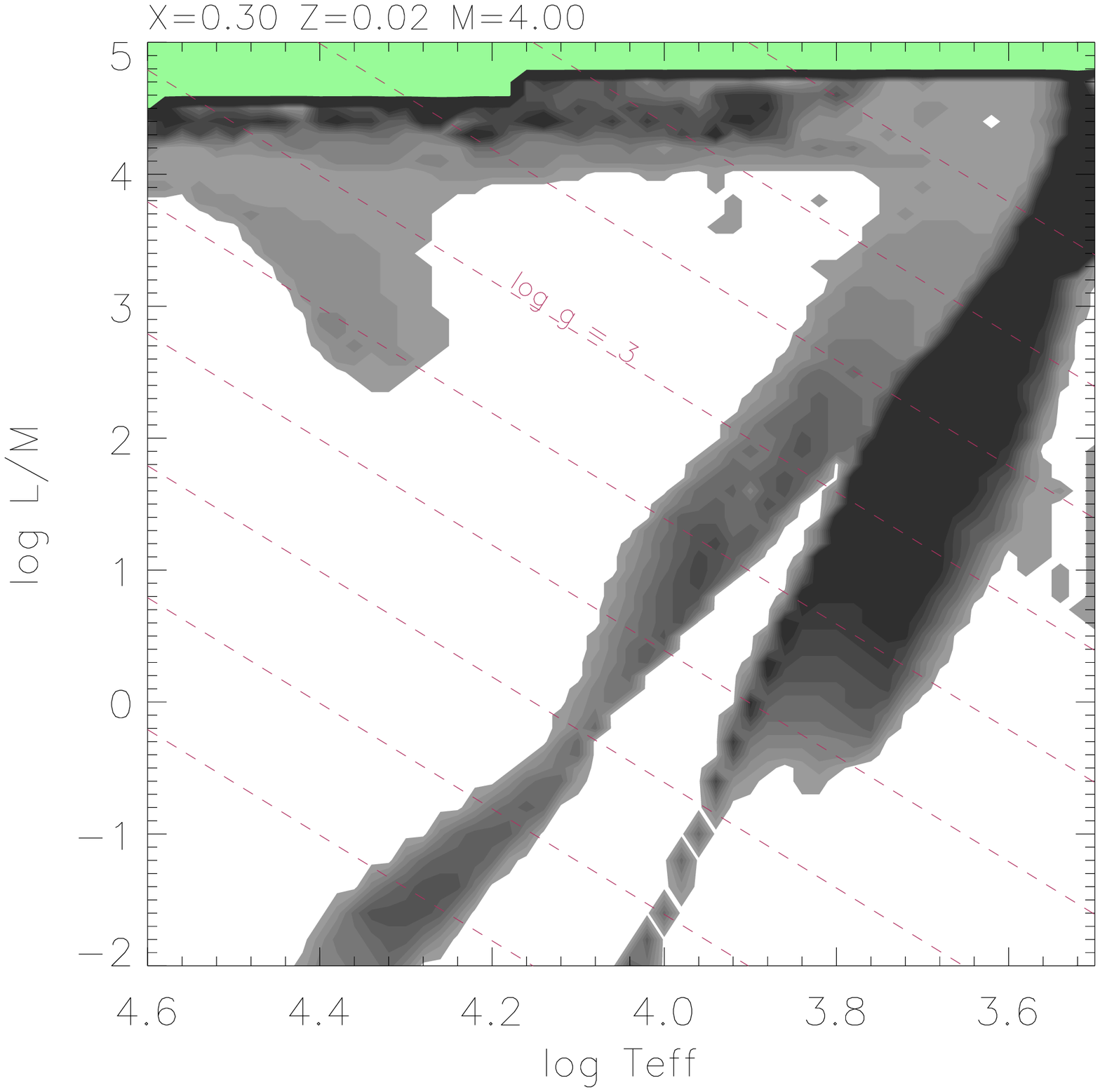,width=4.3cm,angle=0}
\epsfig{file=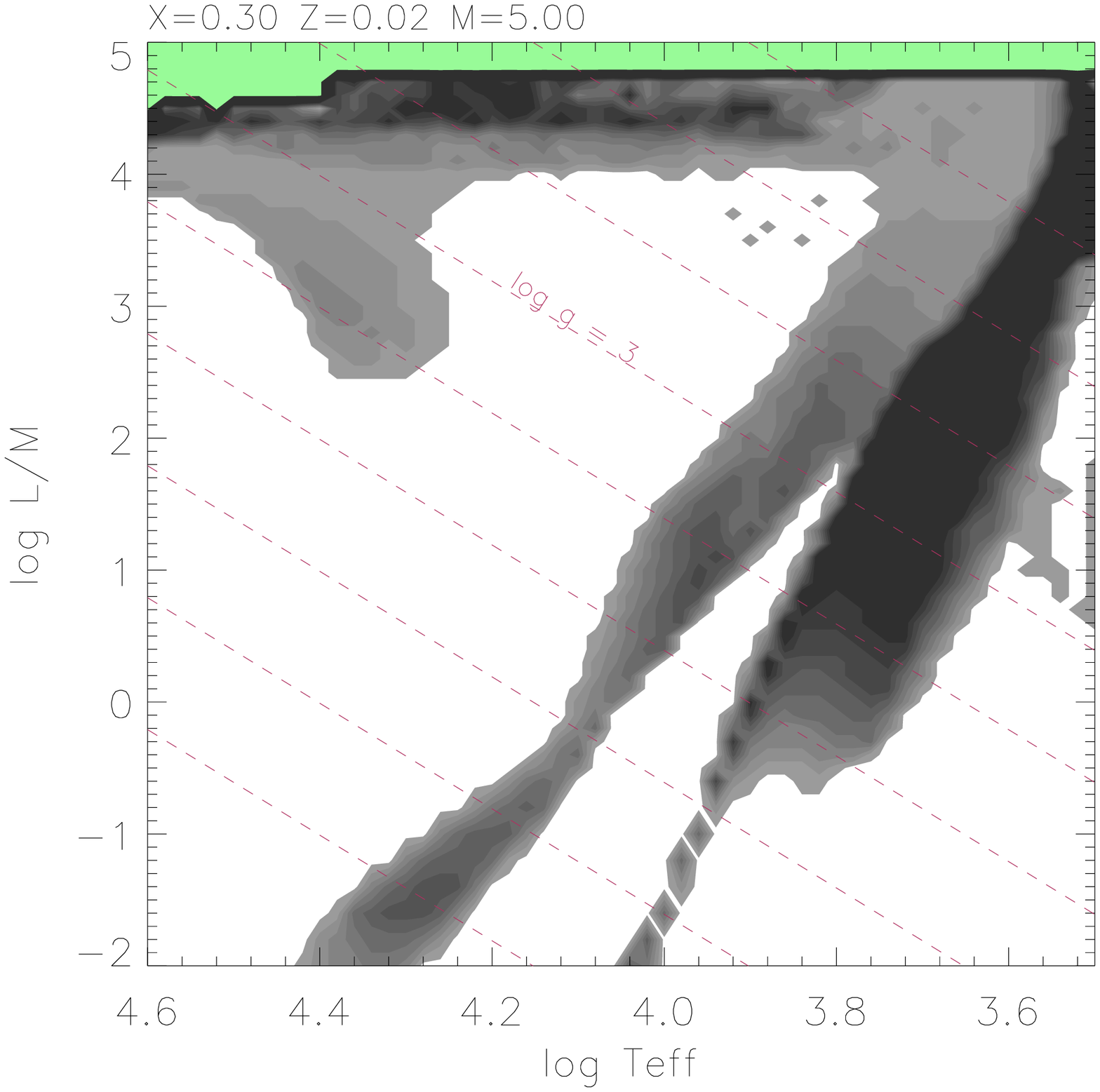,width=4.3cm,angle=0}
\epsfig{file=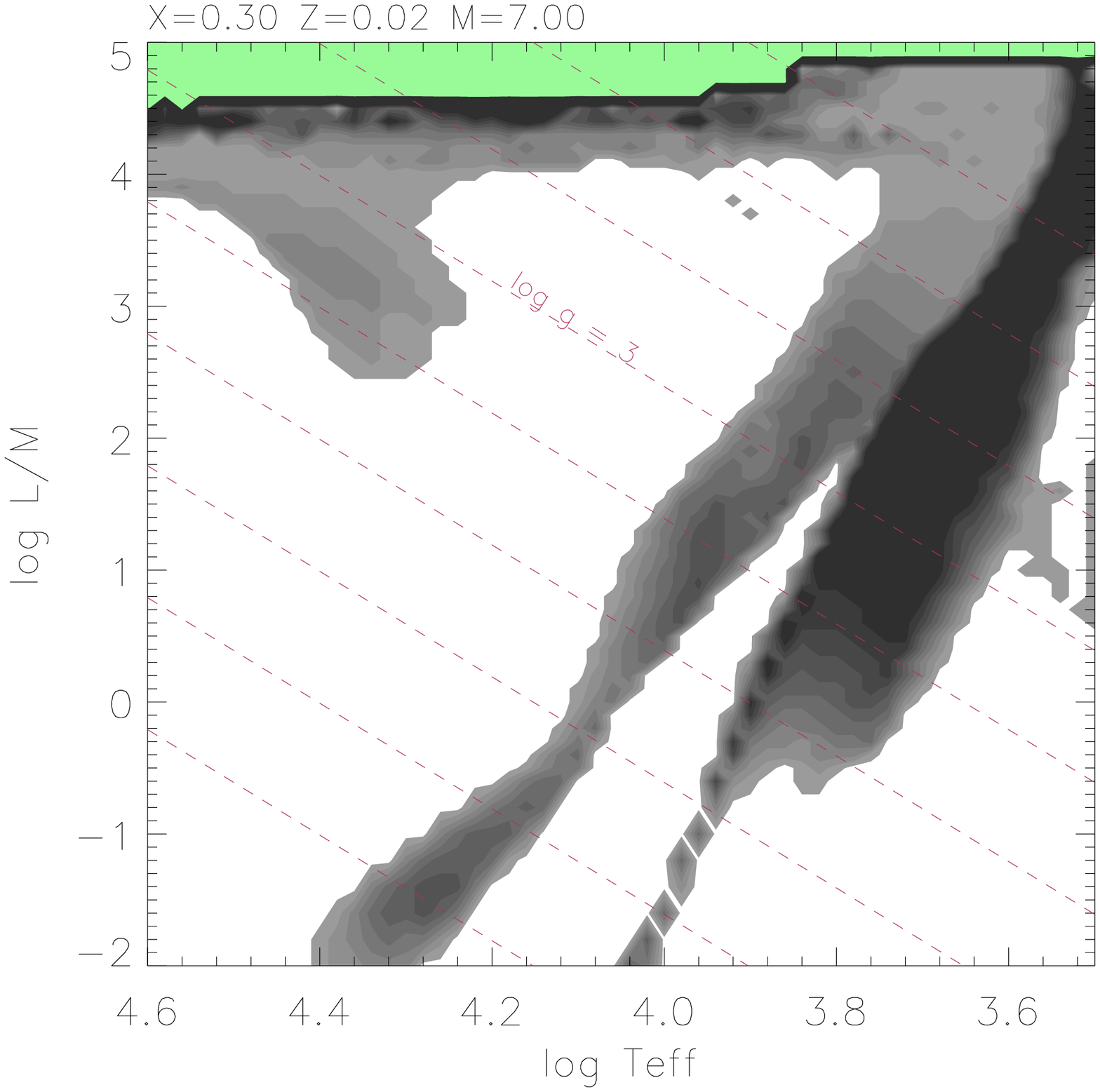,width=4.3cm,angle=0}
\epsfig{file=figs/nmodes_x30z02m10.0_00_opal.eps,width=4.3cm,angle=0}\\
\epsfig{file=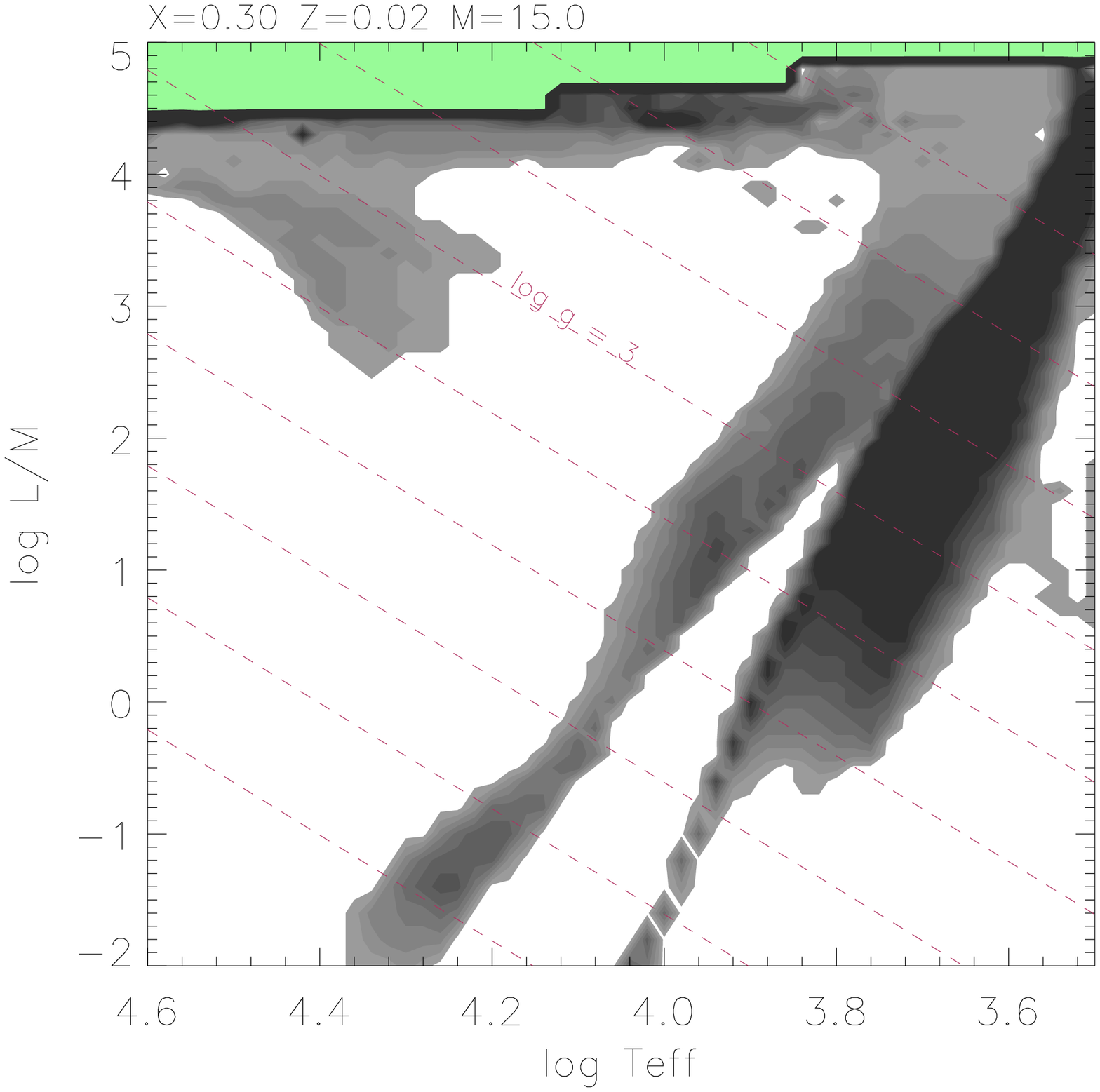,width=4.3cm,angle=0}
\epsfig{file=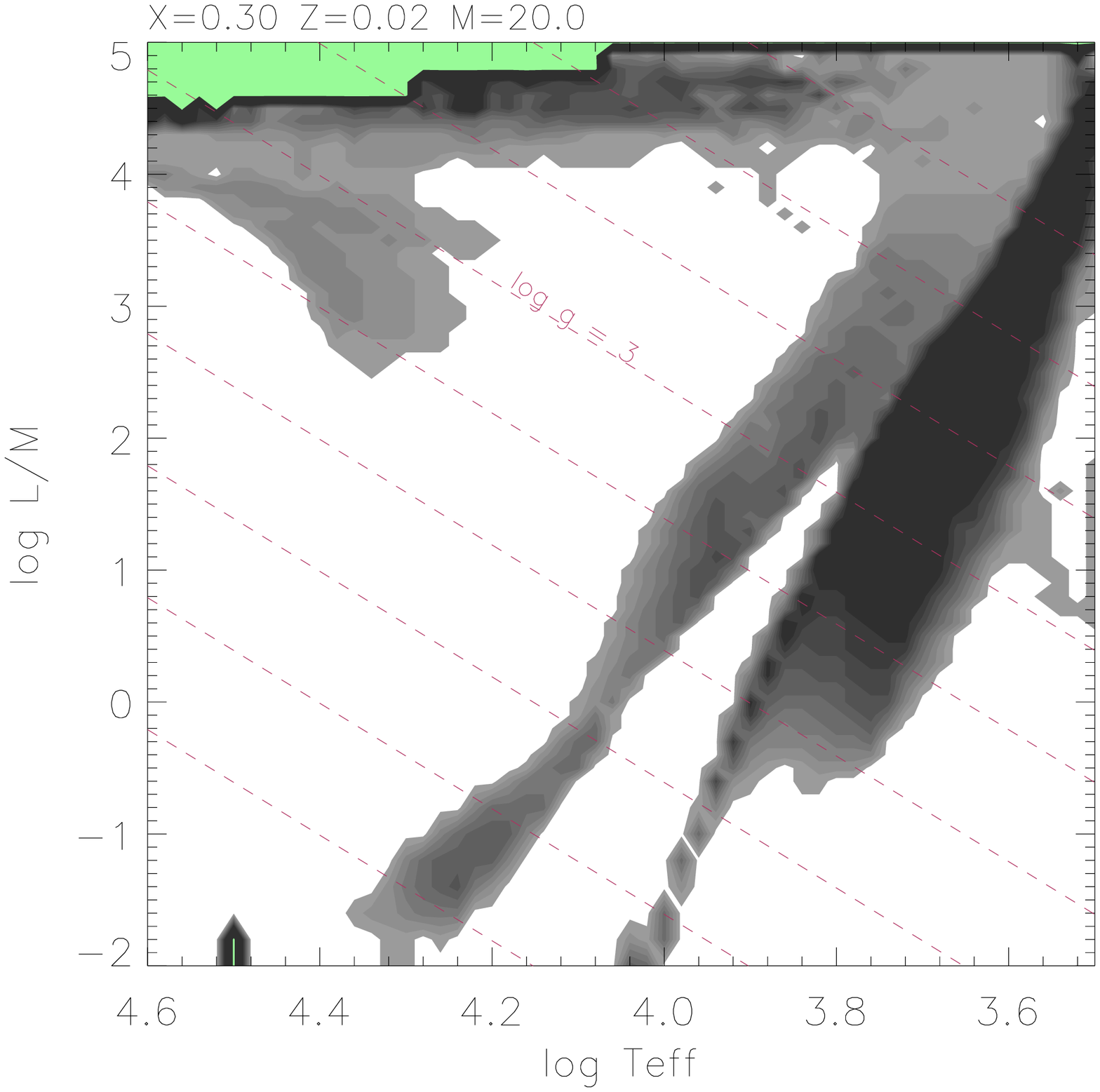,width=4.3cm,angle=0}
\epsfig{file=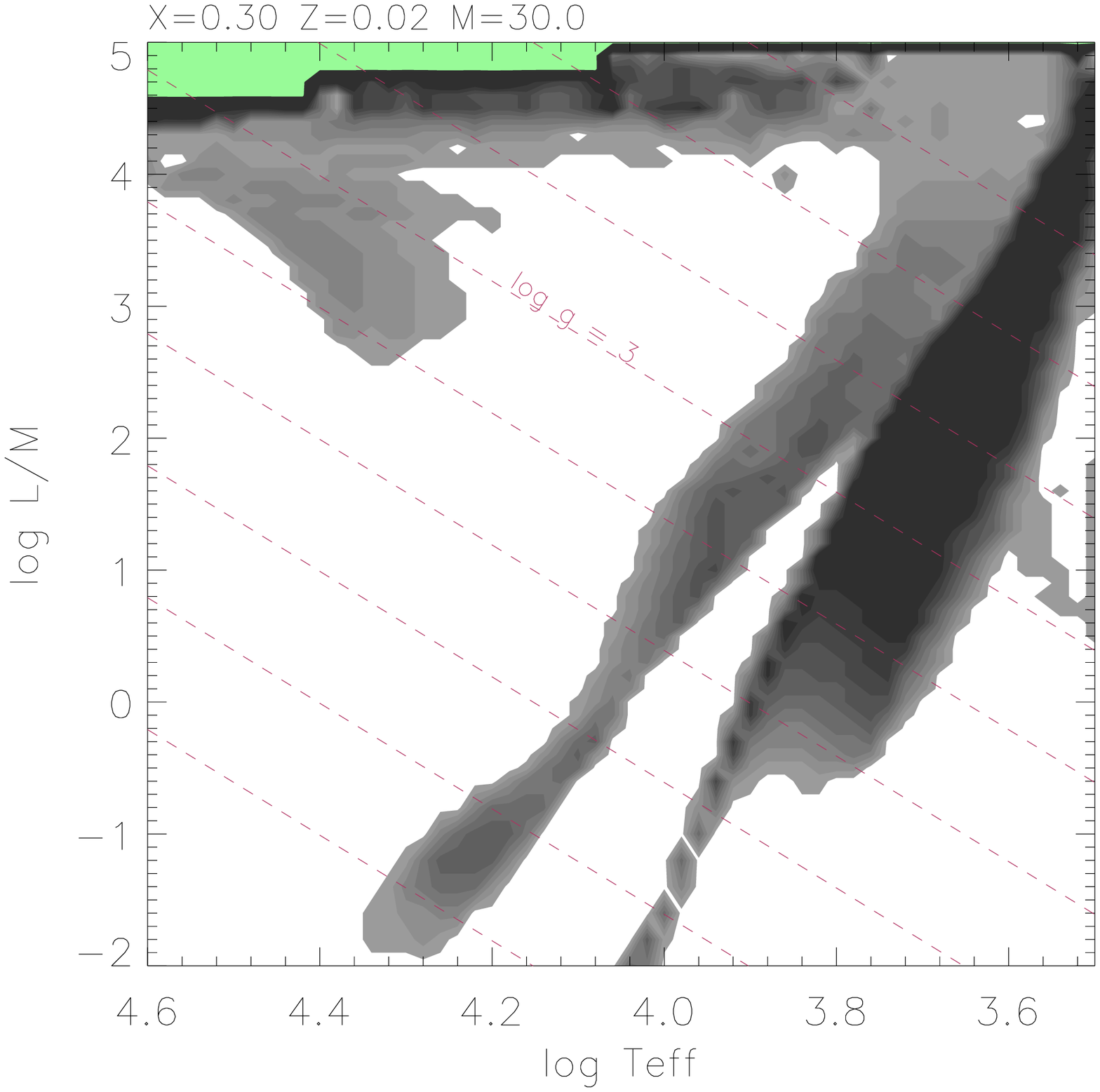,width=4.3cm,angle=0}
\epsfig{file=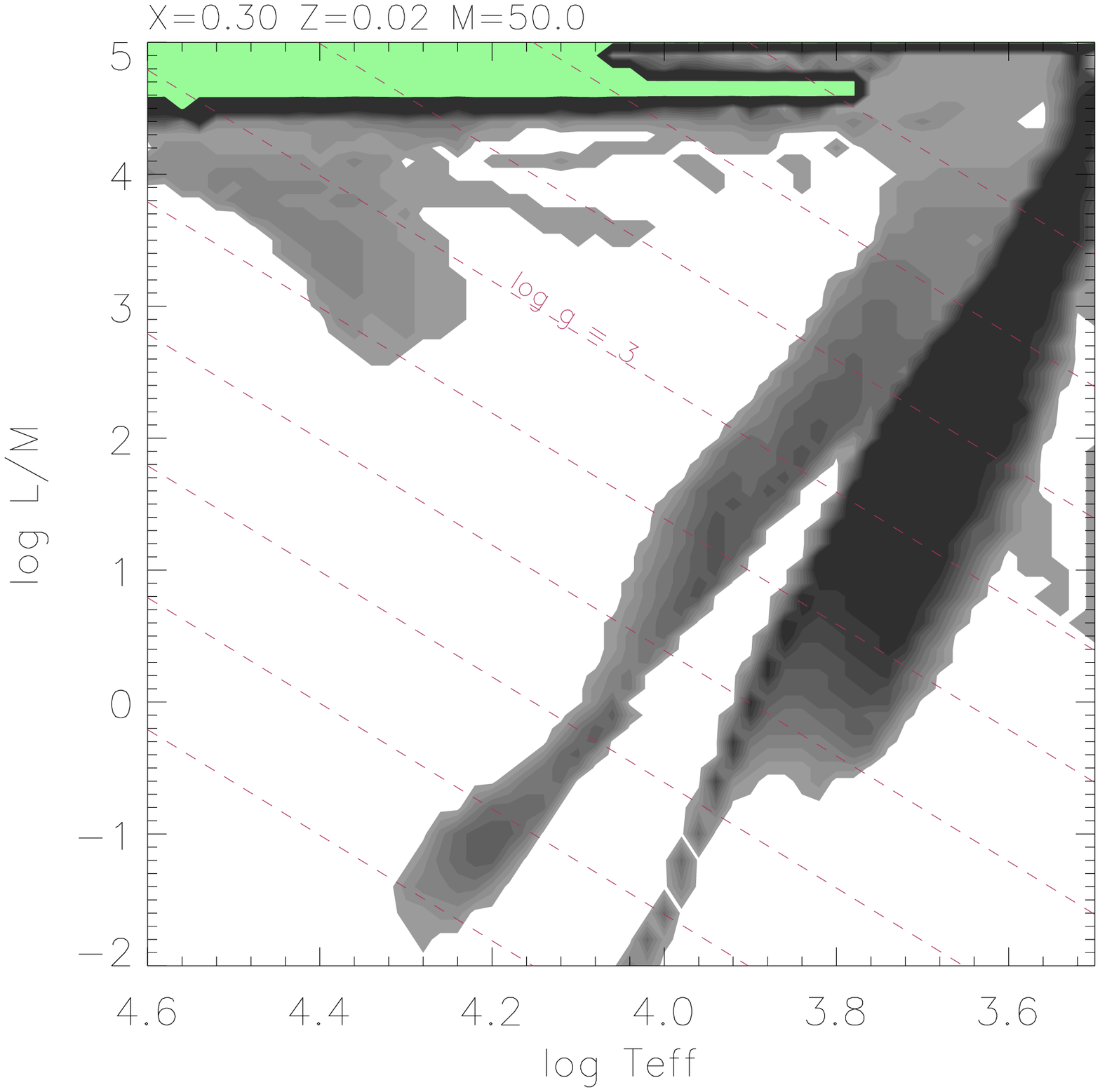,width=4.3cm,angle=0}
\caption[Unstable modes: $X=0.30, Z=0.02$]
{As Fig.~\ref{f:nx70} with $X=0.30, Z=0.02$. 
}
\label{f:nx30}
\end{center}
\end{figure*}

\begin{figure*}
\begin{center}
\epsfig{file=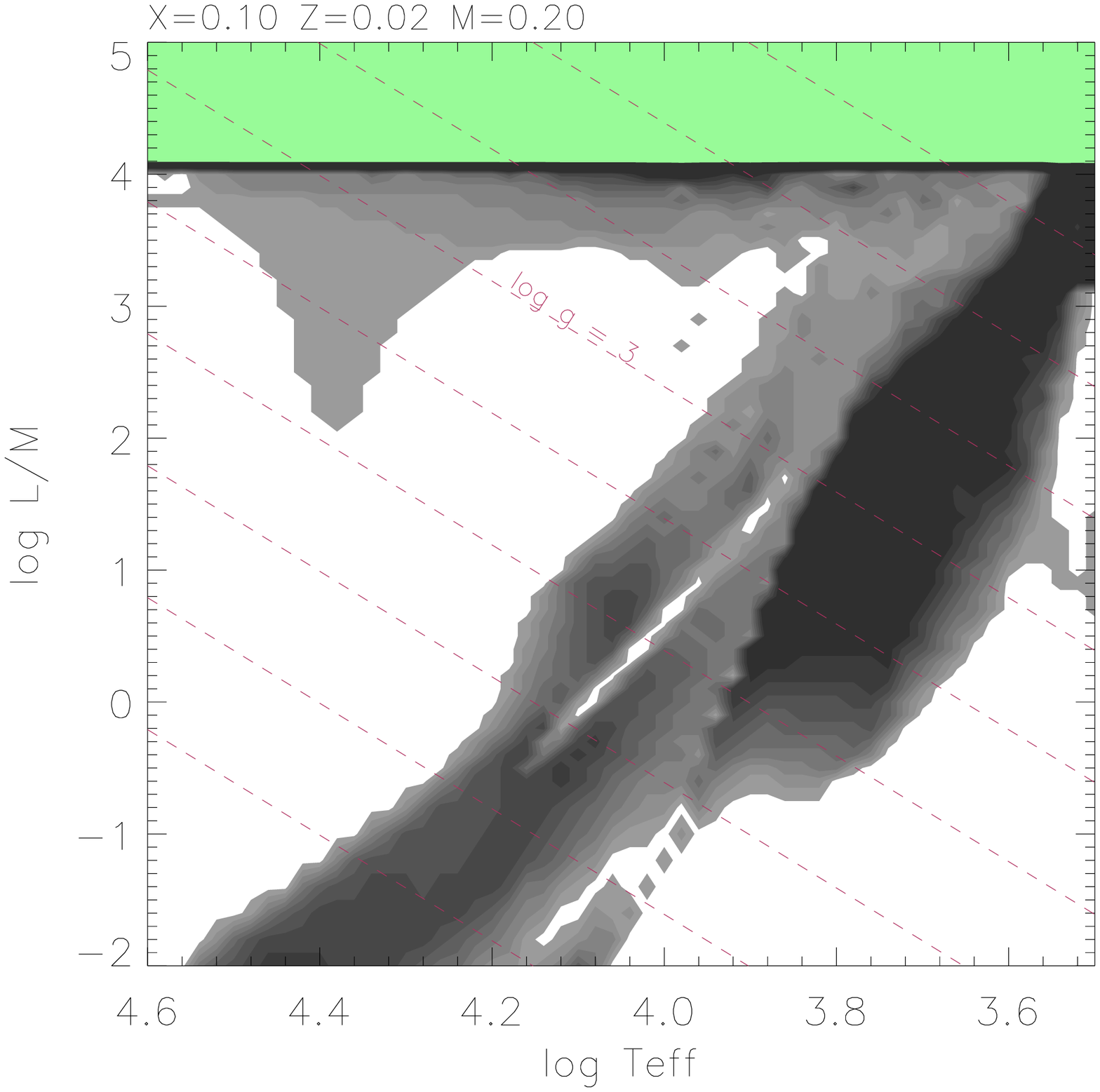,width=4.3cm,angle=0}
\epsfig{file=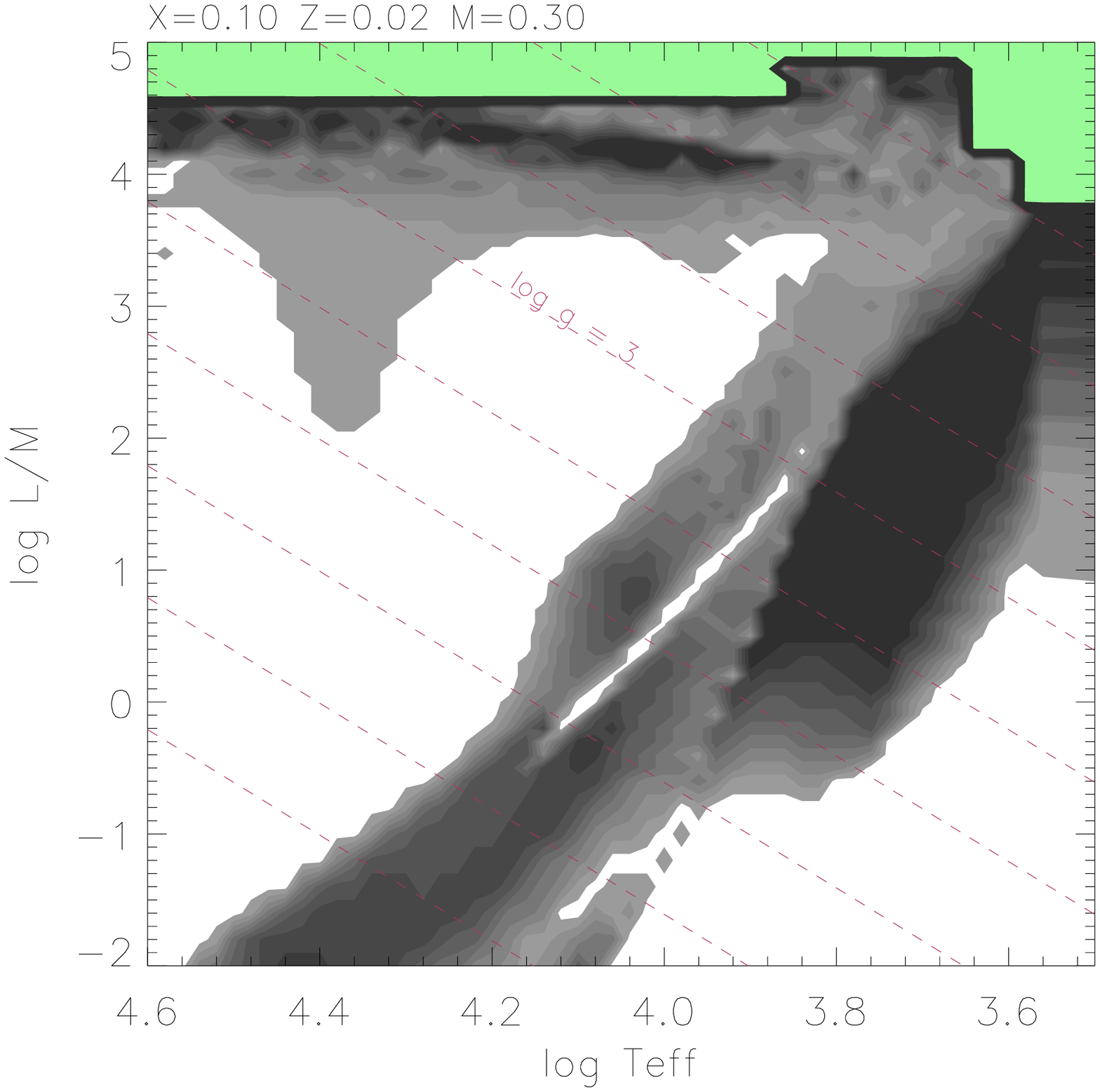,width=4.3cm,angle=0}
\epsfig{file=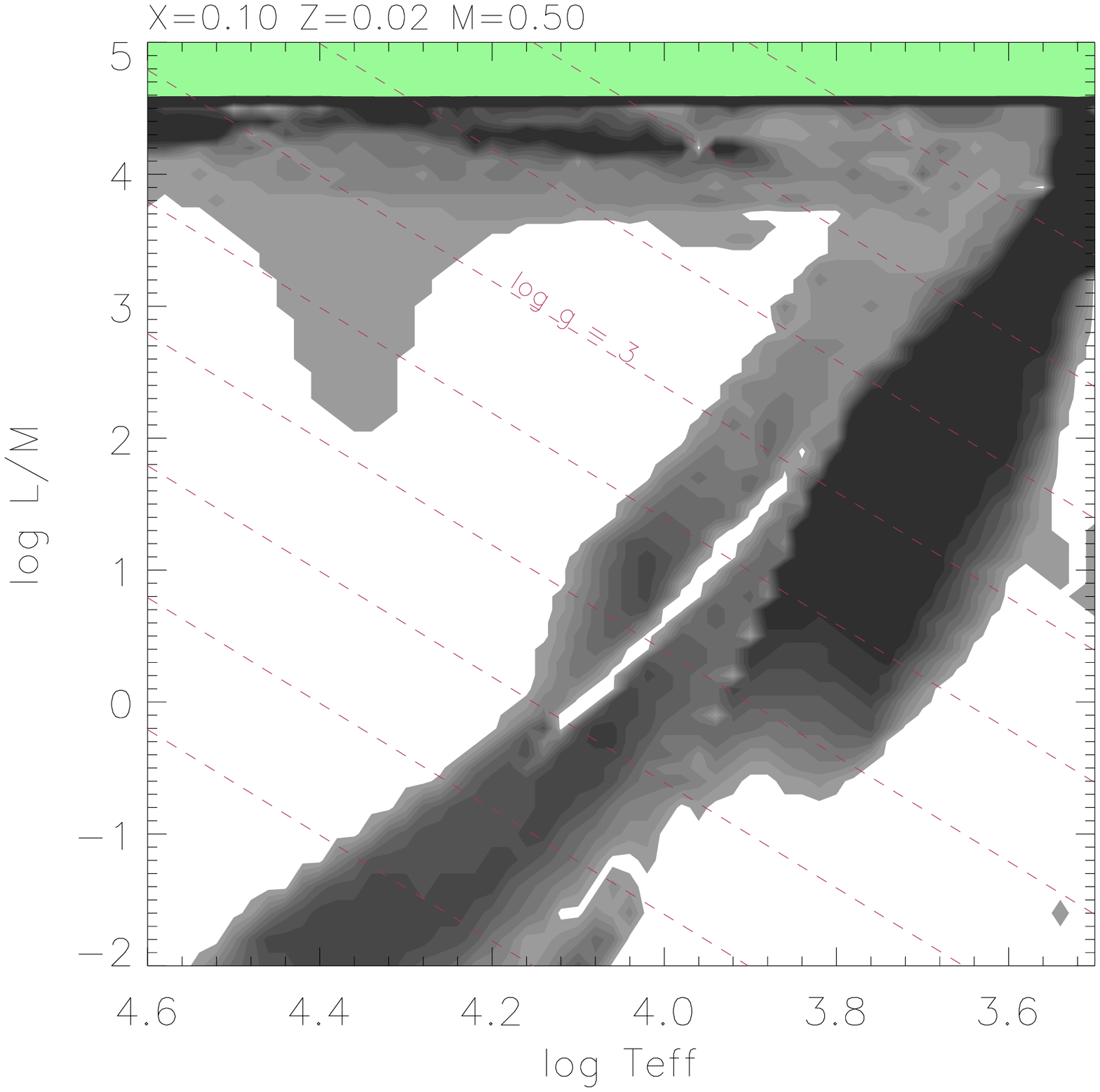,width=4.3cm,angle=0}
\epsfig{file=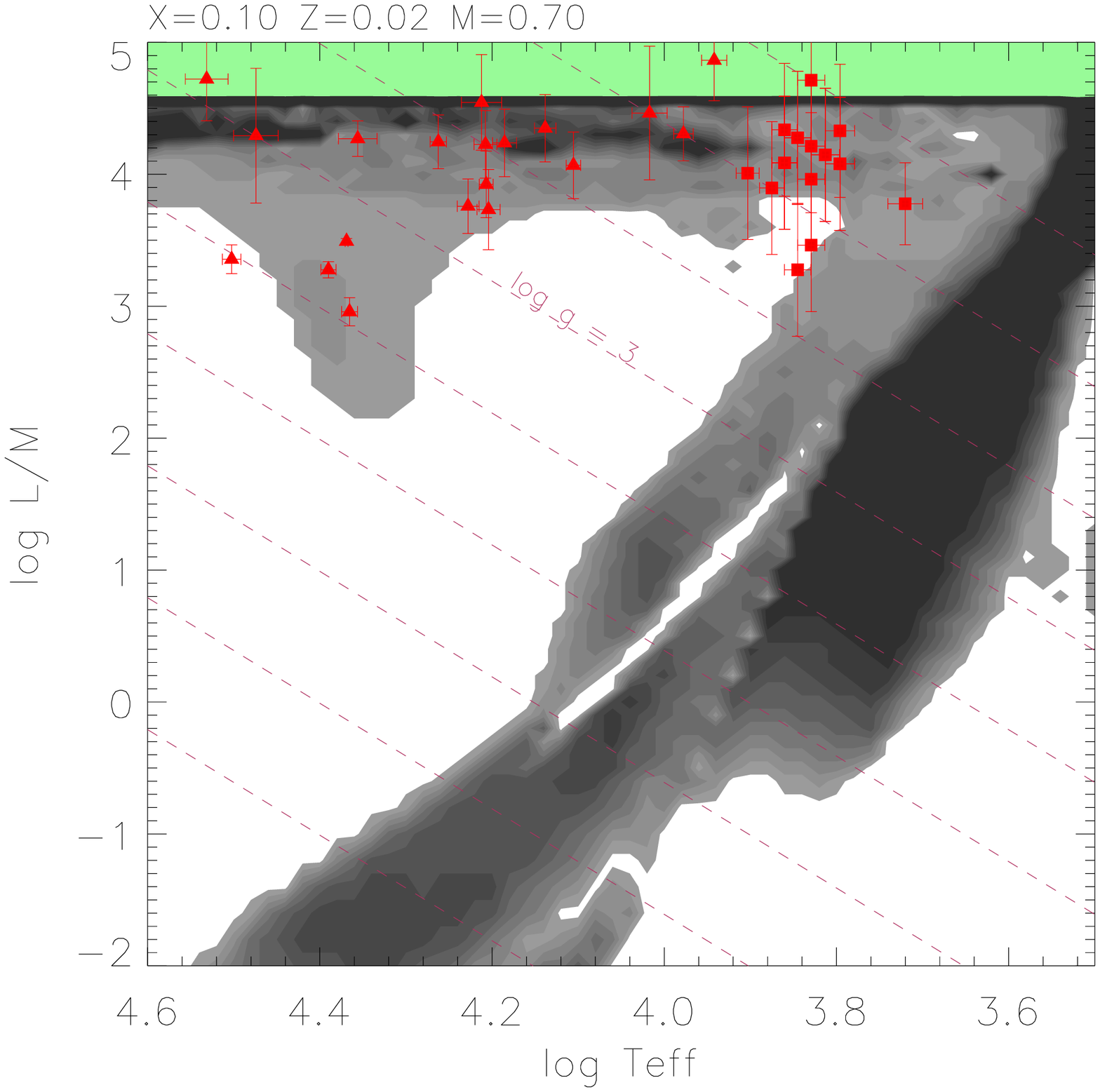,width=4.3cm,angle=0}\\
\epsfig{file=figs/nmodes_x10z02m01.0_00_opal.eps,width=4.3cm,angle=0}
\epsfig{file=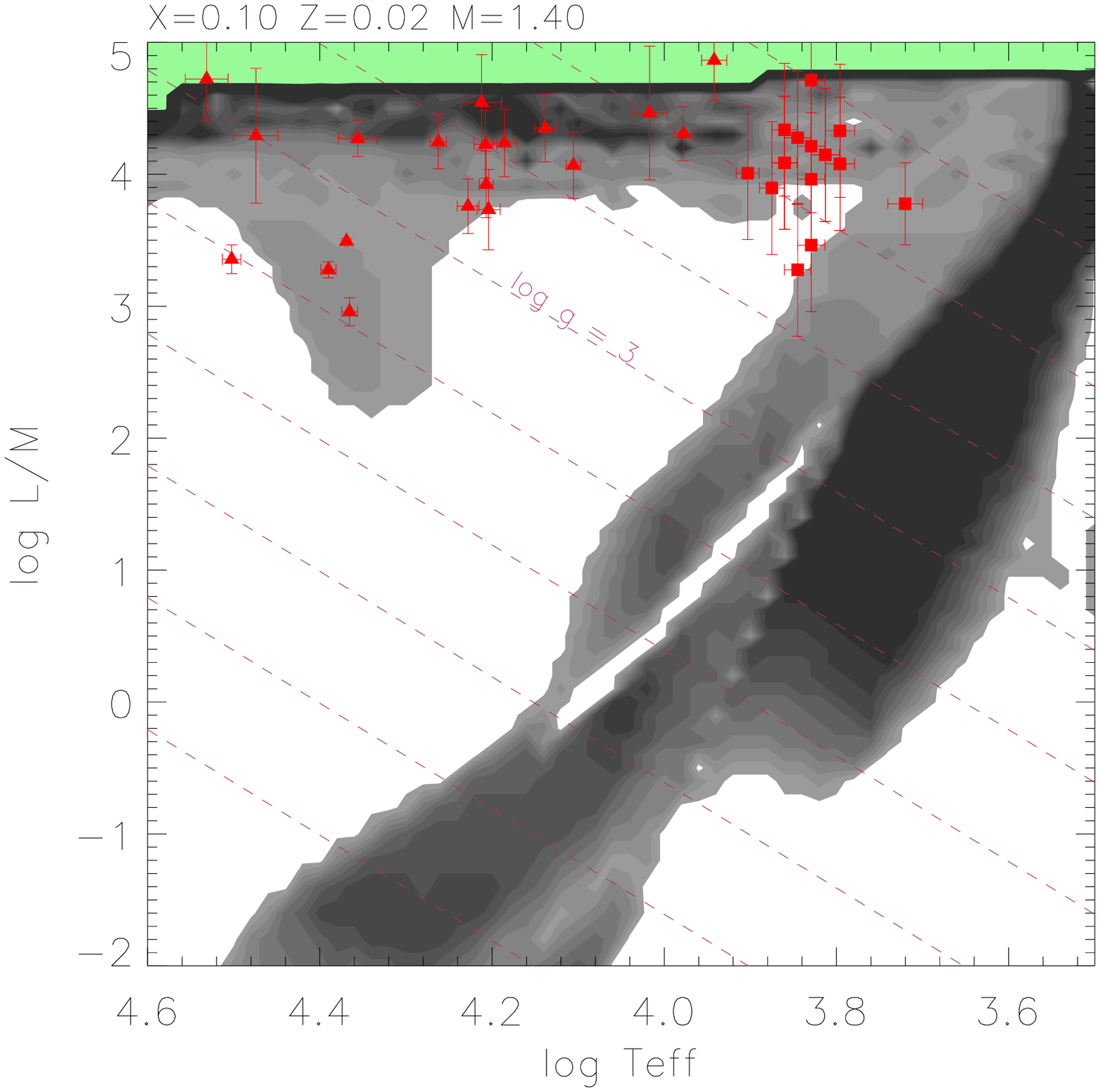,width=4.3cm,angle=0}
\epsfig{file=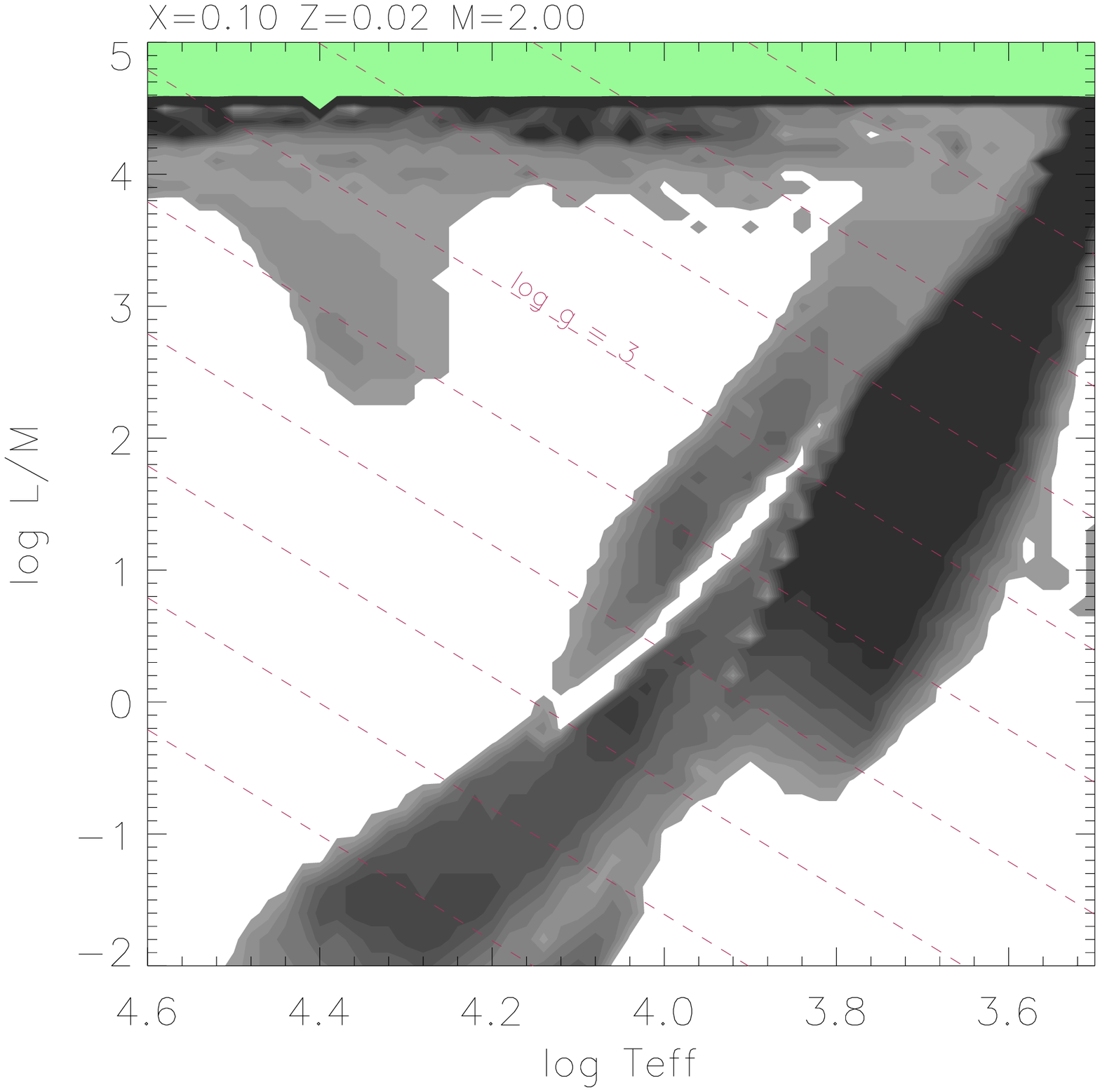,width=4.3cm,angle=0}
\epsfig{file=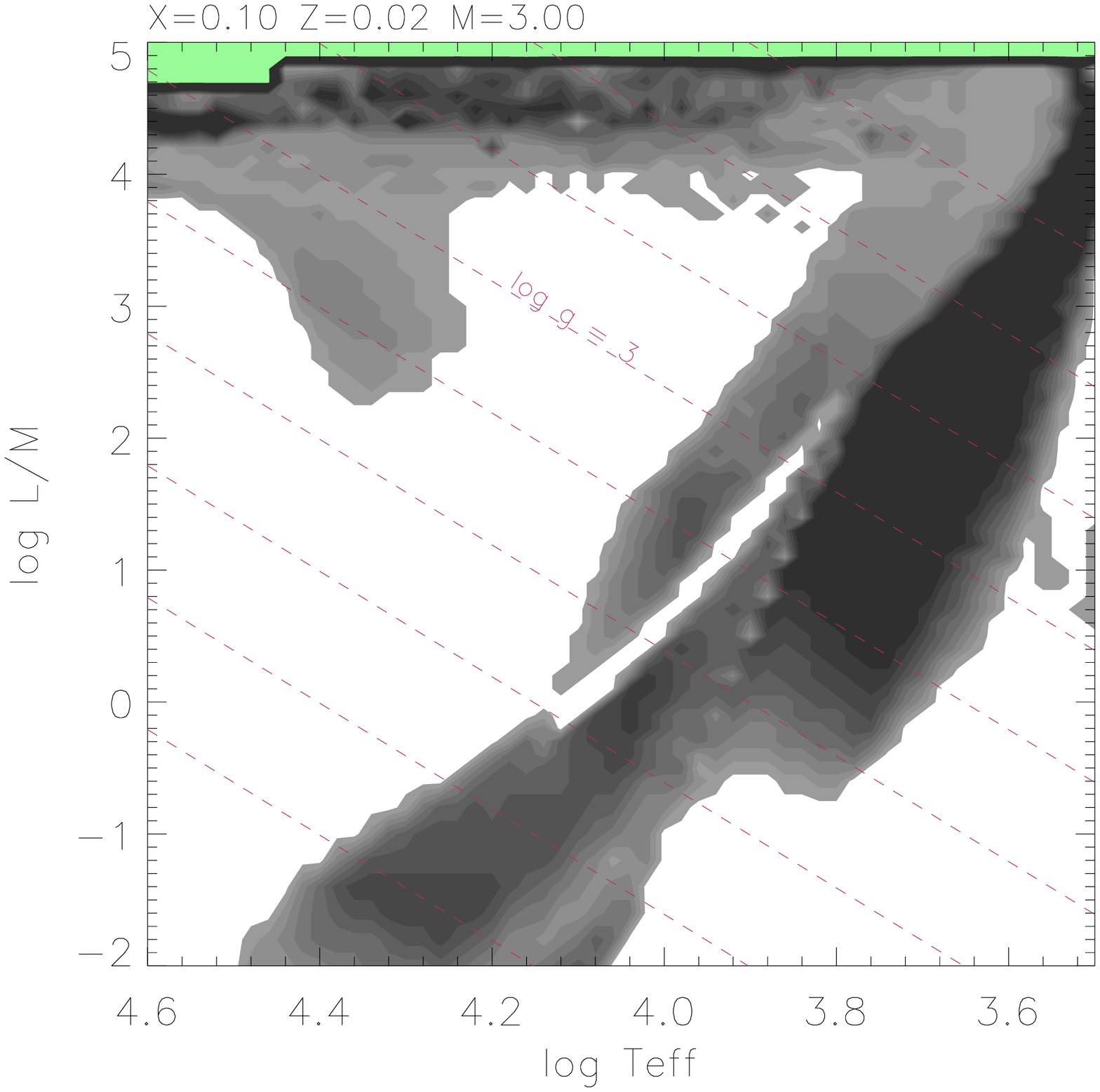,width=4.3cm,angle=0}\\
\epsfig{file=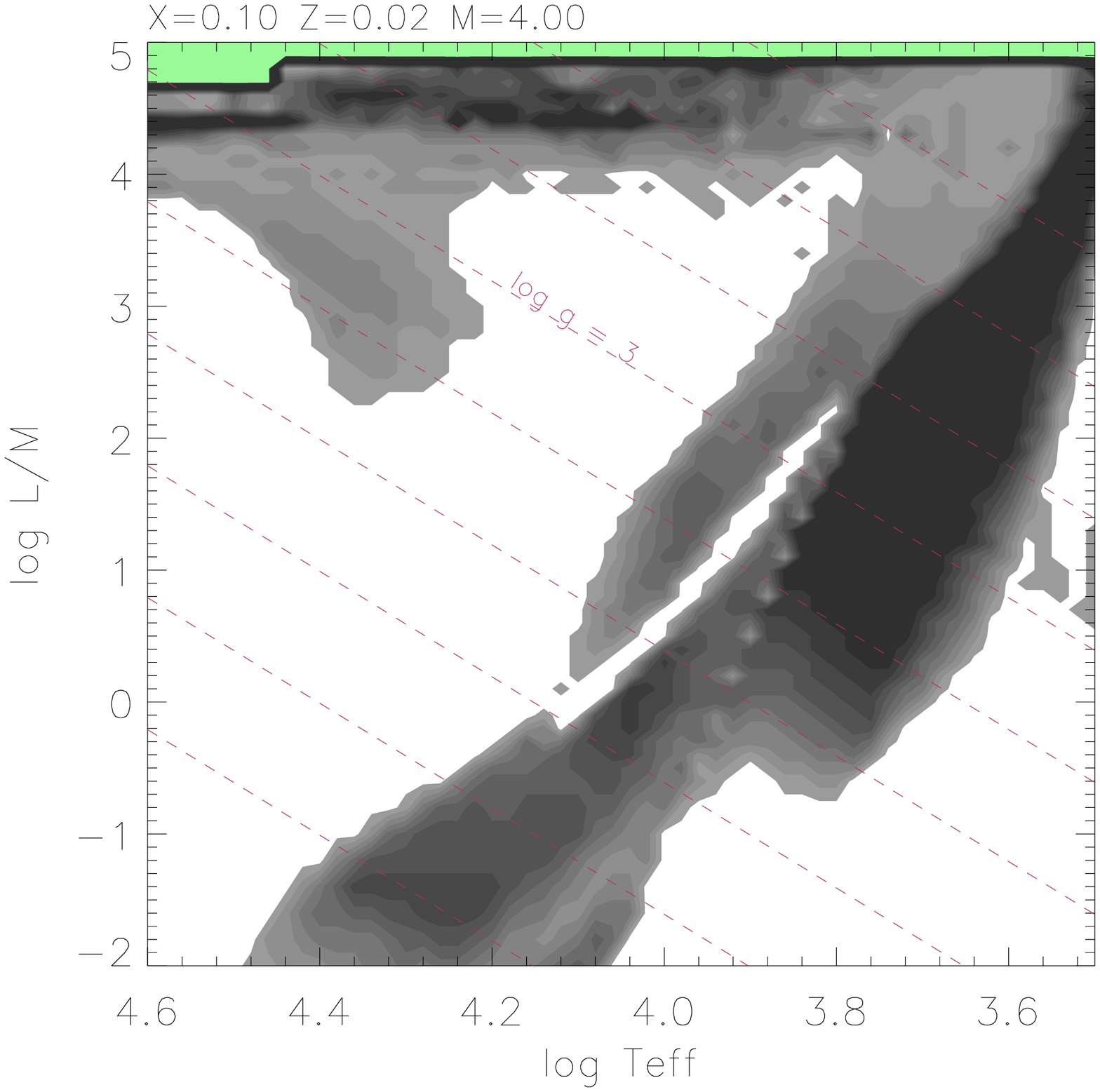,width=4.3cm,angle=0}
\epsfig{file=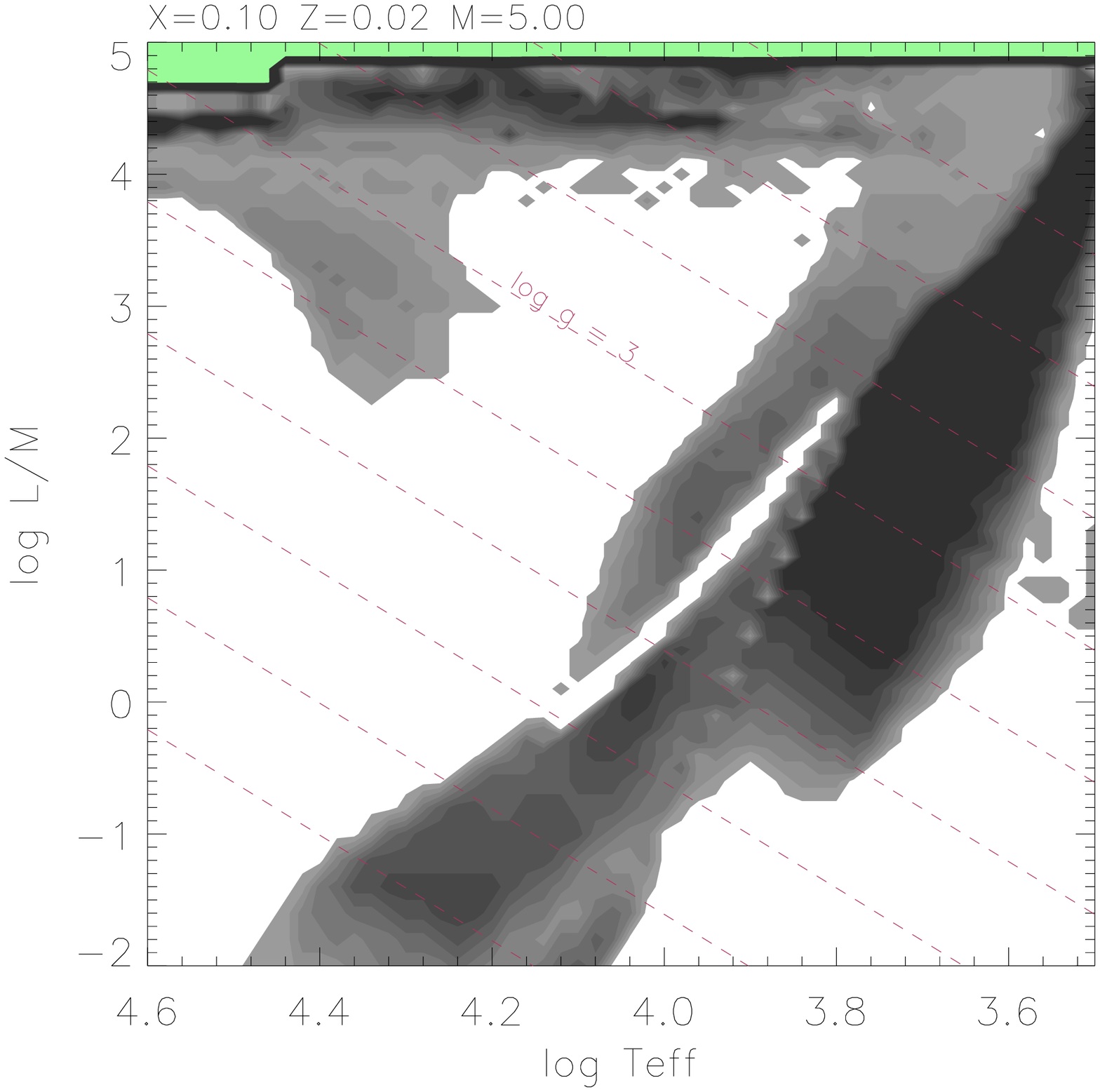,width=4.3cm,angle=0}
\epsfig{file=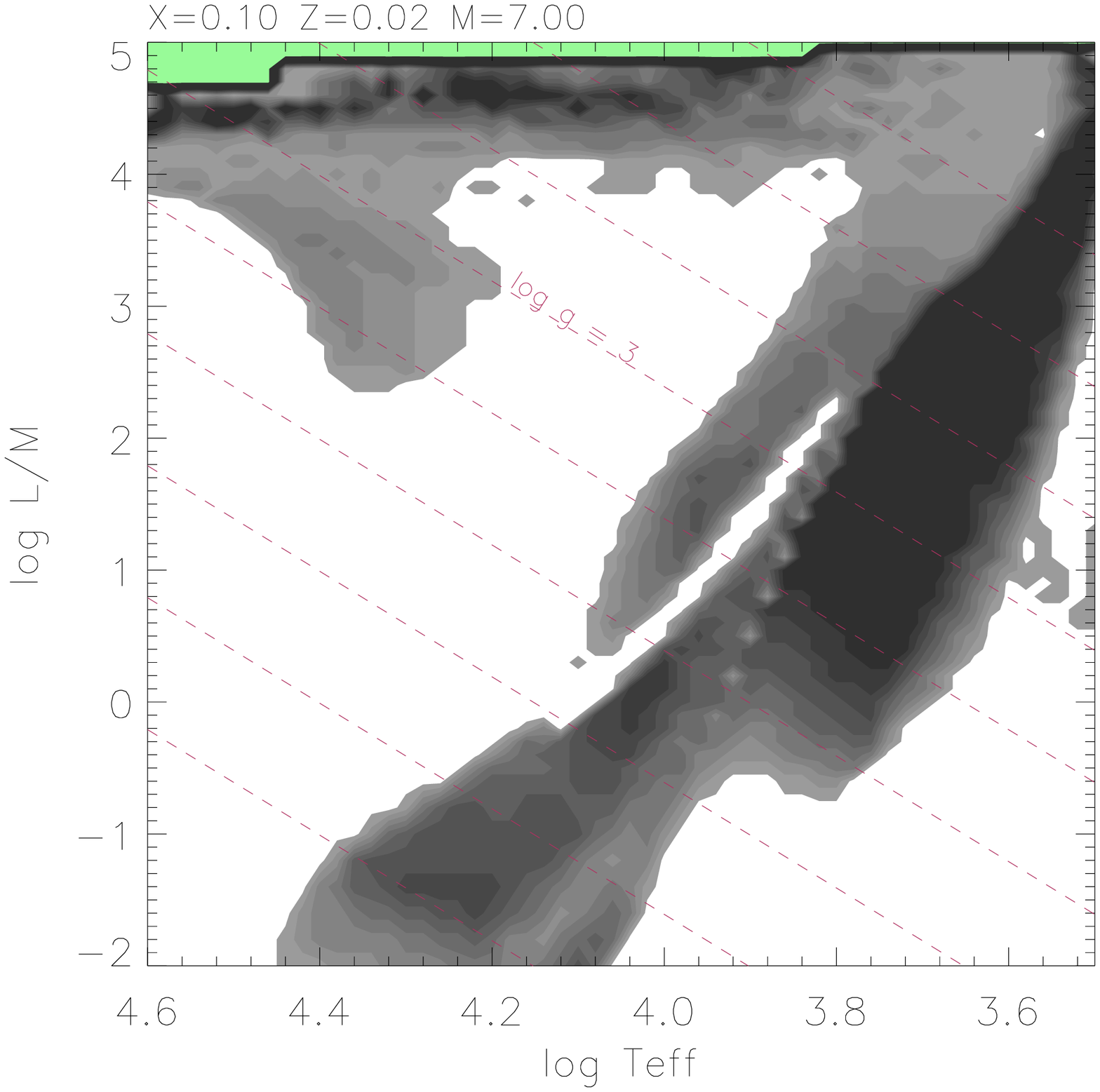,width=4.3cm,angle=0}
\epsfig{file=figs/nmodes_x10z02m10.0_00_opal.eps,width=4.3cm,angle=0}\\
\epsfig{file=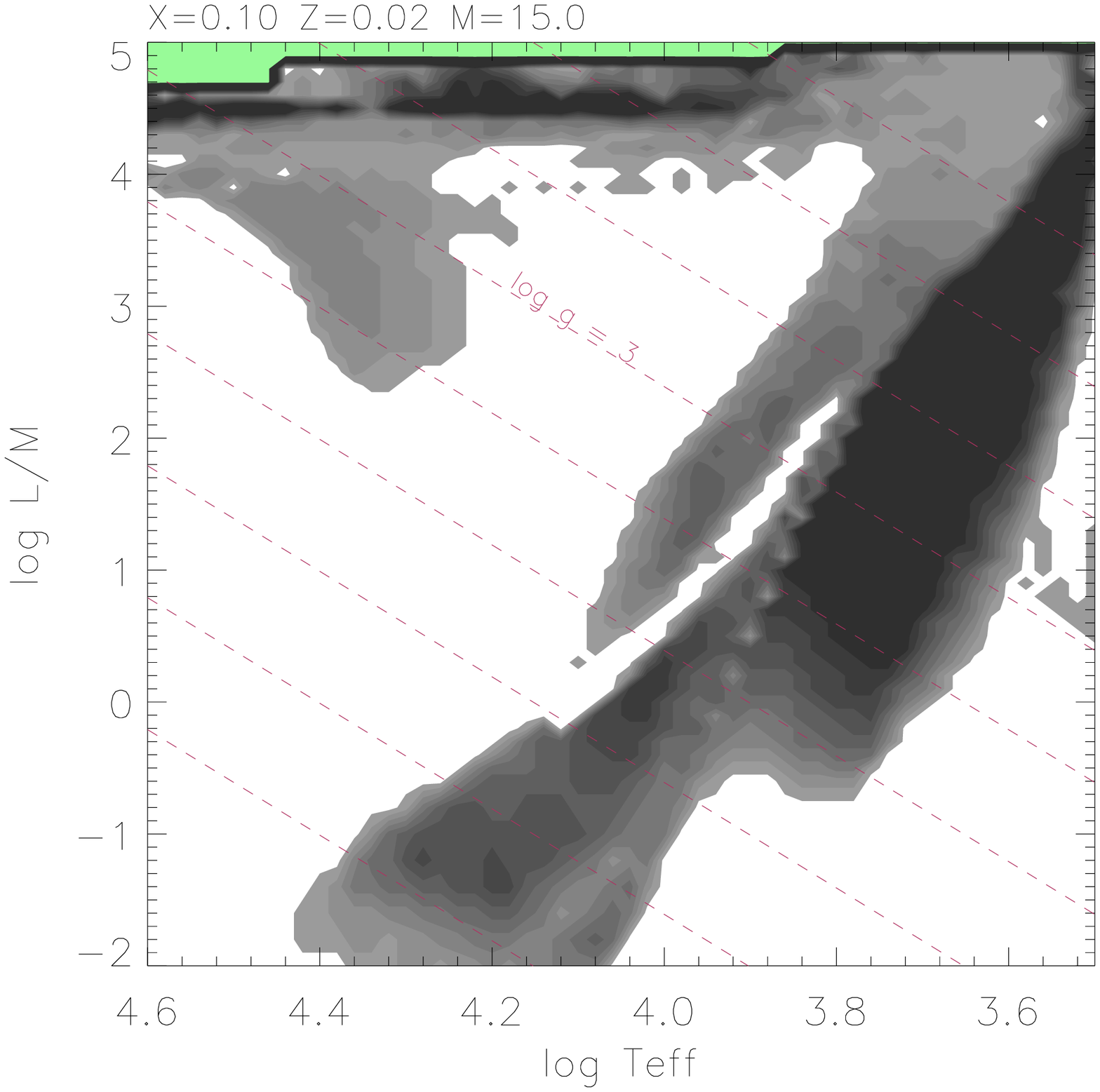,width=4.3cm,angle=0}
\epsfig{file=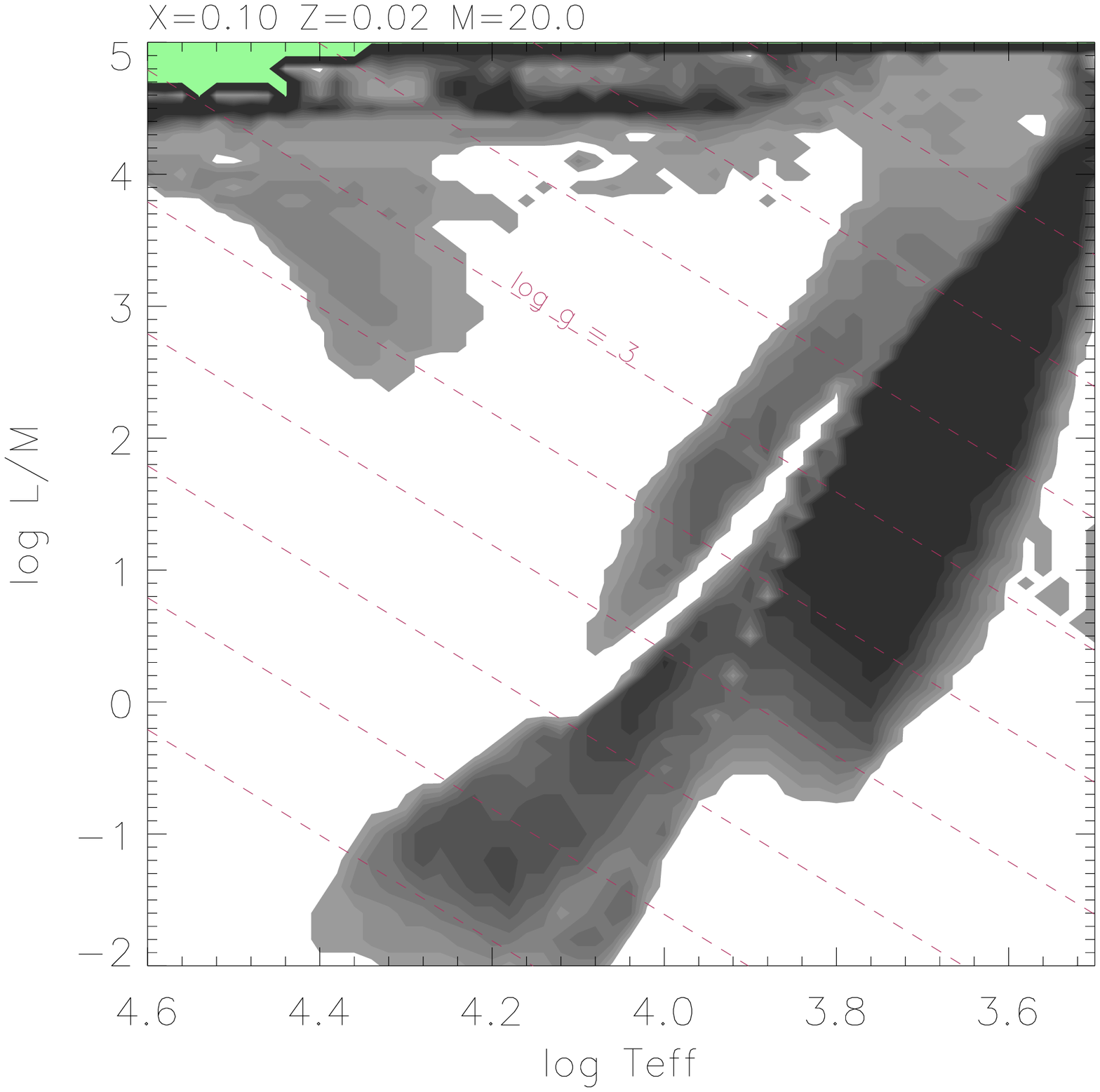,width=4.3cm,angle=0}
\epsfig{file=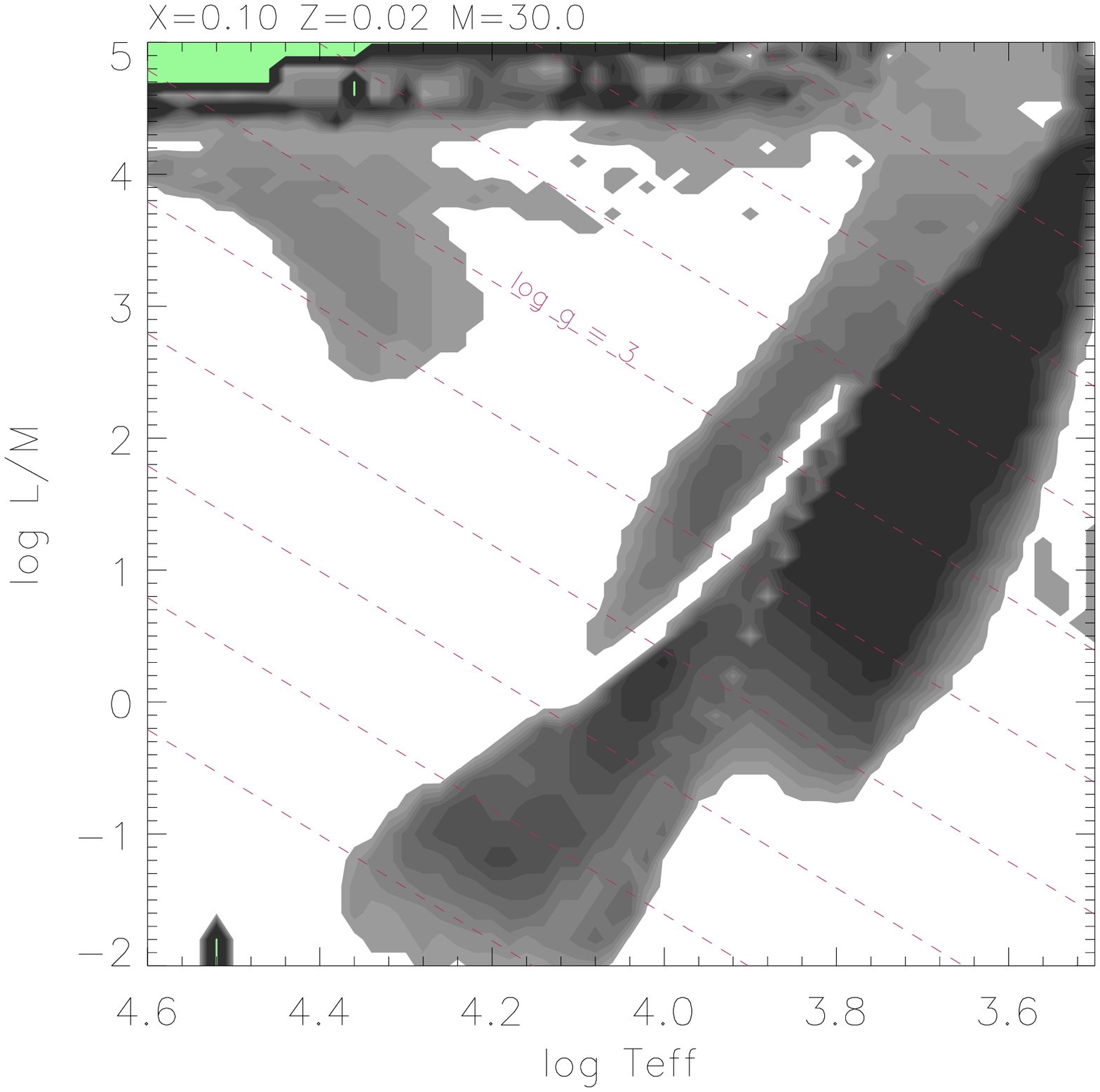,width=4.3cm,angle=0}
\epsfig{file=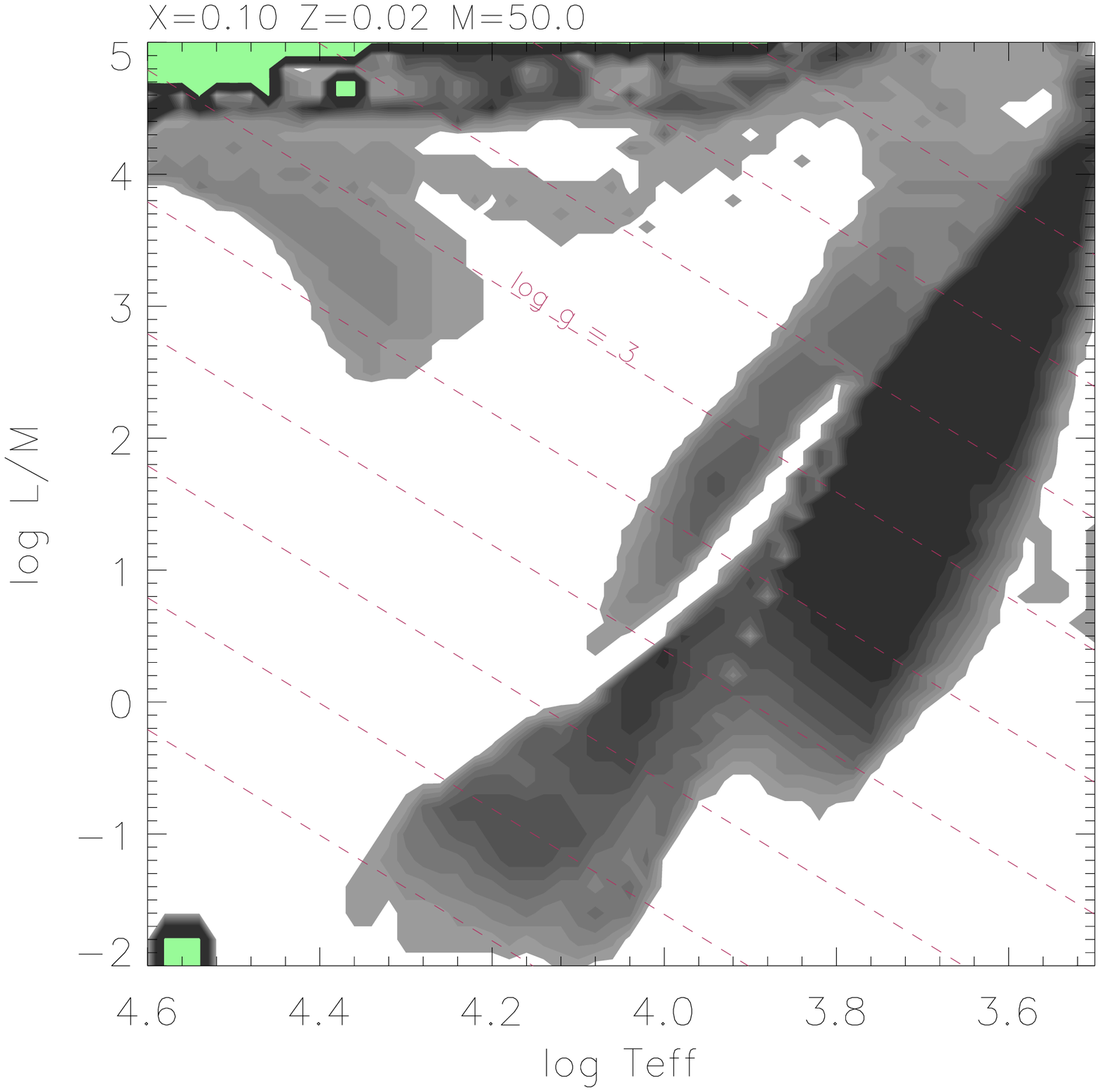,width=4.3cm,angle=0}
\caption[Unstable modes: $X=0.10, Z=0.02$]
{As Fig.~\ref{f:nx70} with $X=0.10, Z=0.02$. Red symbols with error bars 
shown on selected low-mass panels  
represent the observed parameters of pulsating low-mass hydrogen-deficient stars, including extreme
helium stars and R Coronae Borealis variables \citep{jeffery08.ibvs}.
}
\label{f:nx10}
\end{center}
\end{figure*}

\begin{figure*}
\begin{center}
\epsfig{file=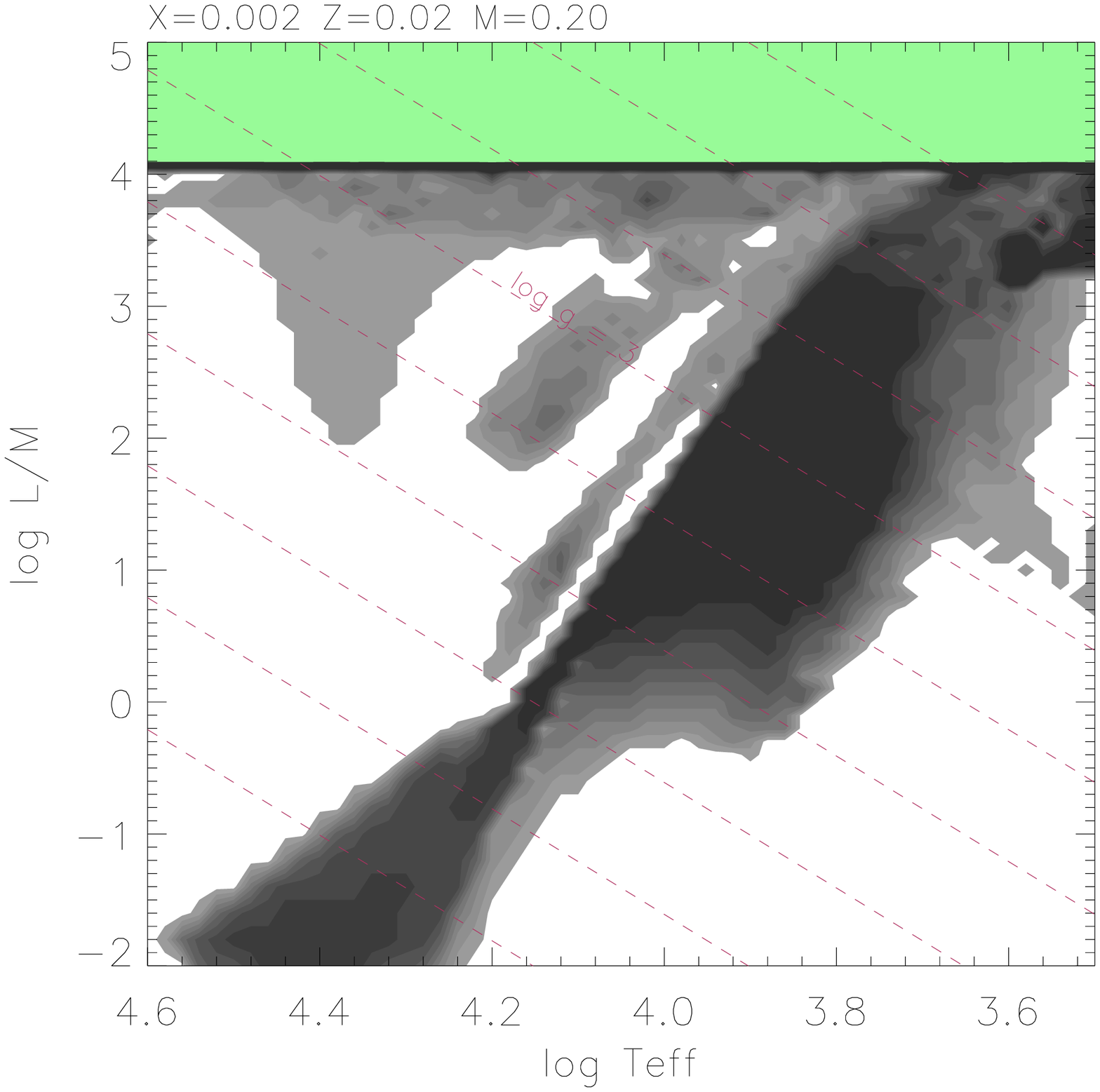,width=4.3cm,angle=0}
\epsfig{file=figs/nmodes_x002z02m00.3_00_opal.eps,width=4.3cm,angle=0}
\epsfig{file=figs/nmodes_x002z02m00.5_00_opal.eps,width=4.3cm,angle=0}
\epsfig{file=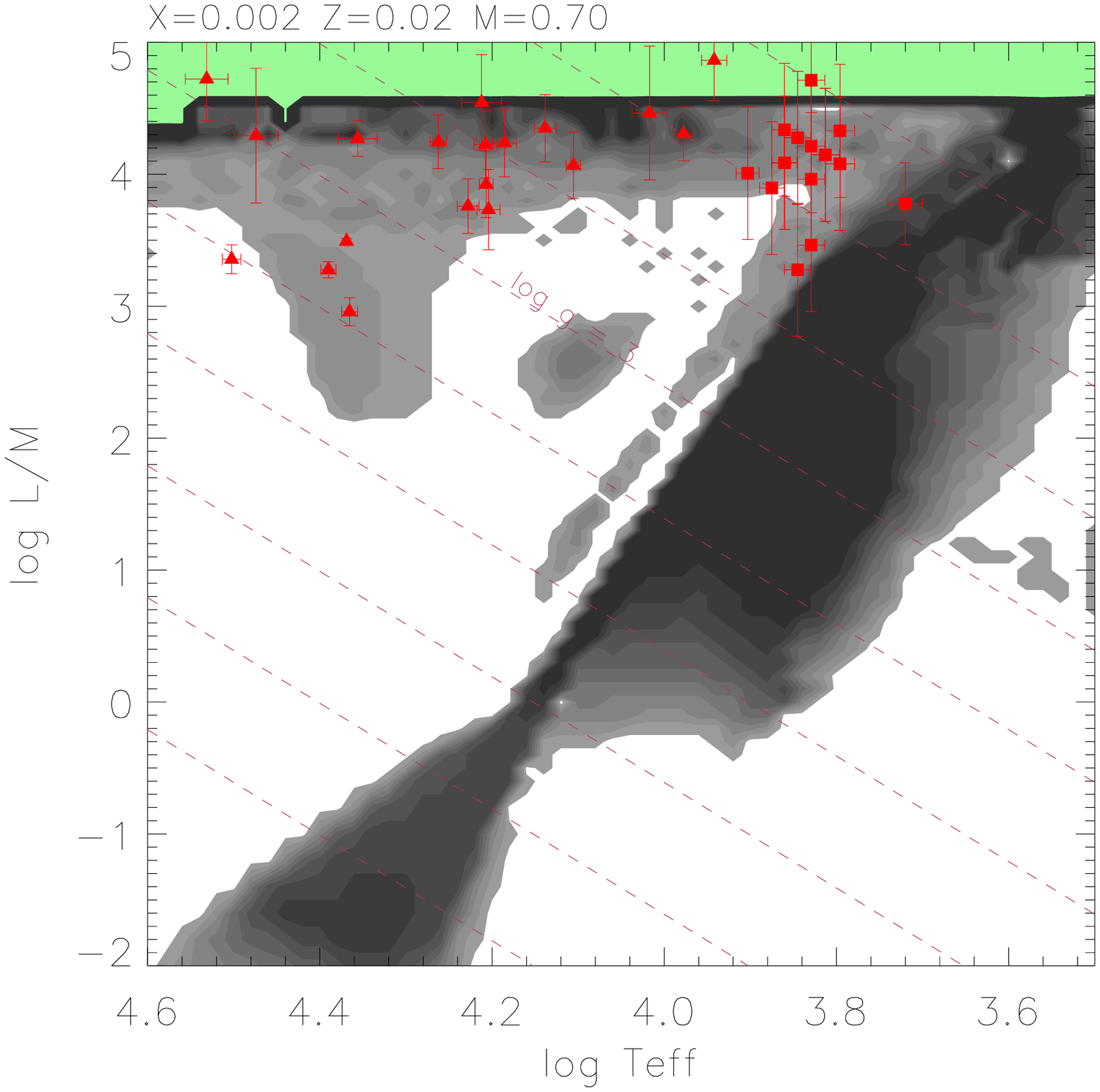,width=4.3cm,angle=0}\\
\epsfig{file=figs/nmodes_x002z02m01.0_00_opal.eps,width=4.3cm,angle=0}
\epsfig{file=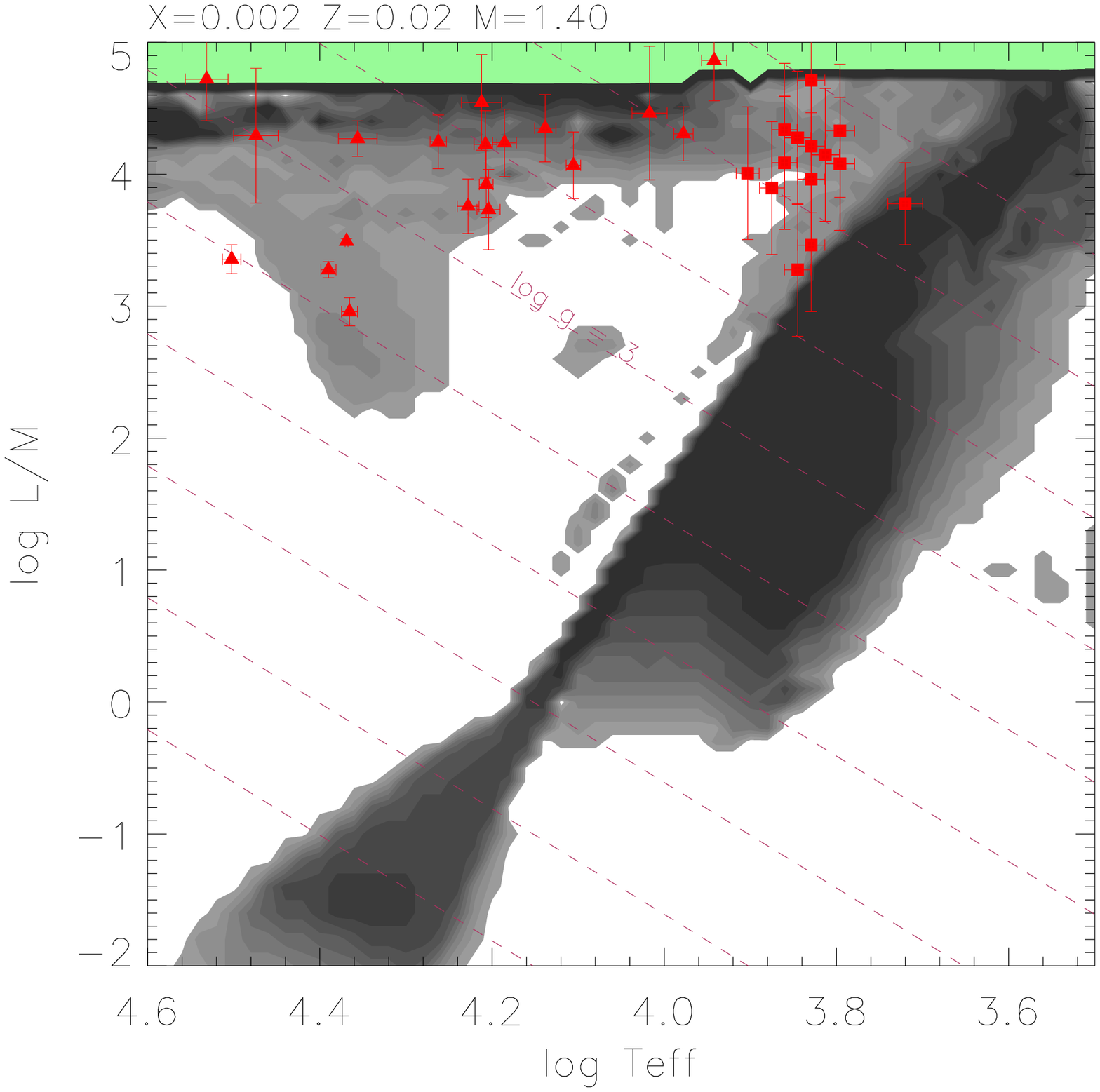,width=4.3cm,angle=0}
\epsfig{file=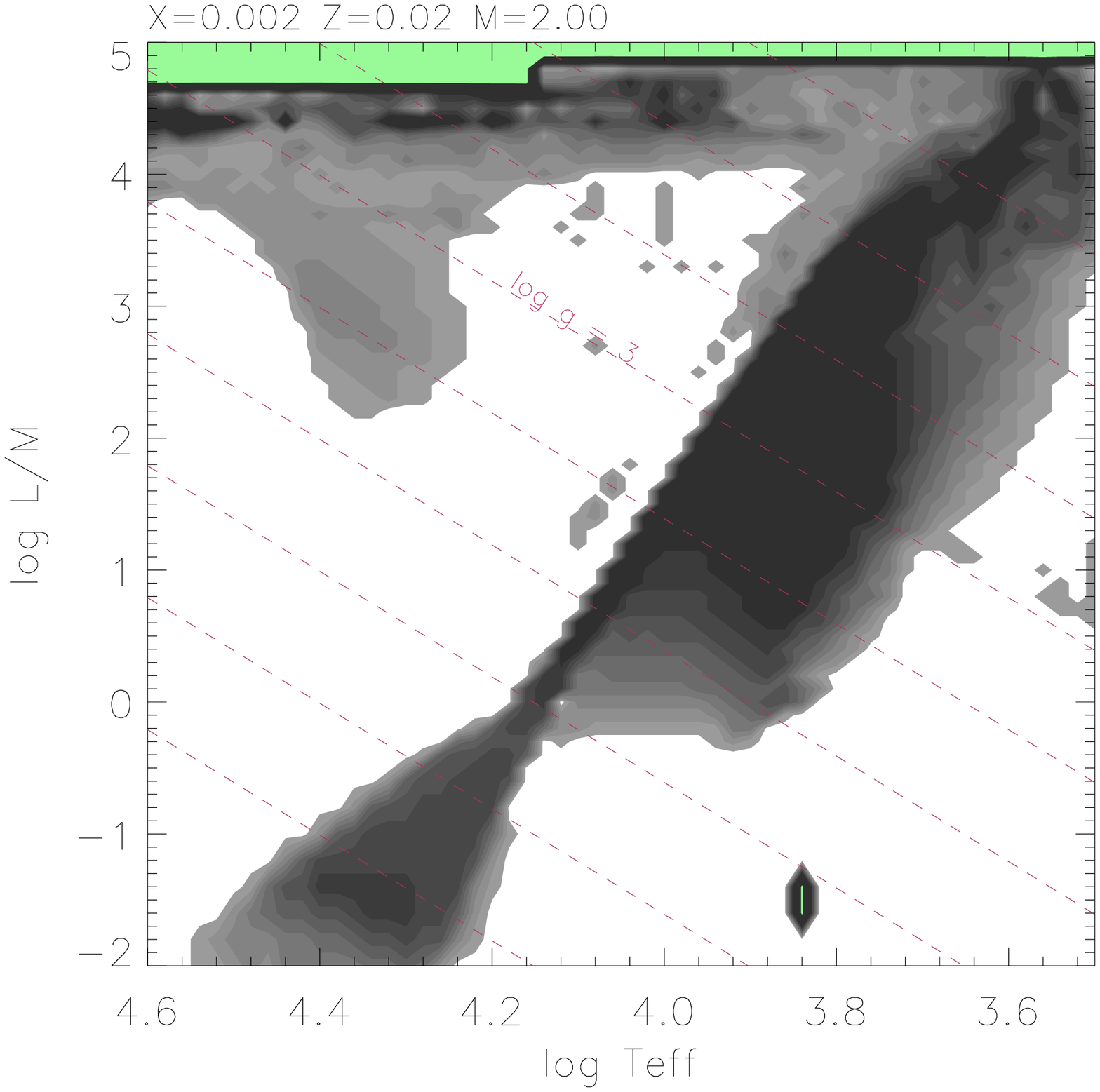,width=4.3cm,angle=0}
\epsfig{file=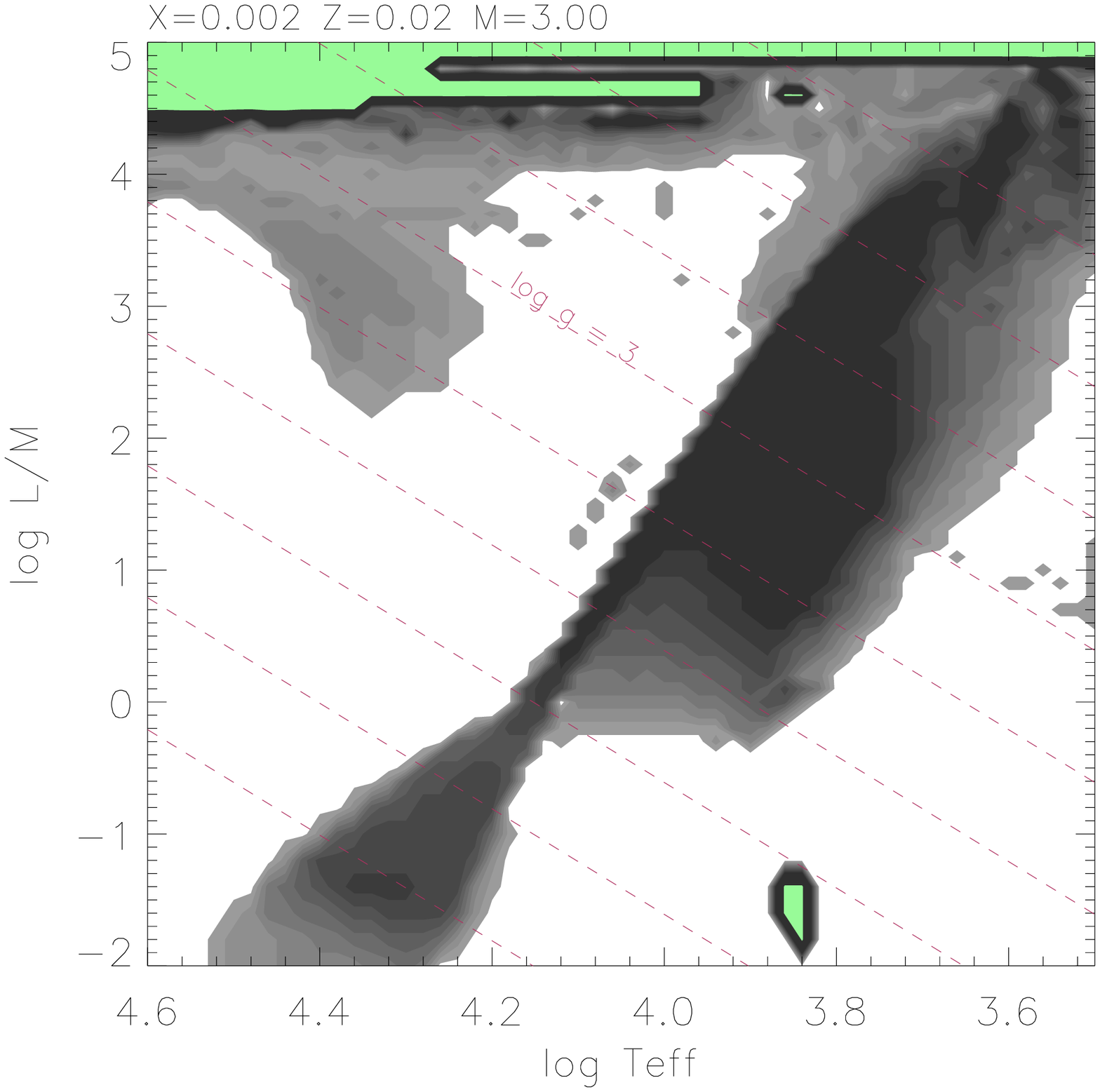,width=4.3cm,angle=0}\\
\epsfig{file=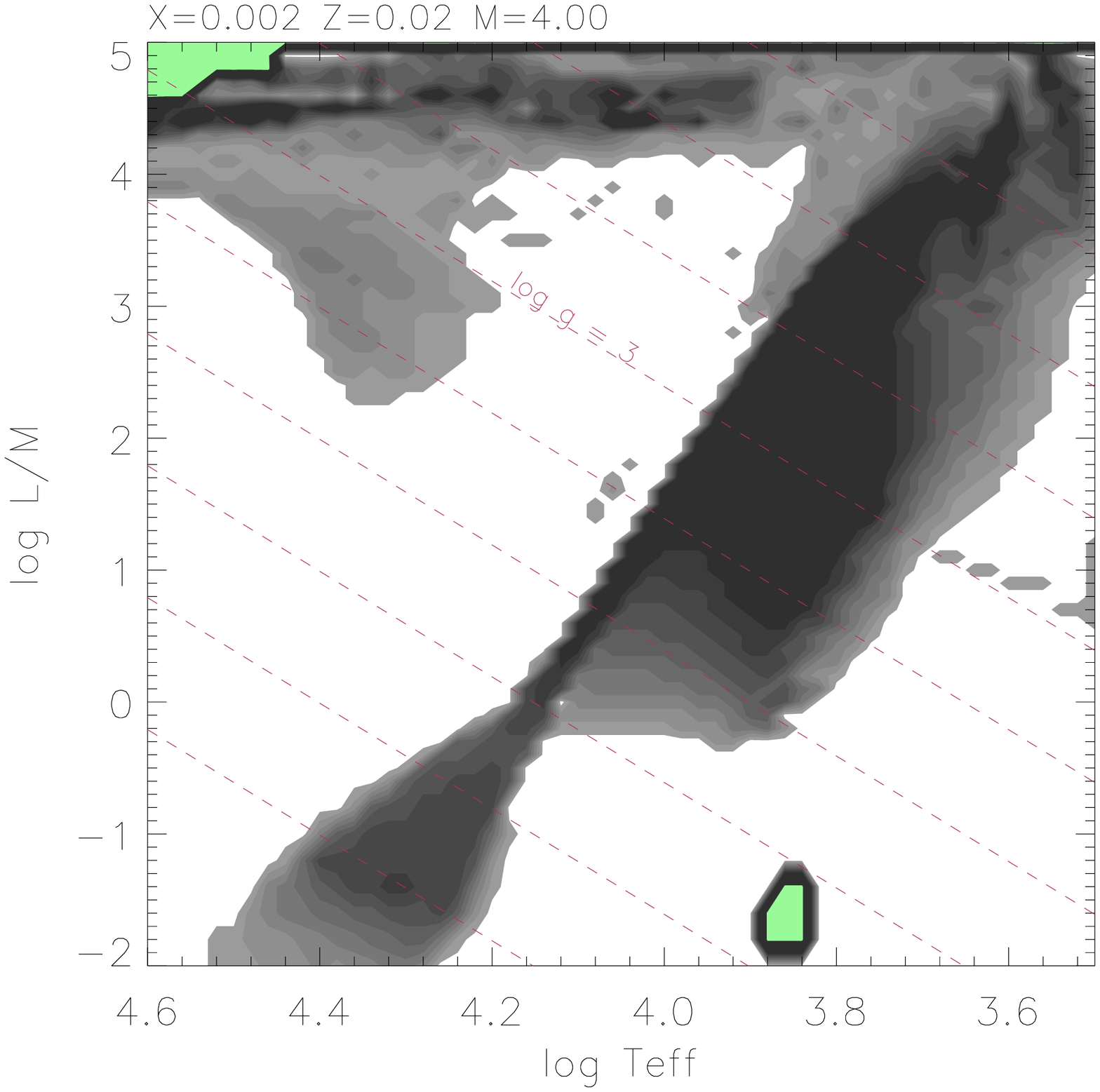,width=4.3cm,angle=0}
\epsfig{file=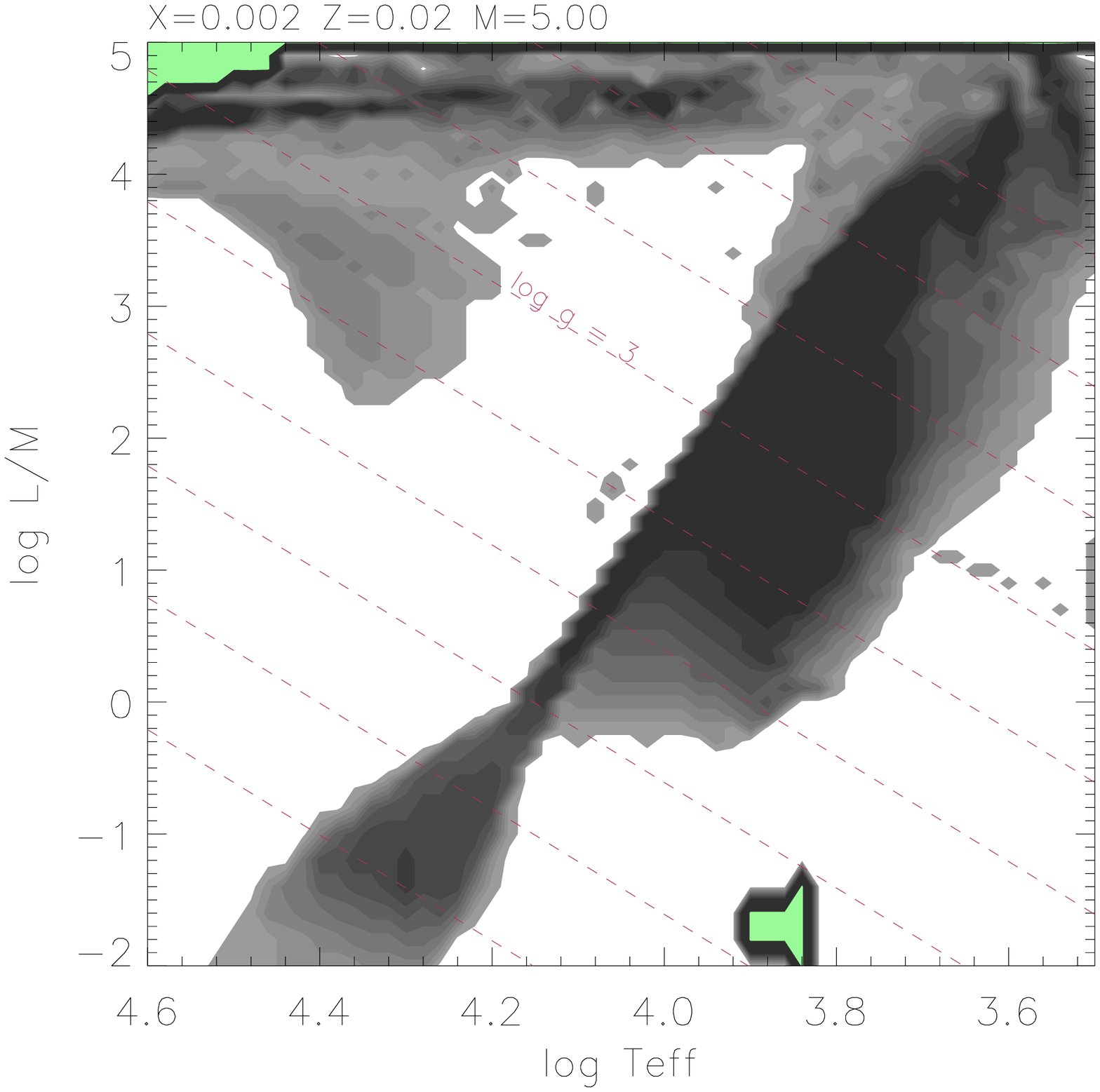,width=4.3cm,angle=0}
\epsfig{file=figs/nmodes_x002z02m07.0_00_opal.eps,width=4.3cm,angle=0}
\epsfig{file=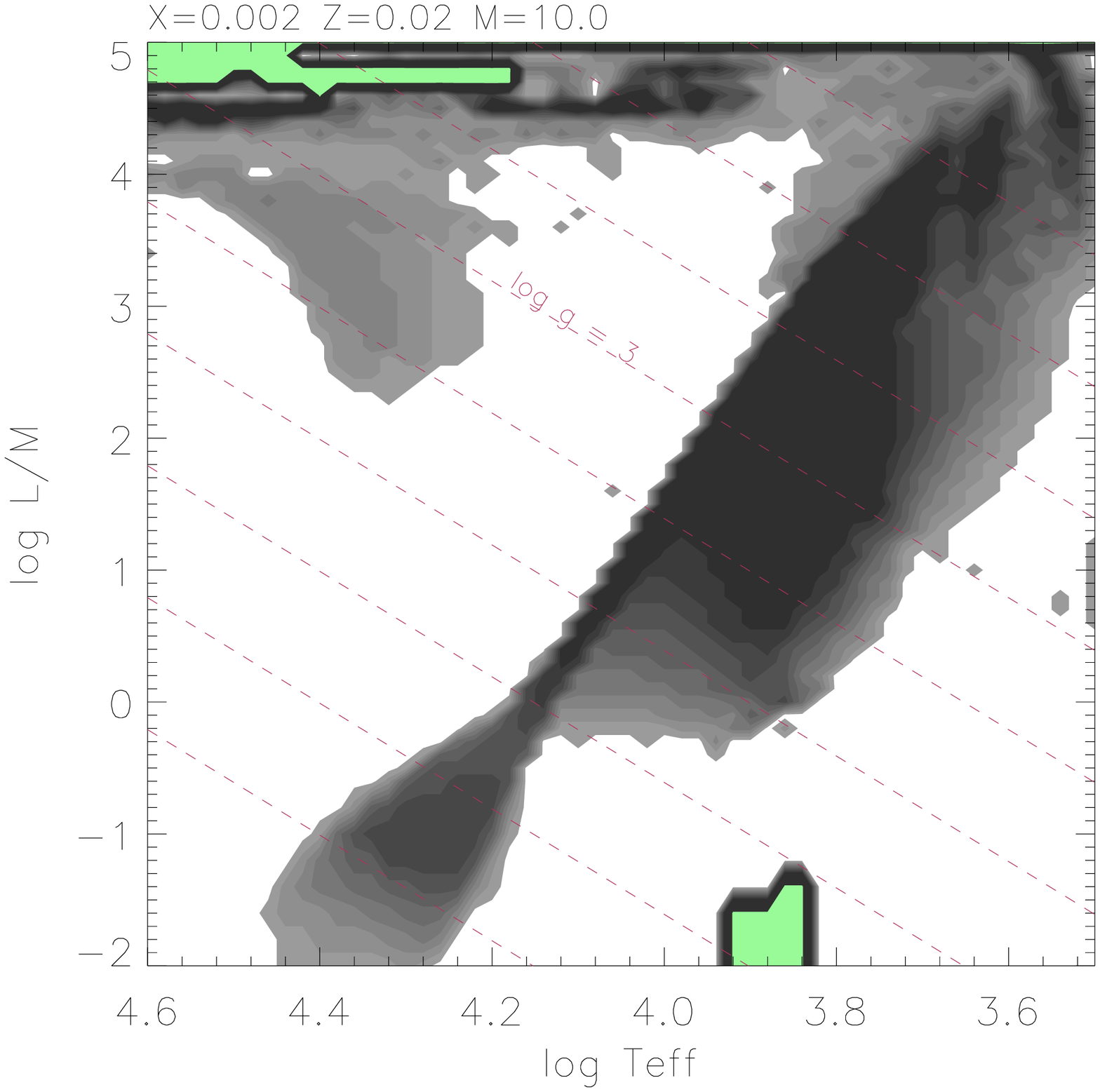,width=4.3cm,angle=0}\\
\epsfig{file=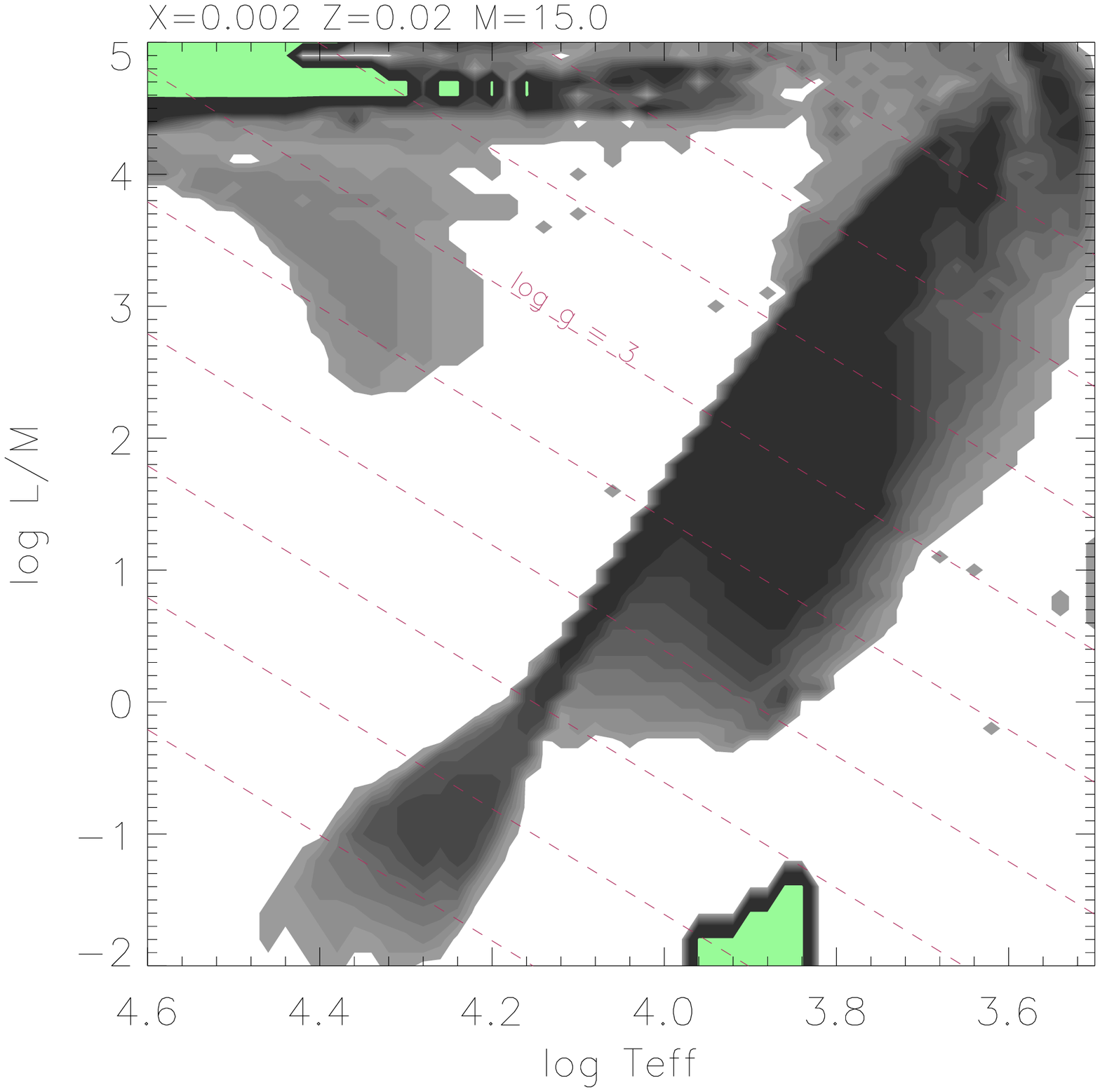,width=4.3cm,angle=0}
\epsfig{file=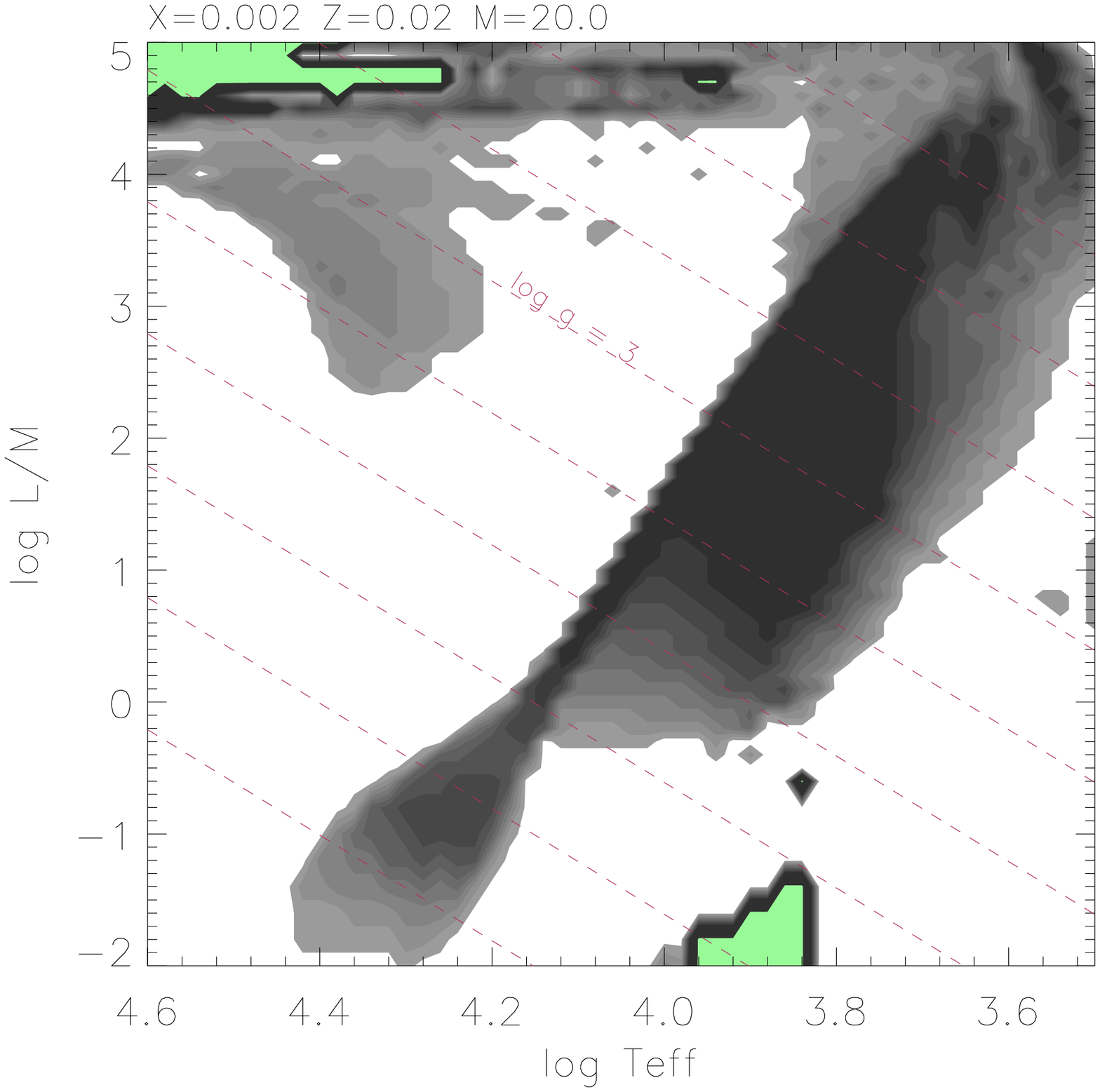,width=4.3cm,angle=0}
\epsfig{file=figs/nmodes_x002z02m30.0_00_opal.eps,width=4.3cm,angle=0}
\epsfig{file=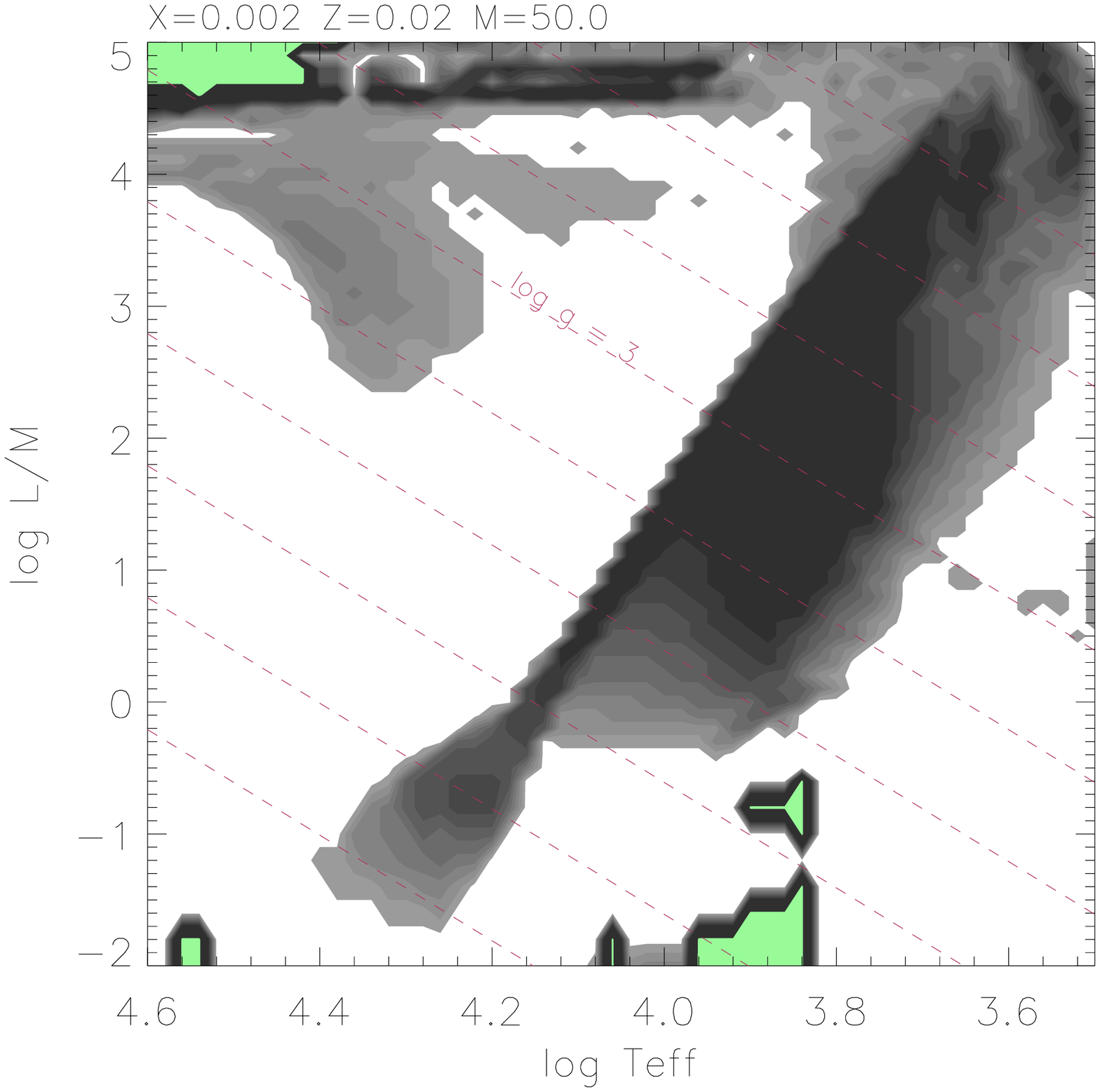,width=4.3cm,angle=0}
\caption[Unstable modes: $X=0.002, Z=0.02$]
{As Fig.~\ref{f:nx10} with $X=0.002, Z=0.02$. 
}
\label{f:nx002}
\end{center}
\end{figure*}

\begin{figure*}
\begin{center}
\epsfig{file=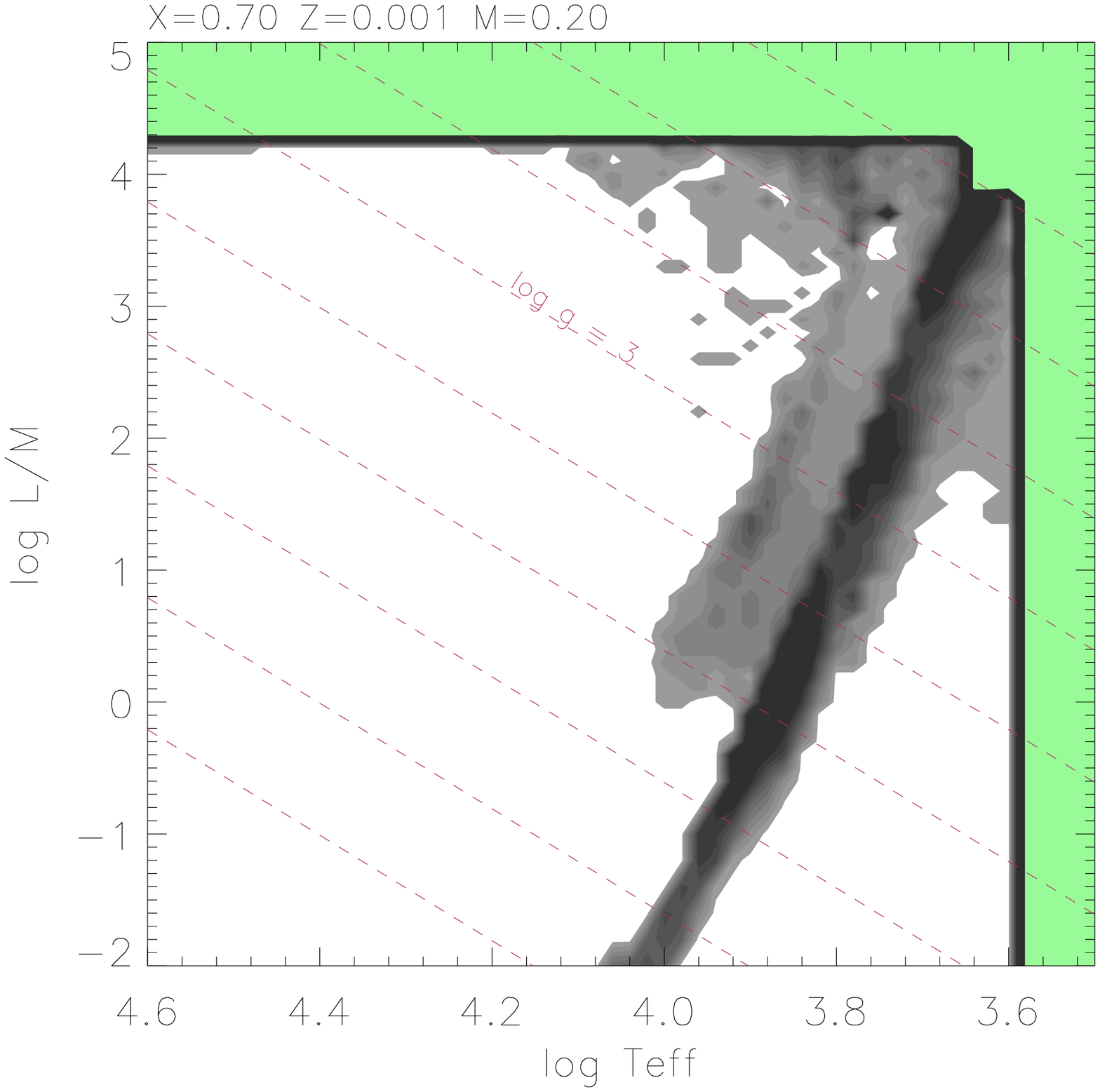,width=4.3cm,angle=0}
\epsfig{file=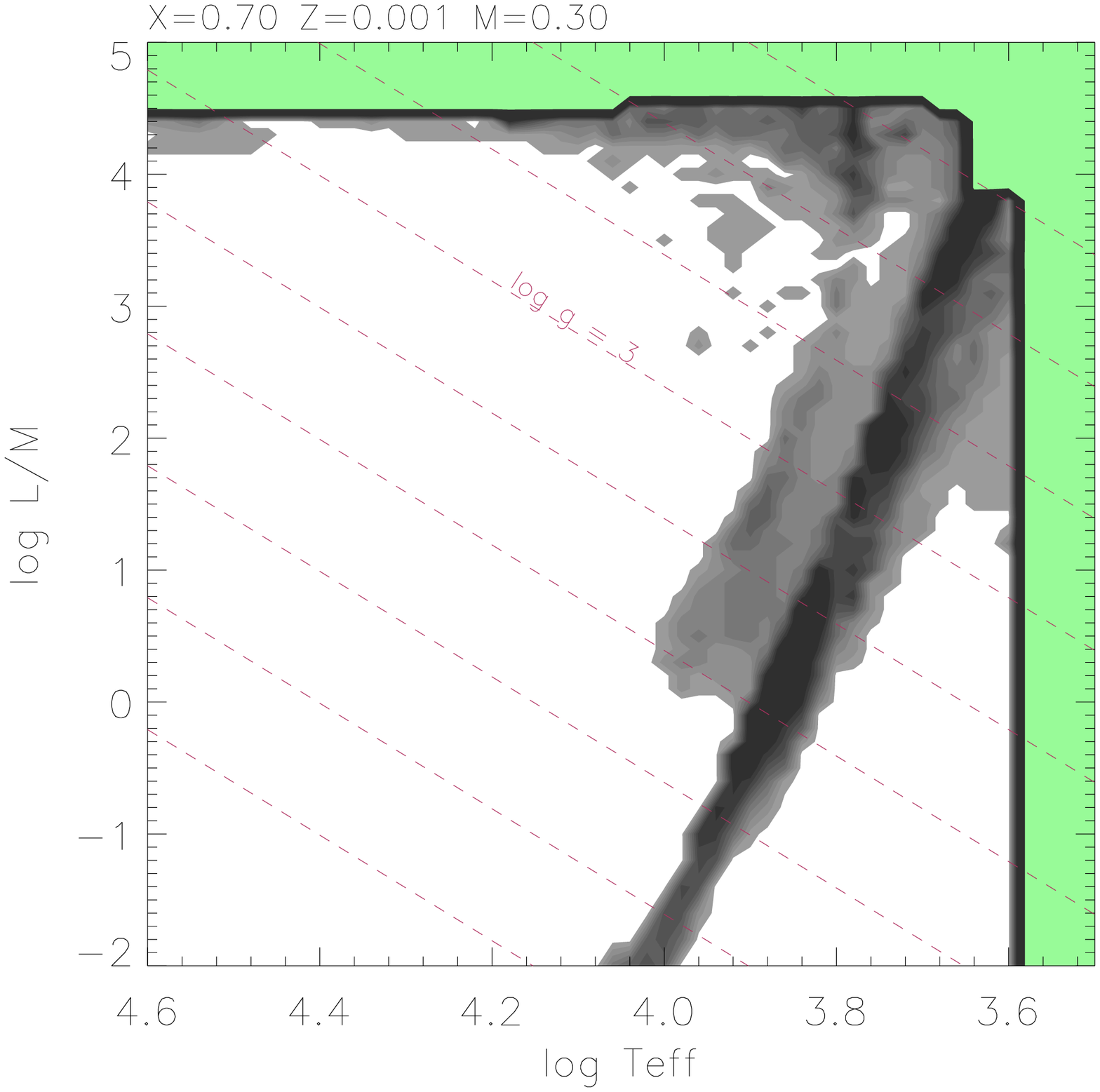,width=4.3cm,angle=0}
\epsfig{file=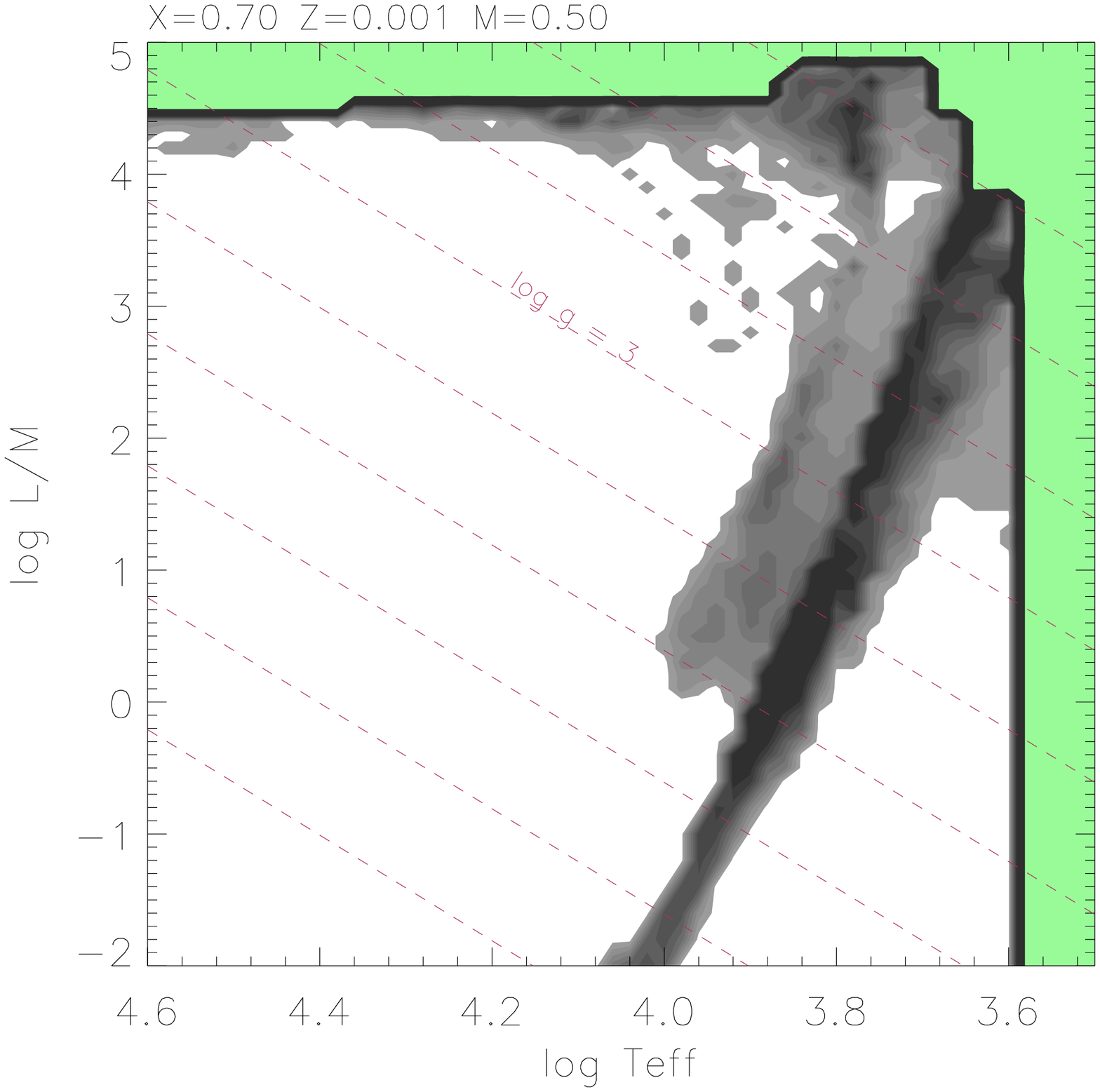,width=4.3cm,angle=0}
\epsfig{file=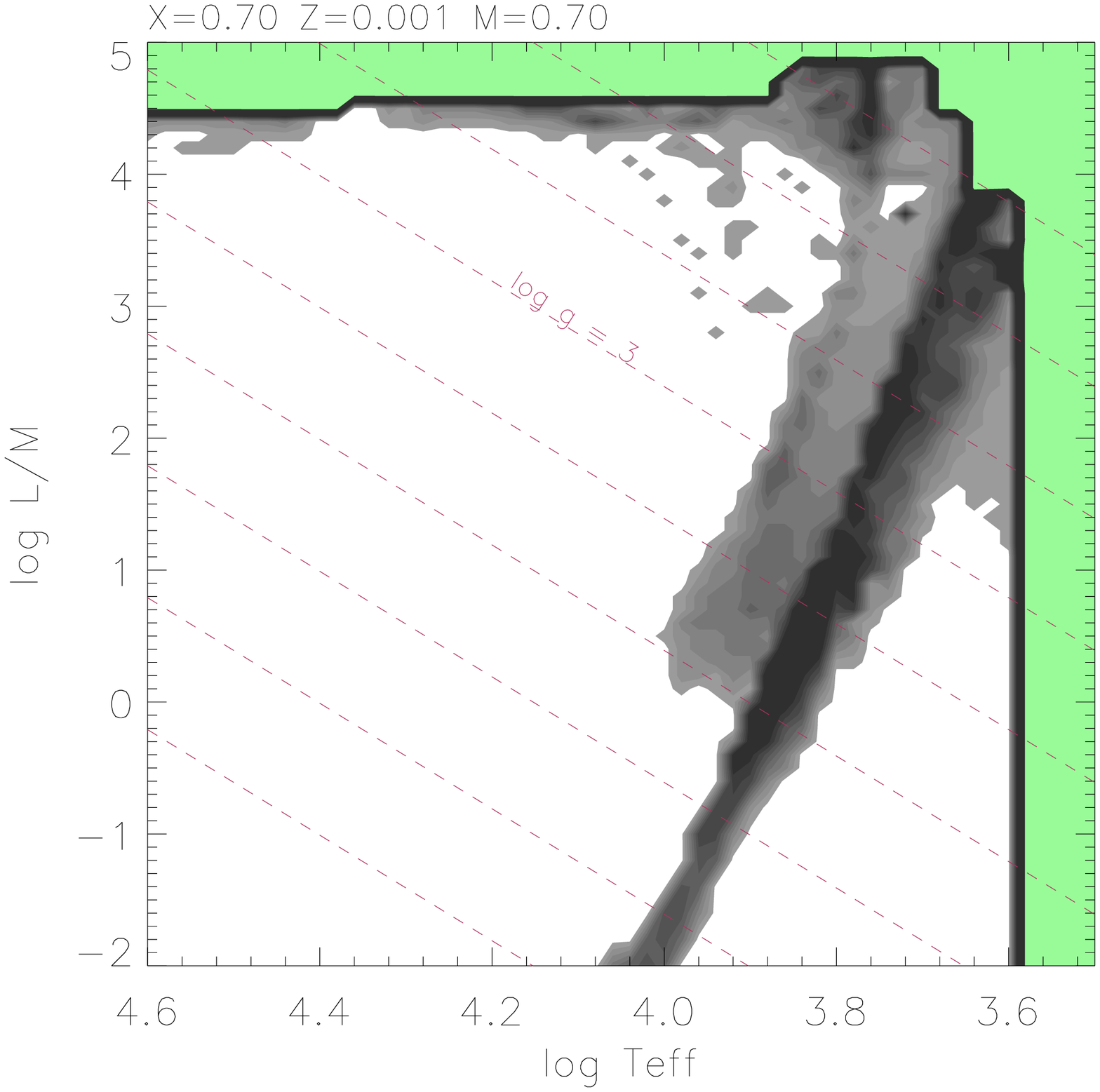,width=4.3cm,angle=0}\\
\epsfig{file=figs/nmodes_x70z001m01.0_00_opal.eps,width=4.3cm,angle=0}
\epsfig{file=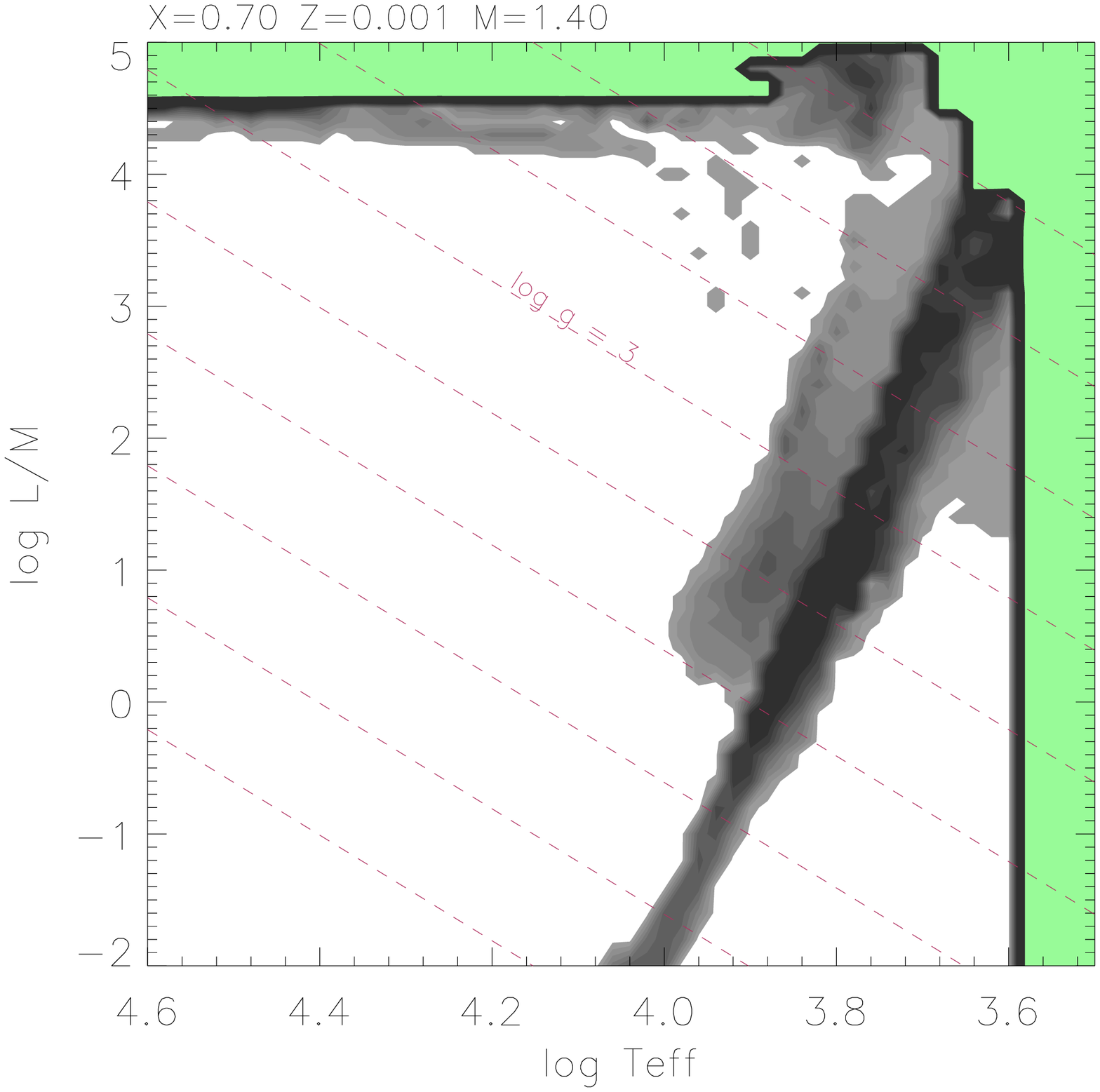,width=4.3cm,angle=0}
\epsfig{file=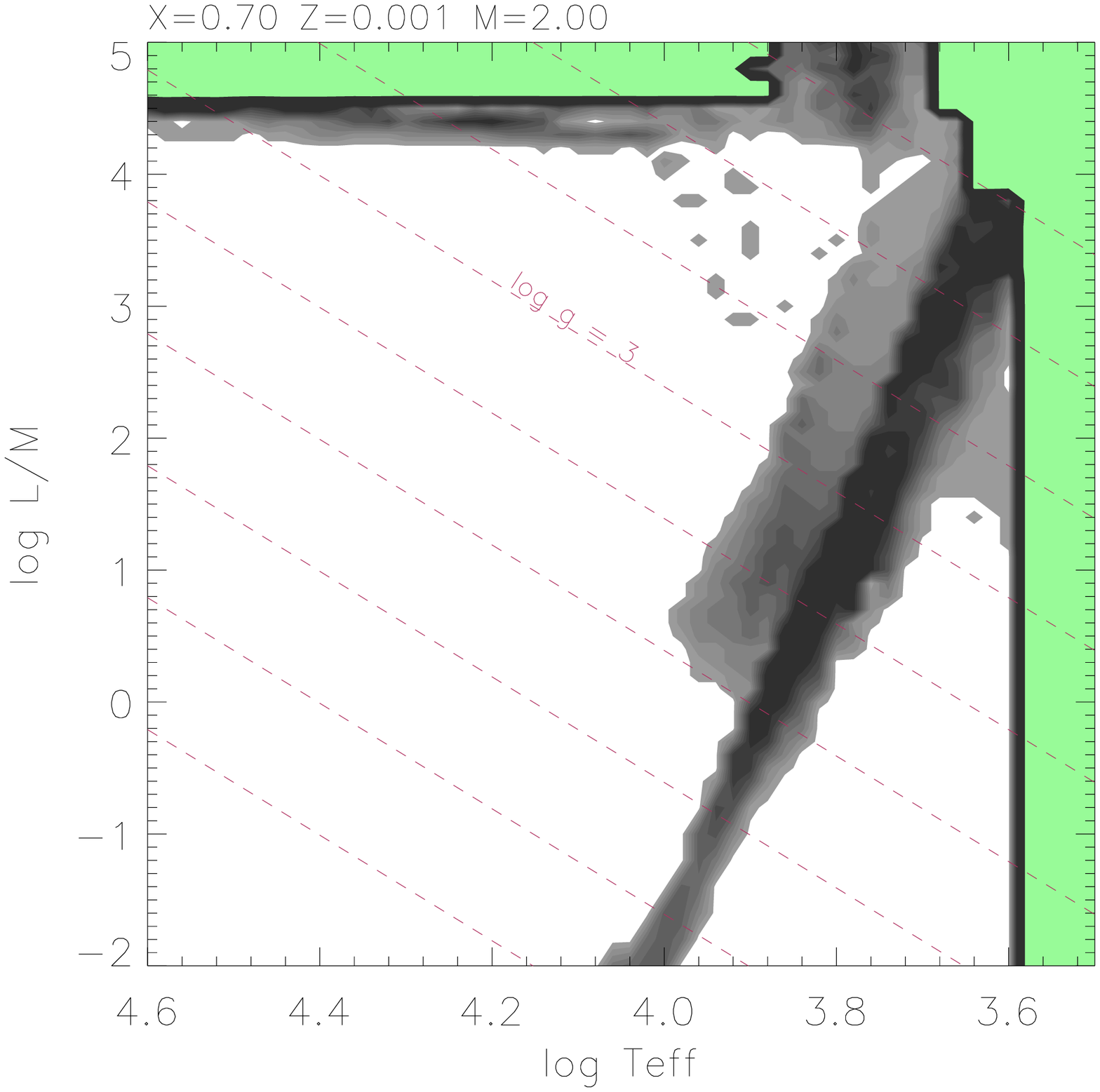,width=4.3cm,angle=0}
\epsfig{file=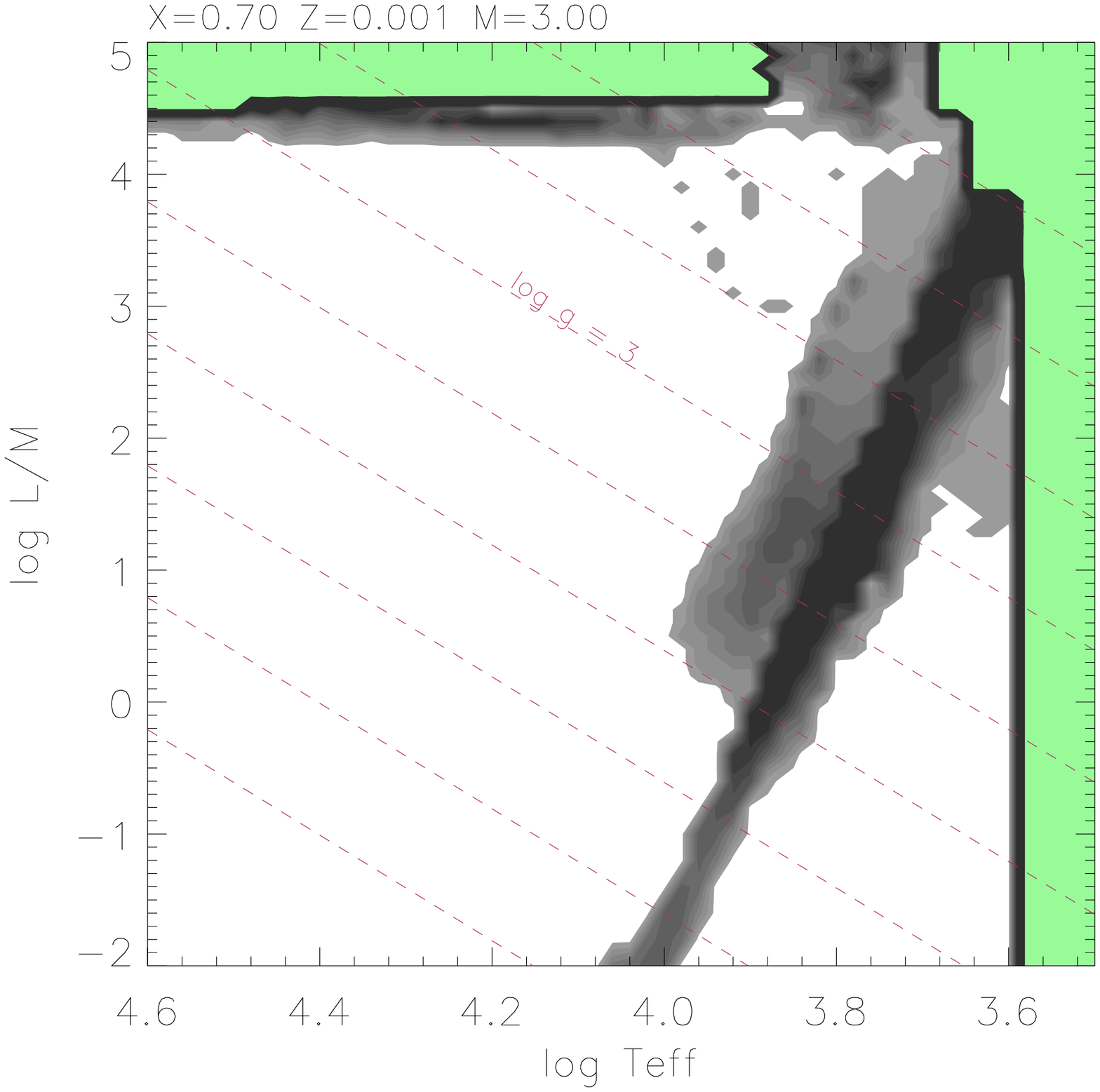,width=4.3cm,angle=0}\\
\epsfig{file=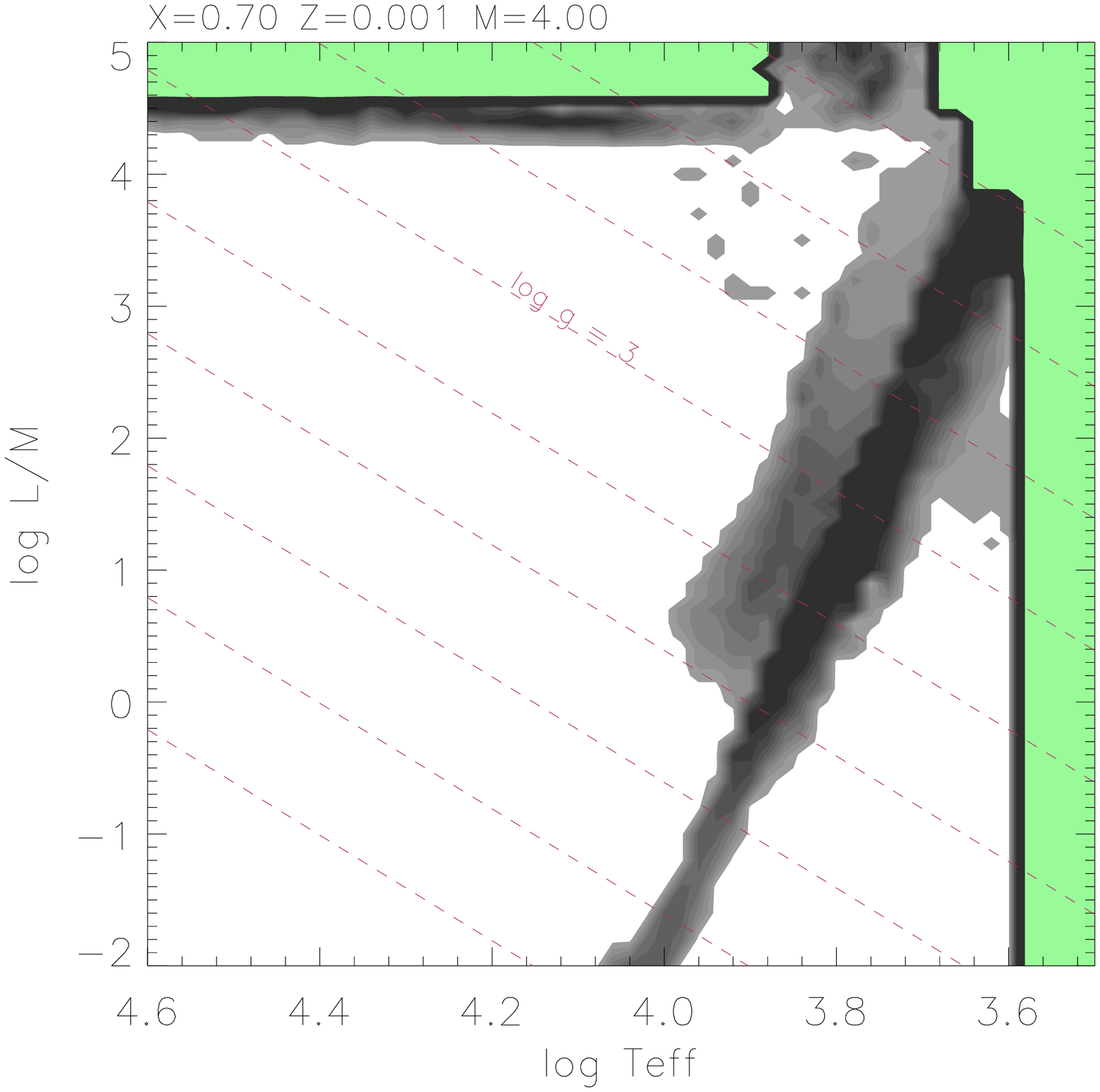,width=4.3cm,angle=0}
\epsfig{file=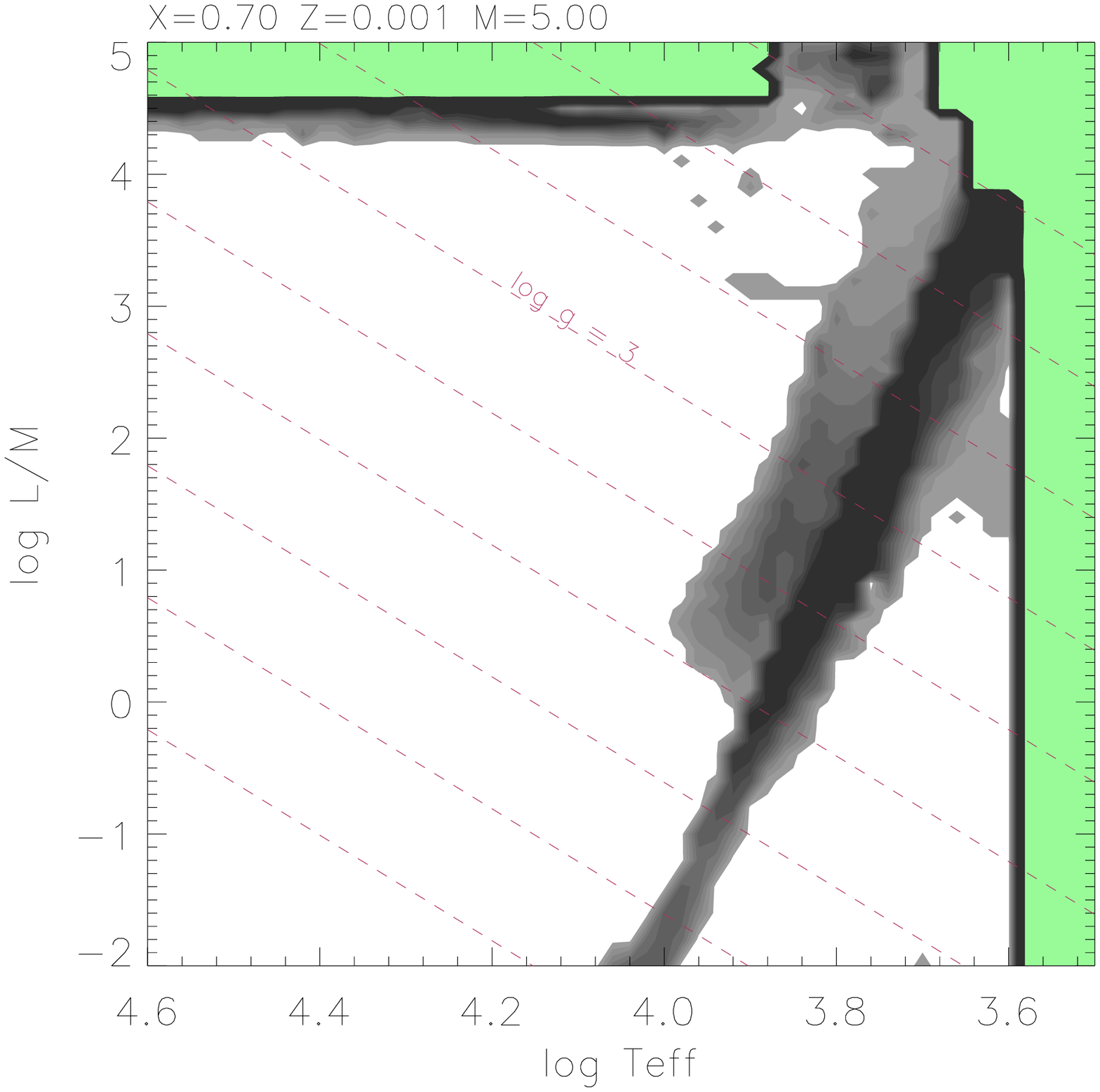,width=4.3cm,angle=0}
\epsfig{file=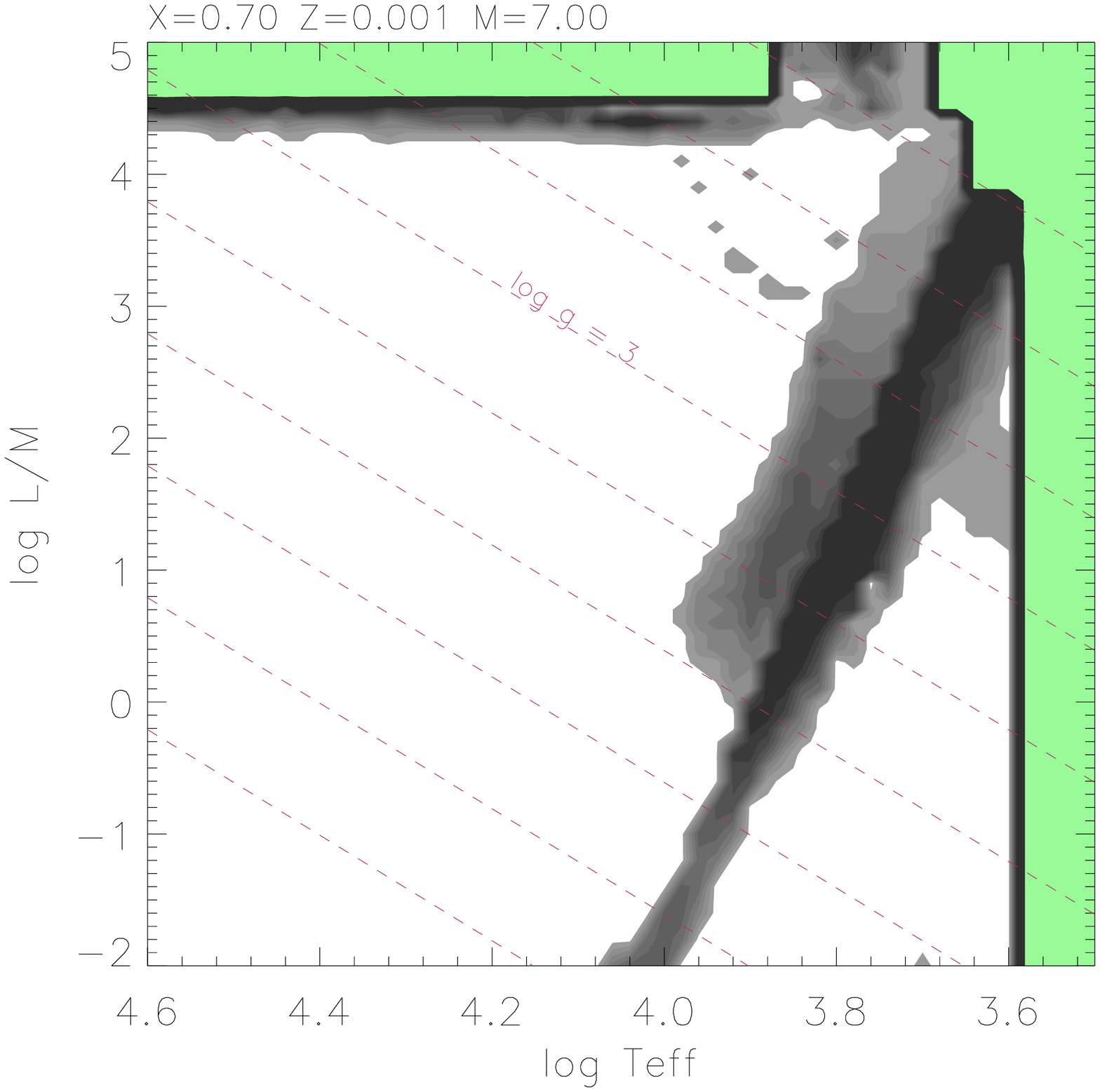,width=4.3cm,angle=0}
\epsfig{file=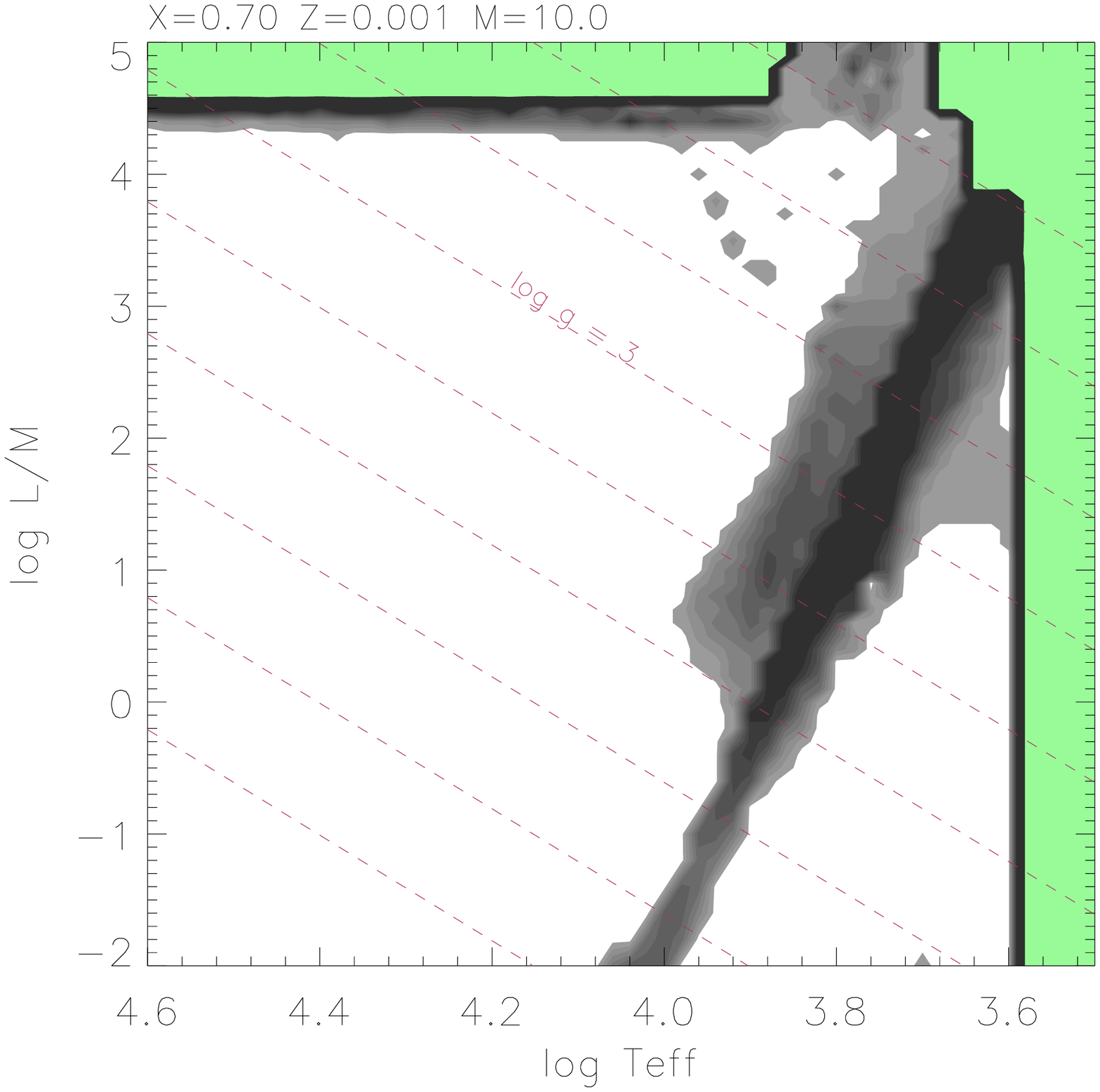,width=4.3cm,angle=0}\\
\epsfig{file=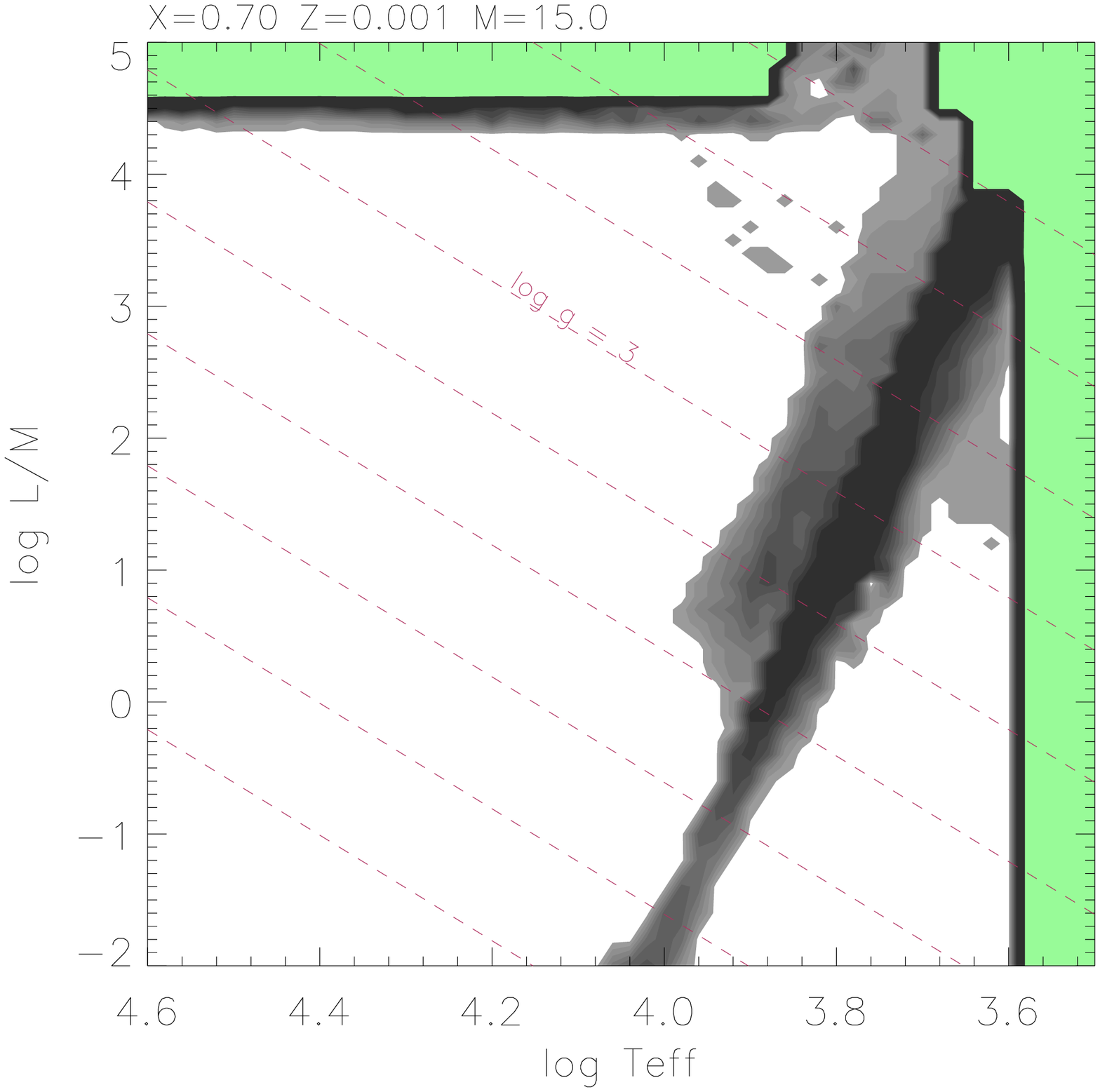,width=4.3cm,angle=0}
\epsfig{file=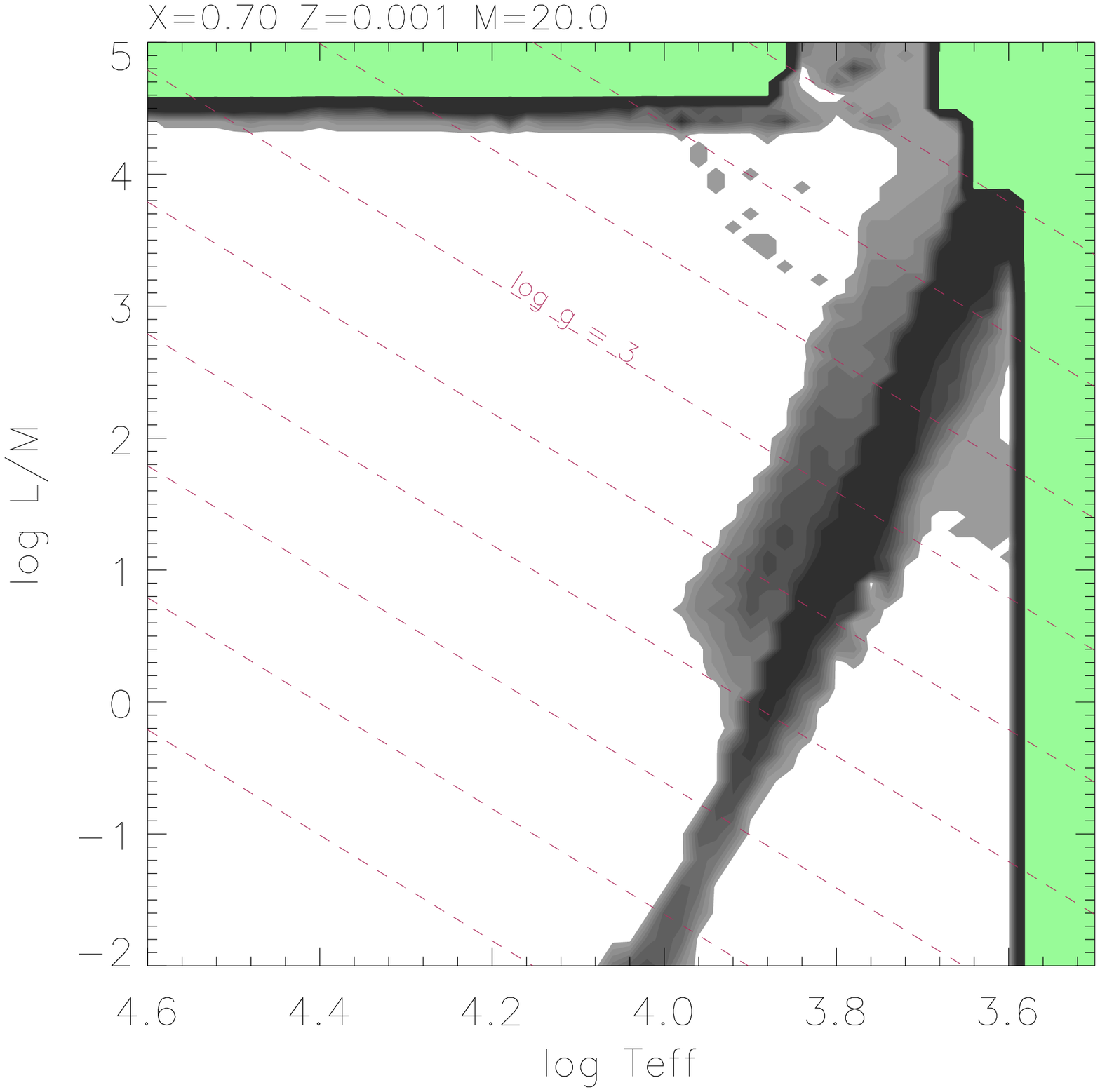,width=4.3cm,angle=0}
\epsfig{file=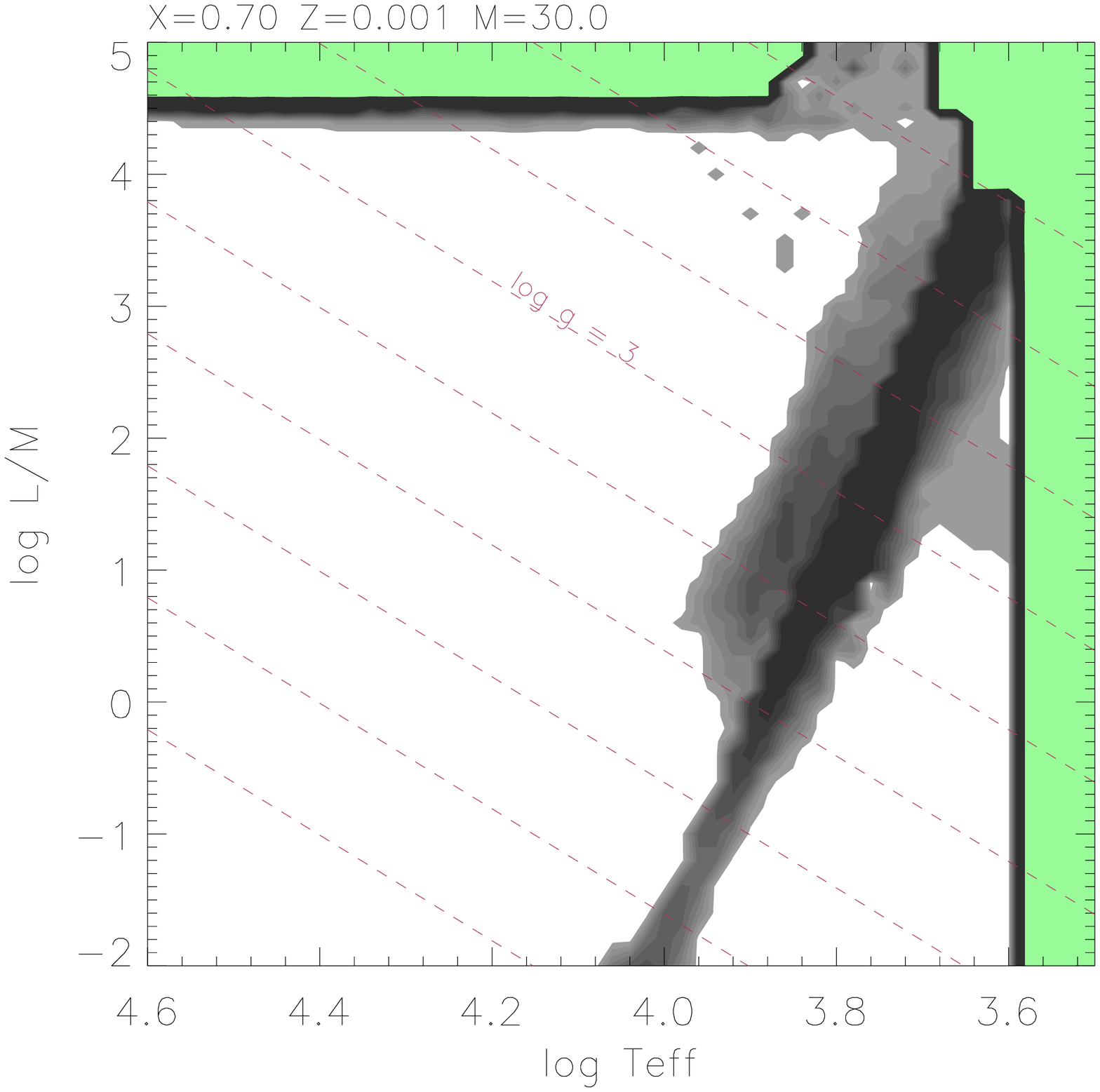,width=4.3cm,angle=0}
\epsfig{file=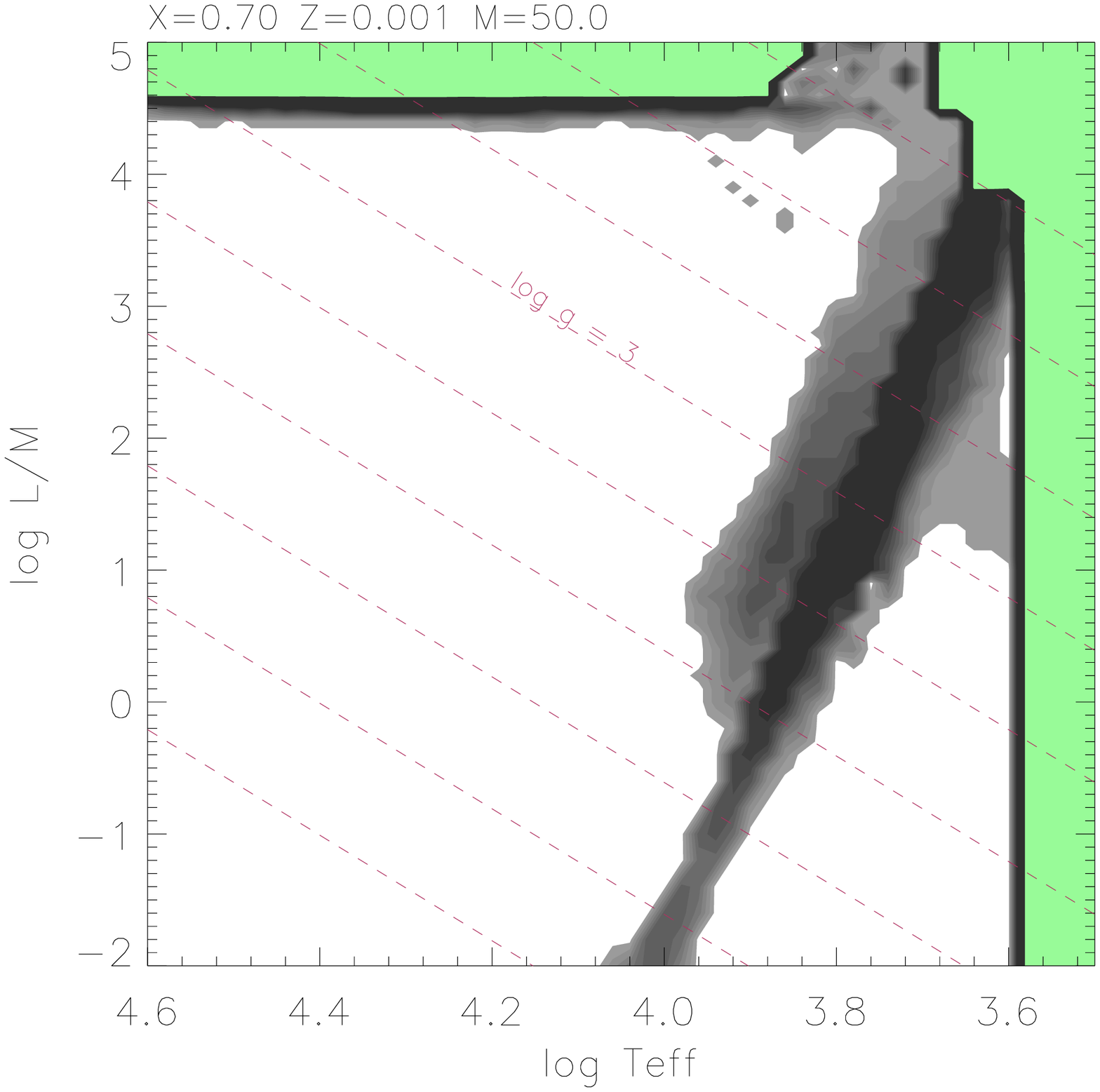,width=4.3cm,angle=0}
\caption[Unstable modes: $X=0.70, Z=0.001$]
{As Fig.~\ref{f:nx70} with $X=0.70, Z=0.001$. 
}
\label{f:nx70z001}
\end{center}
\end{figure*}

\begin{figure*}
\begin{center}
\epsfig{file=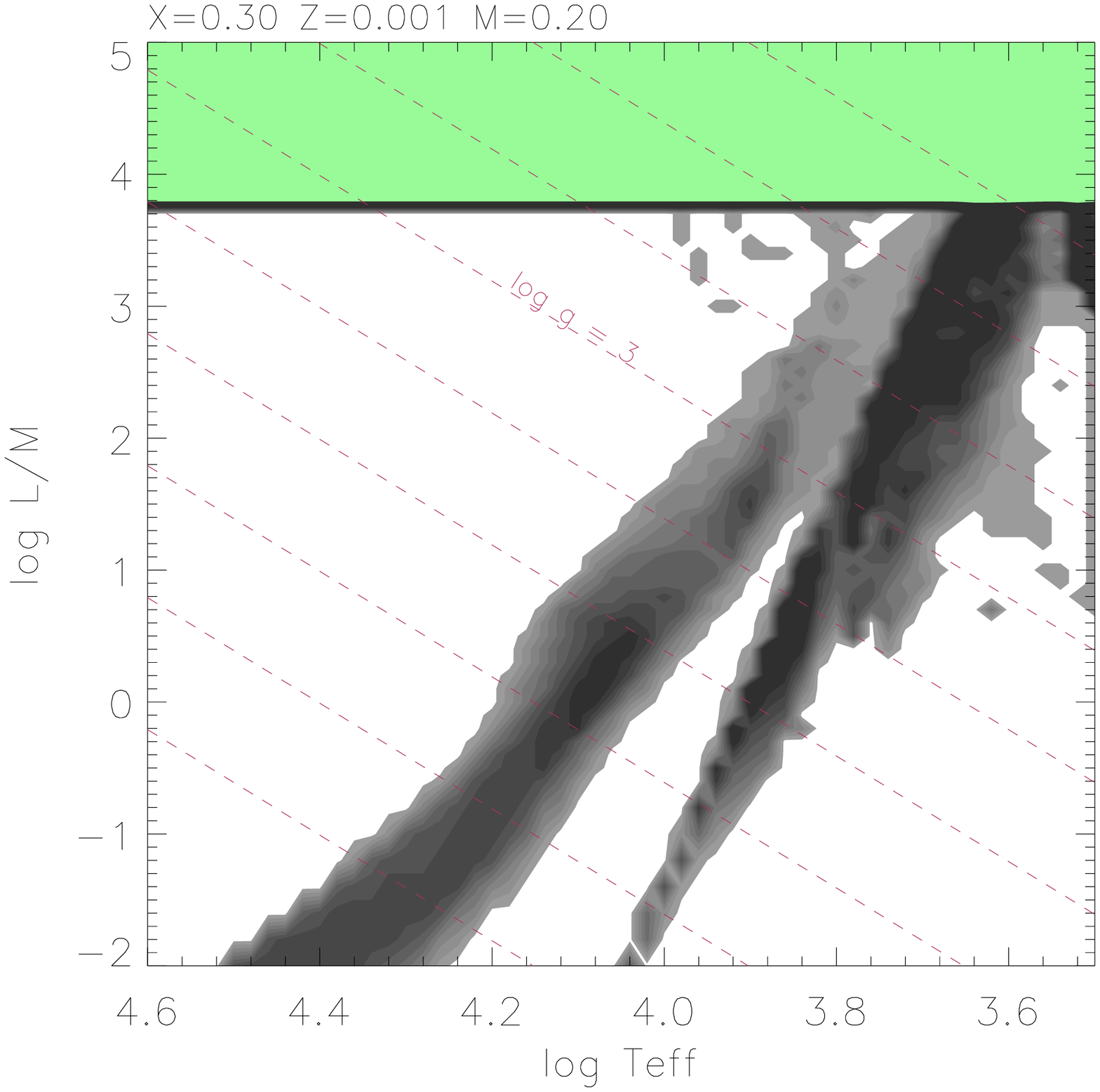,width=4.3cm,angle=0}
\epsfig{file=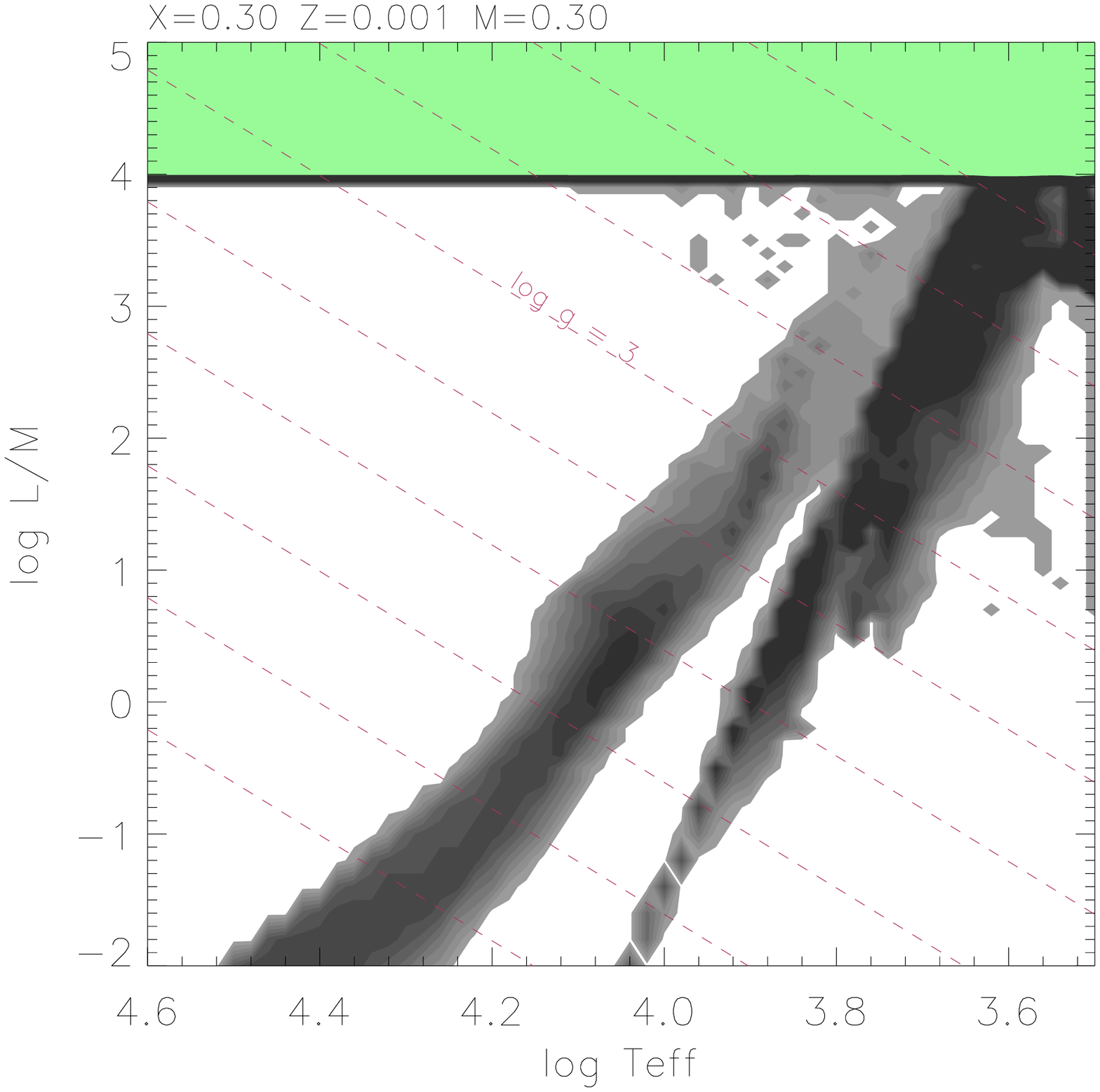,width=4.3cm,angle=0}
\epsfig{file=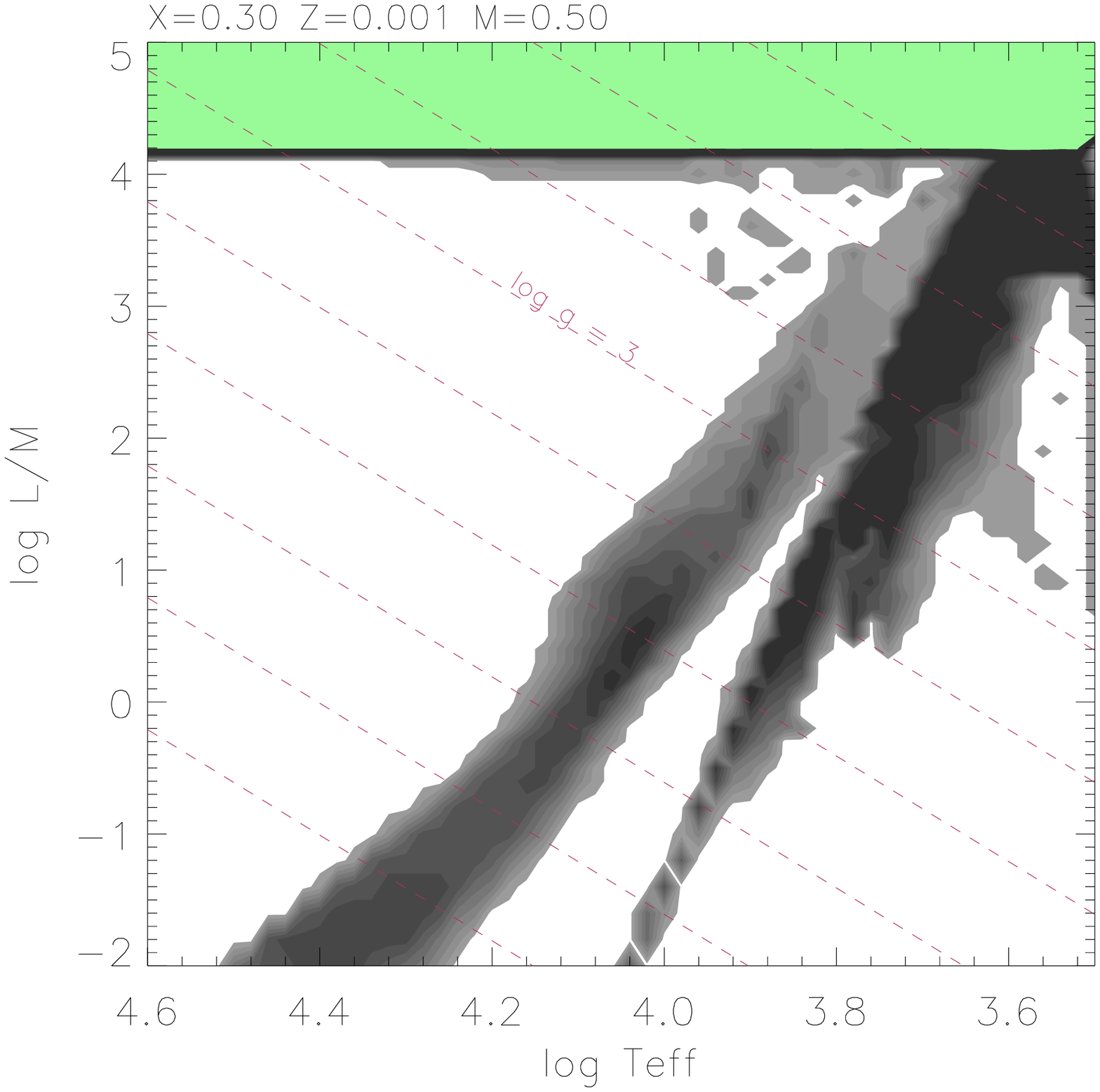,width=4.3cm,angle=0}
\epsfig{file=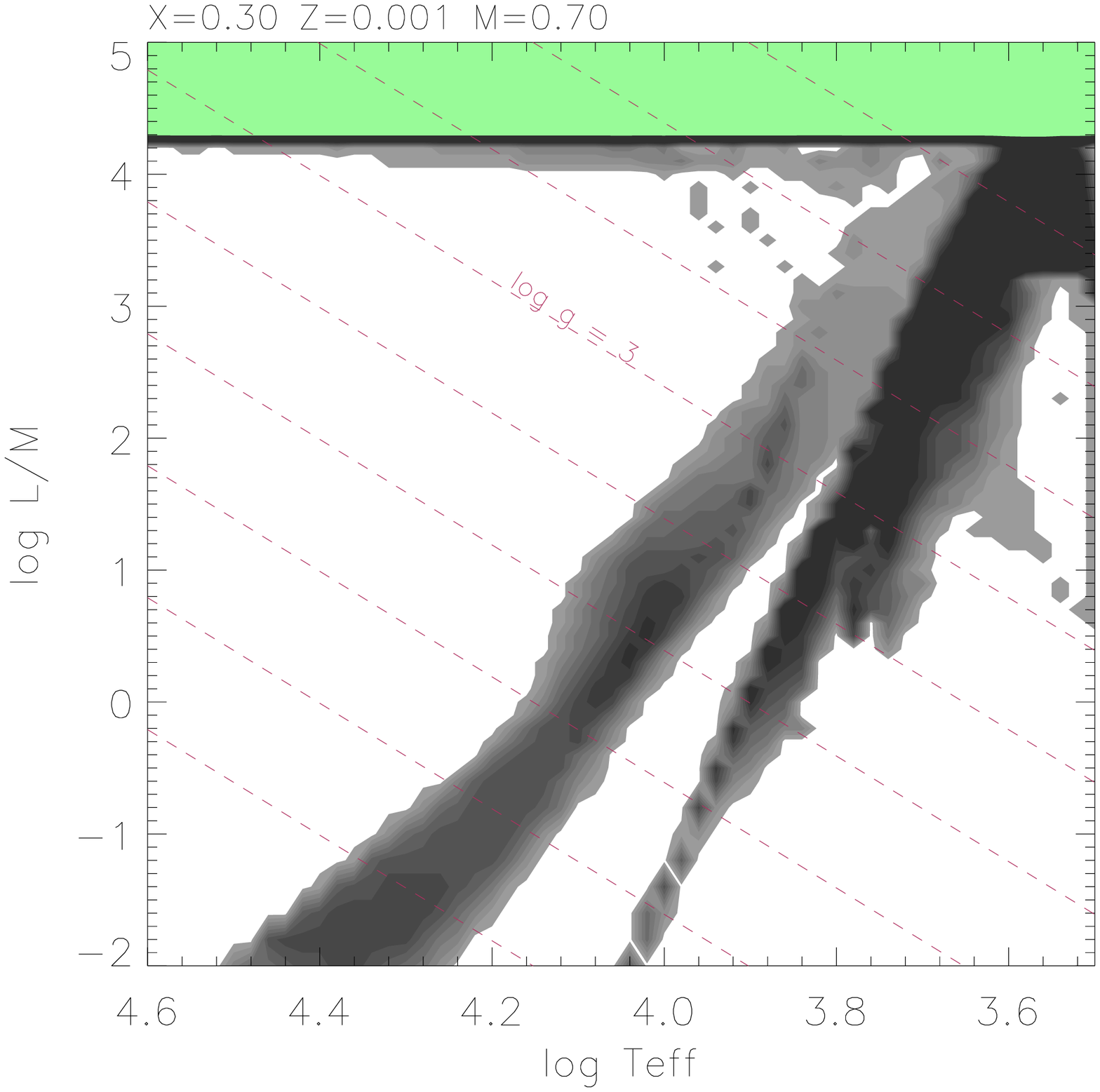,width=4.3cm,angle=0}\\
\epsfig{file=figs/nmodes_x30z001m01.0_00_opal.eps,width=4.3cm,angle=0}
\epsfig{file=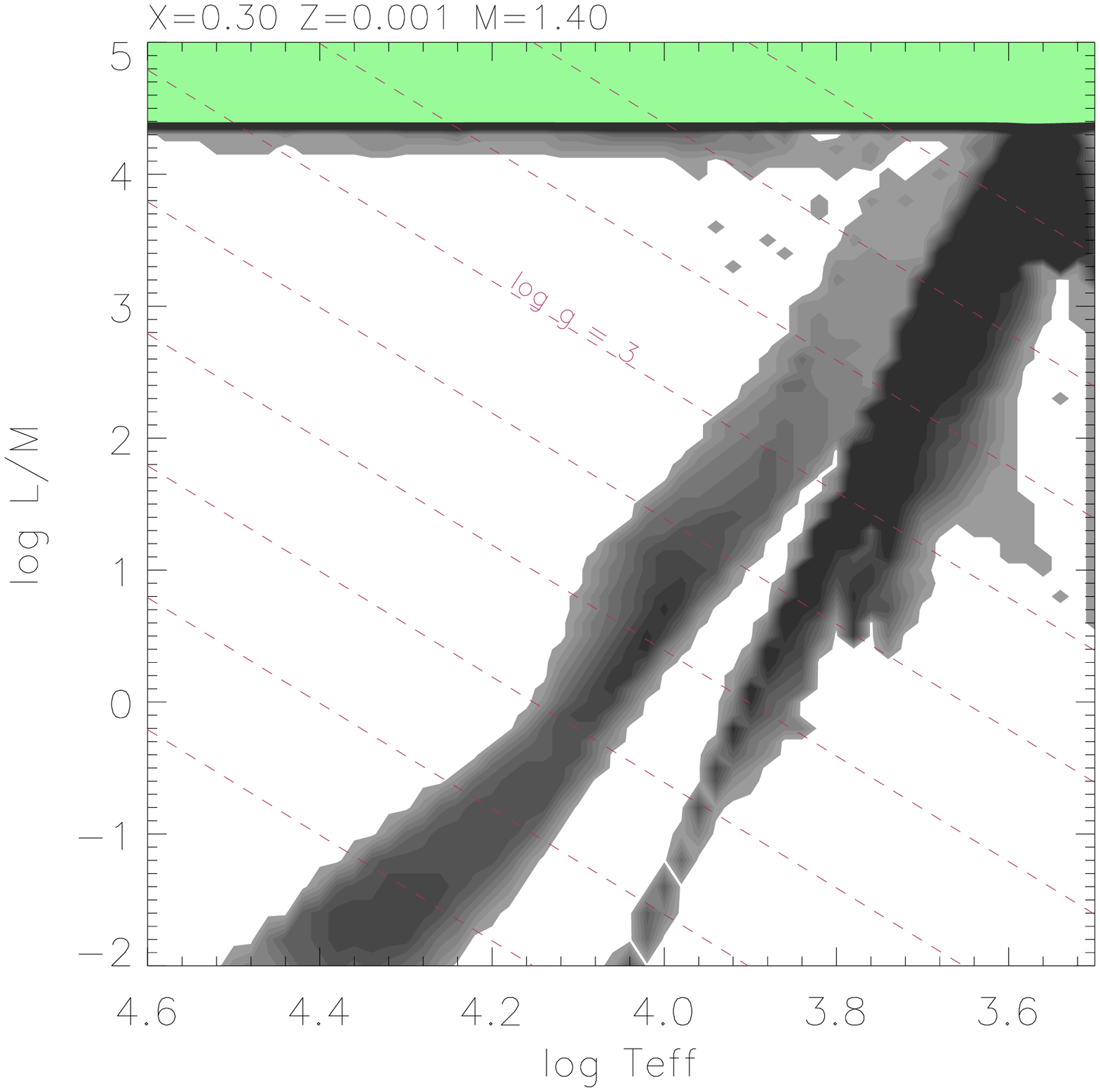,width=4.3cm,angle=0}
\epsfig{file=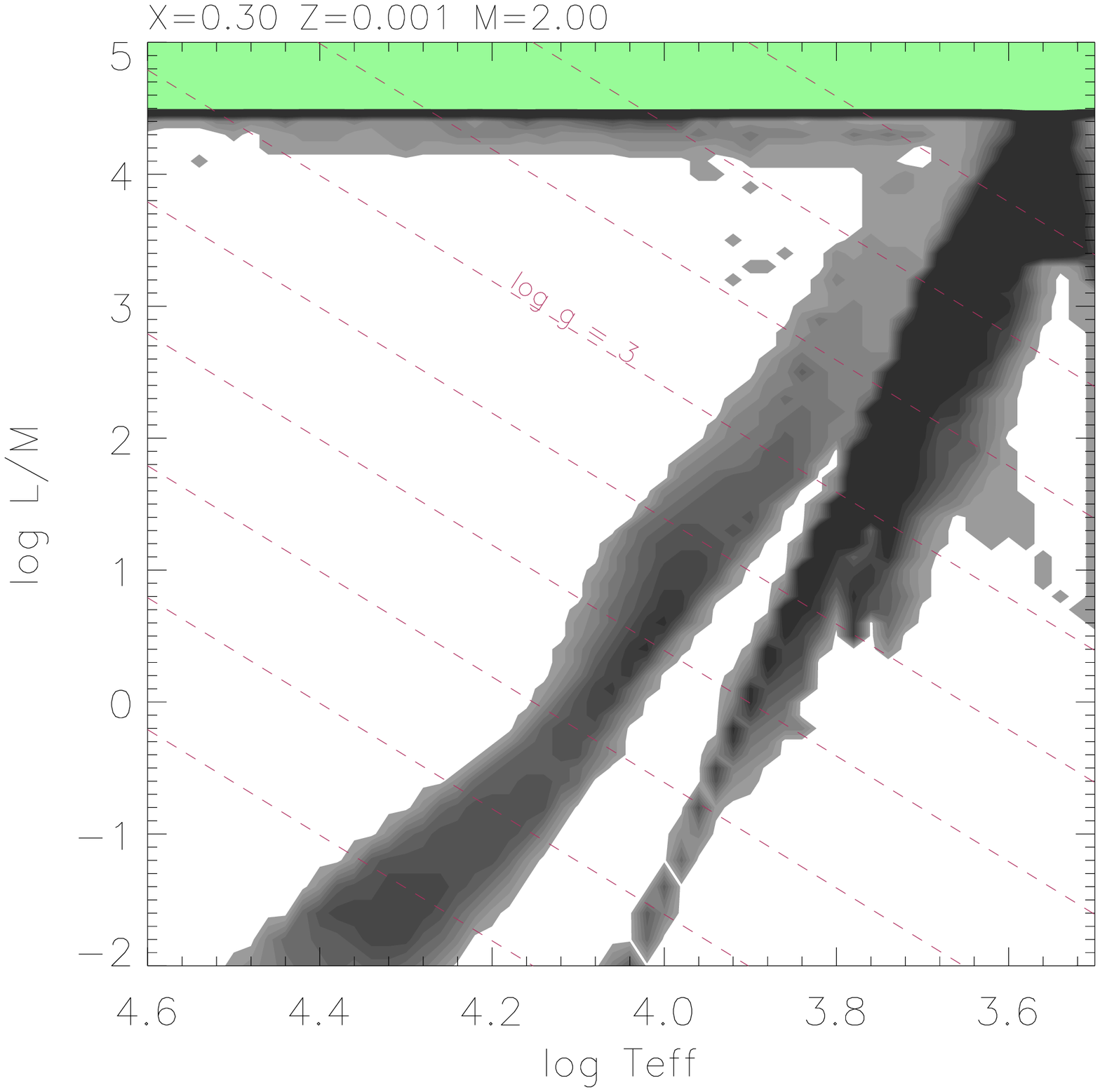,width=4.3cm,angle=0}
\epsfig{file=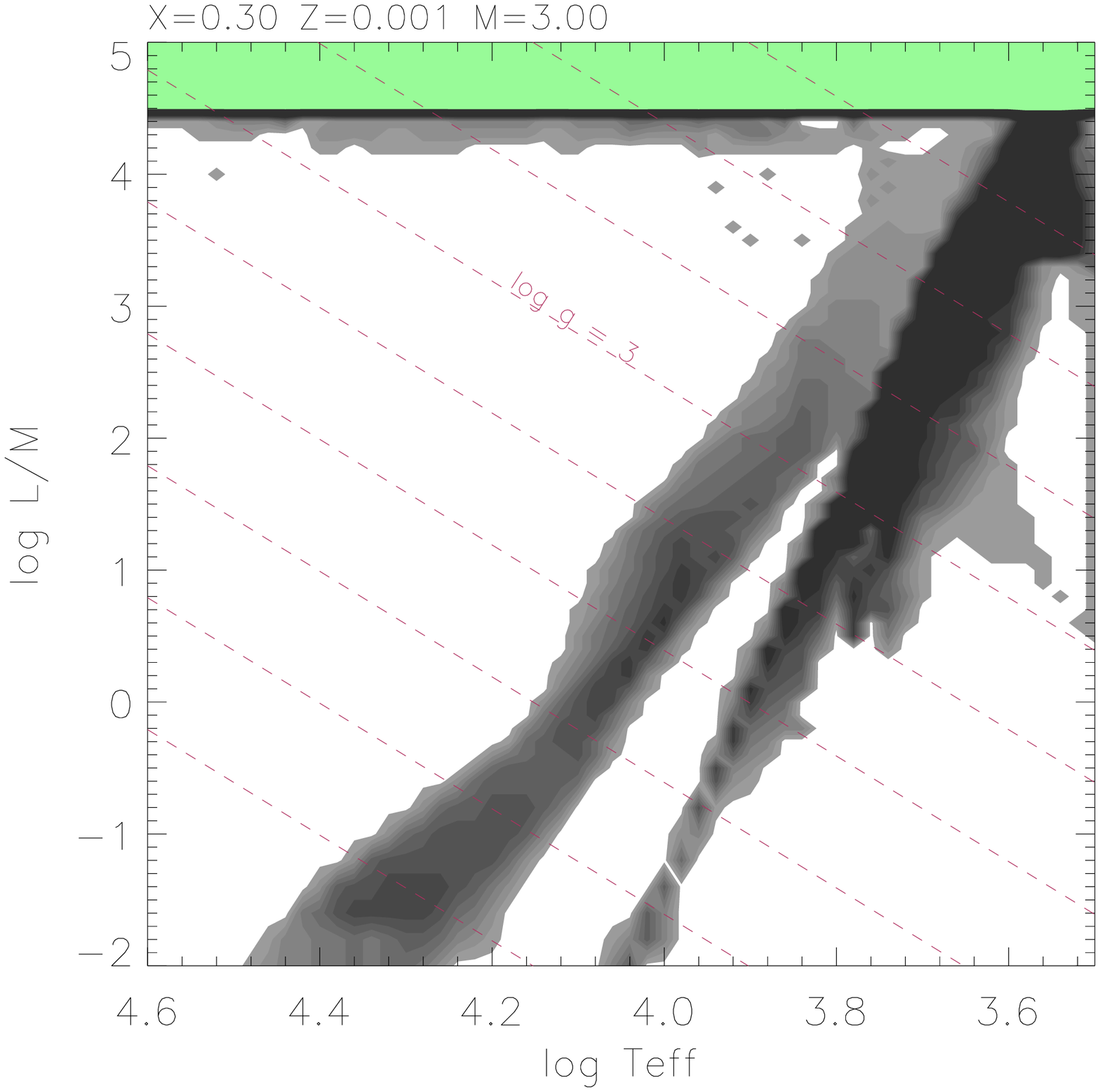,width=4.3cm,angle=0}\\
\epsfig{file=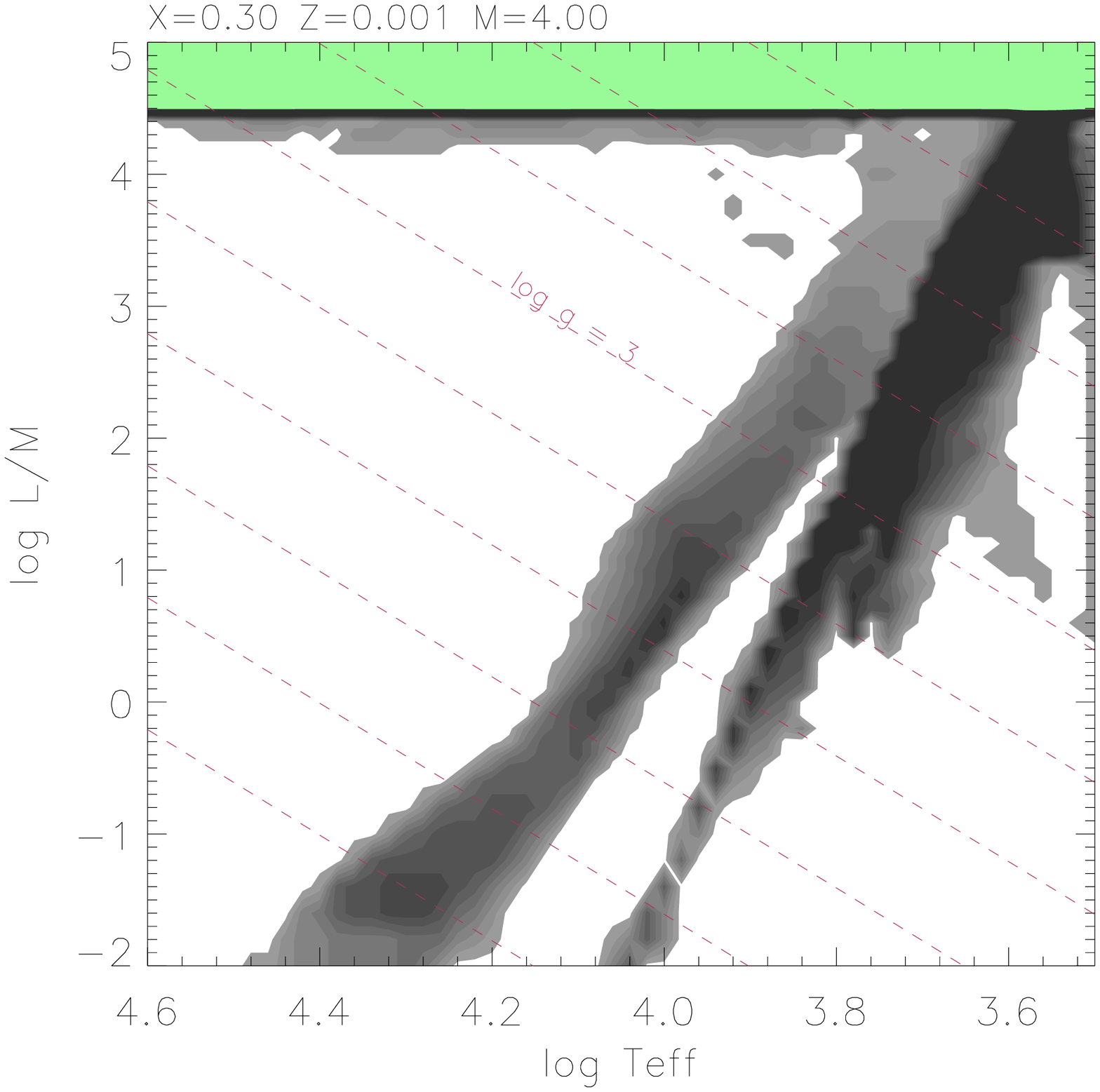,width=4.3cm,angle=0}
\epsfig{file=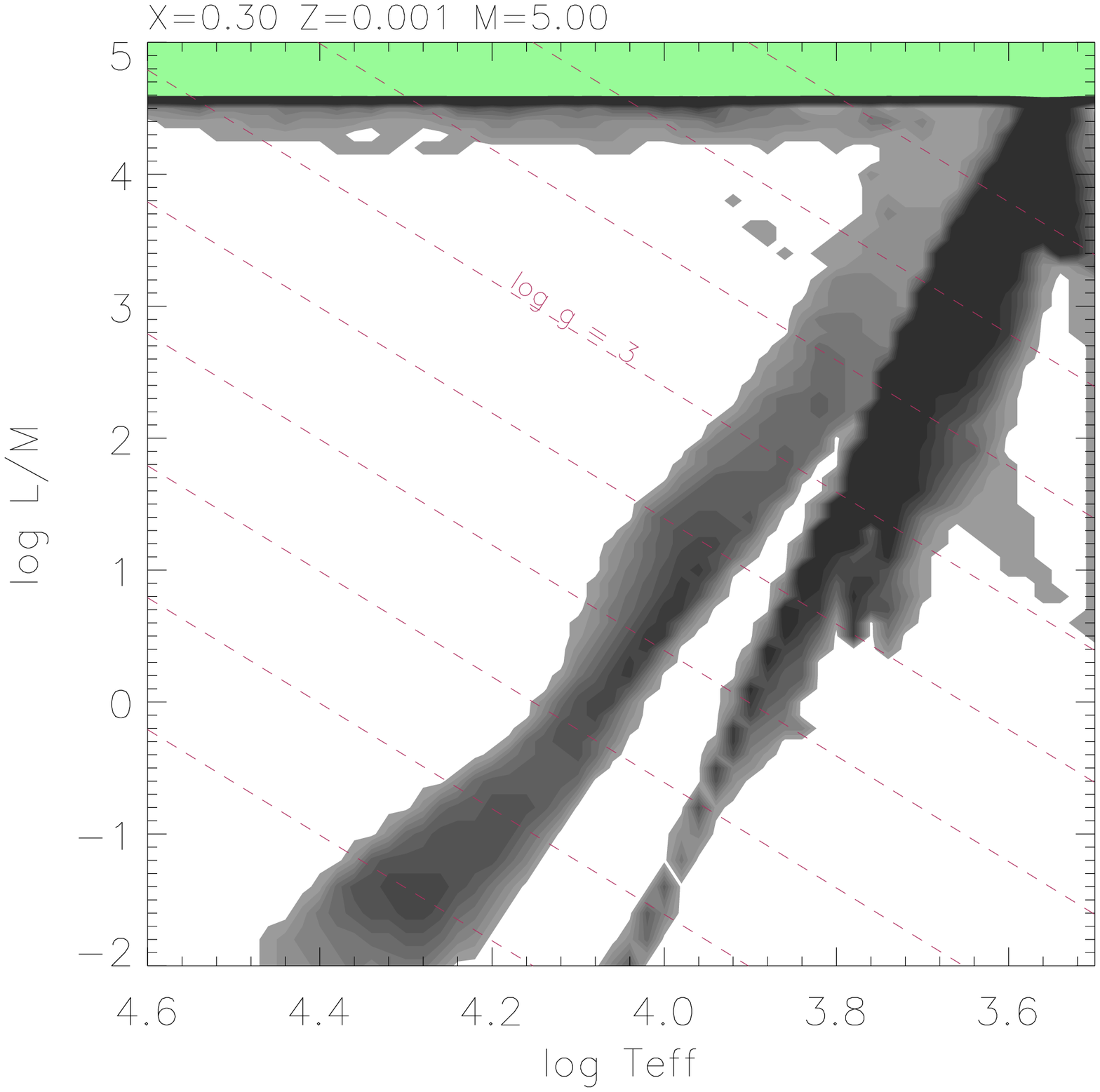,width=4.3cm,angle=0}
\epsfig{file=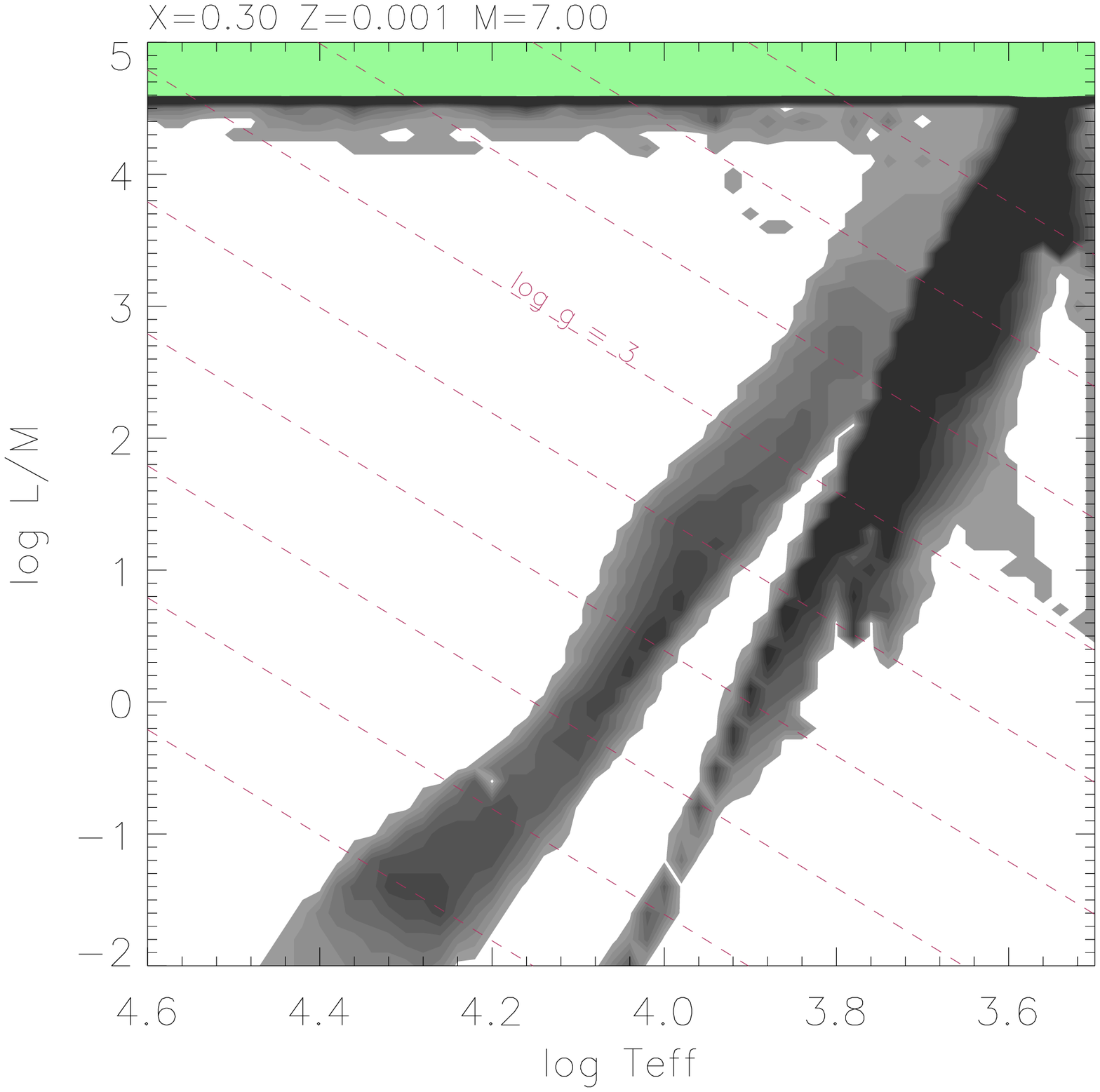,width=4.3cm,angle=0}
\epsfig{file=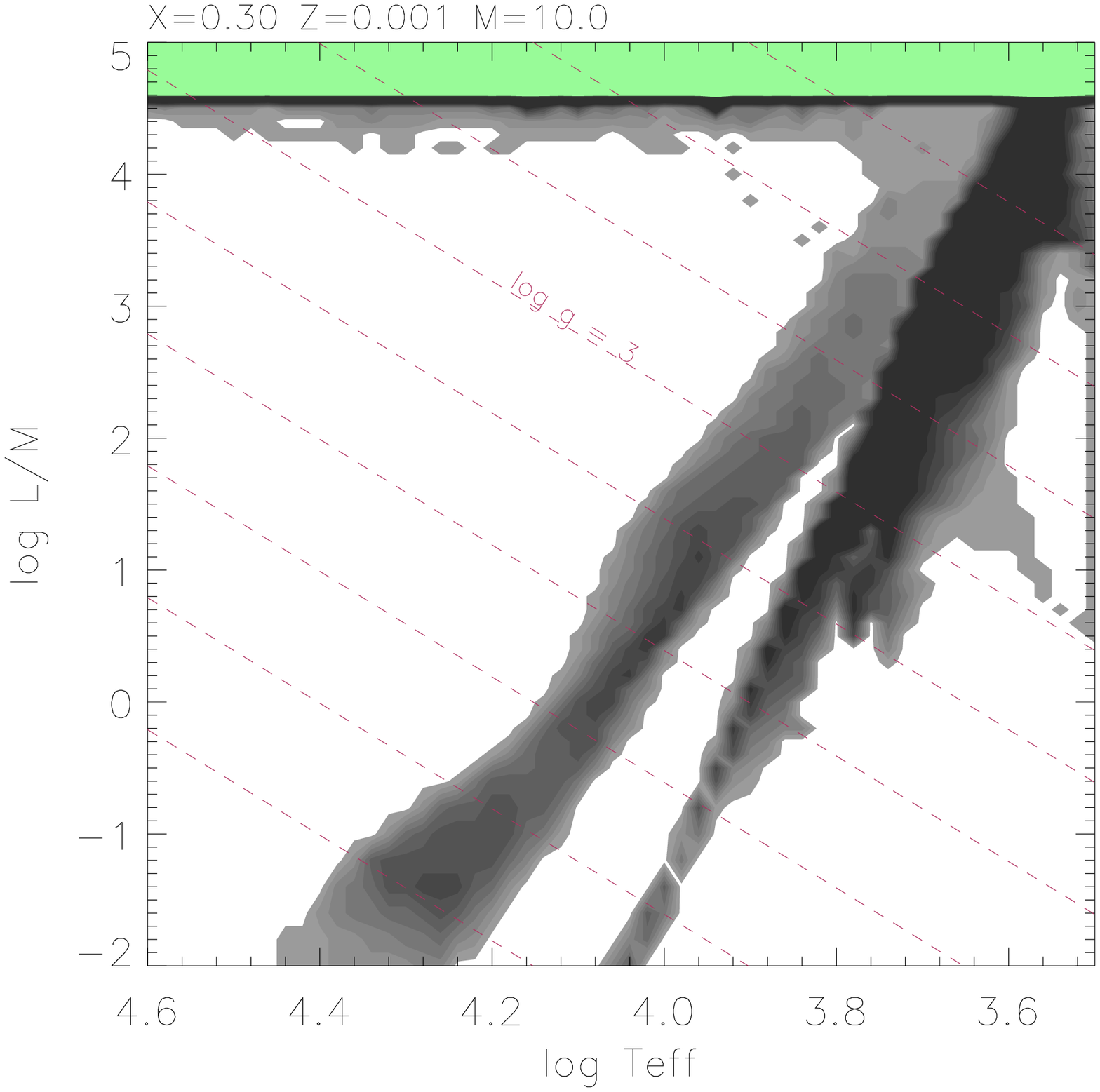,width=4.3cm,angle=0}\\
\epsfig{file=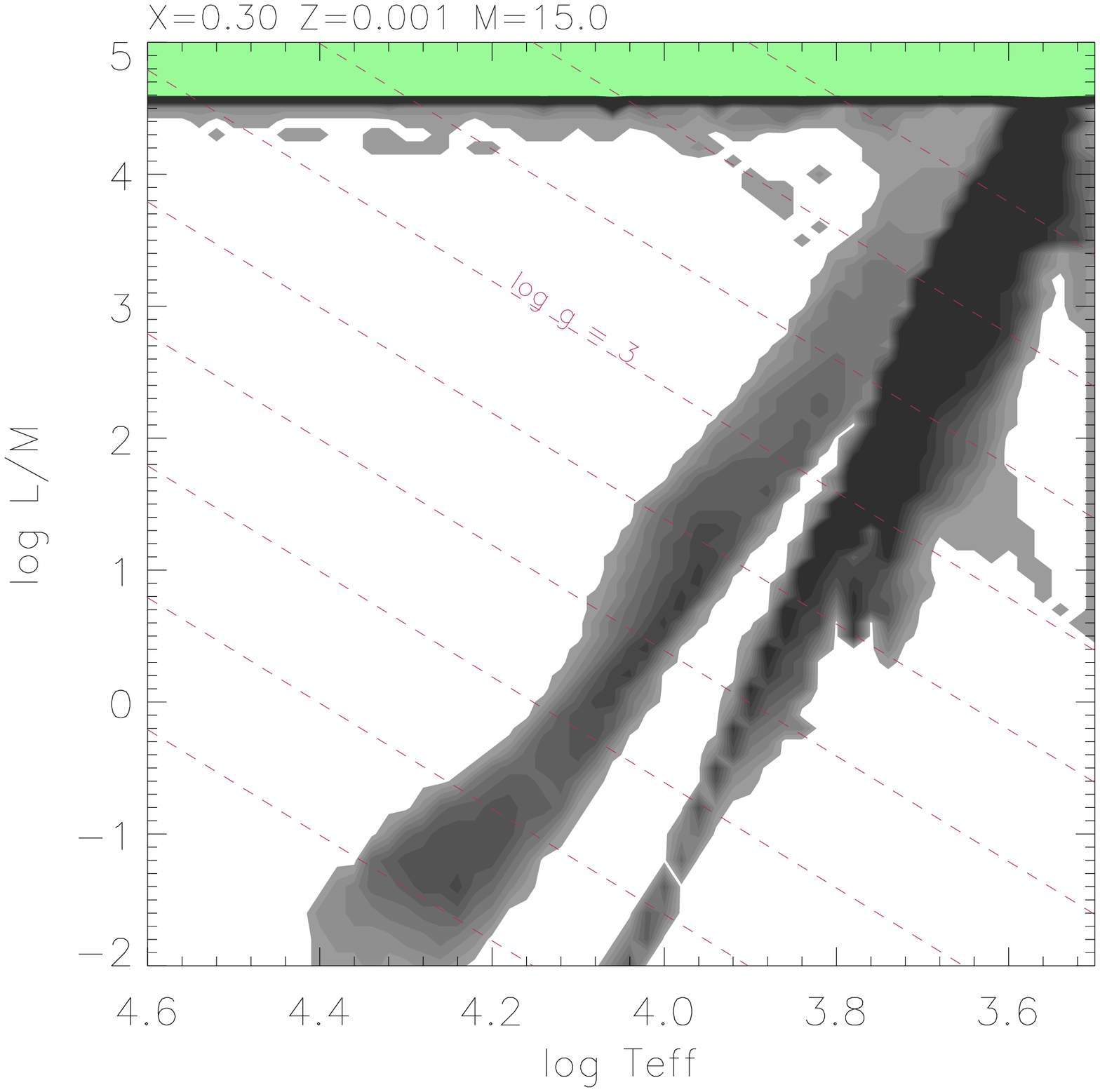,width=4.3cm,angle=0}
\epsfig{file=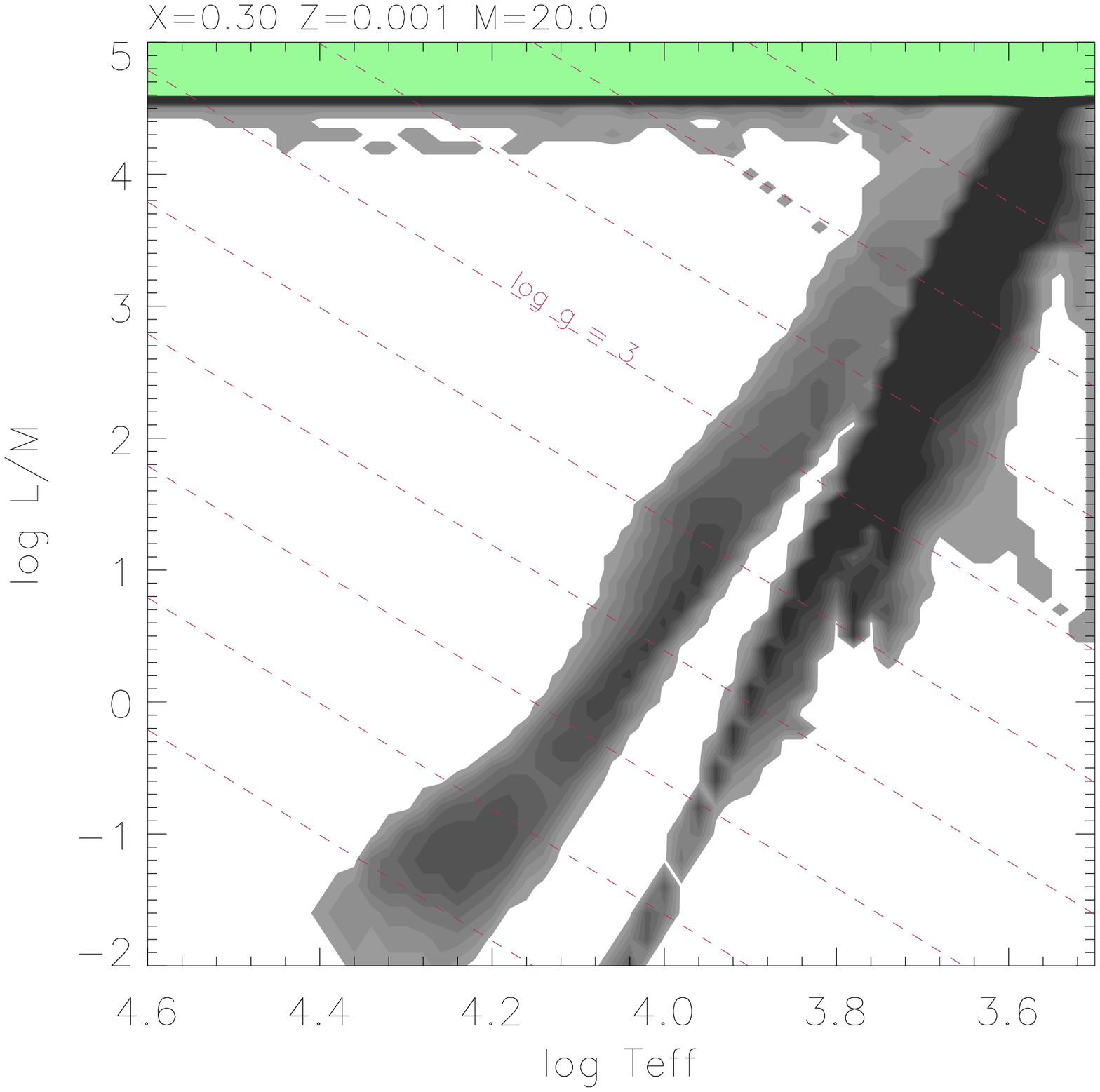,width=4.3cm,angle=0}
\epsfig{file=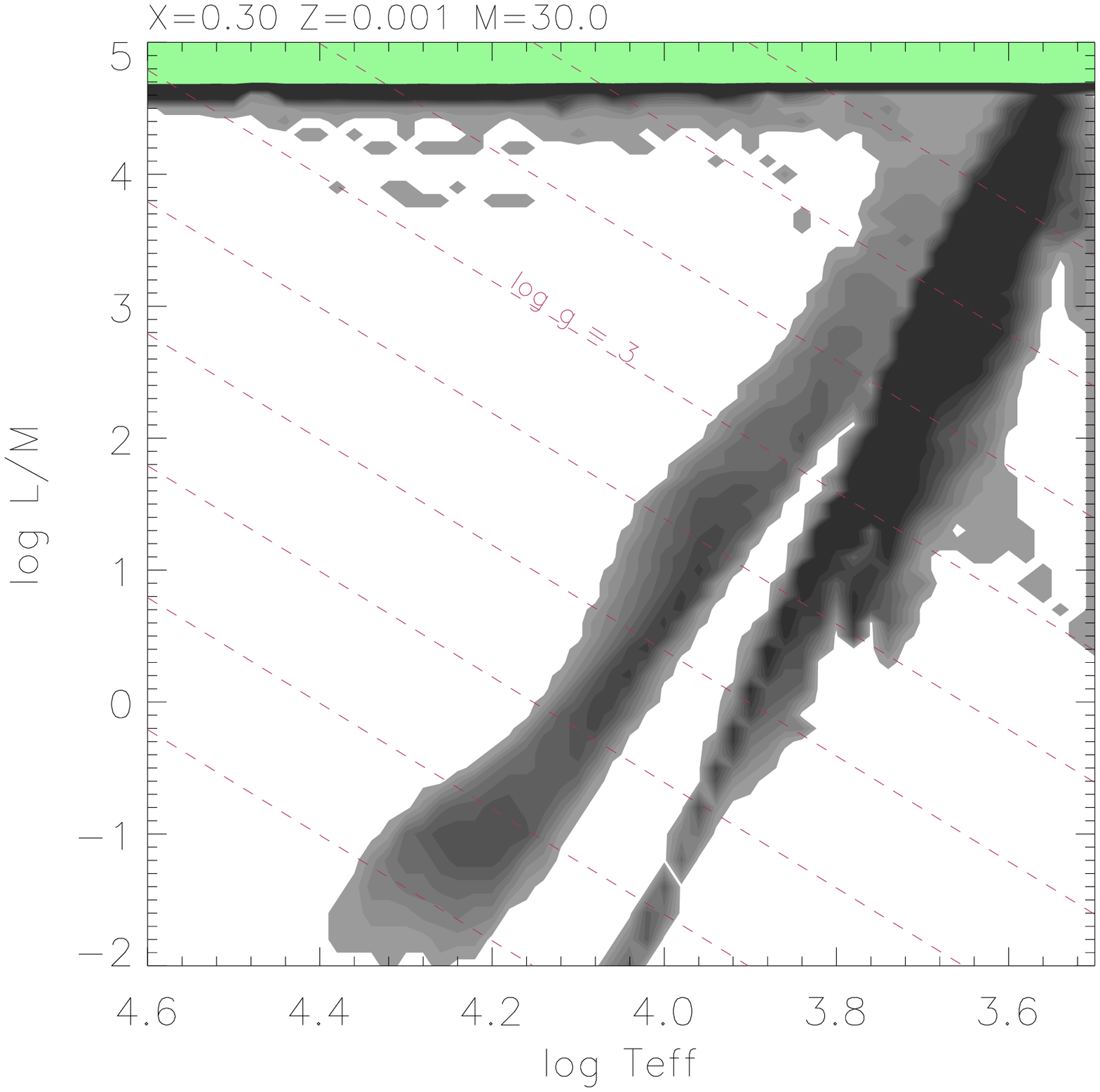,width=4.3cm,angle=0}
\epsfig{file=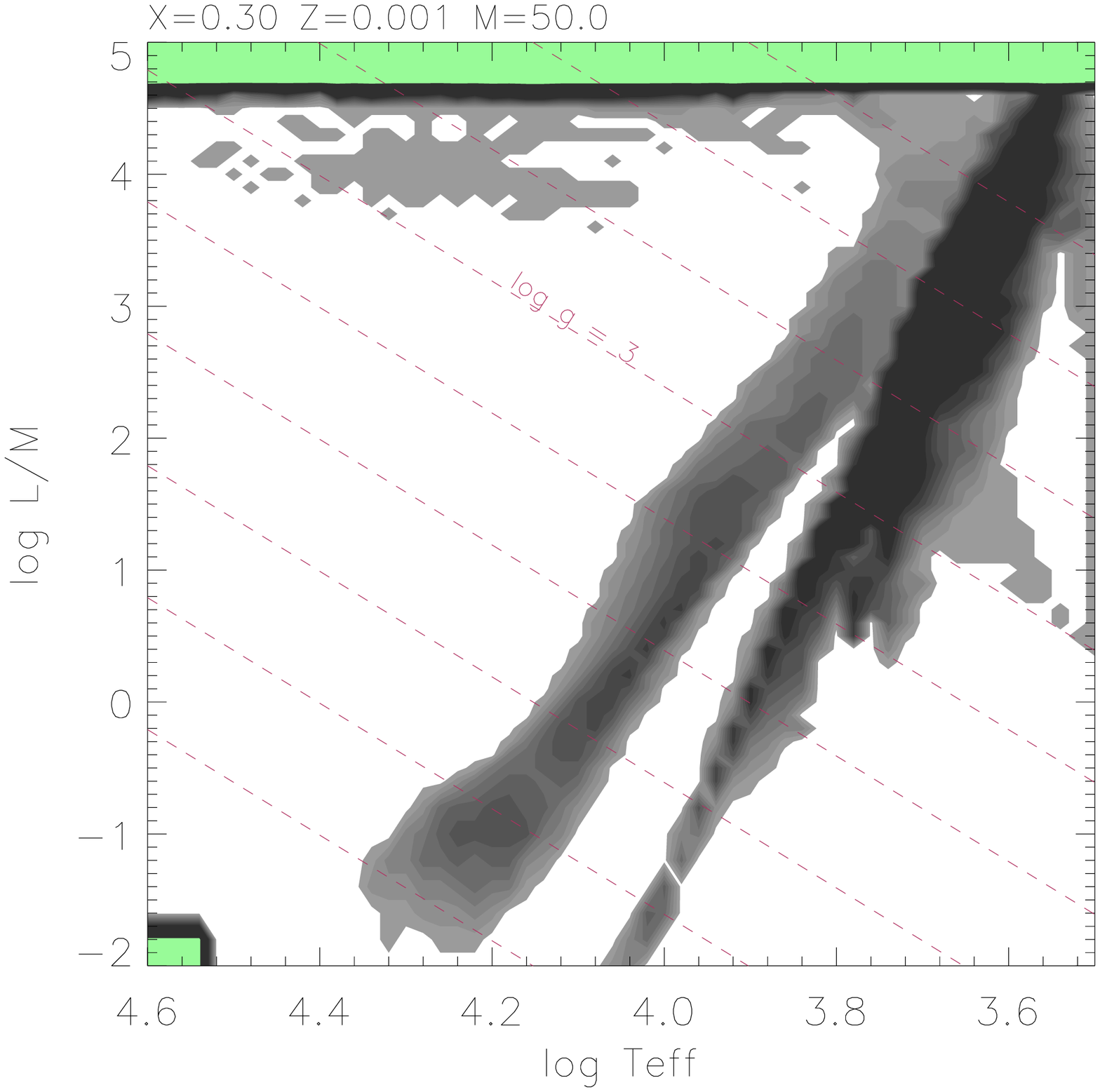,width=4.3cm,angle=0}
\caption[Unstable modes: $X=0.30, Z=0.001$]
{As Fig.~\ref{f:nx70} with $X=0.30, Z=0.001$. 
}
\label{f:nx30z001}
\end{center}
\end{figure*}

\begin{figure*}
\begin{center}
\epsfig{file=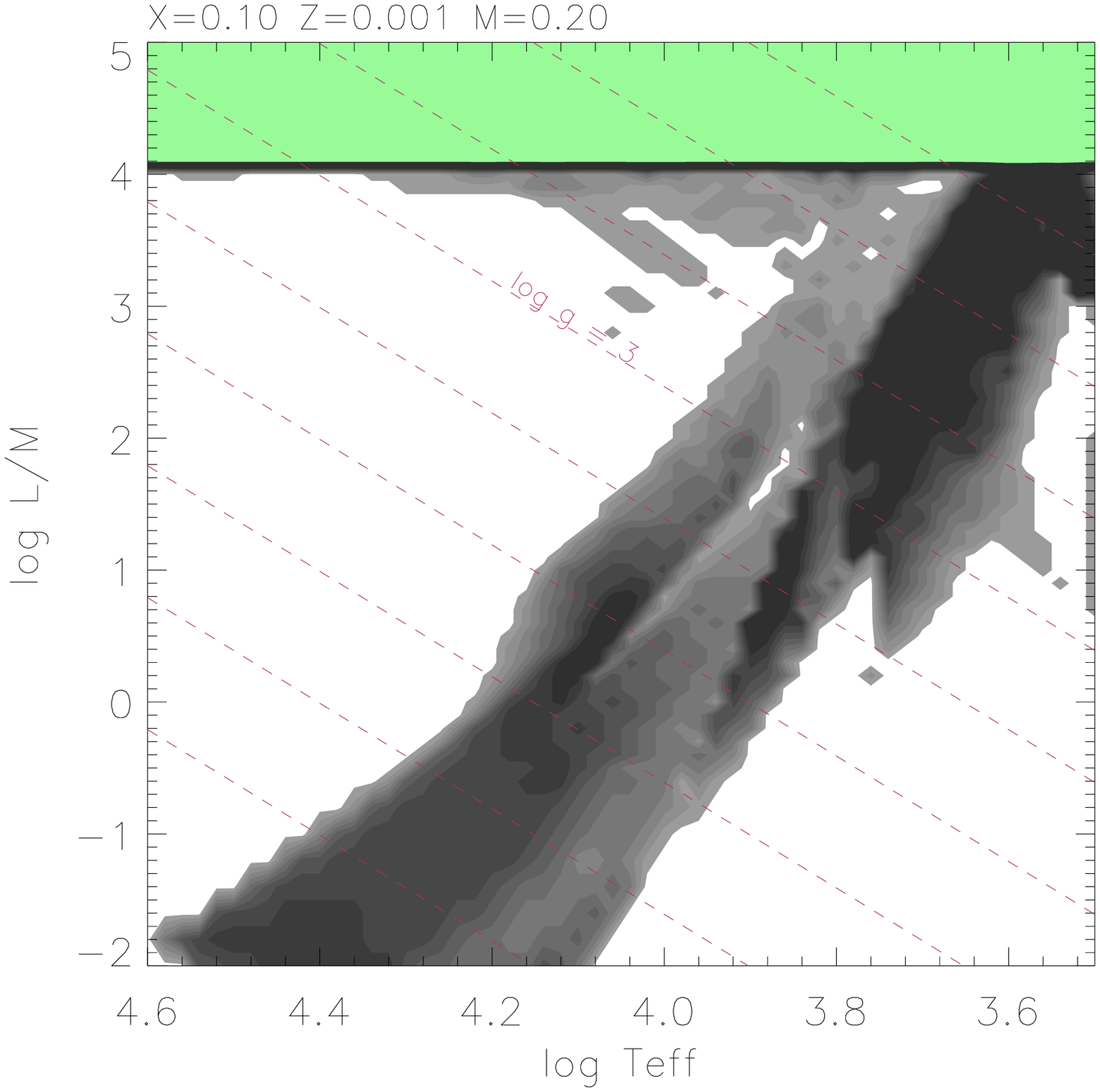,width=4.3cm,angle=0}
\epsfig{file=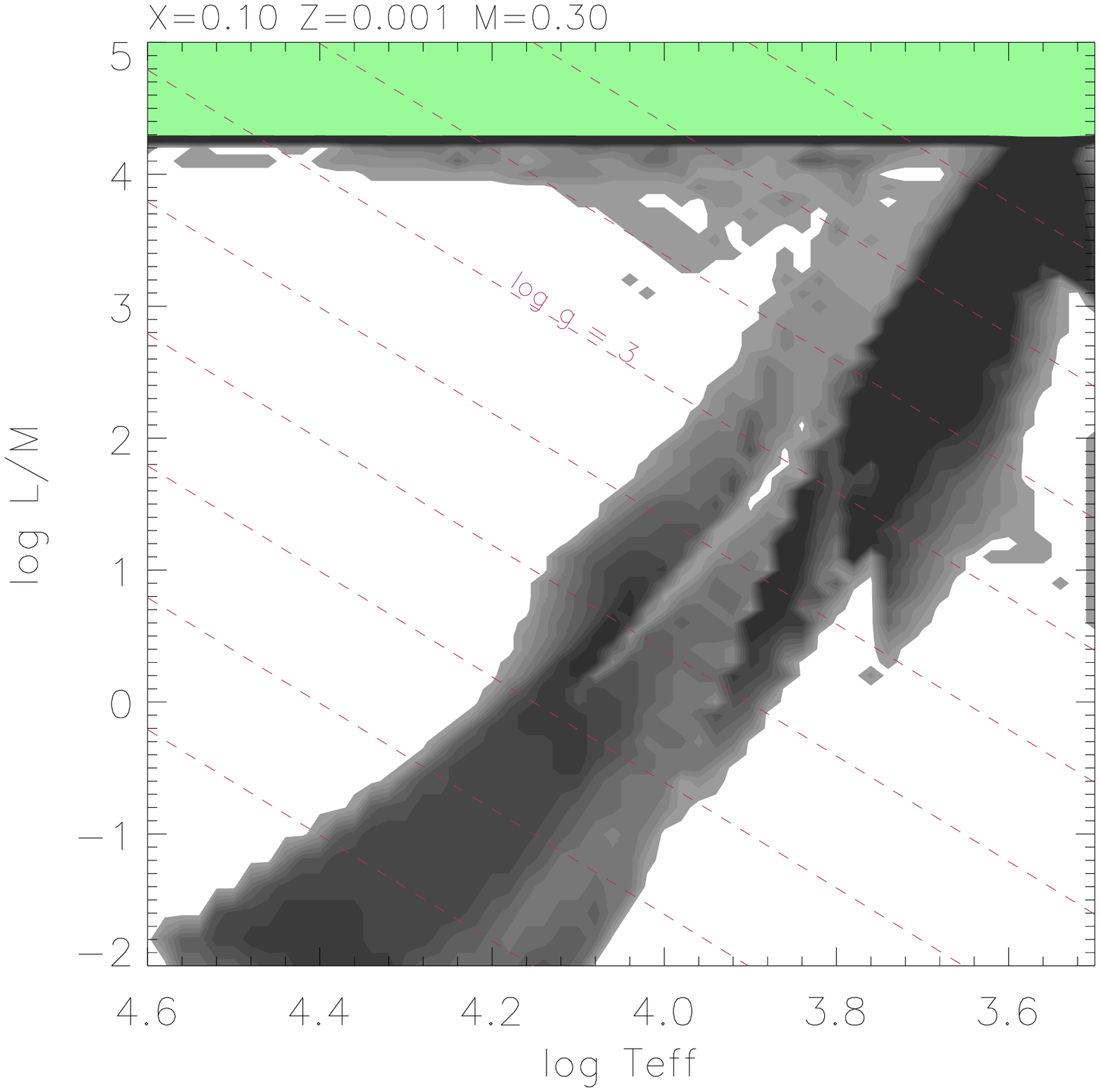,width=4.3cm,angle=0}
\epsfig{file=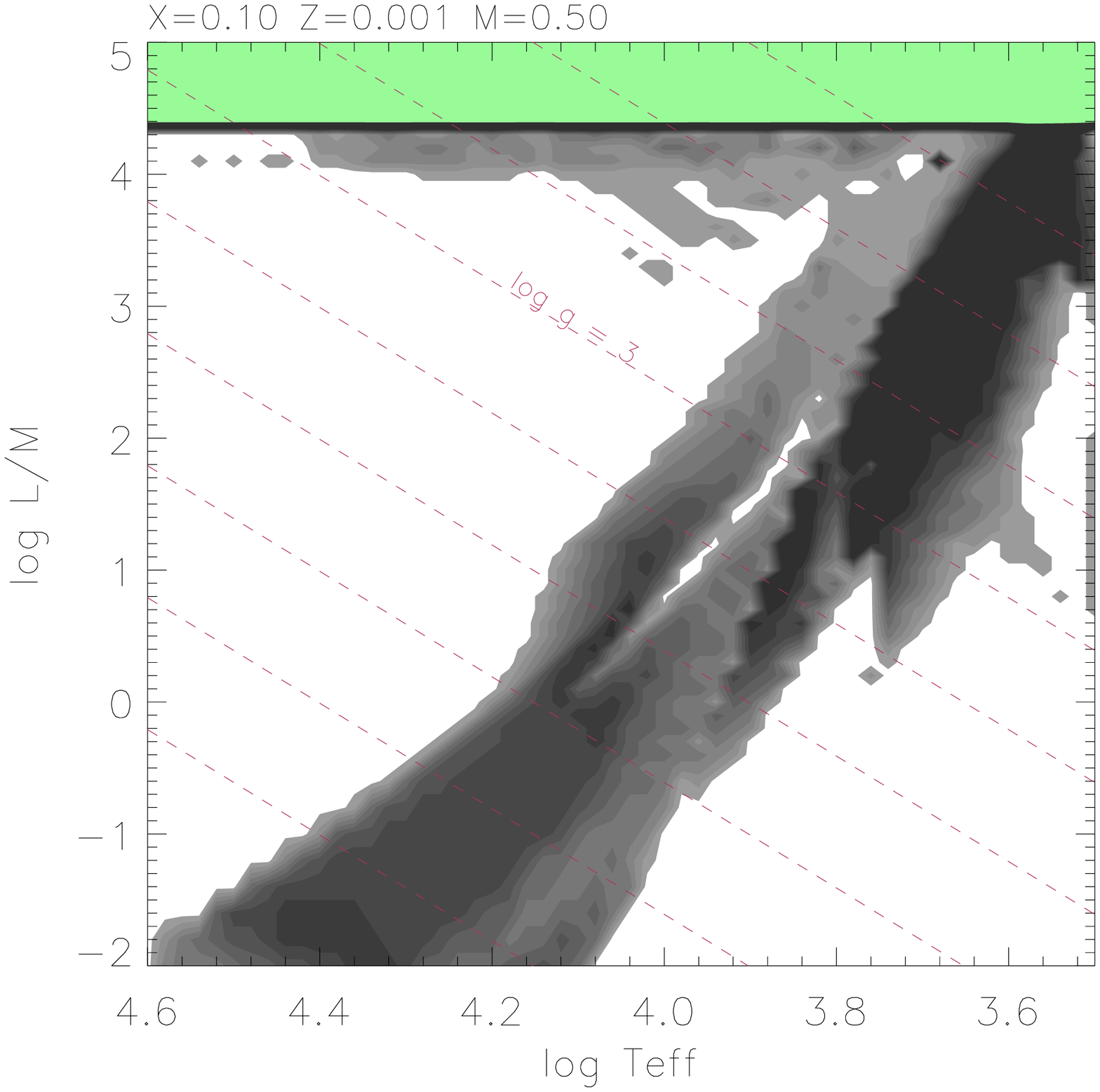,width=4.3cm,angle=0}
\epsfig{file=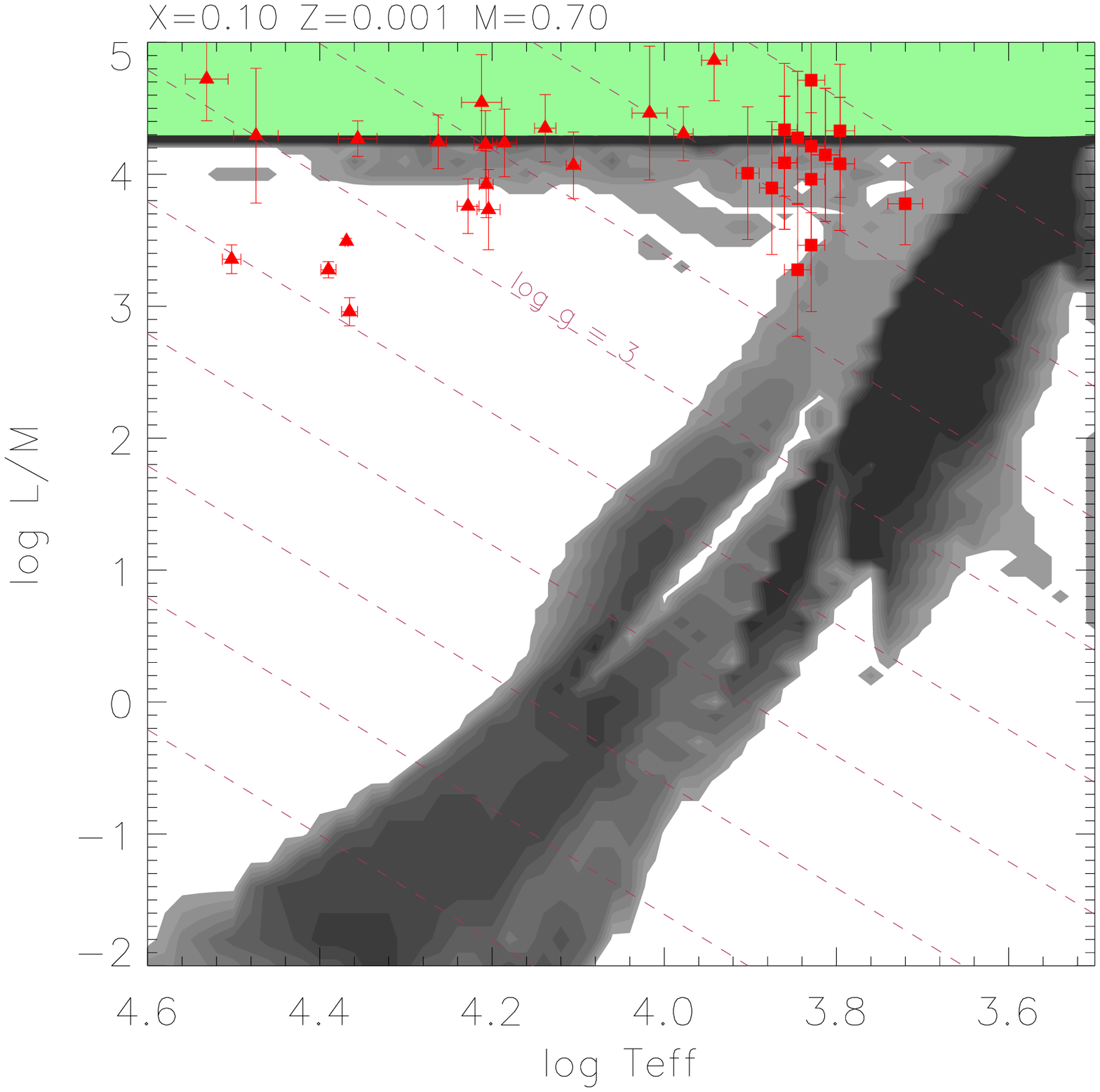,width=4.3cm,angle=0}\\
\epsfig{file=figs/nmodes_x10z001m01.0_00_opal.eps,width=4.3cm,angle=0}
\epsfig{file=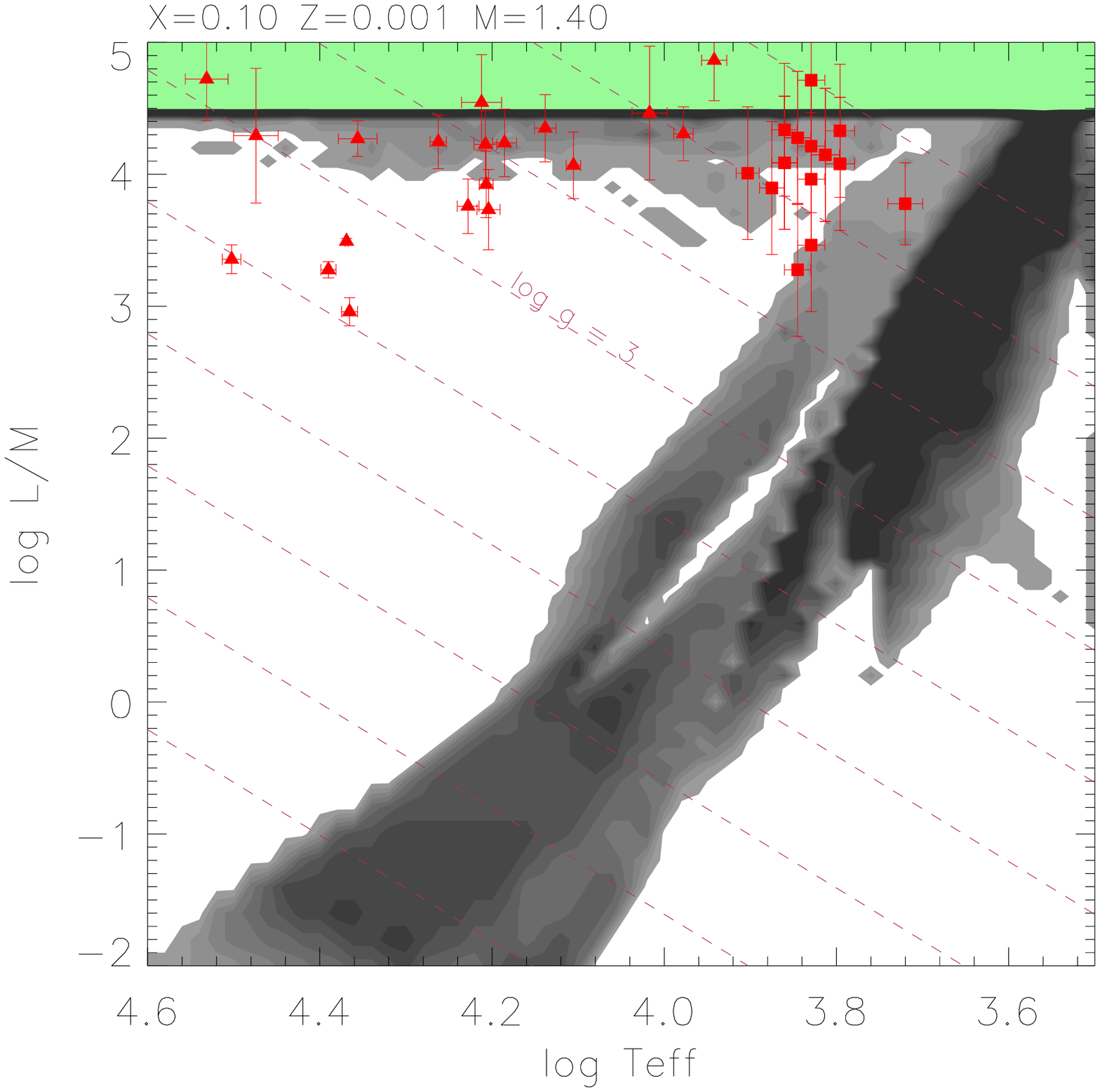,width=4.3cm,angle=0}
\epsfig{file=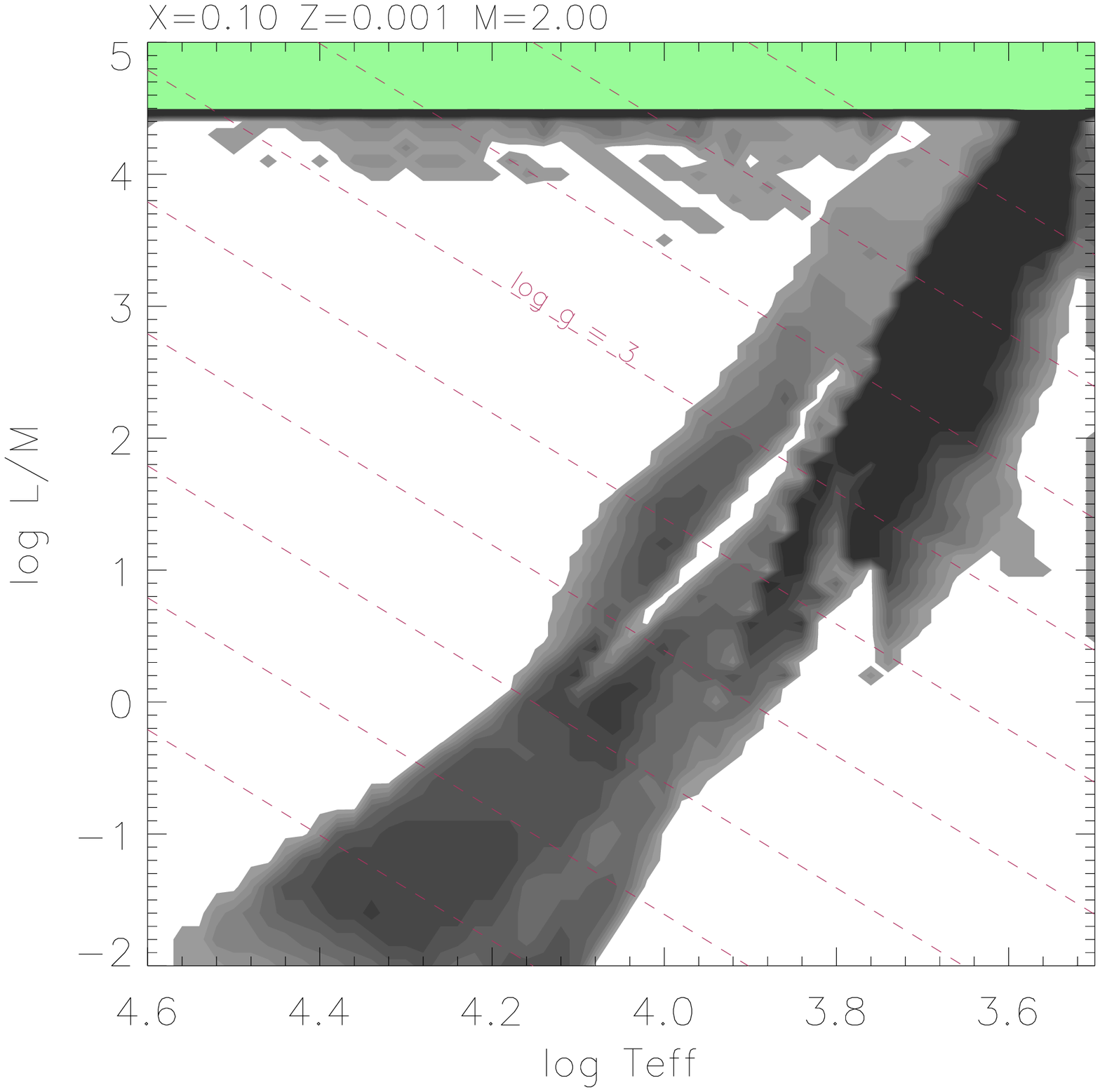,width=4.3cm,angle=0}
\epsfig{file=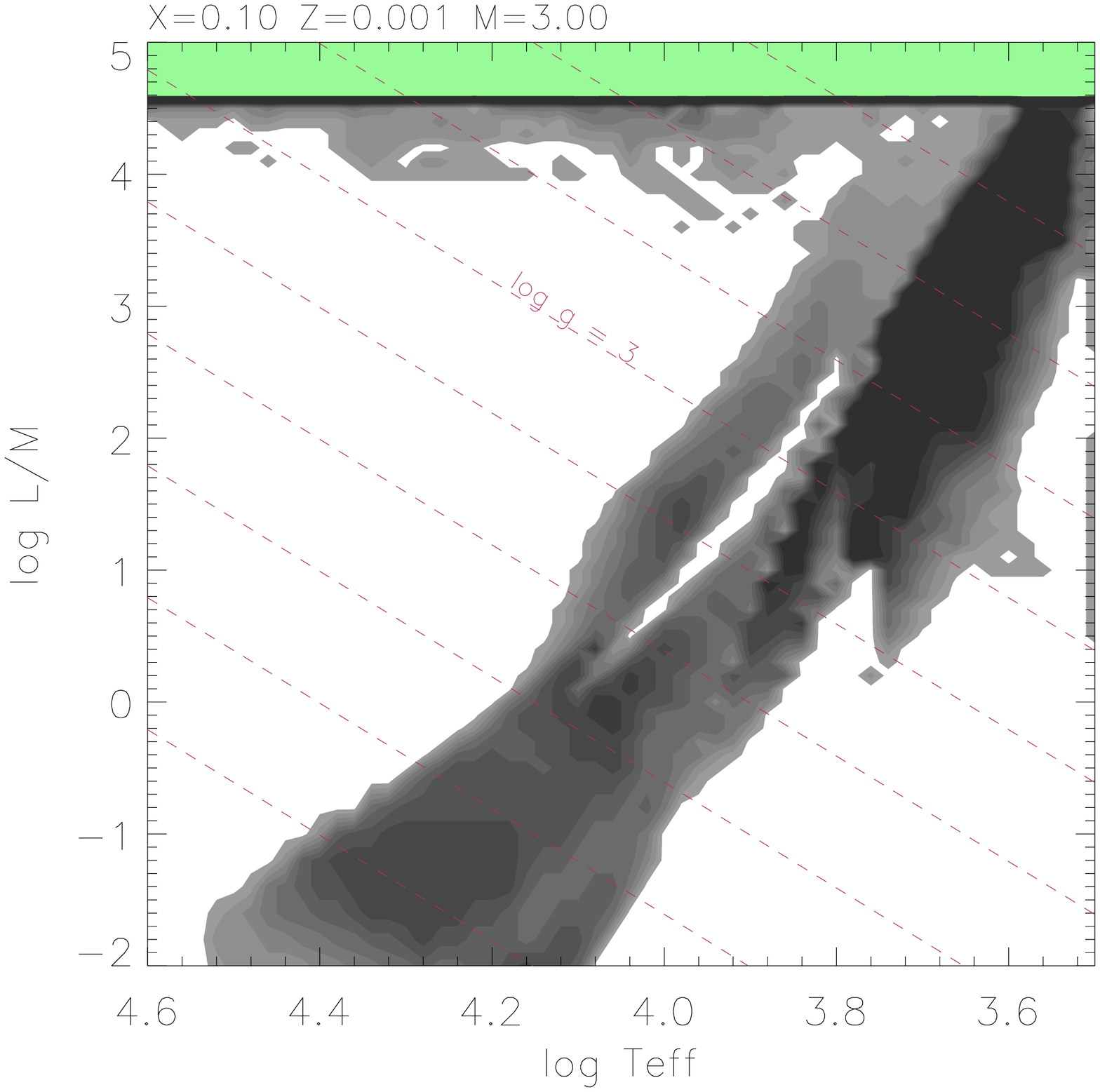,width=4.3cm,angle=0}\\
\epsfig{file=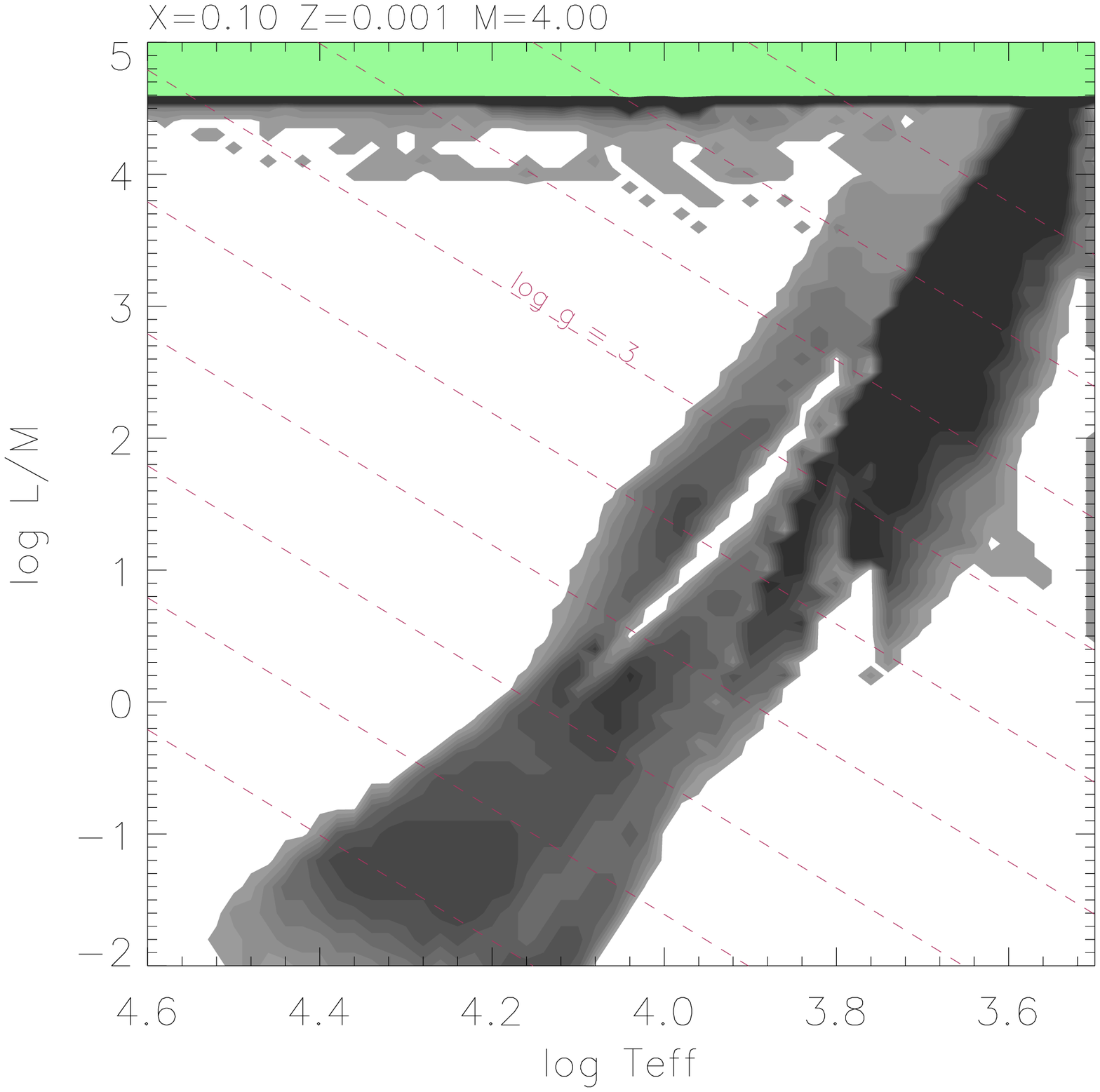,width=4.3cm,angle=0}
\epsfig{file=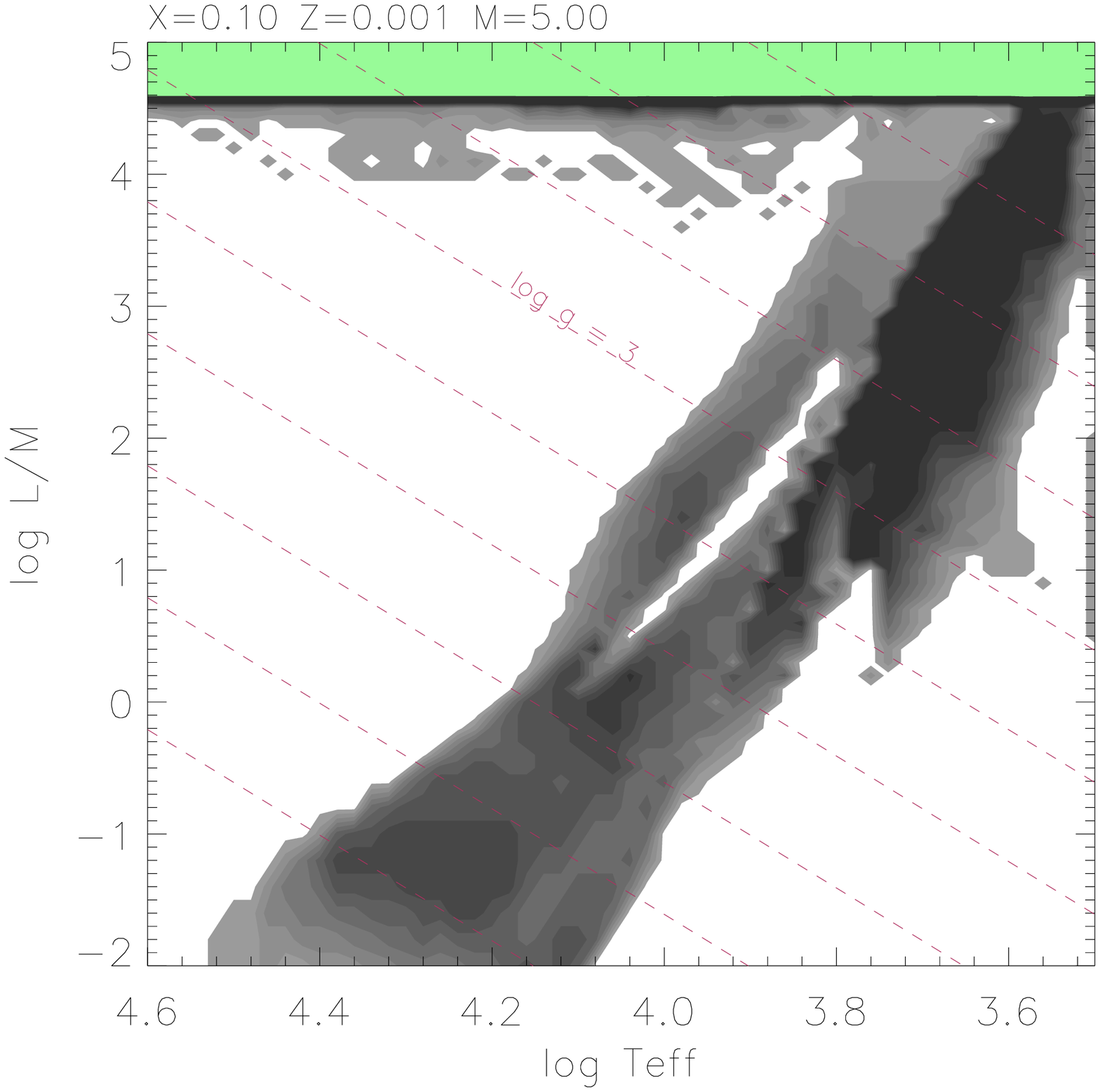,width=4.3cm,angle=0}
\epsfig{file=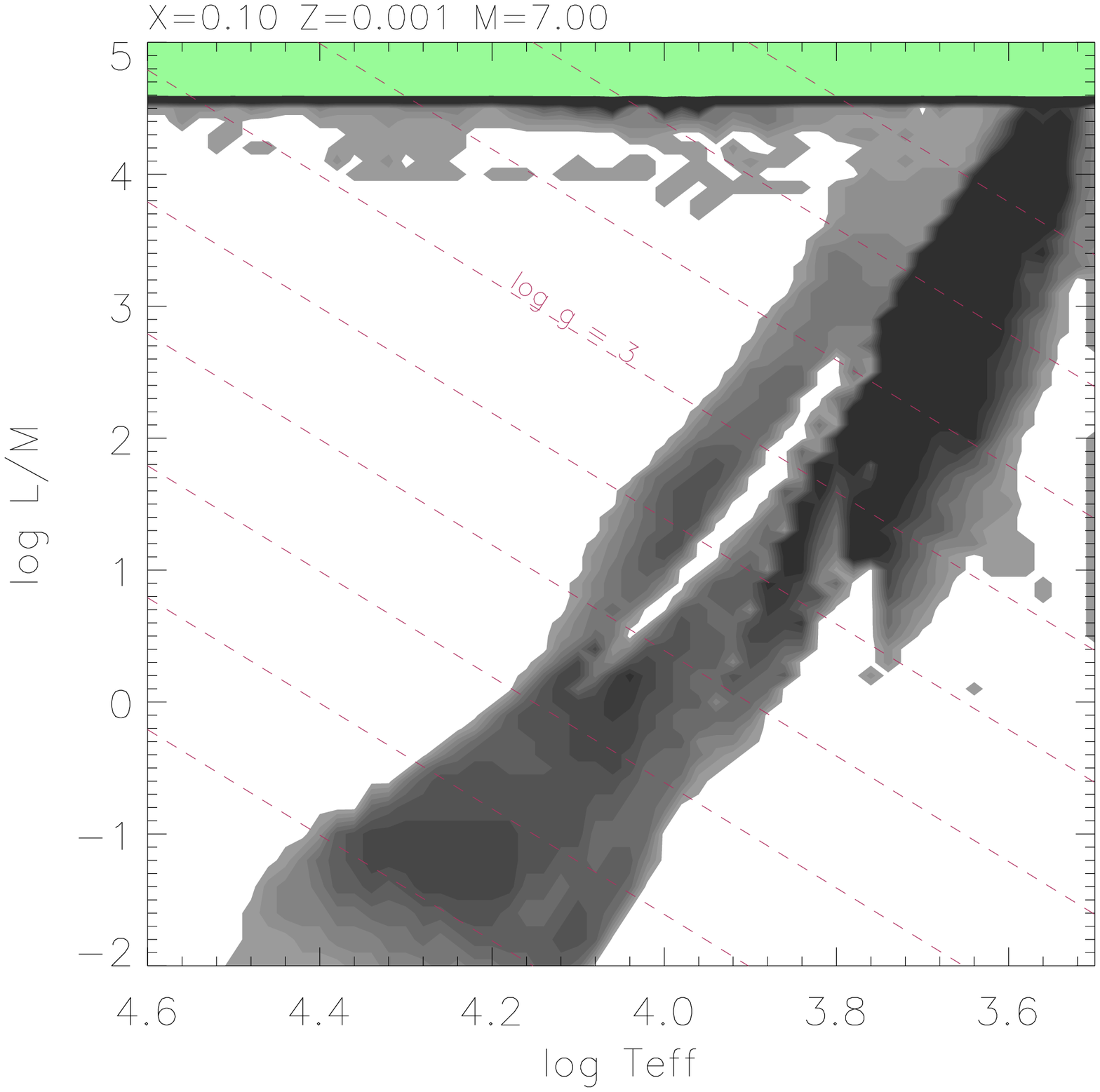,width=4.3cm,angle=0}
\epsfig{file=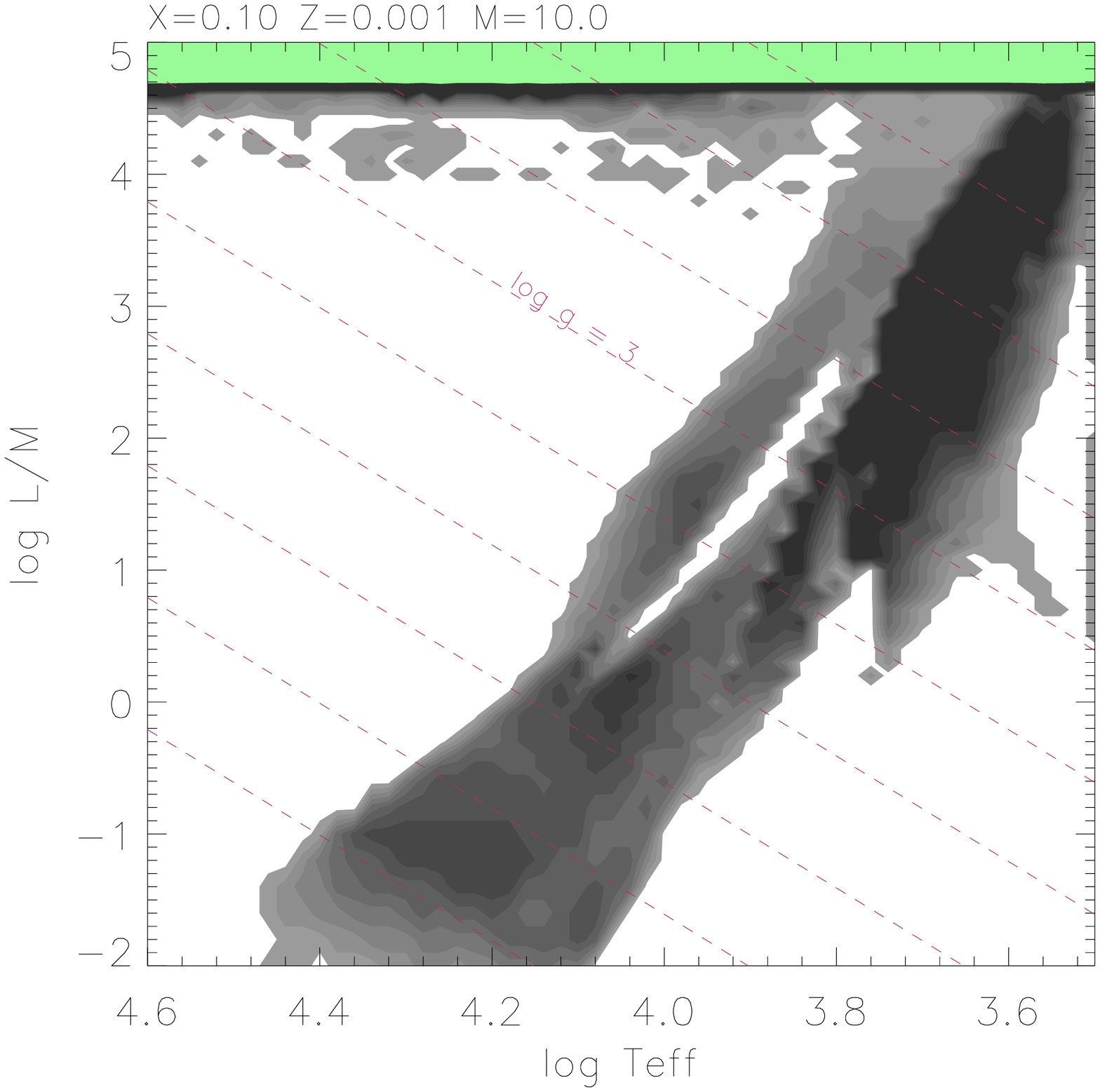,width=4.3cm,angle=0}\\
\epsfig{file=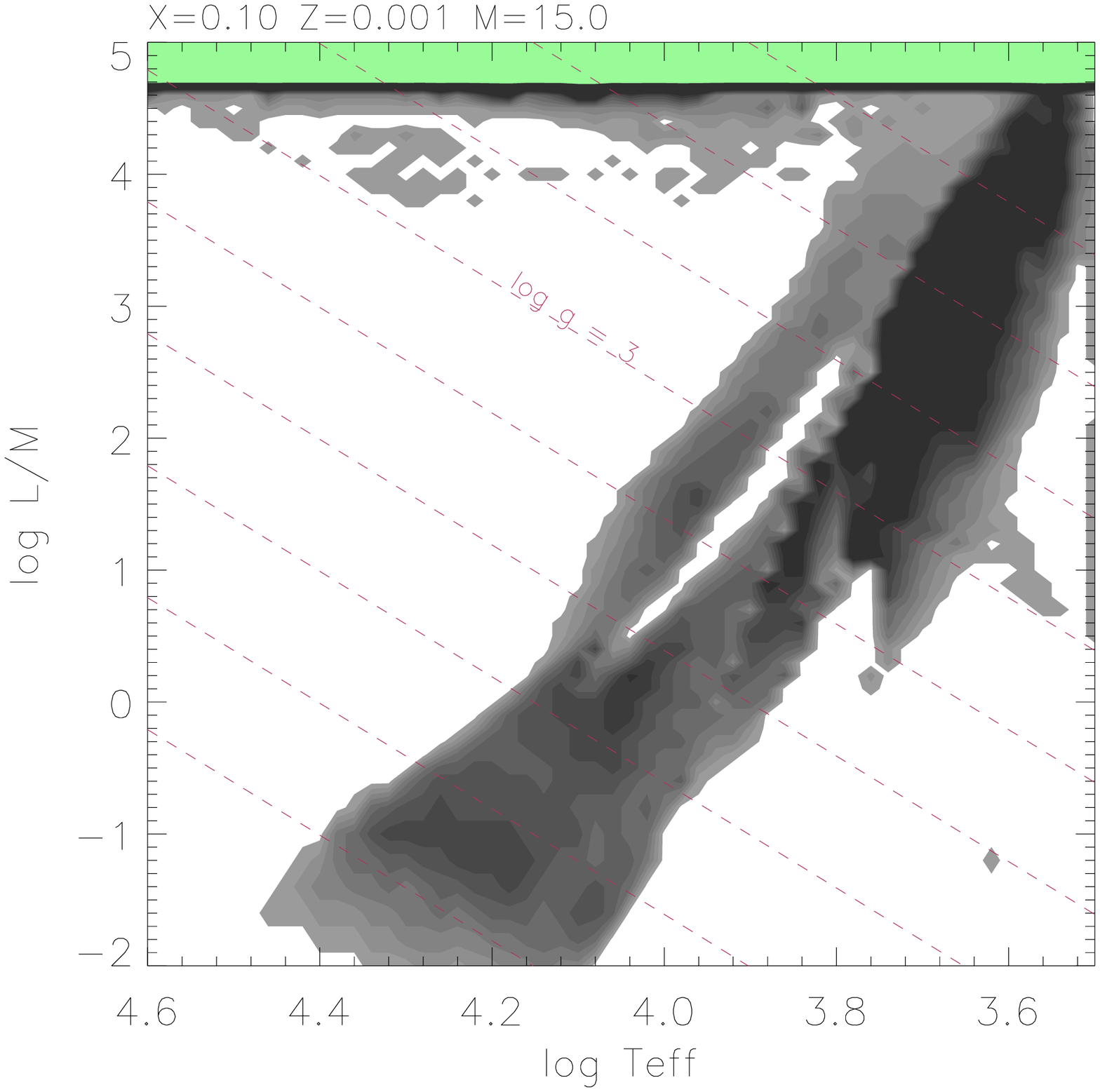,width=4.3cm,angle=0}
\epsfig{file=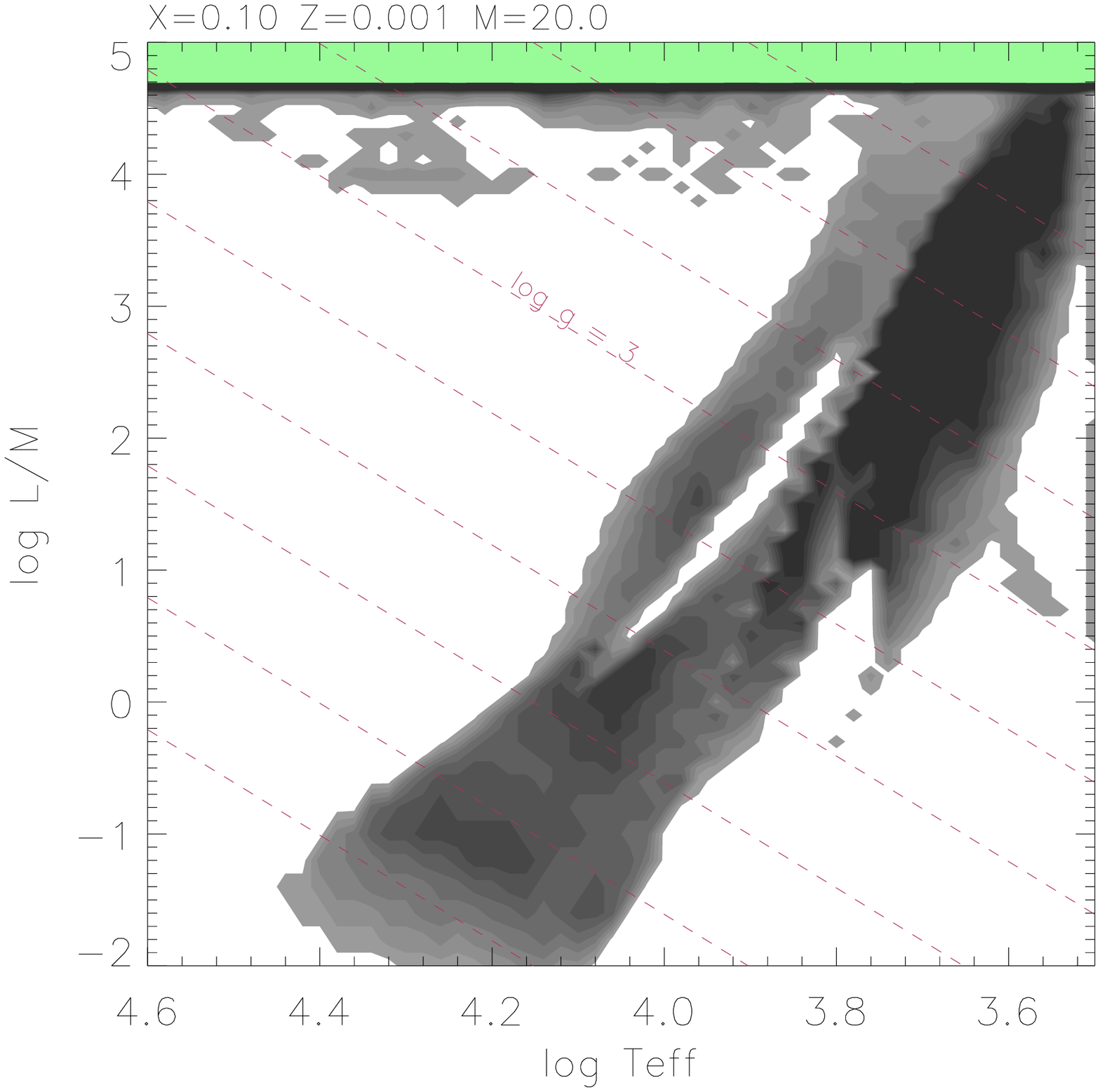,width=4.3cm,angle=0}
\epsfig{file=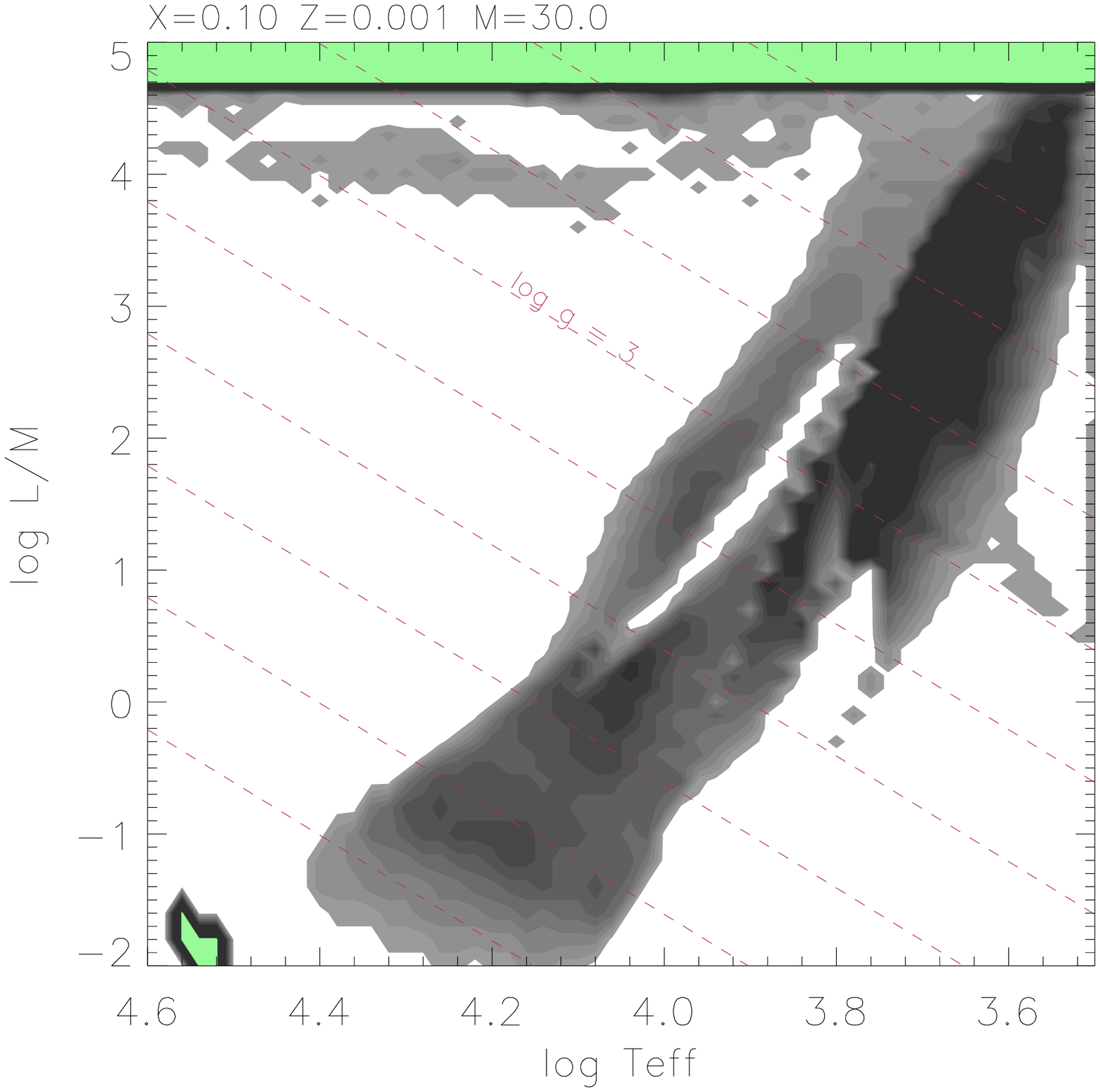,width=4.3cm,angle=0}
\epsfig{file=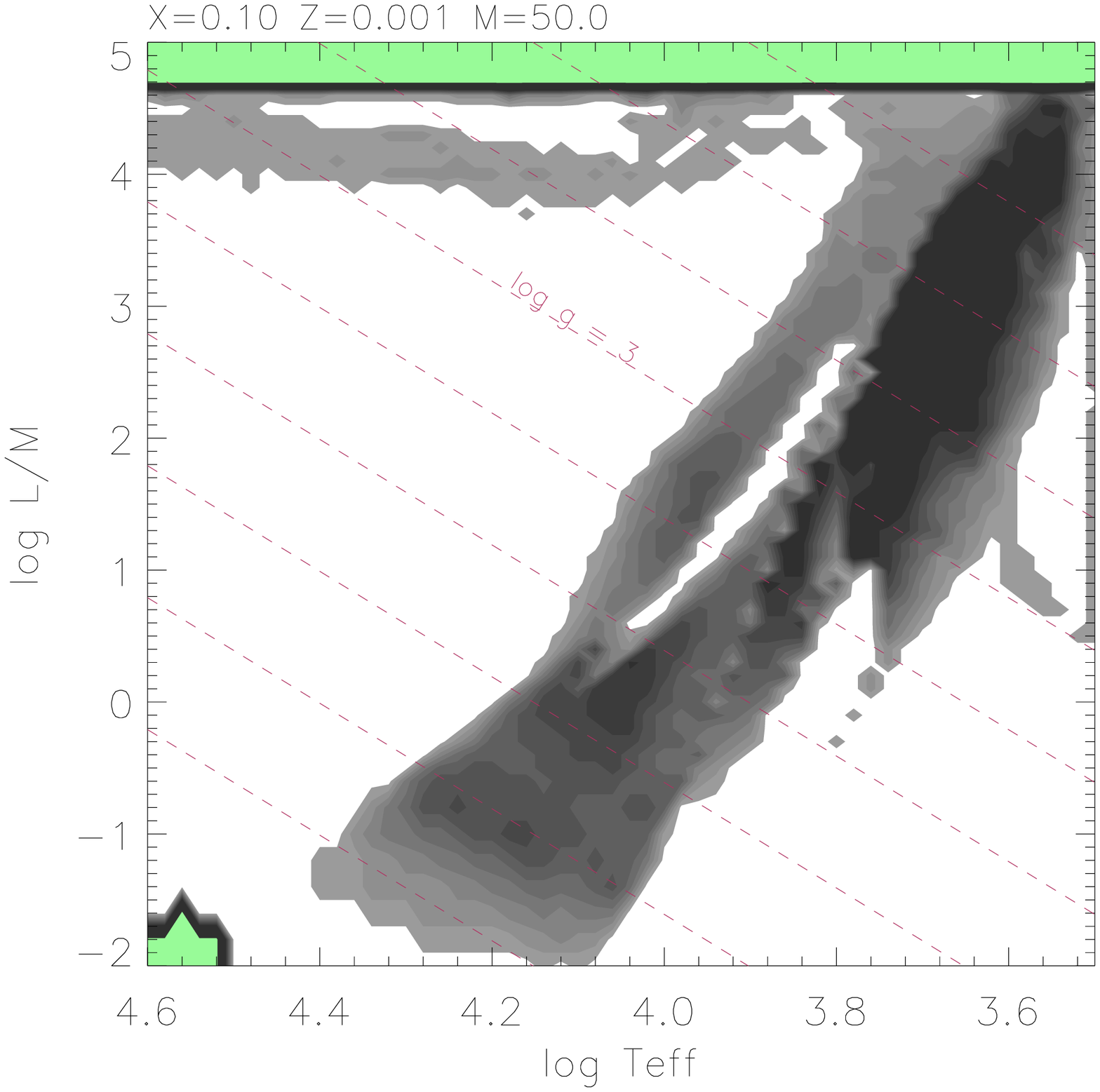,width=4.3cm,angle=0}
\caption[Unstable modes: $X=0.10, Z=0.001$]
{As Fig.~\ref{f:nx70} with $X=0.10, Z=0.001$. 
}
\label{f:nx10z001}
\end{center}
\end{figure*}

\begin{figure*}
\begin{center}
\epsfig{file=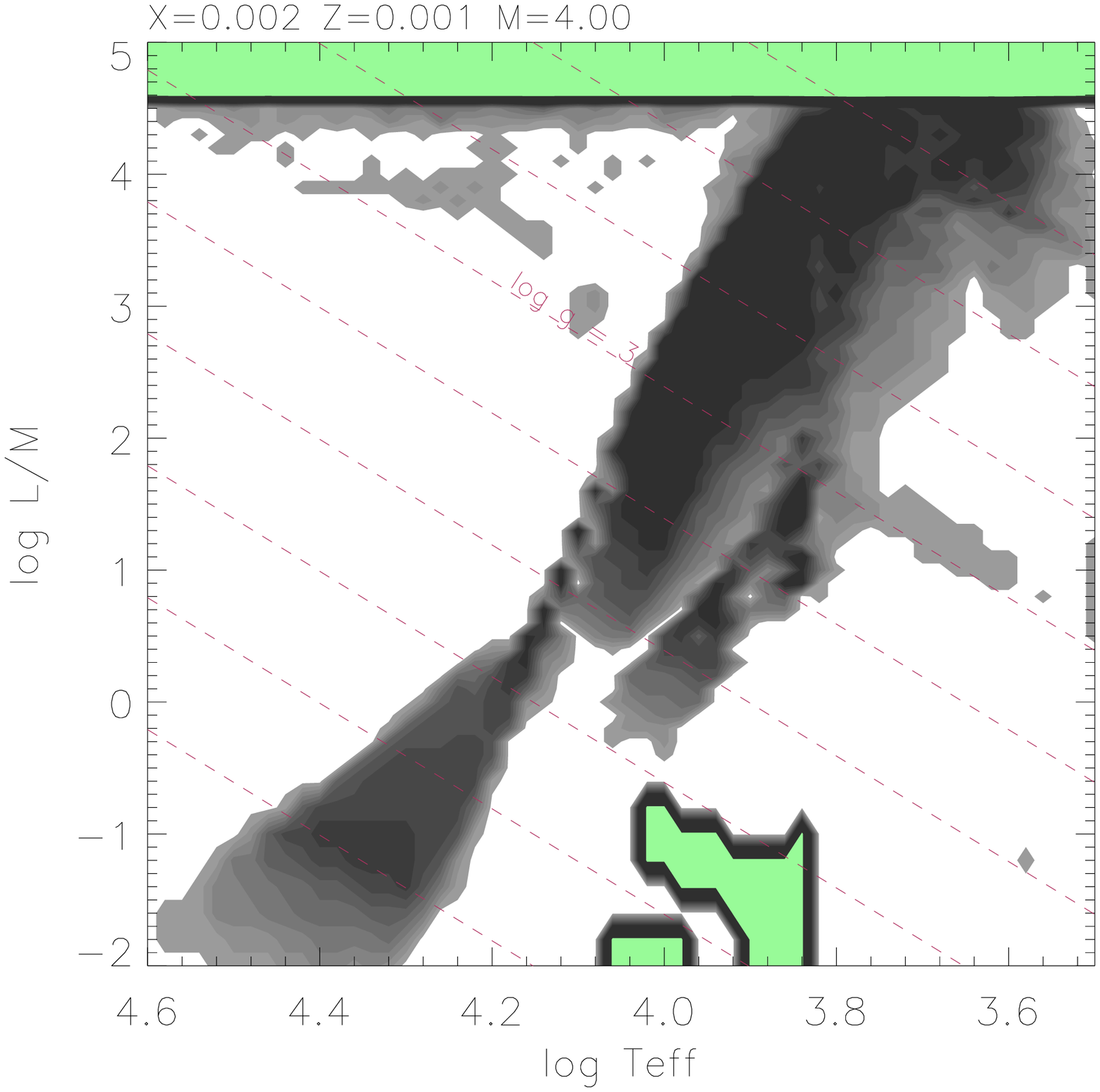,width=4.3cm,angle=0}
\epsfig{file=figs/nmodes_x002z001m05.0_00_opal.eps,width=4.3cm,angle=0}
\epsfig{file=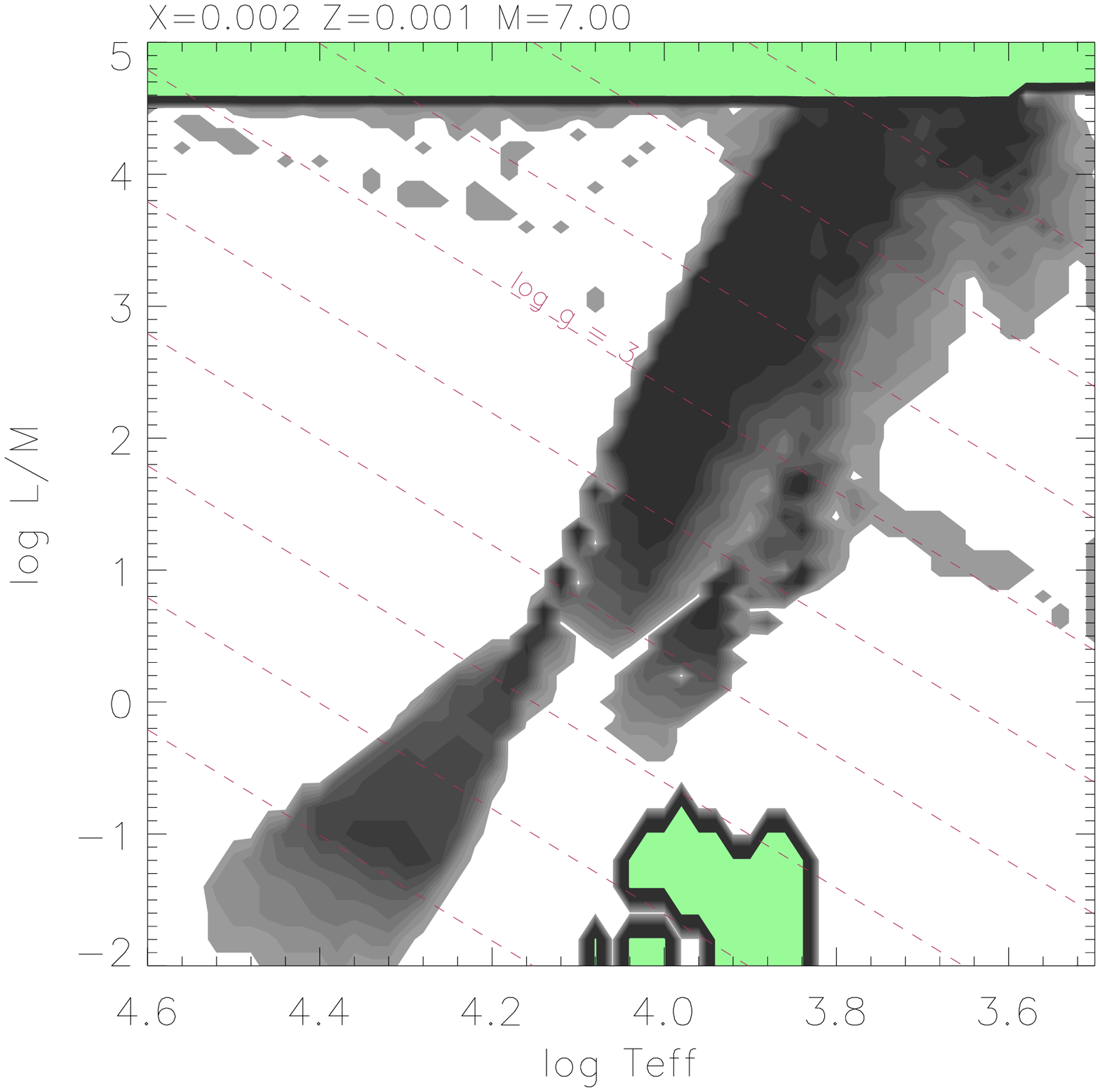,width=4.3cm,angle=0}
\epsfig{file=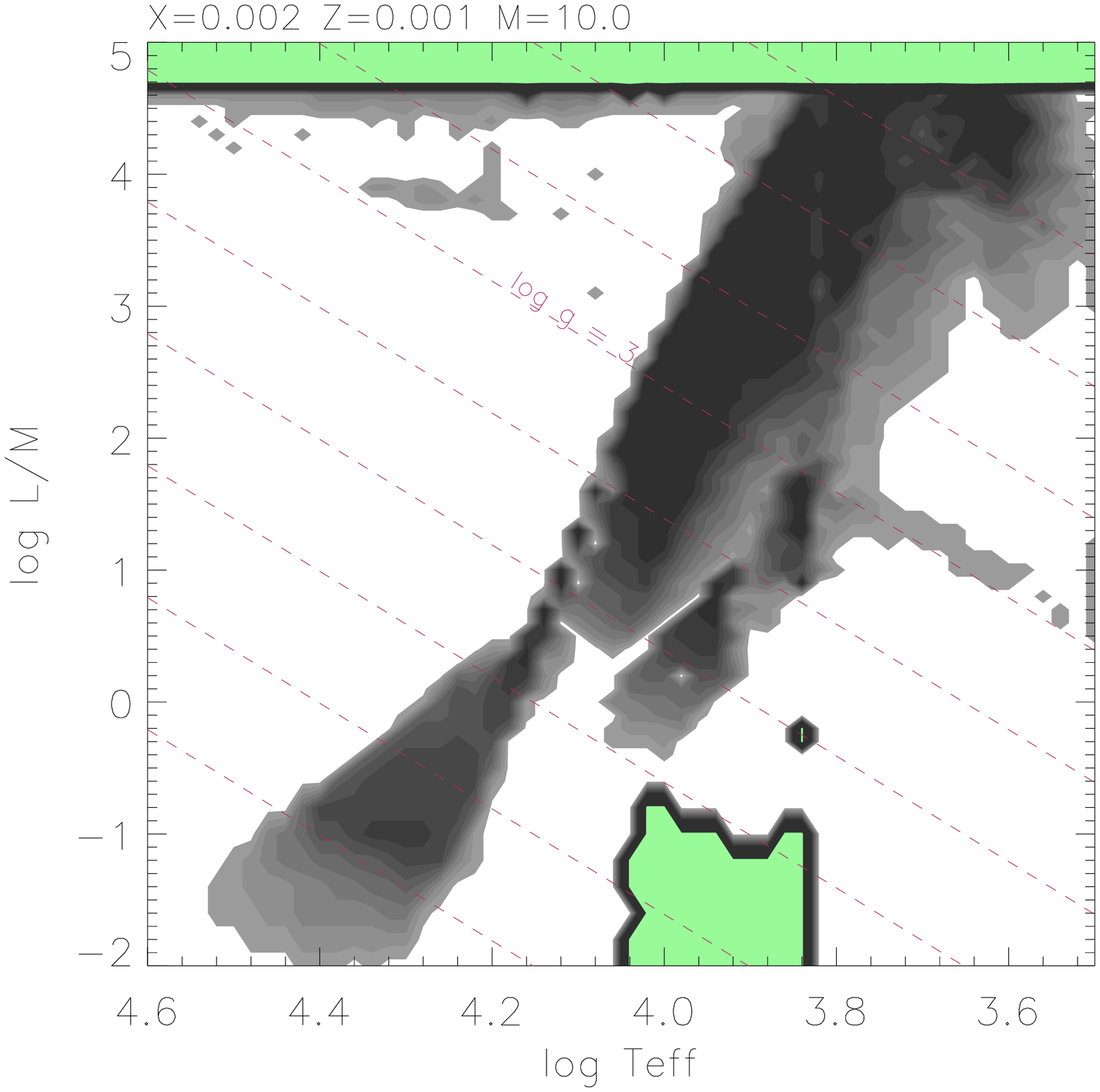,width=4.3cm,angle=0}\\
\epsfig{file=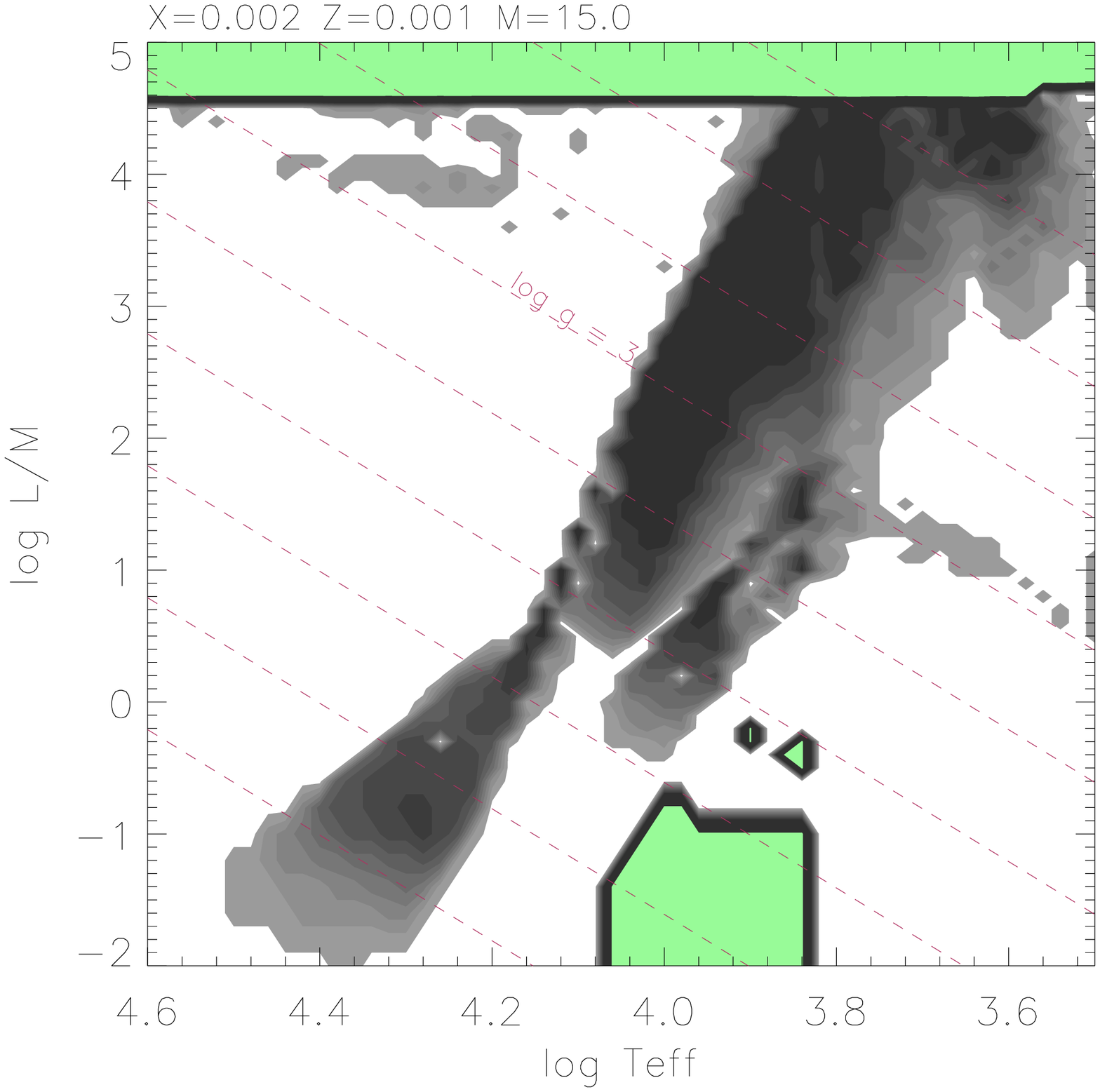,width=4.3cm,angle=0}
\epsfig{file=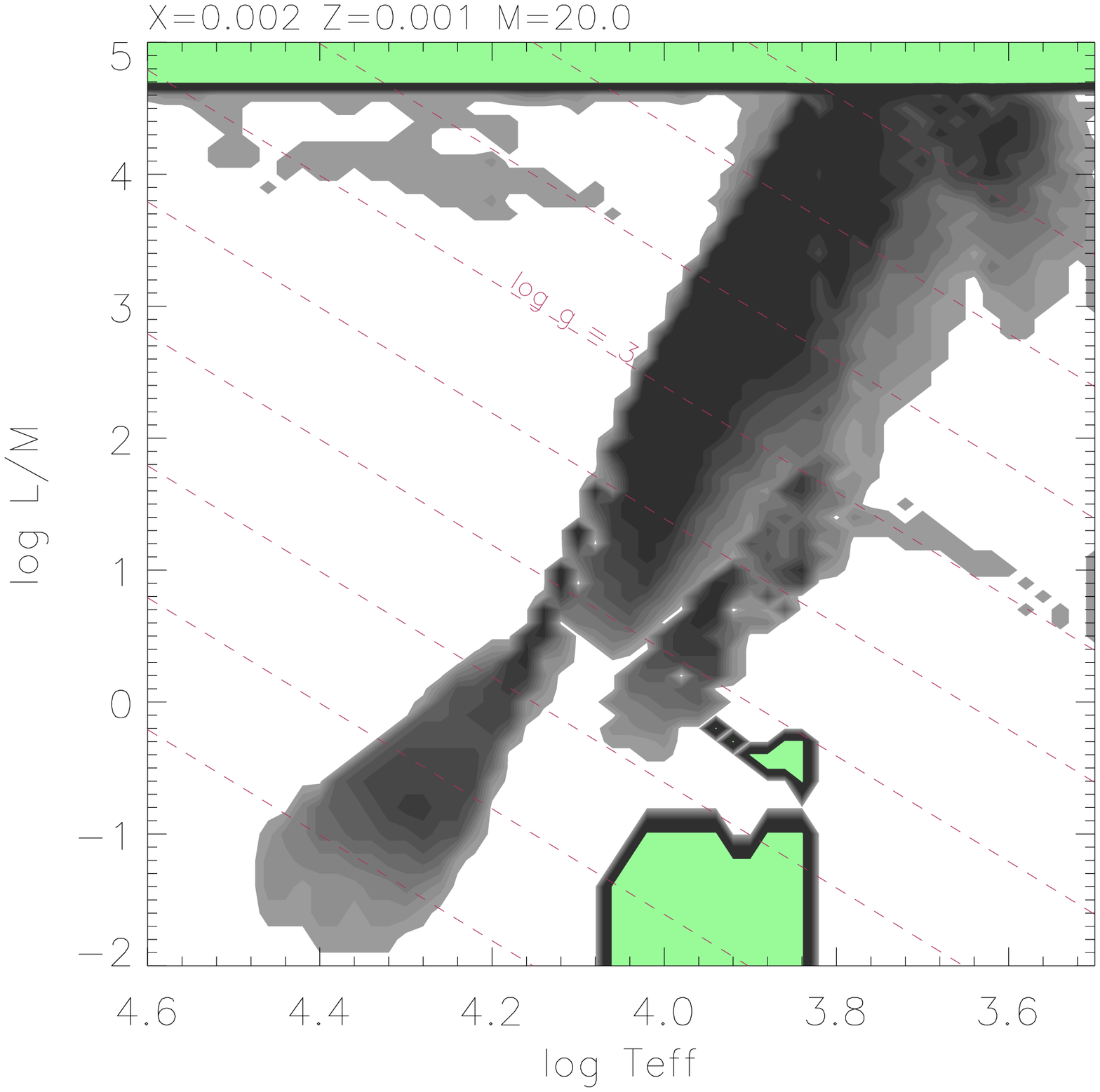,width=4.3cm,angle=0}
\epsfig{file=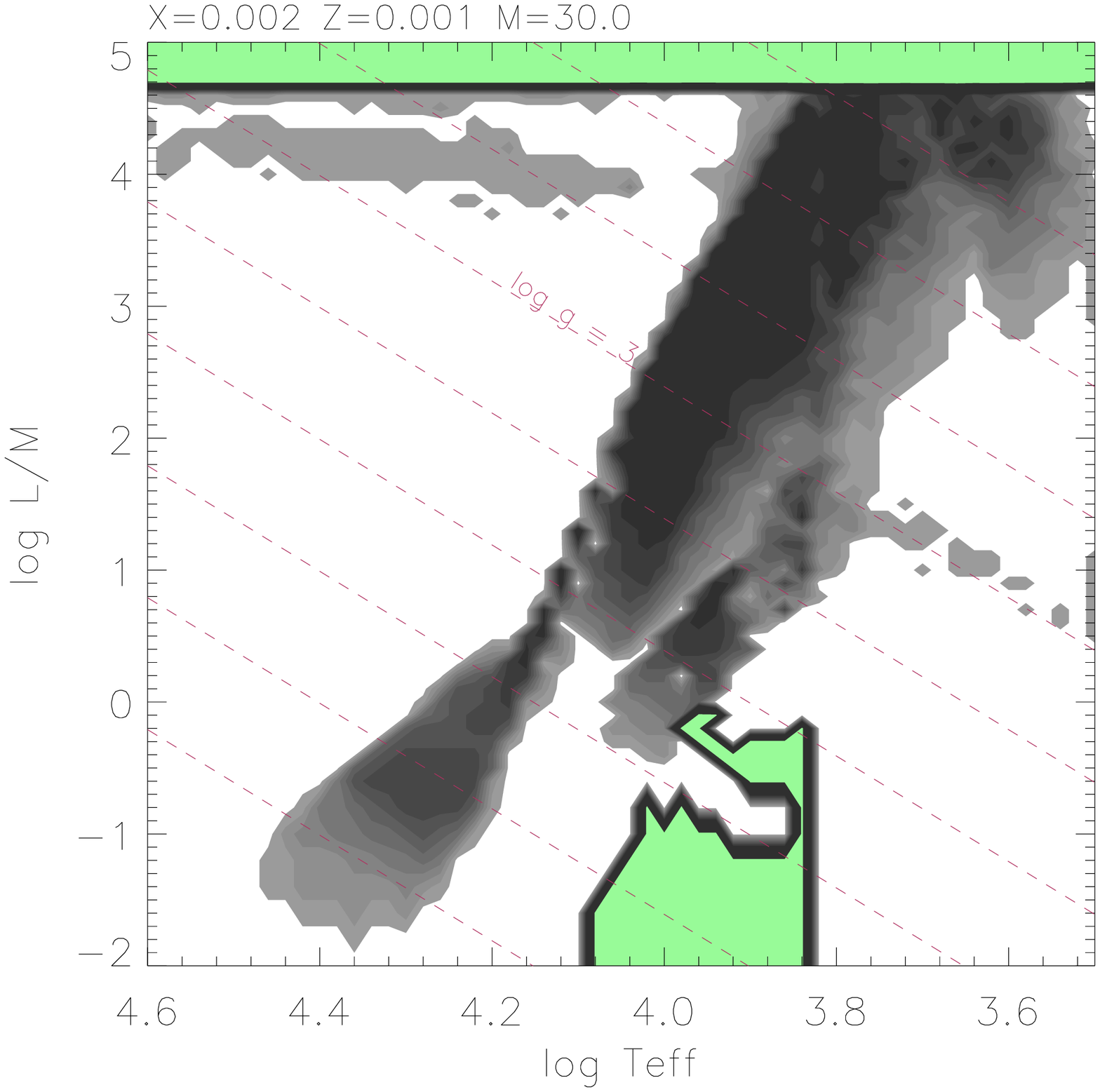,width=4.3cm,angle=0}
\epsfig{file=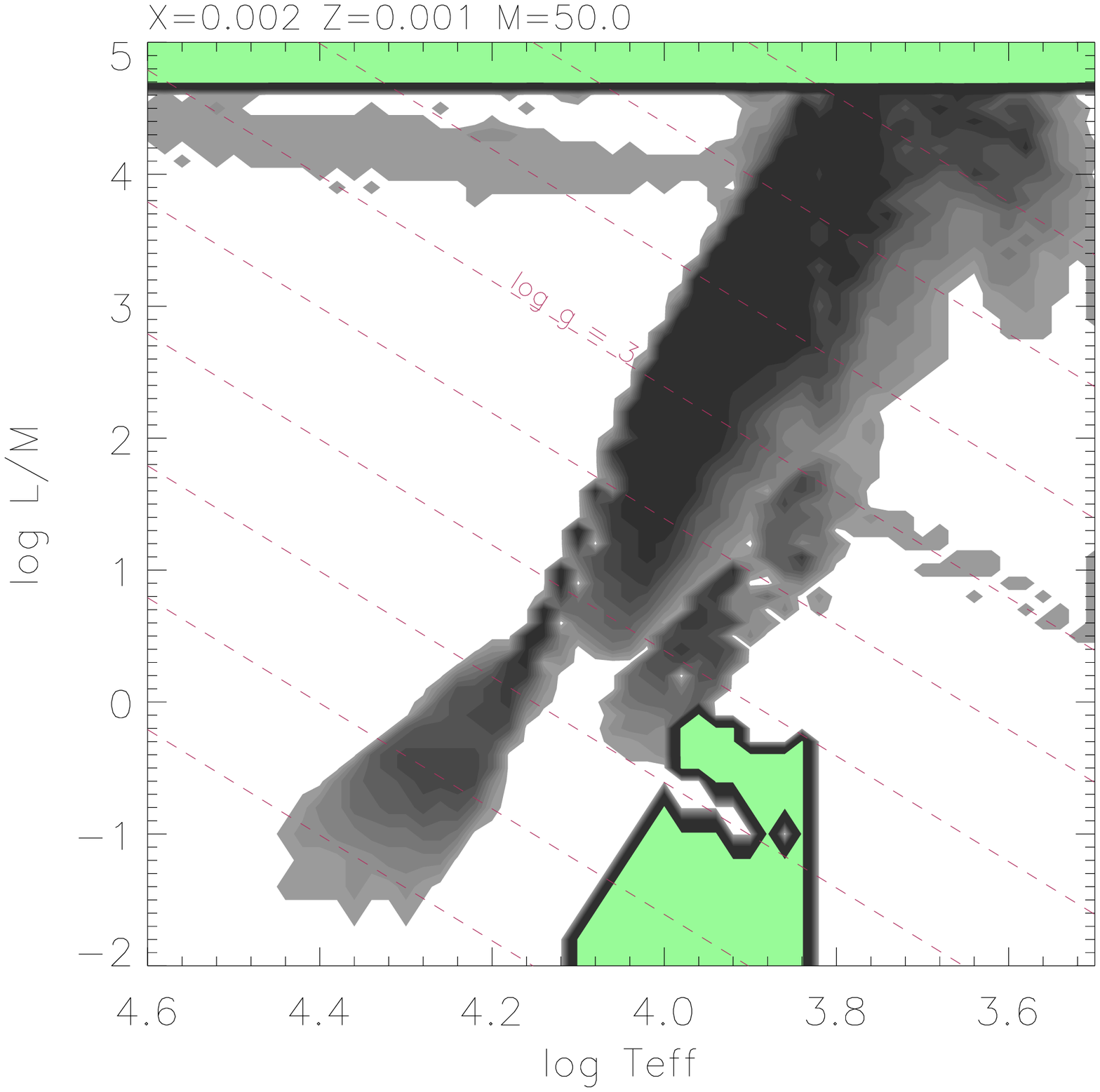,width=4.3cm,angle=0}
\caption[Unstable modes: $X=0.002, Z=0.001$]
{As Fig.~\ref{f:nx70} with $X=0.002, Z=0.001$, for models with $M \geq 4\Msolar$.
Lower-mass models encountered numerical problems at high $L/M$ ratios.  
}
\label{f:nx002z001}
\end{center}
\end{figure*}

\begin{figure*}
\begin{center}
\epsfig{file=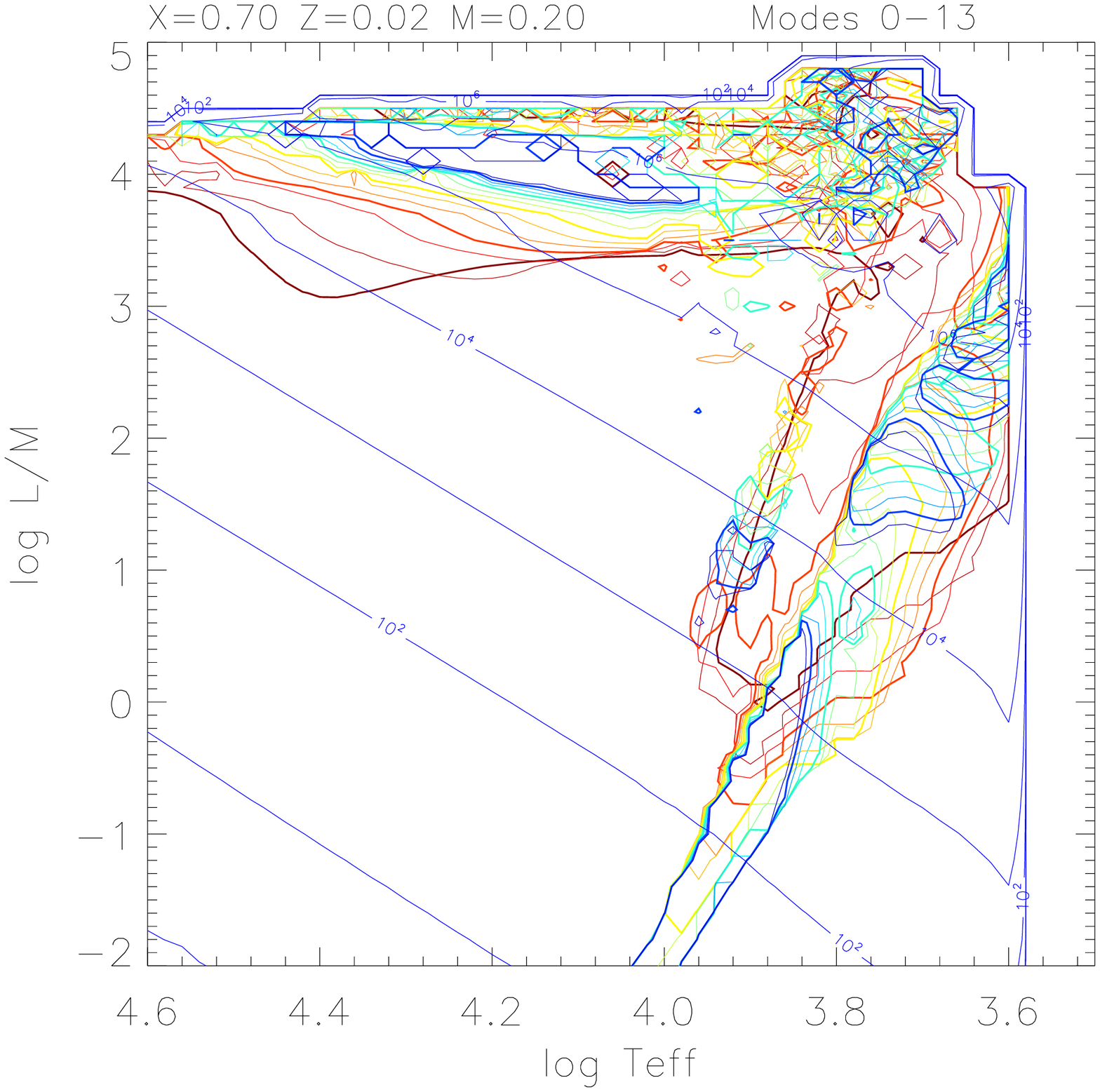,width=4.3cm,angle=0}
\epsfig{file=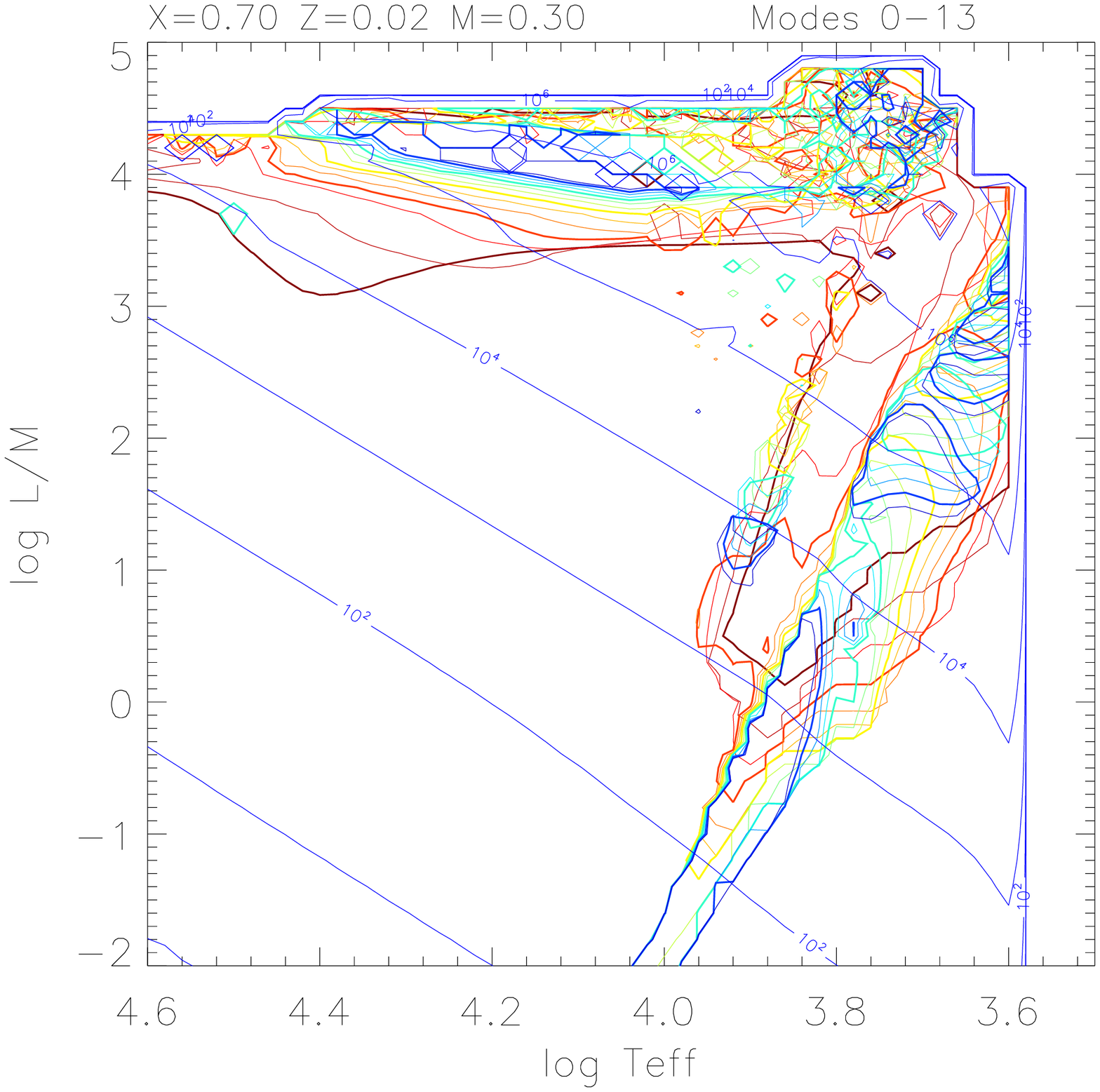,width=4.3cm,angle=0}
\epsfig{file=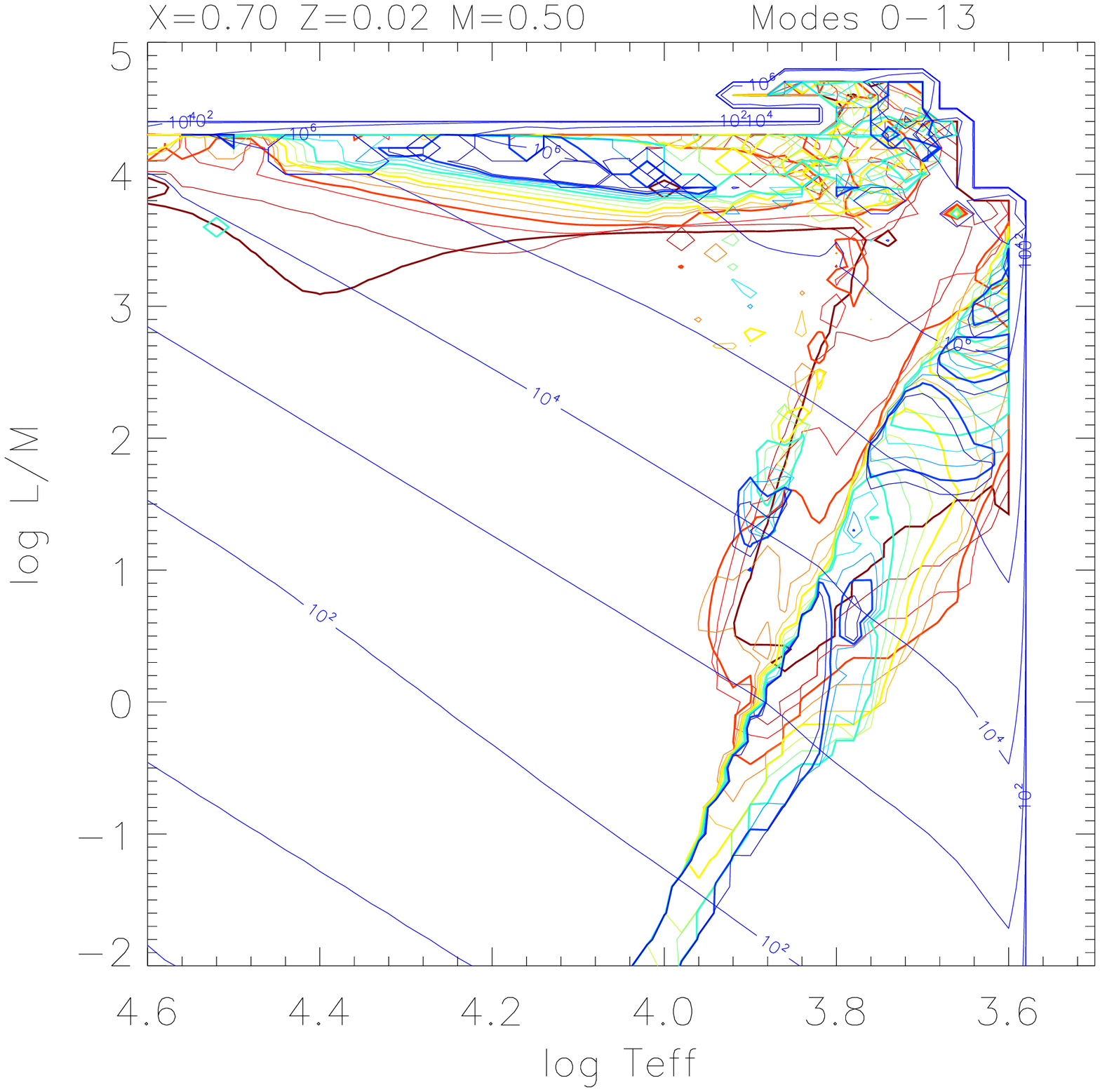,width=4.3cm,angle=0}
\epsfig{file=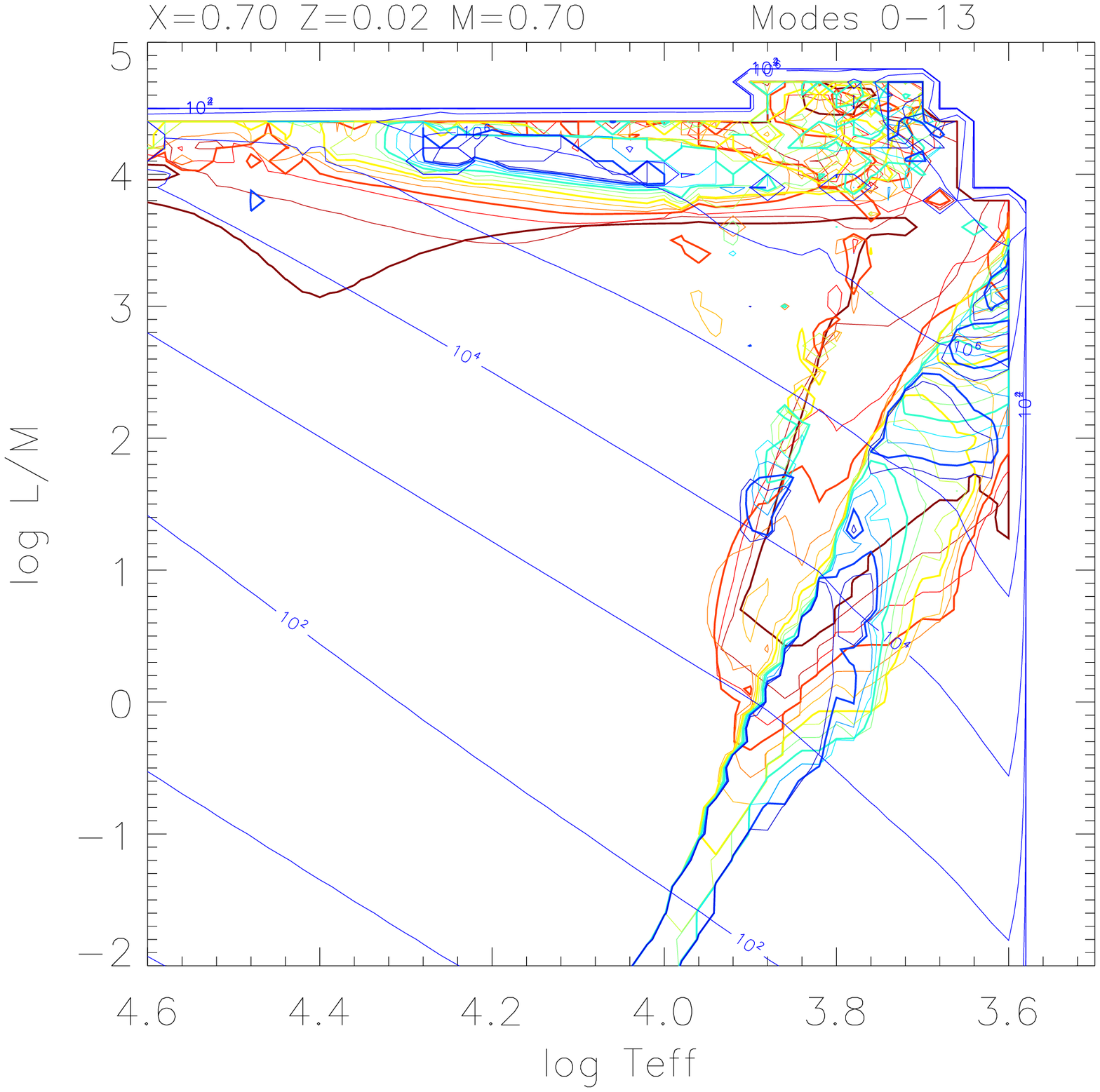,width=4.3cm,angle=0}\\
\epsfig{file=figs/periods_x70z02m01.0_00_opal.eps,width=4.3cm,angle=0}
\epsfig{file=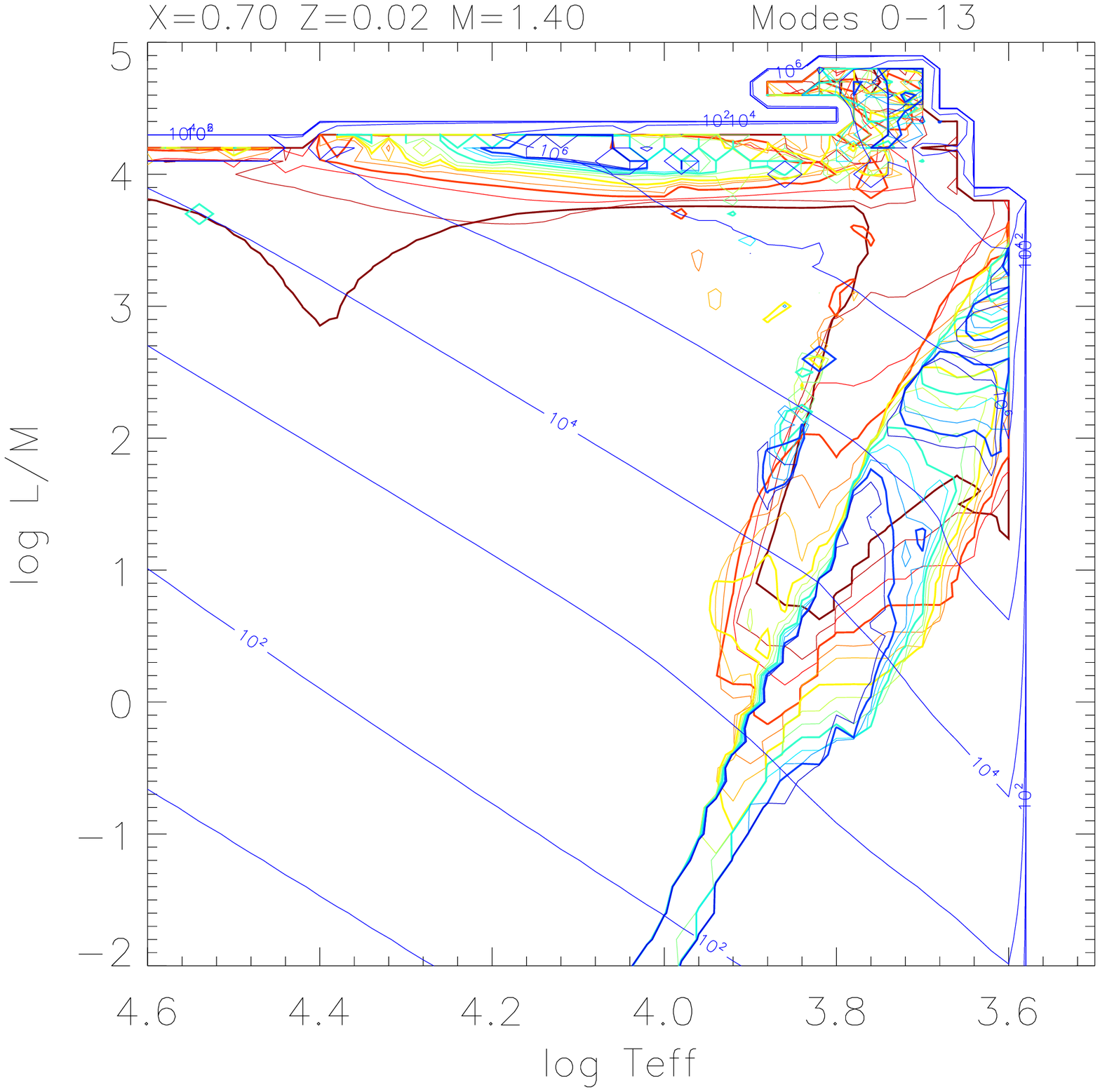,width=4.3cm,angle=0}
\epsfig{file=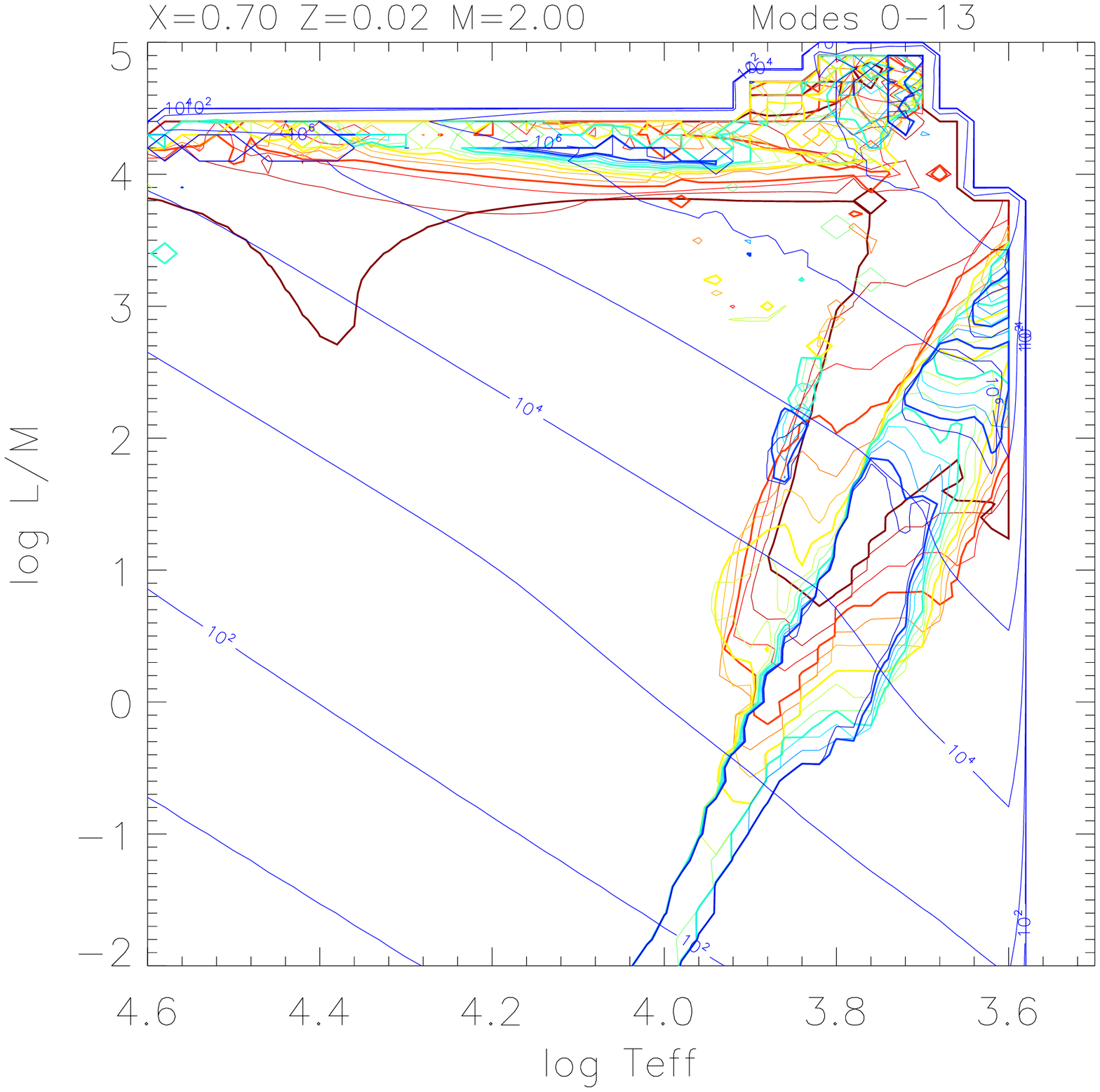,width=4.3cm,angle=0}
\epsfig{file=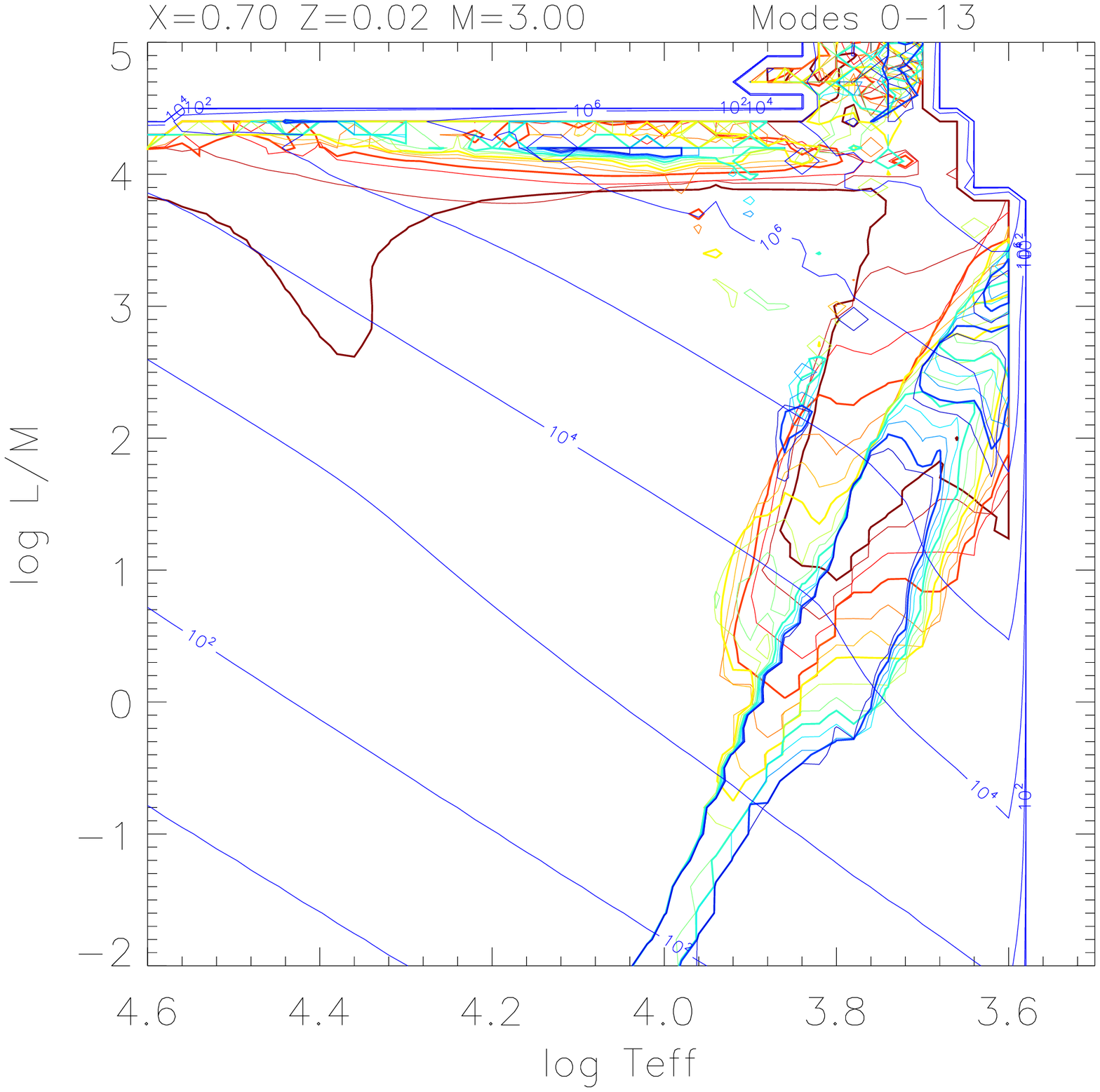,width=4.3cm,angle=0}\\
\epsfig{file=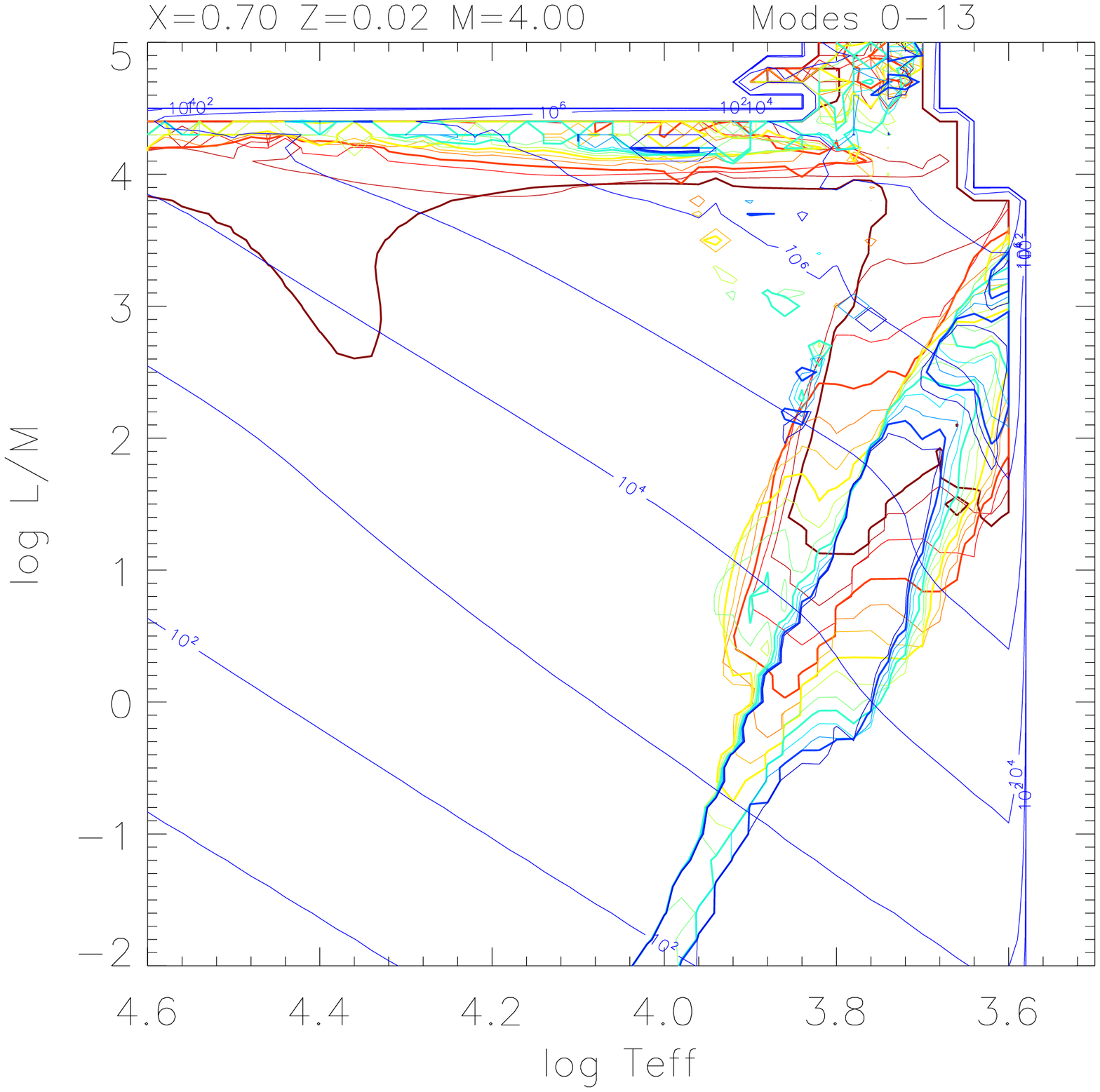,width=4.3cm,angle=0}
\epsfig{file=figs/periods_x70z02m05.0_00_opal.eps,width=4.3cm,angle=0}
\epsfig{file=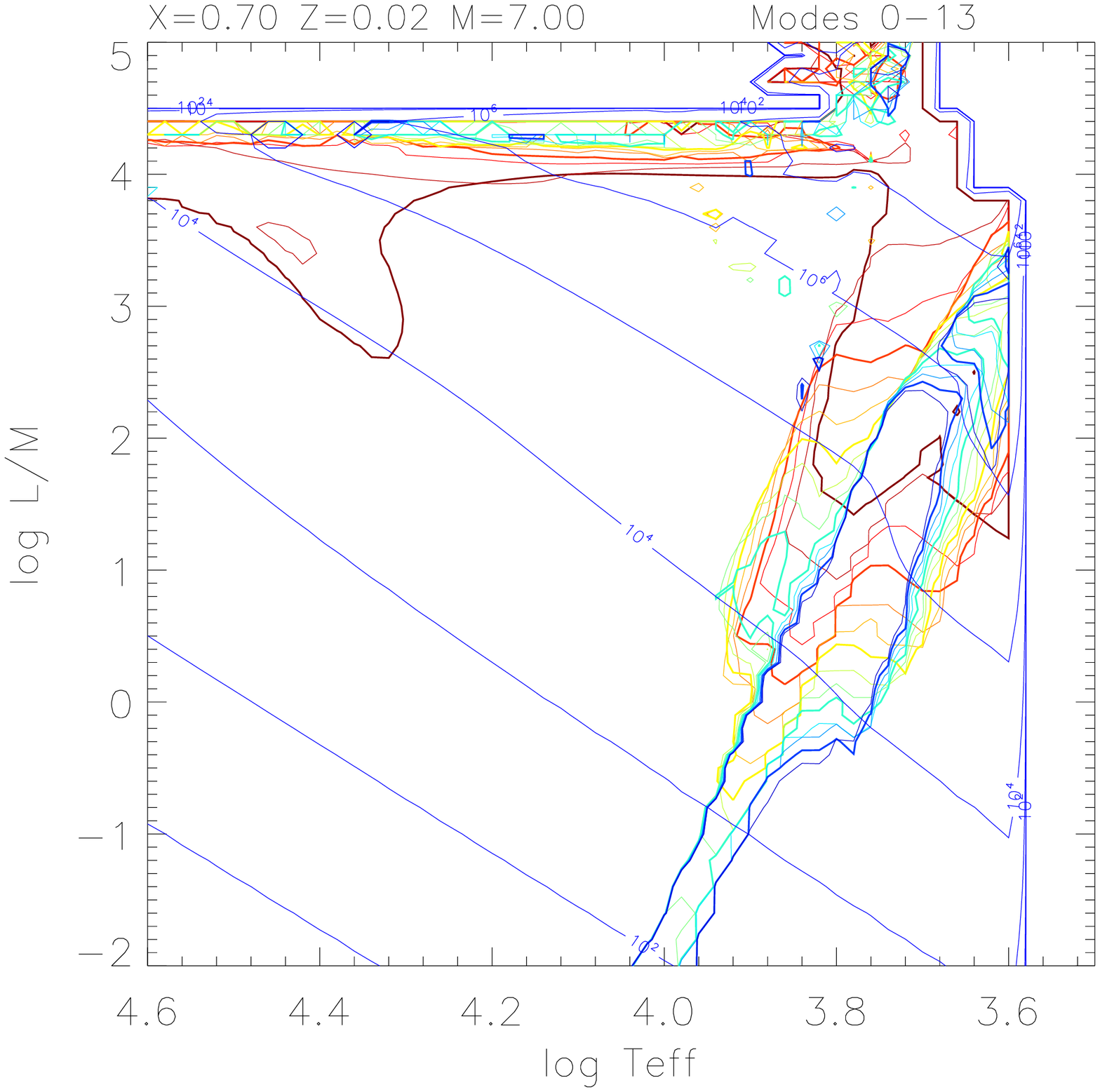,width=4.3cm,angle=0}
\epsfig{file=figs/periods_x70z02m10.0_00_opal.eps,width=4.3cm,angle=0}\\
\epsfig{file=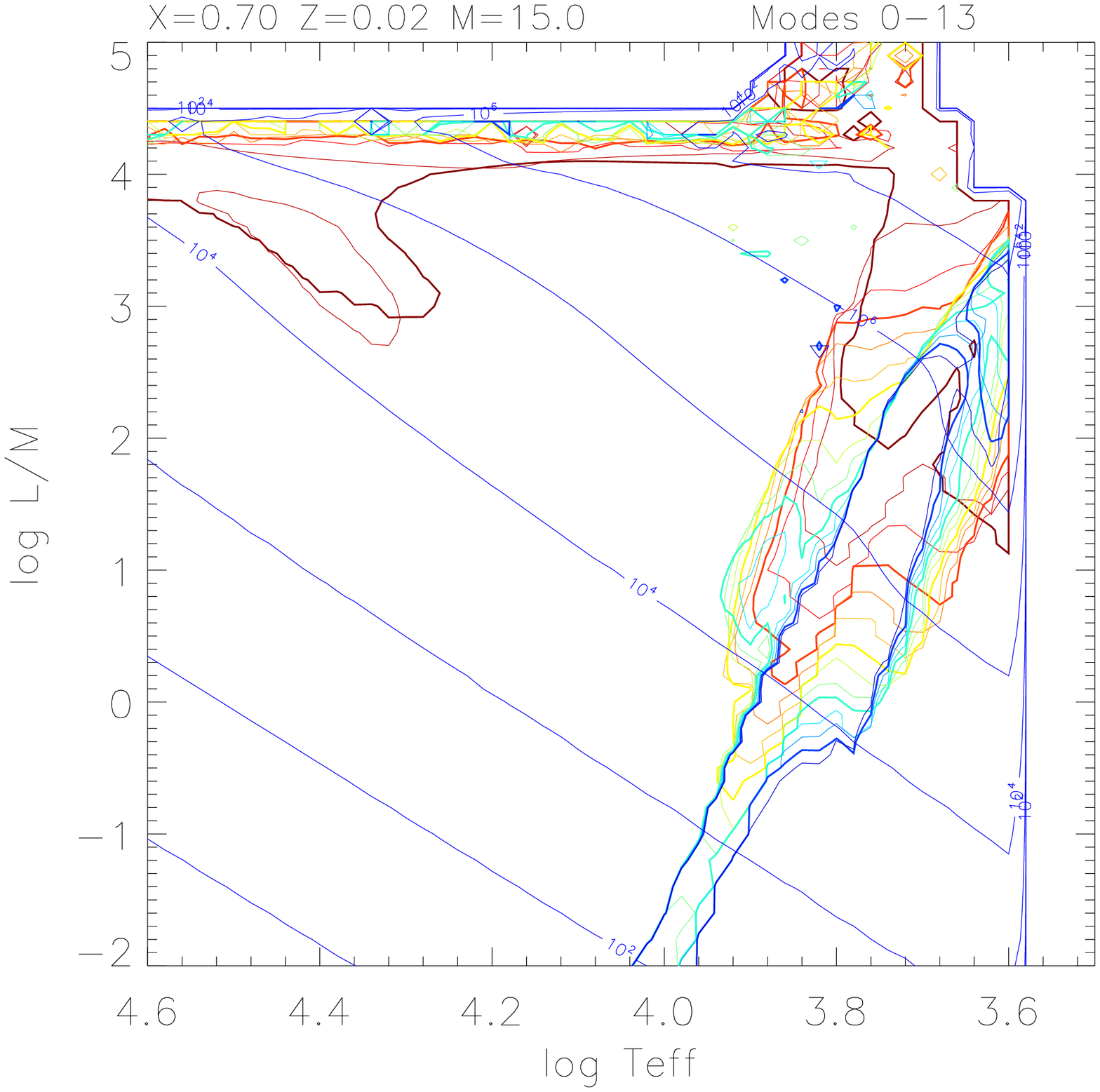,width=4.3cm,angle=0}
\epsfig{file=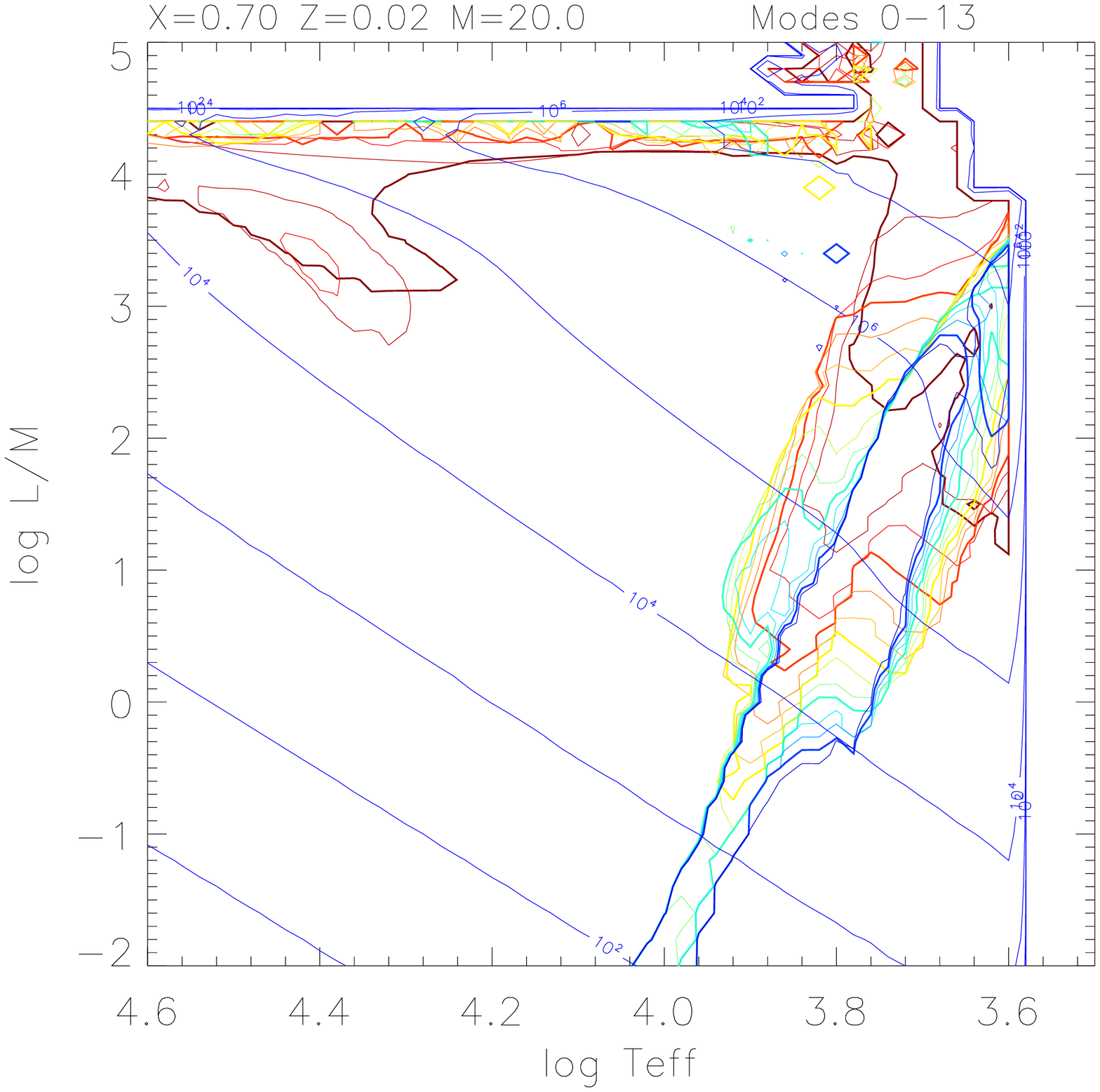,width=4.3cm,angle=0}
\epsfig{file=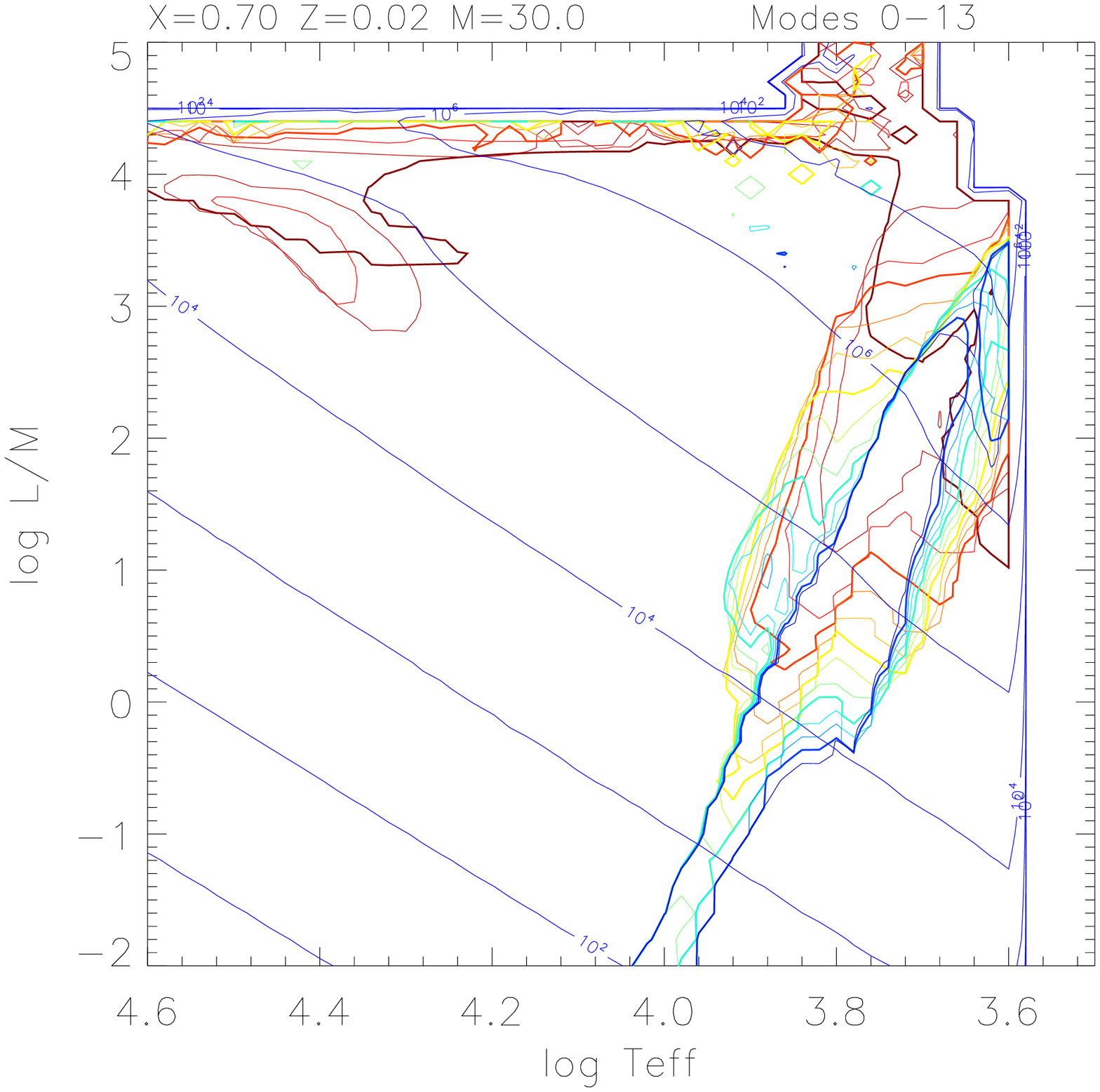,width=4.3cm,angle=0}
\epsfig{file=figs/periods_x70z02m50.0_00_opal.eps,width=4.3cm,angle=0}
\caption[Unstable modes: $X=0.70, Z=0.02$]
{Unstable pulsation mode boundaries in stars with homogeneous
envelopes with hydrogen content $X=0.70, Z=0.02$,  
 and OPAL opacities, and mass $0.20 < M/\Msolar < 50$, as labelled. 
The boundaries of unstable radial modes are  represented by coloured contours, with the darkest red representing
the boundary of the fundamental ($n=0$) mode, with increasing higher orders represented progressively by colours
of increasing frequency (orange, yellow, green, blue \ldots). 
Models with $X=0.7$ and $\log T_{\rm eff} < 3.6$ were excluded.
Solid blue lines represent contours of equal fundamental radial-mode period in seconds spaced at decadal intervals. 
}
\label{f:px70}
\end{center}
\end{figure*}

\begin{figure*}
\begin{center}
\epsfig{file=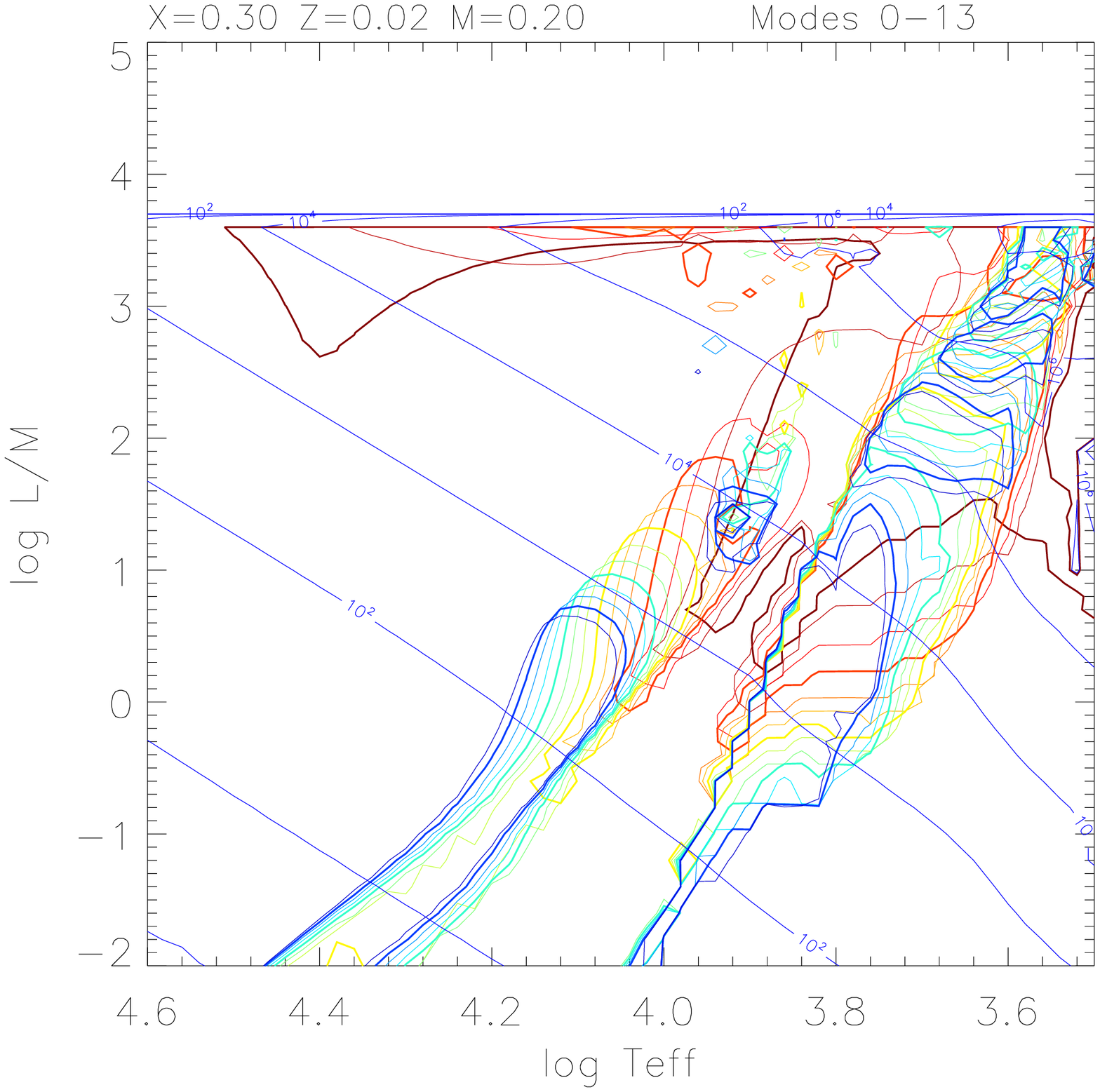,width=4.3cm,angle=0}
\epsfig{file=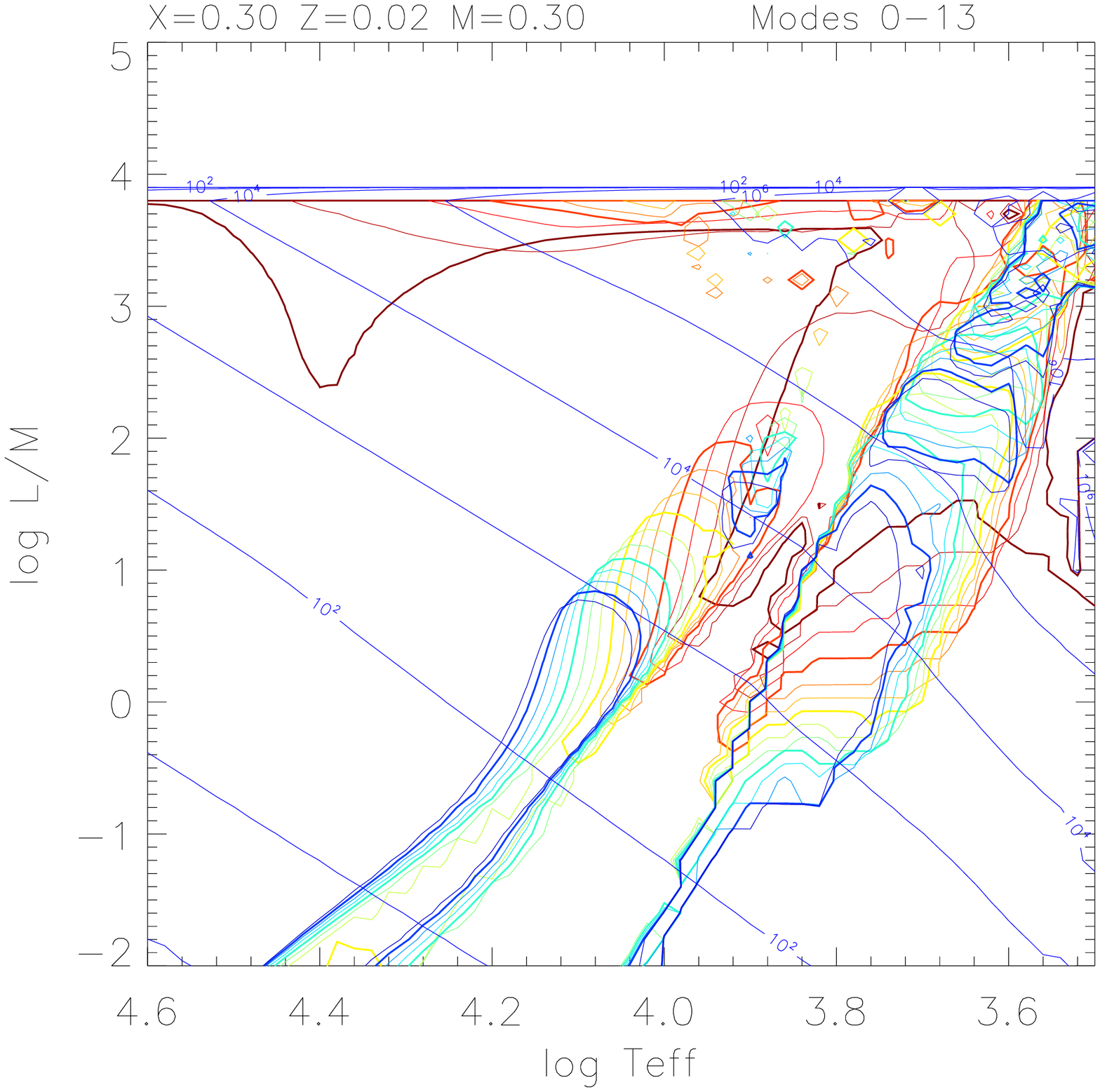,width=4.3cm,angle=0}
\epsfig{file=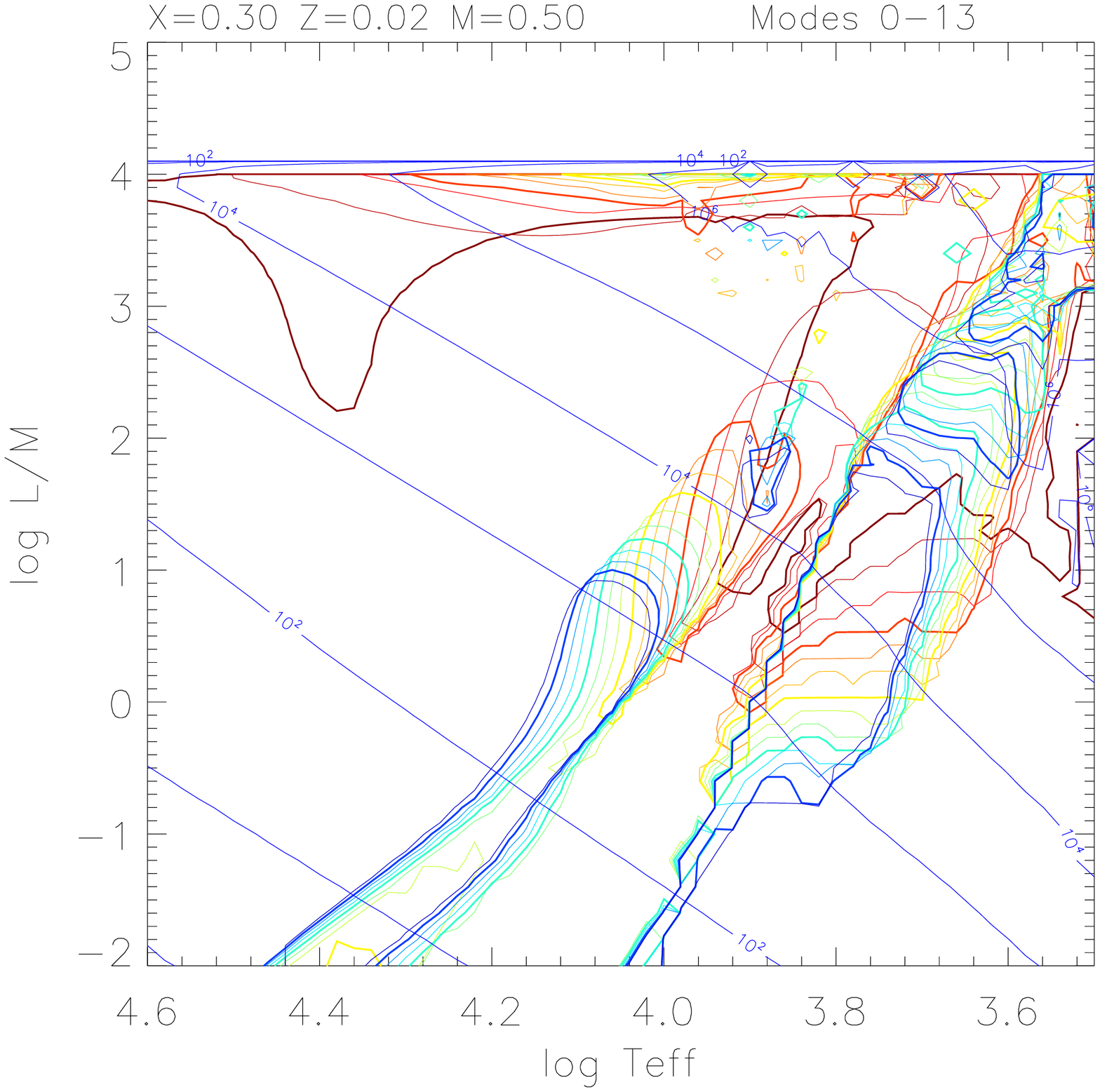,width=4.3cm,angle=0}
\epsfig{file=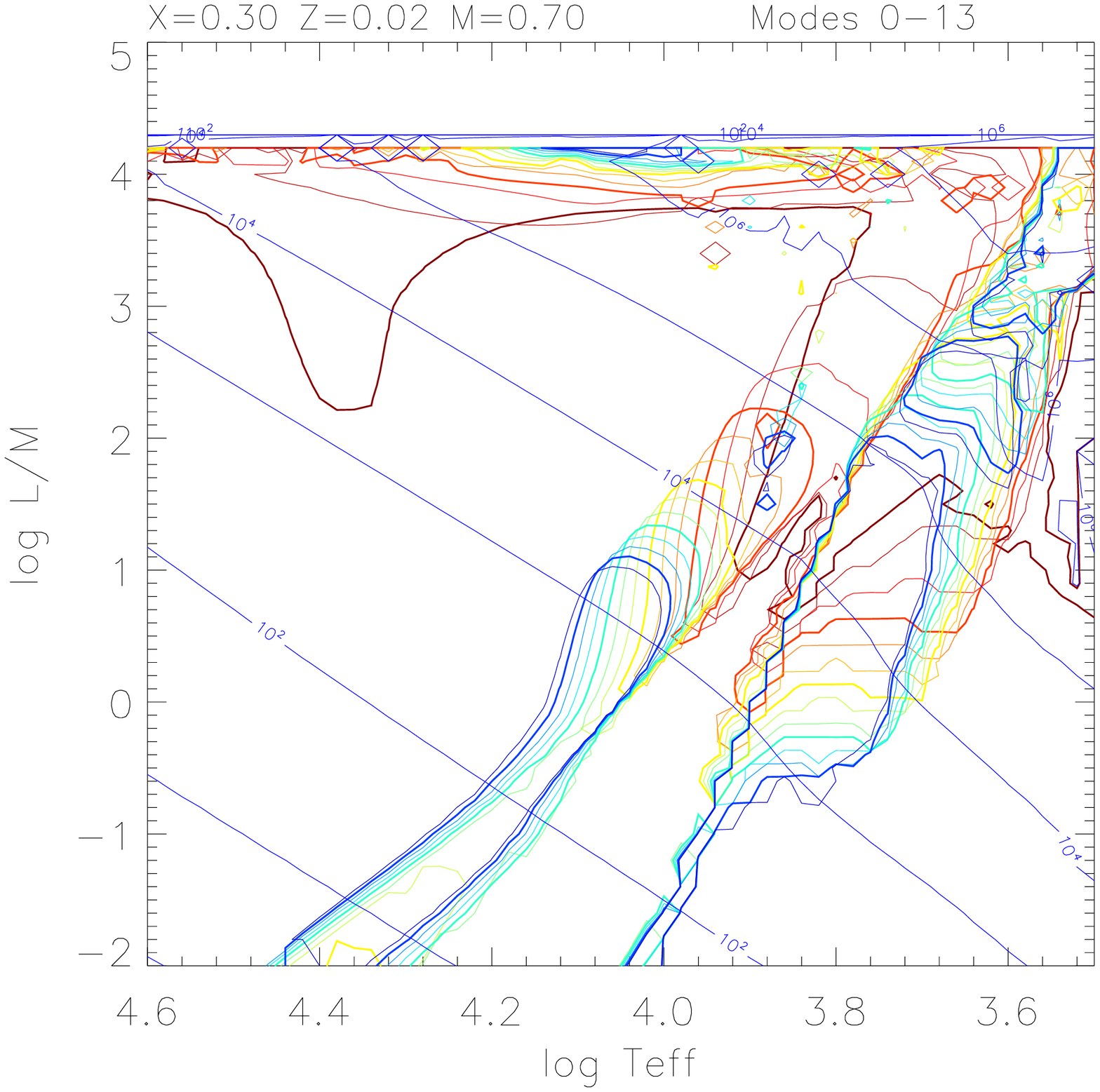,width=4.3cm,angle=0}\\
\epsfig{file=figs/periods_x30z02m01.0_00_opal.eps,width=4.3cm,angle=0}
\epsfig{file=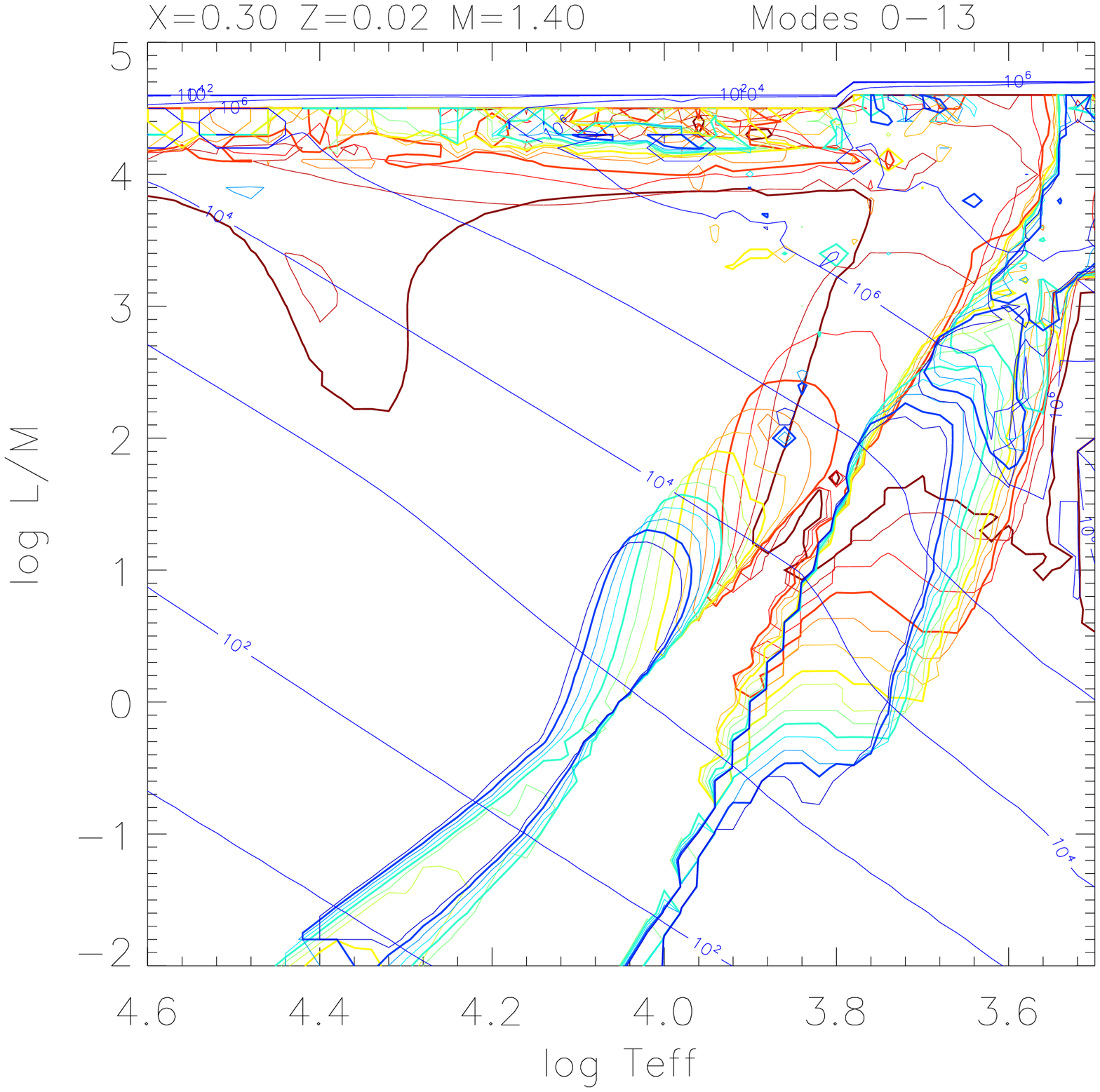,width=4.3cm,angle=0}
\epsfig{file=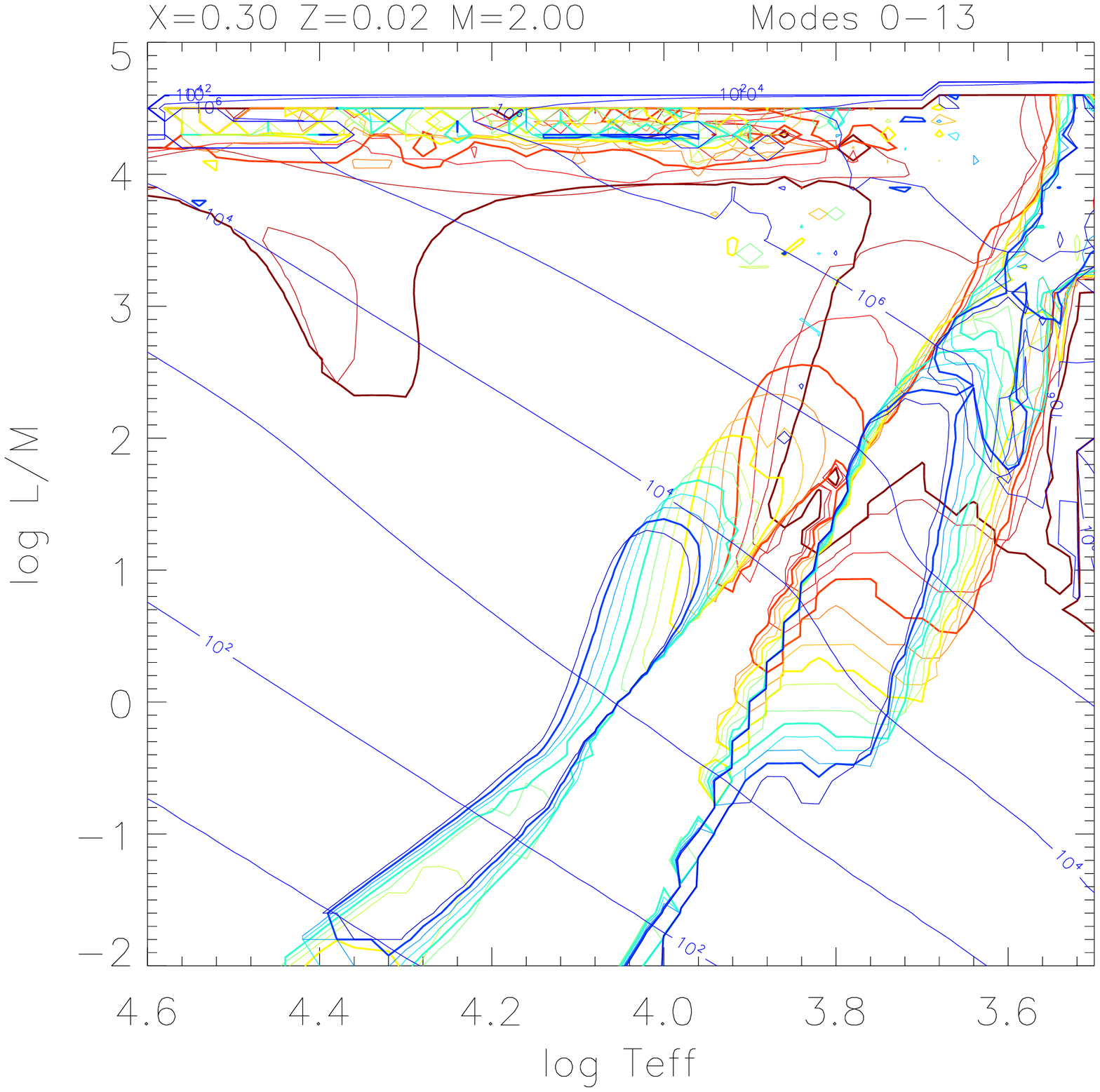,width=4.3cm,angle=0}
\epsfig{file=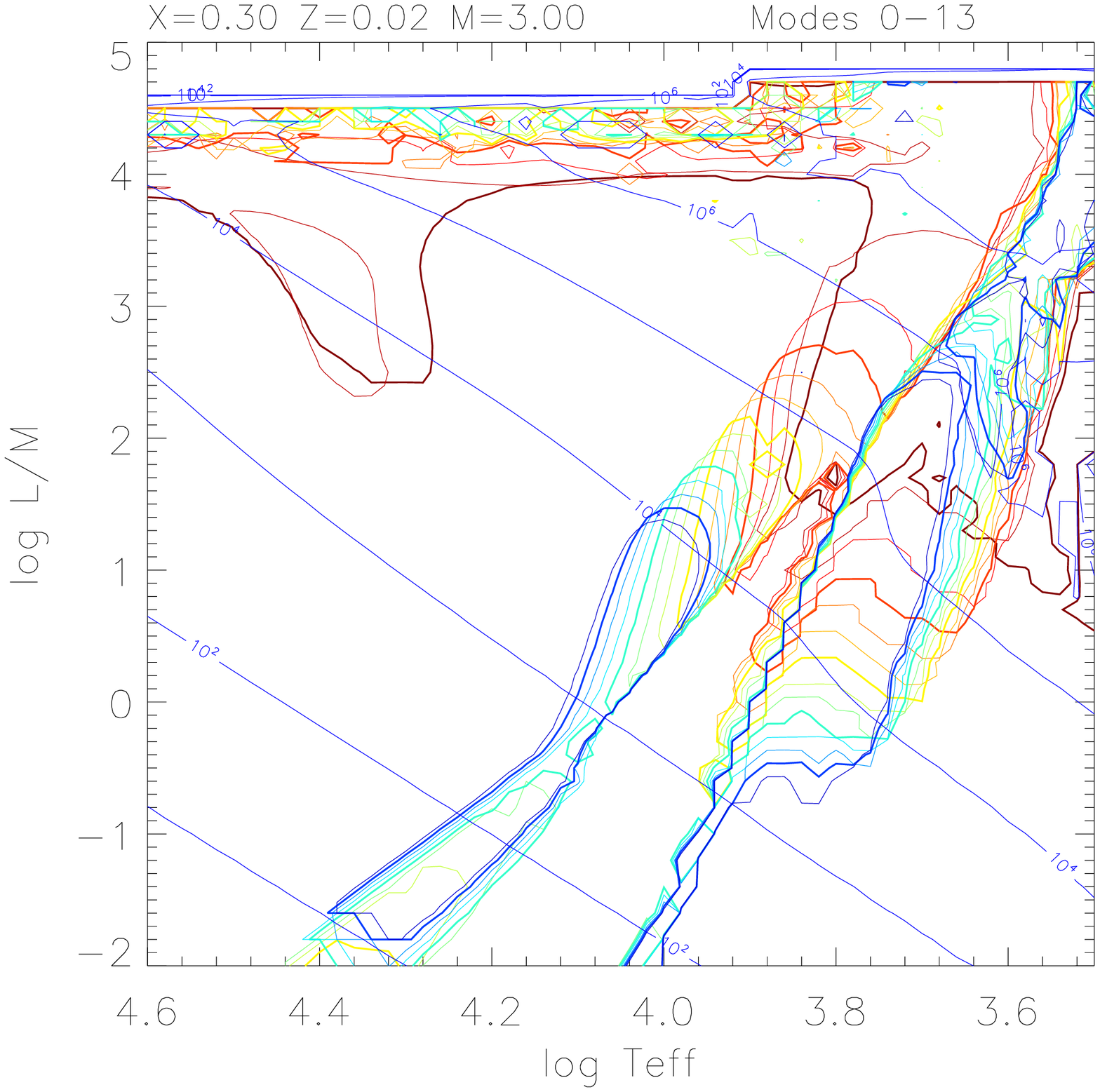,width=4.3cm,angle=0}\\
\epsfig{file=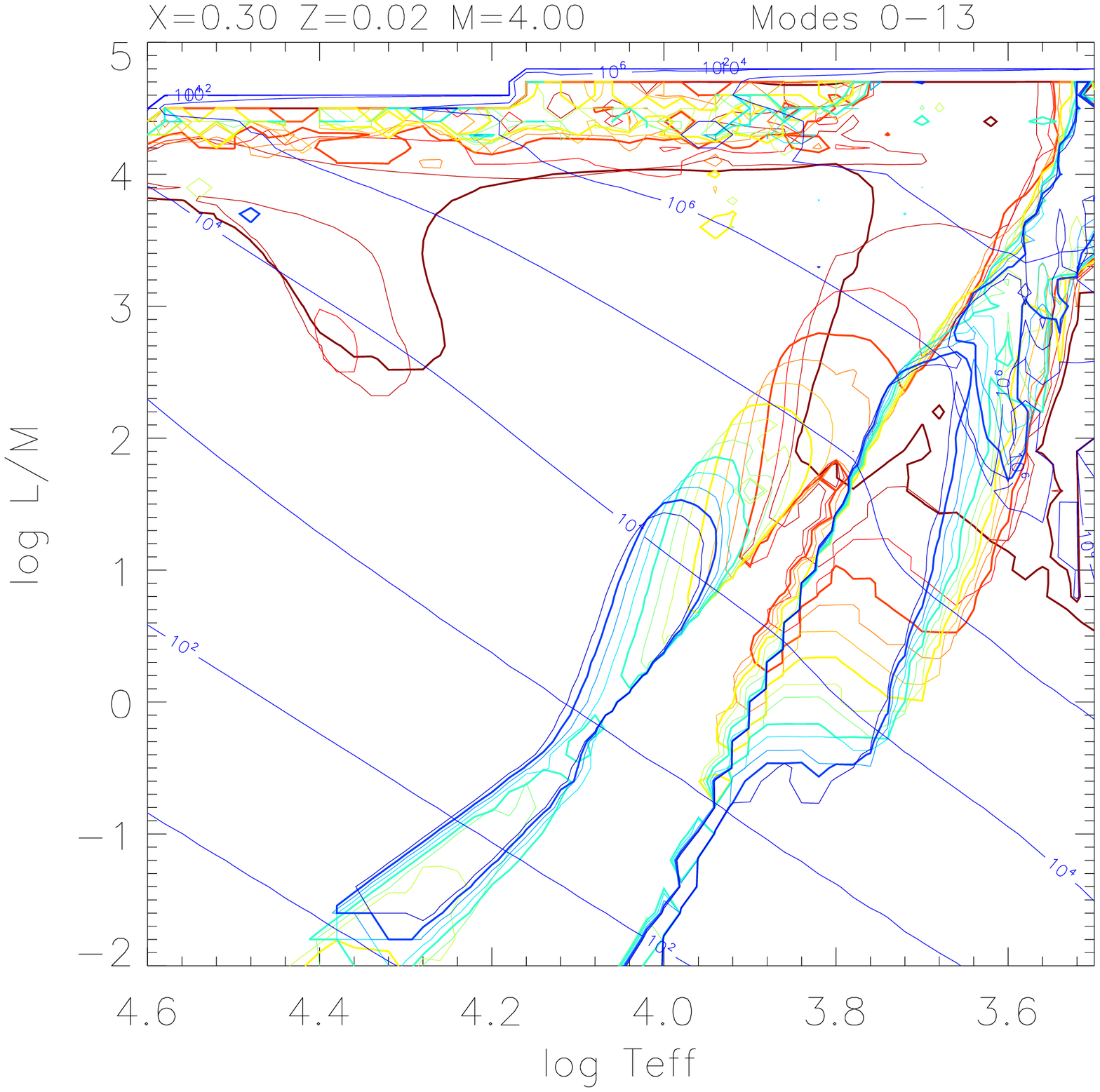,width=4.3cm,angle=0}
\epsfig{file=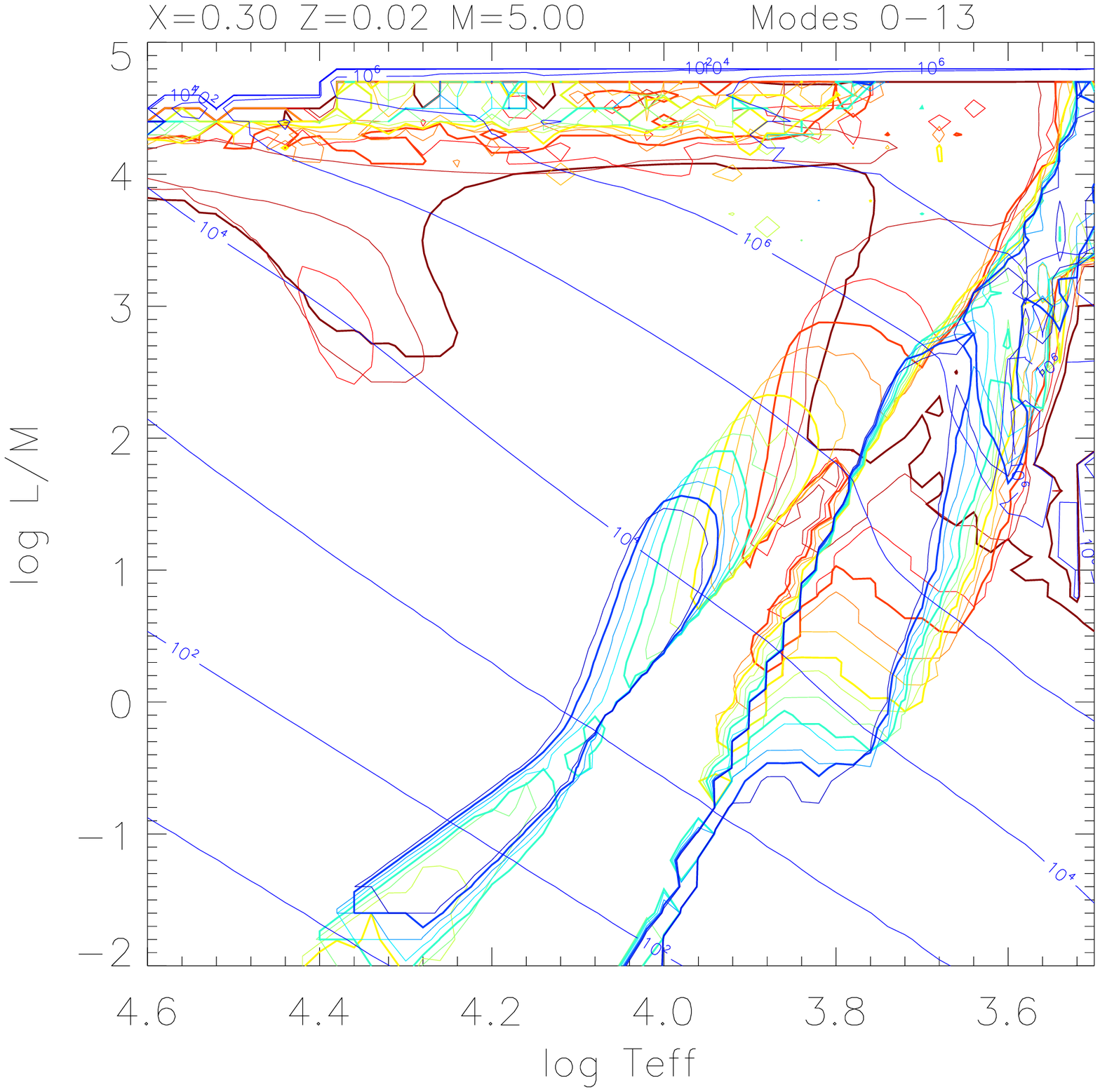,width=4.3cm,angle=0}
\epsfig{file=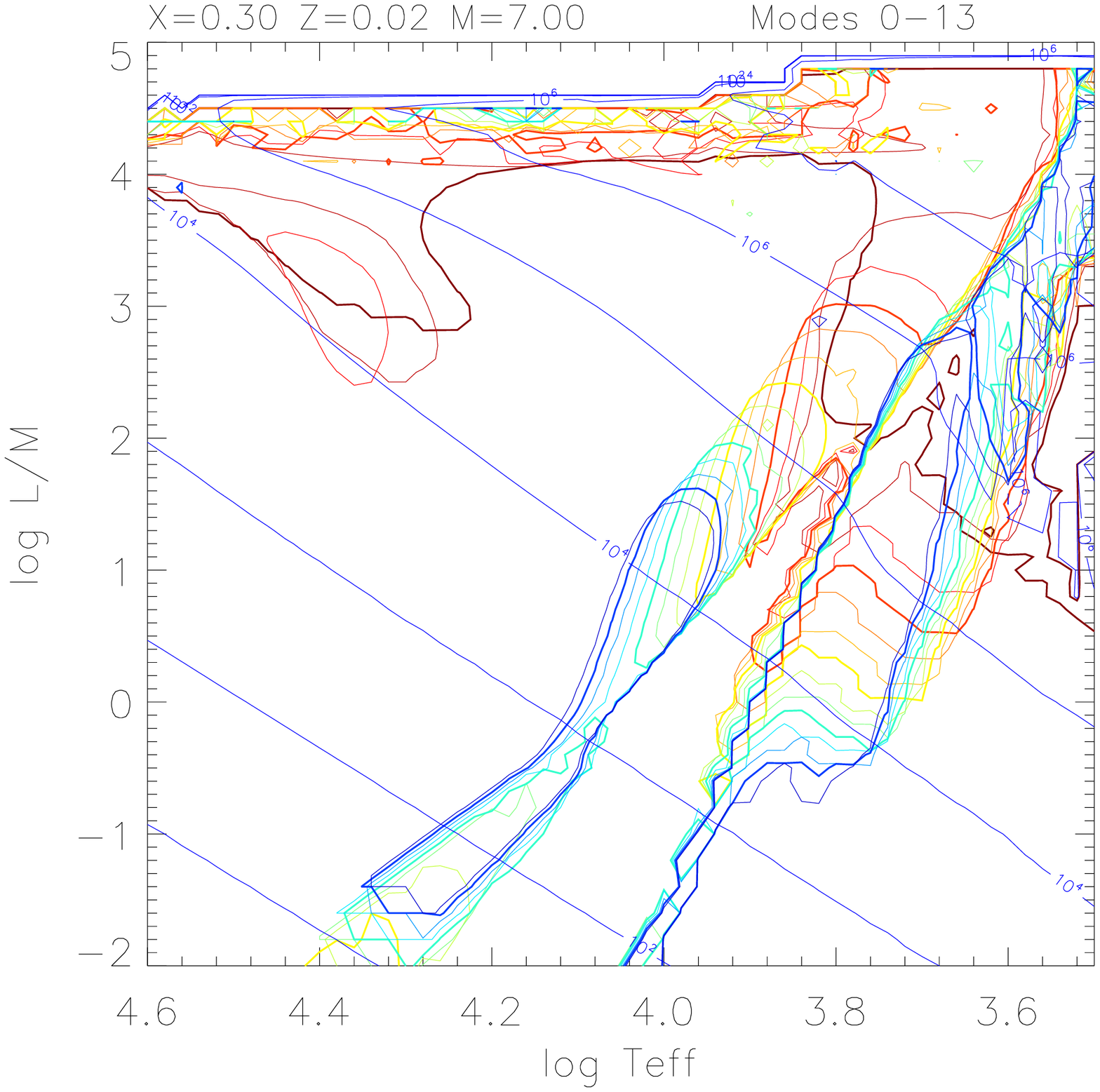,width=4.3cm,angle=0}
\epsfig{file=figs/periods_x30z02m10.0_00_opal.eps,width=4.3cm,angle=0}\\
\epsfig{file=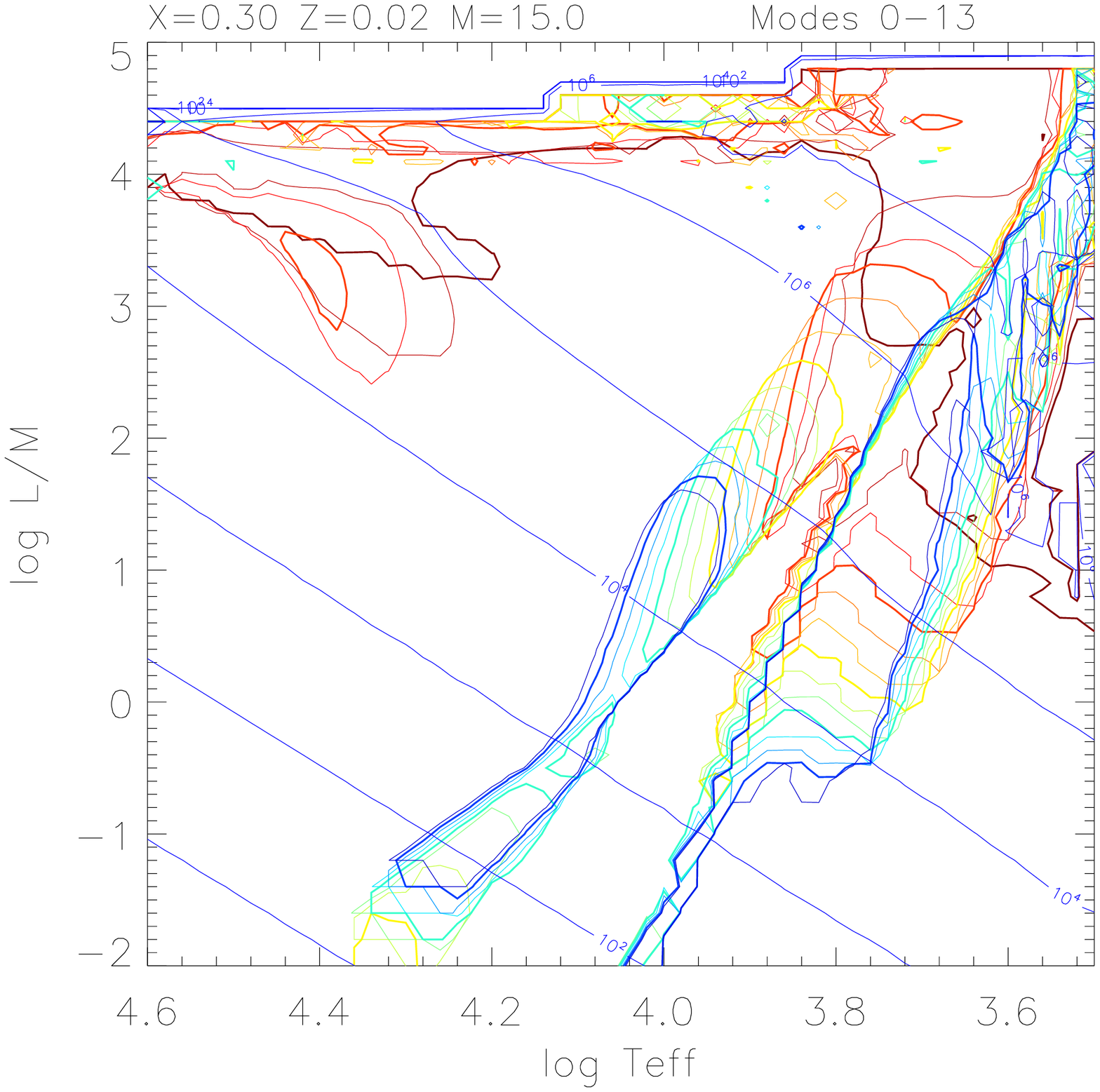,width=4.3cm,angle=0}
\epsfig{file=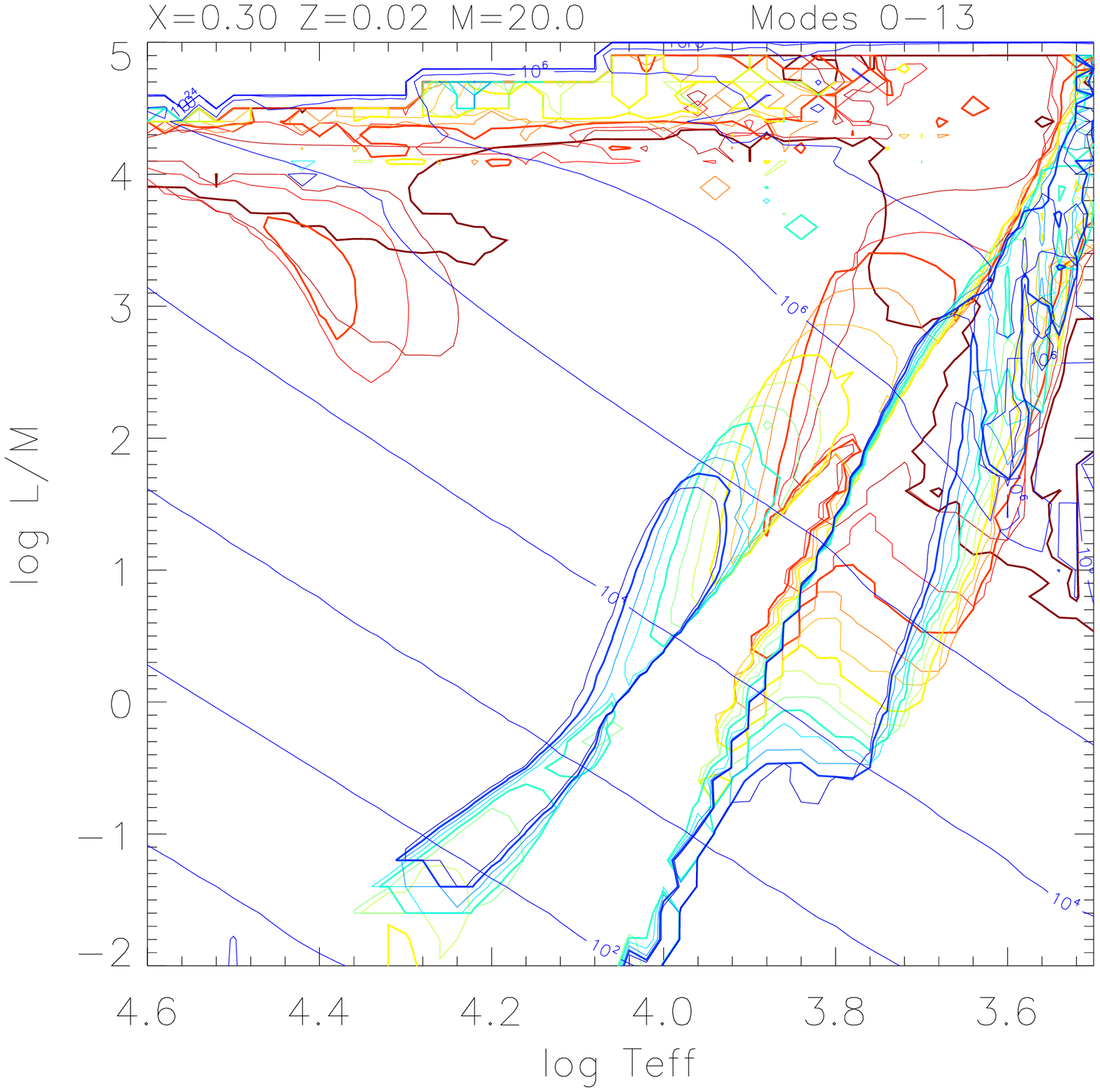,width=4.3cm,angle=0}
\epsfig{file=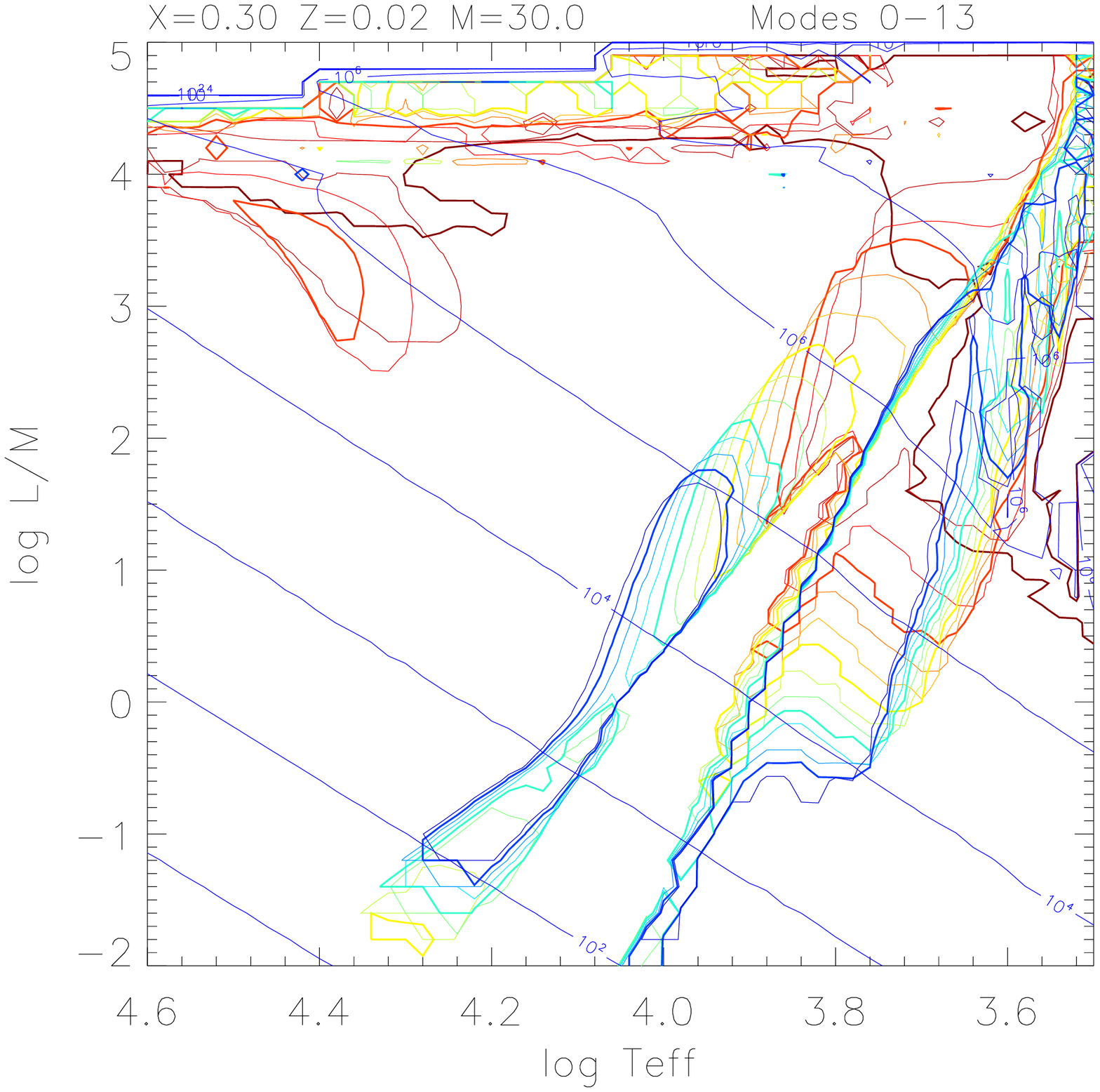,width=4.3cm,angle=0}
\epsfig{file=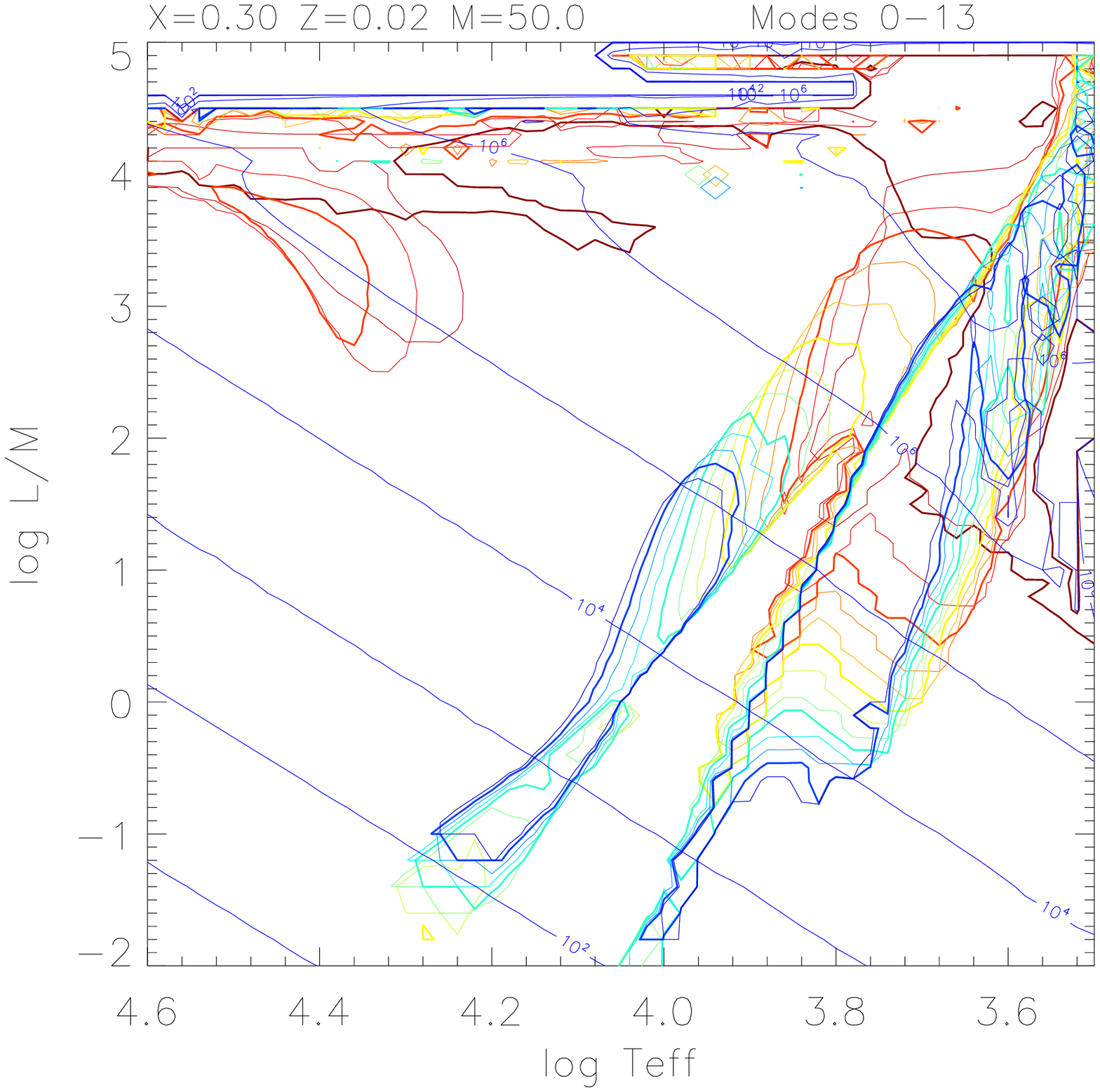,width=4.3cm,angle=0}
\caption[Unstable modes: $X=0.30, Z=0.02$]
{As Fig.~\ref{f:px70} with $X=0.30, Z=0.02$. 
}
\label{f:px30}
\end{center}
\end{figure*}

\begin{figure*}
\begin{center}
\epsfig{file=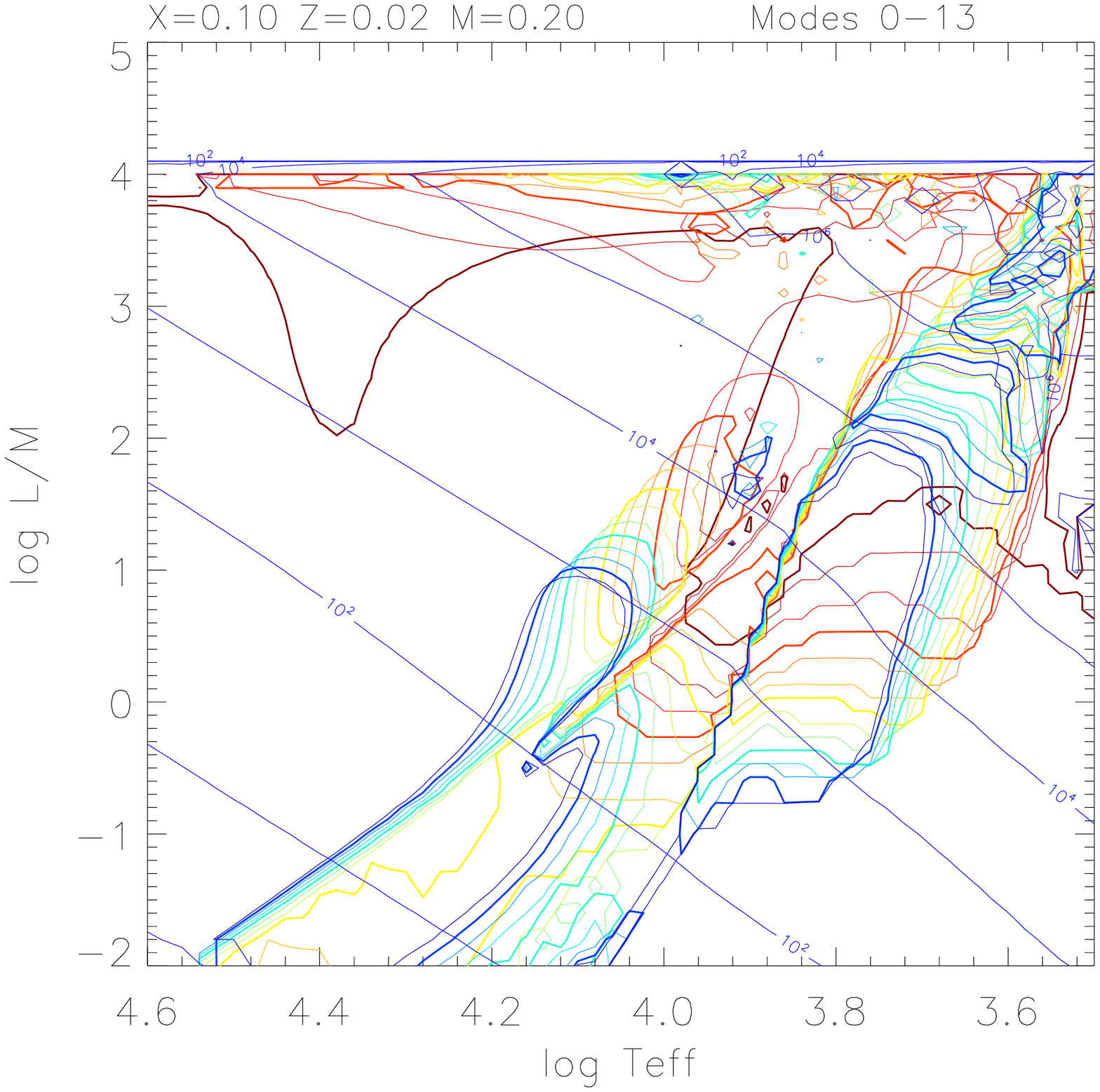,width=4.3cm,angle=0}
\epsfig{file=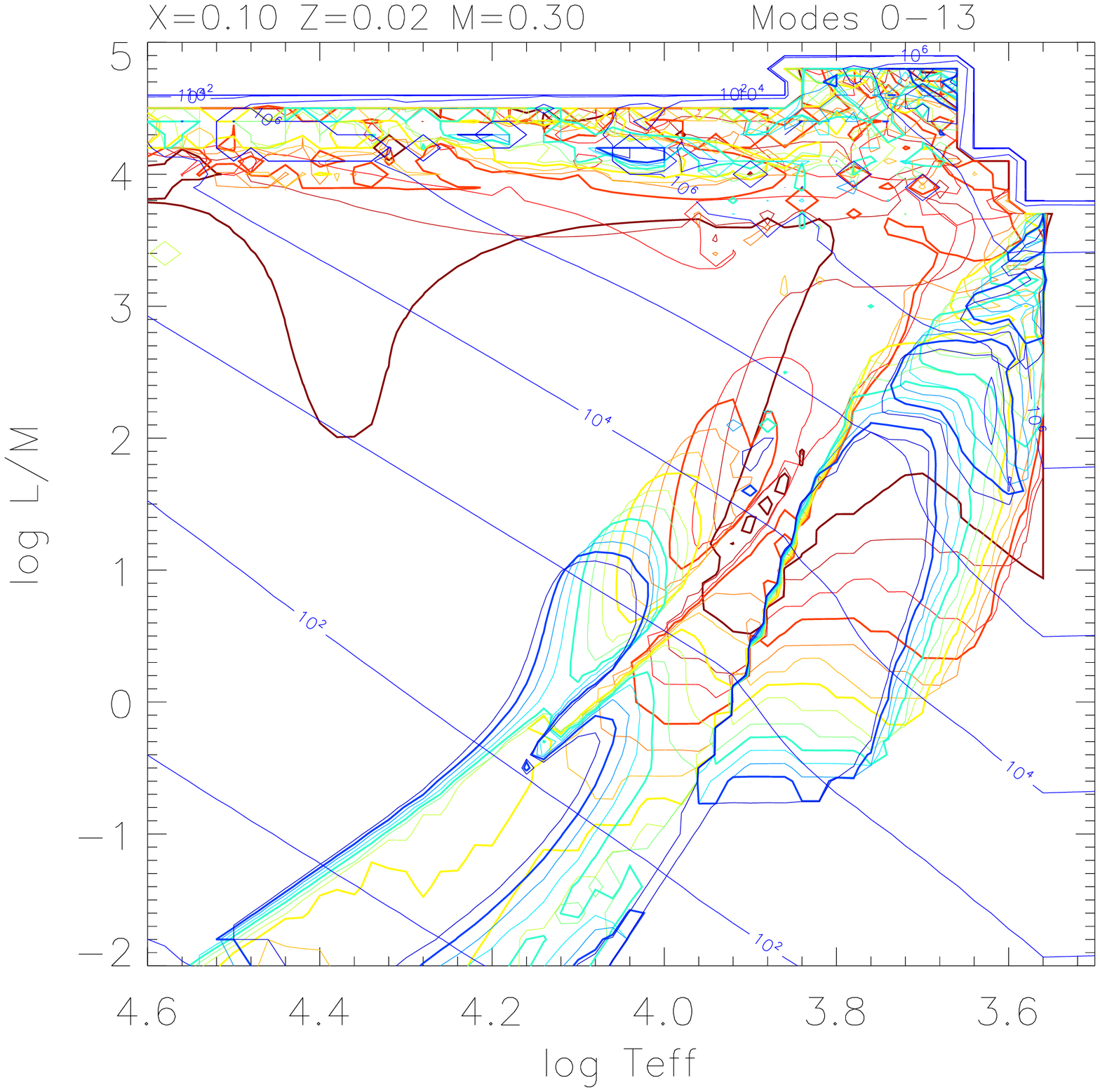,width=4.3cm,angle=0}
\epsfig{file=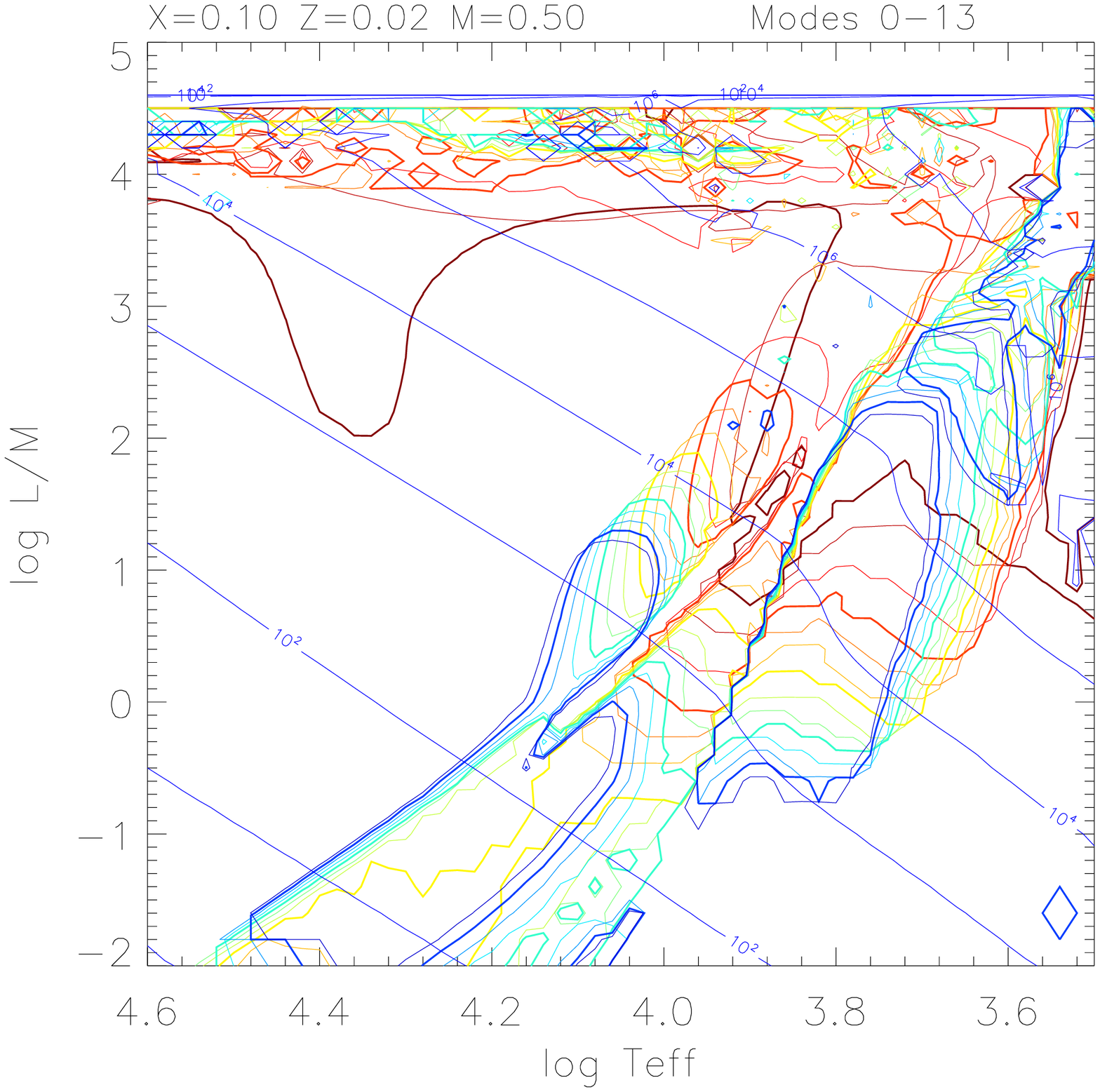,width=4.3cm,angle=0}
\epsfig{file=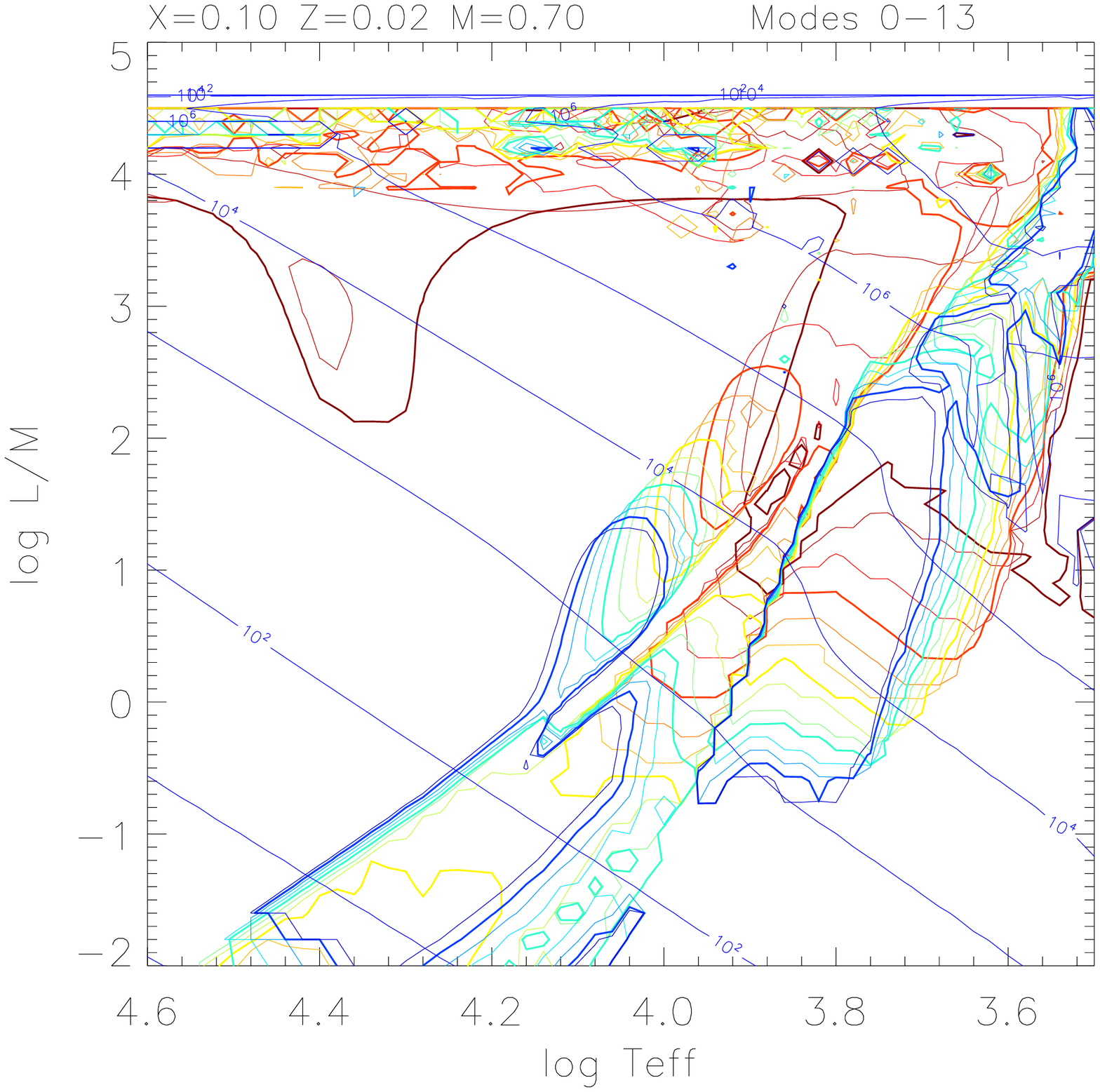,width=4.3cm,angle=0}\\
\epsfig{file=figs/periods_x10z02m01.0_00_opal.eps,width=4.3cm,angle=0}
\epsfig{file=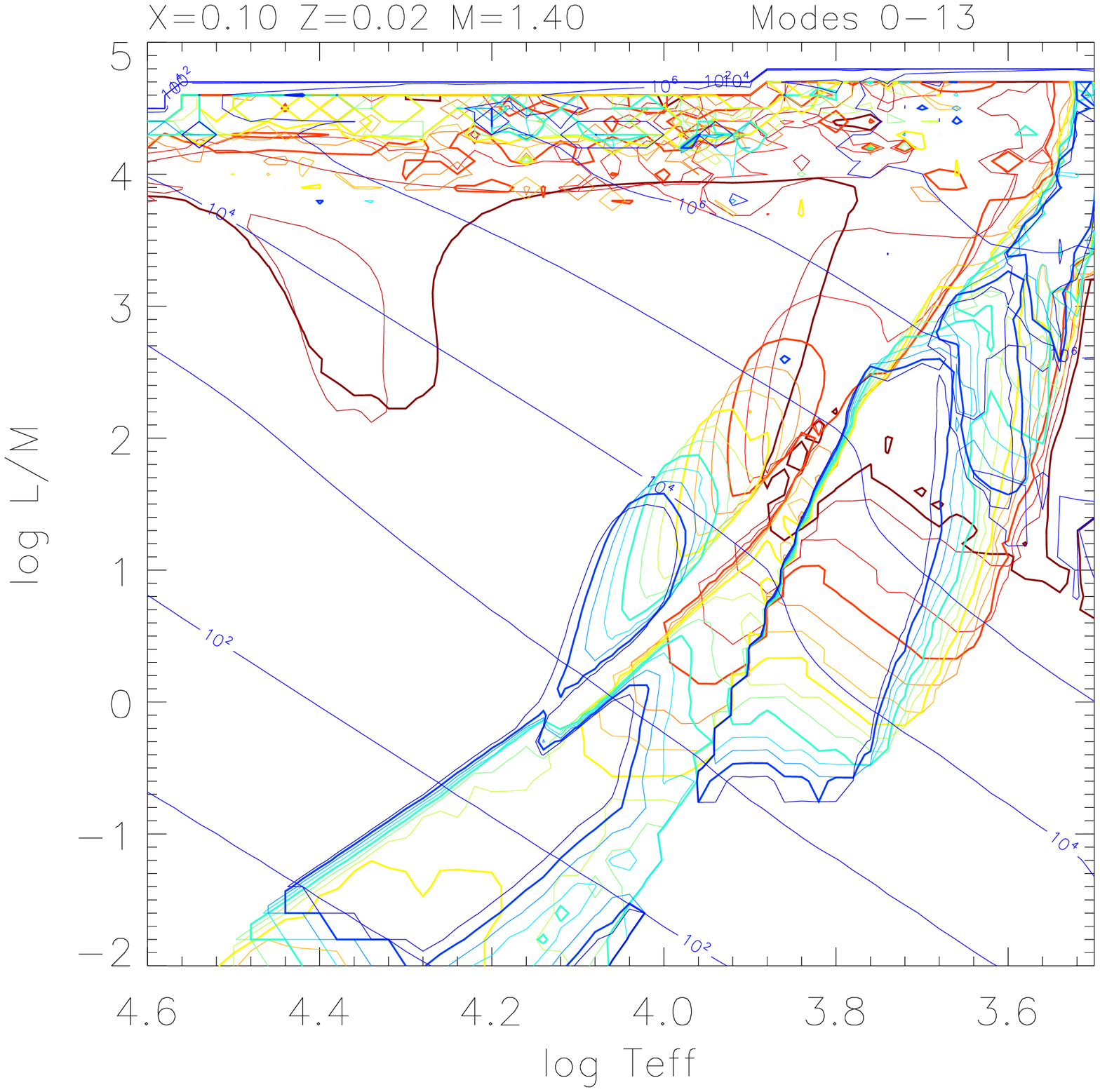,width=4.3cm,angle=0}
\epsfig{file=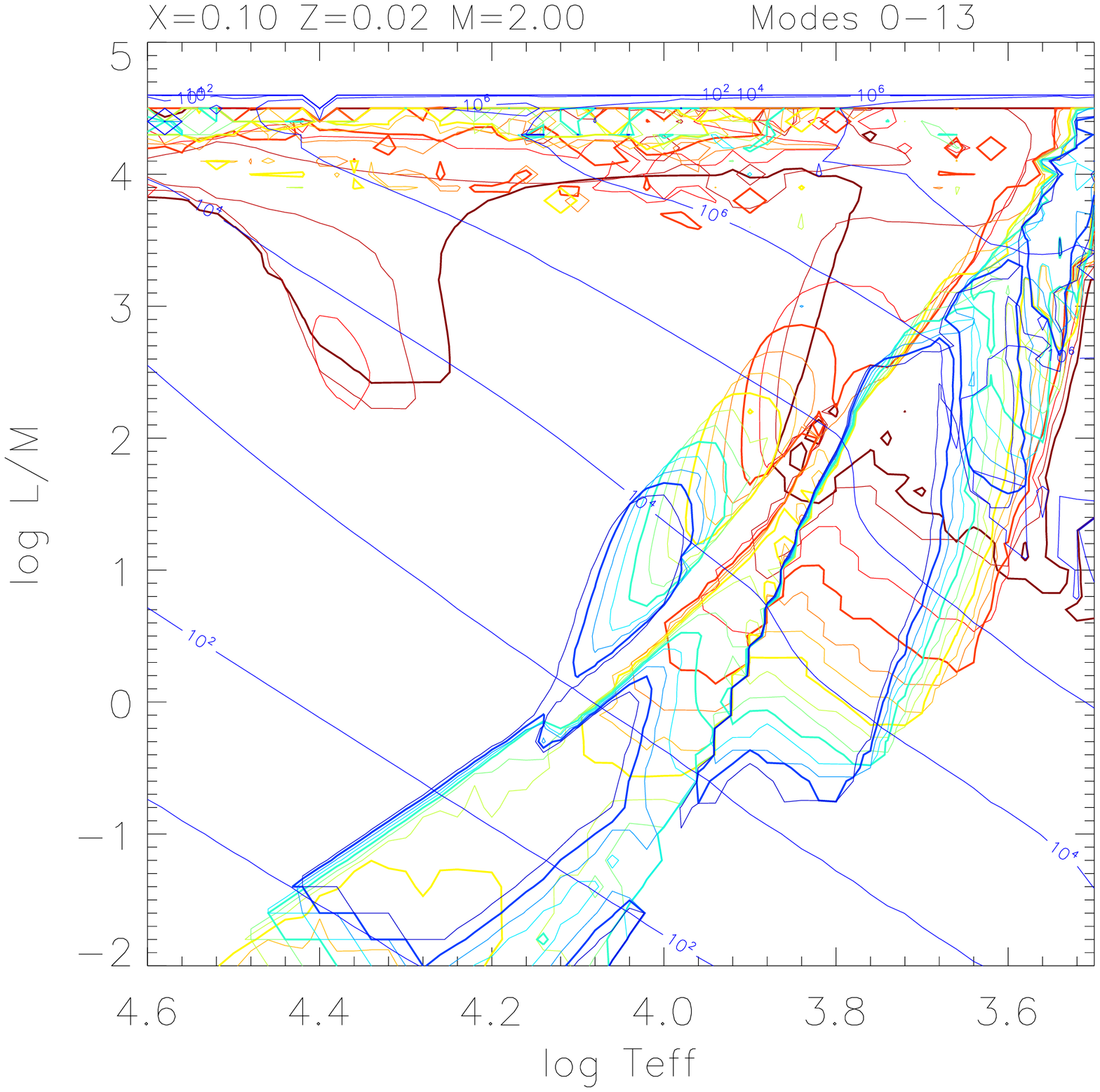,width=4.3cm,angle=0}
\epsfig{file=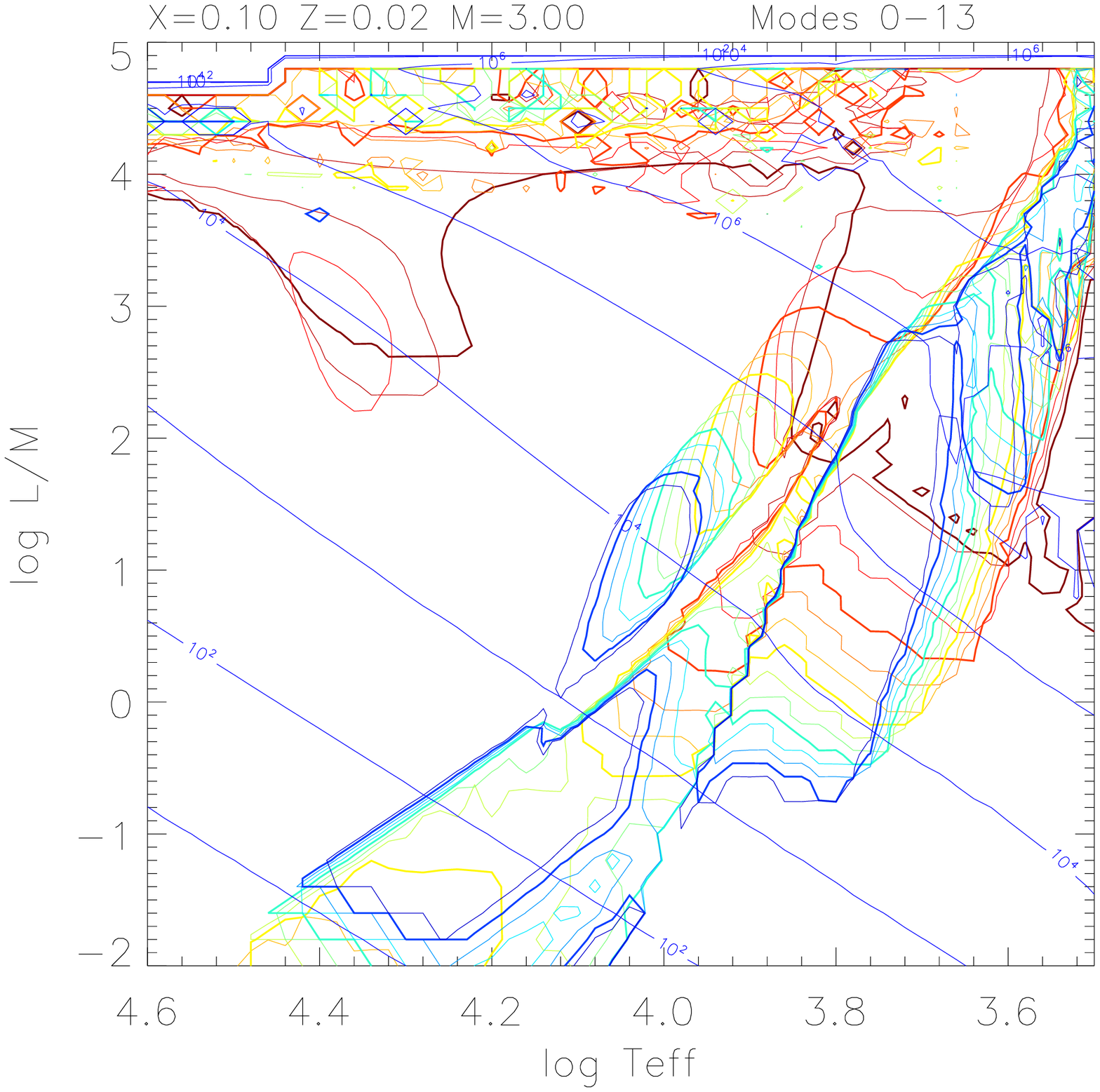,width=4.3cm,angle=0}\\
\epsfig{file=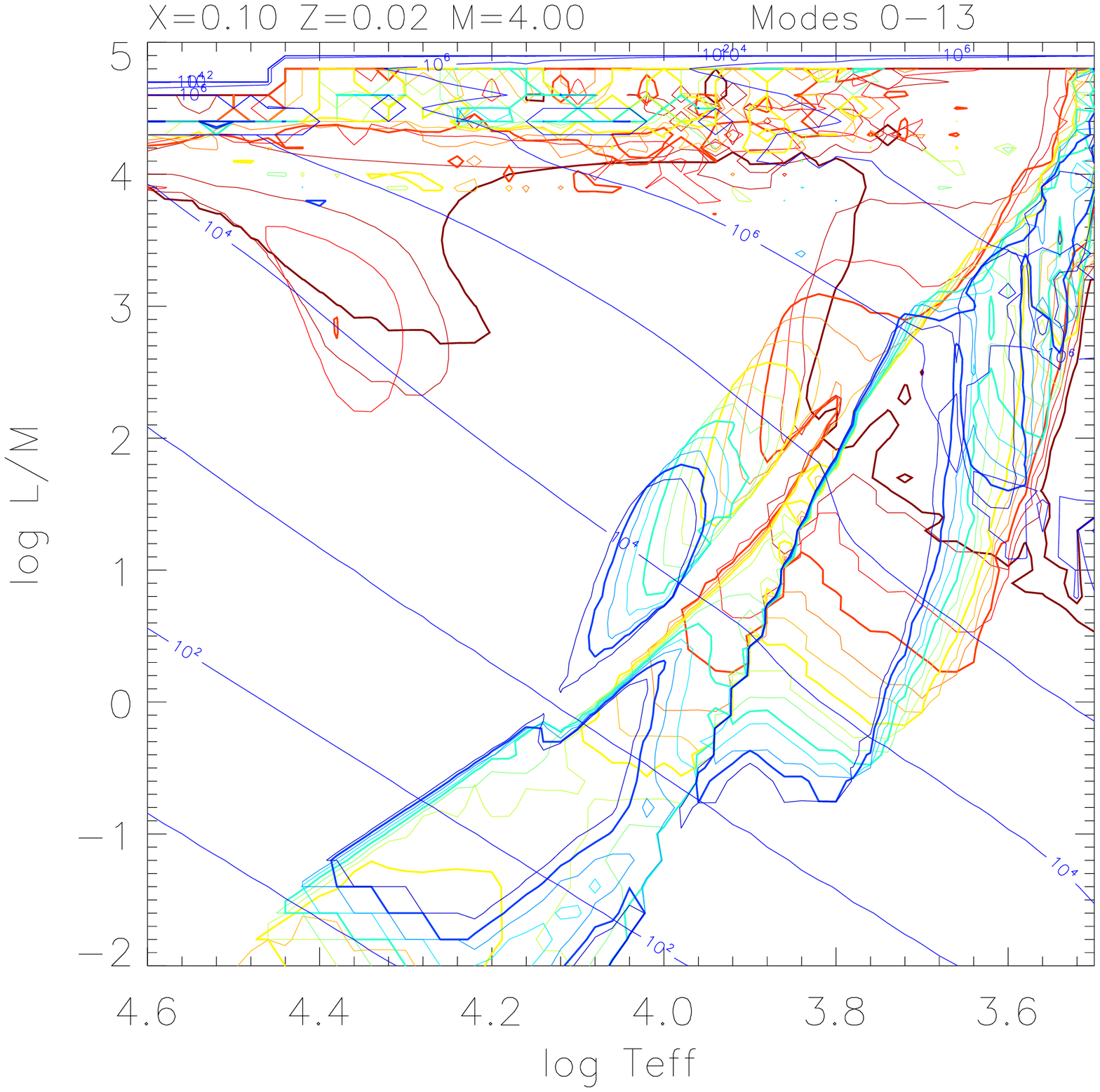,width=4.3cm,angle=0}
\epsfig{file=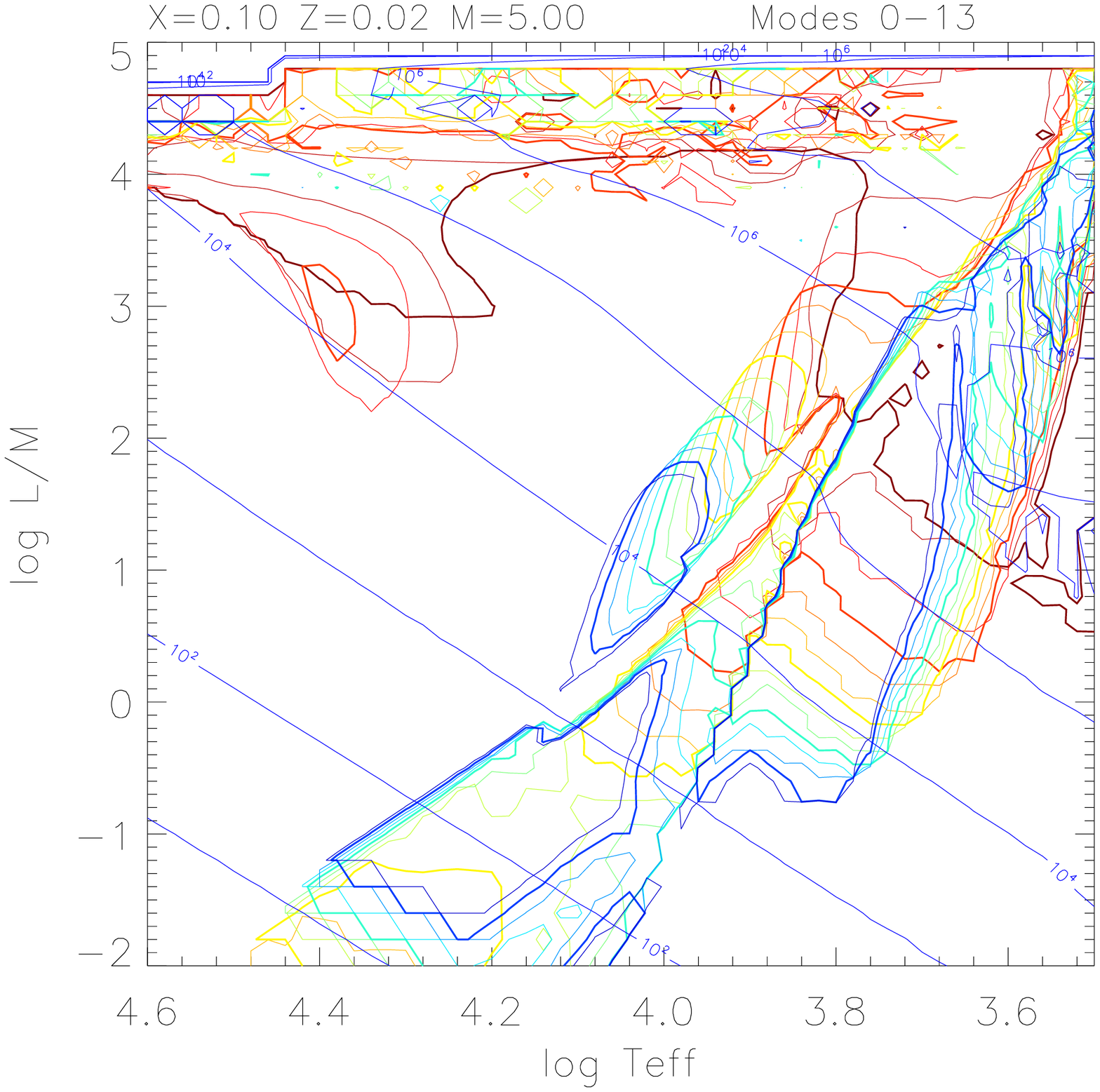,width=4.3cm,angle=0}
\epsfig{file=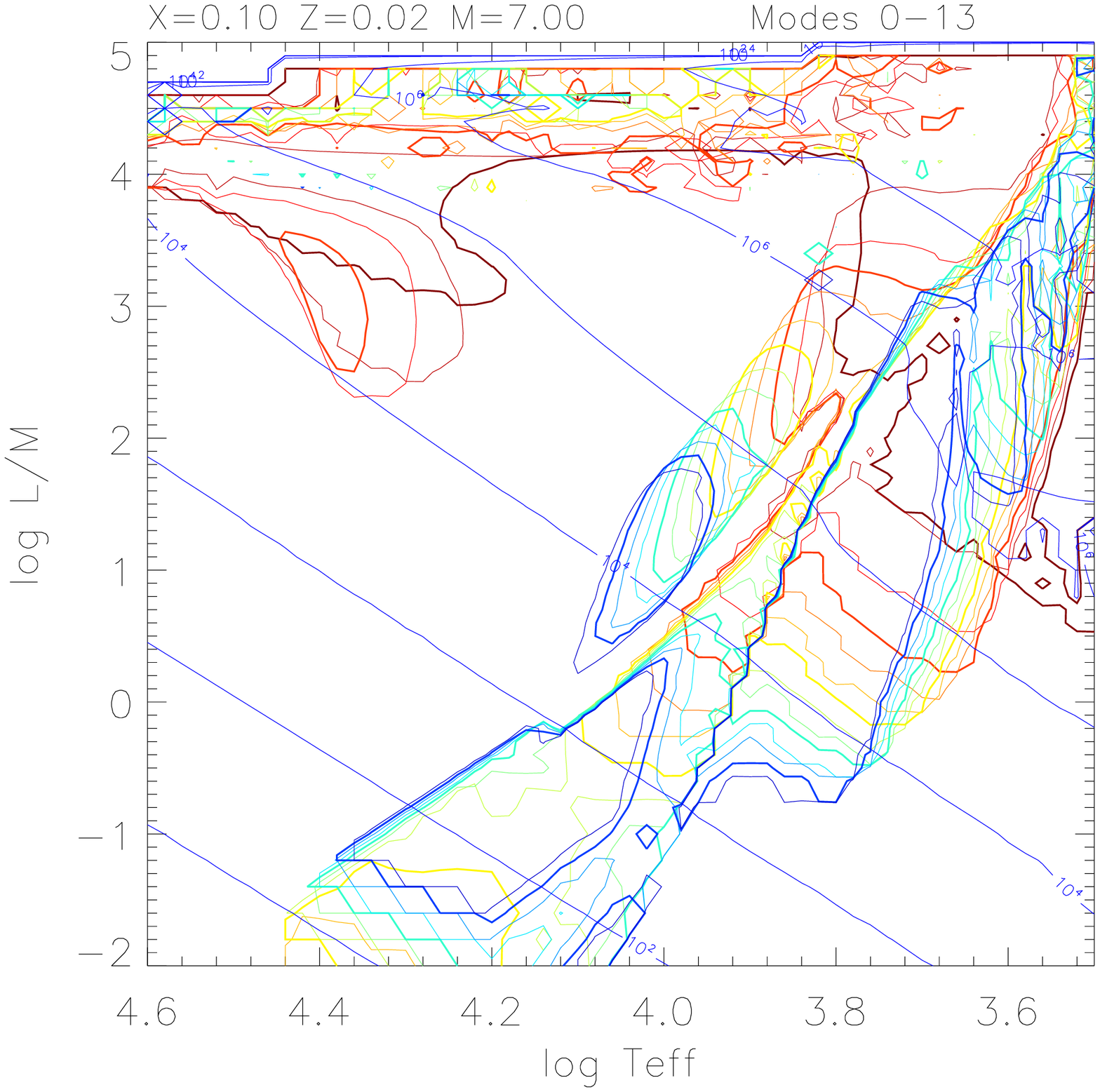,width=4.3cm,angle=0}
\epsfig{file=figs/periods_x10z02m10.0_00_opal.eps,width=4.3cm,angle=0}\\
\epsfig{file=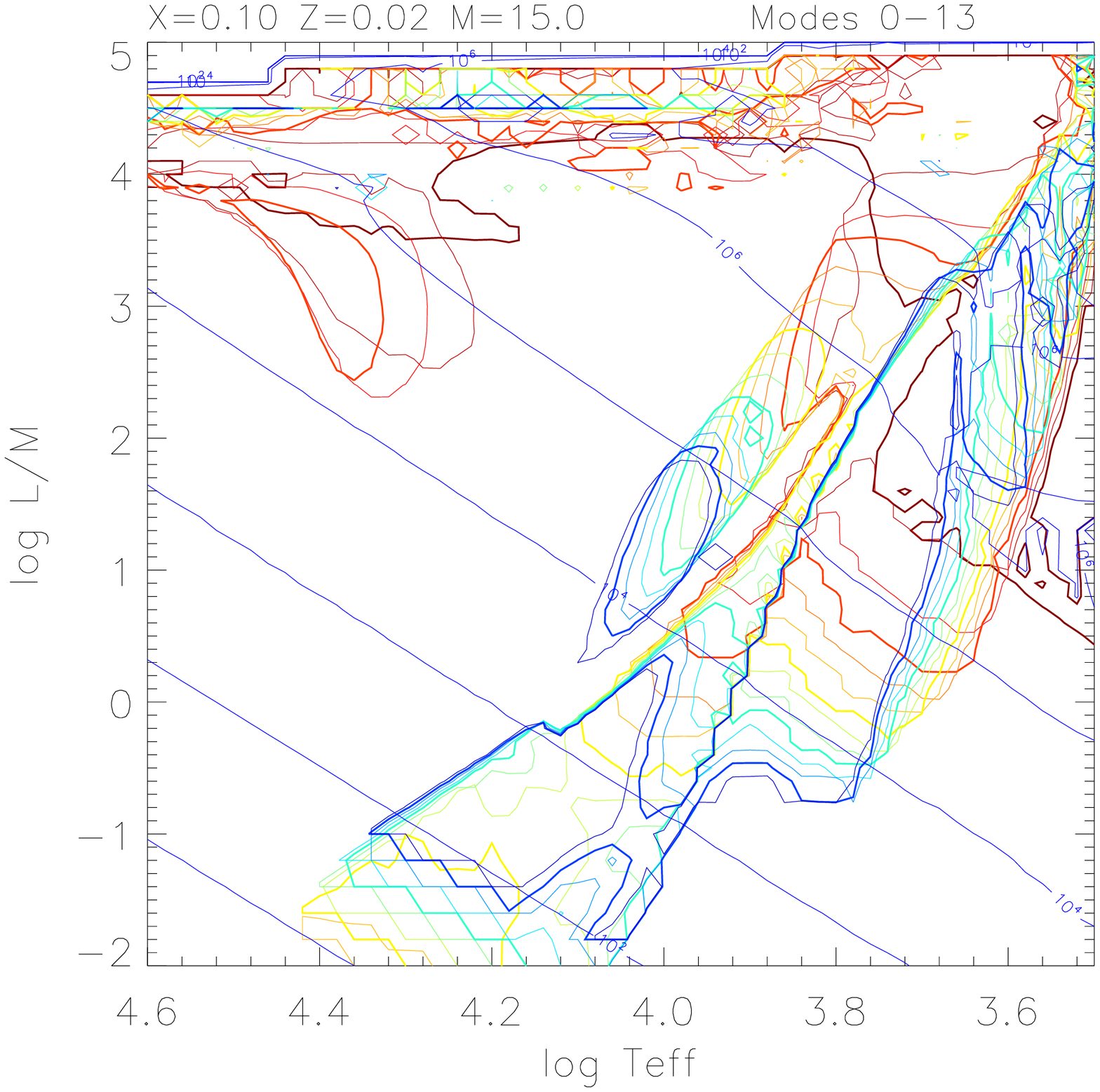,width=4.3cm,angle=0}
\epsfig{file=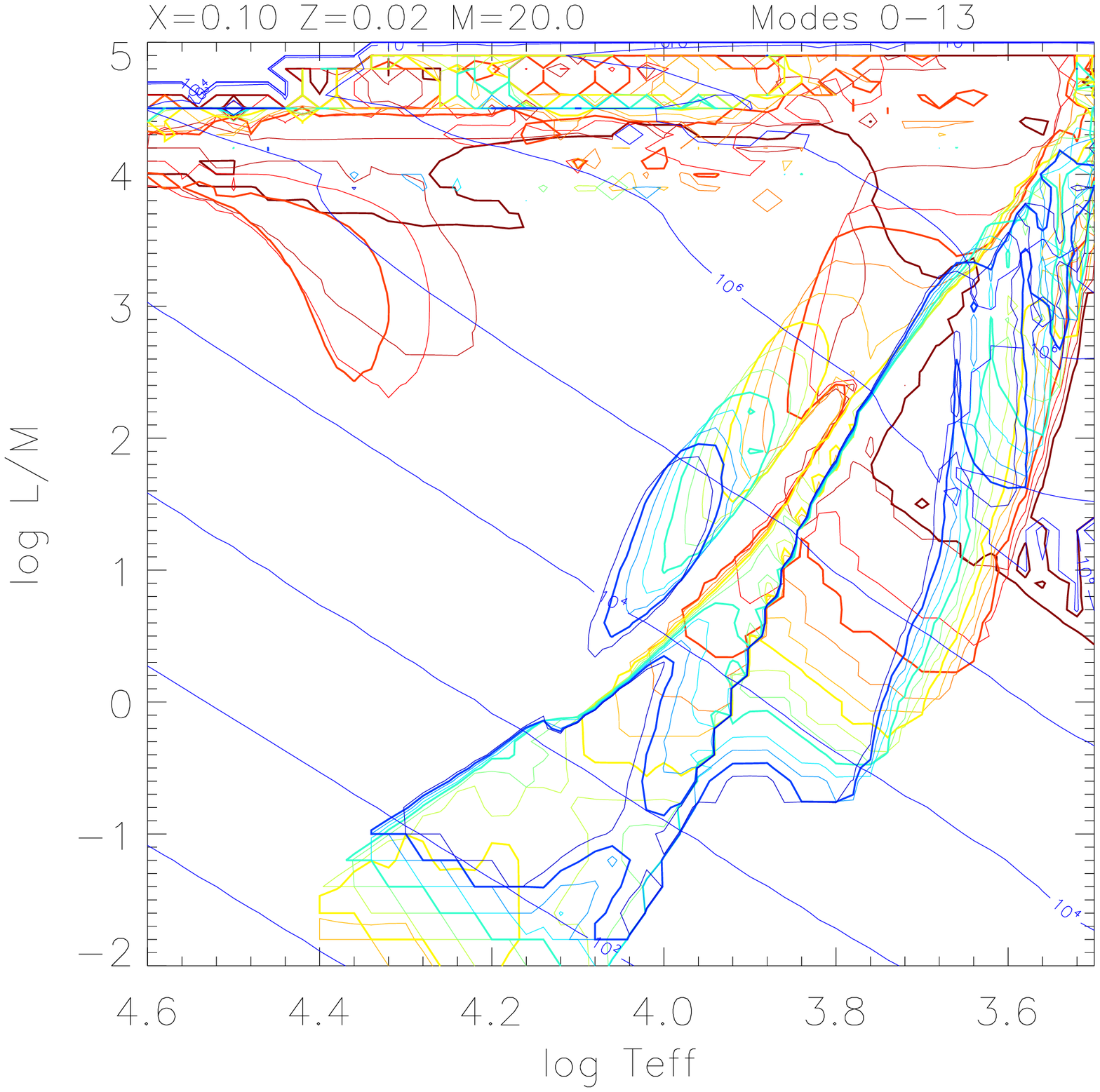,width=4.3cm,angle=0}
\epsfig{file=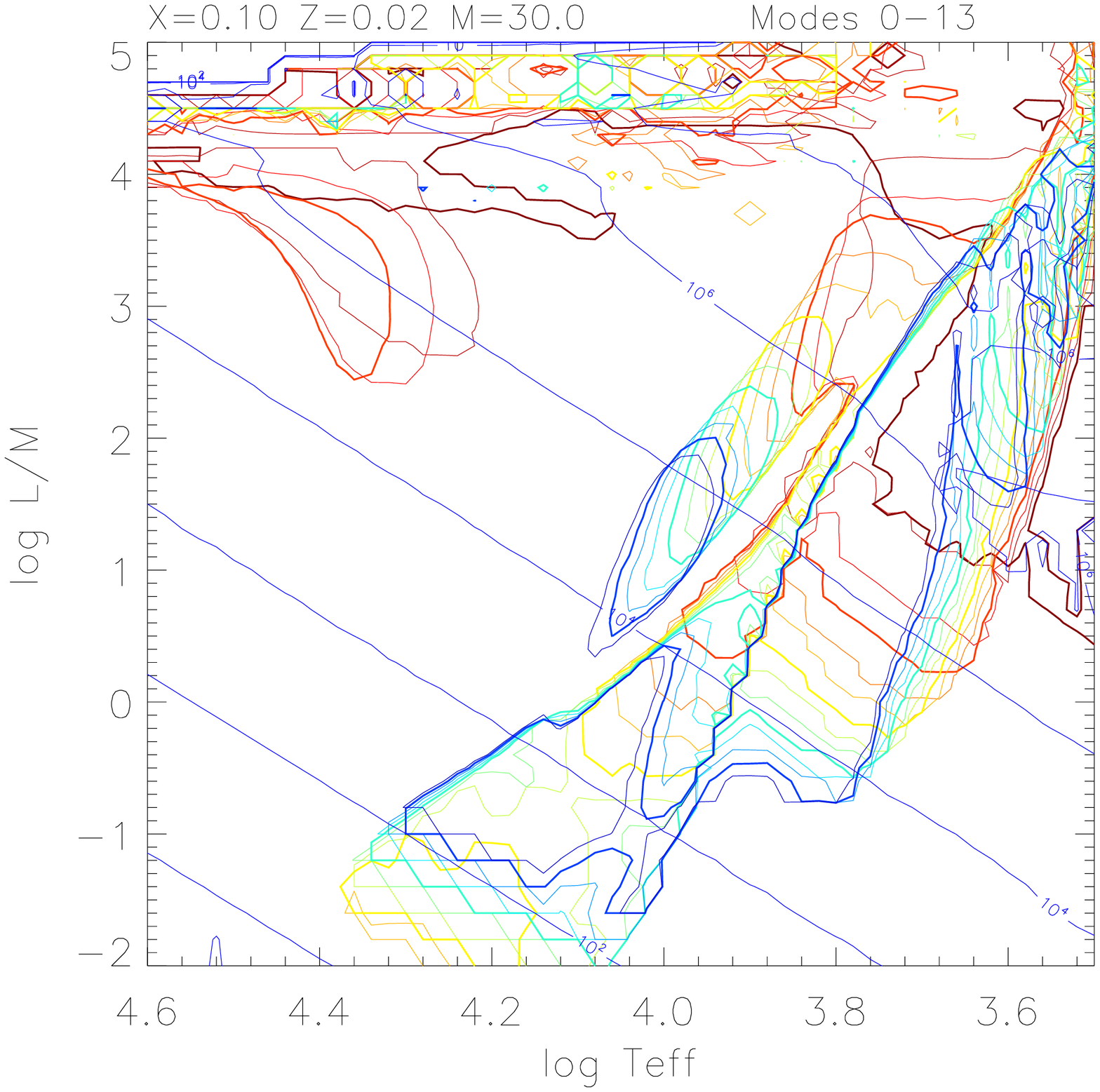,width=4.3cm,angle=0}
\epsfig{file=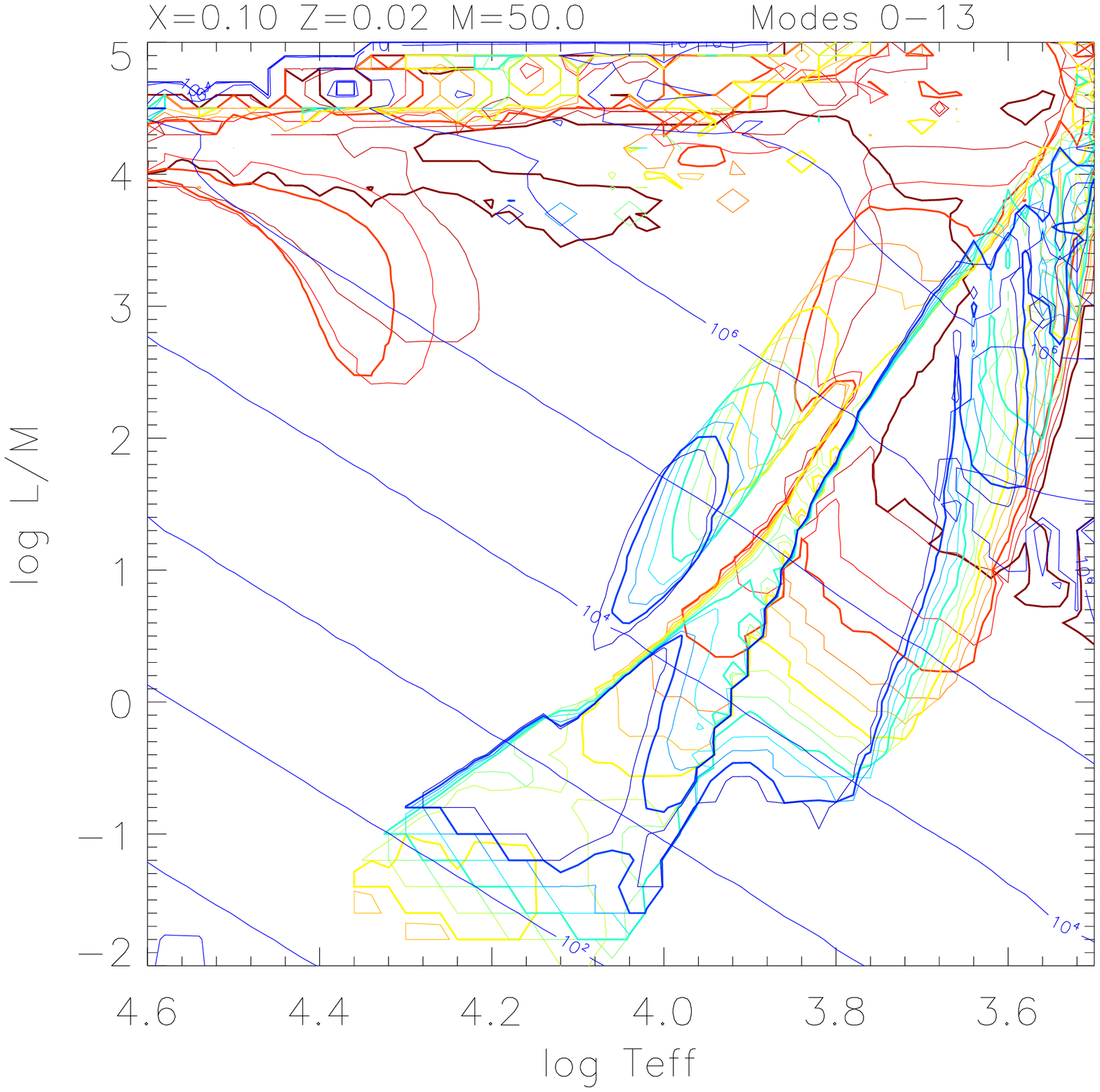,width=4.3cm,angle=0}
\caption[Unstable modes: $X=0.10, Z=0.02$]
{As Fig.~\ref{f:px70} with $X=0.10, Z=0.02$. 
}
\label{f:px10}
\end{center}
\end{figure*}

\begin{figure*}
\begin{center}
\epsfig{file=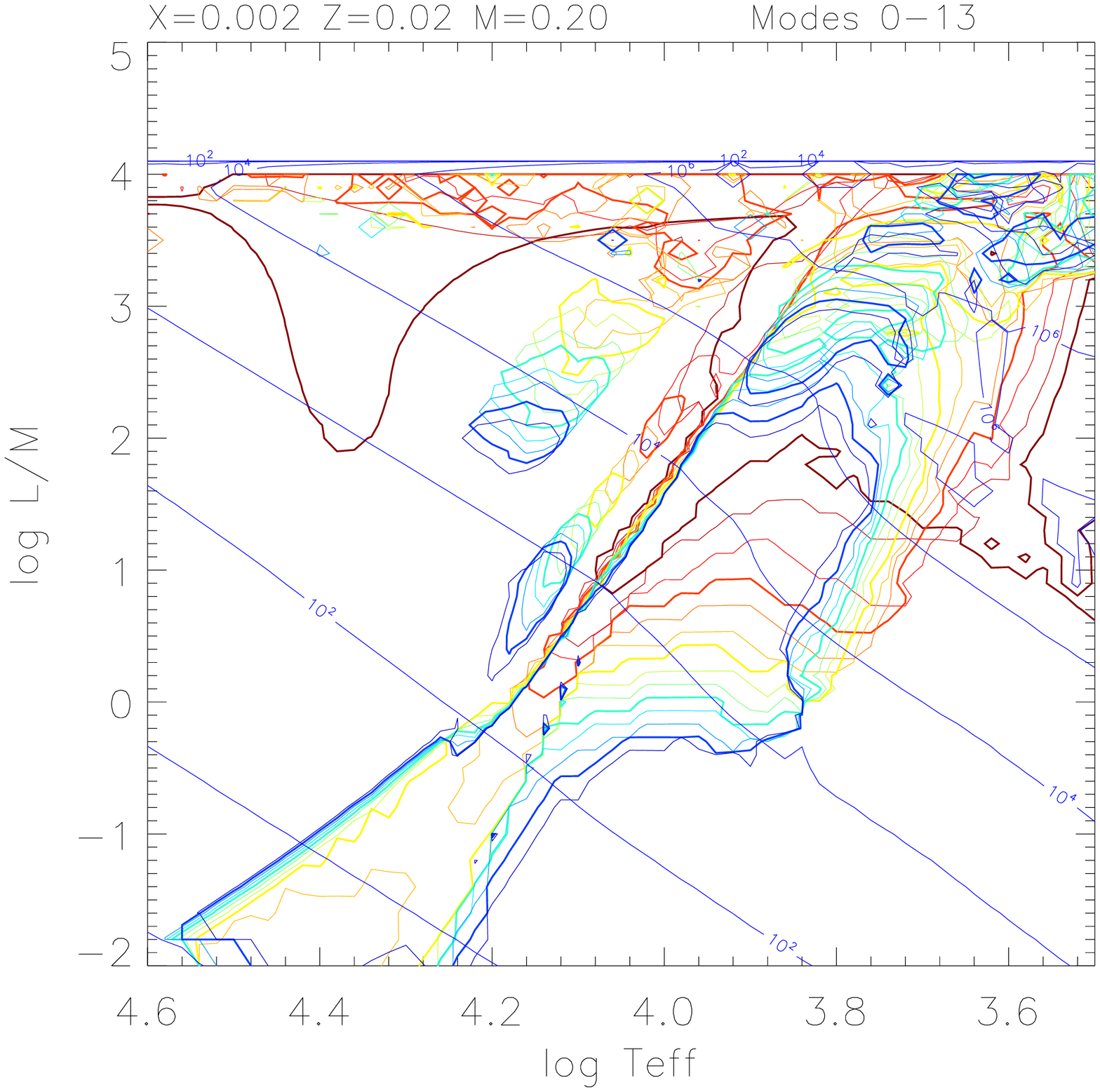,width=4.3cm,angle=0}
\epsfig{file=figs/periods_x002z02m00.3_00_opal.eps,width=4.3cm,angle=0}
\epsfig{file=figs/periods_x002z02m00.5_00_opal.eps,width=4.3cm,angle=0}
\epsfig{file=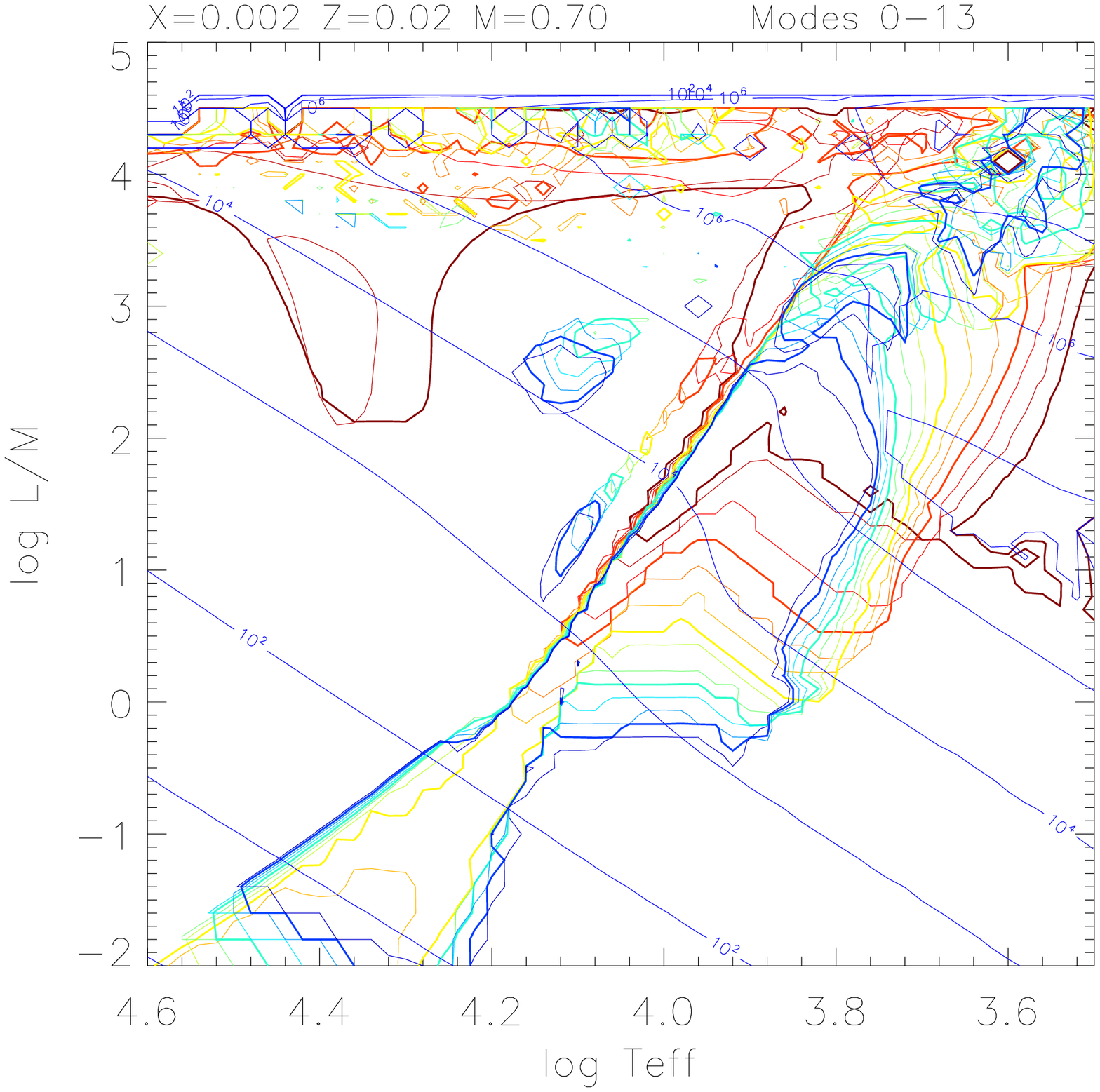,width=4.3cm,angle=0}\\
\epsfig{file=figs/periods_x002z02m01.0_00_opal.eps,width=4.3cm,angle=0}
\epsfig{file=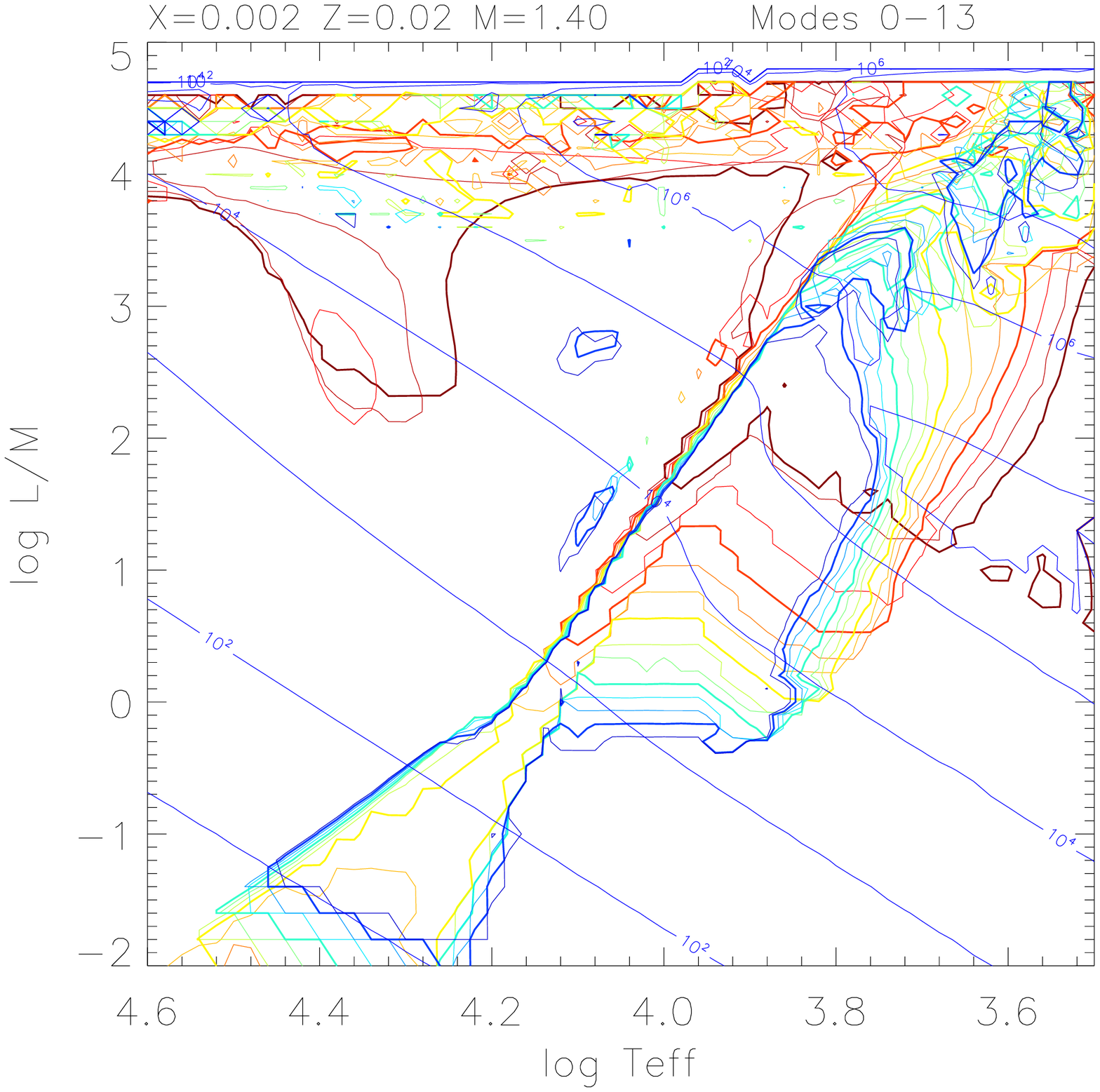,width=4.3cm,angle=0}
\epsfig{file=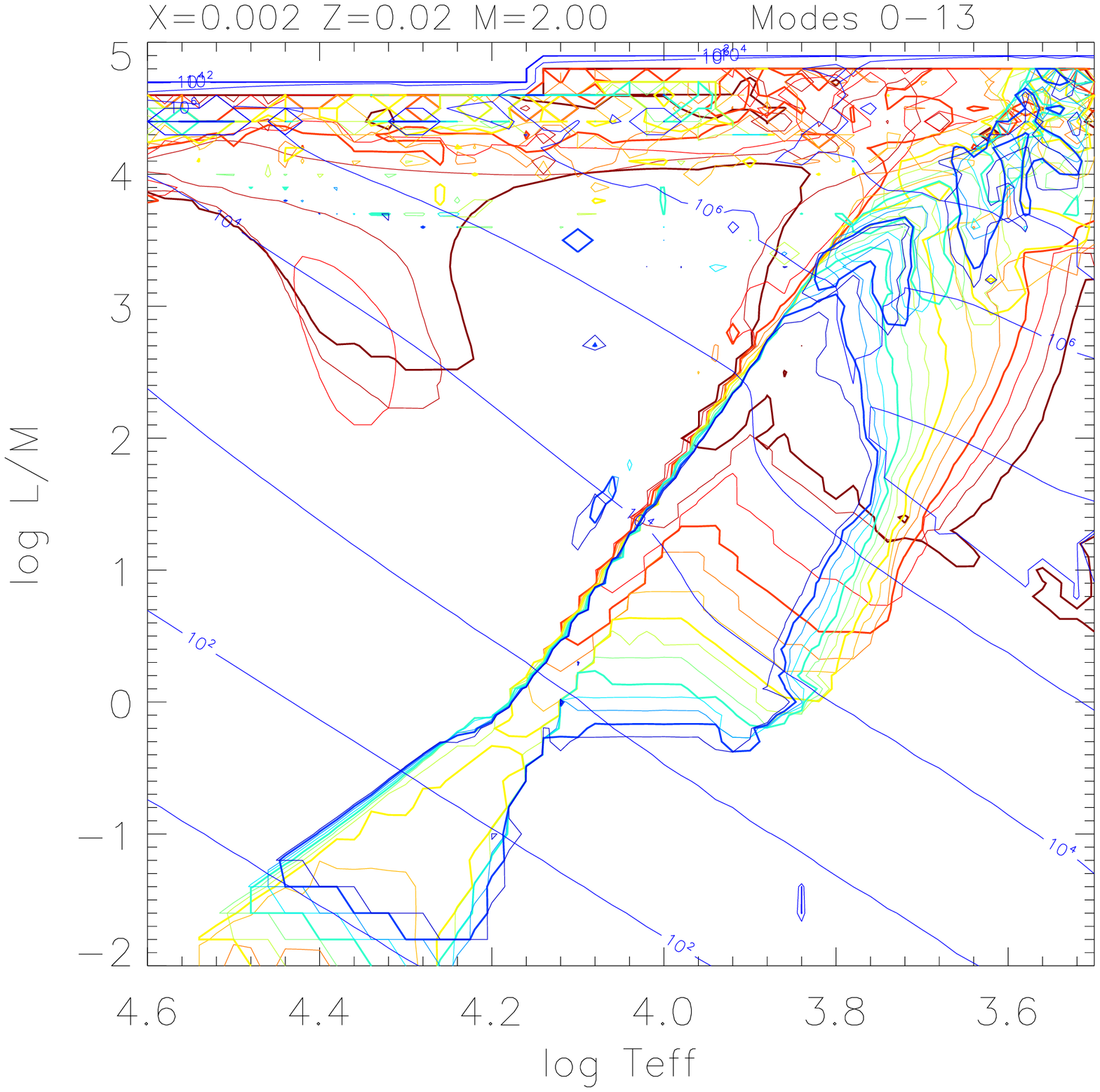,width=4.3cm,angle=0}
\epsfig{file=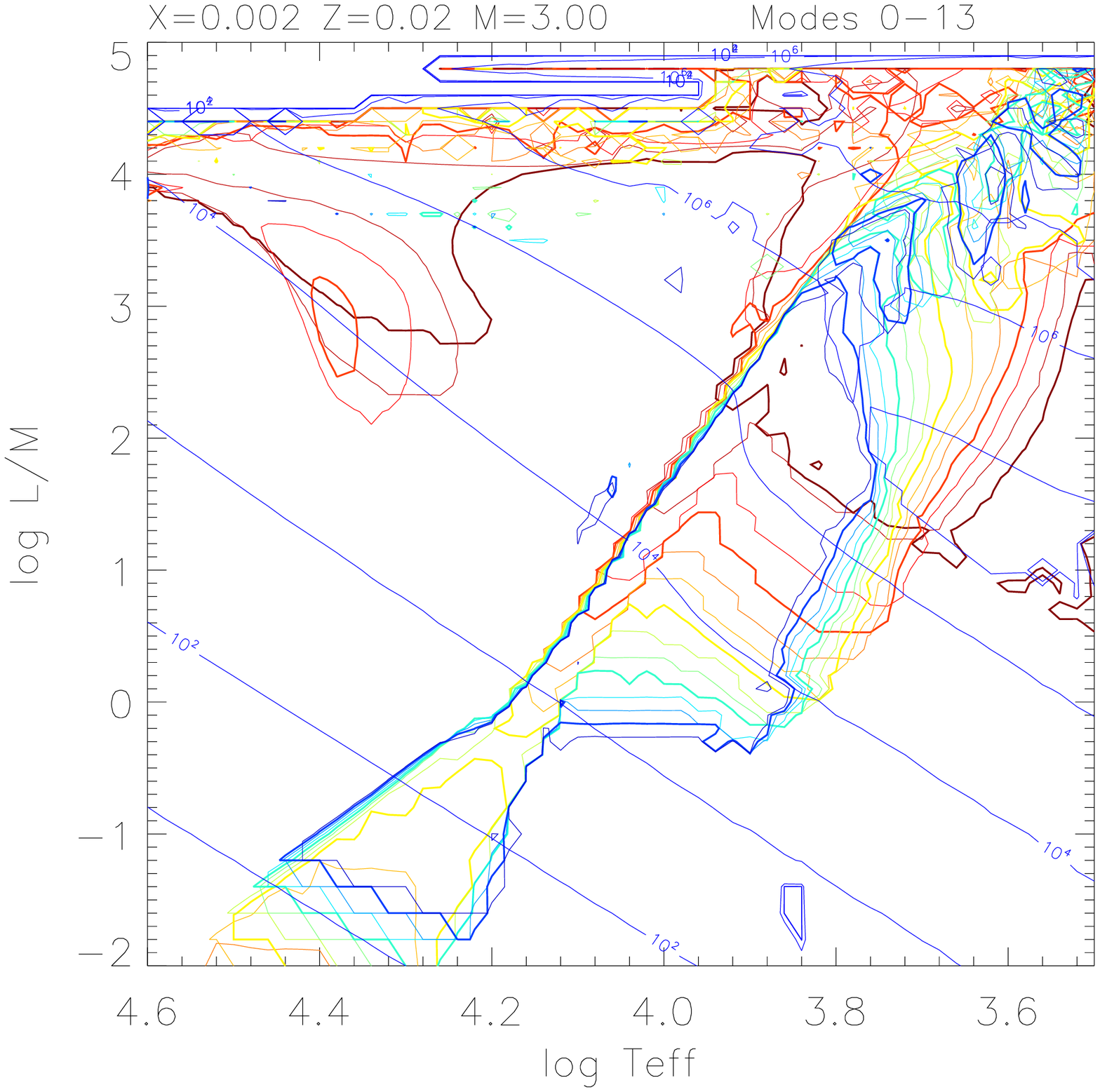,width=4.3cm,angle=0}\\
\epsfig{file=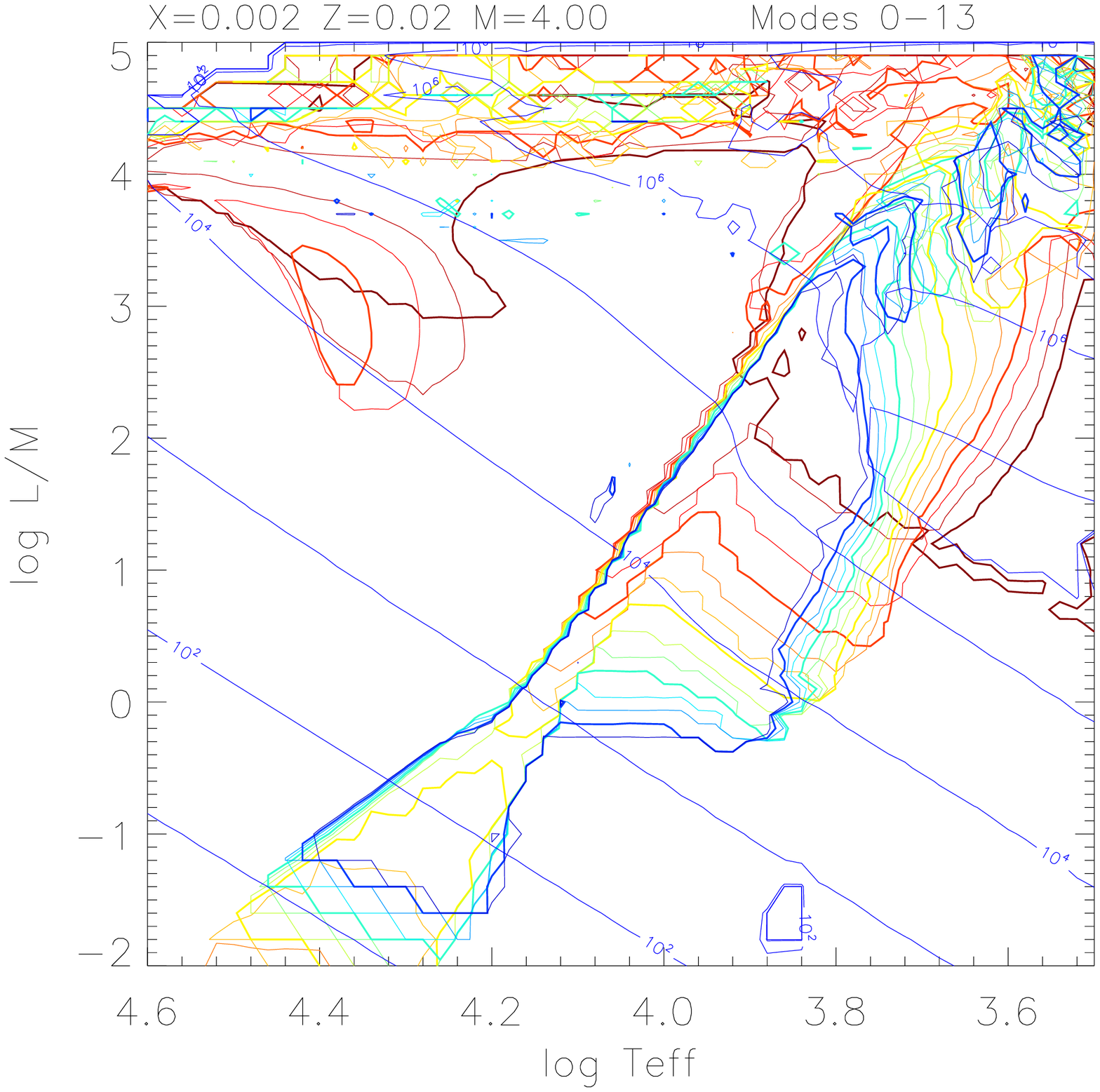,width=4.3cm,angle=0}
\epsfig{file=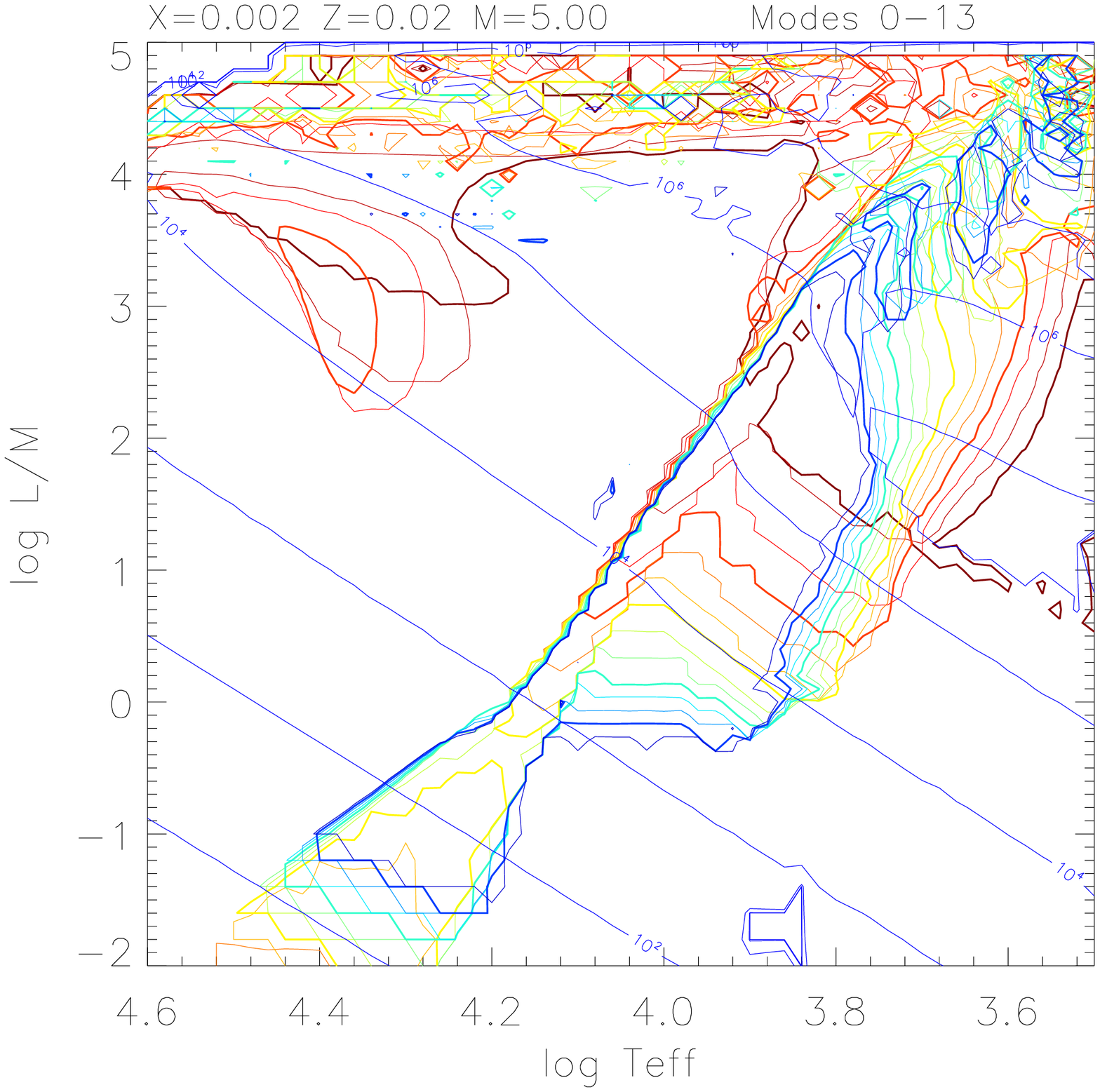,width=4.3cm,angle=0}
\epsfig{file=figs/periods_x002z02m07.0_00_opal.eps,width=4.3cm,angle=0}
\epsfig{file=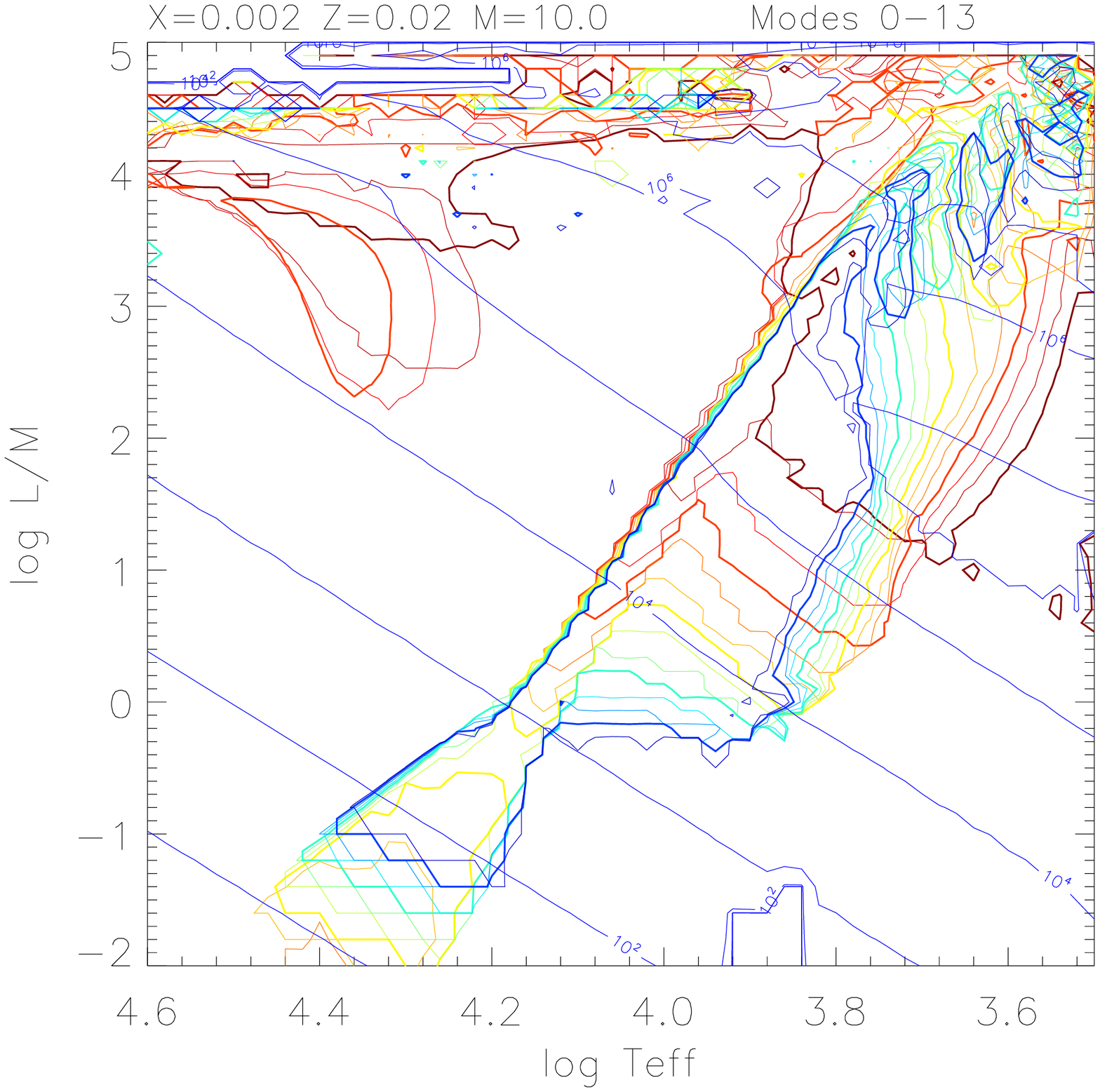,width=4.3cm,angle=0}\\
\epsfig{file=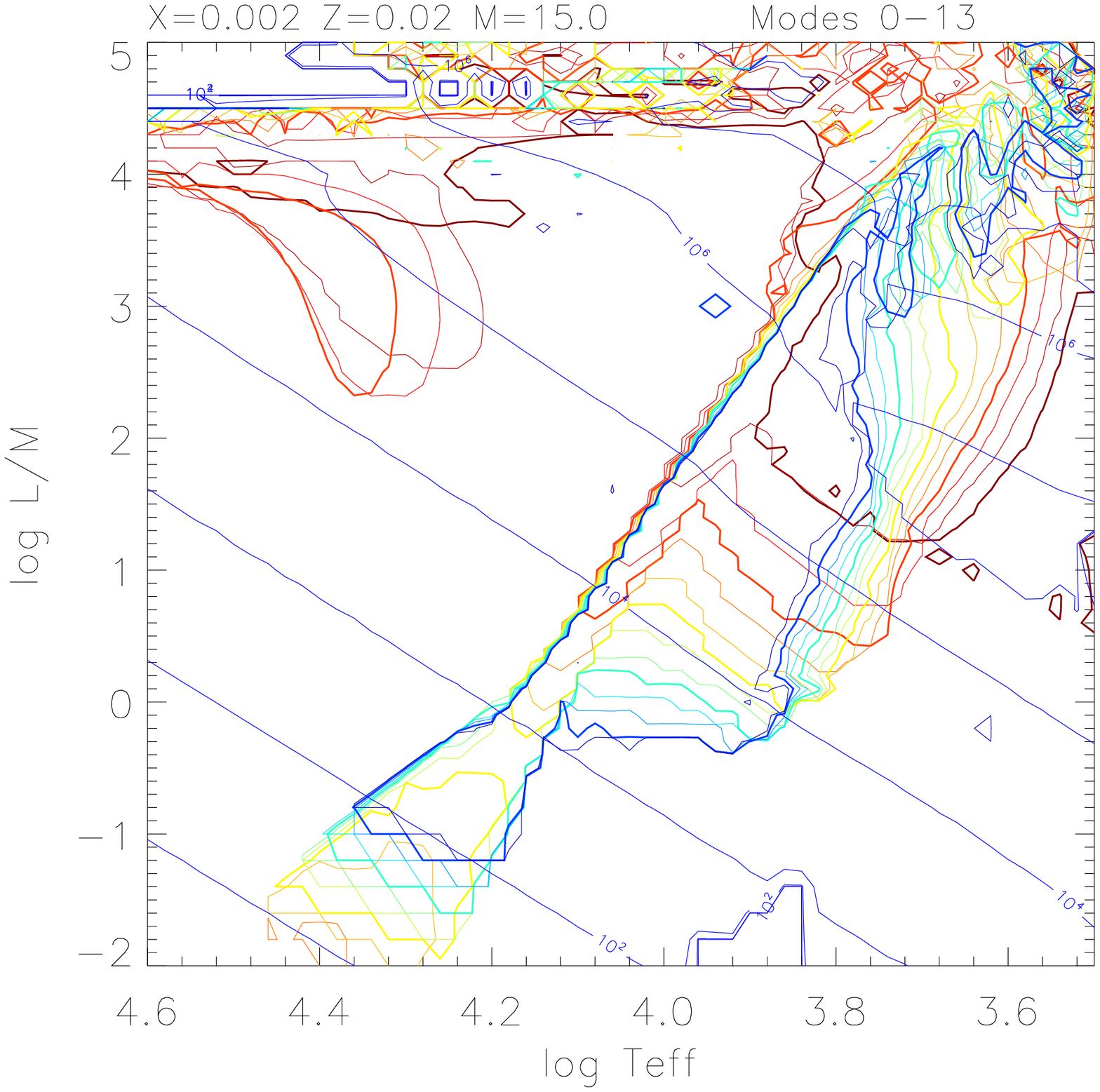,width=4.3cm,angle=0}
\epsfig{file=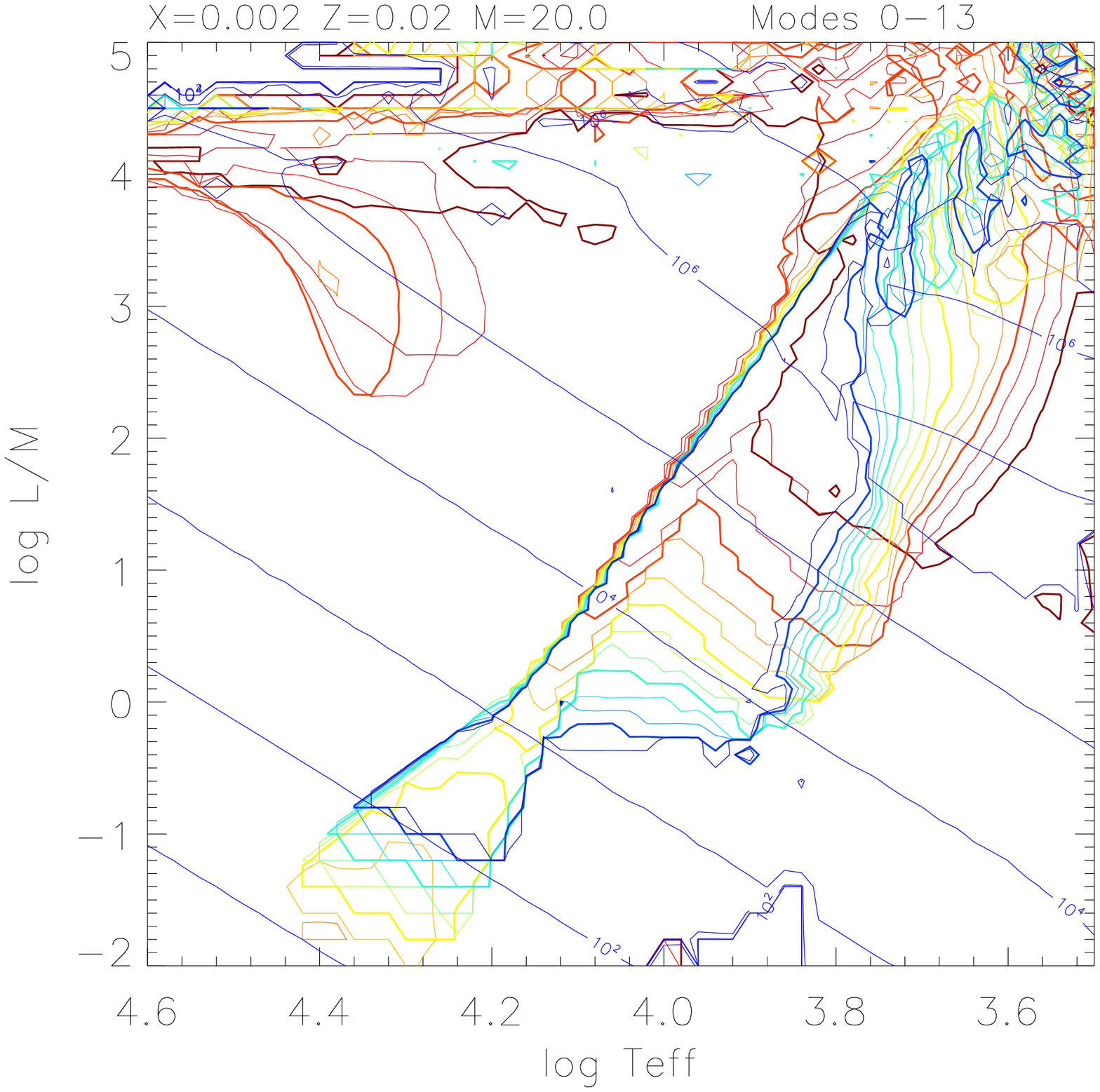,width=4.3cm,angle=0}
\epsfig{file=figs/periods_x002z02m30.0_00_opal.eps,width=4.3cm,angle=0}
\epsfig{file=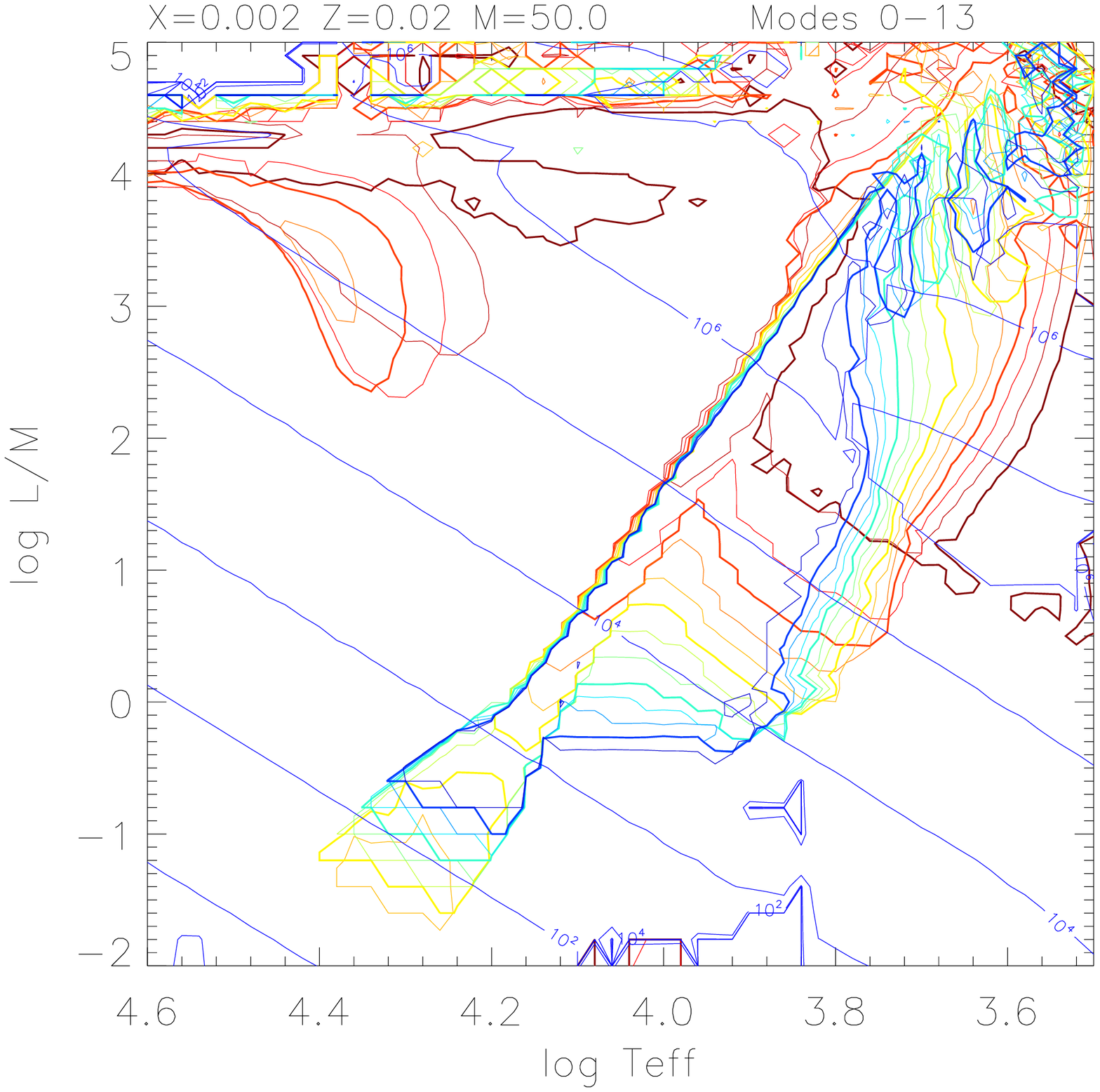,width=4.3cm,angle=0}
\caption[Unstable modes: $X=0.002, Z=0.02$]
{As Fig.~\ref{f:px70} with $X=0.002, Z=0.02$. 
}
\label{f:px002}
\end{center}
\end{figure*}

\begin{figure*}
\begin{center}
\epsfig{file=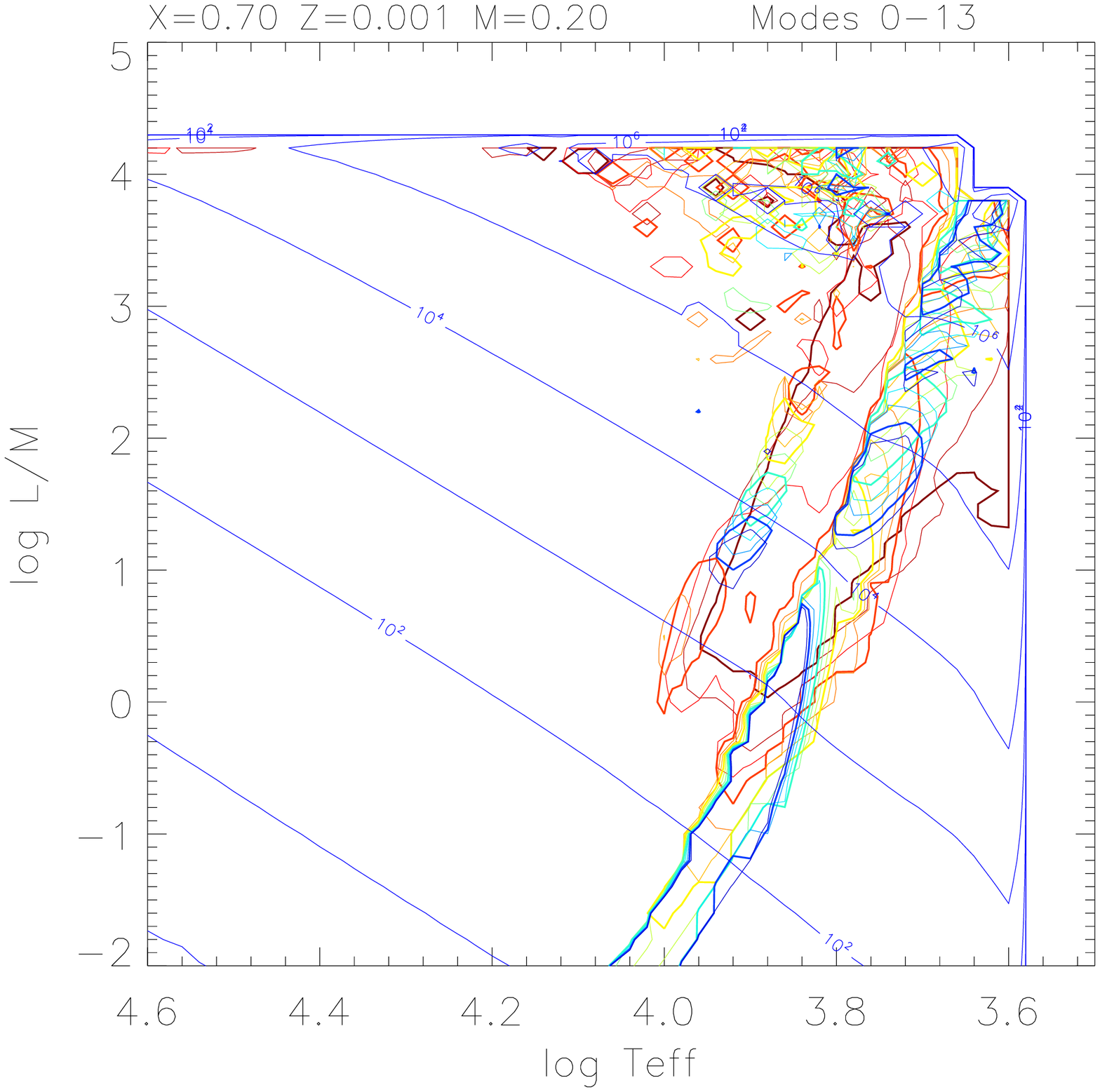,width=4.3cm,angle=0}
\epsfig{file=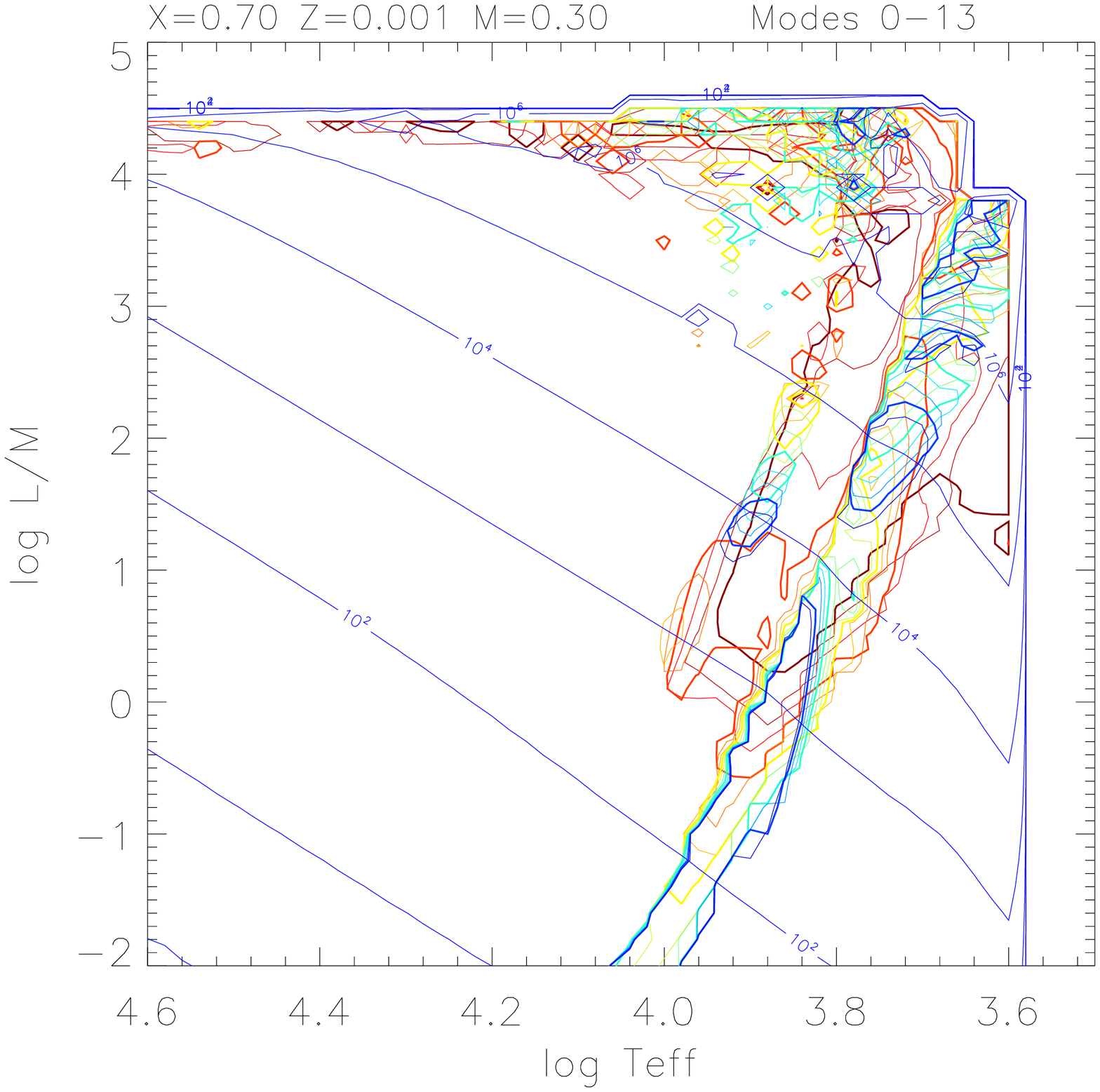,width=4.3cm,angle=0}
\epsfig{file=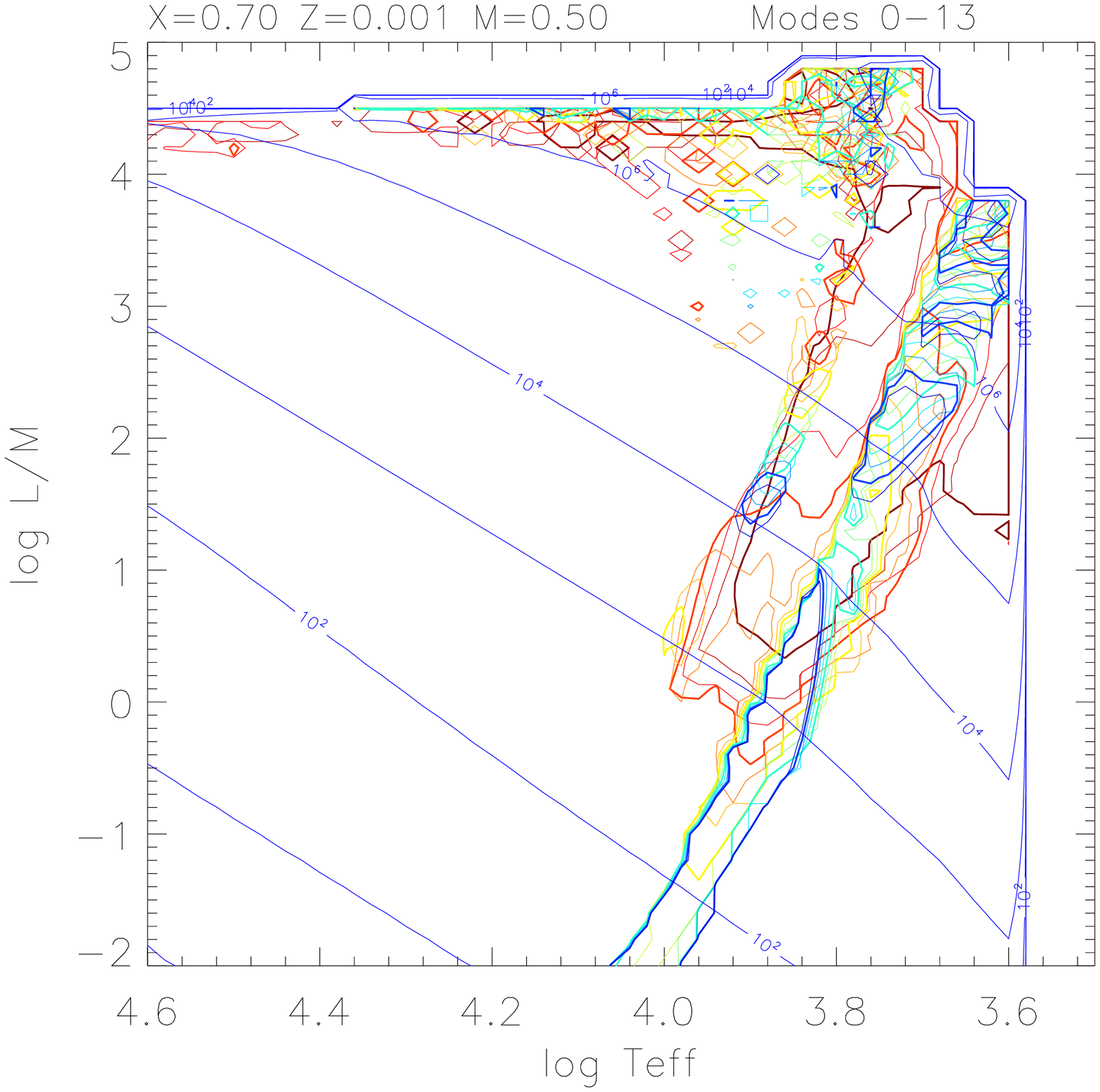,width=4.3cm,angle=0}
\epsfig{file=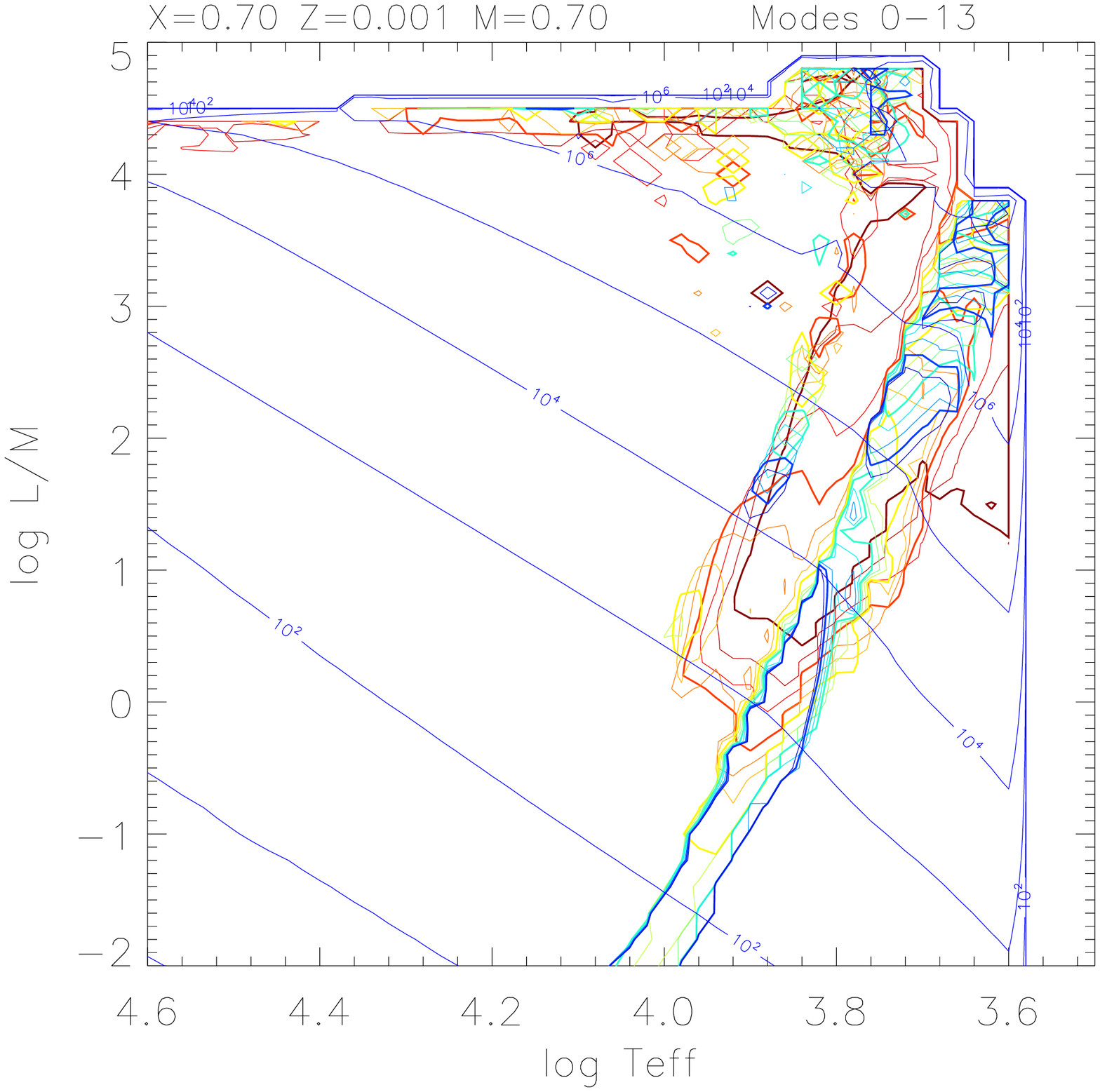,width=4.3cm,angle=0}\\
\epsfig{file=figs/periods_x70z001m01.0_00_opal.eps,width=4.3cm,angle=0}
\epsfig{file=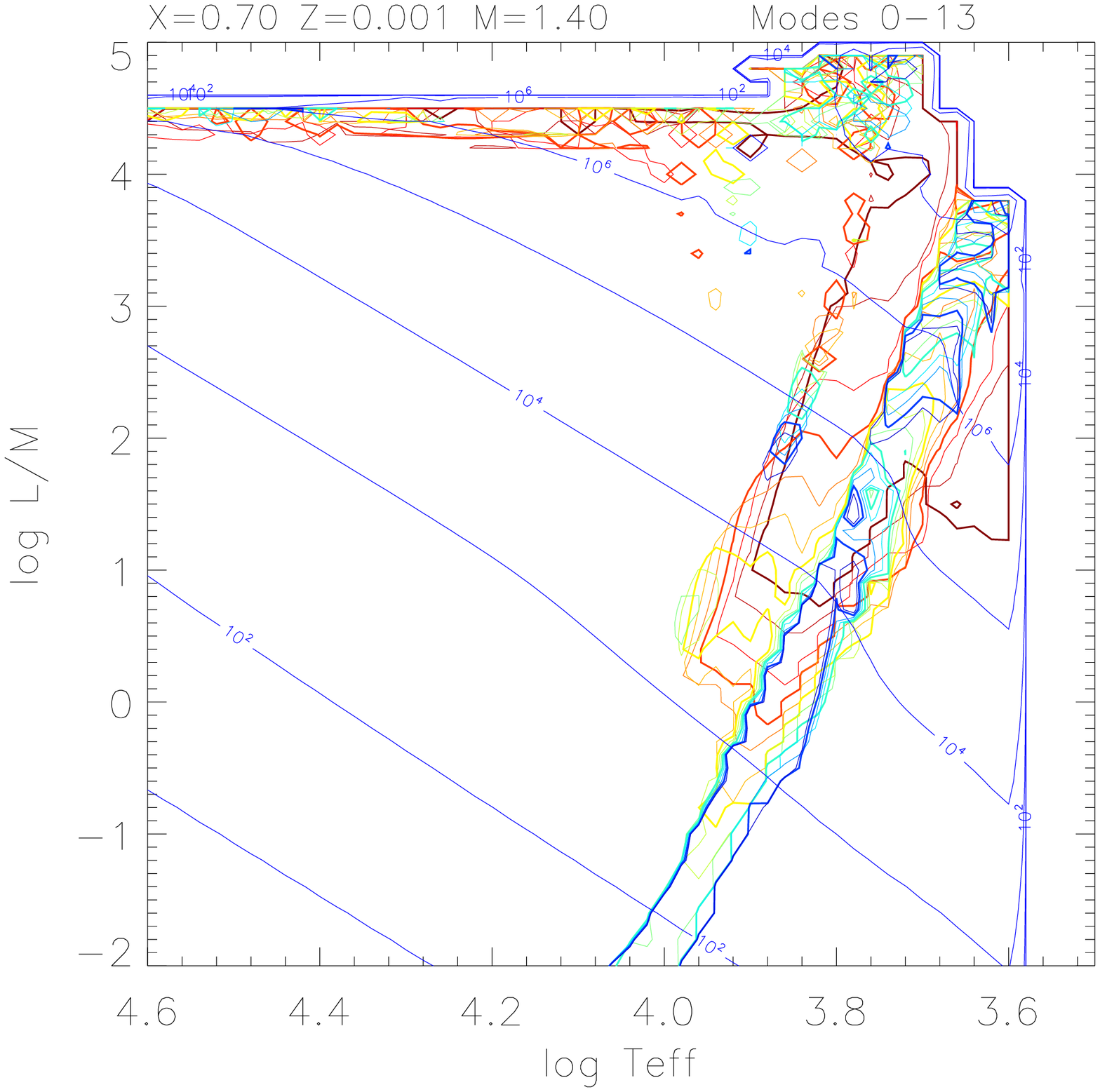,width=4.3cm,angle=0}
\epsfig{file=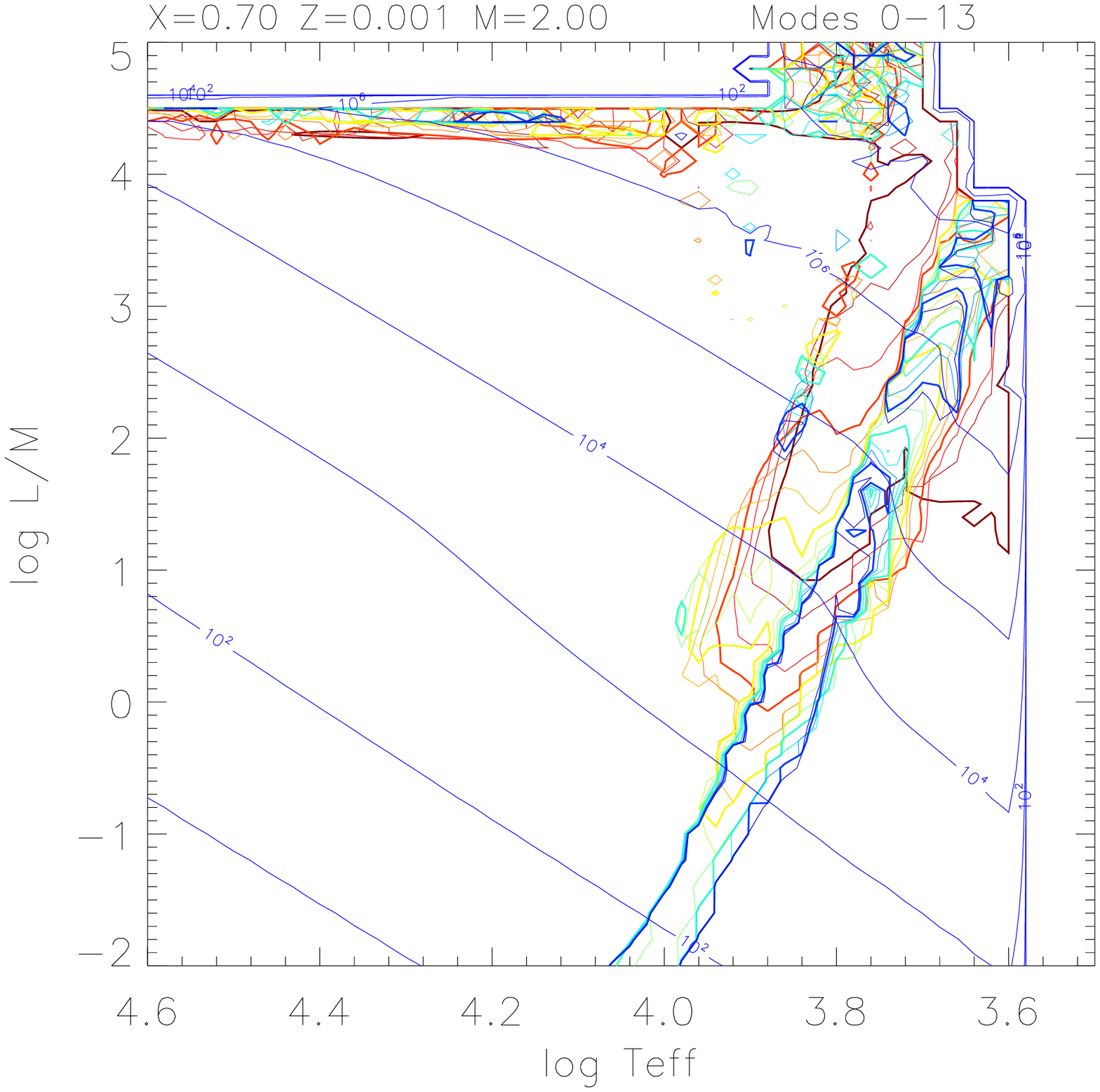,width=4.3cm,angle=0}
\epsfig{file=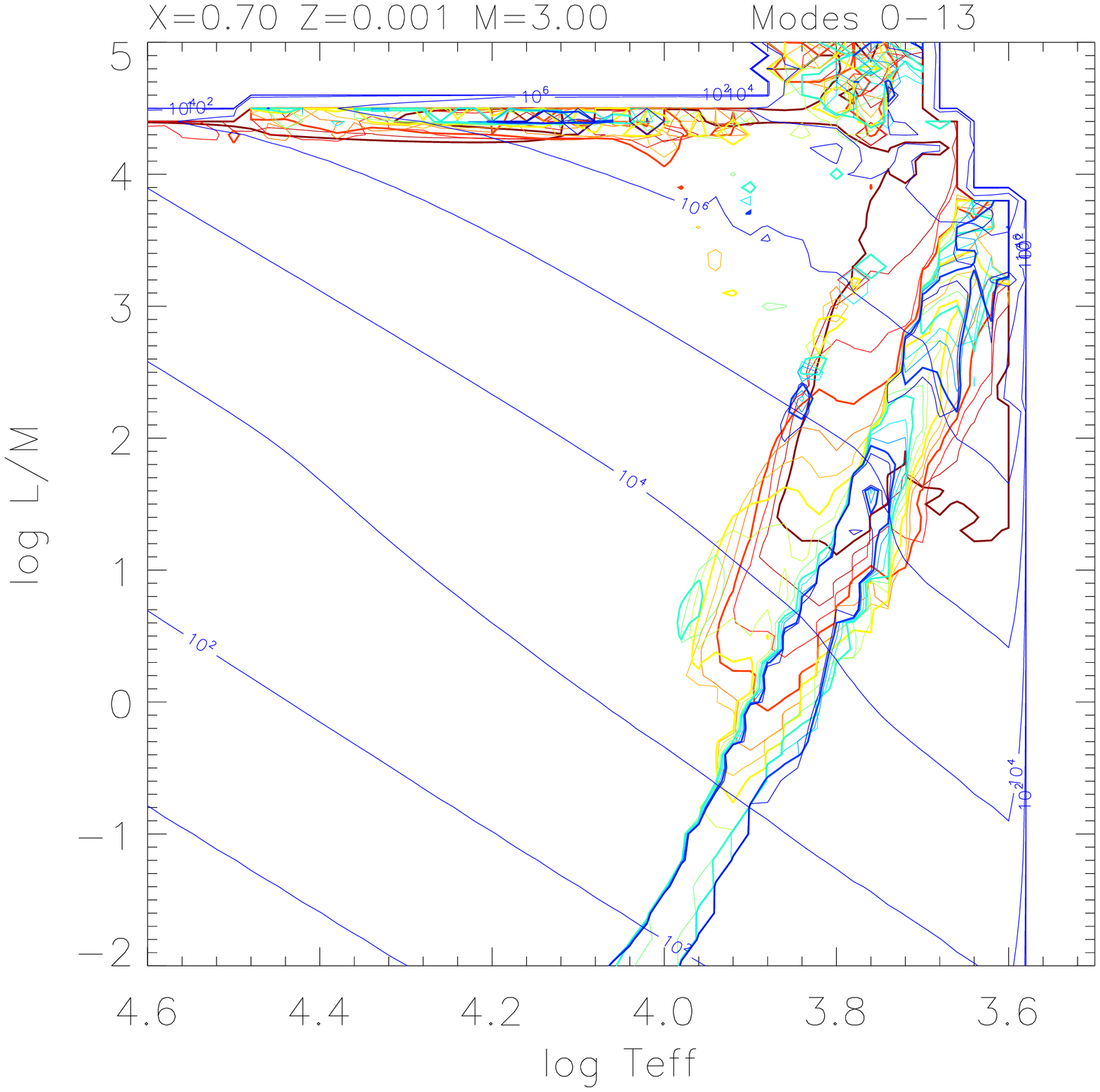,width=4.3cm,angle=0}\\
\epsfig{file=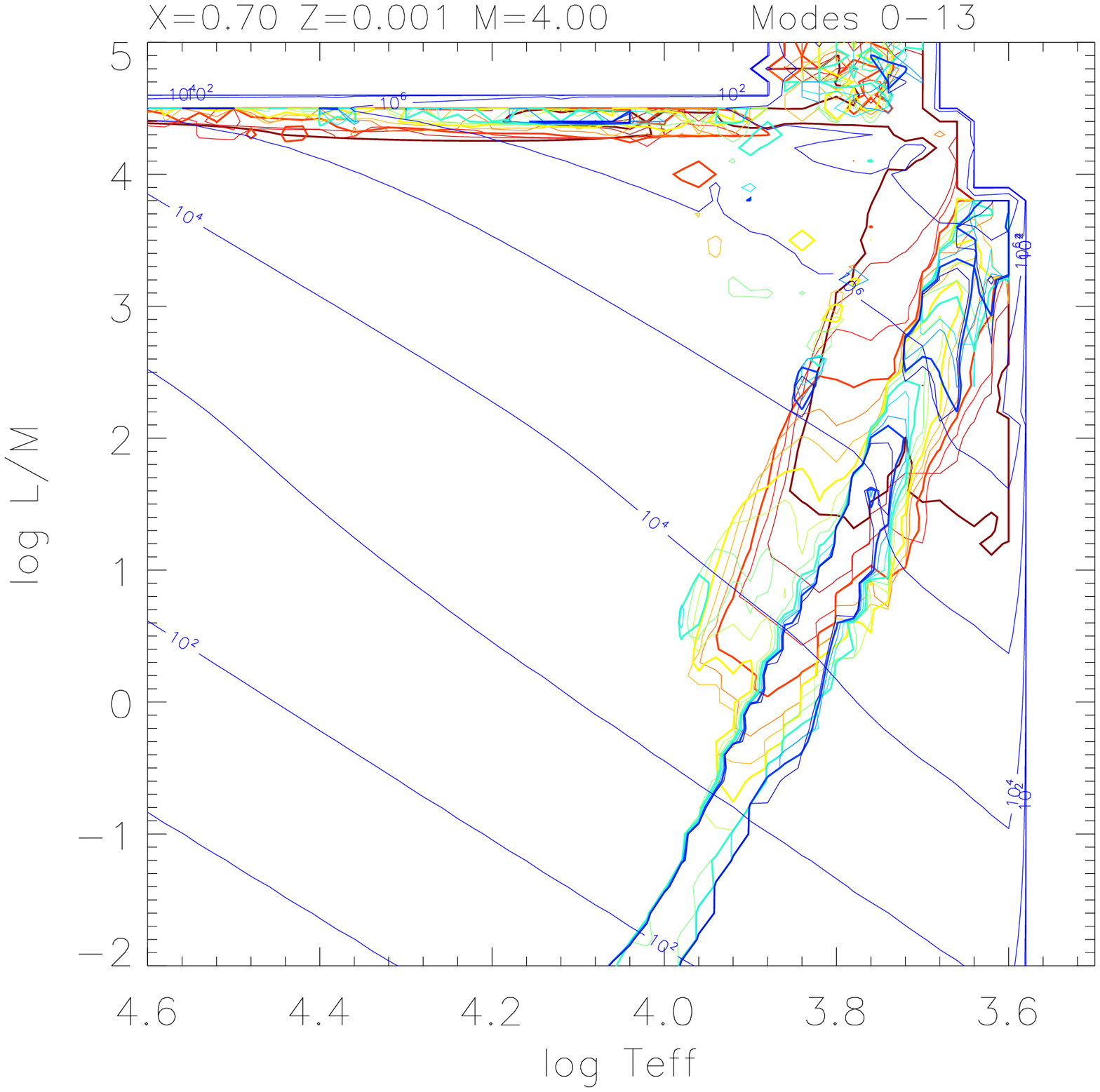,width=4.3cm,angle=0}
\epsfig{file=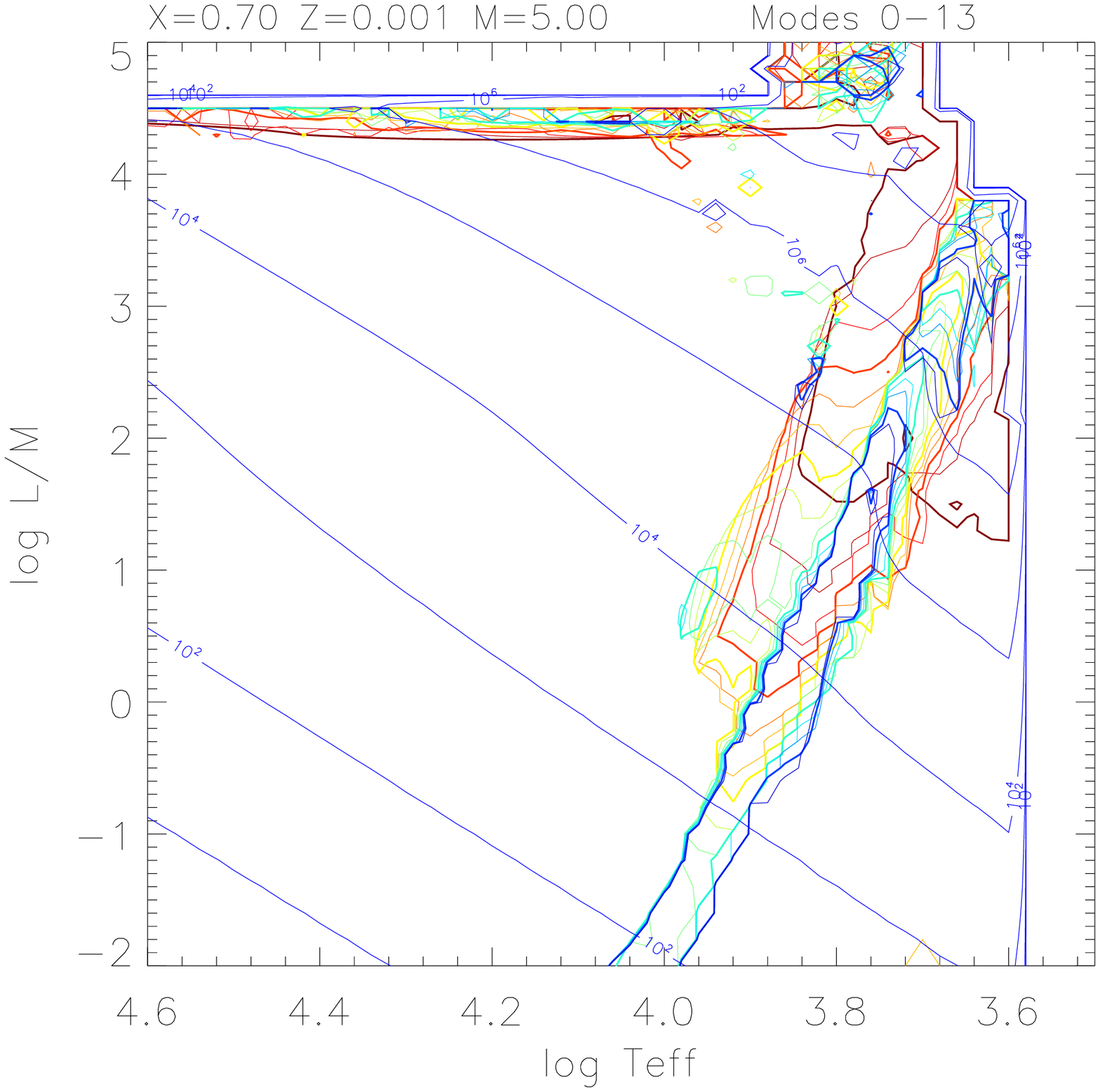,width=4.3cm,angle=0}
\epsfig{file=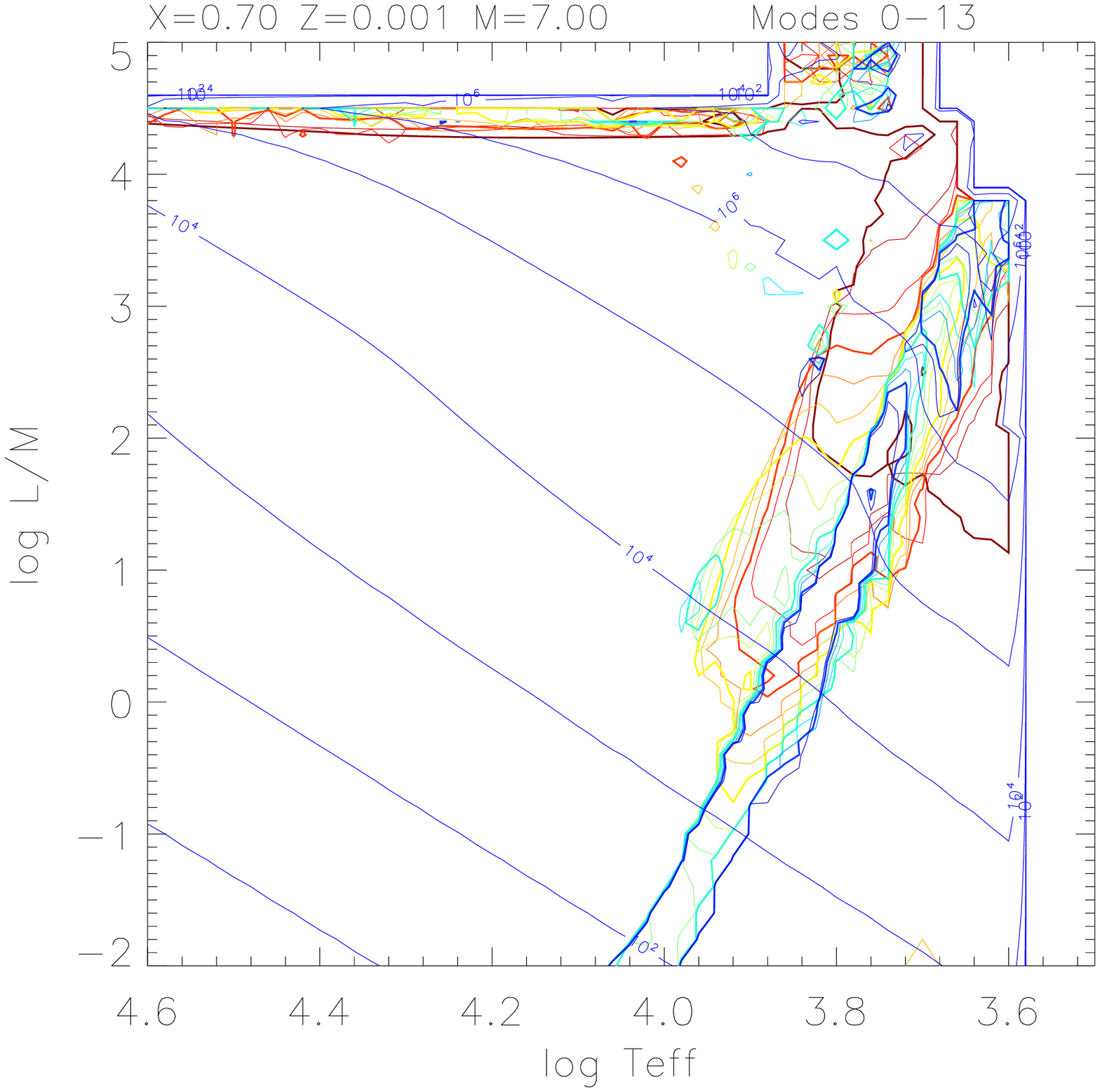,width=4.3cm,angle=0}
\epsfig{file=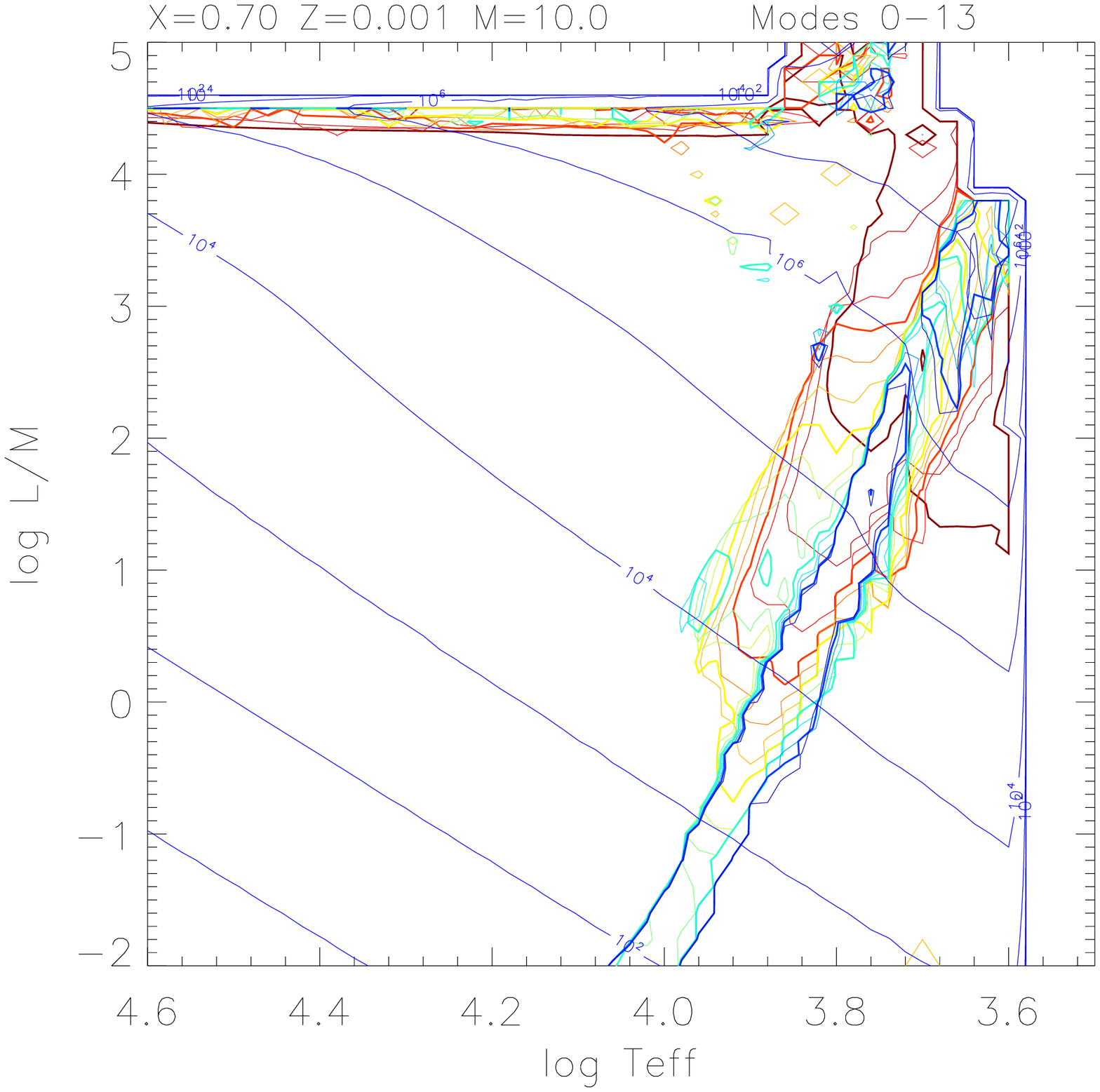,width=4.3cm,angle=0}\\
\epsfig{file=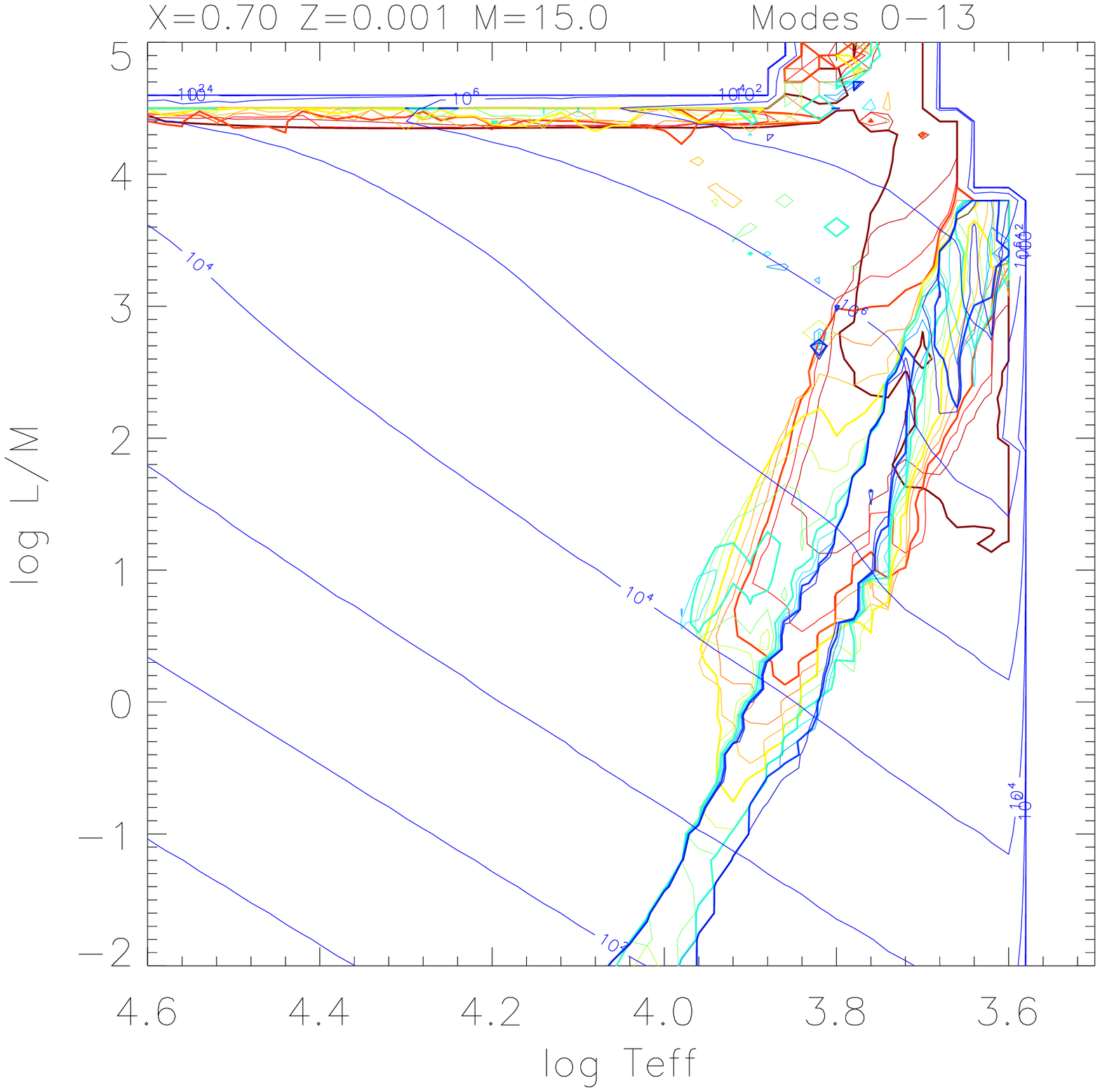,width=4.3cm,angle=0}
\epsfig{file=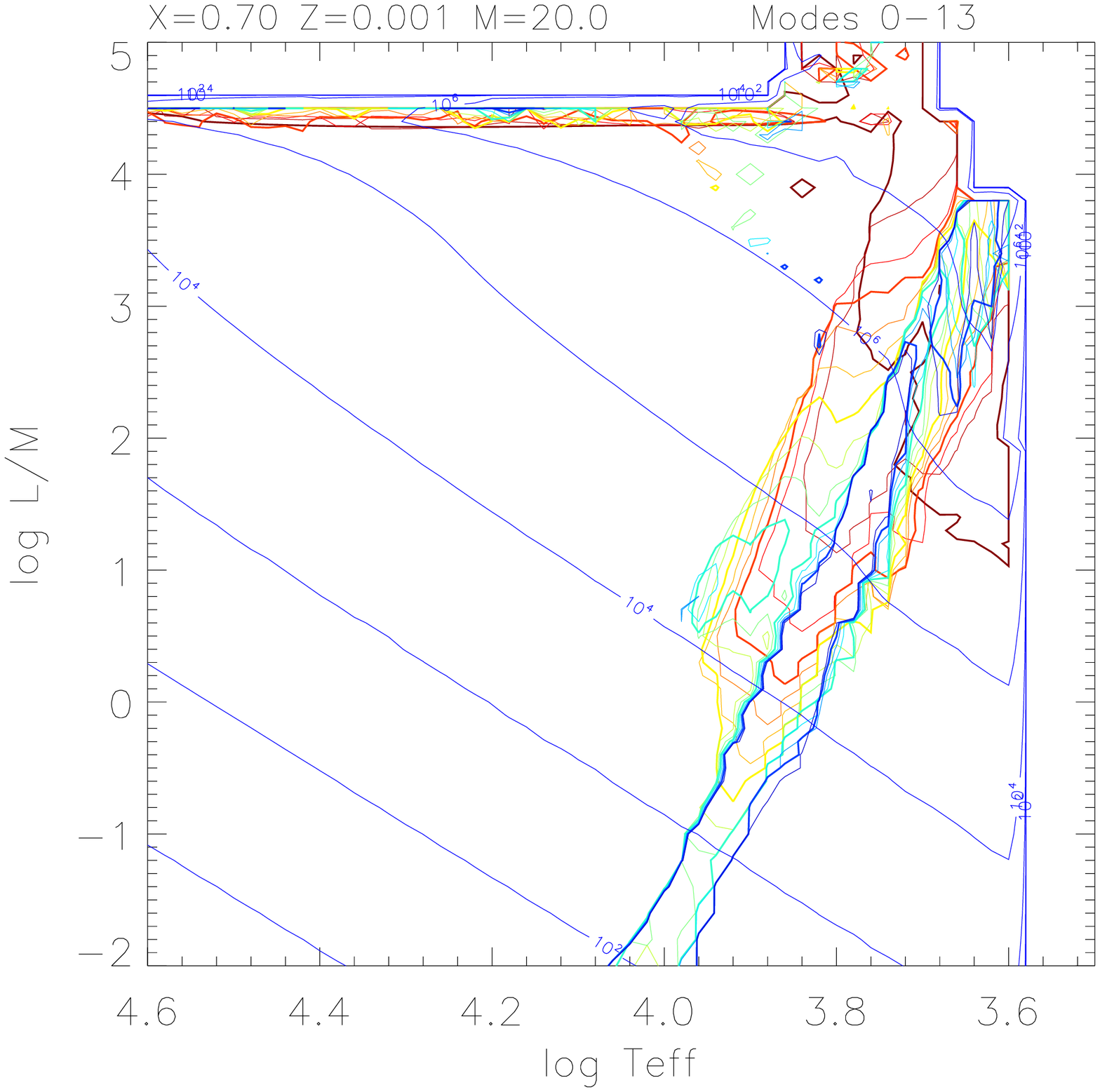,width=4.3cm,angle=0}
\epsfig{file=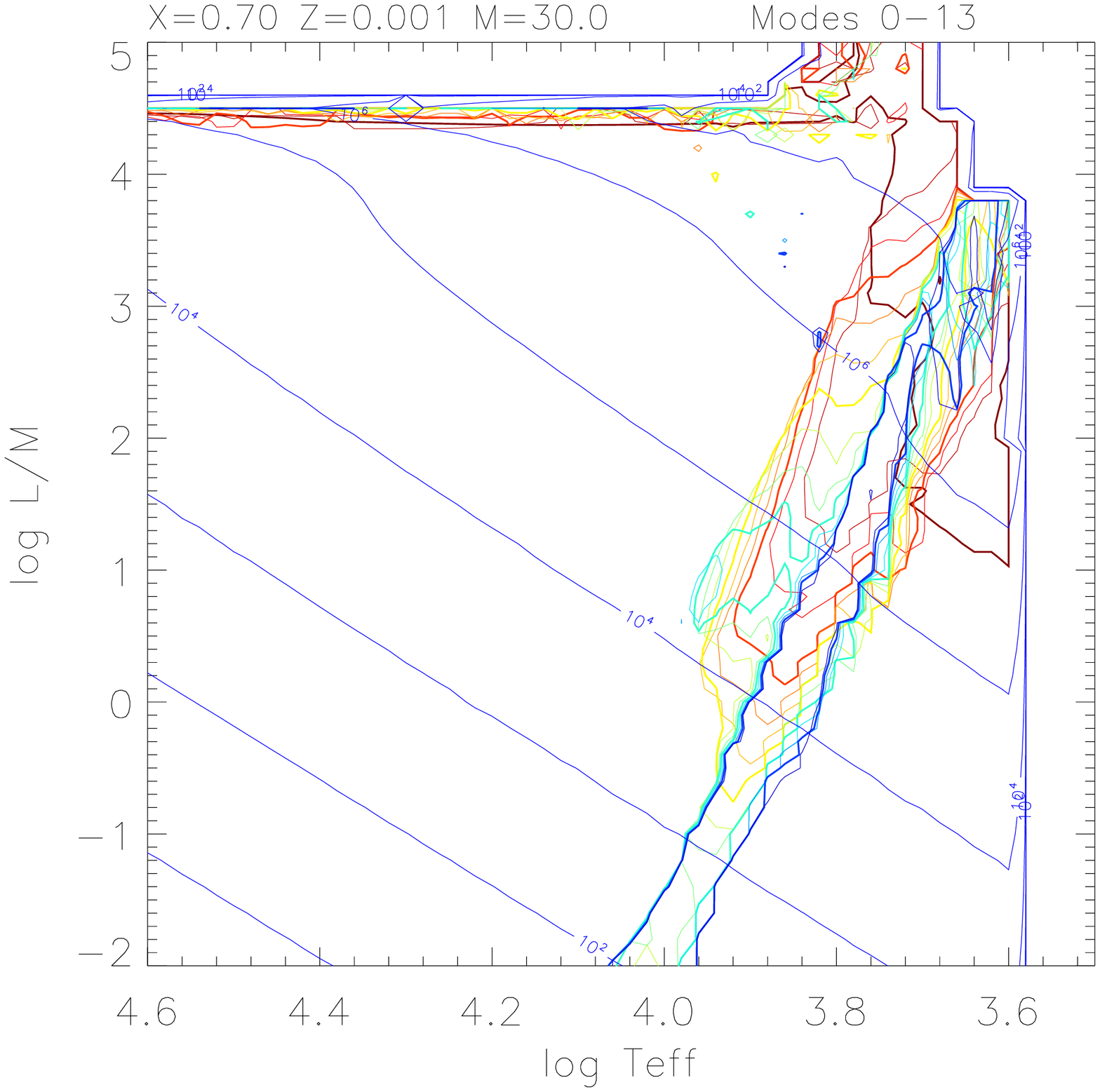,width=4.3cm,angle=0}
\epsfig{file=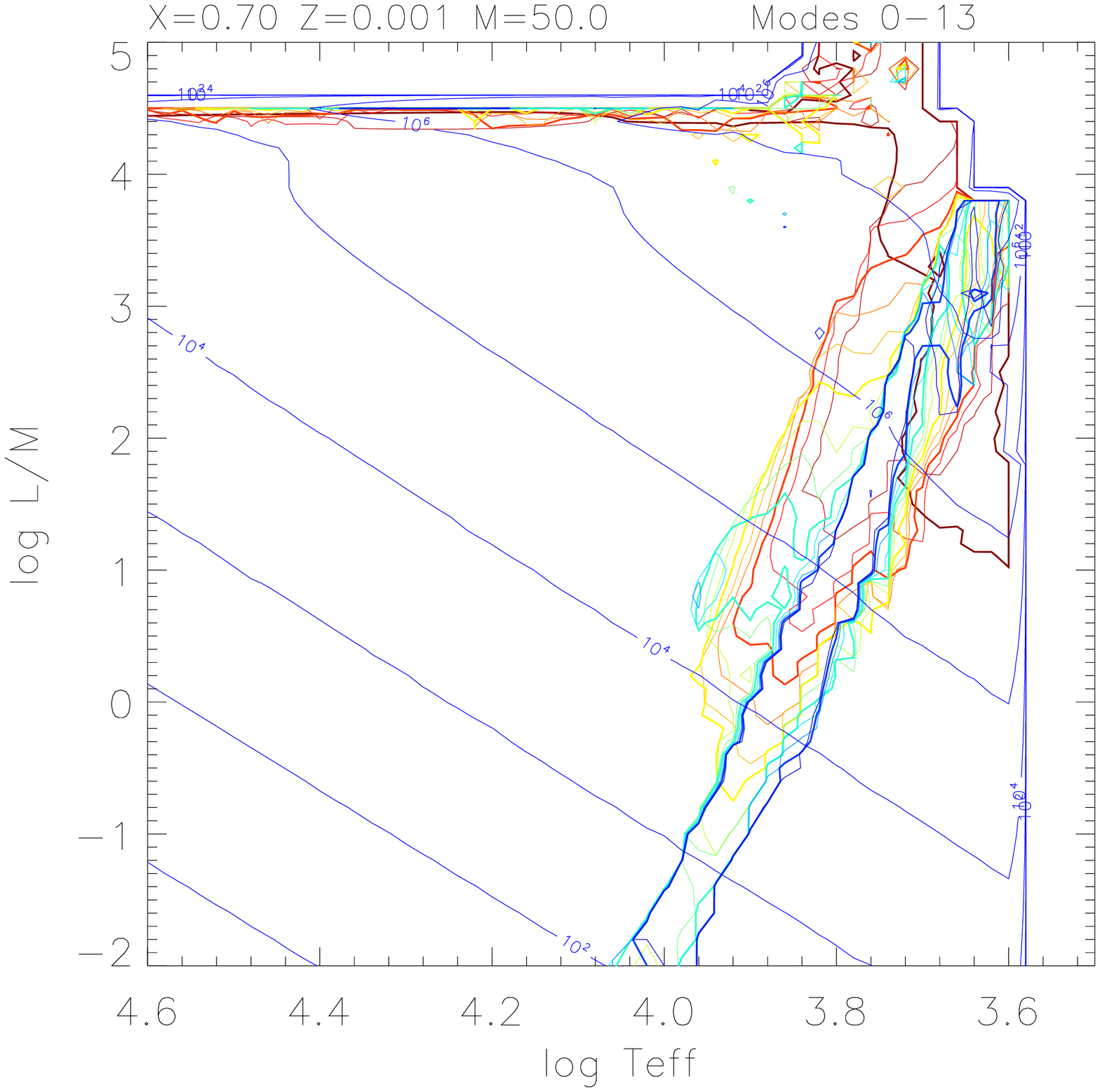,width=4.3cm,angle=0}
\caption[Unstable modes: $X=0.70, Z=0.001$]
{As Fig.~\ref{f:px70} with $X=0.70, Z=0.001$. 
}
\label{f:px70z001}
\end{center}
\end{figure*}

\begin{figure*}
\begin{center}
\epsfig{file=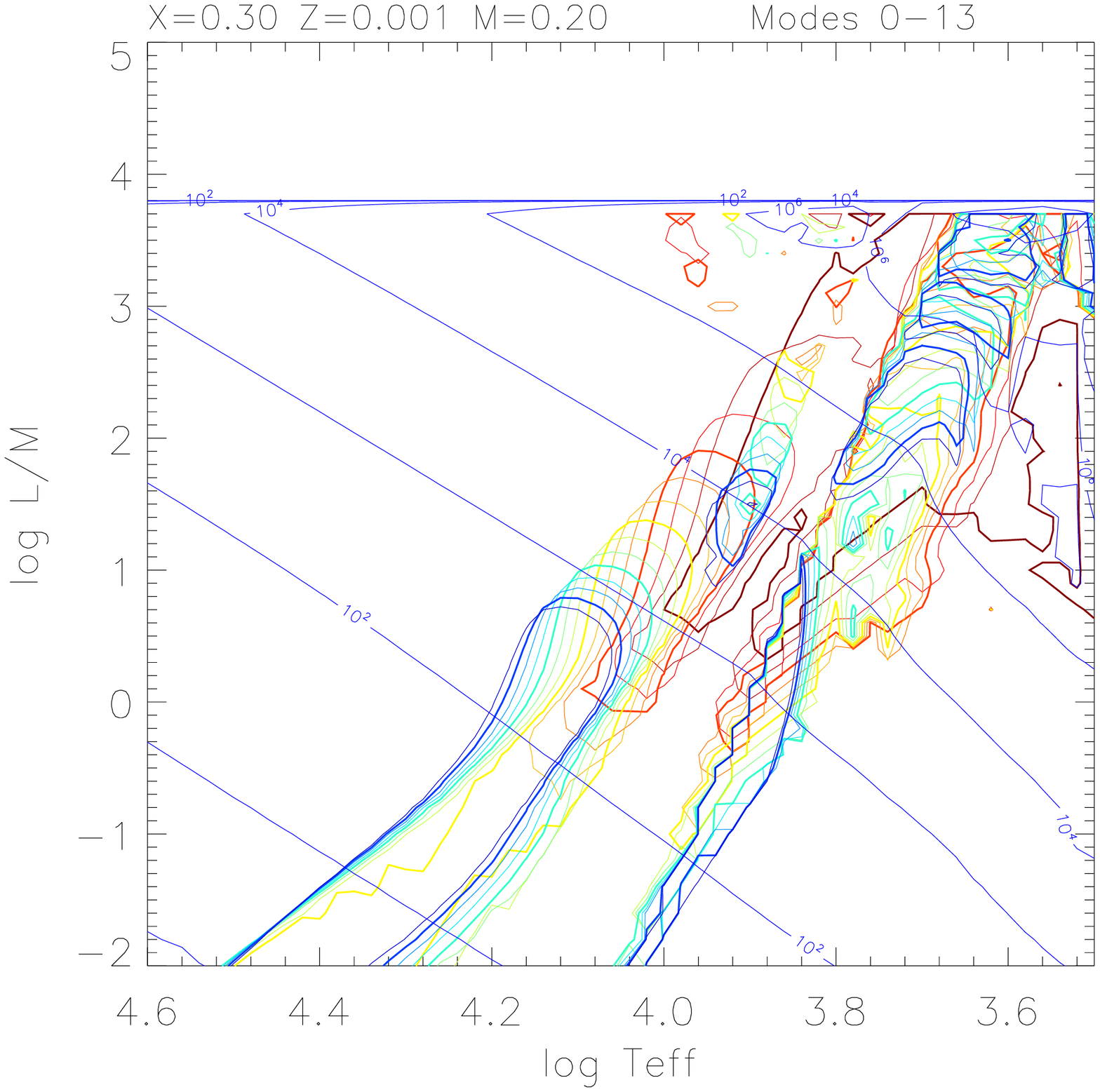,width=4.3cm,angle=0}
\epsfig{file=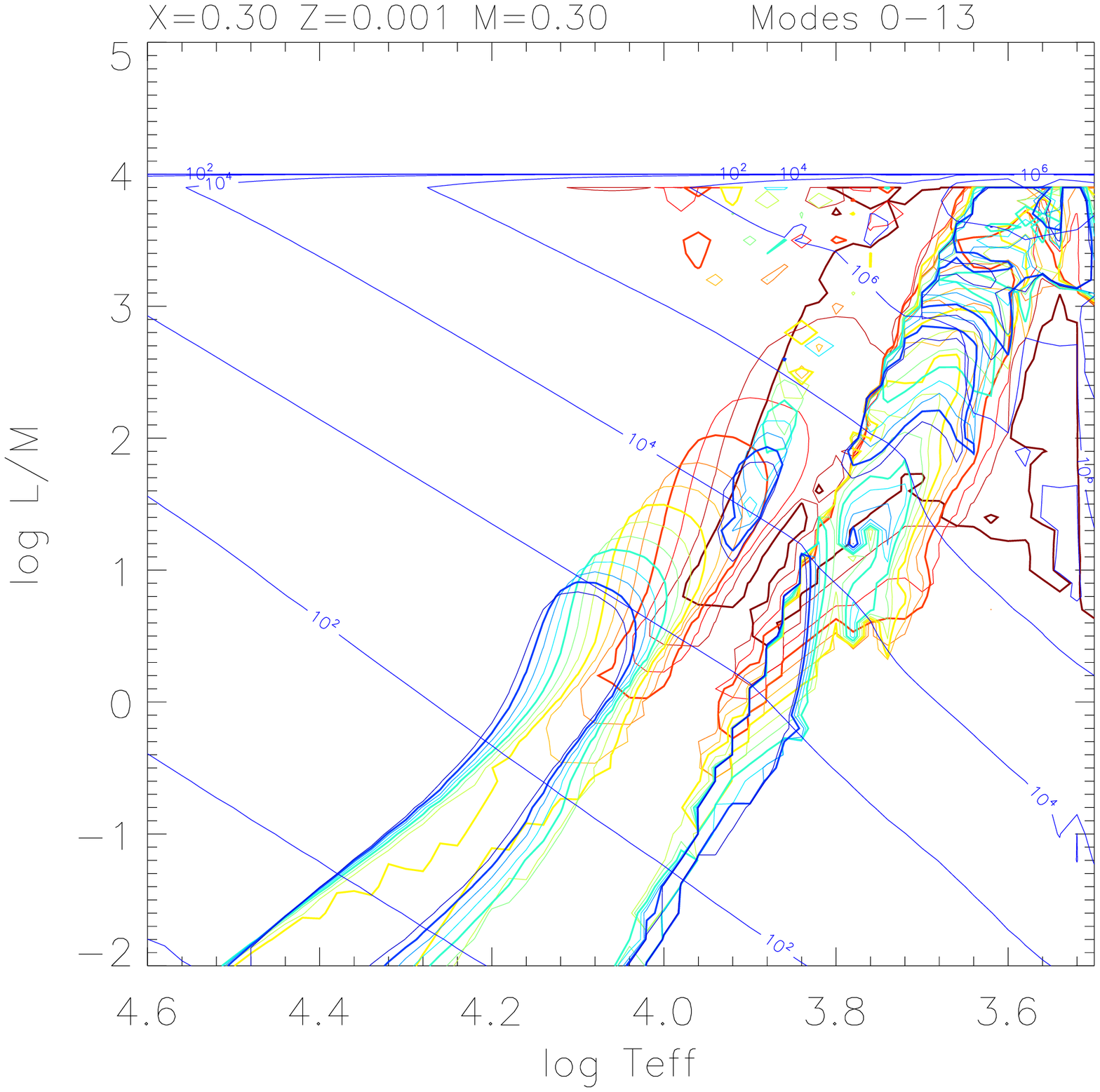,width=4.3cm,angle=0}
\epsfig{file=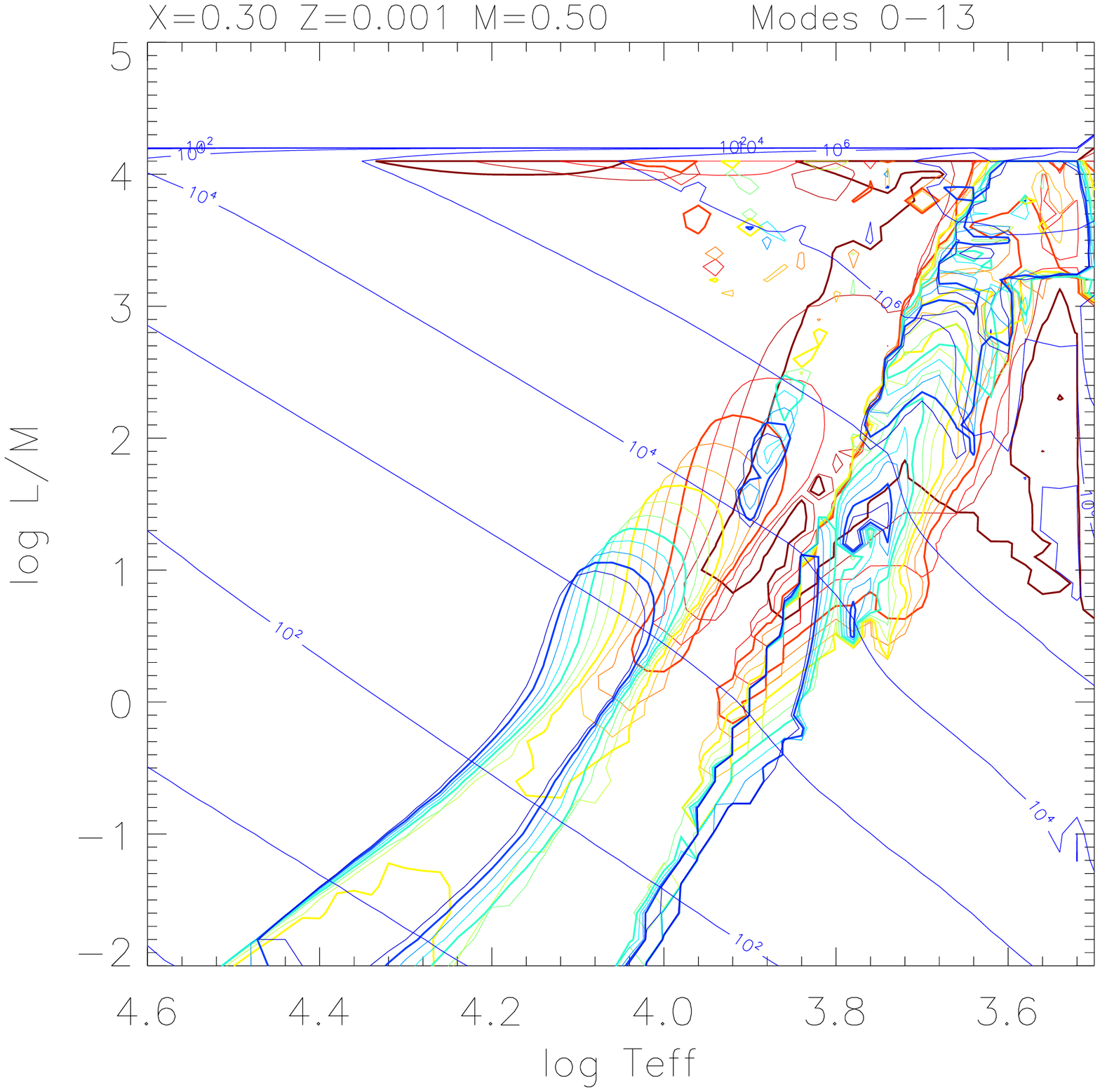,width=4.3cm,angle=0}
\epsfig{file=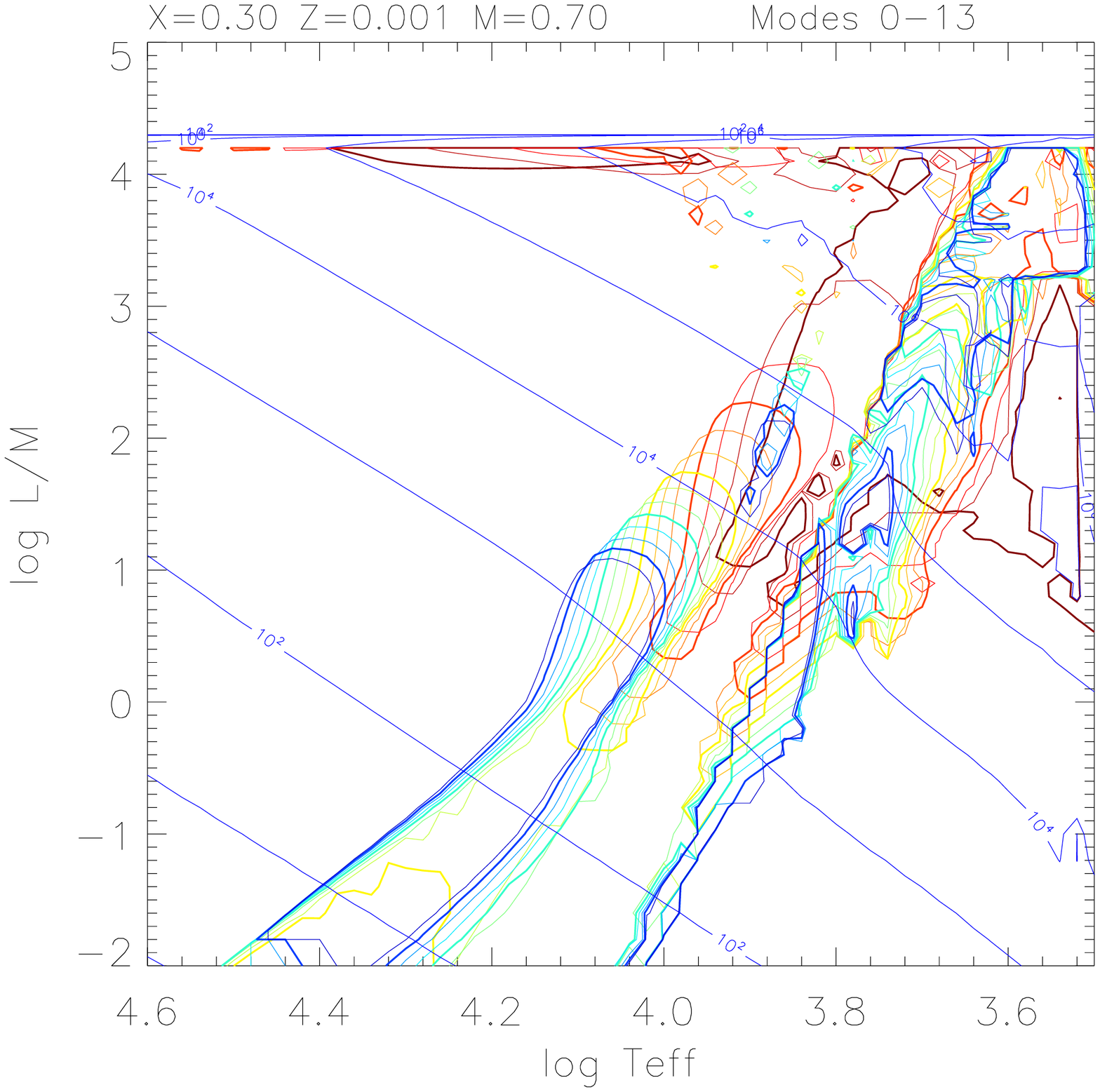,width=4.3cm,angle=0}\\
\epsfig{file=figs/periods_x30z001m01.0_00_opal.eps,width=4.3cm,angle=0}
\epsfig{file=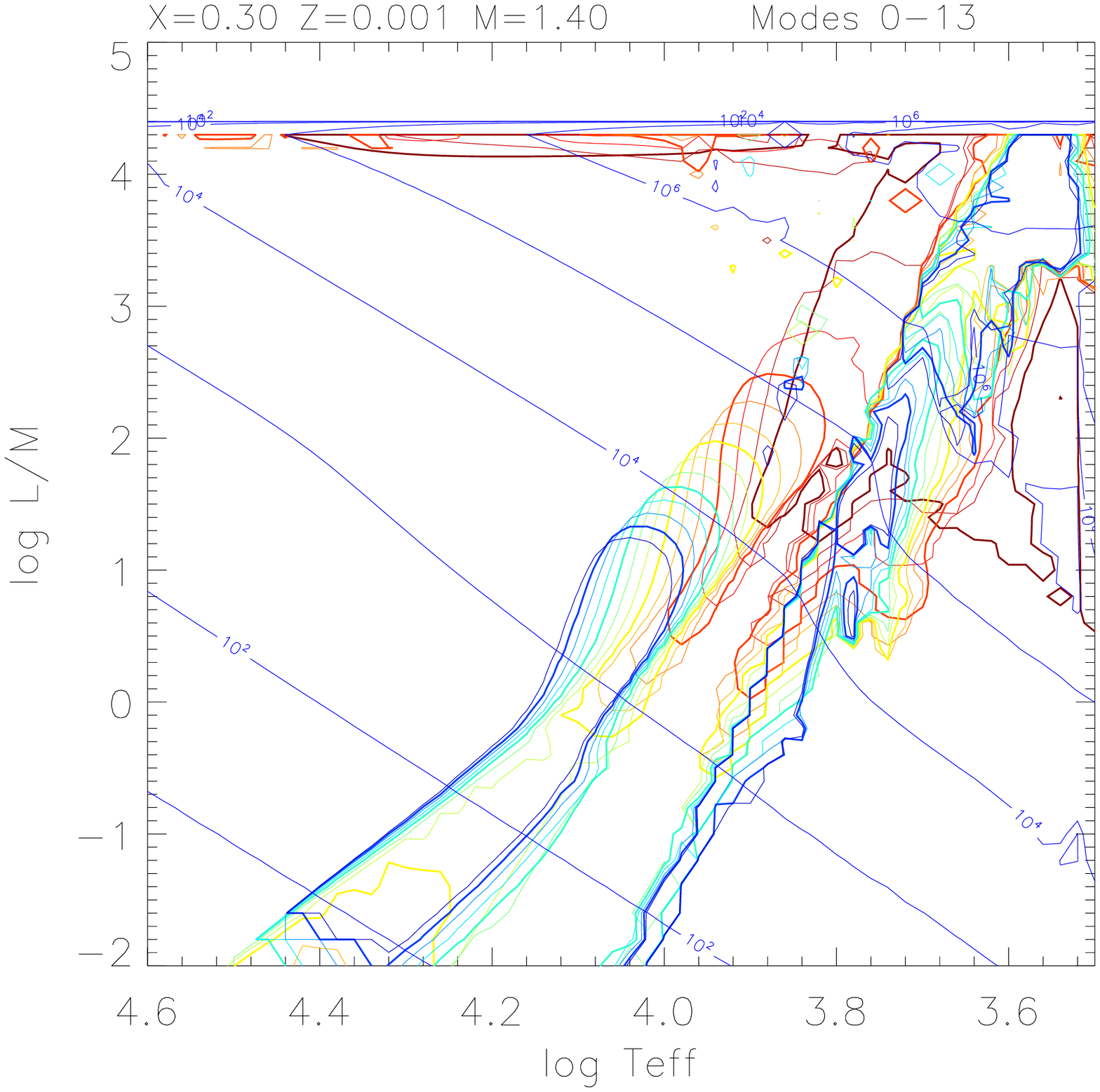,width=4.3cm,angle=0}
\epsfig{file=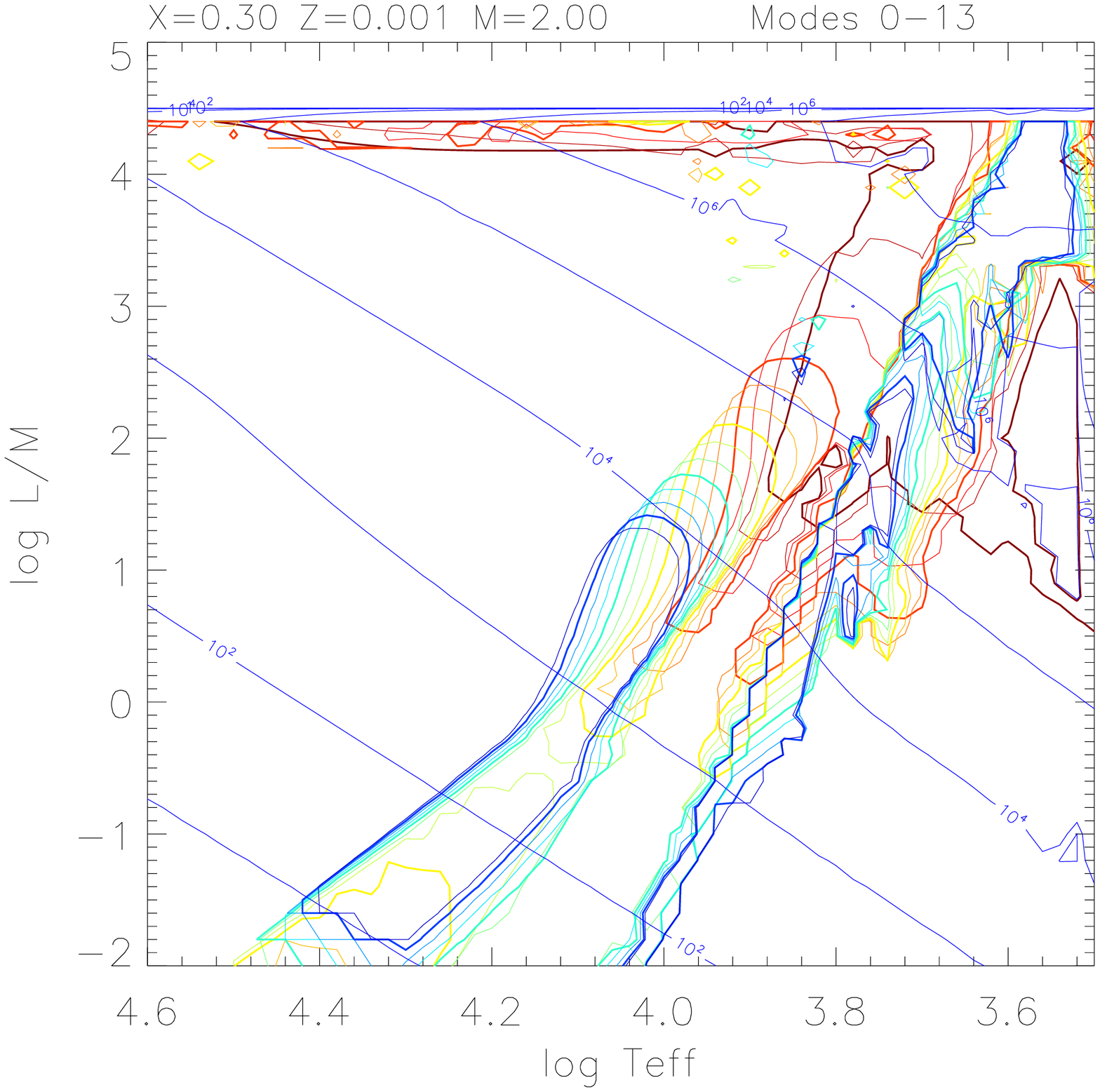,width=4.3cm,angle=0}
\epsfig{file=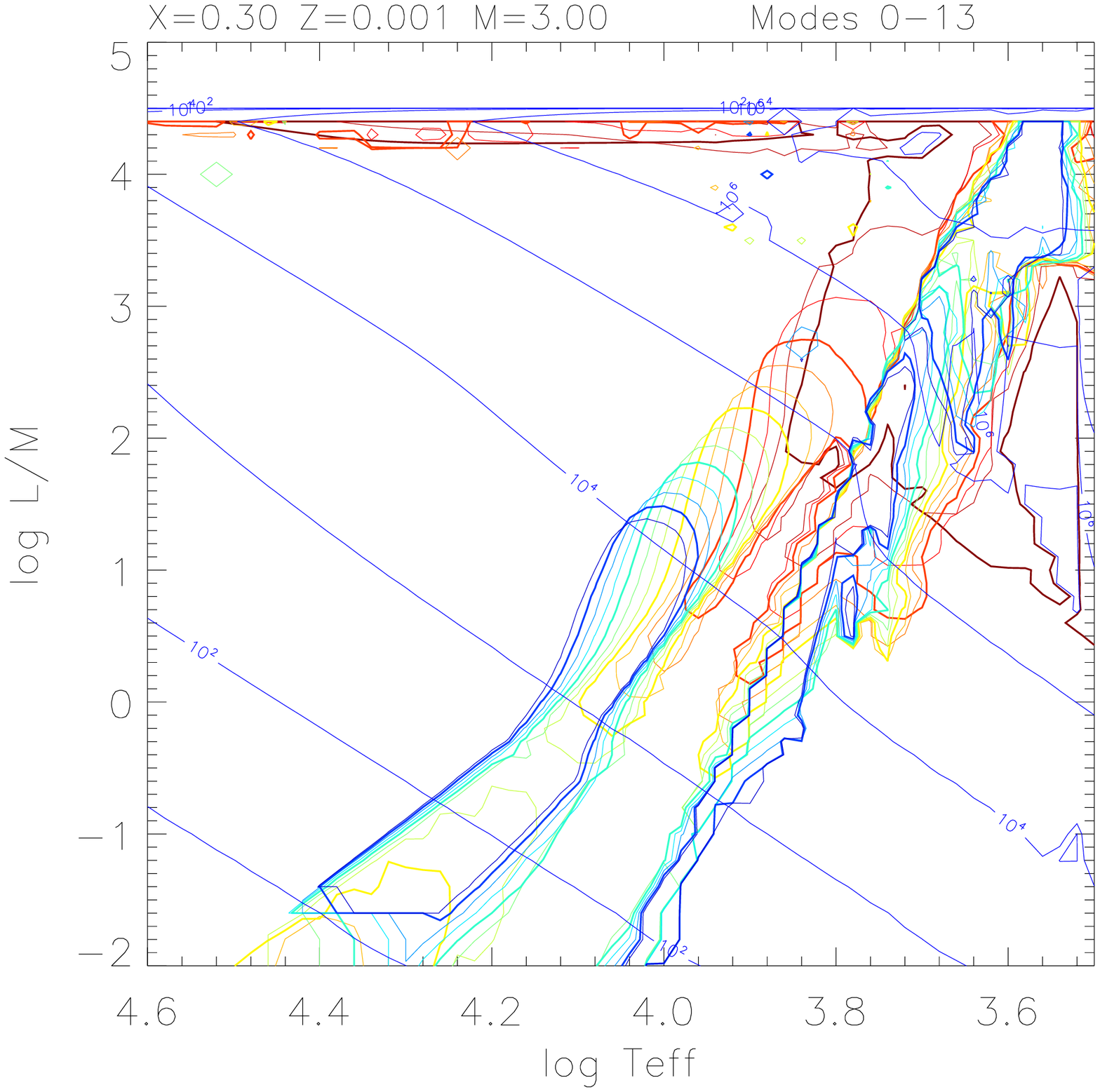,width=4.3cm,angle=0}\\
\epsfig{file=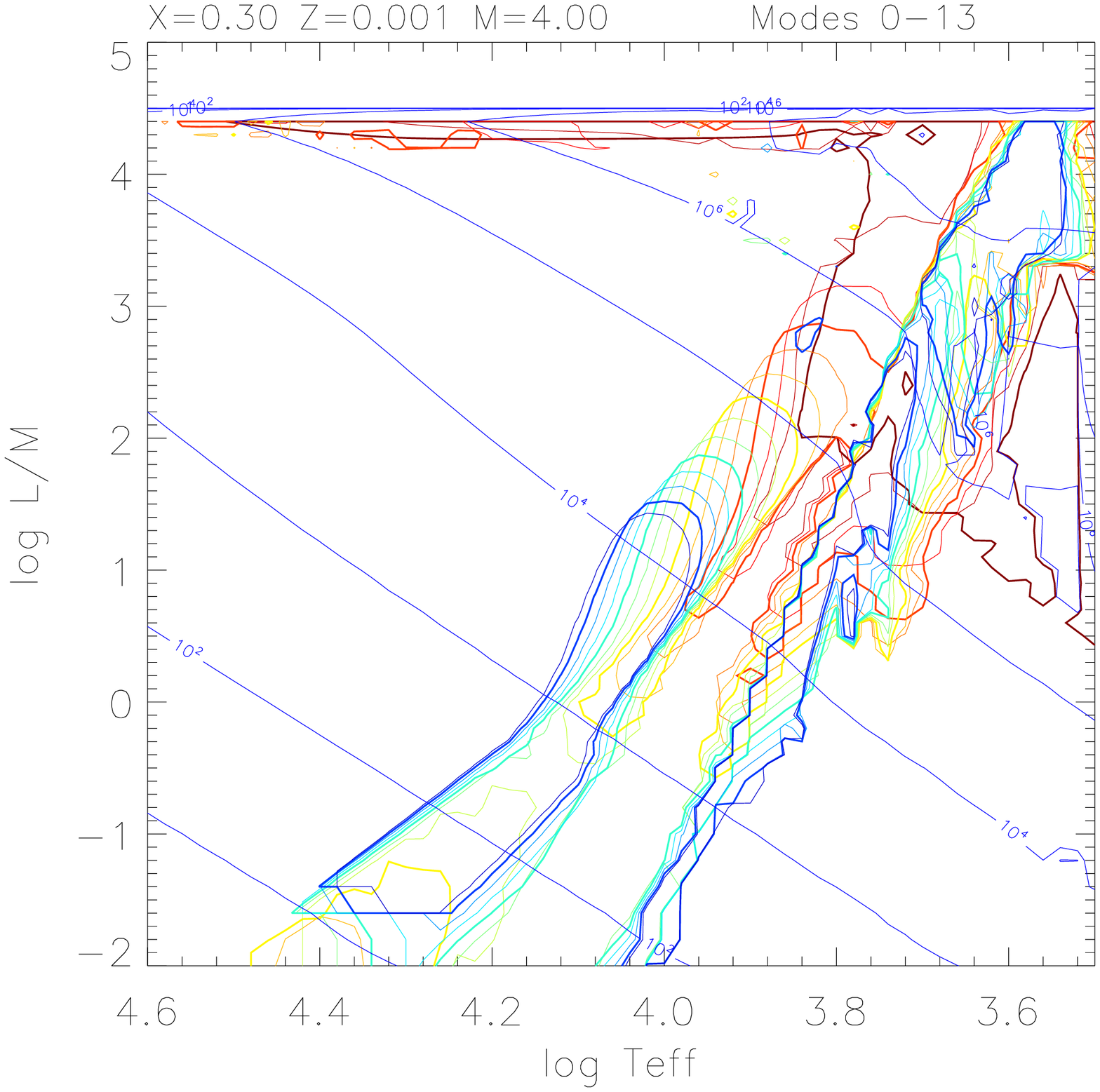,width=4.3cm,angle=0}
\epsfig{file=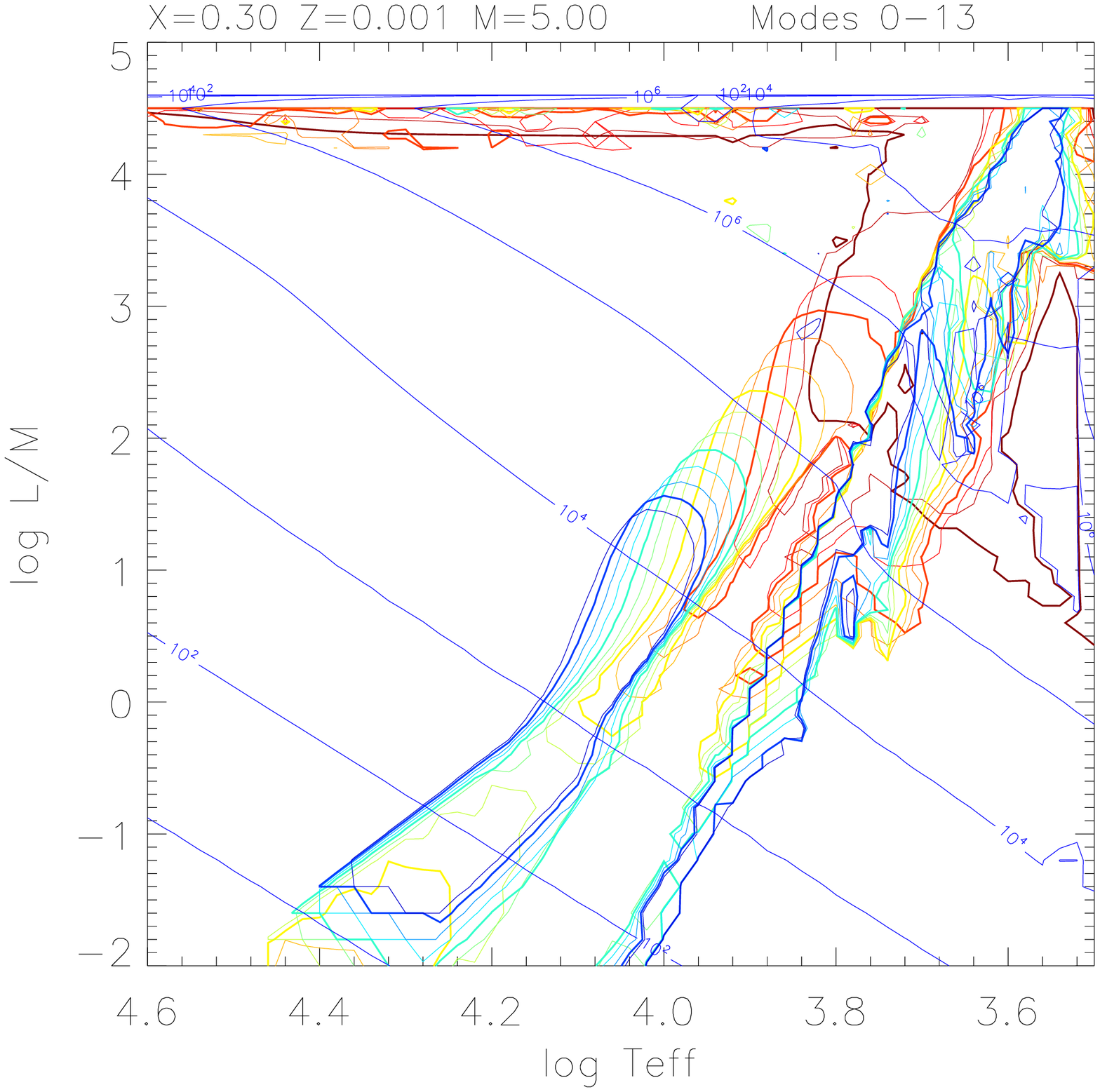,width=4.3cm,angle=0}
\epsfig{file=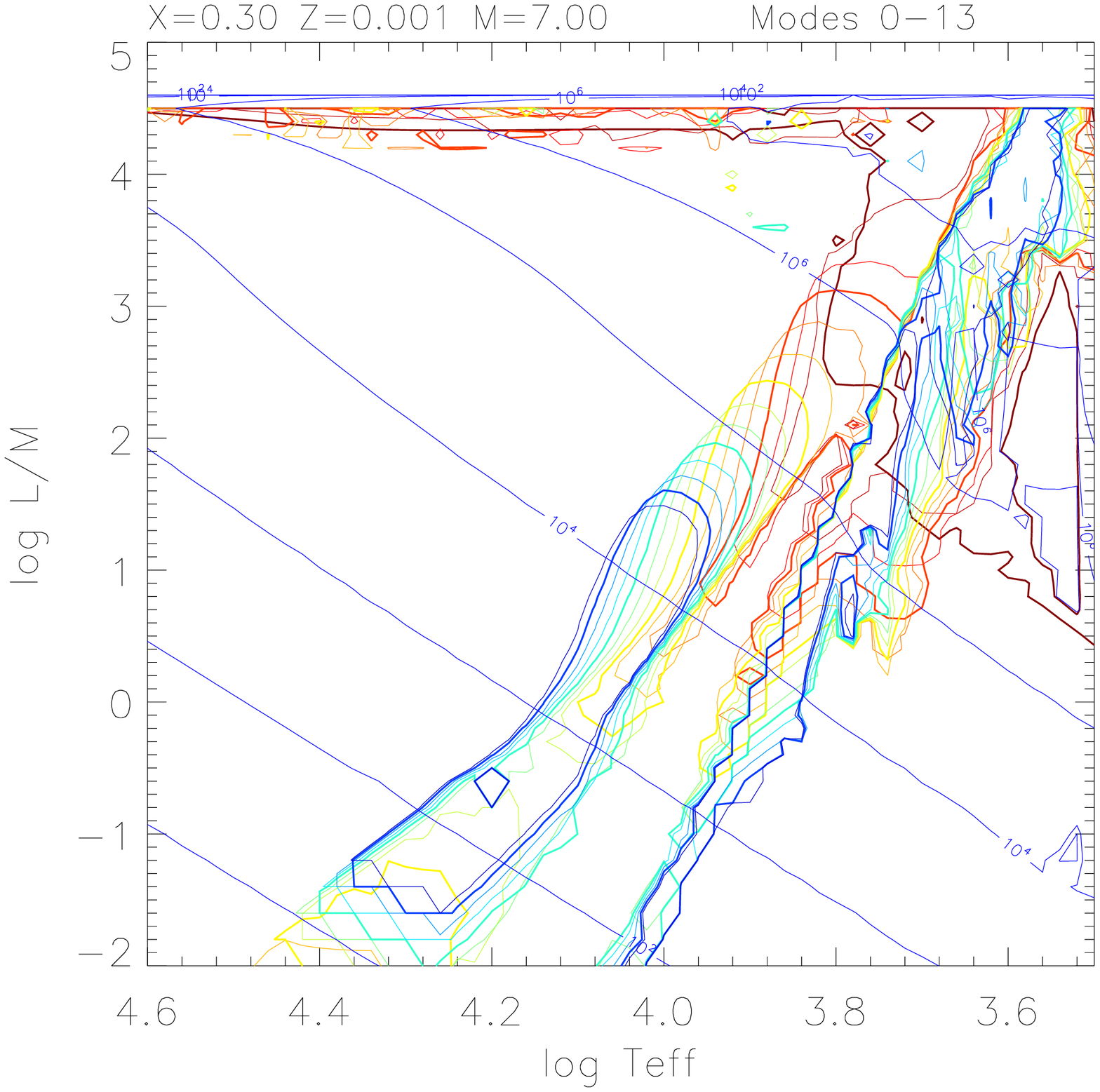,width=4.3cm,angle=0}
\epsfig{file=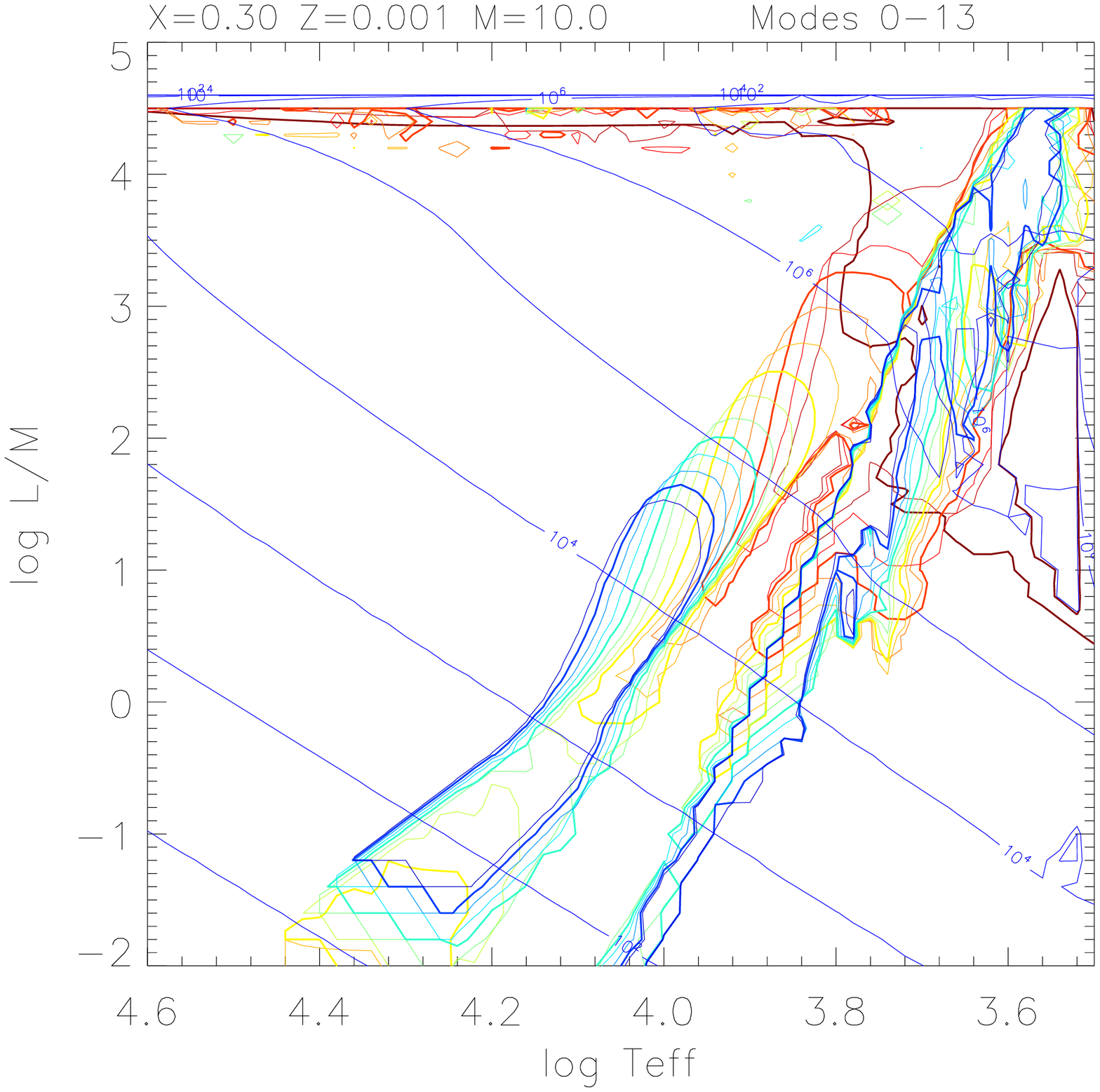,width=4.3cm,angle=0}\\
\epsfig{file=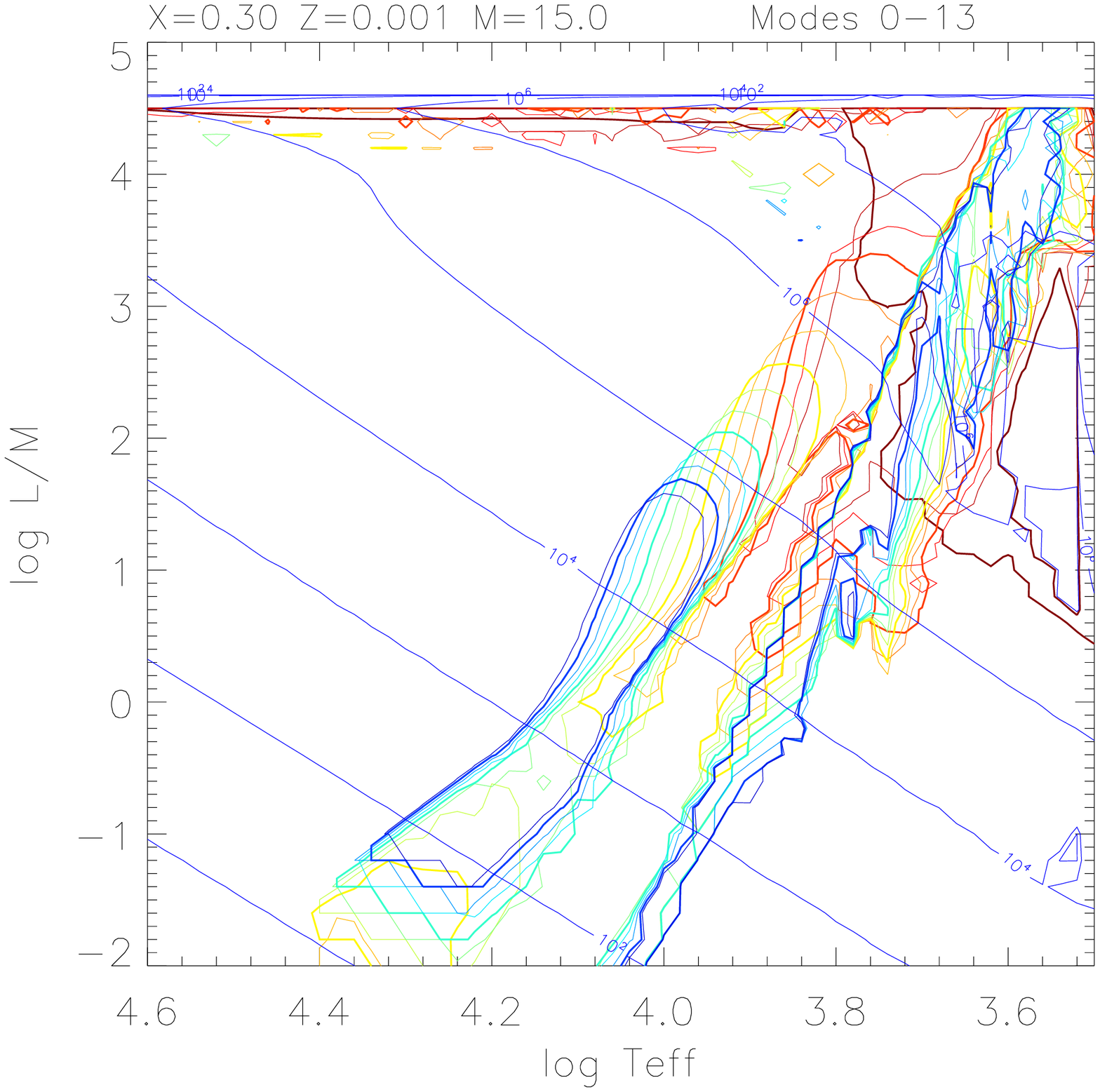,width=4.3cm,angle=0}
\epsfig{file=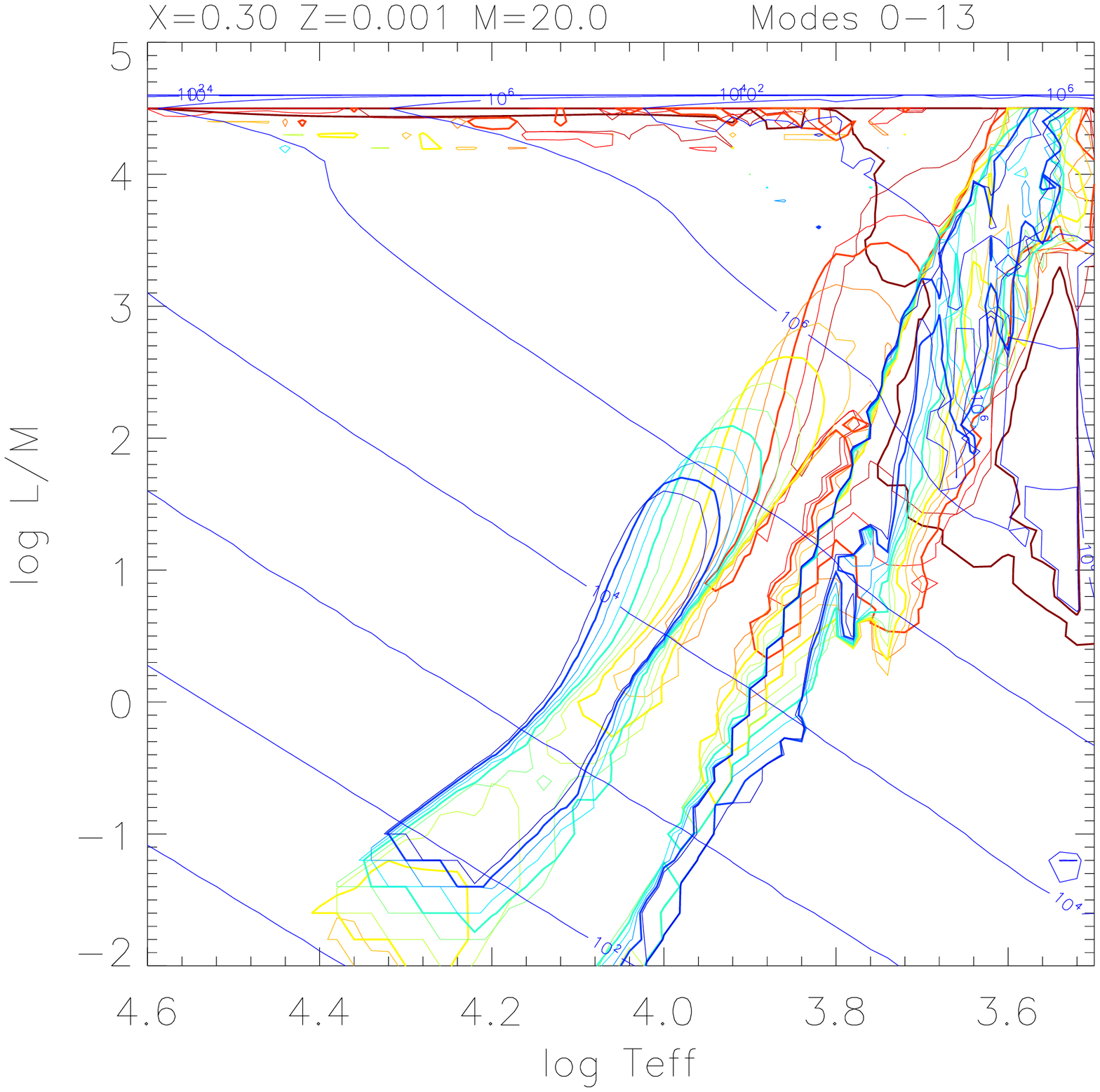,width=4.3cm,angle=0}
\epsfig{file=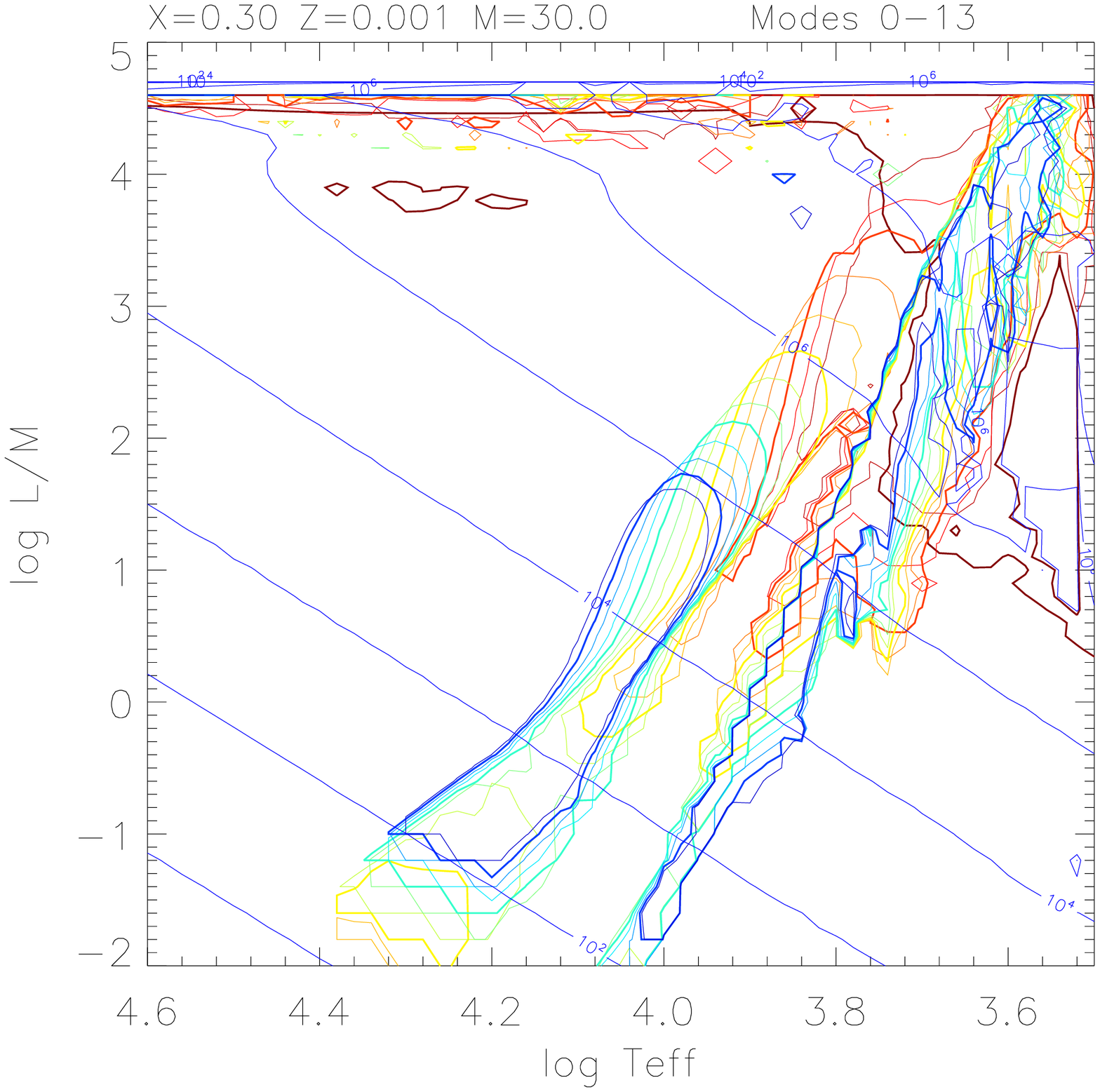,width=4.3cm,angle=0}
\epsfig{file=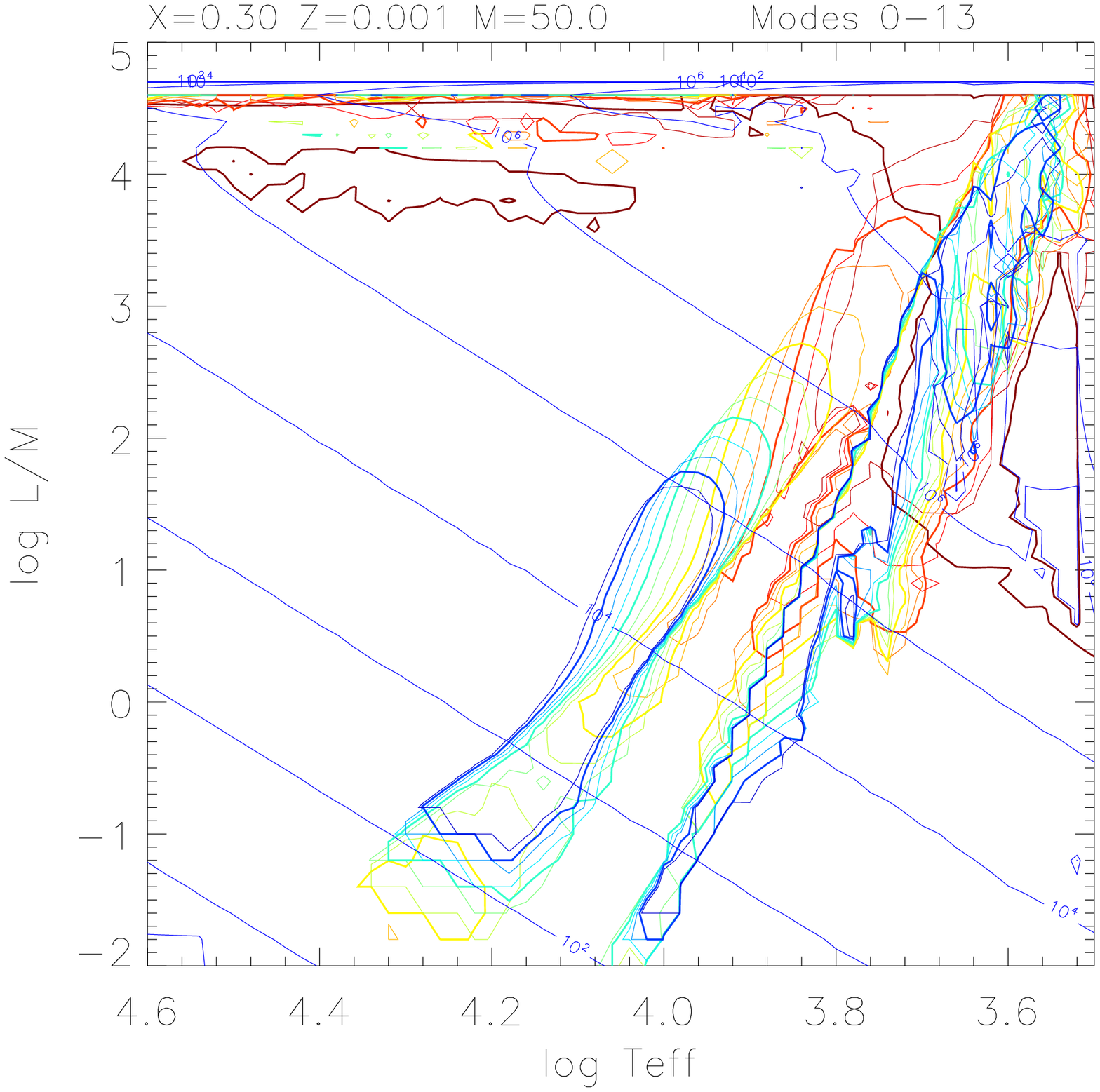,width=4.3cm,angle=0}
\caption[Unstable modes: $X=0.30, Z=0.001$]
{As Fig.~\ref{f:px70} with $X=0.30, Z=0.001$. 
}
\label{f:px30z001}
\end{center}
\end{figure*}

\begin{figure*}
\begin{center}
\epsfig{file=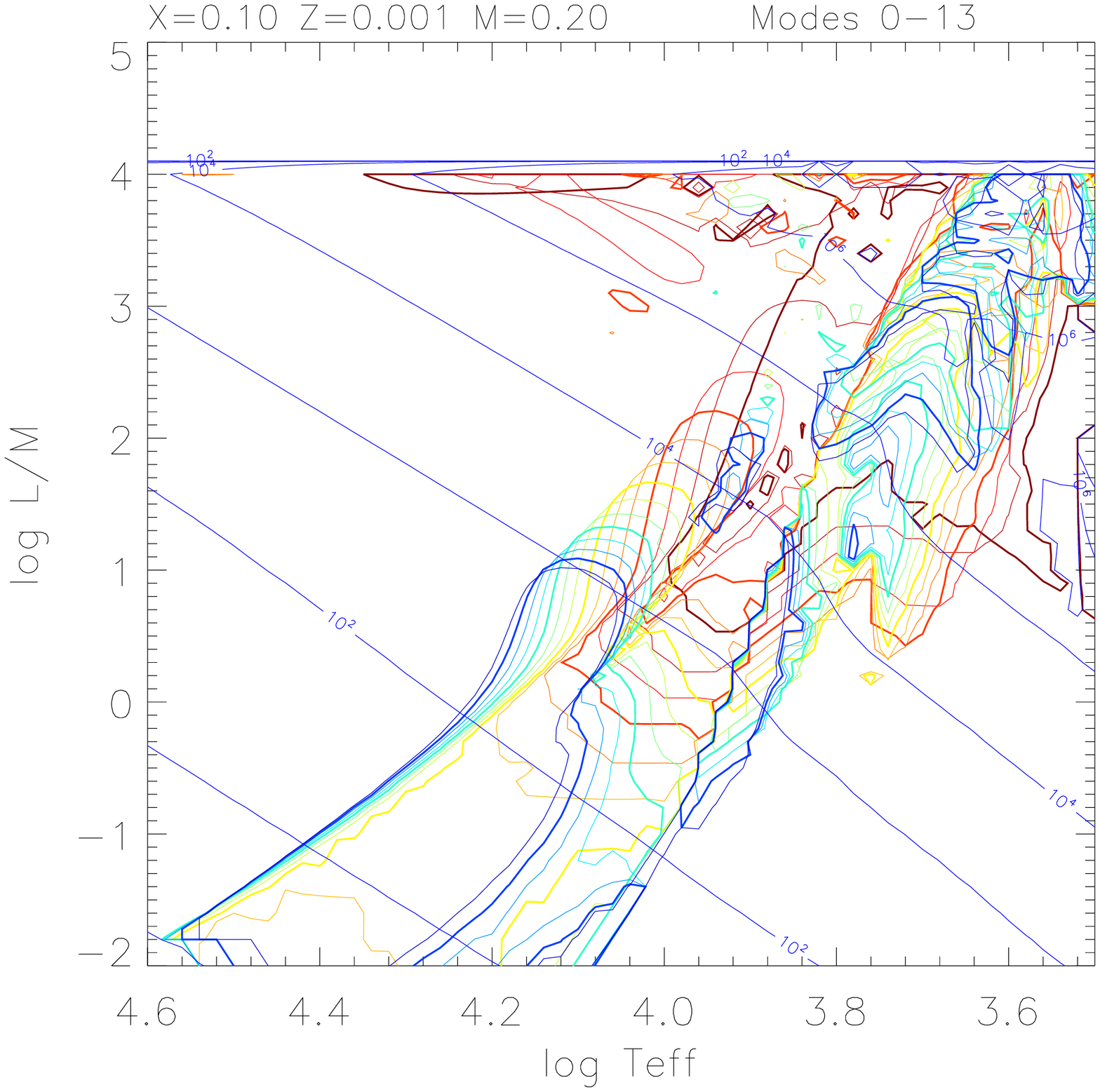,width=4.3cm,angle=0}
\epsfig{file=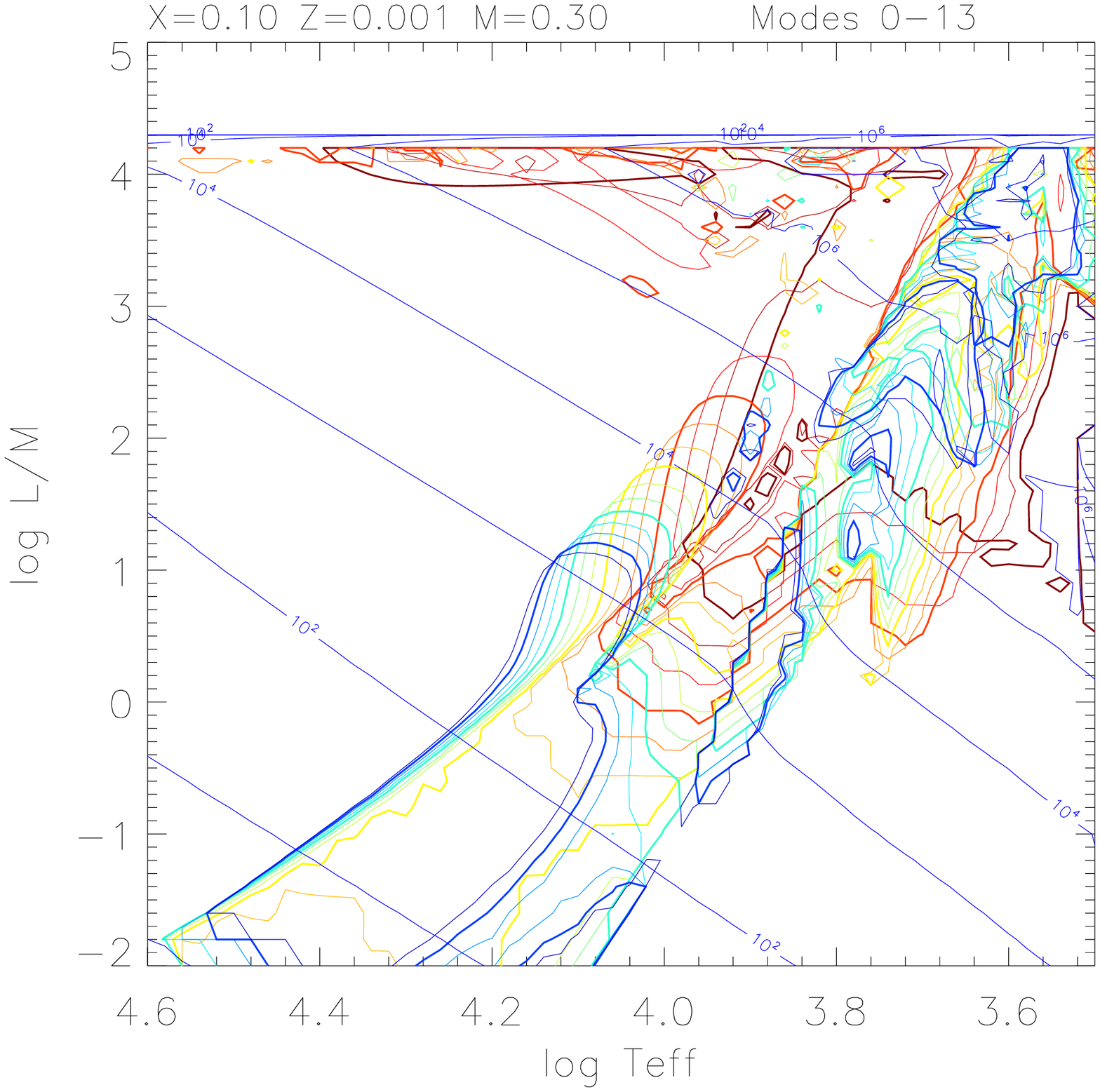,width=4.3cm,angle=0}
\epsfig{file=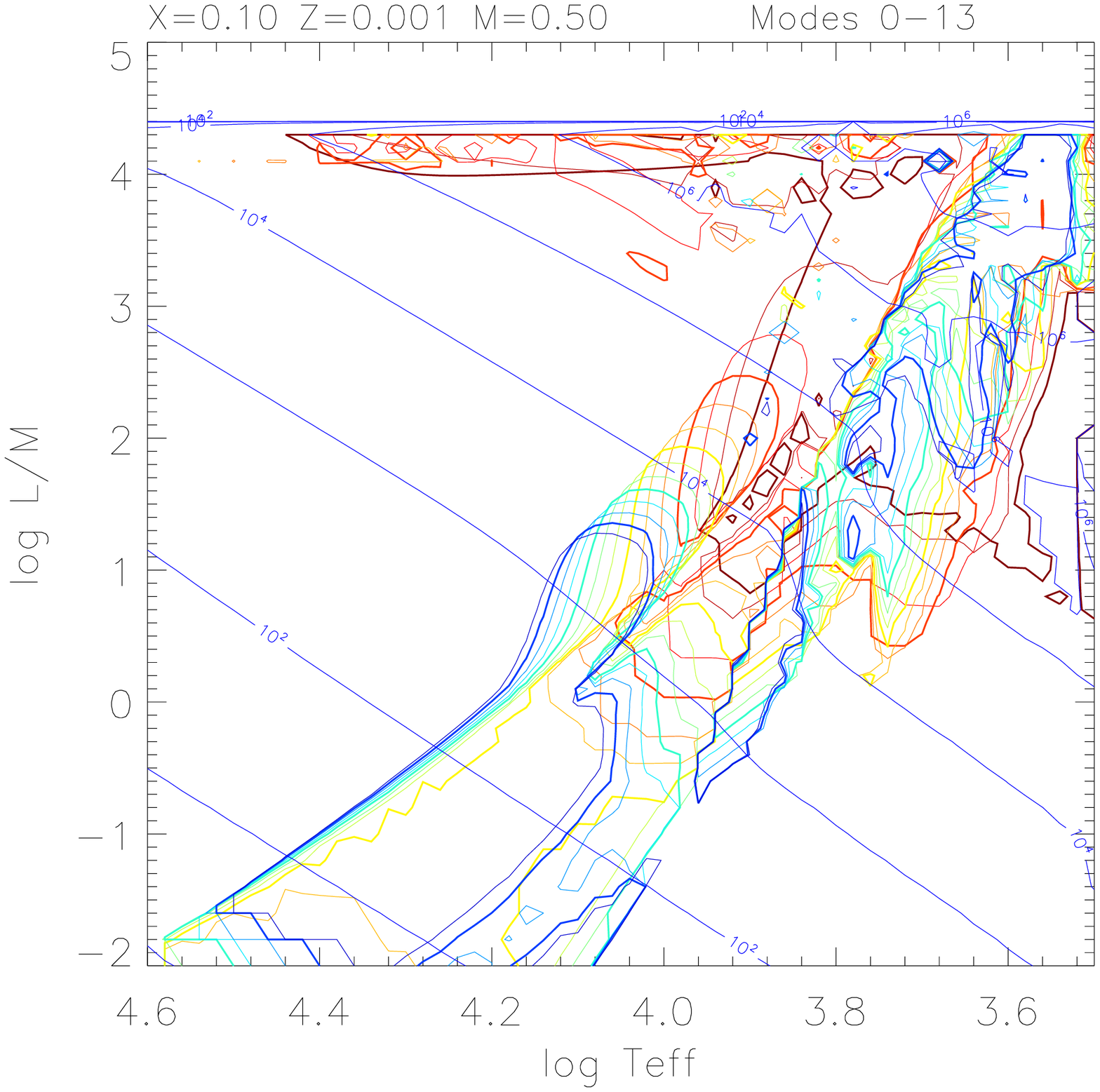,width=4.3cm,angle=0}
\epsfig{file=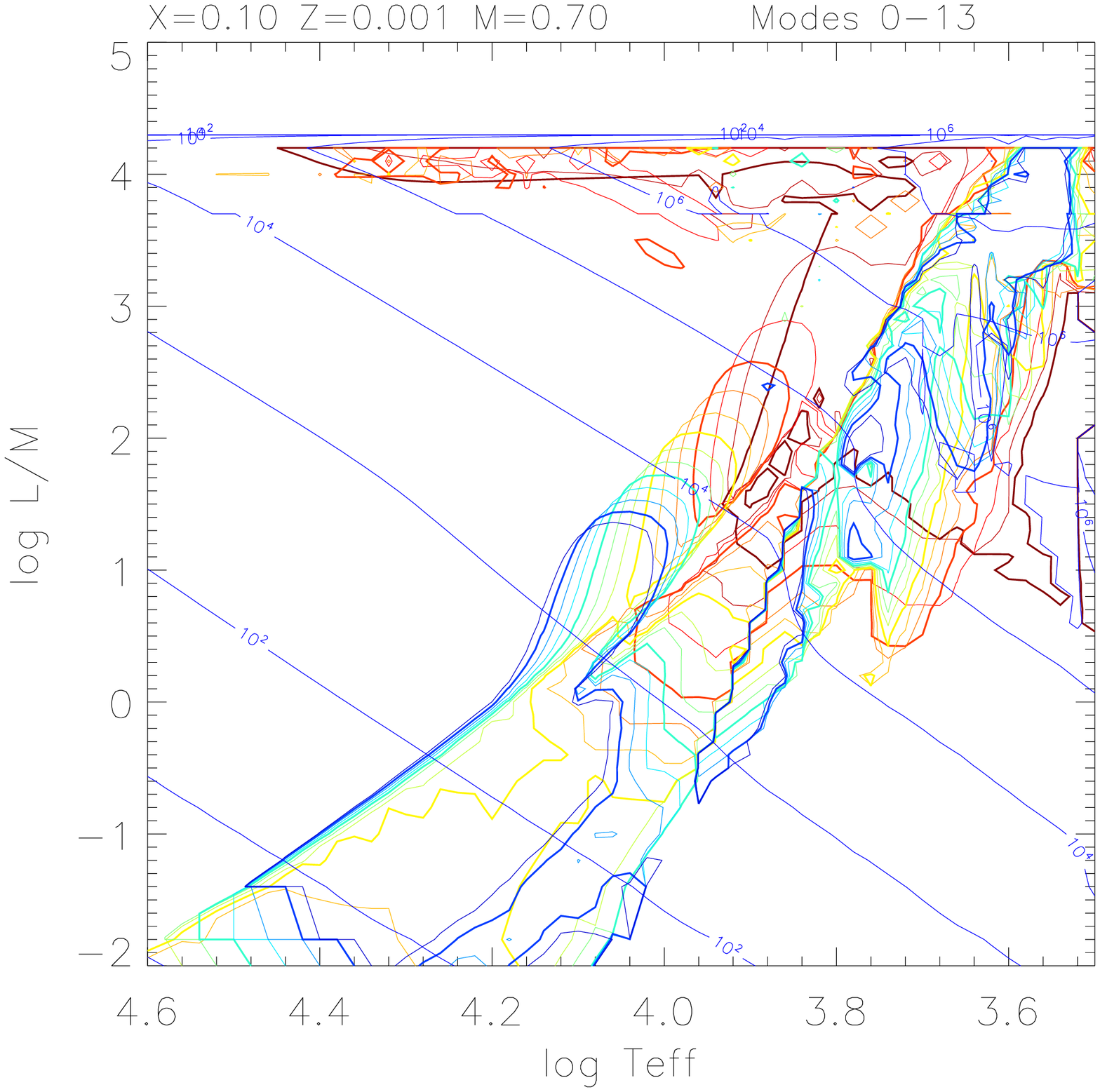,width=4.3cm,angle=0}\\
\epsfig{file=figs/periods_x10z001m01.0_00_opal.eps,width=4.3cm,angle=0}
\epsfig{file=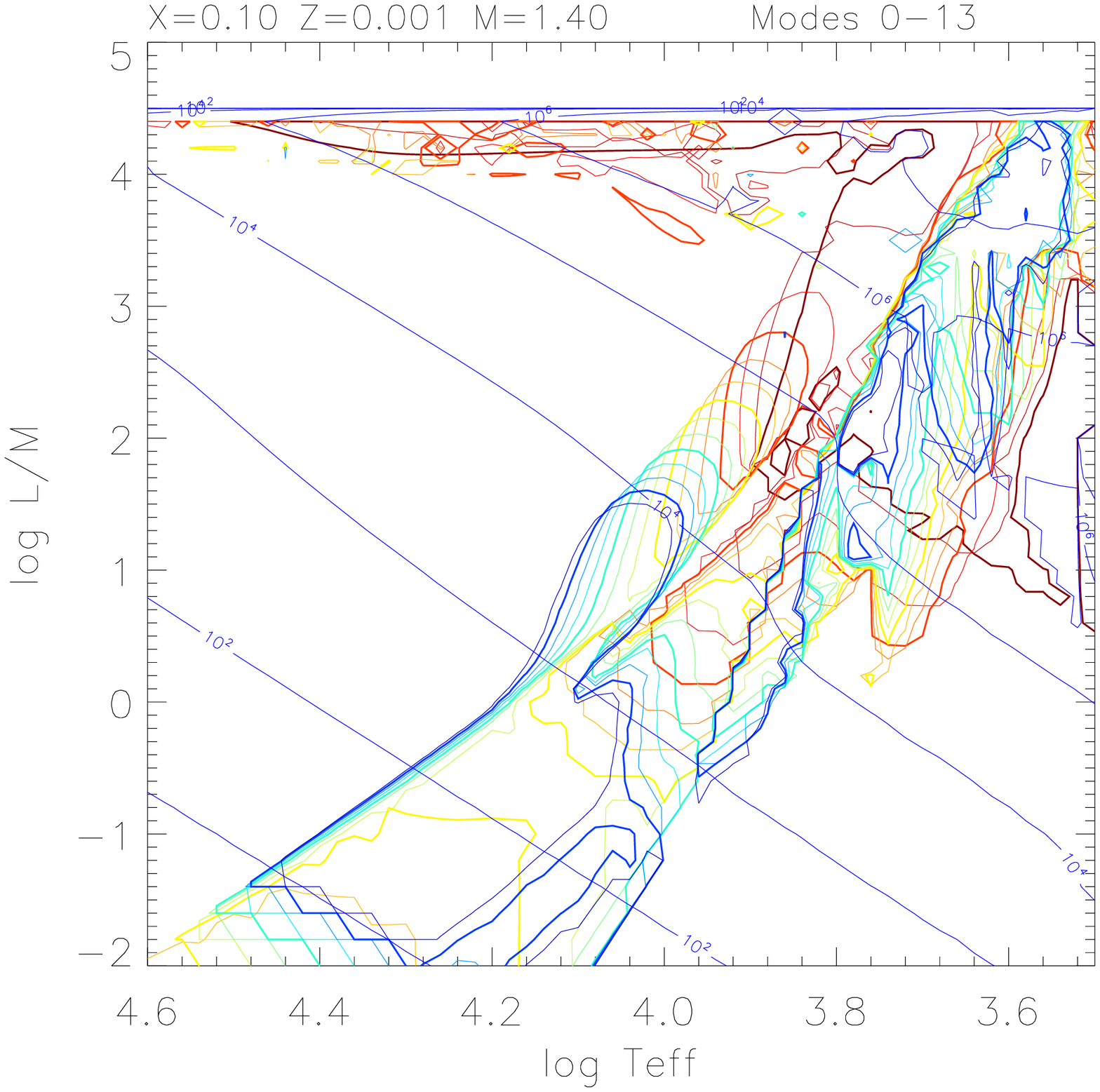,width=4.3cm,angle=0}
\epsfig{file=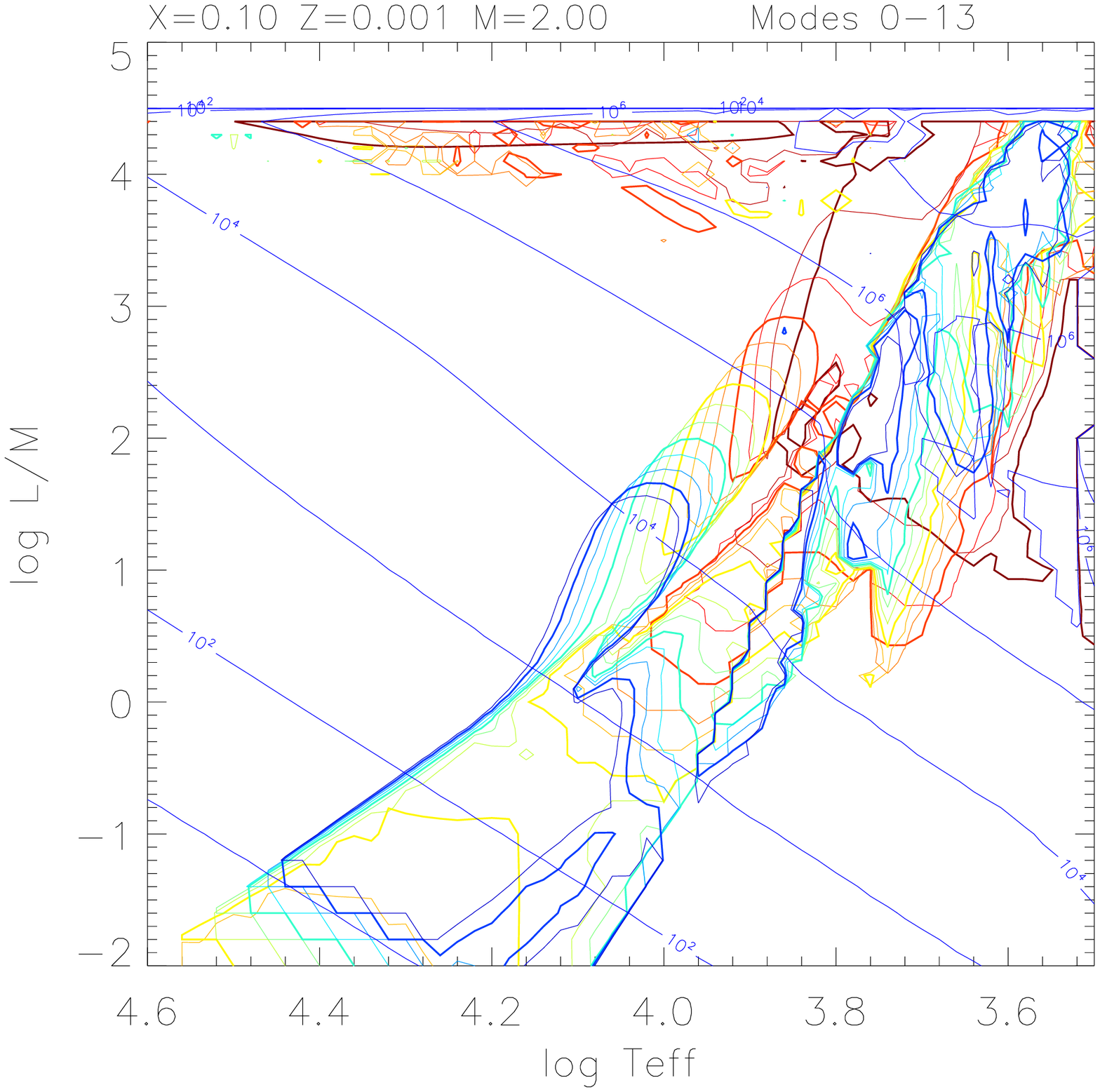,width=4.3cm,angle=0}
\epsfig{file=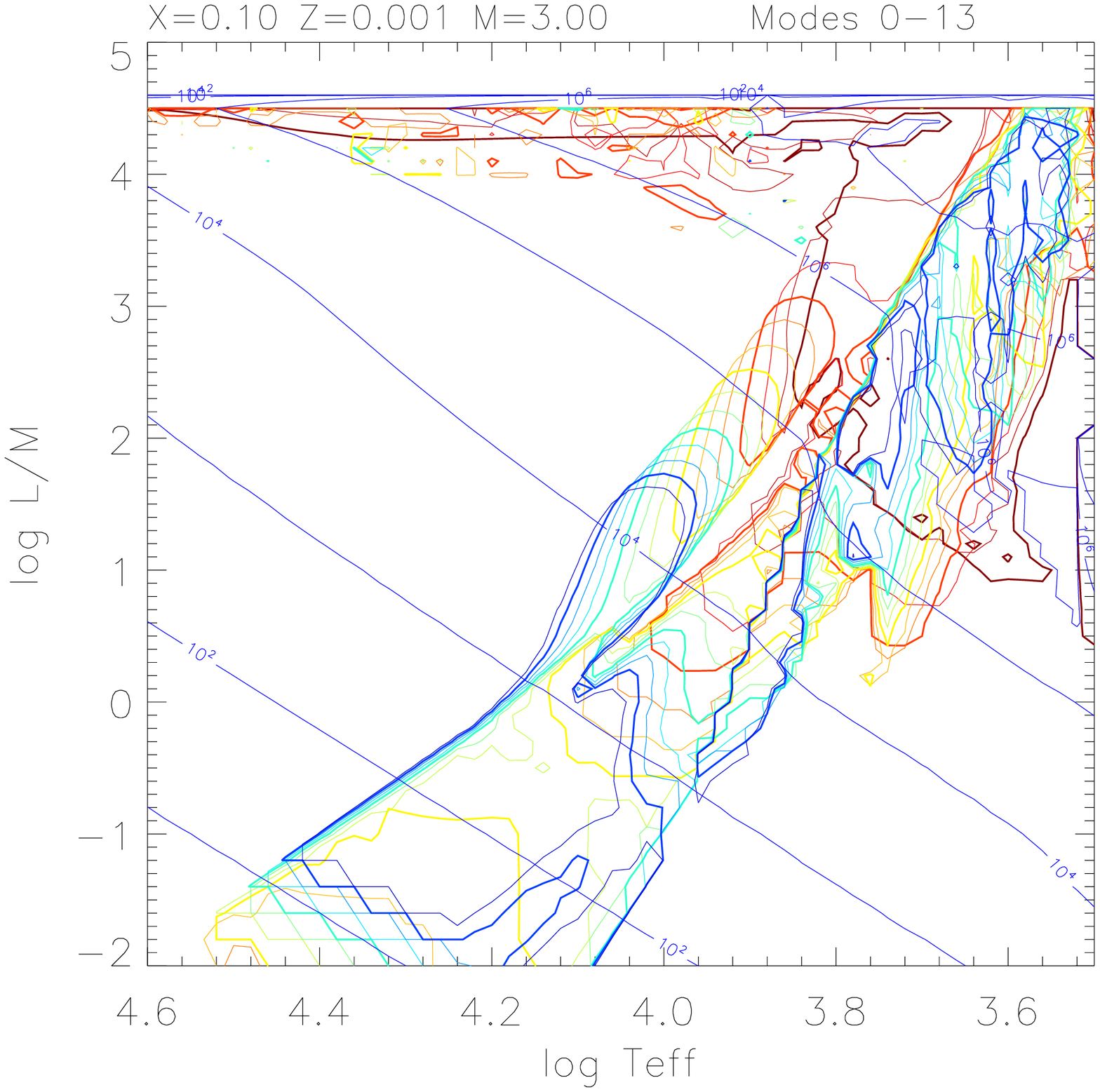,width=4.3cm,angle=0}\\
\epsfig{file=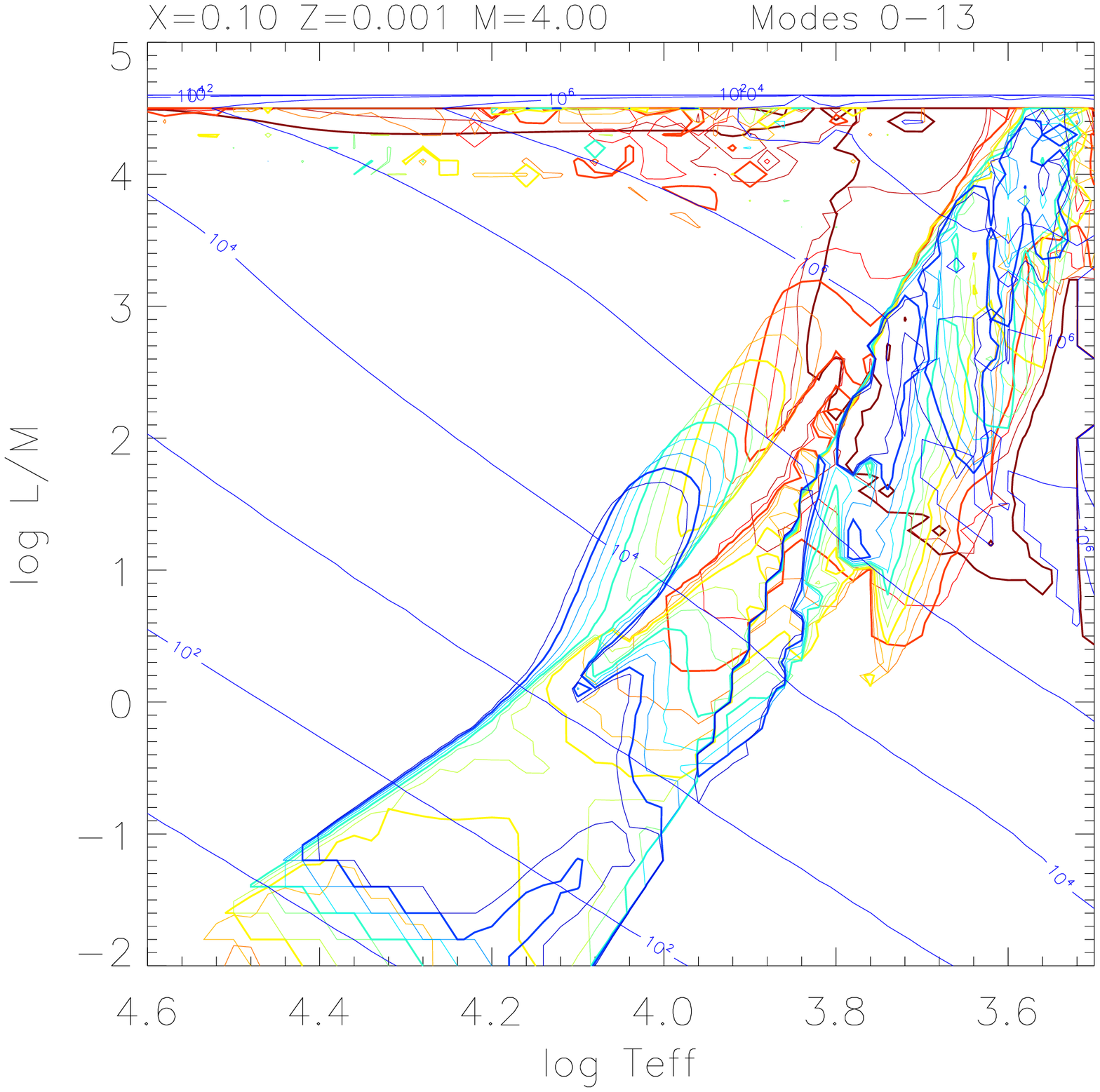,width=4.3cm,angle=0}
\epsfig{file=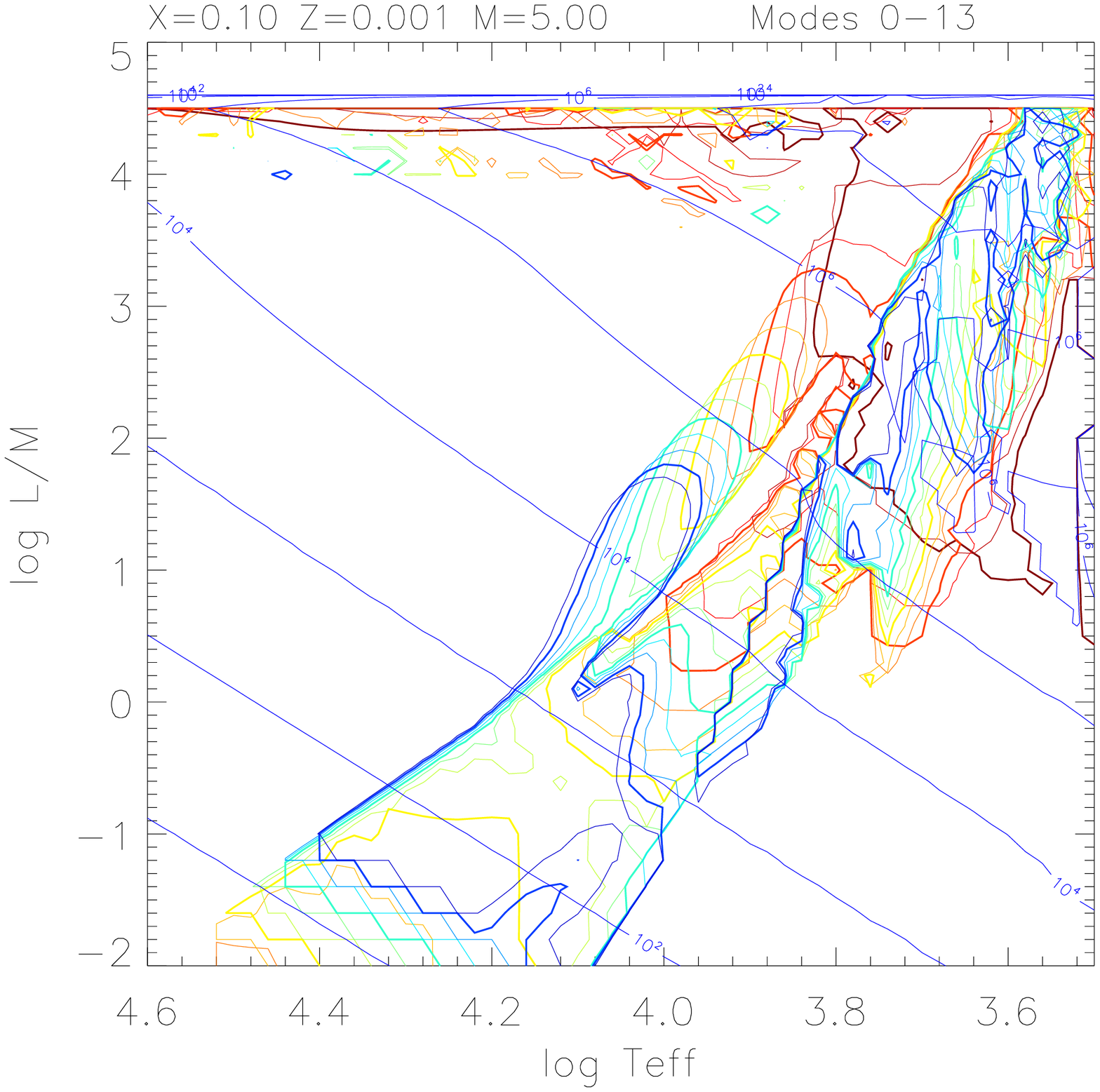,width=4.3cm,angle=0}
\epsfig{file=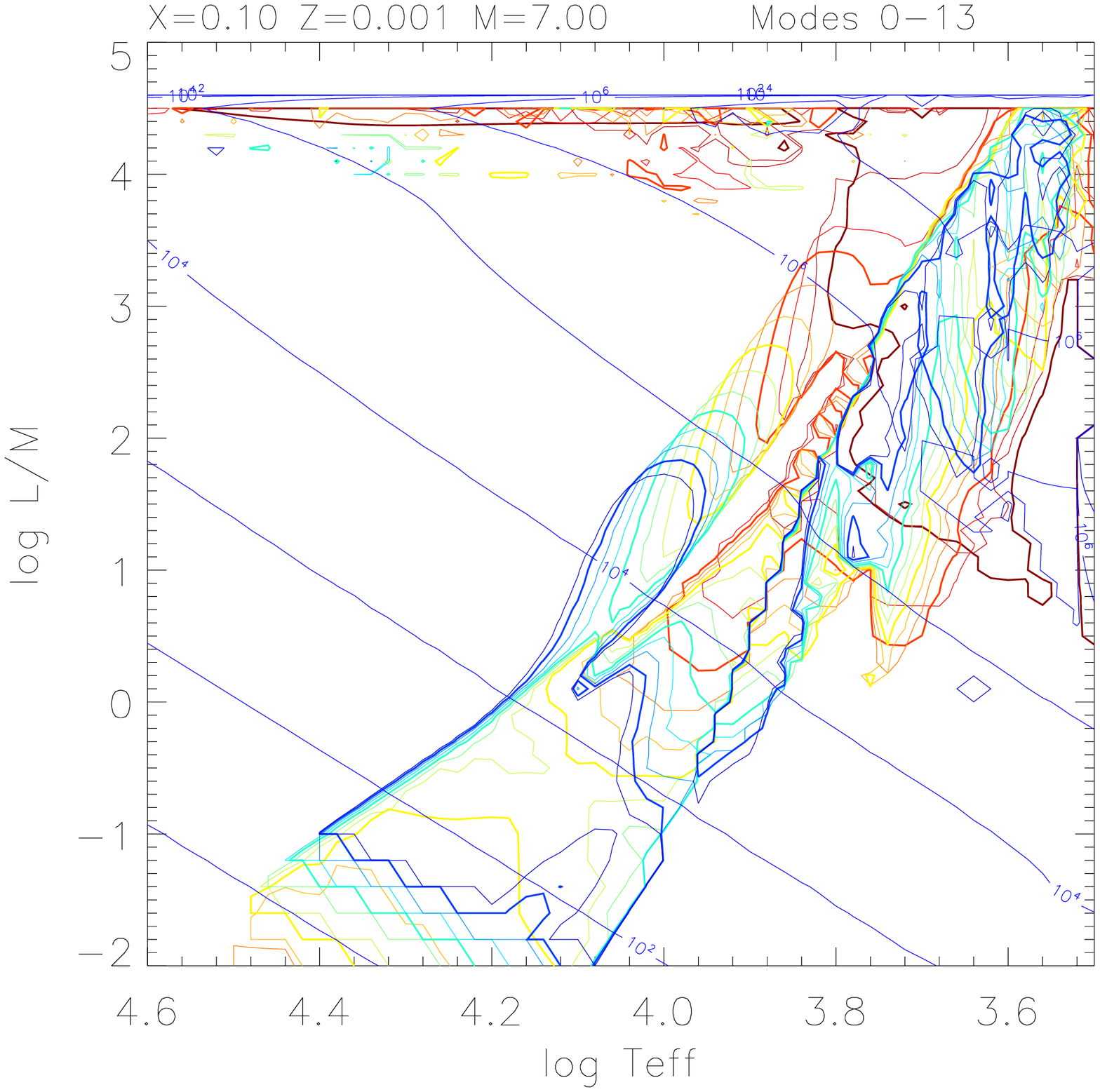,width=4.3cm,angle=0}
\epsfig{file=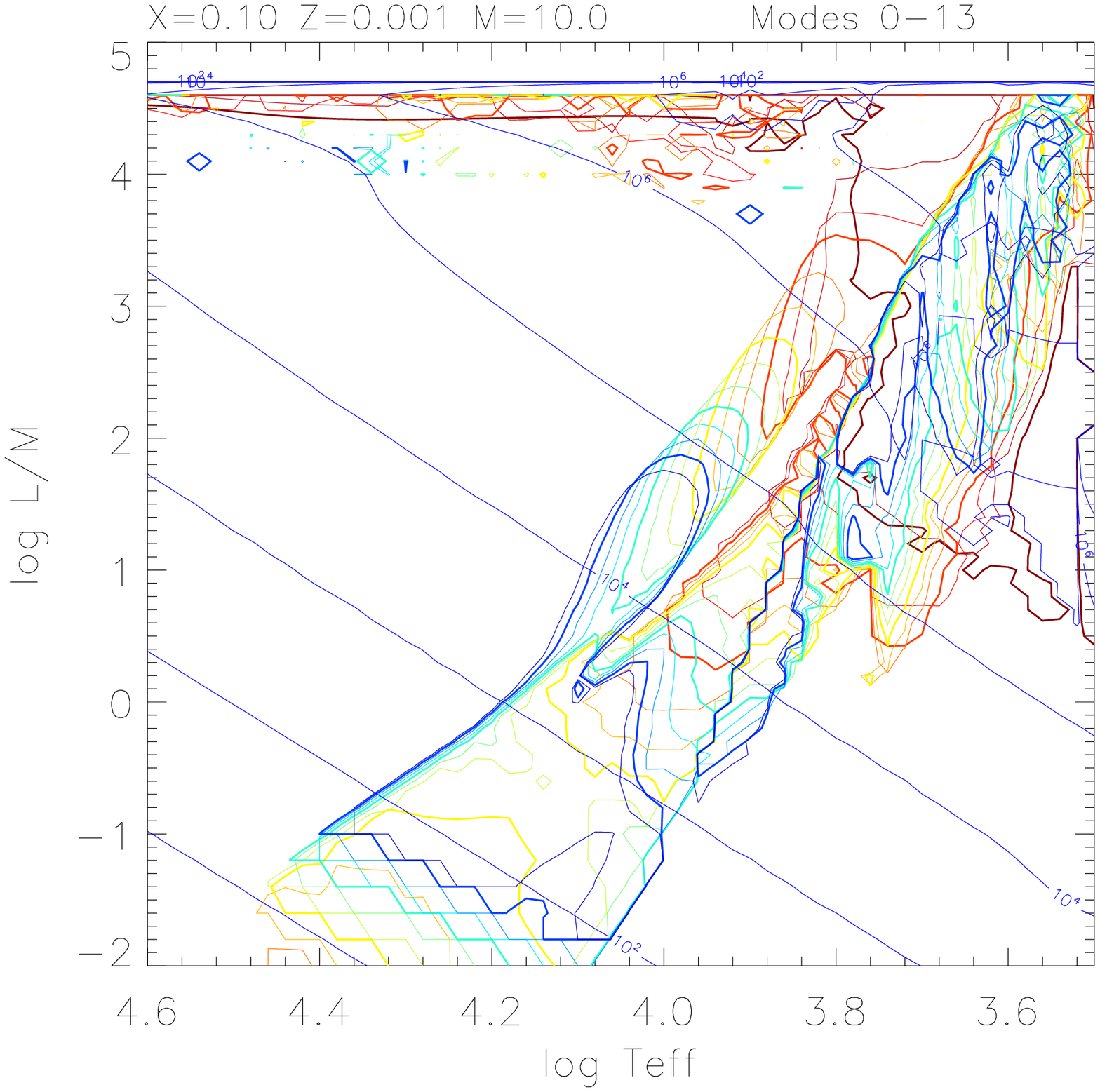,width=4.3cm,angle=0}\\
\epsfig{file=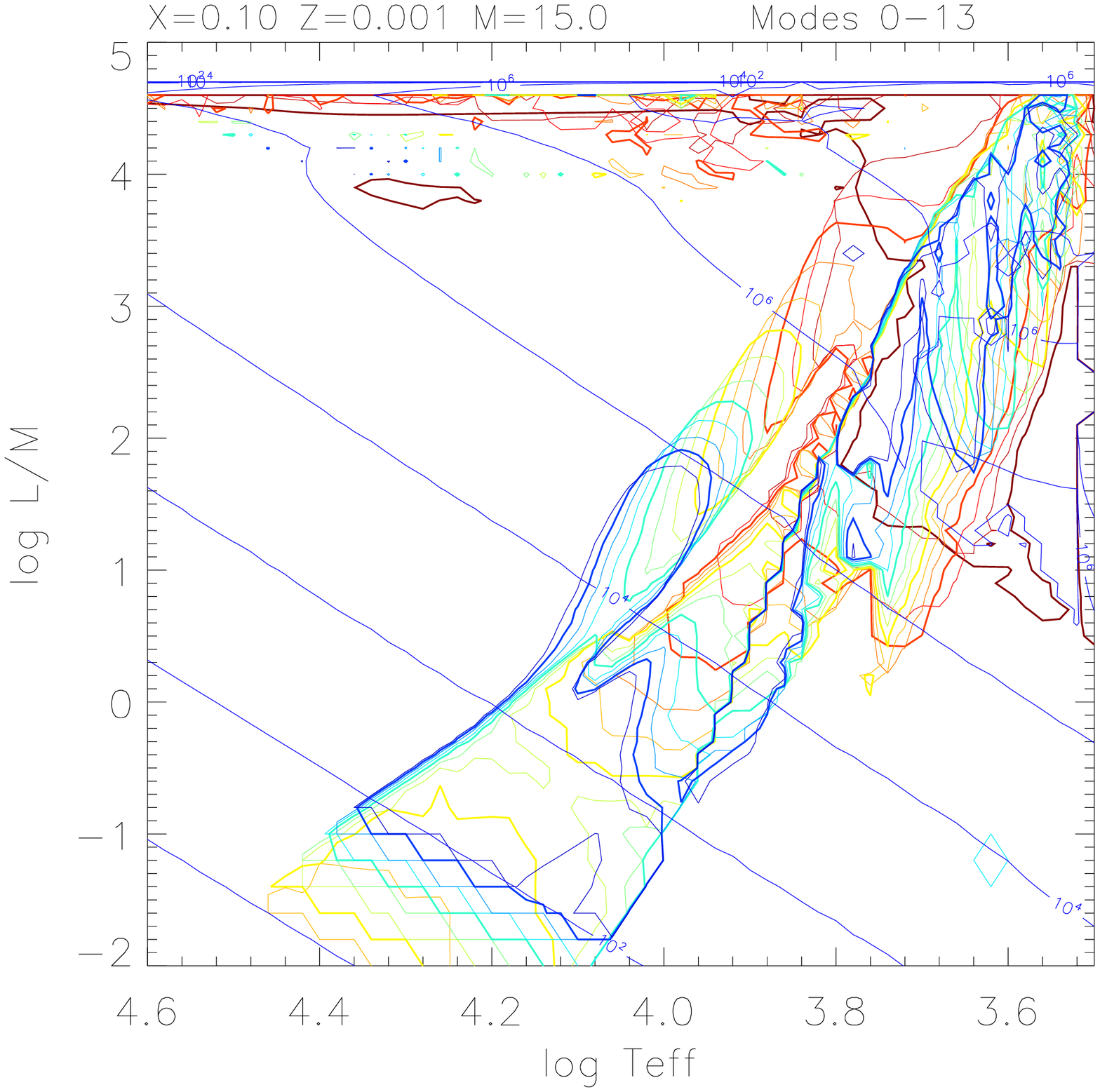,width=4.3cm,angle=0}
\epsfig{file=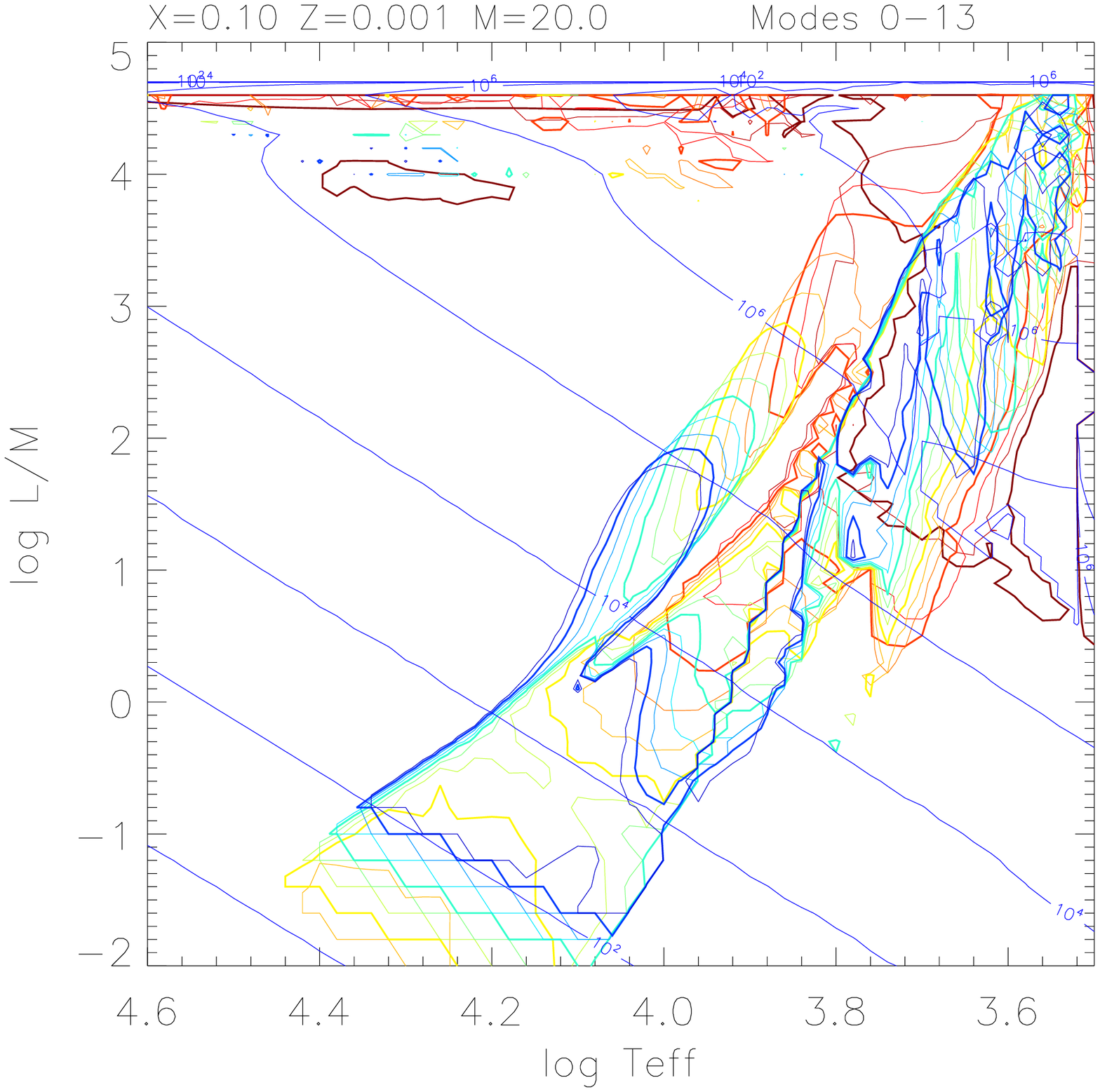,width=4.3cm,angle=0}
\epsfig{file=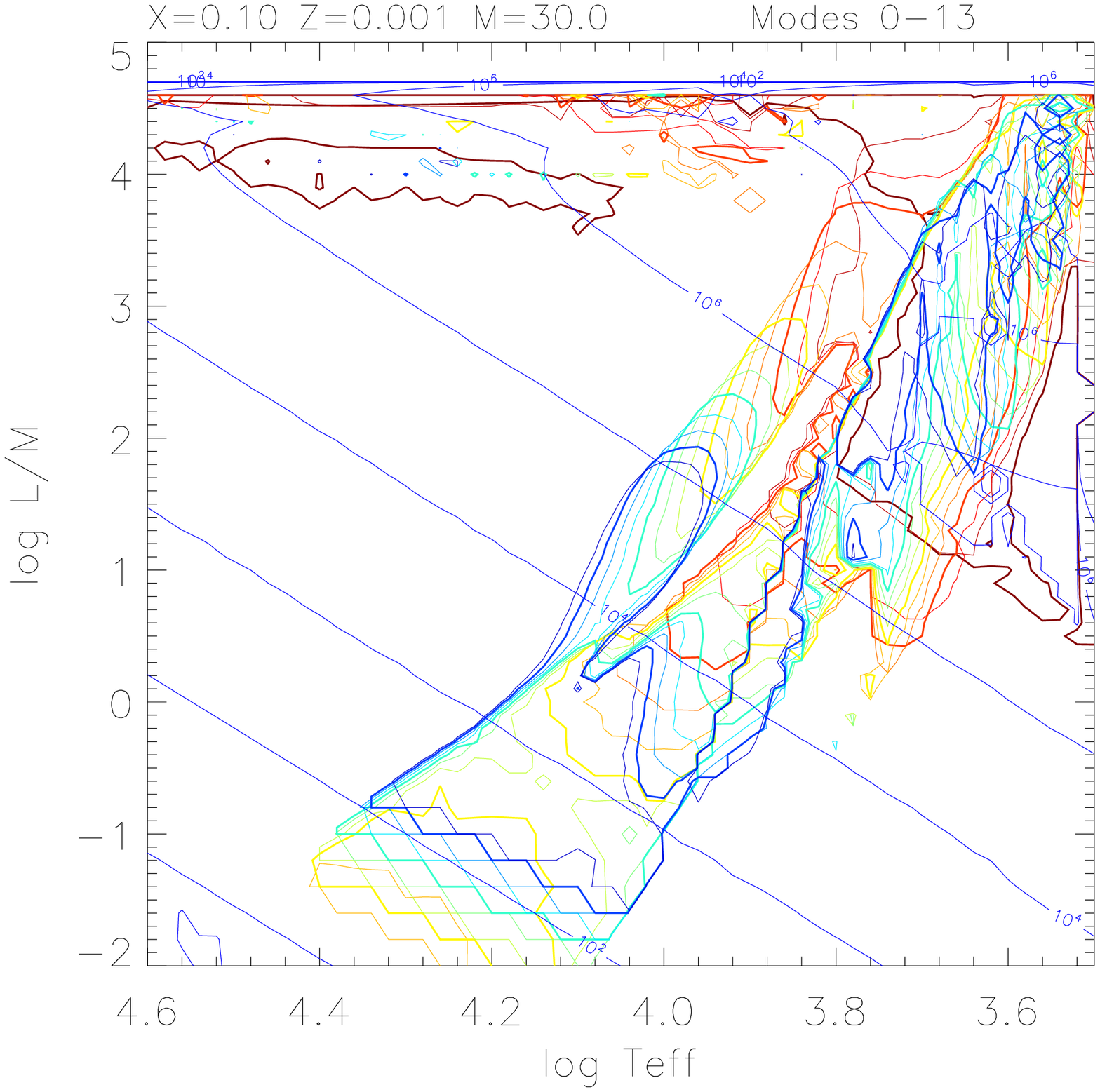,width=4.3cm,angle=0}
\epsfig{file=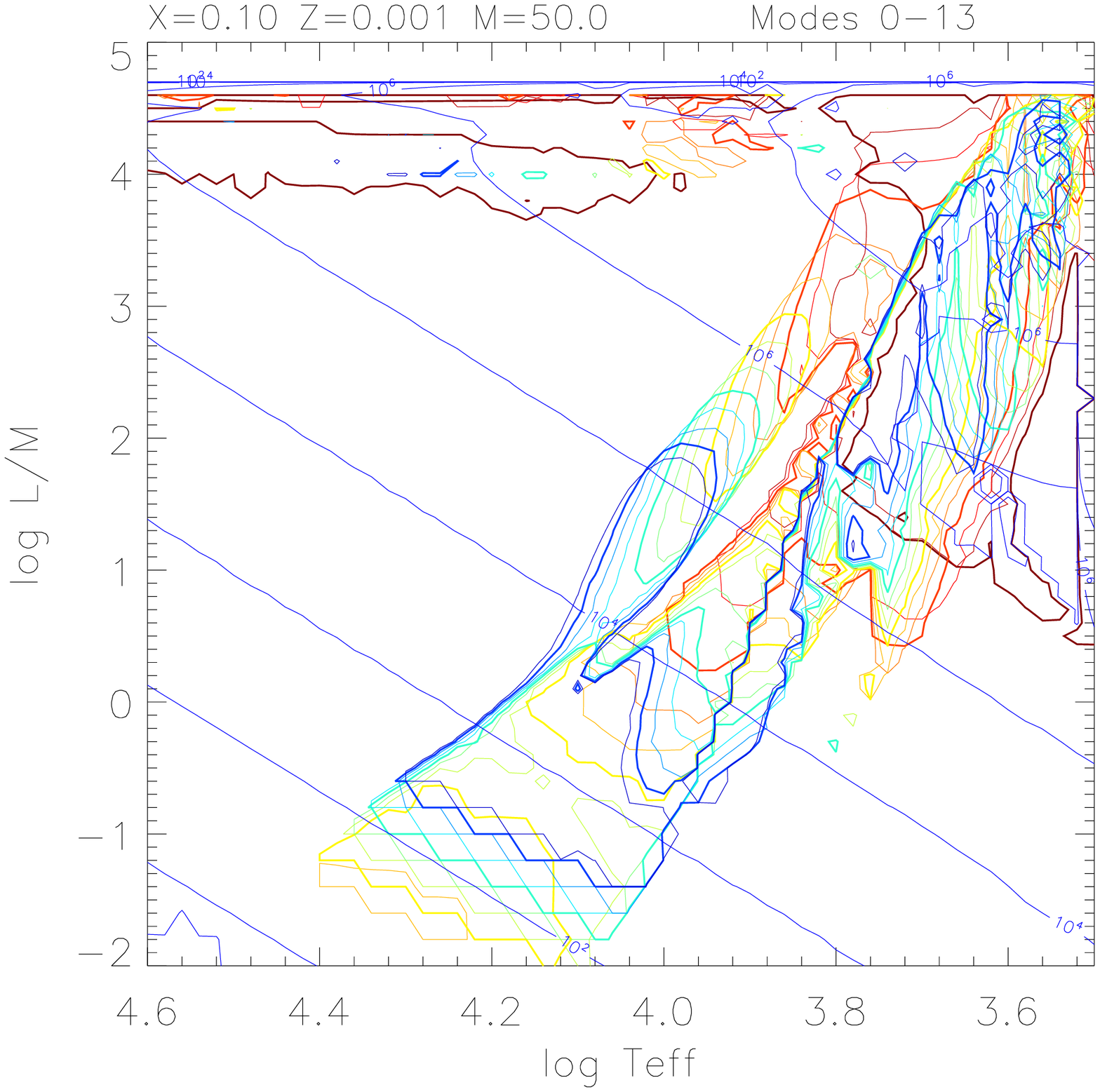,width=4.3cm,angle=0}
\caption[Unstable modes: $X=0.10, Z=0.001$]
{As Fig.~\ref{f:px70} with $X=0.10, Z=0.001$. 
}
\label{f:px10z001}
\end{center}
\end{figure*}

\begin{figure*}
\begin{center}
\epsfig{file=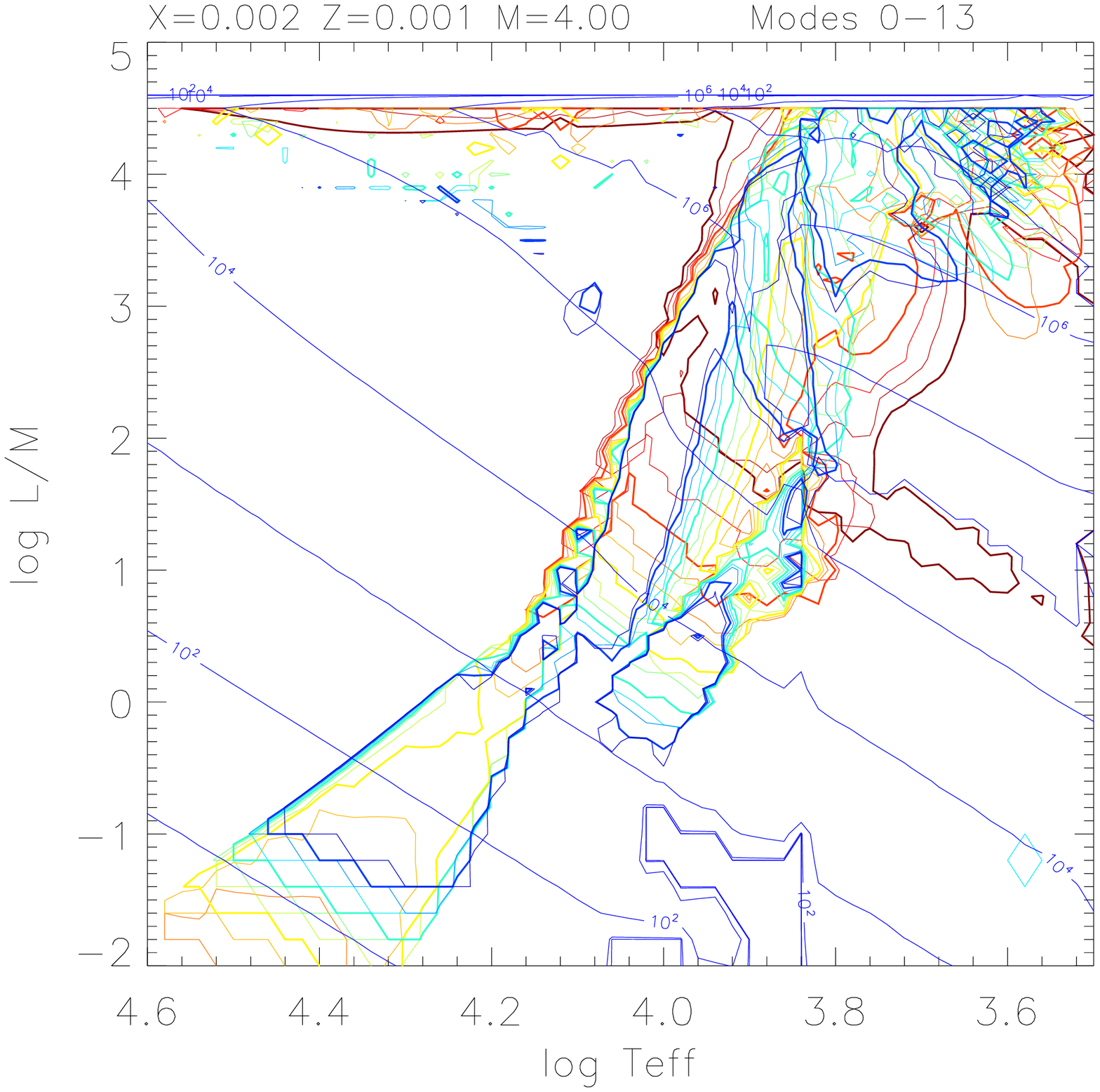,width=4.3cm,angle=0}
\epsfig{file=figs/periods_x002z001m05.0_00_opal.eps,width=4.3cm,angle=0}
\epsfig{file=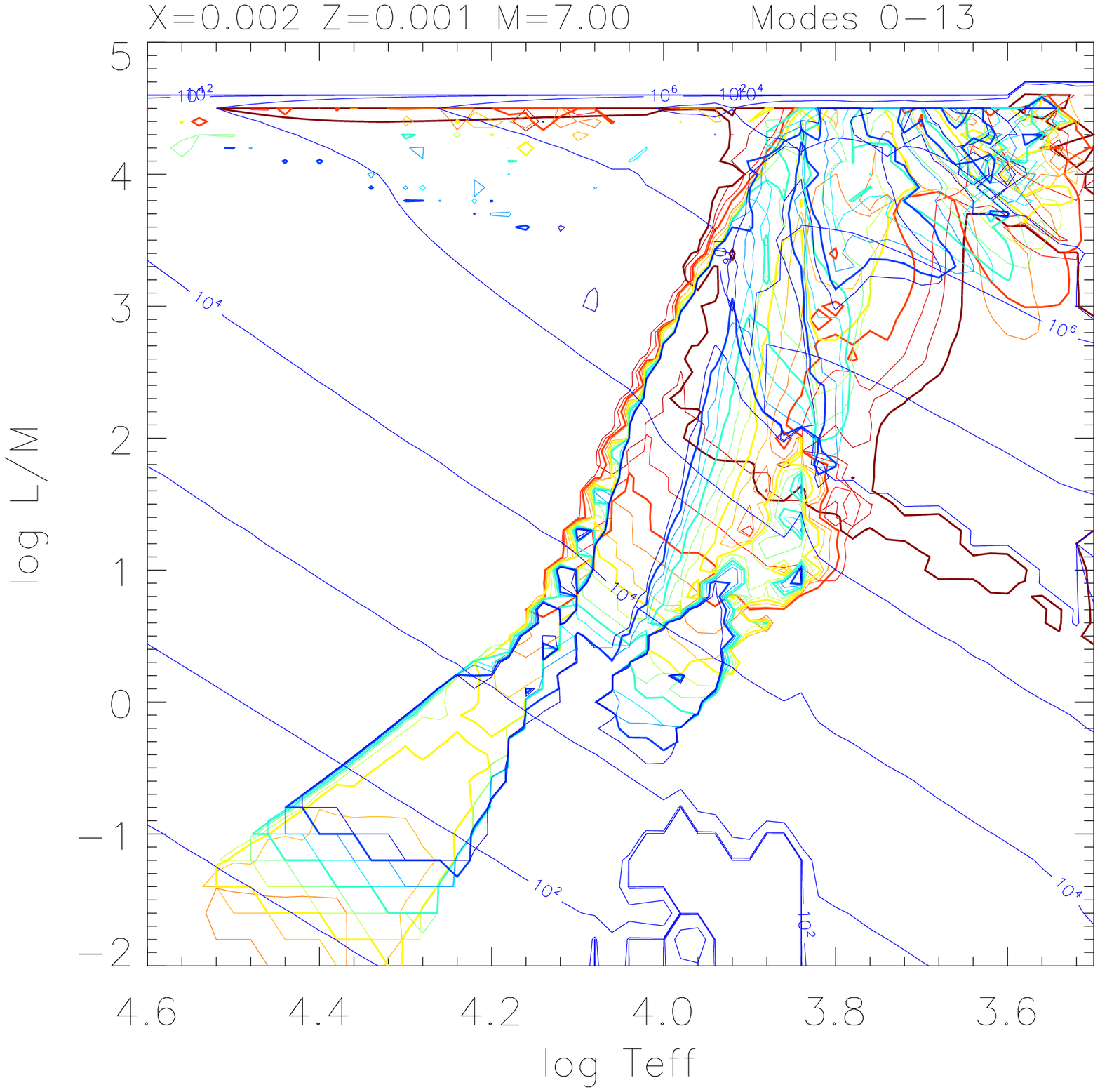,width=4.3cm,angle=0}
\epsfig{file=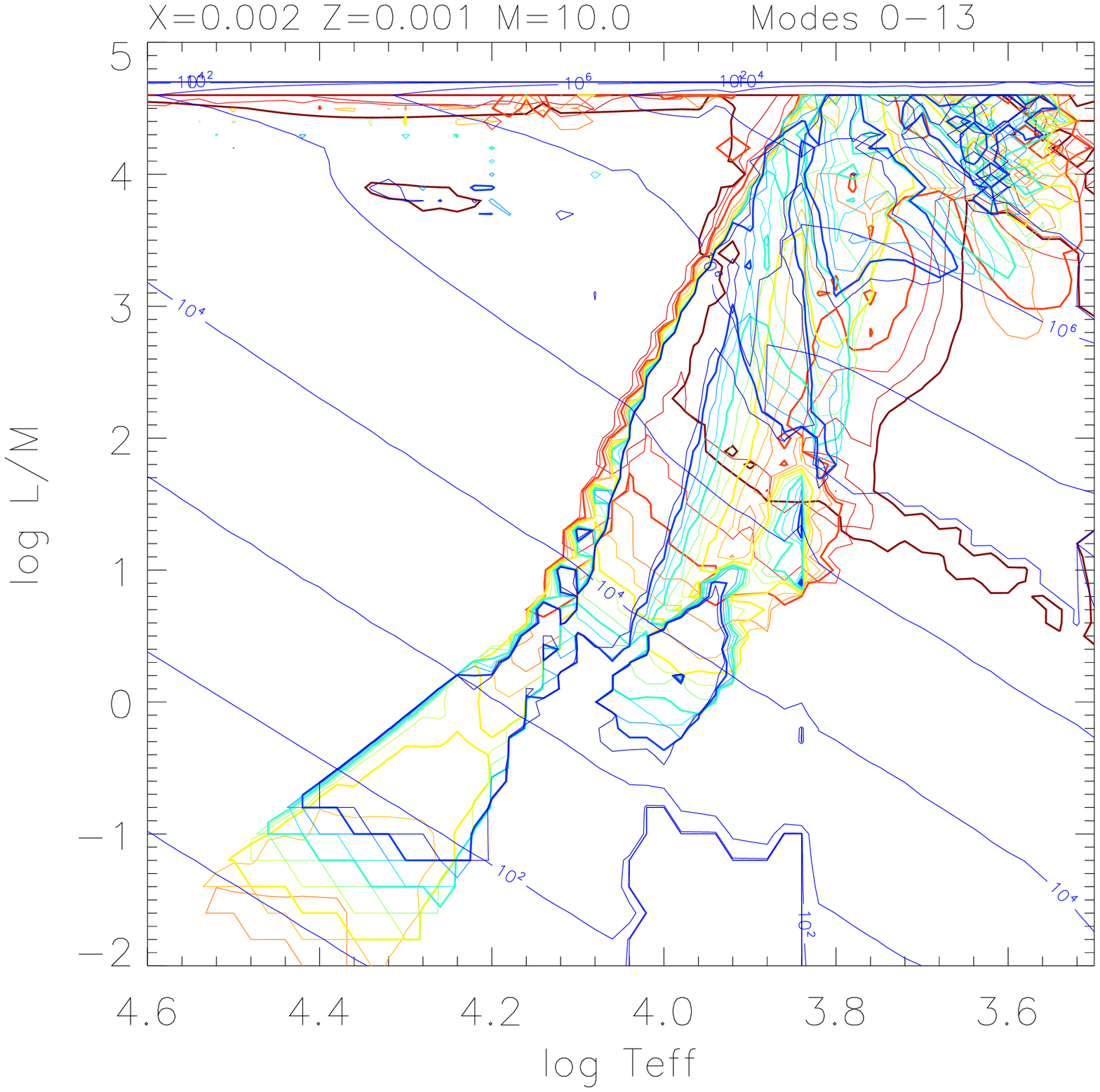,width=4.3cm,angle=0}\\
\epsfig{file=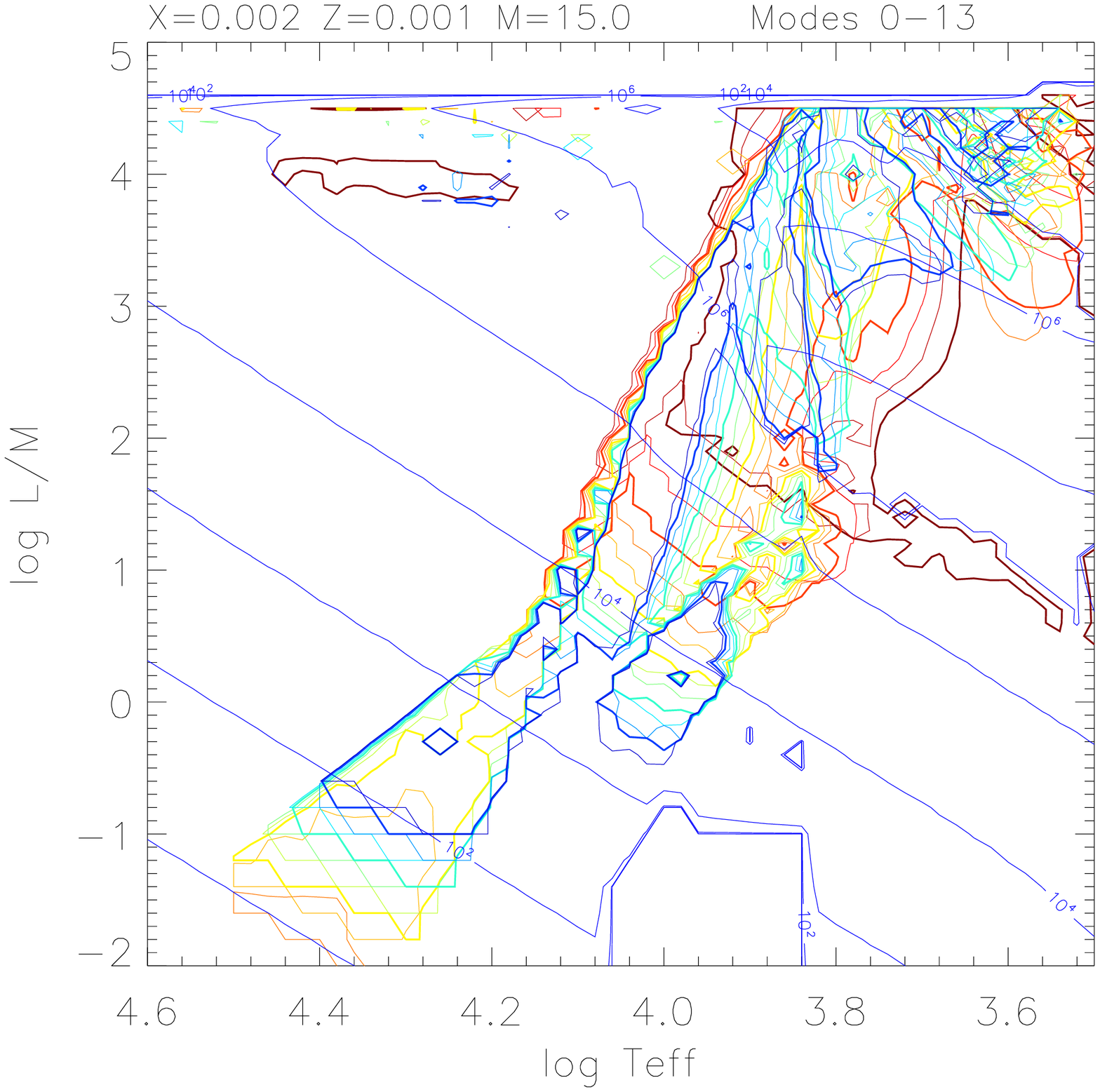,width=4.3cm,angle=0}
\epsfig{file=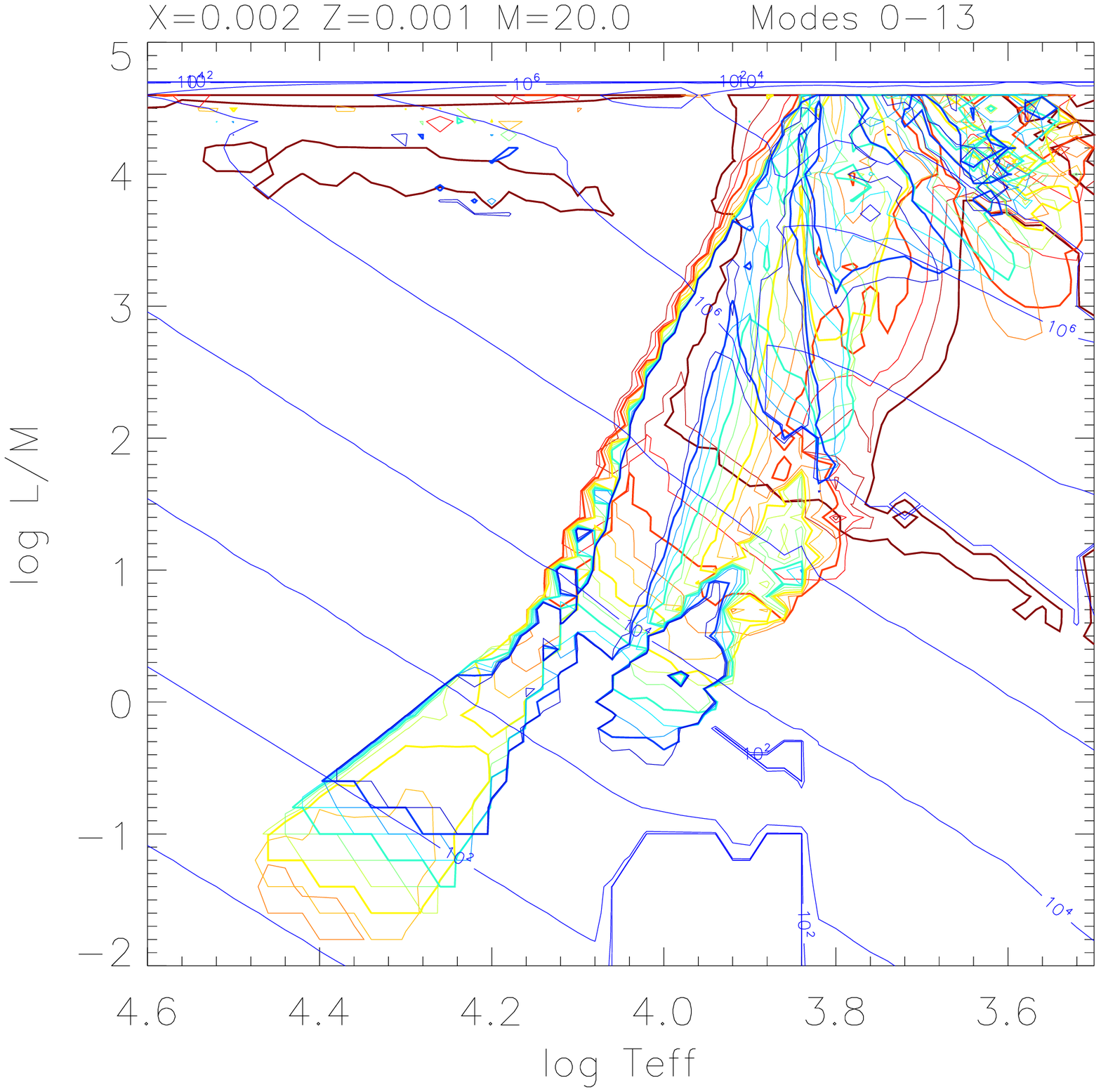,width=4.3cm,angle=0}
\epsfig{file=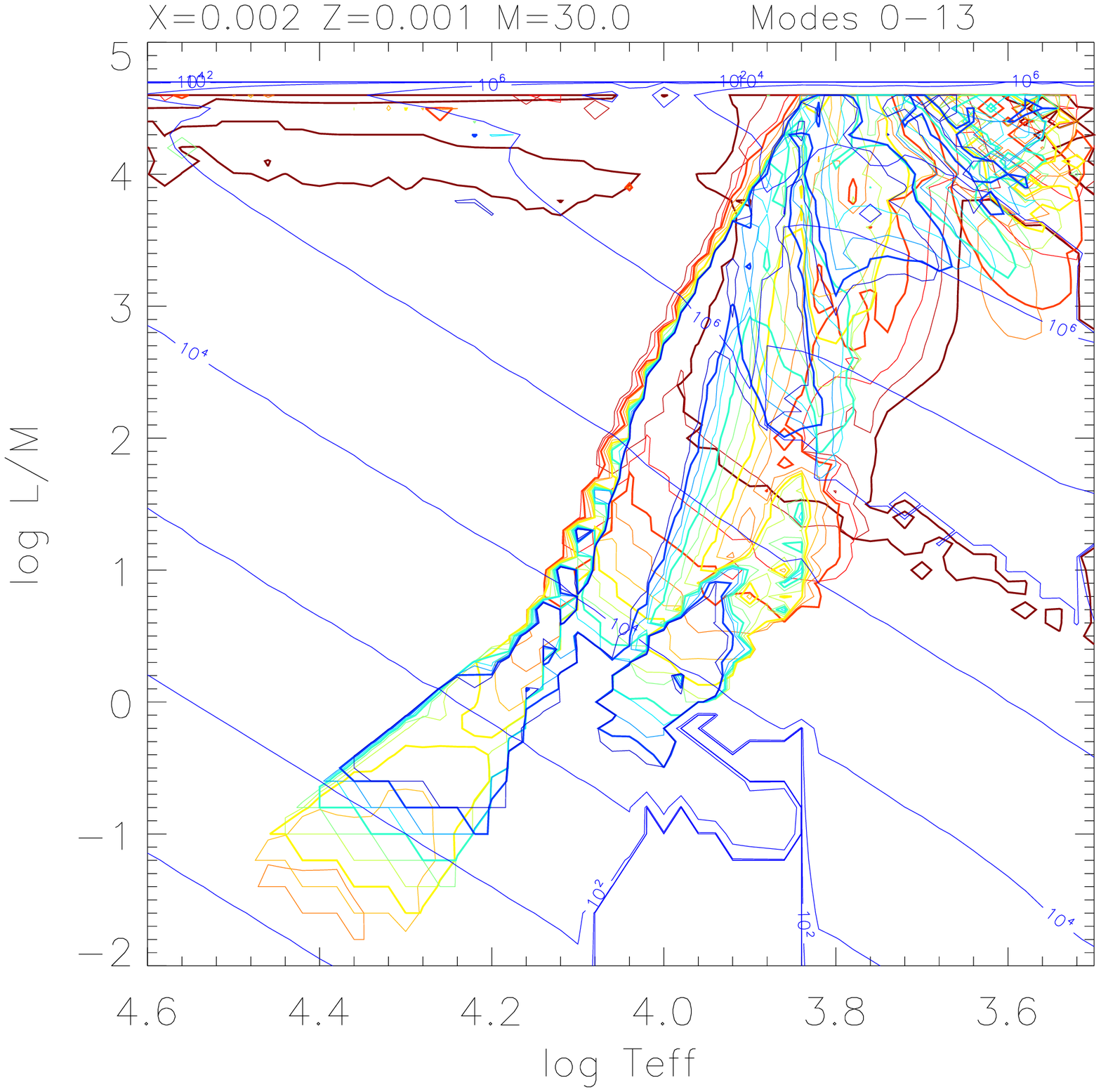,width=4.3cm,angle=0}
\epsfig{file=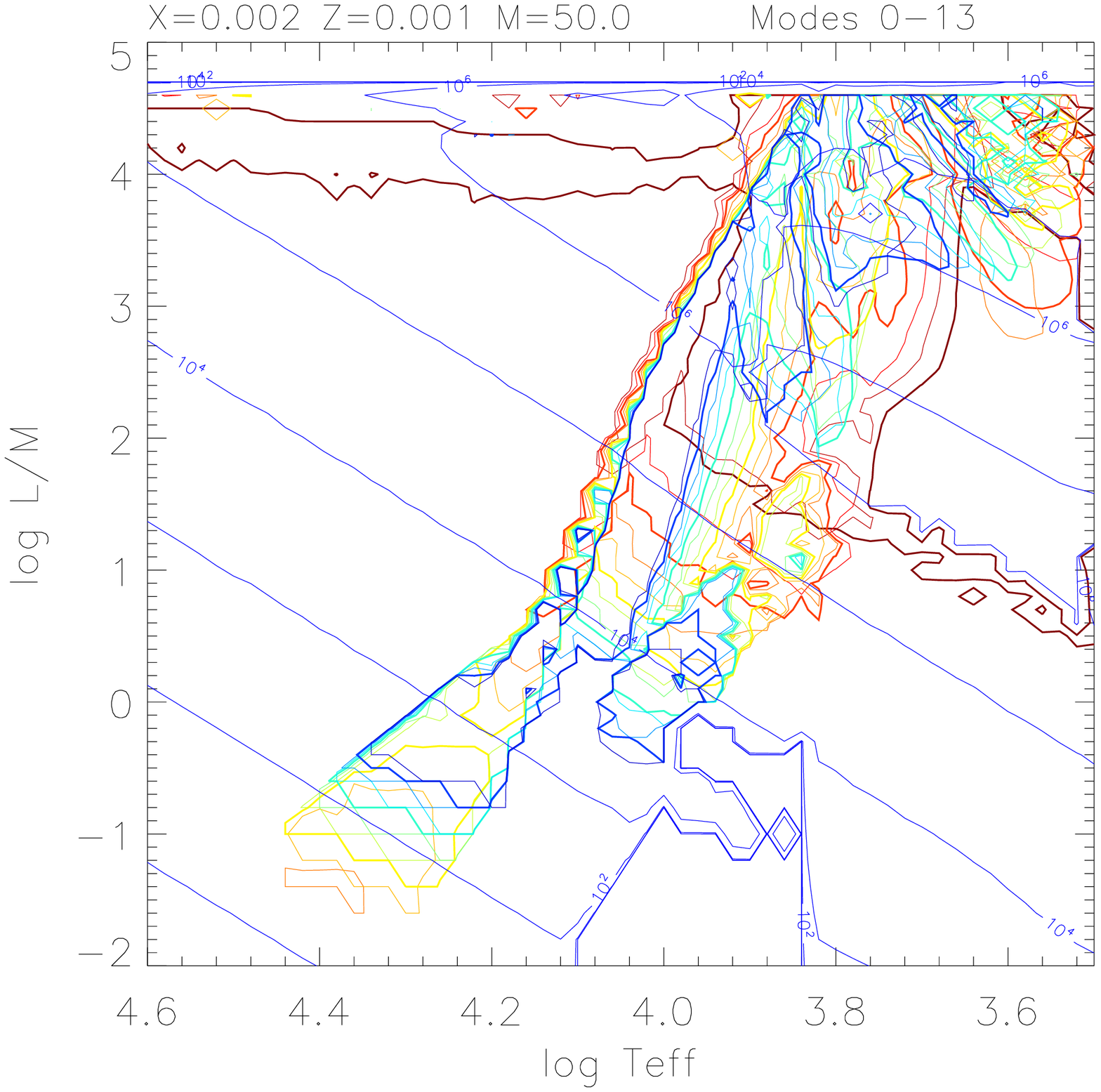,width=4.3cm,angle=0}
\caption[Unstable modes: $X=0.002, Z=0.001$]
{As Fig.~\ref{f:px70} with $X=0.002, Z=0.001$, for models with $M \geq 4\Msolar$.
Lower-mass models encountered numerical problems at high $L/M$ ratios.  
}
\label{f:px002z001}
\end{center}
\end{figure*}

Figures \ref{f:nx70} to \ref{f:px002z001} contain the complete grids showing the 
numbers of unstable radial modes and the instability boundaries for each unstable mode.
\label{lastpage}

\clearpage

\clearpage

\clearpage

\end{document}